# Effectuses in Categorical Quantum Foundations

Kenta Cho

This research has been supported by the European Research Council under the European Union's Seventh Framework Programme (FP7/2007-2013) / ERC grant agreement no. 320571 (Quantum Computation, Logic, and Security).

# Effectuses in Categorical Quantum Foundations

Proefschrift

ter verkrijging van de graad van doctor
aan de Radboud Universiteit Nijmegen
op gezag van de rector magnificus prof. dr. J.H.J.M. van Krieken,
volgens besluit van het college van decanen
in het openbaar te verdedigen op

dinsdag 8 oktober 2019

om 14.30 uur precies

door

Kenta Cho

geboren te Tokio, Japan

**Promotor:**
Prof. dr. B.P.F. Jacobs

**Manuscriptcommissie:**
Prof. dr. J.H. Geuvers
Prof. dr. M. Hasegawa (Kyoto University, Japan)
Prof. dr. D. Kozen (Cornell University, Verenigde Staten)
Dr. A. Wilce (Susquehanna University, Verenigde Staten)
Dr. I. Hasuo (National Institute of Informatics, Japan)

# Acknowledgements

First and foremost, I would like to thank my PhD supervisor Bart Jacobs. I have been impressed and inspired by his vast knowledge of many subjects and his enthusiasm. I enjoyed many stimulating discussions with him and greatly benefited from his insights, advice, and encouragement. Moreover, this thesis is based on a mathematical notion of 'effectus', which was originally introduced by Bart.

This thesis includes results of joint work with Bart, Bas Westerbaan, Abraham (Bram) Westerbaan, and Sean Tull. I am grateful to them for fruitful collaborations. Thanks to Bram also for his help with the Dutch summary of this thesis. I would like to thank Robin Adams, Guillaume Allais, Tobias Fritz, Robert Furber, Robin Kaarsgaard, Aleks Kissinger, Bert Lindenhovius, Mathys Rennela, Jurriaan Rot, Frank Roumen, Sam Staton, Sander Uijlen, John van de Wetering, and Fabio Zanasi for helpful discussions and insights.

I wish to thank the members of the manuscript committee, Herman Geuvers, Masahito Hasegawa, Dexter Kozen, Alexander Wilce, and Ichiro Hasuo, for examining this thesis and for their helpful feedback. In particular, Alexander gave a lot of comments and suggestions.

Ichiro was also my master's supervisor and is a director of the research project I am currently engaged in. I would like to give special thanks to him for all I learned from him (in particular about category theory), for his suggestion that I do a PhD in Nijmegen, and for his support that made it possible to finish this thesis.

I would like to thank my colleagues at Institute for Computing and Information Sciences, in particular those in the Digital Security group, for their help and companionship that made my PhD life in Nijmegen pleasant and enjoyable. Special thanks to Dan Frumin and Sander Uijlen for being my paranymphs and helping me with the organization of my defence.

Finally, I would like to express my gratitude to my parents and family in Japan for their constant support and encouragement.

# Contents

# Chapter 1

# Introduction

In this thesis, we develop the theory of *effectuses* as a categorical approach to quantum theory. An effectus is a category satisfying certain axioms, providing a suitable axiomatic framework for quantum theory, and also for general physical theories including classical probabilistic and deterministic theories. 'Predicates' in an effectus form effect algebras, which are algebraic models of unsharp quantum logic. Effectus theory thus has an aspect of categorical quantum logic.

Axiomatic studies of quantum theory have a long history, forming the background of this thesis. In the first two sections below, we briefly review quantum theory and previous approaches. We then give an overview of effectus theory in Section 1.3, and outline the thesis in Section 1.5.

## 1.1 Quantum theory and foundations

Quantum theory describes physical phenomena at very small scales, for example, behaviours of atoms, electrons, and light. Such quantum physics has many counterintuitive features, which classical physics does not have. For example, a quantum system can be in a *superposition* of states, and thus roughly speaking, in several different states simultaneously. In particular, this means that when we measure a physical quantity (such as position and momentum) on the system, we get an outcome *at random*, according to a probability distribution predicted by quantum theory. It is understood, mainly due to the Bell-Kochen-Specker theorem [15, 173], that such randomness is inherent in quantum physics and cannot be avoided. Another fundamental feature is *Heisenberg's uncertainty principle* [24, 121], which implies that some physical quantities cannot be precisely measured simultaneously, and that measurement on a quantum system necessarily disturbs the state of a system.

Over the last few decades, it has turned out that quantum phenomena can be exploited for computation and communication [212], leading to growing importance of a deeper understanding of quantum physics from a computer science perspective. An example is a quantum key distribution protocol, such as BB84 [16], where eavesdropping can be detected via the fact that measurement on a quantum system disturbs the state. Moreover, quantum computers can solve certain problems, such as prime factorization [238], faster than classical computers. One reason for the speed-up is superposition, which allows us in effect to perform many computations at once.

Since quantum physics is counterintuitive and very different from classical physics, we need a good mathematical framework to understand and utilize quantum physics.

Let us first describe some basic postulates of the standard Hilbert space formulation of quantum theory due to von Neumann [210]. A physical system is represented by a Hilbert space $\mathscr{H}$.[1,2] A state of the system is represented by a *density operator* $\rho$ on $\mathscr{H}$, i.e. a positive operator $\rho\colon \mathscr{H} \to \mathscr{H}$ with trace one: $\mathrm{tr}(\rho) = 1$. A physical quantity that can be measured (observed) on the system — called an *observable* — is represented by a self-adjoint operator $A$ on $\mathscr{H}$. By the spectral theorem (see e.g. [73, 225]), any self-adjoint operator $A$ can be written as $A = \int_{\mathbb{R}} r\, E_A(\mathrm{d}r)$ via a unique projection-valued measure $E_A\colon \Sigma_{\mathbb{R}} \to \mathcal{Pr}(\mathscr{H})$ on the Borel $\sigma$-algebra $\Sigma_{\mathbb{R}}$ of the real line $\mathbb{R}$. The *Born rule* provides us a statistical prediction about measurement: if the observable $A$ is measured on the system of state $\rho$, the probability that the observed value is contained in the Borel subset $U \subseteq \mathbb{R}$ is given by the trace $\mathrm{tr}(\rho E_A(U)) \in [0, 1]$.

Although the Hilbert space formulation is mathematically rigorous and ‘works well’ in the sense that its predictions agree with experimental results, it is unsatisfactory in that the postulates are rather ad hoc and do not admit intuitive interpretations — why is a state of a system represented by a density operator on a Hilbert space, and an observable by a self-adjoint operator? For a fundamental understanding of quantum physics, there is a clear need for an alternative, more insightful axiomatization of quantum theory. Indeed, von Neumann himself was not satisfied by his Hilbert space formulation [224] and made significant contributions to other approaches. Many approaches to quantum theory have been studied, featuring various concepts such as states, observables, propositions, and processes. In the next section we give a brief overview of several approaches.

## 1.2 Approaches to quantum theory

This thesis studies a categorical approach to quantum theory based on *effectuses*. The relevant structures in our approach — such as effect algebras and convex sets — have been studied in prior approaches to quantum theory. Therefore first we give a brief review of several relevant approaches.

### Traditional quantum logic

Projections $P$ on a Hilbert space $\mathscr{H}$ are in bijective correspondence with closed subspaces $U \subseteq \mathscr{H}$ of the Hilbert space. Classical propositional logic is modelled by subsets of a set, which form a Boolean algebra, and intuitionistic propositional logic is modelled by open subsets of a topological space, which form a Heyting algebra. By analogy, we can think of projections on a Hilbert space as a model of ‘quantum logic’. This is the view of the traditional quantum logic initiated by Birkhoff and von Neumann [18]. The set of projections (closed subspaces) $\mathcal{Pr}(\mathscr{H})$ has the following order-theoretic properties.

- There are a greatest element 1 (= ‘truth’) and a least element 0 (= ‘falsity’).

[1] Throughout the thesis, Hilbert spaces are over the complex numbers $\mathbb{C}$.

[2] If the reader is not familiar with Hilbert spaces, consider the complex Euclidean space $\mathscr{H} = \mathbb{C}^n$. Then the operators on $\mathbb{C}^n$ are complex $n \times n$ matrices. Any self-adjoint (Hermitian) matrix can be written as $A = \sum_j r_j P_j$ via diagonalization, where $r_j \in \mathbb{R}$ are the eigenvalues of $A$ and $P_j$ are the projection matrices corresponding to the associated eigenspaces.

- There are meets $P \wedge Q$ (= '$P$ and $Q$') and joins $P \vee Q$ (= '$P$ or $Q$').
- There are orthocomplements $P^\perp$ (= 'not $P$'), which satisfy $P \vee P^\perp = 1$, $P \wedge P^\perp = 0$, and $(P^\perp)^\perp = P$; moreover $P \leq Q$ implies $Q^\perp \leq P^\perp$.

That is, $\mathcal{P}r(\mathscr{H})$ forms an *orthocomplemented lattice*. However, $\mathcal{P}r(\mathscr{H})$ does not satisfy *distributive law*: there exist $P, Q, R \in \mathcal{P}r(\mathscr{H})$ such that

$$P \wedge (Q \vee R) \neq (P \wedge Q) \vee (P \wedge R)\,.$$

The lattice $\mathcal{P}r(\mathscr{H})$ satisfies the following condition called the *orthomodular law*:

$$P \leq Q \quad \text{implies} \quad P \vee (P^\perp \wedge Q) = Q\,,$$

which is weaker than the distributive law. An orthocomplemented lattice satisfying the orthomodular law is called an *orthomodular lattice*. Orthomodular lattice have been studied extensively [17, 129, 168]. They generalize Boolean algebras, which are both orthocomplemented and distributive. Orthomodular lattices contrast with Heyting algebras, which are distributive but not orthocomplemented — the law of excluded middle (or the double negation elimination) fails.

An orthomodular lattice $L$ can be viewed as a representation of a physical system, axiomatizing the set of 'propositions' on the system. We additionally assume that $L$ is a $\sigma$-complete lattice. Then for instance, we can define a *state* of the system as a suitable *probability measure* $\mu\colon L \to [0,1]$ on $L$ (see e.g. [168, 251]). If $L$ is the orthomodular lattice $\mathcal{P}r(\mathscr{H})$ of projections on a Hilbert space $\mathscr{H}$, states (probability measures) on $\mathcal{P}r(\mathscr{H})$ are in bijective correspondence with density operators on $\mathscr{H}$ by Gleason's theorem [95] (when $\mathscr{H}$ is separable and $\dim(\mathscr{H}) > 2$).

### Operational and unsharp quantum logics

An *operational* approach to quantum theory generally demands that primitive concepts and axioms have operational interpretations: for example, we can interpret 'propositions' as (procedures of) measurements answering yes or no. Such operational perspectives led to generalizations of orthomodular lattices such as *orthomodular posets* [79, 200] and *orthoalgebras* [83–85]. In parallel, the convex operational approach (see below) was developed, revealing the importance of *effects*, a concept related to 'unsharp' measurements — those that may not be 'sharp' (ideal or accurate). Concretely in the Hilbert space formulation, effects are positive operators whose spectra are contained in $[0,1]$, generalizing projections. Several authors [64, 82, 94, 178] gave axiomatizations of effects, and hence unsharp quantum logic. The structures introduced there turned out to be the same, and are now called *effect algebras*. Effect algebras generalize orthomodular lattices/posets (hence Boolean algebras), orthoalgebras, and moreover, *MV-algebras* — algebraic models of Łukasiewicz infinitely-many-valued logic. Thus effect algebras provide a general setting for both quantum and classical theories, and also for both sharp and unsharp (fuzzy) logic.

Effect algebras will play an important role in this thesis. We will review basics of effect algebras in Section 2.3. More information about quantum logic approaches can be found e.g. in [56, 86, 206, 222, 239].

### Convex operational approach

The quantum logic approaches feature the structures of 'propositions' (or 'predicates', or 'yes-no measurements') in a physical system. In contrast, the convex operational approach features the structure of *states* of a system, namely the *convex* structure. In other words, the starting point in this approach is a convex set of 'states', which are abstract elements and not assumed to be density operators. Usually one assumes that the set of states can be embedded in an ordered normed vector space in a suitable way — more precisely, the set of states is the *base* (for the cone) of a *base-norm space* [9, 75, 77]. The assumption was justified by Ludwig [195–198] in his axiomatic framework. In the standard Hilbert space formulation, the space of self-adjoint trace-class operators on a Hilbert space $\mathscr{H}$ forms a base-norm space, with the base consisting of density operators. Using base-norm spaces as abstract state spaces, Davies and Lewis [66] developed a framework about measurements, introducing the notion of *instrument*.

Given a base-norm space $\mathscr{V}$ as a state space, suitable elements of the dual space $\mathscr{V}^*$ are called *effects*. They play an important role in the convex operational approach, representing predicates or yes-no measurements in a system. The duality pairing $\langle x, a\rangle = a(x)$ yields the probability of observing effect $a \in \mathscr{V}^*$ in state $x \in \mathscr{V}$. A dual pair of suitable ordered normed vector spaces can be viewed as a model of a physical system, specifying the spaces of states and effects. Such a dual pair is called a *convex operational model* in [11, 258], and also studied in this thesis in Section 7.2.

In general, effects are *unsharp* (or *fuzzy*) — they represent 'unsharp' measurements that may not be perfectly accurate. In the Hilbert space formulation, effects are positive operators whose spectra are contained in $[0, 1]$, generalizing projections that correspond to *sharp* measurements. Unsharp effects are important because they naturally and inevitably occur in sequential measurements, for a reason related to the uncertainty principle. Mathematically this is because if $P, Q$ are incompatible (i.e. non-commuting) projections on a Hilbert space, then $PQP$ (which is intuitively understood as '$P$ and then $Q$') need not be a projection, but only an effect.

We study the relationship of the convex operational approach (in particular, convex operational models) and effectus theory in Section 7.2, where more details about this approach can be found. For further information and references, we refer to [222, Chapter 4] and the recent work of Barnum and others [11–14, 258].

### Algebraic approach

In the algebraic approach, we represent physical systems by *operator algebras*, which are viewed as 'algebras of observables'. Operator algebras refer to both concrete algebras of bounded operators on a Hilbert space and their axiomatizations such as $C^*$*-algebras* and $W^*$*-algebras*. The theory of operator algebras was first developed by Murray and von Neumann in a series of papers starting with [208], motivated by foundational aspects of quantum theory (see [224]). Currently the usefulness of the algebraic formulation of quantum theory is widely known. For example, the algebraic approach has been applied to quantum field theory [6, 111, 112], quantum statistical mechanics [22, 23], and quantum information [171]. One of the advantages of the algebraic formulation is that operator algebras can naturally represent classical

systems — via *commutative* algebras — as well as quantum systems, and moreover mixtures of them.

In this thesis we will use the category of $W^*$-algebras and suitable morphisms as the archetypal example of an effectus, which models quantum systems and processes. We will review some basics of the algebraic formulation in Section 2.6. For further information about the algebraic approach, we refer to [187, 188, 222].

### Categorical approaches

*Category theory* is a very general formalism about objects and morphisms. It provides a suitably abstract language in which we can focus on essential aspects of a subject, and has been used in various fields such as mathematics, logic, computer science, and physics.

A general view in categorical approaches to quantum theory is to see objects $A, B, \ldots$ as types of systems, and to see morphisms $f\colon A \to B$ as processes going from a system of type $A$ to a system of type $B$. Various categorical approaches exist, differing by additional properties and structures assumed on a category.

*Categorical quantum mechanics* [2, 54, 127] was initiated by Abramsky and Coecke [1] and has been developed by many authors associated with Oxford. It is mainly based on a *compact closed category*, a certain type of a monoidal category. The archetypal example is the category of finite-dimensional Hilbert spaces. Morphisms in a compact closed category can be conveniently described by *string diagrams* [164, 234], and thus categorical quantum mechanics emphasizes a formalism of quantum theory as graphical calculus (as in the title of the book [54]).

The *operational probabilistic* framework is another approach based on categories. It was introduced by by Chiribella, D'Ariano, and Perinotti [33–35, 61], aiming at explaining quantum theory from an operational, information-theoretic point of view. The basic notion in the framework is an *operational probabilistic theory* (*OPT*): a monoidal category with the structure of *tests*, which represent physical operations involving measurements. We will review a part of the operational probabilistic framework in Section 6.1.

There are approaches from the perspectives of categorical logic. Heunen and Jacobs [122, 123, 136] studied *dagger kernel categories*, where *kernel subobjects* form orthomodular lattices, capturing the traditional quantum logic. In *topos approaches* to quantum theory [71, 122, 126], one studies a certain *topos* induced from a fixed operator algebra. Toposes are categories that have the structure of *intuitionistic logic* such as Heyting algebras. Thus the topos approaches are radically different from the quantum logic approaches based on orthomodular lattices, effect algebras, etc.

## 1.3 Effectus theory: a new categorical approach

*Effectus theory* is yet another categorical approach to quantum theory, and it is the main topic of this thesis. Here the central notion is *effectus*, a category satisfying certain conditions which provides a suitable axiomatic framework for quantum theory.

An effectus was first introduced by Jacobs in [140][3], and its theory has been developed mainly by him and his colleagues in Nijmegen, including the author of the thesis. In this thesis we aim to give a systematic introduction to effectus theory, and to show the relevance of effectuses in quantum foundations.

One aspect of effectus theory is a new style of categorical logic, as emphasized by Jacobs in [140]. In an effectus 'predicates' form *effect algebras*, capturing the essentials of (unsharp) quantum logic, with probabilistic and Boolean logic as special cases. This contrasts with the traditional categorical logic that often features intuitionistic logic.

Another aspect of effectus theory was revealed by Tull [248–250], who showed that effectuses are closely related to the operational probabilistic framework of Chiribella et al. Specifically, he proved that effectuses are equivalent to a variant of operational probabilistic theories satisfying certain additional properties [248, Corollary 23]. The effect algebra structure of predicates comes from these additional properties, and can be understood as the distinguishing feature of effectus theory. We can thus view effectus theory as the marriage of the operational probabilistic framework and quantum logic.

Though effectus theory uses some assumptions stronger than the operational probabilistic framework, it has good consequences. An effectus admits mathematically clean and reasonably rich structures. As mentioned above, predicates in an effectus form effect algebras. They moreover admit scalar multiplication, forming *effect modules*. *States* in an effectus form *convex sets*. Predicates and states yield a 'state-and-effect' triangle constituted by categories and functors (see § 3.7 and § 4.2.1), capturing the duality between the Schrödinger and Heisenberg pictures. In this thesis we also study the structure of *substates*, which are axiomatized as *weight modules*. The logical structure of predicates further allows us to define notions of *image*, *comprehension*, and *quotients* in an effectus (see Chapter 5).

We usually do not assume finite dimensionality in effectus theory, and indeed we do not in this thesis. In contrast, both the operational probabilistic framework and categorical quantum mechanics focus on the finite-dimensional setting, though in the latter, there are attempts to deal with infinite dimension, e.g. [51, 96]. Indeed, the archetypal example of an effectus is the category of $W^*$-algebras and suitable morphisms, where $W^*$-algebras may be of arbitrary dimension. Thus results obtained abstractly in an effectus are valid for arbitrary $W^*$-algebras, algebraic models of quantum systems. On the other hand, the definition of effectus assumes only finite coproducts, limiting the strength of the results that can be obtained abstractly. This issue will be addressed in Section 7.3 by a notion of $\sigma$-effectus, which is equipped with countable coproducts.

There are two different formulation of effectuses, called *total form* and *partial form*. The equivalence of the two formulations is one of the original contributions in this thesis. An effectus in total form is a category with finite coproducts and a final object that satisfies certain pullback and joint monicity conditions (Definition 4.1.6). It is the original definition of effectus given by Jacobs [140]. The morphisms represent *total* processes between systems. The author of this thesis showed [36] that effectuses can be equivalently defined *in partial form*, i.e. via morphisms representing *partial* processes, in a way that each effectus in total form is a suitable subcategory of the corresponding effectus in partial form. Specifically, an effectus in partial form is defined as a suitable partially additive category equipped with an effect algebra structure (Definition 3.2.1). By the equivalence of the two formulations, whether one starts with an effectus in total form or in partial form is basically a matter of style. Although an effectus in total form admits a simpler definition, an effectus in partial form is in most cases more convenient to work with. Thus in this thesis we decided to develop the theory with effectuses in partial form as a starting point. In particular, the default meaning of 'effectus' in this thesis is 'effectus in partial form'. Effectuses in total form will appear in this thesis as a secondary notion in Chapter 4.

[3]The definition of effectus was established around 2014 and first appeared in an arXiv preprint of [140]. The term 'effectus' was later introduced.

In general, a monoidal structure on a category allows us to express compound systems $A \otimes B$ and processes composed *in parallel*: $f_1 \otimes f_2 \colon A_1 \otimes A_2 \to B_1 \otimes B_2$. There has already been a reasonable definition of an extension of effectuses with a monoidal structure, see *monoidal effectuses* in [40, § 10]. However, the development of monoidal effectuses is still at an early stage and they will not be discussed in this thesis. In other words, this thesis concerns a categorical axiomatization of physical/quantum systems and processes which does not use parallel composition, but uses sequential composition $A \xrightarrow{f} B \xrightarrow{g} C$ of processes and sum $A + B$ of systems (intuitively understood as '$A$ or $B$'). This makes a good contrast with categorical quantum mechanics of the Oxford school, where the monoidal structure plays a central role.

## 1.4 Contributions of this thesis

The contributions of this thesis can be divided into two parts.

First, this thesis provides a comprehensive introduction to effectus theory. This thesis develops the theory based on effectuses in *partial form*, unlike existing introductory papers [40, 140] that are based on effectuses in *total form*. Effectuses in partial form are defined more concretely in terms of partially additive structures, and more directly related to operational probabilistic theories of Chiribella et al., see Chapter 6. Our approach to effectus theory in partial form is probably more accessible to readers who are not very familiar with category theory.

Second, this thesis relates effectus theory to various topics and approaches, in order to reveal the nature and advantages of effectuses. Specifically, it discusses the following topics and approaches in relation to effectus theory.

(i) Effect algebras and orthomodular lattices (Chapter 3 and Section 5.5)

(ii) Partially additive categories (Chapter 3 and Section 7.3)

(iii) State-predicate duality in the form of state-and-effect triangles (Section 3.7)

(iv) Categorical logic in terms of fibrations (Chapter 5)

(v) Janelidze and Weighill's categorical axiomatization of non-abelian algebras (Section 5.6)

(vi) Operational probabilistic framework and measurement theory (Chapter 6)

(vii) Extensive categories (Section 6.6)

(viii) Biproduct categories and ground structures (Section 7.1)

(ix) Convex operational framework (Section 7.2)

The relationships to these topics will demonstrate the mathematical generality and cleanness of effectus theory, and also help to understand the nature of effectuses. The topics (i), (vi), and (ix) concern axiomatization of quantum theory, and (viii) concerns the structures that have been used in categorical quantum mechanics. Thus the relationships to these illustrate the relevance of effectuses in quantum foundations. The topics (ii)–(iv) are closely related to program semantics and logics. Although this thesis does not explicitly deal with programming languages, it discusses concepts of program semantics and logics in an abstract way.

## Original results and publications

A number of people, often jointly, have contributed to effectus theory. Below I will list the author's own results together with relevant references.

(a) The definition of effectus in partial form (Definition 3.2.1; called 'FinPAC with effects' in [36]) and the related results. In particular, the 2-categorical equivalence of effectuses in partial form and total form (§§ 4.1–4.2).

(b) The notion of weight module and the related results (§ 3.5, § 4.4).

(c) The notion of division effect monoid (§ 4.3), and the study of the normalization property on weight modules and general effectuses. In particular, the equivalence of the categories of convex sets and weight modules with the normalization property (Corollary 4.4.9). Normalization in an effectus was first studied by Jacobs et al. in [150], but in a restricted setting with the scalars $[0, 1]$.

(d) The definition of sharp predicates in an effectus via comprehension and images, and the study of sharp predicates based on this definition (§ 5.5).

(e) A systematic study of measurements/instruments in an effectus, via the language from the operational probabilistic framework (§§ 6.3–6.5). Note that the notion of 'instrument' (or 'assert map') in [40, 140] is more restrictive, referring to a fixed family of (canonical or ideal) instruments.

(f) The characterization of Boolean effectuses with comprehension (or quotients) as an extensive category (§ 6.6). This is a joint work with Abraham Westerbaan and the result was included in the preprint [40]. Note that 'Boolean effectus' in [40] is equivalent but differently formulated, due to the difference of the notion of instruments.

(g) Totalization of effectuses and the definition of grounded biproduct categories, which yields an coreflection (§ 7.1). Totalization of effectuses was studied jointly with Tull and related results are also found in his thesis [250, Chapter 3].

(h) The study of the relationship between effectus theory and the convex operational approach. In particular, a categorical equivalence of convex operational models and state-effect models (§ 7.2.4) and an embedding of a real effectus with the order-separation property into the category of convex operational models (§ 7.2.5).

(i) The study of $\sigma$-effectuses (§ 7.3). In particular, we establish state-and-effect triangles over real $\sigma$-effectuses (Corollaries 7.3.42 and 7.3.44), and give an improvement of the embedding into convex operational models for an $\sigma$-effectus (Corollary 7.3.46).

Chapters 3 and 4 are based on [36], but largely expanded and rewritten. The notion of weight module is newly added in this thesis. Chapter 5, except § 5.5 on sharp predicates, originates in [41]. It however discusses only concrete examples of quotients and comprehensions, without using effectuses. Definitions of images, quotients, and comprehension in terms of effectuses have appeared in the preprint [40]. Chapters 6 and 7 have not been published, except that the preprint [40] includes a characterization of Boolean effectuses as extensive categories and the results on grounded biproduct categories (without totalization).

Other publications that the author worked on during his PhD are [38, 39]. To make this thesis focused on effectus theory, the work of [38, 39] is not included here.

Finally let us mention the theses of Abraham and Bas Westerbaan, with whom the author jointly developed effectus theory [40, 41]. Their theses are complementary to the present one: their theses mainly focus more concretely on the category of $W^*$-algebras (the main example of an effectus), whereas the present thesis focuses on abstract theory of effectuses. Abraham's thesis [253] contains a concise yet comprehensive exposition of the theory of operator algebras, and also includes Abraham and the author's results about $W^*$-algebras from preprints [42, 43]. Bas' thesis [256] studies effectuses too, but focuses on the structure of 'canonical' measurements (see Remark 6.3.37) and the *dagger* structure, which are not covered in the present thesis.

## 1.5 Outline

Chapter 2 covers preliminaries for the thesis, and does not contain original results.

Chapter 3 develops basics of effectuses. To define an effectus, first we develop finitely partially additive categories, which are a slight generalization of Arbib and Manes' partially additive categories [7]. We then give a definition of effectus, and describe our leading examples of effectuses. The archetypal effectus for quantum theory is given by $W^*$-algebras. We study structures of predicates and (sub)states in an effectus. Predicates form effect modules, i.e. effect algebras with a scalar multiplication. States form convex sets. In addition, we introduce a new axiomatic structure of substates ('subnormalized states') which we call a *weight module*. There is a dual adjunction between the categories of effect modules and wight modules, formalizing a duality between predicates and substates. These structures in an effectus are neatly summarized as 'state-and-effect' triangles. At the end of the chapter we give a convenient characterization of effectuses.

Chapter 4 is mainly concerned with total morphisms in an effectus. We introduce *effectuses in total form*, which are the original formulation of effectus given by Jacobs [140]. In our setting (where we start with effectuses 'in partial form'), effectuses in total form can be considered as a characterization of the subcategories of effectuses determined by total morphisms. In the other direction, one can recover an effectus 'in partial form' from its subcategory of total morphisms via the lift monad. This gives rise to a 2-equivalence of the 2-categories of effectuses in partial and total form: $\mathbf{Ef} \simeq \mathbf{Eft}$. The chapter continues studying how convex sets and weight modules — axiomatizations of states and substates — are related. It will turn out that under the assumption that the scalars admit division, the category of convex sets is equivalent to the category of weight modules with the *normalization property.* It follows that effectuses with the normalization property admit particularly clean state-and-effect triangles, as diagrams in the 2-categories $\mathbf{Ef} \simeq \mathbf{Eft}$.

In Chapter 5 we study effectuses from a logical point of view, systematically using the language of (Grothendieck) fibrations. The fibrational perspective motivates the notions of *kernel*, *image*, *comprehension*, and *quotient* in an effectus, which are defined by certain universal properties. Via images and comprehension, we define *sharp* predicates (which captures projections in quantum theory), and we prove under a mild assumption that sharp predicates form orthomodular lattices. We then study (bi)fibrations of sharp predicates and sharp morphisms. We conclude the chapter with a comparison to Janelidze and Weighill's theory of non-abelian algebras.

Chapter 6 discusses measurements in an effectus using the language of the operational probabilistic framework (via Tull's result). We study repeatable measurements, side-effect-free measurements, and Boolean measurements, where Boolean is defined to be a property of being both repeatable and side-effect-free. Repeatable measurements are shown to be related to sharp predicates. We abstractly define *Lüders instruments*, certain ideal measurements, and give several characterization of them. Side-effect-freeness will be related to compatibility/commutativity of observables. The study of Boolean measurements leads to a characterization of an extensive category (with a final object) as a 'Boolean' effectus, in which Boolean measurements are possible.

The final chapter, Chapter 7, contains miscellaneous topics in effectus theory. In Section 7.1 we relate effectuses to biproducts or semiadditive structures via a 'totalization' construction. This makes some connection between effectus theory and categorical quantum mechanics. In Section 7.2 we investigate a relation between effectuses and the convex operational approach. A main result here is that a certain class of effectuses can be embedded into the category of *convex operational models* — dual pairs of base-norm and order-unit spaces. In Section 7.3 we study $\sigma$-effectuses, i.e. effectuses with countably partially additive structure. This is a natural extension of effectus which goes back to the setting of Arbib and Manes' partially additive categories. Also from the viewpoint of quantum foundations, it is natural to assume a countably additive structure (cf. Mackey's formulation [200]). State-and-effect triangles and the embedding result to convex operational models will be extended to the setting of $\sigma$-effectuses.

# Chapter 2

# Preliminaries

This chapter covers preliminaries for the thesis.

### Prerequisites

We assume that the reader has a basic knowledge of category theory, including (co)limits, adjunctions, and (co)monads. Concrete 2-categories (e.g. of effectuses) will occasionally appear. For this, it is sufficient to know the definitions of (strict) 2-categories, 2-functors, and 2-natural transformations, see e.g. [199, § XII.3]. A minimal introduction to algebraic quantum theory is included in this chapter (Section 2.6), but further knowledge on the subject will help to understand concrete examples of effectuses that serve as models of quantum theory.

### Notations

We use the following notations.

| | |
|---|---|
| $\mathbb{N} = \{0, 1, 2, \ldots\}$ | natural numbers |
| $\mathbb{Z}$ | integers |
| $\mathbb{Q}$ | rational numbers |
| $\mathbb{R} = (-\infty, \infty)$ | real numbers |
| $\mathbb{N}_{>0} = \{1, 2, \ldots\}$ | nonzero natural numbers |
| $[n] = \{1, 2, \ldots, n\}$ | $n$-element set |
| $\mathbb{R}_+ = [0, \infty)$ | nonnegative real numbers |
| $\mathbb{R}_{>0} = (0, \infty)$ | (strictly) positive real numbers |

## 2.1 Category theory

We assume that the reader is familiar with basic category theory. The standard reference is [199], but see also [10, 19, 20, 191, 227]. Here we fix basic notations, and recall some definitions and results.

Throughout the thesis, **Set** denotes the category of sets and functions.

Let $\mathbf{C}$ be a category. For objects $A, B \in \mathbf{C}$, the *homset* consisting of morphisms $f\colon A \to B$ in $\mathbf{C}$ is denoted by $\mathbf{C}(A, B)$, or sometimes by $\mathrm{Hom}(A, B)$ when the context is clear.

For a family $(A_j)_j$ of objects, the product and coproduct of $(A_j)_j$ are denoted by $\prod_j A_j$ and $\coprod_j A_j$, respectively. Projections and coprojections are denoted by $\pi_j\colon \prod_j A_j \to A_j$ and $\kappa_j\colon A_j \to \coprod_j A_j$. A final object is denoted by 1, and an initial object by 0. Note that 0 also denotes a zero object and zero morphisms.

**Definition 2.1.1.**

(i) A **zero object**, denoted by 0, is an object that is both final and initial.

(ii) A category has **zero morphisms** if there is a family of morphisms $0_{AB}\colon A \to B$ such that $0_{BD} \circ f = 0_{AD} = g \circ 0_{AC}$ for any morphisms $f\colon A \to B$ and $g\colon C \to D$.

A family of zero morphisms is unique if it exists. Indeed, if both $(0_{AB})_{AB}$ and $(0'_{AB})_{AB}$ are families satisfying the condition for zero morphisms, then $0_{AB} = 0_{AB} \circ 0'_{AA} = 0'_{AB}$. A zero object and zero morphisms are closely related:

**Proposition 2.1.2.** *Let* $\mathbf{C}$ *be a category with an initial object* 0*. Then* $\mathbf{C}$ *has zero morphisms if and only if it has a zero object, i.e.* 0 *is final too.*

In particular, a category with finite coproducts has zero morphisms if and only if it has a zero object.

*Proof.* If $\mathbf{C}$ has a zero object 0, it has zero morphisms given by $A \to 0 \to B$. Conversely, suppose that $\mathbf{C}$ has zero morphisms $(0_{AB})_{AB}$. We have $\mathrm{id}_0 = 0_{00}$ by initiality. Then any morphism $f\colon A \to 0$ is equal to $0_{A0}\colon A \to 0$, since $f = \mathrm{id}_0 \circ f = 0_{00} \circ f = 0_{A0}$. Thus 0 is final. ■

We recall two standard constructions involving monads.

**Definition 2.1.3.** Let $T\colon \mathbf{C} \to \mathbf{C}$ be a monad with unit $\eta$ and multiplication $\mu$. The **Kleisli category** $\mathcal{K}\ell(T)$ of $T$ is defined as follows.

- $\mathcal{K}\ell(T)$ has the same objects as $\mathbf{C}$.
- A morphism $f\colon A \twoheadrightarrow B$ in $\mathcal{K}\ell(T)$ is a morphism $f\colon A \to TB$ in $\mathbf{C}$. That is, $\mathcal{K}\ell(T)(A,B) \coloneqq \mathbf{C}(A,TB)$.
- The identities in $\mathcal{K}\ell(T)$ are $\eta_A\colon A \to TA$.
- For morphisms $f\colon A \twoheadrightarrow B$ and $g\colon B \twoheadrightarrow C$ in $\mathcal{K}\ell(T)$, the composite $g \odot f\colon A \twoheadrightarrow C$ is defined to be
$$A \xrightarrow{f} TB \xrightarrow{Tg} TTC \xrightarrow{\mu_C} TC \qquad \text{in } \mathbf{C}\,.$$
The composite $\mu_C \circ Tg\colon TB \to TC$ is called the Kleisli extension of $g$.

**Definition 2.1.4.** Let $T\colon \mathbf{C} \to \mathbf{C}$ be a monad with unit $\eta$ and multiplication $\mu$. The **Eilenberg-Moore category** $\mathcal{EM}(T)$ of $T$ is defined as follows.

- An object in $\mathcal{EM}(T)$ is a pair $(A,\alpha)$ where $A \in \mathbf{C}$ and $\alpha\colon TA \to A$ in $\mathbf{C}$ such that $\alpha \circ \eta_A = \mathrm{id}_A$ and $\alpha \circ \mu_A = \alpha \circ T\alpha$. These objects are called (**Eilenberg-Moore**) **algebras for** $T$, or $T$-algebras.
- A morphism from $(A,\alpha)$ to $(B,\beta)$ in $\mathcal{EM}(T)$ is a morphism $f\colon A \to B$ such that $f \circ \alpha = \beta \circ Tf$.
- The identities and composition in $\mathcal{EM}(T)$ are those in $\mathbf{C}$.

These categories carry adjunctions $\mathbf{C} \rightleftarrows \mathcal{K}\ell(T)$ and $\mathbf{C} \rightleftarrows \mathcal{EM}(T)$, which both recover the monad $T$ on $\mathbf{C}$; see [199, Chapter VI].

The following elementary result will be used several times in the thesis, so we explicitly state it here.

**Lemma 2.1.5.** *Let $T\colon \mathbf{C} \to \mathbf{C}$ be a monad. Then the Kleisli category $\mathcal{K}\ell(T)$ inherits coproducts from $\mathbf{C}$: each coproduct $\coprod_j A_j$ in $\mathbf{C}$, with coprojections $\kappa_j\colon A_j \to \coprod_j A_j$, is also a coproduct in $\mathcal{K}\ell(T)$ with coprojections $\eta \circ \kappa_j\colon A_j \to T(\coprod_j A_j)$.*

*Proof.* Straightforward. ∎

Next we recall the definition of coreflection.

**Definition 2.1.6.** A **coreflection** is an adjunction

$$\mathbf{C} \underset{G}{\overset{F}{\rightleftarrows}} \mathbf{D} \quad (\bot) \tag{2.1}$$

whose unit $\eta\colon \mathrm{id} \Rightarrow GF$ is an isomorphism.

An equivalent definition of coreflection is:

**Lemma 2.1.7.** *An adjunction is a coreflection if and only if the left adjoint functor is full and faithful.*

*Proof.* This is the dual statement of [199, Theorem IV.3.1]. ∎

Thus an example of a coreflection is a full subcategory $\mathbf{C} \hookrightarrow \mathbf{D}$ where the inclusion functor has a right adjoint. Such a subcategory is called a **coreflective subcategory**. Up to equivalence, any coreflection is identified with a coreflective subcategory $F[\mathbf{C}] \hookrightarrow \mathbf{D}$, where $F[\mathbf{C}]$ is the image of the left adjoint $F$.

A coreflection is a 'well-behaved' embedding $F\colon \mathbf{C} \to \mathbf{D}$ where one can transfer certain structures/properties of $\mathbf{D}$ to $\mathbf{C}$. Indeed, $\mathbf{C}$ inherits limits and colimits from $\mathbf{D}$ [146, Theorem 2], and also inherits a monoidal structure from $\mathbf{D}$ under some mild conditions [146, Theorem 5].

The dual notion is called a **reflection**: it is an adjunction such that the counit is an isomorphism, or equivalently, the right adjoint is full and faithful.

The following is a well-known (e.g. [185, Part 0, Proposition 4.2]) result about adjunctions.

**Proposition 2.1.8.** *Consider an adjunction*

$$\mathbf{C} \underset{G}{\overset{F}{\rightleftarrows}} \mathbf{D} \quad (\bot)$$

*with unit $\eta\colon \mathrm{id} \Rightarrow GF$ and counit $\varepsilon\colon FG \Rightarrow \mathrm{id}$. We write*

- *$\mathbf{C}_0 \hookrightarrow \mathbf{C}$ for the full subcategory consisting of objects $A \in \mathbf{C}$ such that $\eta_A\colon A \to GFA$ is an isomorphism, and*

- $\mathbf{D}_0 \hookrightarrow \mathbf{D}$ *for the full subcategory consisting of objects* $B \in \mathbf{D}$ *such that* $\varepsilon_B \colon FGB \to B$ *is an isomorphism.*

*Then the restriction of the adjunction* $\mathbf{C} \rightleftarrows \mathbf{D}$ *to* $\mathbf{C}_0$ *and* $\mathbf{D}_0$ *yields an adjoint equivalence* $\mathbf{C}_0 \simeq \mathbf{D}_0$.

*Proof.* It only has to be shown that the restrictions of the functors $F$ and $G$ to $\mathbf{C}_0$ and $\mathbf{D}_0$ are well-defined. This follows from the zig-zag identities: $\varepsilon_{FA} \circ F\eta_A = \mathrm{id}_{FA}$ and $G\varepsilon_B \circ \eta_{GB} = \mathrm{id}_{GB}$. ■

## 2.2 Partial commutative monoids

In this section we review the notion of partial commutative monoid. Both partially additive categories and effect algebras, introduced in the subsequent sections, are based on partial commutative monoids.

**Definition 2.2.1.** A **partial commutative monoid** (**PCM**, for short) is a set $M$ with a *partial* binary operation $\ovee \colon M \times M \rightharpoonup M$ and an element $0 \in M$ satisfying the three conditions below. We write $x \perp y$ if $x \ovee y$ is defined (i.e. $\perp \subseteq M \times M$ is the domain of definition of $\ovee$).

(a) *Associativity*: $x \perp y$ and $x \ovee y \perp z$ imply $y \perp z$, $x \perp (y \ovee z)$, and $(x \ovee y) \ovee z = x \ovee (y \ovee z)$.

(b) *Commutativity*: $x \perp y$ implies $y \perp x$ and $x \ovee y = y \ovee x$.

(c) *Unit law*: $0 \perp x$ and $0 \ovee x = x$.

We call $x \ovee y$ the **sum** of $x$ and $y$, and 0 the **zero**. We say that elements $x_1, \dots, x_n \in M$ are **summable** if the sum $x_1 \ovee \dots \ovee x_n$ is defined. Summability is well-defined for any finite family (or multiset) by the associativity and commutativity of $\ovee$. By definition, two elements $x$ and $y$ are summable iff $x \perp y$. Note that pairwise summable elements need not be (jointly) summable.

Clearly any commutative monoid is a PCM, whose addition is a total operation. Examples of PCMs with proper partial operation $\ovee$ can be found below in Example 2.3.3, as effect algebras.

**Definition 2.2.2.** Let $M$ and $N$ be PCMs. A **homomorphism of PCMs** $f \colon M \to N$ (or more briefly a **PCM morphism**) is a function satisfying:

(a) $x \perp y$ implies $f(x) \perp f(y)$ and $f(x \ovee y) = f(x) \ovee f(y)$;

(b) $f(0) = 0$.

PCMs and their homomorphisms form a category **PCM**.

For the sake of readability, we use the following convention: when we write an expression containing a sum $x \ovee y$, then (unless stated otherwise) it is assumed that summability $x \perp y$ holds. For example, we simply write $x \ovee y = z$ instead of '$x \perp y$ and $x \ovee y = z$'.

*Categories enriched over PCMs*, defined below, play an important role in this thesis. For PCMs $M, N$, and $L$, a function $f\colon M \times N \to L$ is called a **PCM bimorphism**[1] if it preserves the PCM structure in each argument separately: that is, for all $x \in M$ and $y \in N$, both $f(x, -)\colon N \to L$ and $f(-, y)\colon M \to L$ are PCM morphisms.

**Definition 2.2.3.** We say that a category $\mathbf{C}$ is **enriched over PCMs** if every homset $\mathbf{C}(A, B)$ is a PCM, and for each $A, B, C \in \mathbf{C}$ the composition $\circ\colon \mathbf{C}(B, C) \times \mathbf{C}(A, B) \to \mathbf{C}(A, C)$ is a PCM bimorphism.

The category **PCM** is symmetric monoidal via a tensor product representing PCM bimorphisms [146]. Therefore the definition above may be rephrased more abstractly as a category enriched over the monoidal category **PCM**, see [170]. We will however stick to the concrete definition.

**Proposition 2.2.4.** *Any category enriched over PCMs has zero morphisms.*

*Proof.* The homsets are PCMs and hence contain zeros $0\colon A \to B$, which form zero morphisms. ■

## 2.3 Effect algebras

*Effect algebras* are partial algebraic structures that axiomatize quantum effects. They are a common generalization of Boolean algebras, orthomodular lattices, and MV-algebras. The term 'effect algebra' is due to Foulis and Bennet [82], but the same or equivalent structures were introduced under different names in several papers [64, 94, 101, 178].[2] We refer to [74] for a comprehensive account of the subject.

### 2.3.1 Basics

**Definition 2.3.1.** An **effect algebra** is a PCM $(E, \ovee, 0)$ with an element $1 \in E$, called the **top**, satisfying:

(a) For each $a \in E$, there exists a unique $b \in E$ such that $a \ovee b = 1$.

(b) $a \perp 1$ implies $a = 0$.

The unique element $b \in E$ in condition (a) is written as $a^\perp$ and called the **orthosupplement** of $a$. We note that some authors call sums $\ovee$ in an effect algebra *orthogonal sums* (or *orthosums*), and use *orthogonal* as a synonym for summable.

**Proposition 2.3.2.** *The following hold in an effect algebra.*

(i) $a^{\perp\perp} = a$, *i.e. orthosupplementation is involutive.*

(ii) $0^\perp = 1$ *and* $1^\perp = 0$.

(iii) *Positivity:* $a \ovee b = 0$ *implies* $a = b = 0$.

[1] Be warned that some authors use the term 'bimorphism' differently, referring to a morphism that is both monic and epic.

[2] The names used there are: *weak orthoalgebra* [94], *unsharp orthoalgebra* [64], *D-poset* [178], and *D-algebra* [101].

(iv) *Cancellativity:* $a \ovee c = b \ovee c$ *implies* $a = b$.

(v) *There is a partial order given by* $a \leq b \stackrel{\text{def}}{\iff} \exists c.\, a \ovee c = b$.

(vi) *With respect to the partial order* $\leq$*, the zero* $0$ *is a bottom and the top* $1$ *is indeed a top (i.e. a greatest element).*

The partial order defined in (v) is referred to as the **algebraic ordering**.

*Proof.* Points (i) and (ii) follow immediately from the definition.

(iii) If $a \ovee b = 0$, the sum $(a \ovee b) \ovee 1$ is defined. It follows that both $a \perp 1$ and $b \perp 1$. Hence $a = b = 0$.

(iv) Suppose $a \ovee c = b \ovee c$. Let $d = (a \ovee c)^\perp$ $(= (b \ovee c)^\perp)$. Then $a \ovee c \ovee d = b \ovee c \ovee d = 1$. This implies that $a = (c \ovee d)^\perp = b$.

(v) We have $a \leq a$ for any $a$, since $0 \ovee a = a$. Suppose that $a \leq b$ and $b \leq c$, i.e. that $a \ovee u = b$ and $b \ovee v = c$ for some $u, v$. Then $a \ovee u \ovee v = c$ and thus $a \leq c$. Finally assume that $a \leq b$ and $b \leq a$, i.e. $a \ovee u = b$ and $b \ovee v = a$ for some $u, v$. Then $a \ovee u \ovee v = a$. By cancellativity, $u \ovee v = 0$, and by positivity, $u = v = 0$. Hence $a = b$. Therefore $\leq$ is a partial order.

(vi) Straightforward. ■

From a logical perspective, we view 0 and 1 respectively as the falsity and truth; $a^\perp$ as the negation of $a$; and $a \leq b$ as the entailment. The fact that $a^{\perp\perp} = a$ holds — one can eliminate double negation — shows that effect algebras as a logical structure are quite different from the intuitionistic logical structures such as Heyting algebras. An example below shows that effect algebras generalize Boolean algebras.

**Example 2.3.3.** Here are examples of effect algebras.

(i) Any Boolean algebra is an effect algebra with the obvious top and bottom, and $\ovee$ = 'disjoint sum', i.e. $a \perp b \iff a \wedge b = 0$ and then $a \ovee b = a \vee b$. Clearly the orthosupplement is the complement: $a^\perp = \neg a$. Interestingly, George Boole, the eponym of Boolean algebras, also considered sum/disjunction to be a partially defined operation; see the first paragraph of [82] or the footnote of [140, p. 8].

(ii) Similarly (and more generally), any orthomodular lattice (see Definition 2.3.15) is an effect algebra via $a \perp b \iff a \leq b^\perp$ and $a \ovee b = a \vee b$.

(iii) The unit interval $[0, 1]$ of real numbers is an effect algebra with $r \perp s \iff r + s \leq 1$ and $r \ovee s = r + s$. The bottom is 0 and the top is 1. The orthosupplement is given by $r^\perp = 1 - r$.

(iv) For each set $X$, the set $[0, 1]^X$ of $[0, 1]$-valued functions on $X$ forms an effect algebra in a pointwise manner. These functions $p\colon X \to [0, 1]$ are known as *fuzzy subsets* of the (ordinary, or 'crisp') set $X$ in the theory of fuzzy sets and logic [172, 262]. Two functions $p, q\colon X \to [0, 1]$ are summable when $p(x) + q(x) \leq 1$ for all $x \in X$, and then the sum is $(p \ovee q)(x) = p(x) + q(x)$. The orthosupplement is given by $p^\perp(x) = 1 - p(x)$.

(v) *MV-algebras* — algebraic models of the Łukasiewicz infinitely-many-valued logic — can be viewed as effect algebras, see Section 6.4 for details. In fact, (i), (iii), and (iv) are examples of MV-algebras.

(vi) Let $\mathscr{A}$ be a $C^*$-algebra — it represents a quantum system in the algebraic formulation of quantum theory, see Section 2.6 for a brief introduction. Then the unit interval of $\mathscr{A}$

$$[0,1]_{\mathscr{A}} = \{0 \le x \le 1 \mid x \in \mathscr{A}\}$$

is an effect algebra: $x \perp y$ iff $x + y \le 1$, and then $x \ovee y = x + y$. The orthosupplement is $x^\perp = 1 - x$. The elements in $[0,1]_{\mathscr{A}}$ are called **effects** in $\mathscr{A}$. In particular, when $\mathscr{A} = \mathcal{B}(\mathscr{H})$ is the $C^*$-algebra of bounded operators on a Hilbert space $\mathscr{H}$ (see Example 2.6.2), the effect algebra $[0,1]_{\mathcal{B}(\mathscr{H})}$ is called a *standard effect algebra* [82]. These are the motivating examples of effect algebras, as effects $[0,1]_{\mathscr{A}}$ represent *unsharp* measurements or observations in a quantum system [27, 82]. Note that effects include *projections* ($x \in \mathscr{A}$ with $x^* = x = x^2$), which represent *sharp* measurements.

(vii) More generally, if $G$ is a partially ordered abelian group and $u \in G$ is a positive element, then the interval $[0,u]_G = \{x \in G \mid 0 \le x \le u\}$ forms an effect algebra in the obvious manner. Effect algebras arising in this way (up to isomorphism) are called *interval effect algebras*.

**Definition 2.3.4.** A PCM $M$ is called

(i) **positive** (or conical) if $a \ovee b = 0$ implies $a = b = 0$, for all $a, b \in M$;

(ii) **cancellative** if $a \ovee c = b \ovee c$ implies $a = b$, for all $a, b, c \in M$.

Since the proof of Proposition 2.3.2(v) uses only positivity and cancellativity, any cancellative positive PCM forms a poset via algebraic ordering. Moreover 0 is a bottom. In fact, one has the following characterization of effect algebras.

**Proposition 2.3.5.** *Let $M$ be a PCM and $1 \in M$ an element. Then $(M, 1)$ is an effect algebra if and only if $M$ is positive and cancellative, and $1$ is a greatest element of $M$ with respect to algebraic ordering.*

*Proof.* The 'only if' is proved in Proposition 2.3.2. Suppose that $M$ is positive and cancellative, and 1 is a greatest element of $M$. By the definition of $\le$ and by the assumption that 1 is greatest, for any $a$ there exists $b$ such that $a \ovee b = 1$. By cancellativity such $b$ is unique. Suppose $a \perp 1$. Then $1 \le a \ovee 1$, while $a \ovee 1 \le 1$ since 1 is greatest. By antisymmetry, $a \ovee 1 = 1$, and then $a = 0$ by cancellativity. ∎

**Corollary 2.3.6.** *Let $M$ be a PCM, and $1, 1' \in M$ elements. If both $(M, 1)$ and $(M, 1')$ are effect algebras, then $1 = 1'$.* ∎

Thus a cancellative positive PCM can be thought of as an effect algebra without top 1, and is sometimes called a *generalized effect algebra* [74].

If $a \le b$ in an effect algebra (or more generally, in a cancellative positive PCM), by cancellativity there is a unique $c$ with $a \ovee c = b$. We denote this $c$ by $b \ominus a$ and call the **difference** of $b$ and $a$. In other words, $\ominus$ is a binary partial operation determined by:

$$b \ominus a = c \iff b = a \ovee c\,.$$

We list basic properties of the ordering and the difference, omitting a proof.

**Lemma 2.3.7.** *The following hold in an effect algebra.*

(i) $a \leq b \leq c$ *implies* $c \ominus b \leq c \ominus a$.

(ii) $a \ovee b \leq c$ *iff* $a \leq c \ominus b$.

(iii) $a \leq b$ *iff* $b^\perp \leq a^\perp$.

(iv) $a \perp b$ *iff* $a \leq b^\perp$ *iff* $b \leq a^\perp$.

(v) $a \leq b$ *and* $b \perp c$ *imply* $a \perp c$ *and* $a \ovee c \leq b \ovee c$. ■

**Definition 2.3.8.** Let $E, D$ be effect algebras. A **unital** (resp. **subunital**) **morphism of effect algebras** is a PCM morphism $f\colon E \to D$ such that $f(1) = 1$ (resp. $f(1) \leq 1$). Note that any function $f\colon E \to D$ satisfies $f(1) \leq 1$, since 1 is a greatest element in an effect algebra, and hence subunital morphisms are merely PCM morphisms. Note that in the literature *(homo)morphisms of effect algebras* refer to unital morphisms.

We denote by **EA** the category of effect algebras and unital morphisms; and by $\mathbf{EA}_{\leq}$ the category of effect algebras and subunital morphisms. There are obvious functors:

$$\mathbf{EA} \hookrightarrow \mathbf{EA}_{\leq} \longrightarrow \mathbf{PCM}$$

Here the forgetful functor $\mathbf{EA}_{\leq} \to \mathbf{PCM}$ is full and faithful, and moreover injective on objects by Corollary 2.3.6. Thus one can see $\mathbf{EA}_{\leq}$ as a full subcategory of **PCM**.

**Lemma 2.3.9.** *Let $E$ and $D$ be effect algebras. Let $f\colon E \to D$ be a subunital morphism of effect algebras.*

(i) $f$ *is monotone:* $a \leq b$ *implies* $f(a) \leq f(b)$.

(ii) $f$ *preserves differences:* $f(b \ominus a) = f(b) \ominus f(a)$.

(iii) *If $f$ is unital, it preserves orthosupplements:* $f(a^\perp) = f(a)^\perp$. ■

**Example 2.3.10.** Natural examples of (sub)unital morphisms of effect algebras are found in probability theory. Let $(X, \Sigma_X)$ be a measurable space. Then the $\sigma$-algebra $\Sigma_X$ is a Boolean algebra and hence an effect algebra. Recall that the unit interval $[0, 1]$ is also an effect algebra. Then a probability measure $\mu\colon \Sigma_X \to [0, 1]$ is a unital morphism of effect algebras. Indeed, it must satisfy $\mu(\varnothing) = 0$, $\mu(X) = 1$, and $\mu(A \cup B) = \mu(A) + \mu(B)$ whenever $A$ and $B$ are disjoint. Similarly, a subprobability measure $\mu\colon \Sigma_X \to [0, 1]$ is a subunital morphism. In fact, (sub)probability measures $\mu\colon \Sigma_X \to [0, 1]$ are precisely (sub)unital morphisms of $\sigma$-effect algebras, see Definitions 7.3.4 and 7.3.7. Integration was studied from an effect-algebraic point of view in [151].

## 2.3.2 Ortho-sharpness and orthomodular lattices

**Definition 2.3.11.** Let $E$ be an effect algebra.

(i) Two elements $a, b \in E$ are **disjoint** if $a \wedge b = 0$, that is, if $c \leq a$ and $c \leq b$ implies $c = 0$ for all $c \in E$.

(ii) An element $a \in E$ is **ortho-sharp** if $a$ and $a^\perp$ are disjoint.

In the literature of effect algebras (e.g. [103, 104, 161, 226]), ortho-sharp elements are simply called *sharp* elements. In this thesis we reserve the term 'sharp' for a notion of sharp predicates defined in an effectus, see Section 5.5.

**Lemma 2.3.12.** *Let $a$ be an ortho-sharp element in an effect algebra. Then $a$ is disjoint with any element $b$ such that $a \perp b$.*

*Proof.* Assume $c \leq a$ and $c \leq b$. Then $c \leq b \leq a^{\perp}$. Since $a$ is ortho-sharp, $c = 0$. ∎

The following property of effect algebras is quite useful.

**Proposition 2.3.13.** *Suppose that in an effect algebra, $a \perp b$ and a join $a \vee b$ exists. Then a meet $a \wedge b$ exists too, and $a \ovee b = (a \vee b) \ovee (a \wedge b)$.*

*Proof.* See [100, Theorem 3.5]. ∎

**Corollary 2.3.14.** *Suppose that in an effect algebra, $a \perp b$ and a join $a \vee b$ exists. Then $a \ovee b = a \vee b$ if and only if $a$ and $b$ are disjoint.* ∎

Orthomodular lattices axiomatize the structure of 'sharp' quantum logic, that is, the structure of projections on (i.e. closed subspaces of) a Hilbert spaces. They are the central structure in the traditional quantum logic initiated by Birkhoff and von Neumann [18]. We note that the orthomodular law was not used in [18] but was later discovered independently by Husimi [131], Loomis [193], and Maeda [201]; see [86, § 3] for a historical overview of quantum logic.

**Definition 2.3.15.** An **orthomodular lattice** is a lattice $(L, \vee, \wedge, 0, 1)$ with a unary operation $(-)^{\perp} \colon L \to L$ satisfying the following conditions.

(a) $(-)^{\perp}$ is antitone: $a \leq b$ implies $b^{\perp} \leq a^{\perp}$.

(b) $(-)^{\perp}$ is involutive: $a^{\perp\perp} = a$.

(c) $a^{\perp}$ is a complement of $a$: $a \vee a^{\perp} = 1$ and $a \wedge a^{\perp} = 0$.

(d) The *orthomodular law* holds: $a \leq b$ implies $a \vee (a^{\perp} \wedge b) = b$.

The element $a^{\perp}$ is called the *orthocomplement* of $a$.

The orthomodular law (d) is a weakening of the modular law:

$$a \leq c \implies a \vee (b \wedge c) = (a \vee b) \wedge c\,,$$

which in turn is a weakening of the distributive law:

$$a \vee (b \wedge c) = (a \vee b) \wedge (a \vee c)\,.$$

Thus orthomodular lattices generalize Boolean algebras (= distributive complemented lattices).

**Lemma 2.3.16.** *Every orthomodular lattice is an effect algebra via*

$$a \perp b \iff a \leq b^{\perp} \qquad\qquad a \ovee b = a \vee b$$

*and $a^{\perp}$ as orthosupplements.*

*Proof.* Straightforward. ■

In fact, we can identify orthomodular lattices with effect algebras satisfying special properties. An effect algebra whose partial order is a lattice is called a **lattice effect algebra**.

**Proposition 2.3.17.** *Let $L$ be a lattice effect algebra. The following are equivalent.*

(i) *$L$ is an orthomodular lattice (with orthosupplements $(-)^\perp$ as orthocomplements).*

(ii) *$a \perp a$ implies $a = 0$ for each $a \in L$.*

(iii) *Every element $a \in L$ is ortho-sharp: $a \wedge a^\perp = 0$.*

(iv) *$a \perp b$ implies $a \wedge b = 0$ for each $a, b \in L$.*

(v) *$a \perp b$ implies $a \ovee b = a \vee b$ for each $a, b \in L$.*

*Proof.* (ii) $\implies$ (iii): Let $b \in L$ such that $b \leq a$ and $b \leq a^\perp$. Then $b \perp b$ and hence $b = 0$.

(iii) $\implies$ (ii): Assume $a \perp a$. Then $a \leq a^\perp$, so that $0 = a \wedge a^\perp = a$.

(iv) $\implies$ (iii) is trivial, and the converse (iii) $\implies$ (iv) follows by Lemma 2.3.12.

Equivalence (iv) $\iff$ (v) follows by Corollary 2.3.14.

We have proved that (ii)–(v) are equivalent. Now note that (i) $\implies$ (iii) holds by definition ($a^\perp$ is a complement of $a$). To prove the converse, we assume (iii) and hence all of (ii)–(v). Since $L$ is self-dual via $(-)^\perp$, we have $a \vee a^\perp = (a^\perp \wedge a)^\perp = 0^\perp = 1$. Therefore $a^\perp$ is a complement of $a$. It only remains to prove the orthomodular law. Assume $a \leq b$, i.e. $a \perp b^\perp$. Then

$$b \ominus a = (a \ovee b^\perp)^\perp \overset{\text{(v)}}{=} (a \vee b^\perp)^\perp = a^\perp \wedge b$$

and hence

$$a \vee (a^\perp \wedge b) = a \vee (b \ominus a) \overset{\text{(v)}}{=} a \ovee (b \ominus a) = b\,.$$ ■

**Remark 2.3.18.** Conditions (ii), (iii) and (iv) of Proposition 2.3.17 are equivalent also for any (non-lattice) effect algebras. Effect algebras satisfying these equivalent conditions are called *orthoalgebras*.

### 2.3.3 Effect monoids

We introduce an extension of effect algebras with a (total) multiplication operation. Such an extension, called *effect monoids*, captures abstractly the notion of probabilities [137].

**Definition 2.3.19.** An **effect monoid** is an effect algebra that is at the same time a monoid, in a coherent way: it is an effect algebra $(M, \ovee, 0, 1)$ with a binary operation $\cdot\colon M \times M \to M$ satisfying:

(a) $(M, \cdot, 1)$ is a monoid;

(b) $\cdot\colon M \times M \to M$ is a PCM bimorphism, that is, for each $s \in M$ the mappings $s \cdot (-)$ and $(-) \cdot s$ are PCM morphisms.

An effect monoid is **commutative** if the multiplication $\cdot$ is commutative.

**Remark 2.3.20.** Effect monoids can be abstractly described as monoids [199, § VII.3] in the monoidal category **EA** of effect algebras. In [146] it is shown that **EA** is symmetric monoidal via the tensor product of effect algebras $\otimes$, and the two-element effect algebra $2 = \{0, 1\}$ as tensor unit. The tensor product $\otimes$ has a universal property given by the following natural bijections.

$$\frac{\text{morphisms } E \otimes F \longrightarrow D \text{ in } \mathbf{EA}}{\text{effect algebra bimorphisms } E \times F \longrightarrow D}$$

Here *effect algebra bimorphisms* are PCM bimorphisms $f\colon E \times F \to D$ satisfying $f(1, 1) = 1$. Then it is not hard to see that effect monoids are identified with monoids in **EA**. In fact, this is how effect monoids are first introduced in [137].

**Example 2.3.21.** We give several examples of effect monoids.

(i) The prime example of an effect monoid is the unit interval $[0, 1]$ of real numbers. The multiplication is the ordinary multiplication of real numbers.

(ii) The set $2 = \{0, 1\}$ of Boolean values is an effect monoid. The multiplication is the ordinary conjunction.

(iii) Generalizing (i), for any set $X$ the set $[0, 1]^X$ of fuzzy subsets is an effect monoid in the pointwise manner.

(iv) Generalizing (ii), any Boolean algebra is an effect monoid with conjunction as multiplication.

(v) If $R$ is a partially ordered ring with $0 \leq 1$, the interval $[0, 1]_R$ is an effect monoid. In fact, all of the above examples can be obtained in this way.

Examples (i)–(iv) are all commutative effect monoids. We can find examples of noncommutative effect monoids via (v); see Example 4.3.9. A simpler example can be found in [255, § 2.2.2].

## 2.4 Distribution monads and convex sets

The probability distribution functor/monad $\mathcal{D}\colon \mathbf{Set} \to \mathbf{Set}$, which assigns to each set $X$ the set $\mathcal{D}(X)$ of probability distributions on $X$, is fundamental in modelling probabilistic systems [117, 141, 241]. Algebras for the monad $\mathcal{D}$ are (abstract) convex sets, which have been important structure in a broad context, such as mathematics, physics, computer science, and economics, see e.g. [81, 89, 106, 169, 176, 211, 223, 244, 245]. Here, following [137, 140], we introduce generalizations of these notions by replacing the unit interval $[0, 1]$ with any effect monoid $M$.

**Definition 2.4.1.** Let $M$ be an effect monoid. A **distribution over** $M$ on a set $X$ is a function $\varphi\colon X \to M$ with finite support, i.e. $\mathrm{supp}(\varphi) := \{x \mid \varphi(x) \neq 0\}$ is finite, that satisfies $\bigovee_{x \in X} \varphi(x) = 1$. It is convenient to denote a distribution as a **formal** **convex sum** $r_1|x_1\rangle + \cdots + r_n|x_n\rangle$, where $x_i \in X$ and $r_i \in M$ satisfying $\bigovee_i r_i = 1$.[3] We interpret such an expression as a distribution $\varphi\colon X \to M$ by

$$\varphi(x) = \bigovee_{i \in I(x)} r_i \qquad \text{where } I(x) \coloneqq \{i \mid x_i = x\}\,.$$

Clearly, any distribution $\varphi\colon X \to M$ can be written as the following formal convex sum:

$$\sum_{x \in \mathrm{supp}(\varphi)} \varphi(x)|x\rangle\,.$$

For each set $X$ we write

$$\mathcal{D}_M(X) = \Big\{\varphi\colon X \to M \;\Big|\; \mathrm{supp}(\varphi) \text{ is finite and } \bigovee_{x \in X} \varphi(x) = 1\Big\}$$

for the set of distributions over $M$ on $X$. The assignment $X \mapsto \mathcal{D}_M(X)$ extends to a functor $\mathcal{D}_M\colon \mathbf{Set} \to \mathbf{Set}$. For a function $f\colon X \to Y$, we define $\mathcal{D}_M(f)\colon \mathcal{D}_M(X) \to \mathcal{D}_M(Y)$ by

$$\mathcal{D}_M(f)(\varphi)(y) = \bigovee_{x \in f^{-1}(y)} \varphi(x)\,,$$

or in the formal convex sum notation,

$$\mathcal{D}_M(f)(\textstyle\sum_i r_i|x_i\rangle) = \sum_i r_i|f(x_i)\rangle\,.$$

Moreover $\mathcal{D}_M$ is a monad with unit $\eta_X\colon X \to \mathcal{D}_M(X)$ and multiplication $\mu_X\colon \mathcal{D}_M(\mathcal{D}_M(X)) \to \mathcal{D}_M(X)$ given by:

$$\eta_X(x)(x') = \begin{cases} 1 & \text{if } x = x' \\ 0 & \text{if } x \neq x' \end{cases}$$
$$\mu_X(\Phi)(x) = \bigovee_{\varphi \in \mathcal{D}_M(X)} \Phi(\varphi)\cdot\varphi(x)$$

or in the formal convex sum notation:

$$\eta_X(x) = 1|x\rangle$$
$$\mu_X\Big(\textstyle\sum_i r_i\Big|\sum_j s_{ij}|x_{ij}\rangle\Big\rangle\Big) = \sum_{ij} r_i\cdot s_{ij}|x_{ij}\rangle\,.$$

The monad $\mathcal{D}_M$ is called the **distribution monad over** $M$. Further information can be found in [137, 140].

**Example 2.4.2.** Take $M = [0,1]$, the unit interval of real numbers. Then a distribution over $[0,1]$ on a set $X$ is a finite discrete probability distribution on $X$. The distribution monad $\mathcal{D}_{[0,1]}$ over $[0,1]$ is called the probability distribution monad, or simply, the **distribution monad**, and written as $\mathcal{D} = \mathcal{D}_{[0,1]}$ We will give several variant of $\mathcal{D}$ in the next section.

[3] We use the 'ket' $|-\rangle$ notation to clearly distinguish elements $x \in X$ from scalars $r \in M$, while some authors simply write $r_1 \cdot x_1 + \cdots + r_n \cdot x_n$.

**Definition 2.4.3.** Let $M$ be an effect monoid. A **convex set** over $M$ is an (Eilenberg-Moore) algebra [199, § VI.2] for the distribution monad $\mathcal{D}_M$ over $M$. Explicitly, it is a set $K$ with a function $[\![-]\!]\colon \mathcal{D}_M(K) \to K$, an operation that sends a formal convex sum $\sum_i r_i|x_i\rangle$ to an actual convex sum $[\![\sum_i r_i|x_i\rangle]\!] \in K$. The operation must satisfy the following two axioms:

$$[\![1|x\rangle]\!] = x$$

for $1|x\rangle \in \mathcal{D}_M(K)$, and

$$\Big[\!\Big[\textstyle\sum_i r_i\Big|[\![\sum_j s_{ij}|x_{ij}\rangle]\!]\Big\rangle\Big]\!\Big] = [\![\sum_{ij} r_i \cdot s_{ij}|x_{ij}\rangle]\!] \tag{2.2}$$

for $\sum_i r_i|\sum_j s_{ij}|x_{ij}\rangle\rangle \in \mathcal{D}_M(\mathcal{D}_M(K))$.

An **affine map** $f\colon K \to L$ between convex sets over $M$ is a homomorphism of $\mathcal{D}_M$-algebras. Explicitly, it is a function $f\colon K \to L$ satisfying

$$f\big([\![\textstyle\sum_i r_i|x_i\rangle]\!]\big) = [\![\sum_i r_i|f(x_i)\rangle]\!] \,.$$

We write $M$-**Conv** $= \mathcal{EM}(\mathcal{D}_M)$ for the category of convex sets over $M$ and affine maps.

**Example 2.4.4.** Convex sets over $[0,1]$, i.e. $\mathcal{D}_{[0,1]}$-algebras, are simply called **convex sets**.[4] For the category we write **Conv** $= [0,1]$-**Conv**. Any convex subset of a real vector space forms a convex set. Conversely, it is known [107, 244] that every *cancellative* convex set is isomorphic to a convex subset of a real vector space. In fact, each cancellative convex set can be embedded in a certain ordered vector space. This will be elaborated in Section 7.2.

**Remark 2.4.5.** We give two remarks on the definition of convex sets.

(i) The equation (2.2) is equivalent to the following seemingly weaker one:

$$\Big[\!\Big[\textstyle\sum_i r_i\Big|[\![\sum_j s_{ij}|x_j\rangle]\!]\Big\rangle\Big]\!\Big] = [\![\sum_{ij} r_i \cdot s_{ij}|x_j\rangle]\!] \;\Big(= [\![\sum_j (\bigovee_i r_i \cdot s_{ij})|x_j\rangle]\!]\Big)\,.$$

To see the equivalence, note that $\sum_j s_{ij}|x_{ij}\rangle = \sum_{kj} \delta_{ik} \cdot s_{kj}|x_{kj}\rangle$ in (2.2), where $\delta_{ik}$ is Kronecker's delta, and consider $kj$ as a single index. Some authors define convex sets using this equation (e.g. [21, 176, 177, 223]).

(ii) Convex sets may also be defined in terms of a ternary operation $\langle -; -, - \rangle\colon M \times K \times K \to K$. See Remark 4.4.21 for more details.

## 2.5 Probability monads

We discussed the 'probability' distribution monad $\mathcal{D} = \mathcal{D}_{[0,1]}$ in Example 2.4.2, where $\mathcal{D}(X)$ consists of finite discrete probability distributions on $X$. For the purpose of providing examples of effectuses, we will use several variants of the distribution monad $\mathcal{D}$.

[4] Many synonyms of 'convex set' exist: convex structure [106], semiconvex set [81, 245], convex space [89, 176], convex module [223], abstract convex set [169], barycentric algebra [169]. Historical notes and further references on convex sets can be found in [169, Remark 2.9].

**Definition 2.5.1.** The **subdistribution monad** $\mathcal{D}_{\leq}\colon \mathbf{Set} \to \mathbf{Set}$ is a 'subprobability' variant of the distribution monad $\mathcal{D}$. For a set $X$, the set $\mathcal{D}_{\leq}(X)$ consists of finite discrete subprobability distributions on $X$ (simply called **subdistributions**), namely:

$$\mathcal{D}_{\leq}(X) = \left\{\varphi\colon X \to [0,1] \;\middle|\; \mathrm{supp}(\varphi) \text{ is finite and } \sum_{x\in X} \varphi(x) \leq 1\right\}.$$

The monad structure of $\mathcal{D}_{\leq}$ is very much the same as the distribution monad $\mathcal{D}$, see Definition 2.4.1, and hence not repeated here.

Next we define 'infinite' variants of the distribution and subdistribution monads.

**Definition 2.5.2.** The **infinite distribution monad** $\mathcal{D}^{\infty}\colon \mathbf{Set} \to \mathbf{Set}$ and the **infinite subdistribution monad** $\mathcal{D}^{\infty}_{\leq}\colon \mathbf{Set} \to \mathbf{Set}$ are respectively defined by:

$$\begin{aligned}\mathcal{D}^{\infty}(X) &= \left\{\varphi\colon X \to [0,1] \;\middle|\; \sum_{x\in X} \varphi(x) = 1\right\}\\ \mathcal{D}^{\infty}_{\leq}(X) &= \left\{\varphi\colon X \to [0,1] \;\middle|\; \sum_{x\in X} \varphi(x) \leq 1\right\}.\end{aligned}$$

The only difference from the (sub)distribution monad is that the finite support requirement is dropped. The monad structures of $\mathcal{D}^{\infty}$ and $\mathcal{D}^{\infty}_{\leq}$ are similar to $\mathcal{D}$.

A basic important fact is that $\mathcal{D}^{\infty}_{\leq}(X)$ (hence also $\mathcal{D}^{\infty}(X)$) consists of only countably-supported (sub)distributions.

**Lemma 2.5.3.** *Let $\varphi \in \mathcal{D}^{\infty}_{\leq}(X)$ be an infinite subdistribution on a set $X$. Then the support* $\mathrm{supp}(\varphi) = \{x \in X \mid \varphi(x) \neq 0\}$ *is countable.*

*Proof.* Writing $S_n = \{x \in X \mid \varphi(x) > 1/n\}$ we have $\mathrm{supp}(\varphi) \subseteq \bigcup_{n\in\mathbb{N}_{>0}} S_n$. Because each $S_n$ is finite, $\bigcup_{n\in\mathbb{N}_{>0}} S_n$ is countable, so that $\mathrm{supp}(\varphi)$ is countable. ■

Algebras for the monad $\mathcal{D}^{\infty}$ are known as **superconvex sets** [176, 177], which will be used in Section 7.3.

Finally, we define measure-theoretic probability monads. The monad $\mathcal{G}$ defined below is called the Giry monad after [93]. Its subprobability version $\mathcal{G}_{\leq}$ appeared in [214]. We write **Meas** for the category of measurable spaces and measurable functions. When $X$ is a measurable space, its $\sigma$-algebra is denoted by $\Sigma_X$.

**Definition 2.5.4.** The **Giry monad** $\mathcal{G}\colon \mathbf{Meas} \to \mathbf{Meas}$ and the **subprobability Giry monad** $\mathcal{G}_{\leq}\colon \mathbf{Meas} \to \mathbf{Meas}$ are defined as follows. For a measurable space $X$ with the $\sigma$-algebra $\Sigma_X$, define:

$$\begin{aligned}\mathcal{G}(X) &= \{\mu\colon \Sigma_X \to [0,1] \mid \mu \text{ is } \sigma\text{-additive and } \mu(X) = 1\}\\ \mathcal{G}_{\leq}(X) &= \{\mu\colon \Sigma_X \to [0,1] \mid \mu \text{ is } \sigma\text{-additive}\}.\end{aligned}$$

In other words, $\mathcal{G}(X)$ consists of probability measures on $X$, and $\mathcal{G}_{\leq}(X)$ consists of subprobability measures on $X$. Since the rest of the definitions of $\mathcal{G}$ and $\mathcal{G}_{\leq}$ are basically the same, below we describe $\mathcal{G}$ only.

We need to equip $\mathcal{G}(X)$ with a $\sigma$-algebra. It is defined to be the smallest $\sigma$-algebra such that $\mathrm{ev}_U : \mathcal{G}(X) \to [0,1]$ is measurable for all $U \in \Sigma_X$, where $\mathrm{ev}_U$ is the 'evaluation' map: $\mathrm{ev}_U(\mu) = \mu(U)$. For a measurable function $f : X \to Y$, we define $\mathcal{G}(f) : \mathcal{G}(X) \to \mathcal{G}(Y)$ by

$$\mathcal{G}(f)(\mu)(V) = \mu(f^{-1}(V))$$

for $V \in \Sigma_Y$. The unit $\eta_X : X \to \mathcal{G}(X)$ of the monad is given by the Dirac measures:

$$\eta_X(x)(U) = \begin{cases} 1 & \text{if } x \in U \\ 0 & \text{if } x \notin U\,. \end{cases}$$

The multiplication $\mu_X : \mathcal{G}(\mathcal{G}(X)) \to \mathcal{G}(X)$ is defined by

$$\mu_X(\Phi)(U) = \int_{\mathcal{G}(X)} \mathrm{ev}_U \, \mathrm{d}\Phi \ \left( = \int_{\mathcal{G}(X)} \mu(U)\, \Phi(\mathrm{d}\mu) \right)$$

for $\Phi \in \mathcal{G}(\mathcal{G}(X))$ and $U \in \Sigma_X$.

The verification that these data indeed define monads $\mathcal{G}$ and $\mathcal{G}_{\leq}$ requires some work, based on results from measure theory. For details, we refer to the original work by Giry [93], or [69, 70, 215].

## 2.6 $C^*$-algebras and $W^*$-algebras

In this section, we briefly review the basic definitions and results on $C^*$-algebras and $W^*$-algebras (generally called *operator algebras*). As mentioned in Section 1.2, these operator algebras provide a powerful and convenient 'algebraic' formulation of quantum theory that is alternative to the Hilbert space formulation, see e.g. [6, 112, 187, 188, 228]. In this thesis we use operator algebras to give examples of effectuses that model quantum systems and quantum processes. Specifically, the opposite $\mathbf{Wstar}_{\leq}^{\mathrm{op}}$ of the category of $W^*$-algebras and subunital normal completely positive maps serves as the archetypal example of an effectus. We note that this thesis is mainly focused on abstract theory of effectuses, and not on the category of $W^*$-algebras itself. Complementary to this thesis are Abraham and Bas Westerbaan's theses [253, 256], which are focused more on the category of $W^*$-algebras. Abraham's thesis [253] contains a concise yet comprehensive exposition of operator algebras. More information about operator algebras can be found in the standard textbooks [167, 232, 246].

A general idea in the algebraic approach to quantum theory is that an algebra of observables represents a system. Such algebras are axiomatized as follows.

**Definition 2.6.1.** A **$*$-algebra**[5] $\mathscr{A}$ is a complex unital associative algebra (i.e. a monoid in the category of complex vector spaces) with an 'involution' operation $(-)^* : \mathscr{A} \to \mathscr{A}$ such that for all $a, b \in \mathscr{A}$ and $\lambda \in \mathbb{C}$,

$$(a^*)^* = a \qquad (a+b)^* = a^* + b^* \qquad (\lambda a)^* = \overline{\lambda} a^* \qquad (ab)^* = b^* a^*\,.$$

[5]Note that in this thesis we require $*$-algebras and $C^*$-algebras to be unital.

A $C^*$**-algebra** $\mathscr{A}$ is a complete normed $*$-algebra $\mathscr{A}$ such that $\|ab\| \leq \|a\|\|b\|$ and $\|a^*a\| = \|a\|^2$ for all $a, b \in \mathscr{A}$.

A $*$-algebra is **commutative** if the multiplication is commutative. A $*$**-homomorphism** between $*$-algebras is a linear map that preserves the multiplication and involution. A $*$**-subalgebra** of a $*$-algebra is a linear subspace closed under multiplication and involution. A $*$-homomorphism or $*$-subalgebra is said to be **unital** if it also respects the unit.

One can show that every $*$-homomorphism $f\colon \mathscr{A} \to \mathscr{B}$ between $C^*$-algebras is nonexpansive in the sense that $\|f(a)\| \leq \|a\|$ for all $a \in \mathscr{A}$ [217, Theorem 1.5.7].

**Example 2.6.2.** Let $\mathscr{H}$ be a Hilbert space. We denote by $\mathcal{B}(\mathscr{H})$ the set of bounded operators on $\mathscr{H}$. Then $\mathcal{B}(\mathscr{H})$ forms a $C^*$-algebra with multiplication given by composition of operators, involution given by adjoint operators, and the operator norm. The $C^*$-algebra $\mathcal{B}(\mathscr{H})$ corresponds to the representation of a quantum system by a Hilbert space $\mathscr{H}$ in the standard formalism of quantum theory.

If $\mathscr{A}$ is a $C^*$-algebra, then any norm-closed unital $*$-algebra of $\mathscr{A}$ is a $C^*$-algebra. In particular, any norm-closed unital $*$-algebra of $\mathcal{B}(\mathscr{H})$ is a $C^*$-algebra. In fact, any $C^*$-algebra is of such a form, by the celebrated theorem of Gelfand and Neumark [92].

**Theorem 2.6.3.** *Every $C^*$-algebra is $*$-isomorphic to some norm-closed unital $*$-subalgebra of $\mathcal{B}(\mathscr{H})$ for some Hilbert space $\mathscr{H}$.*

*Proof.* See [246, Theorem I.9.18]. ■

Thus, $C^*$-algebras characterize norm-closed unital $*$-subalgebras of $\mathcal{B}(\mathscr{H})$, without referring to a Hilbert space. The theorem generally justifies the relevance of $C^*$-algebras in quantum theory.

In fact, more than pure quantum systems can be represented by $C^*$-algebras. In particular, we can view *commutative* $C^*$-algebras as classical systems, due to another theorem of Gelfand and Neumark [92].

**Theorem 2.6.4.** *Every commutative $C^*$-algebra is $*$-isomorphic to the $C^*$-algebra $\mathrm{C}(X)$ of continuous functions $\varphi\colon X \to \mathbb{C}$ for some compact Hausdorff space $X$.*

*Proof.* See [246, Theorem I.4.4]. ■

An important subclass of $C^*$-algebras is $W^*$-algebras. They characterize *weakly closed* (i.e. closed under the *weak operator topology* [246, §II.2]) unital $*$-subalgebra of $\mathcal{B}(\mathscr{H})$.

**Definition 2.6.5.** A $W^*$**-algebra** is a $C^*$-algebra $\mathscr{A}$ that has a *predual*, i.e. a Banach space $\mathscr{V}$ with an isometric linear bijection $\mathscr{V}^* \cong \mathscr{A}$.

**Theorem 2.6.6.** *A $C^*$-algebra is a $W^*$-algebra if and only if it is $*$-isomorphic to a weakly closed unital $*$-subalgebra of $\mathcal{B}(\mathscr{H})$ for some Hilbert space $\mathscr{H}$.*

*Proof.* See [246, Theorem III.3.5]. ■

A 'concrete' $W^*$-algebra, i.e. a weakly closed unital $*$-subalgebra of $\mathcal{B}(\mathscr{H})$, is often called a **von Neumann algebra**. The theory of von Neumann algebras, which preceded $C^*$-algebras, was developed by Murray and von Neumann in a series of papers starting with [208]. The abstract characterization above is due to Sakai [231, 232].

**Example 2.6.7.** By Theorem 2.6.6, $\mathcal{B}(\mathscr{H})$ is a $W^*$-algebra, since it is trivially weakly closed. A predual of $\mathcal{B}(\mathscr{H})$ is the space $\mathcal{TC}(\mathscr{H})$ of *trace-class* operators on $\mathscr{H}$. The isomorphism $\Phi\colon \mathcal{B}(\mathscr{H}) \overset{\cong}{\to} \mathcal{TC}(\mathscr{H})^*$ is given by $\Phi(A)(T) = \mathrm{tr}(AT)$. We refer to [246] for the definition of trace-class operators and other details.

A predual of a $W^*$-algebra is unique up to isometric isomorphism [246, Corollary III.3.9]. Therefore a $W^*$-algebra $\mathscr{A}$ has an intrinsic topology, namely the weak* topology induced by the predual. A map $f\colon \mathscr{A} \to \mathscr{B}$ between $W^*$-algebras is said to be **normal** if it is continuous with respect to the weak* topologies. We write $\mathscr{A}_*$ for the set of all normal linear functionals $\varphi\colon \mathscr{A} \to \mathbb{C}$. By the standard theory of dual spaces (see [218, § 2.4] or [58, § V.1]), $\mathscr{A}_*$ is the predual of $\mathscr{A}$, i.e. $(\mathscr{A}_*)^* \cong \mathscr{A}$. Note that any finite-dimensional $C^*$-algebra $\mathscr{A}$ is a $W^*$-algebra since $\mathscr{A} \cong (\mathscr{A}^*)^*$.

We introduce some more terminology and notations.

**Definition 2.6.8.** Let $\mathscr{A}$ be a $C^*$-algebra. An element $a \in \mathscr{A}$ is called

(i) **self-adjoint** if $a^* = a$;

(ii) **positive** if $a = b^*b$ for some $b \in \mathscr{A}$;

(iii) an **effect** if both $a$ and $1 - a$ are positive;

(iv) a **projection** if $a^* = a = a^2$.

We write $\mathscr{A}_{\mathrm{sa}}$, $\mathscr{A}_+$, $[0,1]_{\mathscr{A}}$, and $\mathcal{Pr}(\mathscr{A})$ respectively for the set of self-adjoint elements, positive elements, effects, and projections. It is easy to see that $\mathcal{Pr}(\mathscr{A}) \subseteq [0,1]_{\mathscr{A}} \subseteq \mathscr{A}_+ \subseteq \mathscr{A}_{\mathrm{sa}}$.

Clearly, $\mathscr{A}_{\mathrm{sa}}$ forms a real vector space. Moreover $\mathscr{A}_{\mathrm{sa}}$ is an *ordered vector space* (Definition 7.2.2) with $\mathscr{A}_+$ as the positive cone [246, Theorem I.6.1]. Explicitly, one has a partial order $\leq$ on $\mathscr{A}_{\mathrm{sa}}$ defined by $a \leq b \iff b - a$ is positive. We thus write $a \geq 0$ to mean '$a$ is positive'. Effects are precisely elements $a \in \mathscr{A}_{\mathrm{sa}}$ such that $0 \leq a \leq 1$, which justifies the 'unit interval' notation $[0,1]_{\mathscr{A}}$. Note that if $p \in \mathscr{A}$ is an effect (resp. a projection), $1 - p$ is an effect (resp. a projection) too. We write $p^\perp = 1 - p$, which can intuitively be understood as 'negation of $p$'.

Below we include several results on the partial order on a $C^*$-/$W^*$-algebra.

**Lemma 2.6.9.** *For each self-adjoint element $a \in \mathscr{A}_{\mathrm{sa}}$ of a $C^*$-algebra $\mathscr{A}$,*

$$\|a\| \leq 1 \iff -1 \leq a \leq 1\,.$$

*Proof.* Write $\mathrm{Sp}(a) = \{\lambda \in \mathbb{C} \mid a - \lambda 1 \text{ is not invertible}\}$ for the spectrum of $a \in \mathscr{A}$. Then for each self-adjoint $a \in \mathscr{A}_{\mathrm{sa}}$ one has $\mathrm{Sp}(a) \subseteq \mathbb{R}$ and $\|a\| = \sup_{\lambda \in \mathrm{Sp}(a)} |\lambda|$ [246, Proposition I.4.2 and I.4.3]. Moreover $a \in \mathscr{A}_+$ if and only if $\mathrm{Sp}(a) \subseteq \mathbb{R}_+$ [246,

Theorem I.6.1]. Thus

$$\begin{aligned}\|a\| \leq 1 &\iff -1 \leq \lambda \leq 1 \text{ for all } \lambda \in \mathrm{Sp}(a) \\ &\iff \mathrm{Sp}(1-a) \subseteq \mathbb{R}_+ \text{ and } \mathrm{Sp}(a+1) \subseteq \mathbb{R}_+ \\ &\iff 1-a \in \mathscr{A}_+ \text{ and } a+1 \in \mathscr{A}_+ \\ &\iff -1 \leq a \leq 1\,.\end{aligned}$$ ■

**Corollary 2.6.10.** *For each self-adjoint element $a \in \mathscr{A}_{\mathrm{sa}}$ of a $C^*$-algebra $\mathscr{A}$, one has $-\|a\|1 \leq a \leq \|a\|1$* ■

As mentioned in Example 2.3.3, for each $C^*$-algebra $\mathscr{A}$, the set of effects $[0,1]_{\mathscr{A}}$ forms an effect algebra. Projections can be characterized as ortho-sharp elements there.

**Proposition 2.6.11.** *Let $\mathscr{A}$ be a $C^*$-algebra. Then an effect $p \in [0,1]_{\mathscr{A}}$ is a projection if and only if it is ortho-sharp in the effect algebra $[0,1]_{\mathscr{A}}$*

*Proof.* See [254, Lemma 31]. ■

The order on projections can be characterized in various ways.

**Lemma 2.6.12.** *Let $p, q$ be projections in a $C^*$-algebra. The following are equivalent.*

(i) $p \leq q$.
(ii) $pqp = p$.
(iii) $pq^{\perp}p = 0$.
(iv) $pq = p$.
(v) $pq^{\perp} = 0$.

*Proof.* Conditions (ii) and (iii) are equivalent because

$$p = p(q + q^{\perp})p = pqp + pq^{\perp}p\,.$$

Similarly (iv) and (v) are equivalent.
(i) $\implies$ (iii): If $p \leq q$,

$$0 \leq pq^{\perp}p \leq pp^{\perp}p = p - p = 0\,.$$

(iii) $\implies$ (v): If $pq^{\perp}p = 0$,

$$\|pq^{\perp}\|^2 = \|pq^{\perp}(pq^{\perp})^*\| = \|pq^{\perp}p\| = 0\,,$$

so that $pq^{\perp} = 0$.
(iv) $\implies$ (i): If $pq = p$, then $qp = (pq)^* = p^* = p$ and

$$q - p = q^2 - qp - pq + p^2 = (q-p)^2 \geq 0\,,$$

whence $p \leq q$. ■

We now recap striking order-theoretic properties of $W^*$-algebras.

**Proposition 2.6.13.** *Let $\mathscr{A}$ be a $W^*$-algebra.*

(i) *$\mathscr{A}$ is 'monotone complete' in the following sense: in $\mathscr{A}_{\mathrm{sa}}$, every norm-bounded directed subset has a join (= least upper bound).*

(ii) $\mathcal{P}r(\mathscr{A})$ *is a complete lattice, i.e. all joins and meets exist.*

(iii) $[0,1]_{\mathscr{A}}$ *is directed complete, i.e. all directed joins exist.*

*Proof.* For (i) and (ii), see [232, Lemma 1.7.4 and 1.10.2] respectively. Point (iii) follows from (i) since the set $[0,1]_{\mathscr{A}}$ is bounded in norm. ■

In fact, $W^*$-algebras can be characterized in order-theoretic terms, as monotone complete $C^*$-algebras with a certain requirement; see [246, Theorem III.3.16].

**Lemma 2.6.14.** *Let $\mathscr{A}$ be a $W^*$-algebra and $p \in [0,1]_{\mathscr{A}}$ be an effect. Then there exist a least projection above $p$ and a greatest projection below $p$.*

*Proof.* See [254, Proposition 44] for the existence of a least projection above $p$. Since $[0,1]_{\mathscr{A}}$ is self-dual via $p \mapsto p^\perp \equiv 1-p$, there also exists a greatest projection below $p$ (namely, $\lceil p^\perp \rceil^\perp$ in the notation introduced below). ■

**Definition 2.6.15.** We denote by $\lceil p \rceil$ the least projection above $p \in [0,1]_{\mathscr{A}}$, and by $\lfloor p \rfloor$ the greatest projection below $p$. Note that $\lfloor p \rfloor = \lceil p^\perp \rceil^\perp$ and $\lceil p \rceil = \lfloor p^\perp \rfloor^\perp$.

**Proposition 2.6.16.** *Let $\mathscr{A}$ be a $W^*$-algebra. Joins (resp. meets) of projections in $\mathcal{P}r(\mathscr{A})$ are also joins (resp. meets) in $[0,1]_{\mathscr{A}}$.*

This means that the inclusion $\mathcal{P}r(\mathscr{A}) \hookrightarrow [0,1]_{\mathscr{A}}$ preserves joins and meets.

*Proof.* Let $\bigvee U$ be the join of projections $U \subseteq \mathcal{P}r(\mathscr{A})$. Let $q$ be an effect such that $p \leq q$ for all $p \in U$. Then $p \leq \lfloor q \rfloor$ for all $p \in U$, so that $\bigvee U \leq \lfloor q \rfloor$. Thus we have $\bigvee U \leq \lfloor q \rfloor \leq q$ and conclude that $\bigvee U$ is a join in $[0,1]_{\mathscr{A}}$. The case for meets is similar. ■

We turn to morphisms between $C^*$-algebras. Since we are concerned with linear maps only, we will refer to linear maps between $C^*$-algebras simply as *maps*.

**Definition 2.6.17.** Let $\mathscr{A}, \mathscr{B}$ be $C^*$-algebras.

(i) A map $f\colon \mathscr{A} \to \mathscr{B}$ is **positive** if $f(a) \geq 0$ for all $a \geq 0$.

(ii) A map $f\colon \mathscr{A} \to \mathscr{B}$ is **unital** if $f(1) = 1$; and **subunital** if $f(1) \leq 1$.

(iii) A **state** on $\mathscr{A}$ is a unital positive functional $\omega\colon \mathscr{A} \to \mathbb{C}$. When $\mathscr{A}$ is a $W^*$-algebra, a state is said to be **normal** if it is weak* continuous.

**Proposition 2.6.18.** *Every positive map $f\colon \mathscr{A} \to \mathscr{B}$ between $C^*$-algebras is bounded.*

The proof below shows $\|f\| \leq 2\|f(1)\|$, but in fact, $\|f\| = \|f(1)\|$ holds [216, Corollary 2.9]. The latter is harder to prove.

*Proof.* By Lemma 2.6.9 we have $\|f(a)\| \leq \|f(1)\| \cdot \|a\|$ for each $a \in \mathscr{A}_{\mathrm{sa}}$. Let $a \in \mathscr{A}$ be an arbitrary element. Writing $a_R = (a + a^*)/2$ and $a_I = (a - a^*)/2i$, we have $a_R, a_I \in \mathscr{A}_{\mathrm{sa}}$, $a = a_R + i a_I$, and $\|a_R\|, \|a_I\| \leq \|a\|$. Therefore

$$\|f(a)\| \leq \|f(a_R)\| + \|f(a_I)\| \leq \|f(1)\| \cdot \|a_R\| + \|f(1)\| \cdot \|a_I\| \leq 2\|f(1)\| \cdot \|a\| . \quad ■$$

Note that positive maps are order-preserving on self-adjoint elements. Moreover, normality is characterized by certain order-continuity:

**Proposition 2.6.19.** *A positive map $f\colon \mathscr{A} \to \mathscr{B}$ between $W^*$-algebras is normal (i.e. weak* continuous) if and only if $f$ preserves suprema of norm-bounded directed subsets in $\mathscr{A}_{\mathrm{sa}}$.*

*Proof.* By [232, Theorem 1.13.2], the claim holds for $\mathscr{B} = \mathbb{C}$, i.e. for functionals $\varphi\colon \mathscr{A} \to \mathbb{C}$. The general claim follows because the predual $\mathscr{A}_*$ is spanned by normal positive functionals [232, Theorem 1.14.3]. ∎

When a $C^*$-algebra is viewed as an algebra of observables, a state can be understood as a mapping that sends each observable to the expected value of outcomes. Indeed, by a positivity of states, we have $\omega(x) \in \mathbb{R}$ for any self-adjoint $x \in \mathscr{A}$. The unit $1 \in \mathscr{A}$ is seen as the constant observable whose outcome is always 1, so $\omega(1) = 1$.

**Example 2.6.20.** 'States' in the Hilbert space formulation of quantum theory, i.e. density operators, can be captured as normal states on the $W^*$-algebra $\mathcal{B}(\mathscr{H})$ of bounded operators on a Hilbert space $\mathscr{H}$. Since $\mathcal{TC}(\mathscr{H})$ a predual of $\mathcal{B}(\mathscr{H})$, by the theory of dual spaces, $\mathcal{TC}(\mathscr{H})$ is isomorphic to the space $\mathcal{B}(\mathscr{H})_*$ of normal (= weak* continuous) functionals on $\mathcal{B}(\mathscr{H})$. This yields a bijective correspondence between density operators on $\mathscr{H}$ (i.e. trace-class operators $\rho$ with $\mathrm{tr}(\rho) = 1$) and normal states on $\mathcal{B}(\mathscr{H})$ (i.e. unital positive normal functionals).

If $f\colon \mathscr{A} \to \mathscr{B}$ is a positive unital map, then it sends each state $\omega$ on $\mathscr{B}$ to a state $\omega \circ f$ on $\mathscr{A}$. Therefore $f$ can be understood as a transformation of the system $\mathscr{B}$ to $\mathscr{A}$. However, for the reason explained below, mere positivity is not enough, and *complete* positivity is required. We need some preliminary definitions. We write $\mathcal{M}_n = \mathbb{C}^{n\times n} \cong \mathcal{B}(\mathbb{C}^n)$ for the $C^*$-algebra of complex $n \times n$-matrices. Given a $C^*$-algebra $\mathscr{A}$, let $\mathcal{M}_n \otimes \mathscr{A}$ be the algebraic tensor product (i.e. the tensor product as vector spaces). More concretely, $\mathcal{M}_n \otimes \mathscr{A}$ is isomorphic to the set of $n \times n$-matrices with entries from $\mathscr{A}$. Then $\mathcal{M}_n \otimes \mathscr{A}$ can be equipped with a multiplication and an involution, forming a $C^*$-algebra [246, § IV.3]. We can view $\mathcal{M}_n \otimes \mathscr{A}$ as a compound system of $\mathcal{M}_n$ and $\mathscr{A}$. Let $f\colon \mathscr{A} \to \mathscr{B}$ is a map between $C^*$-algebras. Then it yields a map $\mathcal{M}_n \otimes f\colon \mathcal{M}_n \otimes \mathscr{A} \to \mathcal{M}_n \otimes \mathscr{B}$ between the tensor products. We can view it as a map that transforms only a part of the system, leaving the $\mathcal{M}_n$ part alone. In general, $\mathrm{id} \otimes f$ need not be positive when $f$ is positive. Therefore, for $f$ to be a physically meaningful transformation — so that $f$ can be partially applied to the system — we need the following property:

**Definition 2.6.21.** A map $f\colon \mathscr{A} \to \mathscr{B}$ between $C^*$-algebras is **completely positive** (**CP**, for short) if for each $n \in \mathbb{N}$, the map $\mathcal{M}_n \otimes f\colon \mathcal{M}_n \otimes \mathscr{A} \to \mathcal{M}_n \otimes \mathscr{B}$ is positive.

A more explicit definition for complete positivity can be given, see [246, Corollary 3.4].

Complete positivity is closely related to the notions of tensor products of $C^*$-algebras and $W^*$-algebras. With respect to a suitable notion of tensor product, complete positivity makes tensor product $\otimes$ bifunctorial: for CP maps $f_1\colon \mathscr{A}_1 \to \mathscr{B}_1$ and $f_2\colon \mathscr{A}_2 \to \mathscr{B}_2$, one can construct a CP map $f_1 \otimes f_2\colon \mathscr{A}_1 \otimes \mathscr{A}_2 \to \mathscr{B}_1 \otimes \mathscr{B}_2$ between the tensor products; see [246] and also [37]. Note however that tensor products are not used in this thesis.

We note a convenient result about CP maps.

**Lemma 2.6.22.** *A positive map $f\colon \mathscr{A} \to \mathscr{B}$ between $C^*$-algebras is completely positive whenever at least one of $\mathscr{A}$ and $\mathscr{B}$ is commutative.*

*Proof.* See [246, Corollary IV.3.5 and Proposition IV.3.9]. ■

Thus unital CP maps between $C^*$-algebras can be viewed as 'physical' processes between systems, inducing transformations of states. Similarly, subunital CP maps represent physical processes that are 'incomplete' in some sense. It is possible to explicitly relate normal (sub)unital CP maps between $W^*$-algebras to the notions of *channels* and *operations* used in the Hilbert spaces formulation.

**Example 2.6.23.** Let $\mathscr{H}$ be a Hilbert space. A linear map $f\colon \mathcal{TC}(\mathscr{H}) \to \mathcal{TC}(\mathscr{H})$ on the space of trace-class operators is called a *channel* (resp. an *operation*) if it is completely positive[6] and trace-preserving (resp. trace-decreasing), see e.g. [120, 181, 212]. The channels and operations are known to coincide with classes of transformations on a quantum system that can be realized by preparation, unitary transformation and measurement. Recall that $\mathcal{TC}(\mathscr{H})$ is a predual of $\mathcal{B}(\mathscr{H})$. Using a result form the theory of dual spaces, it is not hard to see that channels (resp. operations) $f\colon \mathcal{TC}(\mathscr{H}) \to \mathcal{TC}(\mathscr{H})$ are in bijective correspondence with normal completely positive unital (resp. subunital) maps $g\colon \mathcal{B}(\mathscr{H}) \to \mathcal{B}(\mathscr{H})$.

In this thesis, we are interested in the following categories, whose objects and morphisms represent quantum systems and processes.

**Definition 2.6.24.** We denote by **Cstar** (resp. $\mathbf{Cstar}_{\leq}$) the category of $C^*$-algebras and unital (resp. subunital) CP maps. We denote by **Wstar** (resp. $\mathbf{Wstar}_{\leq}$) the category of $W^*$-algebras and unital (resp. subunital) normal CP maps.

Note that **Wstar** is a non-full subcategory of **Cstar**.

[6] Defined similarly to completely positive maps between $C^*$-algebras.

# Chapter 3

# Effectuses

An *effectus* is a category satisfying certain conditions which provides a suitable axiomatic framework for quantum theory. In general, objects in an effectus are viewed as types of systems, and morphisms as processes between systems. The archetypal example of an effectus, which models quantum systems and processes, is the opposite $\mathbf{Wstar}_{\leq}^{\mathrm{op}}$ of the category of $W^*$-algebras and subunital normal completely positive maps. Another simple example of an effectus is the category **Pfn** of sets and partial functions, which models classical systems and deterministic processes. Yet another example is the Kleisli category $\mathcal{K}\ell(\mathcal{D}_{\leq})$ of the subdistribution monad, which models classical systems and probabilistic processes. These three effectuses are our leading examples throughout the thesis. Note that the notion of effectus used here is 'effectus in partial form'; in Chapter 4 it will be shown to be equivalent to Jacobs' original formulation [140] of effectus ('in total form').

In this chapter we aim to establish basic notions in effectus theory. The definition of effectus is based on *finitely partially additive category*, which will be covered in Section 3.1. Then in Section 3.2 we define an effectus and related notions, and describe our main examples of effectuses in Section 3.3. An effectus is equipped with a special object $I$, representing the trivial system that has no information, i.e. 'no system'. We call morphisms $p\colon A \to I$ *predicates*, and $\omega\colon I \to A$ *substates*. *States* form a subclass of substates that are *total* ('normalized'). We study structures of them in the subsequent sections (Sections 3.4–3.6). Predicates form *effect modules*, i.e. effect algebras with scalar multiplication, and states form *convex sets*. We will introduce a new axiomatic structure of substates called *weight module*. Similarly to a duality between effect modules and convex sets established by Jacobs [140], there is a duality between weight modules and effect modules. We find that weight modules are a fairly natural and convenient structure, for example because the category of weight modules always forms an effectus, while to prove that the category of convex sets forms an effectus, we need some technical assumption (see Section 4.4).

Section 3.7 summarizes these structures associated with an effectus as 'state-and-effect' triangles of categories and functors. The triangles describe neatly the duality between states and predicates/effects, and thus between the Schrödinger and Heisenberg pictures.

In the last section (Section 3.8) we provide a characterization of effectuses via more elementary, direct conditions that do not mention partially additive structure.

## 3.1 Finitely partially additive categories (finPACs)

We introduce *finitely partially additive categories (finPACs)*, a variant of *partially additive categories (PACs)* studied by Arbib and Manes [7, 8, 202]. As the name suggests, finPACs are equipped with the structure of finite partial sums, i.e. the PCM structure. The definition relaxes the original definition of PACs, which have *countable* partial sums. In this thesis we are mainly concerned with finite partial sums, but the countable structure will also be studied in Section 7.3.

**Definition 3.1.1.** Let $\mathbf{C}$ be a category with zero morphisms.

(i) For each (possibly infinite) coproduct $\coprod_{j\in J} A_j$ that exists in $\mathbf{C}$, we define morphisms $\rhd_j \colon \coprod_{j\in J} A_j \to A_j$ by

$$\rhd_j \circ \kappa_k = \begin{cases} \mathrm{id}_{A_j} & \text{if } k = j \\ 0_{A_k A_j} & \text{if } k \neq j\,. \end{cases}$$

We call the morphisms $\rhd_j$ **partial projections**.[1]

(ii) A family of morphisms $(f_j \colon A \to B_j)_{j\in J}$ in $\mathbf{C}$ is **compatible** if there exist a coproduct $\coprod_j B_j$ and a morphism $f \colon A \to \coprod_j B_j$ such that $f_j = \rhd_j \circ f$ for all $j \in J$.

**Definition 3.1.2** (cf. [7, § 3.3])**.** A **finitely partially additive category** (**finPAC**, for short) is a category $\mathbf{C}$ with finite coproducts $(+, 0)$ that is enriched over PCMs and satisfies the following two axioms.

**(Compatible sum axiom)** If parallel morphisms $f, g \colon A \to B$ are compatible, then $f, g$ are summable in the PCM $\mathbf{C}(A, B)$.

**(Untying axiom)** If $f, g \colon A \to B$ are summable, then $\kappa_1 \circ f, \kappa_2 \circ g \colon A \to B + B$ are summable too.

**Example 3.1.3.**

(i) The definition of finPACs relaxes that of PACs (which have countable addition), see Definition 7.3.27. Hence every PAC is a finPAC.

(ii) Every biproduct category (Definition 7.1.17) is canonically enriched over commutative monoids. It is thus a finPAC with total addition $\ovee = +$.

The theory of finPACs is much the same as that of PACs studied by Arbib and Manes [7, 202]. For the sake of completeness, we elaborate the basic results here. A family of morphisms $(f_j \colon A \to B_j)_{j\in J}$ is **jointly monic** if $f_j \circ g = f_j \circ h$ for all $j \in J$ implies $g = h$ for each pair of morphisms $g, h \colon C \to A$.

**Lemma 3.1.4** (cf. [202, Theorem 3.2.18])**.** *The following hold in a finPAC.*

(i) *The morphisms* $\kappa_1 \circ \rhd_1, \kappa_2 \circ \rhd_2 \colon A + B \to A + B$ *are summable, and* $\kappa_1 \circ \rhd_1 \ovee \kappa_2 \circ \rhd_2 = \mathrm{id}_{A+B}$.

[1] Arbib and Manes [7, 202] call $\rhd_j$ *quasi projections.*

(ii) *Every morphism* $f\colon C \to A+B$ *can be decomposed as* $f = \kappa_1 \circ f_1 \ovee \kappa_2 \circ f_2$*, where* $f_1\colon C \to A$ *and* $f_2\colon C \to B$ *are is given by* $f_j = \rhd_j \circ f$*.*

(iii) *The partial projections* $\rhd_1\colon A+B \to A$ *and* $\rhd_2\colon A+B \to B$ *are jointly monic.*

(iv) *Morphisms* $f_1, f_2\colon A \to B$ *are summable if and only if they are compatible. In that case, for a (unique) morphism* $f\colon A \to B+B$ *with* $\rhd_j \circ f = f_j$*, one has* $f_1 \ovee f_2 = \nabla \circ f$*. Here* $\nabla = [\mathrm{id}, \mathrm{id}]\colon B+B \to B$ *is the codiagonal.*

The last point (iv) shows that the PCM structure of a finPAC is completely determined by its finite coproducts. Clearly the structure is determined without depending a choice of finite coproducts. Therefore if a category forms a finPAC, its structure of a finPAC is unique; see also [202, paragraph after Example 3.2.13]. In §3.8.1 we give a characterization of finPACs, which makes it explicit that finPACs can be defined as categories satisfying certain properties (rather than being equipped with structures).

*Proof.*

(i) Note that the sum $\kappa_1 \circ \rhd_1 \ovee \kappa_2 \circ \rhd_2$ is defined, since the maps are compatible via $\kappa_1 + \kappa_2\colon A+B \to (A+B)+(A+B)$. Then

$$\begin{aligned}(\kappa_1 \circ \rhd_1 \ovee \kappa_2 \circ \rhd_2) \circ \kappa_1 &= \kappa_1 \circ \rhd_1 \circ \kappa_1 \ovee \kappa_2 \circ \rhd_2 \circ \kappa_1 \\ &= \kappa_1 \circ \mathrm{id} \ovee \kappa_2 \circ 0 \\ &= \kappa_1\end{aligned}$$

and similarly $(\kappa_1 \circ \rhd_1 \ovee \kappa_2 \circ \rhd_2) \circ \kappa_2 = \kappa_2$. Hence $\kappa_1 \circ \rhd_1 \ovee \kappa_2 \circ \rhd_2 = \mathrm{id}_{A+B}$.

(ii) This follows from (i) as:

$$\begin{aligned}f = \mathrm{id} \circ f &= (\kappa_1 \circ \rhd_1 \ovee \kappa_2 \circ \rhd_2) \circ f \\ &= \kappa_1 \circ \rhd_1 \circ f \ovee \kappa_2 \circ \rhd_2 \circ f \\ &= \kappa_1 \circ f_1 \ovee \kappa_2 \circ f_2\,.\end{aligned}$$

(iii) Immediate by (ii).

(iv) The 'if' part is the compatible sum axiom. Conversely, if $f_1, f_2\colon A \to B$ are summable, then so are $\kappa_1 \circ f_1, \kappa_2 \circ f_2\colon A \to B+B$ by the untying axiom. Then $f_1$ and $f_2$ are compatible via $f = \kappa_1 \circ f_1 \ovee \kappa_2 \circ f_2$. Thus the maps $f_1, f_2$ are summable if and only if compatible. To verify the latter assertion, assume that $f_1, f_2$ are compatible. If $f\colon A \to B+B$ satisfies $\rhd_j \circ f = f_j$, then $f = \kappa_1 \circ f_1 \ovee \kappa_2 \circ f_2$ by (ii), and we have:

$$\begin{aligned}\nabla \circ f &= \nabla \circ (\kappa_1 \circ f_1 \ovee \kappa_2 \circ f_2) \\ &= \nabla \circ \kappa_1 \circ f_1 \ovee \nabla \circ \kappa_2 \circ f_2 \\ &= \mathrm{id} \circ f_1 \ovee \mathrm{id} \circ f_2 \\ &= f_1 \ovee f_2\,.\end{aligned}$$

∎

The compatible sum and untying axiom mention only pairs of morphisms and hence binary addition. In fact, the $n$-ary versions of the axioms are derivable. We write $[n] = \{1, \dots, n\}$ for the $n$ element set, and $n \cdot A = \coprod_{i \in [n]} A$ for the $n$-fold coproduct of $A$.

**Lemma 3.1.5.** *In a finPAC, the following hold.*

(i) *If a family of morphisms $(f_j \colon A \to B)_{j \in [n]}$ is compatible, then $(f_j)_{j \in [n]}$ is summable.*

(ii) *If a family $(f_j \colon A \to B)_{j \in [n]}$ is summable, then a family $(\kappa_j \circ f_j \colon A \to n \cdot B)_{j \in [n]}$ is summable too.*

*Proof.* (i) We prove the following stronger statement by induction on $n$.

- *If a family $(f_j \colon A \to B)_{j \in [n]}$ is compatible via $f \colon A \to n \cdot B$, then it is summable and $\bigovee_{j \in [n]} f_j = \nabla \circ f$.*

The case $n = 1$ is trivial. Suppose that $(f_j \colon A \to B)_{j \in [n+1]}$ is compatible via $f \colon A \to (n+1) \cdot B$. Then $(f_j)_{j \in [n]}$ is compatible via

$$A \xrightarrow{f} (n+1) \cdot B \xrightarrow{\alpha} n \cdot B + B \xrightarrow{\rhd_1} n \cdot B\,,$$

where $\alpha$ is the associativity isomorphism. By the induction hypothesis, $(f_j)_{j \in [n]}$ is summable and $\bigovee_{j \in [n]} f_j = \nabla \circ \rhd_1 \circ \alpha \circ f$. We claim that $\bigovee_{j \in [n]} f_j$ and $f_{n+1}$ are compatible via

$$A \xrightarrow{f} (n+1) \cdot B \xrightarrow{\alpha} n \cdot B + B \xrightarrow{\nabla + \mathrm{id}} B + B\,.$$

Indeed we have

$$\begin{aligned} \rhd_1 \circ (\nabla + \mathrm{id}) \circ \alpha \circ f &= \nabla \circ \rhd_1 \circ \alpha \circ f = \bigovee_{j \in [n]} f_j \\ \rhd_2 \circ (\nabla + \mathrm{id}) \circ \alpha \circ f &= \rhd_2 \circ \alpha \circ f = \rhd_{n+1} \circ f = f_{n+1}\,. \end{aligned}$$

Hence $\bigovee_{j \in [n]} f_j$ and $f_{n+1}$ are summable, so that the family $(f_j)_{j \in [n+1]}$ is summable. Then

$$\bigovee_{j \in [n+1]} f_j = \Big( \bigovee_{j \in [n]} f_j \Big) \ovee f_{n+1} \stackrel{*}{=} \nabla \circ (\nabla + \mathrm{id}) \circ \alpha \circ f = \nabla \circ f\,.$$

The marked equality $\stackrel{*}{=}$ holds by Lemma 3.1.4(iv).

(ii) By induction on $n$. The case $n = 1$ is trivial. Let $(f_j \colon A \to B)_{j \in [n+1]}$ be a summable family. Then the $n$ morphisms $f_1 \ovee f_{n+1}, f_2, \dots, f_n$ are summable. By the induction hypothesis, the $n$ morphisms

$$\kappa_1 \circ (f_1 \ovee f_{n+1}), \kappa_2 \circ f_2, \dots, \kappa_n \circ f_n \colon A \to n \cdot B$$

are summable. Since $\kappa_1 \circ (f_1 \ovee f_{n+1}) = \kappa_1 \circ f_1 \ovee \kappa_1 \circ f_{n+1}$, it follows that $\bigovee_{j \in [n]} \kappa_j \circ f_j$ and $\kappa_1 \circ f_{n+1}$ are summable. By the untying axiom, $\kappa_1 \circ \bigovee_{j \in [n]} \kappa_j \circ f_j = \bigovee_{j \in [n]} \kappa_1 \circ \kappa_j \circ f_j$ and $\kappa_2 \circ \kappa_1 \circ f_{n+1}$ are summable, which implies that the $n+1$ morphisms

$$\kappa_1 \circ \kappa_1 \circ f_1, \dots, \kappa_1 \circ \kappa_n \circ f_n, \kappa_2 \circ \kappa_1 \circ f_{n+1} \colon A \to n \cdot B + n \cdot B$$

are summable too. Via the associativity $n \cdot B + n \cdot B \cong 2n \cdot B$, the morphisms $\kappa_1 \circ f_1, \ldots, \kappa_{n+1} \circ f_{n+1} \colon A \to 2n \cdot B$ are summable. By postcomposing the obvious 'projection' map $2n \cdot B \to (n+1) \cdot B$, we see that the family $(\kappa_j \circ f_j \colon A \to (n+1) \cdot B)_{j \in [n+1]}$ is summable. ■

Clearly this implies that the compatible sum and untying axiom hold for families $(f_j)_{j \in J}$ indexed by any finite sets $J$. Reasoning as in Lemma 3.1.4, we obtain the following results.

**Lemma 3.1.6.** *Let* $\mathbf{C}$ *be a finPAC. Let* $\coprod_{j \in J} B_j$ *be a finite coproduct.*

(i) *The family* $(\kappa_j \circ \rhd_j \colon \coprod_j B_j \to \coprod_j B_j)_{j \in J}$ *is summable, and* $\bigovee_j \kappa_j \circ \rhd_j = \mathrm{id}$.

(ii) *Every morphism* $f \colon A \to \coprod_{j \in J} B_j$ *can be decomposed as* $f = \bigovee_j \kappa_j \circ f_j$, *where* $f_j \colon A \to B_j$ *is given by* $f_j = \rhd_j \circ f$.

(iii) *The family* $(\rhd_j \colon \coprod_j B_j \to B_j)_{j \in J}$ *of partial projections is jointly monic.*

(iv) *A finite family* $(f_j \colon A \to B)_{j \in J}$ *of morphisms is summable if and only if it is compatible. In the case, there is a unique morphism* $f \colon A \to \coprod_j B$ *such that* $\rhd_j \circ f = f_j$ *for all* $j \in J$, *and one has* $\bigovee_j f_j = \nabla \circ f$. *Here* $\nabla \colon \coprod_j B \to B$ *denotes the codiagonal.* ■

Given the fact that the partial projections $\rhd_j$ are jointly monic (Lemma 3.1.6(iii)), we introduce the following 'partial tuple' notation.

**Definition 3.1.7.** Let $(f_j \colon A \to B_j)_{j \in J}$ be a finite family of morphisms in a finPAC. Then we write $\langle\!\langle f_j \rangle\!\rangle_j \colon A \to \coprod_j B_j$ for a morphism such that $\rhd_j \circ \langle\!\langle f_j \rangle\!\rangle_j = f_j$ for all $j \in J$. The morphism $\langle\!\langle f_j \rangle\!\rangle_j$ may not exist, but if it does, then it is uniquely determined by the joint monicity of $\rhd_j$. We call $\langle\!\langle f_j \rangle\!\rangle_j$ the **partial tuple** of $(f_j)_j$. By definition, the partial tuple $\langle\!\langle f_j \rangle\!\rangle_j$ exists if and only if the family $(f_j)_j$ is compatible.

Specifically for the binary case, a partial tuple $\langle\!\langle f, g \rangle\!\rangle \colon A \to B + C$ of $f \colon A \to B$ and $g \colon A \to C$ is, *if it exists*, the dashed map below making the diagram commute.

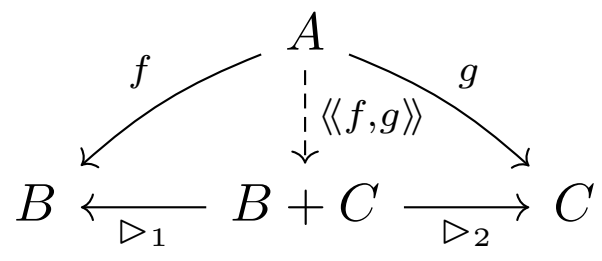

Thus a coproduct in a finPAC behaves 'partially' like a biproduct. In particular, if the finPAC is a biproduct category, the partial tuple $\langle\!\langle f, g \rangle\!\rangle$ is the usual tuple $\langle f, g \rangle \colon A \to B \oplus C$ induced by the universality of products.

The partial tuple notation allows us to rephrase Lemma 3.1.6(iv) more concisely.

**Proposition 3.1.8.** *In a finPAC, a finite family* $(f_j \colon A \to B)_j$ *is summable if and only if* $\langle\!\langle f_j \rangle\!\rangle_j$ *is defined. In that case we have*

$$\bigovee_j f_j = \nabla \circ \langle\!\langle f_j \rangle\!\rangle_j \,. \qquad ■$$

We can also see the partial tuple $\langle\!\langle f_j \rangle\!\rangle_j$ as a shorthand for the sum $\bigovee_j \kappa_j \circ f_j$.

**Proposition 3.1.9.** *Let $(f_j\colon A \to B_j)_{j\in J}$ be a finite family of morphisms in a finPAC. Then*

$$\langle\!\langle f_j \rangle\!\rangle_j = \bigovee_j \kappa_j \circ f_j \,,$$

*where $\langle\!\langle f_j \rangle\!\rangle_j$ is defined if and only if $\bigovee_j \kappa_j \circ f_j$ is defined.*

*Proof.* If $\langle\!\langle f_j \rangle\!\rangle_j$ is defined, then the morphisms $\kappa_j \circ f_j \colon A \to \coprod_j B_j$ are compatible via

$$A \xrightarrow{\langle\!\langle f_j \rangle\!\rangle_j} \coprod_j B_j \xrightarrow{\coprod_j \kappa_j} \coprod_j \coprod_k B_k \,.$$

Therefore the sum $\bigovee_j \kappa_j \circ f_j$ is defined. Conversely, suppose that $\bigovee_j \kappa_j \circ f_j$ is defined. Then one has $\rhd_i \circ (\bigovee_j \kappa_j \circ f_j) = f_i$ for all $i \in J$. By definition, $\langle\!\langle f_j \rangle\!\rangle_j$ exists and $\langle\!\langle f_j \rangle\!\rangle_j = \bigovee_j \kappa_j \circ f_j$. ∎

Finally we show that coproducts in a finPAC are always 'enriched over PCMs' in some appropriate sense (see e.g. [170, § 3.8]).

**Lemma 3.1.10.** *The category* **PCM** *has (small) products, given by cartesian products $\prod_j M_j$ of underlying sets equipped with operations defined pointwise.*

*Proof.* See [146, Proposition 5]. ∎

**Lemma 3.1.11.** *Let* $\mathbf{C}$ *be a finPAC. Let $\coprod_j A_j$ be a coproduct of (possibly infinitely many) objects $A_j$ in $\mathbf{C}$. Then the coproduct $\coprod_j A_j$ is 'enriched over PCMs' in the sense that the canonical bijections*

$$\mathbf{C}\Big(\coprod_j A_j, B\Big) \cong \prod_j \mathbf{C}(A_j, B)\,, \quad f \longmapsto (f \circ \kappa_j)_j \,. \tag{3.1}$$

*are isomorphisms of PCMs, where we interpret the right-hand side as the product of PCMs by Lemma* 3.1.10.

*Proof.* For each $j$, the coprojection $\kappa_j \colon A_j \to \coprod_j A_j$ induces a PCM morphism $\kappa_j^* = (-) \circ \kappa_j \colon \mathbf{C}(\coprod_j A_j, B) \to \mathbf{C}(A_j, B)$, and hence we obtain a PCM morphism

$$\mathbf{C}\Big(\coprod_j A_j, B\Big) \xrightarrow{\langle \kappa_j^* \rangle_j} \prod_j \mathbf{C}(A_j, B) \tag{3.2}$$

by the universality of the product in **PCM**. Clearly the underlying function of (3.2) coincides with the canonical bijection (3.1) for the coproduct $\coprod_j A_j$. We will prove that the map (3.2) is a PCM isomorphism. Since the underlying function is bijective, it suffices to prove that the map reflects summability. Let $f, g \in \mathbf{C}(\coprod_j A_j, B)$ be morphisms such that $(\kappa_j^*(f))_j$ and $(\kappa_j^*(g))_j$ are summable in $\prod_j \mathbf{C}(A_j, B)$, i.e. $\kappa_j^*(f)$ and $\kappa_j^*(g)$ are summable for all $j$. Then there exist tuples $\langle\!\langle \kappa_j^*(f), \kappa_j^*(g) \rangle\!\rangle \colon A_j \to B + B$ for all $j$, and thus we have $[\langle\!\langle \kappa_j^*(f), \kappa_j^*(g) \rangle\!\rangle]_j \colon \coprod_j A_j \to B + B$. Now

$$\rhd_1 \circ [\langle\!\langle \kappa_j^*(f), \kappa_j^*(g) \rangle\!\rangle]_j = [\rhd_1 \circ \langle\!\langle \kappa_j^*(f), \kappa_j^*(g) \rangle\!\rangle]_j = [\kappa_j^*(f)]_j = [f \circ \kappa_j]_j = f$$

and similarly $\rhd_2 \circ [\langle\!\langle \kappa_j^*(f), \kappa_j^*(g) \rangle\!\rangle]_j = g$. Therefore $f$ and $g$ are compatible, and hence summable. ∎

## 3.2 Effectuses

We now give the definition of effectus and the related terminology.

**Definition 3.2.1.** An **effectus** is a finPAC $\mathbf{C}$ with a distinguished 'unit' object $I \in \mathbf{C}$ satisfying the following conditions.

(E1) For each $A \in \mathbf{C}$, the hom-PCM $\mathbf{C}(A, I)$ is an effect algebra. We write $\mathbb{1}_A$ and $\mathbb{0}_A = 0_{AI}$ for the top and bottom in $\mathbf{C}(A, I)$.

(E2) $\mathbb{1}_B \circ f = \mathbb{0}_A$ implies $f = 0_{AB}$ for all $f\colon A \to B$.

(E3) $\mathbb{1}_B \circ f \perp \mathbb{1}_B \circ g$ implies $f \perp g$ for all $f, g\colon A \to B$.

In an effectus, a morphism of the form $p\colon A \to I$ is called a **predicate** on $A \in \mathbf{C}$. We write $\mathrm{Pred}(A) = \mathbf{C}(A, I)$ for the set of predicates. By definition, predicates $\mathrm{Pred}(A)$ form an effect algebra, in which the top predicate $\mathbb{1}_A$ is called the **truth** and the bottom $\mathbb{0}_A$ is called the **falsity**.

A morphism $f\colon A \to B$ is said to be **total** if $\mathbb{1}_B \circ f = \mathbb{1}_A$. The total morphisms form a (wide) subcategory of $\mathbf{C}$, for which we write $\mathrm{Tot}(\mathbf{C}) \hookrightarrow \mathbf{C}$. As usual, the subscripts of $\mathbb{1}_A$ and $\mathbb{0}_A$ may be dropped when it is clear from the context. For $f\colon A \to B$ the predicate $\mathbb{1}_B \circ f \in \mathrm{Pred}(A)$, called the **domain predicate** of $f$, is often simply written as $\mathbb{1}f$.

A **state** on $A$ is a *total* morphism of the form $\omega\colon I \to A$, i.e. a morphism with $\mathbb{1} \circ \omega = \mathbb{1}$, while a **substate** is an arbitrary morphism $\omega\colon I \to A$. We write $\mathrm{St}(A) = \mathrm{Tot}(\mathbf{C})(I, A)$ and $\mathrm{St}_{\leq}(A) = \mathbf{C}(I, A)$ for the set of states and substates, respectively. Given a predicate $p\colon A \to I$ and a (sub)state $\omega\colon I \to A$, the **validity** of $p$ in $\omega$ is defined by composition:

$$\omega \vDash p \;=\; \bigl(I \xrightarrow{\omega} A \xrightarrow{p} I\bigr)\,.$$

The formula may be seen as an *abstract Born rule*, giving the 'probability' that the predicate $p$ holds true in state $\omega$. The endomorphisms on $I$ are called **scalars** and viewed as abstract probabilities. We write $\mathcal{S} = \mathbf{C}(I, I)$ ($= \mathrm{Pred}(I) = \mathrm{St}_{\leq}(I)$) for the set of scalars.

Let us explain some intuition. Objects in an effectus are understood as *types* of systems. Then morphisms $f\colon A \to B$ are viewed as *processes* from a system of type $A$ to a system of type $B$. We primarily interpret 'systems' and 'processes' as physical ones, but sometimes we interpret them in the context of computation: objects/types as *data types* and morphisms/processes as *computations* or *programs*. From an operational perspective, one can understand a morphism $f\colon A \to B$ as an operation on a system of type $A$, which leaves the system in type $B$.

More specifically, in an effectus, morphisms in general represent *partial* processes, which may or may not occur (happen, or succeed). On the other hand, total morphisms represent 'complete' processes, which occur for sure. The difference is clearer in the context of computation: morphisms in general represent possibly non-terminating computation, while total morphisms represent terminating computation. A more precise operational meaning of 'partial processes' can be given via the notion of *tests* from the operational probabilistic framework [33, 35, 61, 248], which we will investigate in Chapter 6. We will see concrete examples of effectuses in Section 3.3 below, which will hopefully give sufficient intuition for now.

The object $I$ is the trivial type that represents a system having no information. In other words, $I$ is the type of ‘no system’. Thus a state $\omega\colon I \to A$ is a total process with no input, i.e. a process that ‘prepares’ a system of type $A$. A predicate $p\colon A \to I$ is a partial process with no output, discarding the system. A predicate can be viewed as a ‘yes-no’ observation/measurement — an *effect* in the terminology of Ludwig [197, 198] — by interpreting the occurrence of the process as answer ‘yes’.

**Remark 3.2.2.** It is often the case that the unit $I$ of an effectus $\mathbf{C}$ is a *monoidal unit*, that is, there is a *monoidal structure* $(\otimes, I)$ [199, Chapter XI] on $\mathbf{C}$ where $I$ coincides with the unit of the effectus. The monoidal structure allows us to compose systems $A$ and $B$ into $A \otimes B$, and also to compose processes *in parallel* as $f_1 \otimes f_2\colon A_1 \otimes A_2 \to B_1 \otimes B_2$. The monoidal unit $I$ satisfies $A \otimes I \cong A \cong I \otimes A$, which formally expresses the idea that $I$ is the type of ‘no system’. In this thesis, however, we do not deal with monoidal structure on an effectus. In other words, our focus here is on sequential composition $\circ$ with sum types $A + B$, rather than parallel composition $\otimes$. We note that an extension of effectuses with monoidal structure has been defined as *monoidal effectuses* in [40, § 10].

Let $f\colon A \to B$ be a morphism. Then pre-composition $f^*(q) = q \circ f$ yields a mapping $f^*\colon \mathrm{Pred}(B) \to \mathrm{Pred}(A)$ between effect algebras. This defines a predicate functor.

**Proposition 3.2.3.** *The mappings* $A \mapsto \mathrm{Pred}(A) = \mathbf{C}(A, I)$ *and* $f \mapsto f^*$ *defines a (contravariant) functor* $\mathrm{Pred}\colon \mathbf{C}^{\mathrm{op}} \to \mathbf{EA}_{\leq}$. *Moreover the functor restricts to* $\mathrm{Pred}\colon \mathrm{Tot}(\mathbf{C})^{\mathrm{op}} \to \mathbf{EA}$ *as follows.*

$$\begin{array}{ccc} \mathbf{C}^{\mathrm{op}} & \xrightarrow{\mathrm{Pred}} & \mathbf{EA}_{\leq} \\ \uparrow & & \uparrow \\ \mathrm{Tot}(\mathbf{C})^{\mathrm{op}} & \xrightarrow{\mathrm{Pred}} & \mathbf{EA} \end{array}$$

*Proof.* For each $f\colon A \to B$, the mapping $f^* = (-) \circ f\colon \mathrm{Pred}(B) \to \mathrm{Pred}(A)$ is a PCM morphism and hence a subunital morphism, since $\mathbf{C}$ is enriched over PCMs. Clearly $\mathrm{Pred}\colon \mathbf{C}^{\mathrm{op}} \to \mathbf{EA}_{\leq}$ is a functor as the underlying mapping is a hom-functor, If $f$ is total, $f^*(\mathbb{1}) = \mathbb{1} \circ f = \mathbb{1}$, i.e. $f^*$ is unital. Thus we obtain $\mathrm{Pred}\colon \mathrm{Tot}(\mathbf{C})^{\mathrm{op}} \to \mathbf{EA}$. ■

Since every effect algebra is partially ordered and every subunital morphism is monotone, we have a composed functor

$$\mathbf{C}^{\mathrm{op}} \xrightarrow{\mathrm{Pred}} \mathbf{EA}_{\leq} \hookrightarrow \mathbf{Poset}\,.$$

This is an instance of *indexed categories*, which are a fundamental structure in categorical logic [133]. The mappings $f^*\colon \mathrm{Pred}(B) \to \mathrm{Pred}(A)$ are called *reindexing maps* in general; *substitution maps* in the context of logic; or **predicate transformers** in program semantics. Here we prefer the latter terminology, in order to emphasize a duality between predicates and (sub)states.

Dually to predicate transformers, a morphism $f\colon A \to B$ induces a **substate transformer** $f_*\colon \mathrm{St}_{\leq}(A) \to \mathrm{St}_{\leq}(B)$ via post-composition $f_*(\omega) = f \circ \omega$. If $f$ is total, the map $f_*\colon \mathrm{St}_{\leq}(A) \to \mathrm{St}_{\leq}(B)$ restricts to a **state transformer** $f_*\colon \mathrm{St}(A) \to \mathrm{St}(B)$, since total morphisms are closed under composition. At this point we can already see them as functors $\mathrm{St}_{\leq}\colon \mathbf{C} \to \mathbf{Set}$ and $\mathrm{St}\colon \mathrm{Tot}(\mathbf{C}) \to \mathbf{Set}$, but later we observe that there are suitable structures on (sub)states.

The following are easy consequences from the definition of an effectus.

**Lemma 3.2.4.** *In an effectus $(\mathbf{C}, I)$, the following hold.*

(i) *For each $f\colon A \to B$, one has $f = 0_{AB}$ if and only if $\mathbb{1}_B \circ f = \mathbb{0}_A$.*

(ii) *For each finite family $(f_j\colon A \to B)_j$ of morphisms, $(f_j)_j$ is summable if and only if $(\mathbb{1}f_j)$ is summable. In that case,*

$$\mathbb{1} \circ \textstyle\bigovee_j f_j = \bigovee_j \mathbb{1} f_j\,.$$

(iii) *Every split mono is total. In particular, every isomorphism is total.*

(iv) *Every coprojection $\kappa_j\colon A_j \to \coprod_j A_j$ is total (in fact, split monic).*

(v) $\mathbb{1}_I = \mathrm{id}_I\colon I \to I$.

*Proof.*

(i) This is immediate by definition.

(ii) This clearly holds if the family consists of two morphisms. The general case can be shown by induction.

(iii) If $g \circ f = \mathrm{id}$, then $\mathbb{1} \circ f \geq (\mathbb{1} \circ g) \circ f = \mathbb{1} \circ \mathrm{id} = \mathbb{1}$. Hence $\mathbb{1} \circ f = \mathbb{1}$.

(iv) A coprojection is split monic as $\rhd_i \circ \kappa_i = \mathrm{id}$, hence total by the previous point.

(v) Note that $\mathbb{1}_I \circ \mathrm{id}_I \perp \mathbb{1}_I \circ \mathrm{id}_I^{\perp}$ and $\mathbb{1}_I \circ \mathrm{id}_I = \mathbb{1}_I$. Then $\mathbb{1}_I \circ \mathrm{id}_I^{\perp} = \mathbb{0}_I$, so that $\mathrm{id}_I^{\perp} = 0_{II} = \mathbb{0}_I$. Hence $\mathrm{id}_I = \mathbb{1}_I$. ∎

Recall that in a finPAC, morphisms $f\colon A \longrightarrow \coprod_j B_j$ can be decomposed as $f = \langle\!\langle f_j \rangle\!\rangle_j$. In an effectus, we can refine the decomposition property as follows.

**Lemma 3.2.5.** *Let $\coprod_j B_j$ be a finite coproduct in an effectus. We have the following bijective correspondence.*

$$\frac{\text{a morphism } f\colon A \longrightarrow \coprod_j B_j}{\text{a family } (f_j\colon A \longrightarrow B_j)_j \text{ with } (\mathbb{1}f_j)_j \text{ summable}}$$

*They are related via $f_j = \rhd_j \circ f$ and $f = \langle\!\langle f_j \rangle\!\rangle_j$. Moreover one has $\mathbb{1}f = \bigovee_j \mathbb{1}f_j$. In particular, $f$ is total if and only if $\bigovee_j \mathbb{1}f_j = \mathbb{1}$.*

*Proof.* Let $f\colon A \to \coprod_j B_j$ be given. Let $f_j = \rhd_j \circ f$. Since the family $(f_j)_j$ is compatible via $\langle\!\langle f_j \rangle\!\rangle_j = f$, it follows that the family $(\kappa_j \circ f_j)_j$ is summable, see Proposition 3.1.9. Then $(\mathbb{1}f_j)_j$ is summable because $\mathbb{1} \circ f_j = \mathbb{1} \circ \kappa_j \circ f_j$. Conversely, let $(f_j\colon A \to B_j)_j$ be a family with $(\mathbb{1}f_j)_j$ summable. Then again by $\mathbb{1} \circ f_j = \mathbb{1} \circ \kappa_j \circ f_j$ the family $(\mathbb{1} \circ \kappa_j \circ f_j)_j$ is summable, and hence so is $(\kappa_j \circ f_j)_j$. Then the partial tuple $\langle\!\langle f_j \rangle\!\rangle_j = \bigovee_j \kappa_j \circ f_j$ is defined. Clearly the correspondence is bijective. Finally,

$$\mathbb{1} \circ f = \mathbb{1} \circ \langle\!\langle f_j \rangle\!\rangle_j = \textstyle\bigovee_j \mathbb{1} \circ \kappa_j \circ f_j = \bigovee_j \mathbb{1} \circ f_j\,.$$ ∎

We note that an effectus is always **Poset**-enriched, i.e. the homsets are partially ordered and the composition is monotone.

**Lemma 3.2.6.** *Let* $\mathbf{C}$ *be an effectus. For each* $A, B$*, the hom-PCM* $\mathbf{C}(A, B)$ *satisfies the following properties.*

(i) *(Positivity)* $f \ovee g = 0$ *implies* $f = g = 0$.

(ii) *(Zero-cancellativity)* $f \ovee g = f$ *implies* $g = 0$.

*Proof.*

(i) If $f \ovee g = 0$ we have

$$\mathbb{1} \circ f \ovee \mathbb{1} \circ g = \mathbb{1} \circ (f \ovee g) = \mathbb{0}\,.$$

By positivity of the effect algebra $\mathrm{Pred}(A)$, we obtain $\mathbb{1} \circ f = \mathbb{1} \circ g = \mathbb{0}$. Therefore $f = g = 0$.

(ii) Assume $f \ovee g = f$. Then

$$\mathbb{1} \circ f \ovee \mathbb{1} \circ g = \mathbb{1} \circ (f \ovee g) = \mathbb{1} \circ f\,.$$

We obtain $\mathbb{1} \circ g = \mathbb{0}$ by cancellation in $\mathrm{Pred}(A)$. Hence $g = 0$. ■

**Proposition 3.2.7.** *Let* $\mathbf{C}$ *be an effectus. For each* $A, B$*, the homset* $\mathbf{C}(A, B)$ *is partially ordered via the algebraic order:*

$$f \leq g \iff \exists h.\, f \ovee h = g\,.$$

*Therefore* $\mathbf{C}$ *is* **Poset***-enriched.*

*Proof.* It is clear that $\leq$ is a preorder. We prove that it is antisymmetric. Assume $f \leq g$ and $g \leq f$. Then there exist $h$ and $k$ such that $f \ovee h = g$ and $g \ovee k = f$. Then

$$f = g \ovee k = f \ovee h \ovee k\,.$$

By zero-cancellation and positivity, $h = k = 0$. Therefore $f = g$. Note that the composition is monotone since it respects $\ovee$. ■

## 3.3 Examples of effectuses

In this section we describe our leading examples of effectuses. In particular, the following three categories are our primary examples.

(i) **Pfn**, the category of sets and partial functions, as a model of deterministic processes.

(ii) $\mathcal{K}\ell(\mathcal{D}_{\leq})$, the Kleisli category of the (finite) subdistribution monad $\mathcal{D}_{\leq}$, as a model of probabilistic processes.

(iii) $\mathbf{Wstar}_{\leq}^{\mathrm{op}}$, the opposite of the category of $W^*$-algebras and subunital normal CP maps, as a model of quantum processes.

We will also describe a few variants of the above examples, such as the Kleisli category $\mathcal{K}\ell(\mathcal{G}_{\leq})$ of the subprobability Giry monad, and the opposite $\mathbf{Cstar}_{\leq}^{\mathrm{op}}$ of the category of $C^*$-algebras.

For each example, we show that it forms an effectus, and describe states, predicates, and validities there.

### 3.3.1 Deterministic example

Let us start with a simple example of an effectus for deterministic processes, namely the category **Pfn** of sets and partial functions. We will write $f\colon X \rightharpoonup Y$ for a partial function from $X$ to $Y$, and $\mathrm{Dom}(f) \subseteq X$ for its domain of definition. The category **Pfn** has coproducts given by disjoint sums $\coprod_j X_j$, with the obvious total functions $\kappa_j\colon X_j \to \coprod_j X_j$ as coprojections, just as in **Set**. (This is because the category **Pfn** is isomorphic to the Kleisli category of the lift monad $1+(-)$ on **Set**, and thus Lemma 2.1.5 can be applied.) In particular, the initial object is the empty set. The sum $f \ovee g$ of partial functions $f, g\colon X \rightharpoonup Y$ is defined by

$$f \perp g \iff \mathrm{Dom}(f) \cap \mathrm{Dom}(g) = \varnothing$$
$$(f \ovee g)(x) = \begin{cases} f(x) & \text{if } x \in \mathrm{Dom}(f) \\ g(x) & \text{if } x \in \mathrm{Dom}(g) \end{cases} \tag{3.3}$$

It is straightforward to see that **Pfn** forms a finPAC with this definition of sum. One might be tempted to relax the definition of summability to $f(x) = g(x)$ for all $x \in \mathrm{Dom}(f) \cap \mathrm{Dom}(g)$. However, the relaxed definition of sum does not satisfy the untying axiom, as noted in [7].

We take the singleton 1 as the unit of **Pfn**. Thus predicates are partial functions $p\colon X \rightharpoonup 1$ and identified with subsets $P \subseteq X$ via

$$x \in P \iff p(x) \text{ is defined}\,.$$

Therefore the set of predicates on $X$ is the powerset of $X$: $\mathrm{Pred}(X) \cong \mathcal{P}(X)$, which is a Boolean algebra and hence an effect algebra. The 'truth' map $\mathbb{1}_X\colon X \rightharpoonup 1$ is the unique total function $X \to 1$, corresponding to the greatest element $X \in \mathcal{P}(X)$.

For each $f\colon X \rightharpoonup Y$, the domain predicate $\mathbb{1}f \in \mathrm{Pred}(X)$ is identified with the domain of definition $\mathrm{Dom}(f) \in \mathcal{P}(X)$. Therefore $\mathbb{1}f = \mathbb{0}$ implies that $f$ is a nowhere defined function, i.e. $f = 0_{XY}$. Since $\mathrm{Pred}(X) \cong \mathcal{P}(X)$ is a 'Boolean' effect algebra, see Example 2.3.3(i), $\mathbb{1}f \perp \mathbb{1}g$ means $\mathrm{Dom}(f) \cap \mathrm{Dom}(g) = \varnothing$, which is precisely the definition of $f \perp g$ above. This shows that **Pfn** is indeed an effectus.

As the domain predicate $\mathbb{1}f$ is the domain of definition, a map $f\colon X \rightharpoonup Y$ in **Pfn** is total if and only if it is a total function. Therefore $\mathrm{Tot}(\mathbf{Pfn}) = \mathbf{Set}$. In particular, states in **Pfn** are (total) functions $\omega\colon 1 \to X$ and are identified with elements $\omega \in X$, i.e. $\mathrm{St}(X) \cong X$. The only non-total substate is the nowhere defined function, i.e. the zero morphism $0_{1X}\colon 1 \rightharpoonup X$. The scalars are Boolean truth values: $\mathbf{Pfn}(1,1) \cong \mathcal{P}(1) \cong \{0,1\}$. Thus, for a state $\omega \in X$ and a predicate $P \in \mathcal{P}(X)$, the validity $\omega \vDash P$ corresponds to the membership $\omega \in P$.

By Proposition 3.2.7, **Pfn** is enriched over posets. For partial functions $f, g\colon X \rightharpoonup Y$, one has $f \le g$ if and only if $\mathrm{Dom}(f) \subseteq \mathrm{Dom}(g)$ and $f$ and $g$ agree on $\mathrm{Dom}(f) \subseteq X$. This is a standard ordering for **Pfn**, which is often used in program semantics, since the ordering is $\omega$-complete.

### 3.3.2 Probabilistic examples

There are examples of effectuses for (classical) probabilistic processes. The simplest one among them is the Kleisli category $\mathcal{K}\ell(\mathcal{D}_{\le})$ of the (finite) subdistribution monad $\mathcal{D}_{\leq}\colon \mathbf{Set} \to \mathbf{Set}$, see Definition 2.5.1. Explicitly, the objects in $\mathcal{K}\ell(\mathcal{D}_{\leq})$ are sets $(X, Y, Z, \dots)$, and the morphisms in $\mathcal{K}\ell(\mathcal{D}_{\leq})$ are functions of the form $f\colon X \to \mathcal{D}_{\leq}(Y)$, where $\mathcal{D}_{\leq}(Y)$ is the set of finite discrete subprobability distributions on $Y$. The composite of two functions $f\colon X \to \mathcal{D}_{\leq}(Y)$ and $g\colon Y \to \mathcal{D}_{\leq}(Z)$, denoted as $g \odot f\colon X \to \mathcal{D}_{\leq}(Z)$, is defined by summing over $Y$:

$$(g \odot f)(x)(z) = \sum_{y \in Y} g(y)(z) \cdot f(x)(y)\,.$$

The identity map $\eta_X\colon X \to \mathcal{D}_{\leq}(X)$ in $\mathcal{K}\ell(\mathcal{D}_{\leq})$ is given by $\eta_X(x)(x) = 1$ and $\eta_X(x)(x') = 0$ for every $x' \neq x$, that is, $\eta_X(x)$ is the Dirac distribution (a.k.a. point-mass) at $x$.

By Lemma 2.1.5, the Kleisli category $\mathcal{K}\ell(\mathcal{D}_{\leq})$ inherits all coproducts from the base category $\mathbf{Set}$. Specifically, coproducts $\coprod_j X_j$ in $\mathcal{K}\ell(\mathcal{D}_{\leq})$ are coproducts of sets, i.e. disjoint sums. The coprojections $\dot{\kappa}_j\colon \coprod_j X_j \to \mathcal{D}_{\leq}(X_j)$ are given by $\dot{\kappa}_j = \eta \circ \kappa_j$ where $\eta\colon X_j \to \mathcal{D}_{\leq}(X_j)$ is the unit of the monad. The sum of maps $f, g\colon X \to \mathcal{D}_{\leq}(Y)$ is given pointwise:

$$(f \ovee g)(x)(y) = f(x)(y) + g(x)(y)\,,$$

where $f$ and $g$ are summable iff $\sum_{y \in Y} f(x)(y) + g(x)(y) \leq 1$ for all $x \in X$. It is straightforward to verify the axioms for finPACs.

The unit object is the singleton 1. Then predicates are functions $p\colon X \to \mathcal{D}_{\leq}(1)$. Since $\mathcal{D}_{\leq}(1) \cong [0,1]$, predicates are identified with $[0,1]$-valued functions $p\colon X \to [0,1]$, that is, *fuzzy subsets* of $X$. We already saw that $\mathrm{Pred}(X) \cong [0,1]^X$ forms an effect algebra in Example 2.3.3(iv). The truth predicate $\mathbb{1}\colon X \to [0,1]$ is the constant function $\mathbb{1}(x) = 1$.

To see that (E2) holds, suppose that $f\colon X \to \mathcal{D}_{\leq}(Y)$ satisfies $\mathbb{1} \odot f = \mathbb{0}$. This means that $\sum_{y \in Y} f(x)(y) = 0$, so that $f(x)(y) = 0$ for all $y \in Y$, whence $f$ is the zero morphism. To see (E3), let $f, g\colon X \to \mathcal{D}_{\leq}(Y)$ be such that $\mathbb{1} \odot f \perp \mathbb{1} \odot g$. Then $\sum_{y \in Y} f(x)(y) + \sum_{y \in Y} g(x)(y) \leq 1$ for all $x \in X$, which means that $f$ and $g$ are summable. Therefore $\mathcal{K}\ell(\mathcal{D}_{\leq})$ is an effectus.

Note that $f\colon X \to \mathcal{D}_{\leq}(Y)$ is total iff $\sum_{y \in Y} f(x)(y) = 1$, i.e. $f(x)$ is a 'proper' probability distribution, for each $x \in X$. Therefore total morphisms in $\mathcal{K}\ell(\mathcal{D}_{\leq})$ are functions $f\colon X \to \mathcal{D}(Y)$, where $\mathcal{D}$ is the distribution monad, and the total subcategory $\mathrm{Tot}(\mathcal{K}\ell(\mathcal{D}_{\leq}))$ equals the Kleisli category $\mathcal{K}\ell(\mathcal{D})$. In particular, states are functions $\omega\colon 1 \to \mathcal{D}(X)$, which can be identified with elements $\omega \in \mathcal{D}(X)$. Hence states in $\mathcal{K}\ell(\mathcal{D}_{\leq})$ are precisely distributions: $\mathrm{St}(X) \cong \mathcal{D}(X)$. Similarly substates are subdistributions: $\mathrm{St}_{\leq}(X) \cong \mathcal{D}_{\leq}(X)$.

Given a state $\omega \in \mathcal{D}(X)$ and a predicate $p\colon X \to [0,1]$, the validity is given as

$$\omega \vDash p \;\;=\;\; \sum_{x \in X} p(x)\omega(x)\,,$$

where we identify scalars $1 \to \mathcal{D}_{\leq}(1) \cong [0,1]$ with probabilities $[0,1]$. The validity $\omega \vDash p$ can be understood as the probability that the predicate $p$ holds in the state/distribution $\omega$.

The canonical partial ordering of morphisms (Proposition 3.2.7) in $\mathcal{K}\ell(\mathcal{D}_{\leq})$ are the obvious pointwise one: for $f\colon X \to \mathcal{D}_{\leq}(Y)$, one has $f \leq g$ if and only if $f(x)(y) \leq g(x)(y)$ for all $x \in X$ and $y \in Y$.

It is instructive to describe states and predicates in $\mathcal{K}\ell(\mathcal{D}_{\leq})$ in the usual language of probability theory and statistics. Let $\omega \in \mathcal{D}(X)$ be a state on a set $X$. We introduce an $X$-valued (discrete) random variable $\mathbf{x}$ that takes a value $x \in X$ with probability $\omega(x)$, that is,

$$\mathrm{P}(\mathbf{x} = x) = \omega(x)\,.$$

For a predicate $p \in [0,1]^X$, we introduce a corresponding event $\mathbf{p}$ that occurs with probability $p(x)$ *given* $\mathbf{x} = x$, that is,

$$\mathrm{P}(\mathbf{p} \mid \mathbf{x} = x) = p(x)\,.$$

In this setting, the validity $\omega \vDash p$ is the probability of the event $\mathbf{p}$, which is verified using the standard probability calculation rules:

$$\begin{aligned}
\mathrm{P}(\mathbf{p}) &= \sum_{x\in X} \mathrm{P}(\mathbf{p}, \mathbf{x} = x) && \text{by marginalization} \\
&= \sum_{x\in X} \mathrm{P}(\mathbf{p} \mid \mathbf{x} = x)\,\mathrm{P}(\mathbf{x} = x) && \text{by the product rule} \\
&= \sum_{x\in X} p(x)\omega(x) \quad \equiv \quad \omega \vDash p\,.
\end{aligned}$$

*Note.* In the above discussion using random variables and events, the underlying probability space $(\Omega, \mu)$ (often called the 'sample space') is made implicit, as is often the case in probability theory and statistics. Precisely speaking, assuming the probability space $(\Omega, \mu)$, an $X$-valued random variable means a measurable function $\mathbf{x} \colon \Omega \to X$. Then the probability that $\mathbf{x}$ takes a value $x \in X$ is defined as $\mathrm{P}(\mathbf{x} = x) \coloneqq \mu(\mathbf{x}^{-1}(\{x\}))$. Similarly, an event $\mathbf{p}$ means a measurable subset $\mathbf{p} \subseteq \Omega$, for which one writes $\mathrm{P}(\mathbf{p}) \coloneqq \mu(\mathbf{p})$.

One can study topics in probability theory and statistics using effectus-theoretic notions such as states and predicates. For example, Jacobs and Zanasi explained Bayesian reasoning in terms of state and predicate transformers [152]. In this thesis we do not pursue this direction, but refer the interested reader to [38, 39, 145, 153, 154].

### Variants of $\mathcal{K}\ell(\mathcal{D}_{\leq})$

The discussions above remains valid if we consider the infinite subdistribution monad $\mathcal{D}^{\infty}_{\leq}$ (Definition 2.5.2) instead of the finite one $\mathcal{D}_{\leq}$. The effectus $\mathcal{K}\ell(\mathcal{D}^{\infty}_{\leq})$ has the same predicates as $\mathcal{K}\ell(\mathcal{D}_{\leq})$, fuzzy subsets $[0,1]^X$. States in $\mathcal{K}\ell(\mathcal{D}^{\infty}_{\leq})$ are infinite discrete probability distributions, i.e. members of $\mathcal{D}^{\infty}(X)$. Since the effectus $\mathcal{K}\ell(\mathcal{D}^{\infty}_{\leq})$ is very much the same as $\mathcal{K}\ell(\mathcal{D}_{\leq})$, in the rest of this thesis, we will mention $\mathcal{K}\ell(\mathcal{D}^{\infty}_{\leq})$ only when there is an interesting difference.

There is a more sophisticated version for measure-theoretic probability — namely, the Kleisli category $\mathcal{K}\ell(\mathcal{G}_{\leq})$ of the subprobability Giry monad $\mathcal{G}_{\leq} \colon \mathbf{Meas} \to \mathbf{Meas}$ on the category $\mathbf{Meas}$ of measurable spaces and measurable functions. See Definition 2.5.4 for details of the monad $\mathcal{G}_{\leq}$. Let us describe the Kleisli category $\mathcal{K}\ell(\mathcal{G}_{\leq})$ concretely. Objects in $\mathcal{K}\ell(\mathcal{G}_{\leq})$ are measurable spaces, and morphisms in $\mathcal{K}\ell(\mathcal{G}_{\leq})$ are measurable functions $f\colon X \to \mathcal{G}_{\leq}(Y)$. Here $\mathcal{G}_{\leq}(Y)$ is the set of subprobability measures on $Y$ equipped with the smallest $\sigma$-algebra such that for each $B \in \Sigma_Y$ the map $\mathcal{G}_{\leq}(Y) \to [0,1], \mu \mapsto \mu(B)$ is measurable. The composite $g \odot f\colon X \to \mathcal{G}_{\leq}(Z)$ of $f\colon X \to \mathcal{G}_{\leq}(Y)$ and $g\colon Y \to \mathcal{G}_{\leq}(Z)$ is defined via integration:

$$(g \odot f)(x)(C) = \int_Y g(y)(C)\, f(x)(\mathrm{d}y)$$

for $x \in X$ and $C \in \Sigma_Z$. The identity map $\eta_X\colon X \to \mathcal{G}_{\leq}(X)$ sends $x \in X$ to the Dirac measure $\eta_X(x) = \delta_x$.

The singleton 1 is the unit in $\mathcal{K}\ell(\mathcal{G}_{\leq})$. We have $\mathcal{G}_{\leq}(1) \cong [0,1]$. Hence predicates are measurable functions $p\colon X \to [0,1]$, which are called *fuzzy events* in [263]. A morphism $f\colon X \to \mathcal{G}_{\leq}(Y)$ in $\mathcal{K}\ell(\mathcal{G}_{\leq})$ is total if and only if $f(x)$ is a probability measure for all $x \in X$. We write $\mathcal{G}(Y) \subseteq \mathcal{G}_{\leq}(Y)$ for the set of probability measures on $Y$. Then $\mathcal{G}$ is the (ordinary) Giry monad, and we have $\mathrm{Tot}(\mathcal{K}\ell(\mathcal{G}_{\leq})) = \mathcal{K}\ell(\mathcal{G})$. States in $\mathcal{K}\ell(\mathcal{G}_{\leq})$ are probability measures $\omega \in \mathcal{G}(X)$. The validity $\omega \models p$ is given by integration:

$$\omega \models p \;=\; \int_X p\, \mathrm{d}\omega\,.$$

Validity defined by integration in this way has appeared in the study of probabilistic programs [180].

### 3.3.3 Quantum examples

The opposite $\mathbf{Wstar}_{\leq}^{\mathrm{op}}$ of the category of $W^*$-algebras and normal subunital CP maps is our archetypal example of an effectus, modelling quantum processes. The opposite category is appropriate since $W^*$-algebras are considered as algebras of observables, which are dual to state spaces. The sum $f \ovee g$ of maps $f, g\colon \mathscr{A} \to \mathscr{B}$ is defined by

$$\begin{aligned} f \perp g \iff f(1) + g(1) \leq 1 \\ (f \ovee g)(x) = f(x) + g(x)\,. \end{aligned}$$

The obvious zero maps $0\colon \mathscr{A} \to \mathscr{B}$ are the neutral elements for addition. The trivial algebra $\{0\}$ is a final object in $\mathbf{Wstar}_{\leq}$, and the *direct sum* $\mathscr{A} \oplus \mathscr{B}$ of $W^*$-algebras $\mathscr{A}$ and $\mathscr{B}$ is a product in $\mathbf{Wstar}_{\leq}$. The direct sum is defined as follows. The underlying set of $\mathscr{A} \oplus \mathscr{B}$ is the cartesian product of $\mathscr{A}$ and $\mathscr{B}$. The operations are defined pointwise, e.g.

$$(x_1, y_1) + (x_2, y_2) = (x_1 + x_2, y_1 + y_2)\,.$$

The norm is given by

$$\|(x,y)\| = \sup\{\|x\|, \|y\|\} = \max\{\|x\|, \|y\|\}\,.$$

It indeed forms a $W^*$-algebra and a product in $\mathbf{Wstar}_{\leq}$; for more details, see [37]. Therefore the opposite $\mathbf{Wstar}_{\leq}^{\mathrm{op}}$ has finite coproducts. Morphisms $f, g\colon \mathscr{B} \to \mathscr{A}$ in $\mathbf{Wstar}_{\leq}^{\mathrm{op}}$ are compatible if and only if there is a map $h\colon \mathscr{B} \to \mathscr{A} \oplus \mathscr{A}$ in $\mathbf{Wstar}_{\leq}^{\mathrm{op}}$ such that $h(x,y) = f(x) + g(y)$. Thus compatibility implies $f(1) + g(1) \leq 1$, i.e. that $f, g$ are summable. It is easy to verify the untying axiom, showing that $\mathbf{Wstar}_{\leq}^{\mathrm{op}}$ is a finPAC.

The unit object of $\mathbf{Wstar}_{\leq}^{\mathrm{op}}$ is the algebra $\mathbb{C}$ of complex numbers. Predicates on $\mathscr{A}$ — morphisms $p\colon \mathscr{A} \to \mathbb{C}$ in $\mathbf{Wstar}_{\leq}^{\mathrm{op}}$ — are normal (completely) positive maps $p\colon \mathbb{C} \to \mathscr{A}$. Since a linear map from $\mathbb{C}$ is determined by its value on 1, predicates are in bijective correspondence with elements $x \in \mathscr{A}$ such that $0 \leq x \leq 1$, that is, *effects* in $\mathscr{A}$. As we saw in Example 2.3.3(vi), the predicates $\mathrm{Pred}(\mathscr{A}) \cong [0,1]_{\mathscr{A}}$ form an effect algebra. In particular, the truth map $\mathbb{1}\colon \mathscr{A} \to \mathbb{C}$ in $\mathbf{Wstar}_{\leq}^{\mathrm{op}}$ corresponds to the unit $1 \in \mathscr{A}$.

Let us verify the axioms (E2) and (E3) of effectuses from Definition 3.2.1. Suppose that $f\colon \mathscr{A} \to \mathscr{B}$ in $\mathbf{Wstar}_{\leq}^{\mathrm{op}}$ satisfies $\mathbb{1} \circ f = \mathbb{0}$. Then $f$ is a normal subunital CP map $f\colon \mathscr{B} \to \mathscr{A}$ such that $f(1) = 0$. We have $f(x) = 0$ for any self-adjoint $x \in \mathscr{B}$ since $-\|x\| \cdot 1 \leq x \leq \|x\| \cdot 1$ (Corollary 2.6.10). This implies $f(x) = 0$ for arbitrary $x \in \mathscr{B}$ as $x$ can be written as a linear combination of self-adjoint elements. Thus $f = 0$. Next assume that $f, g\colon \mathscr{A} \to \mathscr{B}$ in $\mathbf{Wstar}_{\leq}^{\mathrm{op}}$ satisfy $\mathbb{1} \circ f \perp \mathbb{1} \circ g$. By definition of the partially additive structure, it follows that $f(1) + g(1) \leq 1$ and hence $f \perp g$. Therefore $\mathbf{Wstar}_{\leq}^{\mathrm{op}}$ is an effectus.

In the effectus $\mathbf{Wstar}_{\leq}^{\mathrm{op}}$, total morphisms in $\mathbf{Wstar}_{\leq}^{\mathrm{op}}$ are precisely unital ones, so that $\mathrm{Tot}(\mathbf{Wstar}_{\leq}^{\mathrm{op}}) = \mathbf{Wstar}^{\mathrm{op}}$. States in $\mathbf{Wstar}_{\leq}^{\mathrm{op}}$ are thus normal unital (completely) positive maps $\omega\colon \mathscr{A} \to \mathbb{C}$, which are exactly normal states on $\mathscr{A}$, see Definition 2.6.17. Therefore for a Hilbert space $\mathscr{H}$, states on $\mathcal{B}(\mathscr{H})$ are density operators on $\mathscr{H}$. Substates are normal subunital positive maps $\omega\colon \mathscr{A} \to \mathbb{C}$.

Given a state $\omega \in \mathrm{St}(\mathscr{A})$ and a predicate $p \in [0,1]_{\mathscr{A}} \cong \mathrm{Pred}(\mathscr{A})$, the validity is given by

$$\omega \vDash p \quad = \quad \omega(p) \in [0,1]\,.$$

Here scalars $\mathbf{Wstar}_{\leq}^{\mathrm{op}}(\mathbb{C}, \mathbb{C})$ are identified with probabilities in the unit interval $[0,1] \subseteq \mathbb{R}$. This agrees with the Born rule in quantum theory, calculating the probability of a certain observation in a quantum measurement. In particular, if $\mathscr{A} = \mathcal{B}(\mathscr{H})$, (normal) states $\omega\colon \mathcal{B}(\mathscr{H}) \to \mathbb{C}$ are of the form $\omega(A) = \mathrm{tr}(A\rho)$ for some density operator $\rho$ on $\mathscr{H}$. Given an effect $P$ on $\mathscr{H}$ the validity is

$$\mathrm{tr}(-\rho) \vDash P \quad = \quad \mathrm{tr}(P\rho)\,,$$

which is a more common form of the Born rule.

A variant of the effectus $\mathbf{Wstar}_{\leq}^{\mathrm{op}}$ is the opposite $\mathbf{Cstar}_{\leq}^{\mathrm{op}}$ of the category of $C^*$-algebras. The effectus $\mathbf{Cstar}_{\leq}^{\mathrm{op}}$ is more general than $\mathbf{Wstar}_{\leq}^{\mathrm{op}}$ in the sense that $\mathbf{Wstar}_{\leq}^{\mathrm{op}}$ is a subcategory of $\mathbf{Cstar}_{\leq}^{\mathrm{op}}$. States in $\mathbf{Cstar}_{\leq}^{\mathrm{op}}$ are (not necessarily normal) states on a $C^*$-algebra $\mathscr{A}$, and predicates in $\mathbf{Cstar}_{\leq}^{\mathrm{op}}$ are identified with effects $[0,1]_{\mathscr{A}}$. We take $\mathbf{Wstar}_{\leq}^{\mathrm{op}}$, rather than $\mathbf{Cstar}_{\leq}^{\mathrm{op}}$, as our main example, because extra properties of $W^*$-algebras allow us to show that the effectus $\mathbf{Wstar}_{\leq}^{\mathrm{op}}$ admits extra structures and properties, such as *images*, *comprehension*, and *quotients*, which will be discussed in Chapter 5.

Note that $\mathbf{Wstar}_{\leq}^{\mathrm{op}}$ is a *non-full* subcategory $\mathbf{Cstar}_{\leq}^{\mathrm{op}}$, since morphisms in $\mathbf{Wstar}_{\leq}^{\mathrm{op}}$ are required to be normal. The full subcategory of $\mathbf{Cstar}_{\leq}^{\mathrm{op}}$ consisting of $W^*$-algebras, whose morphisms are arbitrary subunital CP maps, forms an effectus too. However, proofs that $\mathbf{Wstar}_{\leq}^{\mathrm{op}}$ admits extra structures use normality of morphisms and hence are not valid when we consider arbitrary subunital CP maps.

Further details about the category of $W^*$-algebras can be found in [253, 254, 256].

**Remark 3.3.1.** The partial orders on the homsets in $\mathbf{Wstar}_{\le}^{\mathrm{op}}$ (or $\mathbf{Cstar}_{\le}^{\mathrm{op}}$), given by Proposition 3.2.7, are stronger than what one might expect. Indeed, it is easy to see that two subunital (normal) CP maps $f, g\colon \mathscr{A} \to \mathscr{B}$ satisfy $f \le g$ in $\mathbf{Wstar}_{\le}^{\mathrm{op}}$ (or $\mathbf{Cstar}_{\le}^{\mathrm{op}}$) if and only if $g - f$ is CP. By the definition of CP (Definition 2.6.21), this means for any $n \in \mathbb{N}$ the map

$$\mathcal{M}_n \otimes (g - f) = \mathcal{M}_n \otimes g - \mathcal{M}_n \otimes f \quad : \mathcal{M}_n \otimes \mathscr{A} \longrightarrow \mathcal{M}_n \otimes \mathscr{B}$$

is positive. It is equivalent to saying that $(\mathcal{M}_n \otimes f)(x) \le (\mathcal{M}_n \otimes g)(x)$ for all positive $x \in (\mathcal{M}_n \otimes \mathscr{A})_+$.

The ordering $\le$ is strictly stronger than the 'pointwise' ordering $f \le' g$ defined as $f(a) \le g(a)$ for all $a \in \mathscr{A}_+$. To show this, we give an example from [261]. We consider endomaps on the $n \times n$-matrix algebra $\mathcal{M}_n$, and thus the example works both for $\mathbf{Wstar}_{\le}^{\mathrm{op}}$ and $\mathbf{Cstar}_{\le}^{\mathrm{op}}$. Write $(|i\rangle)_{i\in[n]}$ for the standard basis of $\mathbb{C}^n$ and define $f\colon \mathcal{M}_n \to \mathcal{M}_n$ by

$$f(A) = \sum_{i,j\in[n]} S_{ij} A S_{ij}^* \qquad \text{where} \quad S_{ij} = \frac{1}{\sqrt{2}}(|i\rangle\langle j| - |j\rangle\langle i|)\,.$$

The map $f$ is CP because it is of the *Kraus form*, see [120, § 4.2.3]. Then

$$\begin{aligned}
f(A) &= \sum_{i,j} S_{ij} A S_{ij}^* \\
&= \frac{1}{2}\sum_{i,j}\bigl(|i\rangle\langle j| - |j\rangle\langle i|\bigr)A\bigl(|j\rangle\langle i| - |i\rangle\langle j|\bigr) \\
&= \frac{1}{2}\sum_{i,j}\Bigl(|i\rangle\langle j|A|j\rangle\langle i| - |i\rangle\langle j|A|i\rangle\langle j| - |j\rangle\langle i|A|j\rangle\langle i| + |j\rangle\langle i|A|i\rangle\langle j|\Bigr) \\
&= \sum_{i,j}|i\rangle\langle j|A|j\rangle\langle i| - \sum_{i,j}|i\rangle\langle j|A|i\rangle\langle j| \\
&= \mathrm{tr}(A)\mathcal{I} - A^T\,.
\end{aligned}$$

We define $g(A) = \mathrm{tr}(A)\mathcal{I}$ and $h(A) = A^T$, so that $f + h = g$. Here $g$ is clearly CP. It is well known (see e.g. [120, Example 4.3]) that the transpose $h(A) = A^T$ is positive but not CP. Thus, $\frac{1}{n}f$ and $\frac{1}{n}g$ are subunital CP maps such that $\frac{1}{n}f \le' \frac{1}{n}g$ but not $\frac{1}{n}f \le \frac{1}{n}g$, since $\frac{1}{n}g - \frac{1}{n}f = \frac{1}{n}h$ is positive but not CP.

The strong order $\le$ is reasonably well-behaved. For example, with the order $\le$ the homset $\mathbf{Wstar}_{\le}(\mathscr{A}, \mathscr{B})$ is directed complete, so that the category $\mathbf{Wstar}_{\le}$ is enriched over dcpos; see [37].

## 3.4 Structure of predicates: effect modules

Since scalars are predicates, i.e. $\mathcal{S} = \mathbf{C}(I, I) = \mathrm{Pred}(I)$, they form an effect algebra. At the same time there is another (total) monoid structure on $\mathbf{C}(I, I)$ given by composition $\circ$ of morphisms. Therefore we immediately obtain the following result.

**Proposition 3.4.1.** *Scalars $\mathcal{S} = \mathbf{C}(I, I)$ form an effect monoid via composition $\circ$ as multiplication.*

*Proof.* The only nontrivial point, $\mathbb{1}_1 = \mathrm{id}_1$, is shown in Lemma 3.2.4(v). ■

The effect algebra $\mathrm{Pred}(A) = \mathbf{C}(A, I)$ of predicates carries an extra action $s \cdot p = s \circ p$ of scalars $s\colon I \to I$ via composition, like modules over a ring or vector spaces. We call the operation $(s, p) \mapsto s \cdot p$ **scalar multiplication**. The structure of predicates can be axiomatized as follows.

**Definition 3.4.2.** Let $M$ be an effect monoid. A **partial module**[2] over $M$ is a PCM $X$ equipped with an operation $\cdot\colon M \times X \to X$ satisfying:

(a) $\cdot$ is a monoid action, i.e. $(s \cdot t) \cdot x = s \cdot (t \cdot x)$ and $1 \cdot x = x$;

(b) $\cdot\colon M \times X \to X$ is a PCM bimorphism.

Let $X, Y$ be partial modules over $M$. A **module map** $f\colon X \to Y$ is a PCM morphism that preserves the scalar multiplication: $f(s \cdot x) = s \cdot f(x)$. Partial modules over $M$ and module maps form a category $M$-**PMod**.

**Definition 3.4.3.** An **effect module** over an effect monoid $M$ is a partial module over $M$ that is at the same time an effect algebra. Similarly to effect algebras (see Definition 2.3.8), we use two types of morphisms $f\colon E \to D$ between effect modules: **unital module maps**, i.e. module maps with $f(1) = 1$; and **subunital module maps**, i.e. module maps with $f(1) \le 1$. Since the latter condition is trivial, subunital module maps are merely module maps.

We write $M$-**EMod** and $M$-$\mathbf{EMod}_{\le}$ for the categories of effect modules over $M$ with unital and subunital module maps, respectively. We simply write $\mathbf{EMod} = [0, 1]$-**EMod** for the standard effect monoid $M = [0, 1]$ of probabilities.

**Remark 3.4.4.** Effect modules over $[0, 1]$ are also known as *convex effect algebra* [105, 109] (as they are indeed convex, see § 3.6.1). Our terminology 'effect module', introduced in [137, 147], is due to the following categorical fact. Recall from Remark 2.3.20 that effect monoids are monoids in the symmetric monoidal category **EA**. Similarly, an effect module over an effect monoid $M$ is identified with a *module over a monoid* $M$ in **EA**, that is, an object $E \in \mathbf{EA}$ with a morphism $M \otimes E \to E$ making certain diagrams commute. For the complete definition, see [199, § VII.4] (where it is called an *action of a monoid*).

With the notion of effect modules, we can more properly capture the structure of predicates.

**Proposition 3.4.5.** *For each $A \in \mathbf{C}$ predicates* $\mathrm{Pred}(A) = \mathbf{C}(A, I)$ *form an effect module over scalars $\mathcal{S} = \mathbf{C}(I, I)$, with scalar multiplication given by composition, i.e. $s \cdot p = s \circ p$. Moreover, the functors in Proposition* 3.2.3 *can be lifted to the categories of effect modules:*

$$\begin{array}{ccc} \mathbf{C}^{\mathrm{op}} & \xrightarrow{\mathrm{Pred}} & \mathcal{S}\text{-}\mathbf{EMod}_{\le} \\ \uparrow & & \uparrow \\ \mathrm{Tot}(\mathbf{C})^{\mathrm{op}} & \xrightarrow{\mathrm{Pred}} & \mathcal{S}\text{-}\mathbf{EMod} \end{array}$$

[2] Called a partial commutative module in [36, 139].

*Proof.* It is clear that Pred($A$) is an effect module over scalars. To prove the latter claim it suffices to show that reindexing maps $f^*\colon \mathrm{Pred}(B) \to \mathrm{Pred}(A)$ preserve scalars. Indeed we have $f^*(s \cdot q) = s \circ q \circ f = s \cdot f^*(q)$. ■

**Example 3.4.6.** We look at predicates and predicate transformers in effectuses, continuing Section 3.3.

(i) In **Pfn** the predicates on a set $X$ are subsets: $\mathrm{Pred}(X) \cong \mathcal{P}(X)$. The scalars in **Pfn** are Boolean values $2 = \{0, 1\}$. However, the scalar multiplication over 2 is trivial, and hence effect modules over 2 are merely effect algebras: 2-**EMod** $\cong$ **EA**. Therefore the predicate functor is simply $\mathrm{Pred}\colon \mathbf{Pfn} \to \mathbf{EA}_{\leq}$, which restricts to $\mathrm{Pred}\colon \mathbf{Set} \to \mathbf{EA}$. The predicate transformer $f^*\colon \mathcal{P}(Y) \to \mathcal{P}(X)$ for $f\colon X \rightharpoonup Y$ is given by, for $Q \subseteq Y$,

$$f^*(Q) = \{x \in X \mid f(x) \text{ is defined and } f(x) \in Q\}\,.$$

In particular, if $f$ is total then $f^*(Q)$ is the inverse image. Note that in this case it is clear that we have functors $\mathrm{Pred}\colon \mathbf{Pfn} \to \mathbf{BA}_{\leq}$ and $\mathrm{Pred}\colon \mathbf{Set} \to \mathbf{BA}$ into the categories of Boolean algebras. Later we will show that this holds generally for so-called Boolean effectuses, and in particular any extensive category with a final object, see Sections 6.5 and 6.6.

(ii) In the effectus $\mathcal{K}\ell(\mathcal{D}_{\leq})$, the predicates on a set $X$ are fuzzy subsets $p\colon X \to [0, 1]$. The fuzzy subsets $[0, 1]^X$ form an effect module over $[0, 1]$ via the pointwise scalar multiplication: $(r \cdot p)(x) = r \cdot p(x)$. For a morphism $f\colon X \to \mathcal{D}_{\leq}(Y)$, the predicate transformer $f^*\colon [0, 1]^Y \to [0, 1]^X$ is given by:

$$f^*(q)(x) = \sum_{y \in Y} q(y) \cdot f(x)(y)\,.$$

This yields functors $\mathrm{Pred}\colon \mathcal{K}\ell(\mathcal{D}_{\leq})^{\mathrm{op}} \to \mathbf{EMod}_{\leq}$ and $\mathrm{Pred}\colon \mathcal{K}\ell(\mathcal{D})^{\mathrm{op}} \to \mathbf{EMod}$. If $f$ is total, then $f(x)$ is a probability distribution on $Y$ and the value $f^*(q)(x)$ can be understood as the expected value of $q$ with respect to $f(x)$.

We can also explain predicate transformation $f^*(q)$ in the language of probability theory, following the discussion in § 3.3.2. For a predicate $q \in [0, 1]^Y$, we introduce a corresponding event $\mathbf{q}$ and a $Y$-valued random variable $\mathbf{y}$ such that

$$\mathrm{P}(\mathbf{q} \mid \mathbf{y} = y) = q(y)\,.$$

We also introduce an $X$-valued random variable $\mathbf{x}$ and view a total morphism $f\colon X \to \mathcal{D}(Y)$ as a conditional probability distribution:

$$\mathrm{P}(\mathbf{y} = y \mid \mathbf{x} = x) = f(x)(y)\,.$$

We assume that the event $\mathbf{q}$ is conditionally independent of $\mathbf{x}$ given $\mathbf{y}$, i.e.

$\mathrm{P}(\mathbf{q} \mid \mathbf{y} = y) = \mathrm{P}(\mathbf{q} \mid \mathbf{y} = y, \mathbf{x} = x)$. Then

$$\begin{aligned} f^*(q)(x) &= \sum_{y \in Y} q(y) \cdot f(x)(y) \\ &= \sum_{y \in Y} \mathrm{P}(\mathbf{q} \mid \mathbf{y} = y) \, \mathrm{P}(\mathbf{y} = y \mid \mathbf{x} = x) \\ &= \sum_{y \in Y} \mathrm{P}(\mathbf{q} \mid \mathbf{y} = y, \mathbf{x} = x) \, \mathrm{P}(\mathbf{y} = y \mid \mathbf{x} = x) \\ &= \sum_{y \in Y} \mathrm{P}(\mathbf{q}, \mathbf{y} = y \mid \mathbf{x} = x) \\ &= \mathrm{P}(\mathbf{q} \mid \mathbf{x} = x) \,, \end{aligned}$$

that is, $f^*(q)(x)$ is the probability of the event $\mathbf{q}$ given $\mathbf{x} = x$. The last two equalities hold by the usual probability calculation rules, assuming $\mathrm{P}(\mathbf{x} = x) \neq 0$.

Predicate transformers in $\mathcal{K}\ell(\mathcal{G}_{\leq})$ are similar. For a Kleisli map $f \colon X \to \mathcal{G}_{\leq}(Y)$, the predicate transformer $f \colon \mathrm{Pred}(Y) \to \mathrm{Pred}(X)$ is given by integration:

$$f^*(q)(x) = \int_Y q \, \mathrm{d} f(x) \,.$$

(iii) In the effectus $\mathbf{Wstar}_{\leq}^{\mathrm{op}}$ of $W^*$-algebras, the predicates on a $W^*$-algebra $\mathscr{A}$ are effects $[0,1]_{\mathscr{A}}$. The effect algebra $[0,1]_{\mathscr{A}}$ is an effect module over $[0,1]$ via the obvious scalar multiplication, i.e. the restriction of the scalar multiplication on $\mathscr{A}$ over $\mathbb{C}$. If $f \colon \mathscr{A} \to \mathscr{B}$ is a morphism in $\mathbf{Wstar}_{\leq}^{\mathrm{op}}$, the predicate transformer $f^* \colon [0,1]_{\mathscr{B}} \to [0,1]_{\mathscr{A}}$ is simply given by restricting the map $f \colon \mathscr{B} \to \mathscr{A}$, i.e. $f^*(a) = f(a)$. This defines a functor $\mathrm{Pred} \colon \mathbf{Wstar}_{\leq} \to \mathbf{EMod}_{\leq} \equiv [0,1]\text{-}\mathbf{EMod}_{\leq}$ and also a unital variant $\mathrm{Pred} \colon \mathbf{Wstar} \to \mathbf{EMod}$. Predicates and predicate transformers for $\mathbf{Cstar}_{\leq}^{\mathrm{op}}$ are basically the same.

Note that the effectuses $\mathbf{Wstar}_{\leq}^{\mathrm{op}}$, $\mathbf{Cstar}_{\leq}^{\mathrm{op}}$, $\mathcal{K}\ell(\mathcal{D}_{\leq})$, and $\mathcal{K}\ell(\mathcal{G}_{\leq})$ have the real unit interval $[0,1] \subseteq \mathbb{R}$ as scalars. These form an important class of effectuses, since there we can interpret scalars — hence validities $\omega \vDash p$ — as probabilities in the usual sense. Therefore we introduce the following terminology.

**Definition 3.4.7.** A **real effectus** is an effectus $(\mathbf{C}, I)$ such that the effect monoid $\mathcal{S} = \mathbf{C}(I, I)$ of scalars is isomorphic to $[0,1]$.

One may think that an isomorphism $\varphi \colon \mathbf{C}(I, I) \to [0,1]$ should be specified as a structure of a real effectus. This is unnecessary, however, since the isomorphism is unique by the following lemma.

**Lemma 3.4.8.** *The effect algebra $[0,1] \subseteq \mathbb{R}$ satisfies the following properties.*

(i) *If $\varphi \colon [0,1] \to [0,1]$ is a unital morphism of effect algebras, then $\varphi = \mathrm{id}_{[0,1]}$.*

(ii) *For each effect algebra $E$, if $\varphi, \psi \colon E \to [0,1]$ are isomorphisms of effect algebras, then $\varphi = \psi$.*

*Proof.*

(i) Let $\varphi\colon [0,1] \to [0,1]$ be a homomorphism of effect algebras. For each $n \in \mathbb{N}_{>0}$,

$$1 = \varphi(1) = \varphi(\tfrac{1}{n} + \cdots + \tfrac{1}{n}) = \varphi(\tfrac{1}{n}) + \cdots + \varphi(\tfrac{1}{n}) = n \cdot \varphi(\tfrac{1}{n}),$$

so that $\varphi(\frac{1}{n}) = \frac{1}{n}$. Then for each $m \in \{0, \ldots, n\}$,

$$\varphi(\tfrac{m}{n}) = \varphi(\tfrac{1}{n} + \cdots + \tfrac{1}{n}) = \varphi(\tfrac{1}{n}) + \cdots + \varphi(\tfrac{1}{n}) = \tfrac{1}{n} + \cdots + \tfrac{1}{n} = \tfrac{m}{n}.$$

Therefore $\varphi(q) = q$ for all $q \in [0,1] \cap \mathbb{Q}$. Let $r \in [0,1]$ be given. For any $\varepsilon > 0$, there are $q, q' \in [0,1] \cap \mathbb{Q}$ such that $q \leq r \leq q'$ and $q' - q < \varepsilon$. Since $\varphi$ is monotone, $q \leq \varphi(r) \leq q'$ and hence $|r - \varphi(r)| < \varepsilon$. Since $\varepsilon > 0$ is arbitrary, we conclude that $\varphi(r) = r$.

(ii) By the previous point, $\varphi \circ \psi^{-1} = \mathrm{id}_{[0,1]}$ and hence $\varphi = \psi$. ∎

We study the category $M$-$\mathbf{EMod}_{\leq}$ of effect modules over an effect monoid $M$. The (opposite) category turns out to form an effectus.

**Lemma 3.4.9.** *The category $M$-$\mathbf{EMod}_{\leq}$ has all small products and a zero object.*

*Proof.* Let $(E_j)_j$ be a (small) family of effect modules over $M$. It is straightforward to verify that the set theoretic product $\prod_j E_j$ with the operations given pointwise, i.e.

$$\begin{gathered}
(a_j)_j \perp (b_j)_j \iff \forall j.\ a_j \perp b_j \\
(a_j)_j \ovee (b_j)_j = (a_j \ovee b_j)_j \qquad s \cdot (a_j)_j = (s \cdot a_j)_j \\
0 = (0)_j \qquad\qquad 1 = (1)_j
\end{gathered}$$

forms a product in $M$-$\mathbf{EMod}_{\leq}$. In particular, the trivial effect module $1 = \{0\}$ is a final object. It is also initial: for any $E \in M$-$\mathbf{EMod}_{\leq}$ there is a unique subunital module map $\mathsf{i}\colon 1 \to E$ given by $\mathsf{i}(0) = 0$. ∎

The opposite category $M$-$\mathbf{EMod}_{\leq}^{\mathrm{op}}$ is a category with finite coproducts and zero morphisms. Note that $M$ itself can be seen as an effect module over $M$ in the obvious way. Taking $M \in M$-$\mathbf{EMod}_{\leq}^{\mathrm{op}}$ as the unit object, we define the 'truth' maps $\mathbb{1}_E\colon E \to M$ in $M$-$\mathbf{EMod}_{\leq}^{\mathrm{op}}$, i.e. a subunital module map $\mathbb{1}_E\colon M \to E$, by $\mathbb{1}_E(s) = s \cdot 1$.

**Proposition 3.4.10.** *Let $M$ be an effect monoid. The category $M$-$\mathbf{EMod}_{\leq}^{\mathrm{op}}$ is an effectus with $M$ as unit, and the truth maps $\mathbb{1}_E$ defined above. The total maps in $M$-$\mathbf{EMod}_{\leq}^{\mathrm{op}}$ are unital maps:* $\mathrm{Tot}(M\text{-}\mathbf{EMod}_{\leq}^{\mathrm{op}}) = M\text{-}\mathbf{EMod}^{\mathrm{op}}$.

Though it is not hard to prove this directly, we defer a proof to § 3.8.3, to give a more concise proof via a characterization of an effectus. Let us quickly describe the structure of the effectus $M$-$\mathbf{EMod}_{\leq}^{\mathrm{op}}$. For $f, g \in M\text{-}\mathbf{EMod}_{\leq}^{\mathrm{op}}(E, D) = M\text{-}\mathbf{EMod}_{\leq}(D, E)$ we define sum $f \ovee g$ pointwise, namely:

$$\begin{aligned}
f \perp g &\iff \forall a \in D.\, f(a) \perp g(a) \\
(f \ovee g)(a) &\coloneqq f(a) \ovee g(a)
\end{aligned}$$

**Proposition 3.4.11.** *Let $M$ be an effect monoid. Consider the effectus $M$-$\mathbf{EMod}_{\le}^{\mathrm{op}}$.*

(i) *For each $E \in M$-$\mathbf{EMod}_{\le}^{\mathrm{op}}$, the set of predicates $\mathrm{Pred}(E) \equiv M$-$\mathbf{EMod}_{\le}^{\mathrm{op}}(E, M)$ is isomorphic to $E$ as an effect algebra.*

(ii) *The set of scalars $\mathcal{S} \equiv M$-$\mathbf{EMod}_{\le}^{\mathrm{op}}(M, M)$ is isomorphic to $M$ as an effect monoid.*

(iii) *Via the isomorphism $\mathcal{S} \cong M$ of the previous point,* $\mathrm{Pred}(E)$ *can be seen as an effect module over $M$. Then* $\mathrm{Pred}(E)$ *is isomorphic to $E$ as an effect module over $M$.*

*Proof.*

(i) There is a bijection between $M$-$\mathbf{EMod}_{\le}^{\mathrm{op}}(E, M) = M$-$\mathbf{EMod}_{\le}(M, E)$ and $E$ sending $p\colon M \to E$ to $p(1) \in E$, with inverse taking $a \in E$ to $(-) \cdot a\colon M \to E$. This preserves the structure of an effect algebra.

(ii) We show that the mapping $M$-$\mathbf{EMod}_{\le}^{\mathrm{op}}(M, M) \to M, \varphi \mapsto \varphi(1)$ preserves the multiplication. Let $\varphi, \psi \in M$-$\mathbf{EMod}_{\le}^{\mathrm{op}}(M, M)$. Then

$$\begin{aligned}(\varphi \cdot \psi)(1) &= (\varphi \circ^{\mathrm{op}} \psi)(1)\\ &= (\psi \circ \varphi)(1)\\ &= \psi(\varphi(1))\\ &= \psi(\varphi(1) \cdot 1)\\ &= \varphi(1) \cdot \psi(1)\,.\end{aligned}$$

(iii) The isomorphism $\mathcal{S} \cong M$ induces an action of $M$ on $\mathrm{Pred}(E)$ as follows: for $s \in M$ and $p \in \mathrm{Pred}(E)$, one defines $s \cdot p \in \mathrm{Pred}(E)$ by

$$(s \cdot p)(t) \coloneqq p(t \cdot s) \qquad \text{for each} \quad t \in M\,.$$

Then, for all $s \in M$ and $p \in \mathrm{Pred}(E)$,

$$(s \cdot p)(1) = p(1 \cdot s) = p(s \cdot 1) = s \cdot p(1)\,.$$

Therefore the bijection $\mathrm{Pred}(E) \to E, p \mapsto p(1)$ preserves the scalar multiplication. ■

**Remark 3.4.12.** Since $2$-$\mathbf{EMod}_{\le} \cong \mathbf{EA}_{\le}$, the opposite $\mathbf{EA}_{\le}^{\mathrm{op}}$ of the category of effect algebras and subunital maps is an effectus. By Proposition 3.4.11(i), each effect algebra $E$ is isomorphic to the effect algebra $\mathrm{Pred}(E)$ of predicates on $E$ in $\mathbf{EA}_{\le}^{\mathrm{op}}$. Therefore every effect algebra may appear as the effect algebra of predicates in an effectus.

**Proposition 3.4.13.** *The predicate functor* $\mathrm{Pred}\colon \mathbf{C} \to \mathcal{S}$-$\mathbf{EMod}_{\le}^{\mathrm{op}}$ *preserves all small coproducts that exist in* $\mathbf{C}$*. This means that* Pred *sends a coproduct in* $\mathbf{C}$ *to a product in* $\mathcal{S}$-$\mathbf{EMod}_{\le}$*. Explicitly, if* $\coprod_j A_j$ *is a coproduct of objects* $(A_j)_j$ *in* $\mathbf{C}$*, the canonical morphism*

$$\mathrm{Pred}\Big(\coprod\nolimits_j A_j\Big) \xrightarrow[\cong]{\langle \kappa_j^* \rangle_j} \prod\nolimits_j \mathrm{Pred}(A_j) \;\; \textit{in } \mathcal{S}\text{-}\mathbf{EMod}_{\le} \tag{3.4}$$

*is an isomorphism, where* $\kappa_j^* \colon \mathrm{Pred}(\coprod_j A_j) \to \mathrm{Pred}(A_j)$ *is the predicate transformer for the* $j$*th coprojection* $\kappa_j \colon A_j \to \coprod_j A_j$*. The inverse is given by cotupling* $(p_j)_j \mapsto [p_j]_j$*.*

*Proof.* By Lemma 3.4.9, the product $\prod_j \mathrm{Pred}(A_j)$ indeed exists. By the universality of the coproduct $\coprod_j A_j$, there is a canonical bijection between sets:

$$\mathbf{C}\Big(\coprod_j A_j, I\Big) \cong \prod_j \mathbf{C}(A_j, I)\,.$$

It is easy to see that this bijection coincides with the map (3.4). Moreover the bijection is a PCM isomorphism by Lemma 3.1.11. Therefore, the morphism (3.4) is bijective and reflects summability, so that it is an isomorphism in $\mathcal{S}$-$\mathbf{EMod}_{\leq}$. ∎

Note that $\mathrm{Pred}(I) = \mathbf{C}(I, I) = \mathcal{S}$, where $\mathcal{S}$ is the unit of the effectus $\mathcal{S}$-$\mathbf{EMod}_{\leq}$. Thus the functor $\mathrm{Pred} \colon \mathbf{C} \to \mathcal{S}$-$\mathbf{EMod}_{\leq}$ preserves finite coproducts and the unit object, which basically says that Pred is a *morphism of effectuses*; see Section 4.2 for the precise definition of this term.

## 3.5 Structure of substates: weight modules

For each $A \in \mathbf{C}$, the set of substates $\mathrm{St}_{\leq}(A) = \mathbf{C}(I, A)$ is a PCM since an effectus is enriched over PCMs. The substates $\omega \colon I \to A$ also carry an action of scalars $s \colon I \to I$ via composition $s \cdot \omega = \omega \circ s$. However, $\mathrm{St}_{\leq}(A)$ is not quite a partial module over $\mathcal{S} = \mathbf{C}(I, I)$, because the action is defined by composition from the right, and hence $(s \cdot t) \cdot \omega = s \cdot (t \cdot \omega)$ may fail if the effect monoid $\mathcal{S}$ of scalars is noncommutative. A remedy is to use the *opposite* $M^{\mathrm{op}}$ of an effect monoid $M$. The effect monoid $M^{\mathrm{op}}$ has the same structure as $M$ except that the multiplication is given by $s \cdot^{\mathrm{op}} t = t \cdot s$. Then:

**Proposition 3.5.1.** *For each* $A \in \mathbf{C}$*, the substates* $\mathrm{St}_{\leq}(A) = \mathbf{C}(I, A)$ *form a partial module over* $\mathcal{S}^{\mathrm{op}}$*. The mappings* $A \mapsto \mathrm{St}_{\leq}(A)$ *and* $f \mapsto f_*$ *yield a functor* $\mathrm{St}_{\leq} \colon \mathbf{C} \to \mathcal{S}^{\mathrm{op}}$-$\mathbf{PMod}$*.*

*Proof.* To see that $\mathrm{St}_{\leq}(A)$ is a partial module over $\mathcal{S}^{\mathrm{op}}$, note that

$$(s \cdot^{\mathrm{op}} t) \cdot \omega = \omega \circ (s \cdot^{\mathrm{op}} t) = \omega \circ t \circ s = (t \cdot \omega) \circ s = s \cdot (t \cdot \omega)\,.$$

For $f \colon A \to B$ the state transformer $f_*(\omega) = f \circ \omega$ preserves the PCM structure, since $\mathbf{C}$ is enriched over PCMs, and preserves the scalar multiplication:

$$f_*(s \cdot \omega) = f \circ (s \cdot \omega) = f \circ \omega \circ s = f_*(\omega) \circ s = s \cdot f_*(\omega)\,.$$ ∎

In other words, $\mathrm{St}_{\leq}(A)$ is a partial *right* module over $\mathcal{S}$, if the partial modules in Definition 3.4.2 are called *left* ones. Clearly there is no such distinction if the scalars are commutative, i.e. $\mathcal{S}^{\mathrm{op}} = \mathcal{S}$.

**Remark 3.5.2.** It is quite common that an effectus has commutative scalars. Indeed, the scalars in our main examples of an effectus are commutative, because they are either $\{0, 1\}$ or $[0, 1]$. Moreover, it is well known (see e.g. [2, § 3.2]) that for any monoidal category $(\mathbf{A}, \otimes, I)$ the homset $\mathbf{A}(I, I)$ forms a commutative monoid. Thus whenever the unit $I$ of an effectus $\mathbf{C}$ is a monoidal unit of some monoidal structure on $\mathbf{C}$ (see Remark 3.2.2), the scalars are commutative.

To every substate $\omega\colon I \to A$ there is an associated scalar $|\omega| := \mathbb{1} \circ \omega$. We call $|\omega|$ the **weight** of $\omega$. This leads the following axiomatization of the structure of substates.

**Definition 3.5.3.** Let $M$ be an effect monoid. A **weight module** over $M$ is a partial module $X$ over $M$ equipped with a **weight** map $|-|\colon X \to M$ satisfying:

(a) $|-|\colon X \to M$ is a module map (over $M$);

(b) $|x| = 0$ implies $x = 0$;

(c) $|x| \perp |y|$ implies $x \perp y$.

Let $X, Y$ be weight modules over $M$. A module map $f\colon X \to Y$ is said to be **weight-preserving** if $|f(x)| = |x|$ for all $x \in X$, and **weight-decreasing** if $|f(x)| \leq |x|$ for all $x \in X$.

We denote by $M$-**WMod** (resp. $M$-$\mathbf{WMod}_{\leq}$) the category of weight modules over $M$ and weight-preserving (resp. weight-decreasing) module maps. We simply write $\mathbf{WMod} = [0,1]$-$\mathbf{WMod}$ for the effect monoid $M = [0,1]$.

It will turn out that weight modules are closely related with convex sets; see Sections 3.6 and 4.4. Weight modules (over $[0,1]$) are also closely related to *base-norm spaces*, see Section 7.2.

**Proposition 3.5.4.** *For each $A \in \mathbf{C}$ the substates $\mathrm{St}_{\leq}(A) = \mathbf{C}(A, I)$ form a weight module over $\mathcal{S}^{\mathrm{op}}$, with weight $|\omega| = \mathbb{1} \circ \omega$. Moreover the mappings $A \mapsto \mathrm{St}_{\leq}(A)$ and $f \mapsto f_*$ define a functor $\mathrm{St}_{\leq}\colon \mathbf{C} \to \mathcal{S}^{\mathrm{op}}\text{-}\mathbf{WMod}_{\leq}$, which restricts to subcategories as in the following diagram.*

$$\begin{array}{ccc} \mathbf{C} & \xrightarrow{\mathrm{St}_{\leq}} & \mathcal{S}^{\mathrm{op}}\text{-}\mathbf{WMod}_{\leq} \\ \uparrow & & \uparrow \\ \mathrm{Tot}(\mathbf{C}) & \xrightarrow{\mathrm{St}_{\leq}} & \mathcal{S}^{\mathrm{op}}\text{-}\mathbf{WMod} \end{array}$$

*Proof.* We already showed that $\mathrm{St}_{\leq}(A)$ is a partial module over $\mathcal{S}^{\mathrm{op}}$. The weight map $|-|\colon \mathrm{St}_{\leq}(A) \to \mathcal{S}^{\mathrm{op}}$ given by $|\omega| = \mathbb{1} \circ \omega$ is a module map; in particular, it preserves the scalar multiplication:

$$|s \cdot \omega| = \mathbb{1} \circ \omega \circ s = |\omega| \circ s = s \cdot^{\mathrm{op}} |\omega| \,.$$

Conditions (b) and (c) in Definition 3.5.3 follow immediately from the axioms of an effectus.

Let $f\colon A \to B$ be a morphism in $\mathbf{C}$. Then the state transformer $f_*\colon \mathrm{St}_{\leq}(A) \to \mathrm{St}_{\leq}(B)$ is a module map. It is moreover weight-decreasing:

$$|f_*(\omega)| = \mathbb{1}_B \circ f \circ \omega \leq \mathbb{1}_A \circ \omega = |\omega| \,,$$

so that we obtain a functor $\mathrm{St}_{\leq}\colon \mathbf{C} \to \mathcal{S}^{\mathrm{op}}\text{-}\mathbf{WMod}_{\leq}$. If $f$ is total, i.e. $\mathbb{1}_B \circ f = \mathbb{1}_A$, then the inequality above becomes equality, and thus $f$ is weight-preserving. Hence the functor $\mathrm{St}_{\leq}$ restricts to $\mathrm{Tot}(\mathbf{C}) \to \mathcal{S}^{\mathrm{op}}\text{-}\mathbf{WMod}$. ∎

**Example 3.5.5.** Continuing Section 3.3 we review substates in effectuses.

(i) Substates on a set $X$ in **Pfn** are partial functions $\omega\colon 1 \rightharpoonup X$. Clearly $\mathrm{St}_{\le}(X) \cong X + \{0\}$, i.e. a substate on $X$ is either an element $x \in X$ or the zero 0. It is a weight module over the Boolean values $2 = \{0, 1\}$, with weight $|x| = 1$ for $x \in X$ and $|0| = 0$. For a partial function $f\colon X \rightharpoonup Y$, the substate transformer $f_*\colon \mathrm{St}_{\le}(X) \to \mathrm{St}_{\le}(Y)$ is given by $f_*(x) = f(x)$ for $x \in X$ and $f_*(0) = 0$. This yields a functor $\mathrm{St}_{\le}\colon \mathbf{Pfn} \to 2\text{-}\mathbf{WMod}_{\le}$. Similarly to effect modules over 2, weight modules over 2 are degenerate and turn out to be equivalent to pointed sets, as shown in Proposition 3.5.6 below.

(ii) Substates in $\mathcal{K}\ell(\mathcal{D}_{\le})$ are subdistributions: $\mathrm{St}_{\le}(X) \cong \mathcal{D}_{\le}(X)$. The $[0, 1]$-module structure on substates $\omega\colon X \to [0, 1]$ are given pointwise. The weight is given by $|\omega| = \sum_{x \in X} \omega(x)$, i.e. the total mass of $\omega$. A Kleisli map $f\colon X \to \mathcal{D}_{\le}(Y)$ induces a substate transformer $f_*\colon \mathcal{D}_{\le}(X) \to \mathcal{D}_{\le}(Y)$ by

$$f_*(\omega)(y) = \sum_{x \in X} f(x)(y) \cdot \omega(x)\,.$$

This yields a functor $\mathrm{St}_{\le}\colon \mathcal{K}\ell(\mathcal{D}_{\le}) \to \mathbf{WMod}_{\le}$.

In the effectus $\mathcal{K}\ell(\mathcal{G}_{\le})$ for measure-theoretic probability, substates are subprobability measures: $\mathrm{St}_{\le}(X) \cong \mathcal{G}_{\le}(X)$. For $f\colon X \to \mathcal{G}_{\le}(Y)$, a substate transformer is now given by integration:

$$f_*(\omega)(B) = \int_X f(x)(B)\,\omega(\mathrm{d}x)\,,$$

for measurable subsets $B \in \Sigma_Y$. We then obtain $\mathrm{St}_{\le}\colon \mathcal{K}\ell(\mathcal{G}_{\le}) \to \mathbf{WMod}_{\le}$.

(iii) In the effectus $\mathbf{Wstar}_{\le}^{\mathrm{op}}$ of $W^*$-algebras, substates on $\mathscr{A}$ are normal subunital positive map $\omega\colon \mathscr{A} \to \mathbb{C}$. They form a weight module over $[0, 1]$ via $|\omega| = \omega(1)$ and the obvious pointwise operations. A morphism $f\colon \mathscr{A} \to \mathscr{B}$ in $\mathbf{Wstar}_{\le}^{\mathrm{op}}$, i.e. a normal unital CP map $f\colon \mathscr{B} \to \mathscr{A}$, defines a substate transformer $f_*\colon \mathrm{St}_{\le}(\mathscr{A}) \to \mathrm{St}_{\le}(\mathscr{B})$ by post-composition $f_*(\omega) = \omega \circ f$, yielding a functor $\mathrm{St}_{\le}\colon \mathbf{Wstar}_{\le}^{\mathrm{op}} \to \mathbf{WMod}_{\le}$. Substates and substate transformers in $\mathbf{Cstar}_{\le}^{\mathrm{op}}$ are the same, except that they need not be normal.

**Proposition 3.5.6.** *The category* $2\text{-}\mathbf{WMod}_{\le}$ *of weight module over* 2 *is isomorphic to the coslice category* $1/\mathbf{Set}$*, i.e. the category of pointed sets.*

*Proof.* Any weight module $X$ over 2 forms a pointed set $(X, 0)$. Conversely, if $(X, x_0)$ is a pointed set, define the PCM structure on $X$ by $x_0 \ovee x = x = x \ovee x_0$ where $x \perp y$ iff $x = x_0$ or $y = x_0$. With the zero element $x_0$, the trivial scalar multiplication over 2, and the weight given by $|x_0| = 0$ and $|x| = 1$ for $x \neq x_0$, the set $X$ forms a weight module over 2. It is straightforward to see that any weight module over 2 is defined in this way, so that the constructions above yields a bijective correspondence between weight modules over 2 and pointed sets. Moreover, a function $f\colon X \to Y$ between weight modules over 2 is a weight-decreasing module map if and only if $f(0) = 0$, i.e. $f$ is a morphism of pointed sets. We conclude that $2\text{-}\mathbf{WMod}_{\le} \cong 1/\mathbf{Set}$. ■

We study the category $M\text{-}\mathbf{WMod}_{\le}$ for an effect monoid $M$. For $X, Y \in M\text{-}\mathbf{WMod}_{\le}$ define

$$X + Y = \{(x, y) \in X \times Y \mid |x| \perp |y|\}$$

and a weight map $|-|\colon X+Y \to M$ by $|(x,y)| = |x| \ovee |y|$. We define operations on $X+Y$ as follows.

$$\begin{aligned}(x,y) \perp (x',y') &\iff |(x,y)| \perp |(x',y')| \\ (x,y) \ovee (x',y') &= (x \ovee x', y \ovee y') \\ s \cdot (x,y) &= (s\cdot x, s\cdot y) \\ 0 &= (0,0)\,.\end{aligned}$$

Moreover there are coprojections $\kappa_1\colon X \to X+Y$ and $\kappa_2\colon Y \to X+Y$ given by $\kappa_1(x) = (x,0)$ and $\kappa_2(y) = (0,y)$ respectively. Then:

**Lemma 3.5.7.** *$X+Y$ with the coprojections given above is a coproduct in $M$-$\mathbf{WMod}_{\le}$.*

*Proof.* We first verify that $X+Y$ with the weight map $|-|$ form a weight module over $M$.

*(The operations are well-defined.)* Let $(x,y),(x',y') \in X+Y$ be elements with $(x,y) \perp (x',y')$, i.e. $|(x,y)| \perp |(x',y')|$. Then $|x|,|y|,|x'|,|y'|$ are summable, and in particular we have $|x| \perp |x'|$ and $|y| \perp |y'|$, so that $x \perp x'$ and $y \perp y'$. We have $(x \ovee x', y \ovee y') \in X+Y$ since

$$|x| \ovee |y| \ovee |x'| \ovee |y'| = |x \ovee x'| \ovee |y \ovee y'|$$

is defined. For any $(x,y) \in X+Y$ and $s \in M$, we have $|s\cdot x| = s\cdot |x| \le |x|$ and similarly $|s\cdot y| \le |y|$. Thus $|s\cdot x| \perp |s \cdot y|$, and $(s\cdot x, s\cdot y) \in X+Y$. It is clear that $(0,0) \in X+Y$.

*(The operations satisfy the axioms.)* Before proving that $X+Y$ is a partial module, we first verify that the weight map $|-|\colon X+Y \to M$ satisfies conditions (a), (b) and (c) in Definition 3.5.3. Indeed, $(x,y) \perp (x',y') \iff |(x,y)| \perp |(x',y')|$ by definition, and

$$|(x,y) \ovee (x',y')| = |(x \ovee x', y \ovee y')| = |x| \ovee |y| \ovee |x'| \ovee |y'| = |(x,y)| \ovee |(x',y')|\,.$$

Clearly $|(0,0)| = 0$. Conversely, if $|(x,y)| = 0$, then $|x| \ovee |y| = 0$. By the cancellativity of an effect algebra, $|x| = |y| = 0$, so that $(x,y) = (0,0)$. It preserves the scalar multiplication:

$$\begin{aligned}|s\cdot (x,y)| &= |(s\cdot x, s\cdot y)| \\ &= |s\cdot x| \ovee |s \cdot y| \\ &= s\cdot |x| \ovee s\cdot |y| \\ &= s\cdot (|x| \ovee |y|) \\ &= s \cdot |(x,y)|\end{aligned}$$

We prove that $X+Y$ is a partial module over $M$. For the associativity, using $|(x,y) \ovee (x',y')| = |(x,y)| \ovee |(x',y')|$ we have

$$\begin{aligned}&(x,y) \perp (x',y') \;\;\&\;\; (x,y) \ovee (x',y') \perp (x'',y'') \\ &\implies |(x,y)| \perp |(x',y')| \;\;\&\;\; |(x,y)| \ovee |(x',y')| \perp |(x'',y'')| \\ &\implies |(x',y')| \perp |(x'',y'')| \;\;\&\;\; |(x,y)| \perp |(x',y')| \ovee |(x'',y'')| \\ &\implies (x',y') \perp (x'',y'') \;\;\&\;\; (x,y) \perp (x',y') \ovee (x'',y'')\,.\end{aligned}$$

It is clear that

$$((x,y) \ovee (x',y')) \ovee (x'',y'') = (x,y) \ovee ((x',y') \ovee (x'',y''))$$

when both sides are defined, since $\ovee$ is given pointwise. The commutativity and the zero law are shown similarly. It is easy to check that $s \cdot (x,y) = (s \cdot x, s \cdot y)$ is a monoid action. The mapping $s \cdot (-)$ is a PCM morphism. Indeed, $(x,y) \perp (x',y')$ implies $s \cdot (x,y) \perp s \cdot (x',y')$, since

$$|s \cdot (x,y)| = s \cdot |(x,y)| \leq |(x,y)|$$

and similarly $|s \cdot (x',y')| \leq |(x',y')|$, hence $|s \cdot (x,y)| \perp |s \cdot (x',y')|$. It is clear that $s \cdot ((x,y) \ovee (x',y')) = s \cdot (x,y) \ovee s \cdot (x',y')$, and $s \cdot (0,0) = (0,0)$. For each $(x,y) \in X+Y$ the mapping $(-) \cdot (x,y)$ is a PCM morphism too. If $s \perp t$ then $s \cdot (x,y) \perp t \cdot (x,y)$ since

$$|s \cdot (x,y)| = s \cdot |(x,y)| \leq s$$

and similarly $|t \cdot (x,y)| \leq t$. We clearly have $(s \ovee t) \cdot (x,y) = s \cdot (x,y) \ovee t \cdot (x,y)$ and $0 \cdot (x,y) = (0,0)$.

Finally, we show that $X+Y$ is a coproduct. It is straightforward to check that the coprojections $\kappa_1 \colon X \to X+Y$ and $\kappa_2 \colon Y \to X+Y$ are morphisms in $M$-$\mathbf{WMod}_{\leq}$. In fact, they are weight-preserving:

$$|\kappa_1(x)| = |(x,0)| = |x| \ovee |0| = |x| \,.$$

To show the universality of the coproduct, let $f \colon X \to Z$ and $g \colon Y \to Z$ be morphisms in $M$-$\mathbf{WMod}_{\leq}$. We define a cotuple $[f,g] \colon X+Y \to Z$ by $[f,g](x,y) = f(x) \ovee g(y)$. The sum is defined since $|f(x)| \leq |x|$, $|g(y)| \leq |y|$ and $|x| \perp |y|$, so $|f(x)| \perp |g(x)|$. The map $[f,g]$ is weight-decreasing since

$$|[f,g](x,y)| = |f(x) \ovee g(y)| = |f(x)| \ovee |g(y)| \leq |x| \ovee |y| \equiv |(x,y)| \,.$$

From this it follows that $(x,y) \perp (x',y')$ implies $[f,g](x,y) \perp [f,g](x',y')$. The map $[f,g]$ preserves sum $\ovee$:

$$\begin{aligned} [f,g]((x,y) \ovee (x',y')) &= [f,g](x \ovee x', y \ovee y') \\ &= f(x \ovee x') \ovee g(y \ovee y') \\ &= f(x) \ovee f(x') \ovee g(y) \ovee g(y') \\ &= f(x) \ovee g(y) \ovee f(x') \ovee g(y') \\ &= [f,g](x,y) \ovee [f,g](x',y') \,. \end{aligned}$$

Moreover

$$[f,g](0,0) = f(0) \ovee g(0)0 \ovee 0 = 0 \,,$$

and

$$\begin{aligned} [f,g](s \cdot (x,y)) &= [f,g](s \cdot x, s \cdot y) \\ &= f(s \cdot x) \ovee g(s \cdot y) \\ &= s \cdot f(x) \ovee s \cdot g(y) \\ &= s \cdot (f(x) \ovee g(y)) \\ &= s \cdot [f,g](x,y) \,, \end{aligned}$$

so that $[f, g]$ is a morphism in $M$-**WMod**$_{\leq}$. To show that $[f, g]$ is a unique mediating map, assume that $h\colon X + Y \to Z$ satisfies $h \circ \kappa_1 = f$ and $h \circ \kappa_2 = g$. Because $(x, 0) \perp (0, y)$ for each $(x, y) \in X + Y$, we have

$$\begin{aligned} h(x, y) &= h((x, 0) \ovee (0, y)) \\ &= h(x, 0) \ovee h(0, y) \\ &= (h \circ \kappa_1)(x) \ovee (h \circ \kappa_2)(y) \\ &= f(x) \ovee g(y) \\ &= [f, g](x, y)\,. \end{aligned}$$

■

The singleton $1 = \{0\}$ forms a weight module over $M$ in a trivial way. It is a zero object.

**Lemma 3.5.8.** *The trivial weight module* $1 = \{0\}$ *is a zero object in* $M$-**WMod**$_{\leq}$. ■

*Proof.* For each $X \in M$-**WMod**$_{\leq}$ there is a unique function $!\colon X \to 1$, which is a weight-decreasing morphism. On the other hand, the function $¡\colon 1 \to X$ given by $¡(0) = 0$ is a unique homomorphism of partial modules. The function $¡$ is trivially weight-preserving. ■

Thus $M$-**WMod**$_{\leq}$ is a category with finite coproducts and zero morphisms. Note that $M$ itself can be seen as a weight module over $M$, via a weight $|s| = s$. Then for each $X \in M$-**WMod**$_{\leq}$ the 'truth' map $\mathbb{1}_X\colon X \to M$ given by $\mathbb{1}_X(x) = |x|$ is a morphism in $M$-**WMod**$_{\leq}$, since $\mathbb{1}_X = |{-}|\colon X \to M$ is a module map by definition, and weight-preserving: $|\mathbb{1}_X(x)| = \mathbb{1}_X(x) = |x|$.

**Proposition 3.5.9.** *Let $M$ be an effect monoid. The category* $M$-**WMod**$_{\leq}$ *is an effectus, with the unit object $M$ and the truth maps $\mathbb{1}_X\colon X \to M$ defined above. The total maps in* $M$-**WMod**$_{\leq}$ *are weight-preserving maps:* $\mathrm{Tot}(M\text{-}\mathbf{WMod}_{\leq}) = M\text{-}\mathbf{WMod}$.

We defer a proof to § 3.8.3. We here describe the PCM structure in $M$-**WMod**$_{\leq}$. Let $f, g\colon X \to Y$ be morphisms in $M$-**WMod**$_{\leq}$. They are summable, $f \perp g$, if and only if $|f(x)| \ovee |g(x)| \leq |x|$ for all $x \in X$. In that case the sum $f \ovee g$ is defined pointwise: $(f \ovee g)(x) = f(x) \ovee g(x)$. It is weight-decreasing by construction.

**Proposition 3.5.10.** *Let $M$ be an effect monoid. Consider the effectus* $M$-**WMod**$_{\leq}$.

(i) *For each* $X \in M$-**WMod**$_{\leq}$, *the set of substates* $\mathrm{St}_{\leq}(X) \equiv M\text{-}\mathbf{WMod}_{\leq}(M, X)$ *is isomorphic to $X$ as PCMs.*

(ii) *The set of scalars* $\mathcal{S} \equiv M\text{-}\mathbf{WMod}_{\leq}(M, M)$ *is isomorphic to* $M^{\mathrm{op}}$ *as effect monoids.*

(iii) *Via the isomorphism* $\mathcal{S} \cong M^{\mathrm{op}}$, *or* $\mathcal{S}^{\mathrm{op}} \cong M$, $\mathrm{St}_{\leq}(X)$ *can be seen as a weight module over $M$. Then* $\mathrm{St}_{\leq}(X)$ *is isomorphic to $X$ as weight modules over $M$.*

*Proof.*

(i) The bijection between $M$-**WMod**$_{\leq}(M, X)$ and $X$ is given as follows: it sends $\omega\colon M \to X$ to $\omega(1)$, and $x \in X$ to $(-) \cdot x\colon M \to X$. The bijection preserves the structure of PCMs.

(ii) The bijection preserves the top 1. It preserves the multiplication (contravariantly) since for $\varphi, \psi \in M\text{-}\mathbf{WMod}_{\leq}(M, M)$,

$$\begin{aligned}(\varphi \cdot \psi)(1) &= (\varphi \circ \psi)(1) \\ &= \varphi(\psi(1)) \\ &= \varphi(\psi(1) \cdot 1) \\ &= \psi(1) \cdot \varphi(1) = \varphi(1) \cdot^{\mathrm{op}} \psi(1) .\end{aligned}$$

(iii) The isomorphism $\mathcal{S}^{\mathrm{op}} \cong M$ induces an action of $M$ on $\mathrm{St}_{\leq}(X)$ by:

$$(s \cdot \omega)(t) = \omega(t \cdot s)$$

The $M$-valued weight of $\omega$ is given by $|\omega(1)|$. Then the mapping $\omega \mapsto \omega(1)$ preserves the weight by definition: $|\omega(1)| = |\omega|$. It also preserves the scalar multiplication: for $\omega \in \mathrm{St}_{\leq}(X)$ and $s \in M$,

$$\begin{aligned}(s \cdot \omega)(1) &= \omega(1 \cdot s) \\ &= \omega(s \cdot 1) \\ &= s \cdot \omega(1) .\end{aligned}$$

Thus $\mathrm{St}_{\leq}(X) \cong X$ as weight modules. ■

**Proposition 3.5.11.** *The substate functor* $\mathrm{St}_{\leq} \colon \mathbf{C} \to \mathcal{S}^{\mathrm{op}}\text{-}\mathbf{WMod}_{\leq}$ *preserves finite coproducts.*

*Proof.* We have $\mathrm{St}_{\leq}(0) = \mathbf{C}(I, 0) \cong 1 = \{0\}$ and hence it preserves the initial object. For objects $A, B \in \mathbf{C}$ the canonical map

$$\mathrm{St}_{\leq}(A) + \mathrm{St}_{\leq}(B) \xrightarrow{[(\kappa_1)_*, (\kappa_2)_*]} \mathrm{St}_{\leq}(A + B)$$

is given by

$$[(\kappa_1)_*, (\kappa_2)_*](\sigma, \tau) = (\kappa_1)_*(\sigma) \ovee (\kappa_2)_*(\tau) = \kappa_1 \circ \sigma \ovee \kappa_2 \circ \tau$$

This mapping is bijective by Lemma 3.2.5. Since $[(\kappa_1)_*, (\kappa_2)_*]$ is weight-preserving, the inverse $[(\kappa_1)_*, (\kappa_2)_*]^{-1}$ is also a weight-preserving module map. Hence $[(\kappa_1)_*, (\kappa_2)_*]$ is an isomorphism. ■

## 3.6 Structure of states: convex sets

States in an effectus are morphisms $\omega \colon I \to A$ with $\mathbb{1} \circ \omega = \mathbb{1}$. Since $|\omega| \equiv \mathbb{1} \circ \omega$, states can be characterized by the weight module structure of substates $\mathrm{St}_{\leq}(A)$: states are substates $\omega$ of weight one. This motivates the following definition:

**Definition 3.6.1.** Let $X$ be a weight module over an effect monoid $M$. The **base** of $X$ is the subset of elements of weight one, i.e.

$$\mathrm{B}(X) = \big\{ x \in X \bigm| |x| = 1 \big\} .$$

Then it is quite straightforward to see that the base $\mathrm{B}(X)$ forms a convex set in the sense of Definition 2.4.3.

**Proposition 3.6.2.** *If $X$ is a weight module over an effect monoid $M$, the base* $\mathrm{B}(X)$ *is a convex set over $M$.*

*Proof.* Let $\sum_i r_i|x_i\rangle \in \mathcal{D}_M(\mathrm{B}(X))$ be a distribution over $M$ on $\mathrm{B}(X)$. We define the convex sum by

$$[\![\textstyle\sum_i r_i|x_i\rangle]\!] = \bigovee_i r_i \cdot x_i\,.$$

The sum $\bigovee_i r_i \cdot x_i$ is defined since

$$|r_i \cdot x_i| = r_i \cdot |x_i| = r_i \cdot 1 = r_i$$

and $\bigovee_i r_i$ is defined. We indeed have $\bigovee_i r_i \cdot x_i \in \mathrm{B}(X)$ since

$$\big|\bigovee_i r_i \cdot x_i\big| = \bigovee_i |r_i \cdot x_i| = \bigovee_i r_i = 1\,.$$

The convex sum satisfies the axioms of convex sets:

$$[\![1|x\rangle]\!] = 1 \cdot x = x$$

and

$$\begin{aligned}\Big[\!\!\Big[\textstyle\sum_i r_i\Big|[\![\textstyle\sum_j s_{ij}|x_{ij}\rangle]\!]\Big\rangle\Big]\!\!\Big] &= \bigovee_i r_i \cdot (\bigovee_j s_{ij} \cdot x_{ij}) \\ &= \bigovee_{ij} r_i \cdot (s_{ij} \cdot x_{ij}) \\ &= \bigovee_{ij} (r_i \cdot s_{ij}) \cdot x_{ij} \\ &= [\![\textstyle\sum_{ij} r_i \cdot s_{ij}|x_{ij}\rangle]\!]\,.\end{aligned}$$

■

Let $f\colon X \to Y$ be a weight-preserving module map between weight modules. If $|x| = 1$, then $|f(x)| = |x| = 1$ and hence $f$ restricts to $f\colon \mathrm{B}(X) \to \mathrm{B}(Y)$.

**Proposition 3.6.3.** *If $f\colon X \to Y$ is a weight-preserving module map between weight modules, the restriction $f\colon \mathrm{B}(X) \to \mathrm{B}(Y)$ is an affine map.*

*Proof.* For $\sum_i r_i|x_i\rangle \in \mathcal{D}_M(X)$,

$$f\big([\![\textstyle\sum_i r_i|x_i\rangle]\!]\big) = f\big(\bigovee r_i \cdot x_i\big) = \bigovee r_i \cdot f(x_i) = [\![\textstyle\sum_i r_i|f(x_i)\rangle]\!]\,.$$

■

**Corollary 3.6.4.** *The assignment $X \mapsto \mathrm{B}(X)$ defines a functor $\mathrm{B}\colon M\text{-}\mathbf{WMod} \to M\text{-}\mathbf{Conv}$.* ■

Now we turn back to states $\mathrm{St}(A)$ in an effectus $\mathbf{C}$. We have $\mathrm{St}(A) = \mathrm{B}(\mathrm{St}_\le(A))$. For a total morphism $f\colon A \to B$, moreover, the state transformer $f_*\colon \mathrm{St}(A) \to \mathrm{St}(B)$ agrees with the morphism $\mathrm{B}(\mathrm{St}_\le(f))$ obtained via

$$\mathrm{Tot}(\mathbf{C}) \xrightarrow{\mathrm{St}_\le} \mathcal{S}^{\mathrm{op}}\text{-}\mathbf{WMod} \xrightarrow{\mathrm{B}} \mathcal{S}^{\mathrm{op}}\text{-}\mathbf{Conv}\,.$$

To summarize, we obtain the following result.

**Corollary 3.6.5.** *For each $A \in \mathbf{C}$ the states* $\mathrm{St}(A)$ *form a convex set over scalars* $\mathcal{S}$*. Moreover, the mappings* $A \mapsto \mathrm{St}(A)$ *and* $f \mapsto f_*$ *define a functor* $\mathrm{St}\colon \mathrm{Tot}(\mathbf{C}) \to \mathcal{S}^{\mathrm{op}}\text{-}\mathbf{Conv}$*, which makes the following diagram commute.*

$$\begin{array}{ccc}
\mathrm{Tot}(\mathbf{C}) & \xrightarrow{\mathrm{St}_{\leq}} & \mathcal{S}^{\mathrm{op}}\text{-}\mathbf{WMod} \\
& \searrow^{\mathrm{St}} & \downarrow \mathrm{B} \\
& & \mathcal{S}^{\mathrm{op}}\text{-}\mathbf{Conv}
\end{array}$$

■

**Example 3.6.6.** Continuing Example 3.5.5, we briefly review states in our examples of effectuses.

(i) In the effectus **Pfn**, states on a set $X$ are elements $x \in X$. By the general theory, states form a convex set over $2 = \{0, 1\}$. However, convex structure over 2 is completely trivial and we simply have $2\text{-}\mathbf{Conv} \cong \mathbf{Set}$. Therefore the state functor $\mathrm{St}\colon \mathrm{Tot}(\mathbf{Pfn}) = \mathbf{Set} \to \mathbf{Set}$ is simply the identity functor.

(ii) States in the effectus $\mathcal{K}\ell(\mathcal{D}_{\leq})$ are distributions $\omega \in \mathcal{D}(X)$. The convex structure over $[0, 1]$ is obvious and we get a functor $\mathrm{St}\colon \mathcal{K}\ell(\mathcal{D}) \to \mathbf{Conv}$, where the state transformer $f_*\colon \mathcal{D}(X) \to \mathcal{D}(Y)$ for $f\colon X \to \mathcal{D}(Y)$ is given by:

$$f_*(\omega)(y) = \sum_{x \in X} f(x)(y) \cdot \omega(x)\,.$$

Similarly to Example 3.4.6(ii), we can explain state transformation in the usual language of probability. Recall that a state $\omega \in \mathcal{D}(X)$ can be viewed as a probability distribution for an $X$-valued random variable via $\mathrm{P}(\mathbf{x} = x) = \omega(x)$, and a total morphism $f\colon X \to \mathcal{D}(Y)$ as a conditional probability distribution via $\mathrm{P}(\mathbf{y} = y \mid \mathbf{x} = x) = f(x)(y)$. Then

$$\begin{aligned}
f_*(\omega)(y) &= \sum_{x \in X} f(x)(y) \cdot \omega(x) \\
&= \sum_{x \in X} \mathrm{P}(\mathbf{y} = y \mid \mathbf{x} = x)\, \mathrm{P}(\mathbf{x} = x) \\
&= \sum_{x \in X} \mathrm{P}(\mathbf{y} = y, \mathbf{x} = x) \\
&= \mathrm{P}(\mathbf{y} = y)\,.
\end{aligned}$$

Therefore the state transformation computes a probability distribution of $\mathbf{y}$, via what is often called the *law of total probability*.

Similarly, states in $\mathcal{K}\ell(\mathcal{G}_{\leq})$ are probability measures and state transformers are given by integration, yielding $\mathrm{St}\colon \mathcal{K}\ell(\mathcal{G}) \to \mathbf{Conv}$.

(iii) States in the effectus $\mathbf{Wstar}_{\leq}^{\mathrm{op}}$ are normal unital positive maps $\omega\colon \mathscr{A} \to \mathbb{C}$, i.e. normal states of a $W^*$-algebra. These clearly forms a convex set over $[0, 1]$, yielding a functor $\mathrm{St}\colon \mathbf{Wstar}^{\mathrm{op}} \to \mathbf{Conv}$ from the subcategory of total morphisms. Similarly there is $\mathrm{St}\colon \mathbf{Cstar}^{\mathrm{op}} \to \mathbf{Conv}$ for $C^*$-algebras.

### 3.6.1 Convexity of predicates and substates

Convexity is an important property in quantum foundations. Sometimes not only states but also predicates/effects are assumed to form a convex set, see e.g. [197, 198]. Such assumption is compatible with our setting: every effect module is a convex set. This explains why effect modules were called *convex effect algebras* in [105, 109]. In parallel, we also show that weight modules are convex sets. In this subsection, $M$ denotes an effect monoid.

**Definition 3.6.7.** A partial module $X$ over $M$ is said to be **convex** if for each formal convex sum $\sum_i r_i|x_i\rangle \in \mathcal{D}_M(X)$, the sum $\bigovee_i r_i \cdot x_i$ is defined in $X$.

**Lemma 3.6.8.** *If $X$ is a convex partial module over an effect monoid $M$, then $X$ is a convex set over $M$ by*

$$[\![\textstyle\sum_i r_i|x_i\rangle]\!] = \bigovee_i r_i \cdot x_i \,.$$

*Moreover, if $f\colon X \to Y$ is a module map between convex partial modules, then $f$ is affine.*

*Proof.* Straightforward. ■

**Proposition 3.6.9.** *Any effect module over $M$ is convex, and any weight module over $M$ is convex. Therefore both of them form convex sets.*

*Proof.* Let $E$ be an effect module over $M$ and let $\sum_i r_i|a_i\rangle \in \mathcal{D}_M(E)$. Then the sum $(\bigovee_i r_i) \cdot 1 = \bigovee_i r_i \cdot 1$ is defined. Since $r_i \cdot a_i \le r_i \cdot 1$ for each $i$, it follows, by using Lemma 2.3.7(v) repeatedly, that the sum $\bigovee_i r_i \cdot a_i$ is defined.

Next, let $X$ be a weight module over $M$, and let $\sum_i r_i|x_i\rangle \in \mathcal{D}_M(X)$. Because $|r_i \cdot x_i| = r_i \cdot |x_i| \le r_i$ and the sum $\bigovee_i r_i$ is defined, the sum $\bigovee_i |r_i \cdot x_i|$ is defined too by a reasoning similar to the above. By the definition of weight module, it follows that $\bigovee_i r_i \cdot x_i$ is defined. ■

## 3.7 State-and-effect triangles

In the previous sections, we studied the structures of predicates, substates, and states in an effectus. Note that the notions of predicates and (sub)states are formally dual in the effectus in the sense that predicates are morphisms $A \to I$, while (sub)states are morphisms $I \to A$. In this section we further show that there are dualities between their algebraic structures, namely, dual adjunctions between the categories of effect modules and weight modules, and between the categories of effect modules and convex sets. To put the pieces all together, it is shown that every effectus induces *state-and-effect triangles*, which summarizes all the results so far.

### 3.7.1 Triangles with substates

We have seen that for each effectus $\mathbf{C}$, there are a predicate functor $\mathrm{Pred}\colon \mathbf{C} \to \mathcal{S}\text{-}\mathbf{EMod}_{\le}^{\mathrm{op}}$ and a substate functor $\mathrm{St}_{\le}\colon \mathbf{C} \to \mathcal{S}^{\mathrm{op}}\text{-}\mathbf{WMod}_{\le}$. We fix an effect monoid $M$. Then the categories $M\text{-}\mathbf{EMod}_{\le}^{\mathrm{op}}$ and $M\text{-}\mathbf{WMod}_{\le}$ are effectuses, and scalars in both effectuses are identified with $M$:

$$M\text{-}\mathbf{EMod}_{\le}^{\mathrm{op}}(M,M) \cong M \cong M^{\mathrm{op}}\text{-}\mathbf{WMod}_{\le}(M,M)\,,$$

see Propositions 3.4.11 and 3.5.10. Therefore we have the following contravariant functors.

$$M\text{-}\mathbf{EMod}_{\le}^{\mathrm{op}} \underset{\mathrm{Pred}}{\overset{\mathrm{St}_{\le}}{\rightleftarrows}} M^{\mathrm{op}}\text{-}\mathbf{WMod}_{\le}$$

Note that both functors are given by 'homming into $M$':

$$\begin{aligned} \mathrm{St}_{\le}(E) &= M\text{-}\mathbf{EMod}_{\le}^{\mathrm{op}}(M,E) = M\text{-}\mathbf{EMod}_{\le}(E,M) \\ \mathrm{Pred}(X) &= M^{\mathrm{op}}\text{-}\mathbf{WMod}_{\le}(X,M)\,. \end{aligned}$$

This is a typical situation where a dual adjunction exists, with $M$ being a *dualizing object* (see e.g. [162, § VI.4]). This is indeed the case:

**Proposition 3.7.1.** *For each effect monoid $M$, one has the following adjunction given by 'homming into $M$'.*

$$M\text{-}\mathbf{EMod}_{\le}^{\mathrm{op}} \underset{\mathrm{Hom}(-,M)}{\overset{\mathrm{Hom}(-,M)}{\rightleftarrows}} M^{\mathrm{op}}\text{-}\mathbf{WMod}_{\le} \quad (\top)$$

*Proof.* As shown above, the hom functors respectively coincide with the state and the predicate functor, and thus are well-defined. To simplify the notation, in the proof we will write:

$$\mathbf{EMod}_{\le} = M\text{-}\mathbf{EMod}_{\le} \qquad\qquad \mathbf{WMod}_{\le} = M^{\mathrm{op}}\text{-}\mathbf{WMod}_{\le}\,.$$

To give an adjunction, we establish the following natural bijective correspondence.

$$\frac{X \xrightarrow{f} \mathbf{EMod}_{\le}(E,M) \quad \text{in } \mathbf{WMod}_{\le}}{E \xrightarrow{g} \mathbf{WMod}_{\le}(X,M) \text{ in } \mathbf{EMod}_{\le}} \tag{3.5}$$

The correspondence is given by 'swapping arguments', i.e. $f(x)(a) = g(a)(x)$ for $x \in X$ and $a \in E$. If $f\colon X \to \mathbf{EMod}_{\le}(E,M)$ is a morphism in $\mathbf{WMod}_{\le}$, then $g\colon E \to \mathbf{WMod}_{\le}(X,M)$ given by $g(a)(x) \coloneqq f(x)(a)$ is indeed a morphism in $\mathbf{EMod}_{\le}$ as follows. Note that $|x| \ge |f(x)| = f(x)(1)$ because $f$ is weight-decreasing. If $a \perp b$ in $E$, then $g(a) \perp g(b)$ since

$$|x| \ge f(x)(1) \ge f(x)(a \ovee b) = f(x)(a) \ovee f(x)(b) = g(a)(x) \ovee g(b)(x)\,.$$

We then have $g(a \ovee b) = g(a) \ovee g(b)$ as

$$\begin{aligned} g(a \ovee b)(x) &= f(x)(a \ovee b) \\ &= f(x)(a) \ovee f(x)(b) \\ &= g(a)(x) \ovee g(b)(x)\,. \end{aligned}$$

We have $g(0) = 0$ since $g(0)(x) = f(x)(0) = 0$, and $g(s \cdot a) = s \cdot g(a)$ since

$$g(s \cdot a)(x) = f(x)(s \cdot a) = s \cdot f(x)(a) = s \cdot g(a)(x)\,.$$

Thus $g\colon E \to \mathbf{WMod}_{\leq}(X, E)$ is a morphism in $\mathbf{EMod}_{\leq}$. Similarly a morphism in $\mathbf{EMod}_{\leq}$ defines one in $\mathbf{WMod}_{\leq}$, establishing the bijective correspondence. It is straightforward to check the naturality of the correspondence. ■

We can now put a predicate and a substate functor into one picture.

**Corollary 3.7.2.** *Every effectus* $(\mathbf{C}, I)$ *induces the following 'state-and-effect' triangle.*

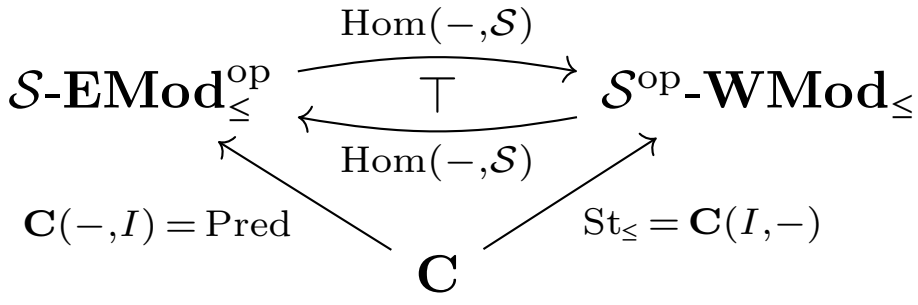

*where* $\mathcal{S} = \mathbf{C}(I, I)$ *is the effect monoid of scalars.* ■

The functors in the state-and-effect triangle need not commute, but there are canonical natural transformations to fill in the triangle. Recall that for a predicate $p$ and a substate $\omega$, the validity $(\omega \models p) = p \circ \omega$ is defined via composition. The validity gives a 'bimorphism'

$$\models\colon \mathrm{St}_{\leq}(A) \times \mathrm{Pred}(A) \to \mathcal{S}$$

in the sense that $\omega \models -\colon \mathrm{Pred}(A) \to \mathcal{S}$ is a subunital module map, and $- \models p\colon \mathrm{St}_{\leq}(A) \to \mathcal{S}$ is a weight-decreasing module map. Then we can 'curry' the bimorphism in two ways:

$$\mathrm{St}_{\leq}(A) \xrightarrow{\alpha_A} \mathcal{S}\text{-}\mathbf{EMod}_{\leq}(\mathrm{Pred}(A), \mathcal{S}) \quad \text{via} \quad \alpha_A(\omega)(p) = (\omega \models p)$$

$$\mathrm{Pred}(A) \xrightarrow{\beta_A} \mathcal{S}^{\mathrm{op}}\text{-}\mathbf{WMod}_{\leq}(\mathrm{St}_{\leq}(A), \mathcal{S}) \quad \text{via} \quad \beta_A(p)(\omega) = (\omega \models p)\,.$$

**Proposition 3.7.3.** *The 'curried' validities* $\alpha_A$ *and* $\beta_A$ *defined above are the natural transformations below:*

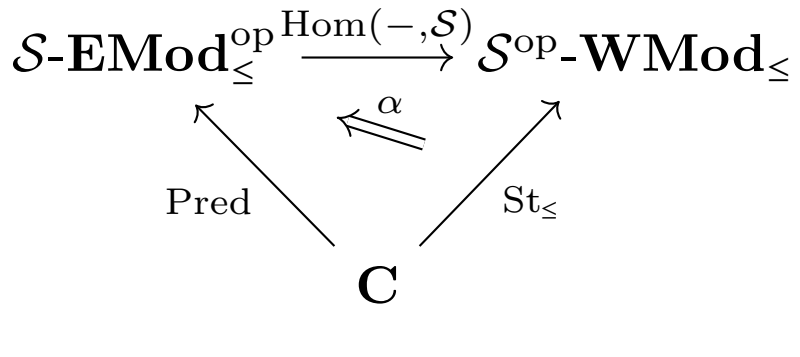

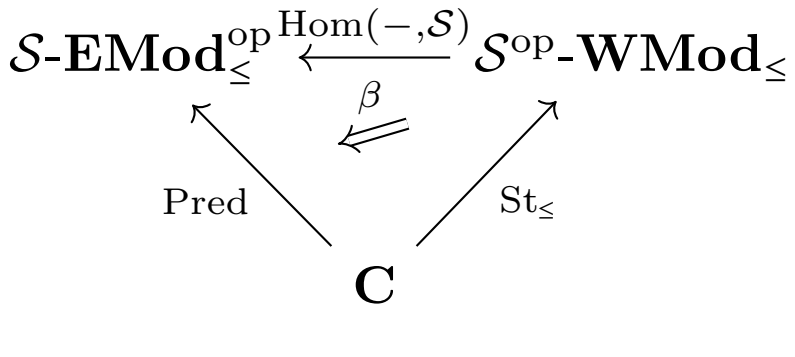

*The* $\alpha$ *and* $\beta$ *are (a simple special case of) 'mates' with respect to the adjunction* $\mathcal{S}$-$\mathbf{EMod}_{\leq} \leftrightarrows \mathcal{S}^{\mathrm{op}}$-$\mathbf{WMod}_{\leq}$ *— namely, for each* $A \in \mathbf{C}$ *the components* $\alpha_A$ *and* $\beta_A$ *correspond in the natural bijection* (3.5) *of the adjunction.*

*Proof.* We need to show that

$$\mathrm{St}_{\leq}(A) \xrightarrow{\alpha_A} \mathcal{S}\text{-}\mathbf{EMod}_{\leq}(\mathrm{Pred}(A), \mathcal{S})$$

is a well-defined morphism in $\mathcal{S}^{\mathrm{op}}$-$\mathbf{WMod}$. This boils down to the fact that $\models$ is a 'bimorphism'. The naturality of $\alpha$ amounts to the formula $f_*(\omega) \models p = \omega \models f^*(p)$. Clearly $\alpha$ and $\beta$ are mates, from this it follows that $\beta$ is well-defined too. ■

We show that the triangle can be restricted to total morphisms.

**Proposition 3.7.4.** *For any effect monoid $M$, the adjunction $M$-$\mathbf{EMod}_{\leq}^{\mathrm{op}} \rightleftarrows M^{\mathrm{op}}$-$\mathbf{WMod}_{\leq}$ of Proposition* 3.7.1 *restricts to the adjunction*

$$M\text{-}\mathbf{EMod}^{\mathrm{op}} \underset{\longleftarrow}{\overset{\longrightarrow}{\top}} M^{\mathrm{op}}\text{-}\mathbf{WMod}$$

*between the subcategories of total morphisms.*

*Proof.* The claim boils down to the fact that the bijective correspondence (3.5) restricts to weight-preserving and unital maps. ∎

**Corollary 3.7.5.** *By restricting all the functors in Corollary* 3.7.2*, we obtain the state-and-effect triangle over* $\mathrm{Tot}(\mathbf{C})$*:*

$$\begin{array}{ccc} \mathcal{S}\text{-}\mathbf{EMod}^{\mathrm{op}} & \underset{\longleftarrow}{\overset{\longrightarrow}{\top}} & \mathcal{S}^{\mathrm{op}}\text{-}\mathbf{WMod} \\ {\scriptstyle \mathbf{C}(-,I)=\mathrm{Pred}} \nwarrow & & \nearrow {\scriptstyle \mathrm{St}_{\leq}=\mathbf{C}(I,-)} \\ & \mathrm{Tot}(\mathbf{C}) & \end{array} \tag{3.6}$$

*Moreover, the natural transformations in Proposition* 3.7.3 *given by 'currying' validity* $\models$ *also restricts in the triangle above.*

*Proof.* The first claim follows immediately from what we have shown. The latter claim amounts to the fact that the components of the 'validity' natural transformations

$$\mathrm{St}_{\leq}(A) \xrightarrow{\alpha_A} \mathcal{S}\text{-}\mathbf{EMod}_{\leq}(\mathrm{Pred}(A), \mathcal{S})$$
$$\mathrm{Pred}(A) \xrightarrow{\beta_A} \mathcal{S}^{\mathrm{op}}\text{-}\mathbf{WMod}_{\leq}(\mathrm{St}(A), \mathcal{S})$$

are weight-preserving and unital, respectively. ∎

### 3.7.2 Triangles with states

Let $M$ be an effect monoid. Then for the effectus $M$-$\mathbf{EMod}_{\leq}^{\mathrm{op}}$ one has the state functor:

$$M\text{-}\mathbf{EMod}^{\mathrm{op}} = \mathrm{Tot}(M\text{-}\mathbf{EMod}_{\leq}^{\mathrm{op}}) \xrightarrow{\mathrm{St}} M^{\mathrm{op}}\text{-}\mathbf{Conv}\,,$$

via the identification of scalars: $M$-$\mathbf{EMod}_{\leq}^{\mathrm{op}}(M, M) \cong M$. Explicitly, the functor is defined by 'homming into $M$':

$$\mathrm{St}(E) = M\text{-}\mathbf{EMod}^{\mathrm{op}}(M, E) = M\text{-}\mathbf{EMod}(E, M)\,.$$

For the other direction, note that $M$ itself is a convex set over $M^{\mathrm{op}}$ via convex sum:

$$[\![ \textstyle\sum_i r_i |s_i\rangle ]\!] = \textstyle\bigovee_i s_i \cdot r_i\,.$$

Then homsets $M^{\mathrm{op}}$-$\mathbf{Conv}(K, M)$ yield a functor in the other way, forming an adjunction:

**Theorem 3.7.6** (Jacobs [140]). *Let $M$ be an effect monoid. By 'homming into $M$' we have an adjunction:*

$$M\text{-}\mathbf{EMod}^{\mathrm{op}} \underset{\mathrm{Hom}(-,M)}{\overset{\mathrm{Hom}(-,M)}{\rightleftarrows}} M^{\mathrm{op}}\text{-}\mathbf{Conv} \qquad (\top)$$

*Proof.* One can verify that the set $M^{\mathrm{op}}\text{-}\mathbf{Conv}(K, M)$ of affine maps forms an effect module in a pointwise manner, and then can establish bijective correspondence of morphisms by 'swapping arguments', similarly to Proposition 3.7.1. See [140, Proposition 2.6] for more details. ■

**Corollary 3.7.7.** *For every effectus $\mathbf{C}$, there is the following 'state-and-effect' triangle:*

$$\begin{array}{ccc} \mathcal{S}\text{-}\mathbf{EMod}^{\mathrm{op}} & \underset{\mathrm{Hom}(-,\mathcal{S})}{\overset{\mathrm{Hom}(-,\mathcal{S})}{\rightleftarrows}} \; (\top) & \mathcal{S}^{\mathrm{op}}\text{-}\mathbf{Conv} \\ & {\scriptstyle \mathbf{C}(-,I) = \mathrm{Pred}} \nwarrow \quad \nearrow {\scriptstyle \mathrm{St} = \mathrm{Tot}(\mathbf{C})(I,-)} & \\ & \mathrm{Tot}(\mathbf{C}) & \end{array} \tag{3.7}$$

■

In the same way as Proposition 3.7.3, the validity $\vDash \colon \mathrm{St}(A) \times \mathrm{Pred}(A) \to \mathcal{S}$ can be 'curried' into natural transformations

$$\mathrm{St}(A) \xrightarrow{\alpha_A} \mathcal{S}\text{-}\mathbf{EMod}(\mathrm{Pred}(A), \mathcal{S}) \qquad \mathrm{Pred}(A) \xrightarrow{\beta_A} \mathcal{S}^{\mathrm{op}}\text{-}\mathbf{Conv}(\mathrm{St}(A), \mathcal{S})$$

that fill in the triangle.

### 3.7.3 Examples

We summarize the examples we saw in Section 3.3 and Examples 3.4.6, 3.5.5 and 3.6.6 in state-and-effect triangles.

**Example 3.7.8.** For the effectus **Pfn** of sets and partial functions, we have the following triangles:

$$\begin{array}{ccc} \mathbf{EA}_{\leq}^{\mathrm{op}} & \underset{\mathrm{Hom}(-,2)}{\overset{\mathrm{Hom}(-,2)}{\rightleftarrows}} \; (\top) & 1/\mathbf{Set} \\ & {\scriptstyle \mathcal{P} = \mathrm{Pred}} \nwarrow \quad \nearrow {\scriptstyle \mathrm{St}_{\leq} = (-)+1} & \\ & \mathbf{Pfn} & \end{array} \qquad \begin{array}{ccc} \mathbf{EA}^{\mathrm{op}} & \underset{\mathrm{Hom}(-,2)}{\overset{\mathrm{Hom}(-,2)}{\rightleftarrows}} \; (\top) & \mathbf{Set} \\ & {\scriptstyle \mathcal{P} = \mathrm{Pred}} \nwarrow \quad \nearrow {\scriptstyle \mathrm{St} = \mathrm{id}} & \\ & \mathbf{Set} & \end{array}$$

Here we use the isomorphisms of categories $2\text{-}\mathbf{EMod}_{\leq} \cong \mathbf{EA}_{\leq}$, $2\text{-}\mathbf{WMod}_{\leq} \cong 1/\mathbf{Set}$, $2\text{-}\mathbf{EMod} \cong \mathbf{EA}$, and $2\text{-}\mathbf{Conv} \cong \mathbf{Set}$ for the case where scalars are Boolean values: $2 = \{0, 1\}$. These triangles capture the duality between state and predicate transformers in the standard deterministic setting.

**Example 3.7.9.** For the Kleisli category $\mathcal{K}\ell(\mathcal{D}_\leq)$ of the subdistribution monad, we have the following triangles:

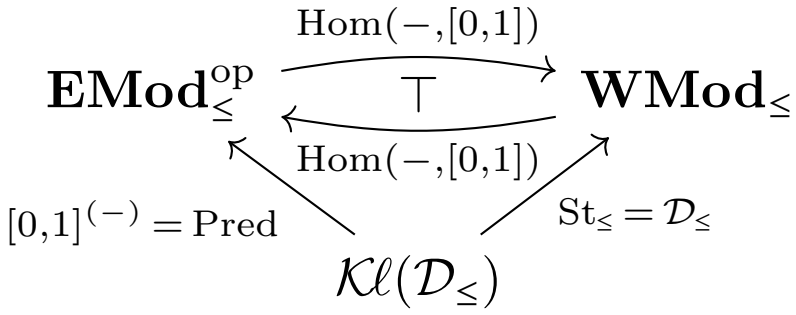

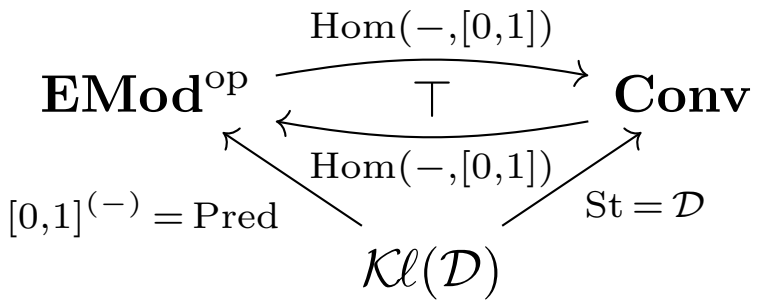

Similar triangles exist over $\mathcal{K}\ell(\mathcal{G}_\leq)$, for measure-theoretic probability, and they capture the duality between state and predicate transformer semantics of probabilistic programs by Kozen [179, 180].

**Example 3.7.10.** For the effectus $\mathbf{Wstar}_\leq^{\mathrm{op}}$ of $W^*$-algebras, we have the following triangles, for a substate functor and a state functor:

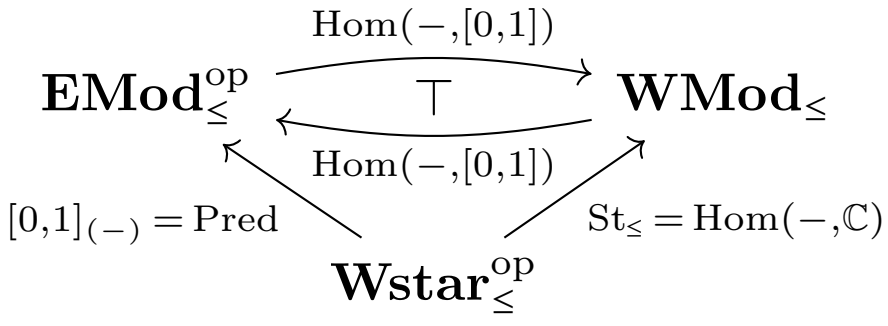

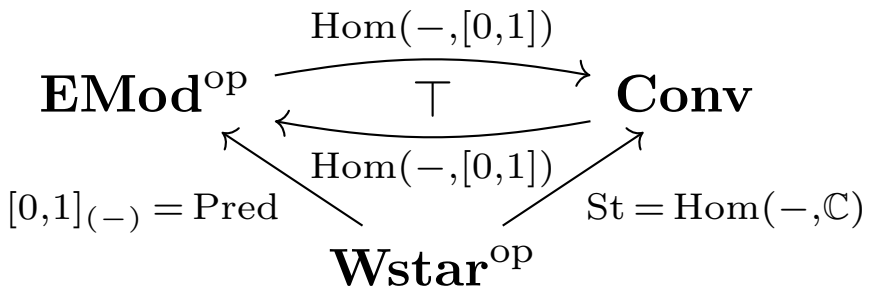

Similar triangles also exist for the effectus $\mathbf{Cstar}_\leq^{\mathrm{op}}$ of $C^*$-algebras. These triangles capture the duality between state and predicate transformer semantics of quantum programs [62, 235]. These triangles may also be seen as a concise presentation of the duality between the Schrödinger and Heisenberg picture for quantum processes.

## 3.8 A characterization of effectuses

In § 3.2 we defined an effectus as a finPAC with a suitable structure of effect algebras. On the one hand, the definition is reasonable in the sense that both (fin)PACs and effect algebras are well-established notions. On the other hand, it is not the most convenient definition in order to check if a certain category is an effectus, since the definition involves a lot of structures. In this section, we give a characterization of effectuses that is simpler than the original definition; see Proposition 3.8.6. Using the characterization, we prove that the categories of effect modules and weight modules are effectuses.

### 3.8.1 A characterization of finPACs

By definition, an effectus is a finPAC with additional structures and properties. Therefore we start with a characterization of finPACs, which is similar to a characterization of PACs given by Arbib and Manes [7, § 5.3]. Recall from Lemma 3.1.4(iii) that in a finPAC, partial projections are jointly monic. We use this property to characterize finPACs. First we show that several possible definitions of 'partial projections are jointly monic' agree.

**Lemma 3.8.1.** *Let* $\mathbf{C}$ *be a category with finite coproducts and zero morphisms. The following are equivalent.*

(i) *For each object* $A \in \mathbf{C}$*, the following two partial projections are jointly monic.*

$$A + A \underset{\rhd_2}{\overset{\rhd_1}{\rightrightarrows}} A$$

(ii) *For each pair of objects* $A, B \in \mathbf{C}$*, the following two partial projections are jointly monic.*

$$\begin{array}{ccc} & \overset{\rhd_1}{\nearrow} & A \\ A + B & & \\ & \underset{\rhd_2}{\searrow} & B \end{array}$$

(iii) *For each* $n$*-tuple of objects* $A_1, \ldots, A_n \in \mathbf{C}$*, the following* $n$ *partial projections are jointly monic.*

$$A_1 + \cdots + A_n \xrightarrow{\rhd_j} A_j \qquad (j = 1, \ldots, n)$$

*Proof.* The implication (iii) $\Longrightarrow$ (i) is trivial, and (ii) $\Longrightarrow$ (iii) follows by induction. We prove (i) $\Longrightarrow$ (ii)

Let $f, g \colon C \to A + B$ be morphisms with $\rhd_1 \circ f = \rhd_1 \circ g$ and $\rhd_2 \circ f = \rhd_2 \circ g$. For $j \in \{1, 2\}$ we have

$$\rhd_j \circ (\kappa_1 + \kappa_2) \circ f = \kappa_j \circ \rhd_j \circ f = \kappa_j \circ \rhd_j \circ g = \rhd_j \circ (\kappa_1 + \kappa_2) \circ g\,.$$

By the joint monicity of

$$(A + B) + (A + B) \underset{\rhd_2}{\overset{\rhd_1}{\rightrightarrows}} A + B$$

we obtain $(\kappa_1 + \kappa_2) \circ f = (\kappa_1 + \kappa_2) \circ g$. It follows that $f = g$, since $\kappa_1 + \kappa_2$ is a split mono as $\nabla \circ (\kappa_1 + \kappa_2) = \mathrm{id}$. ■

In the rest of this section, we use categories with this joint monicity property repeatedly. For convenience, we introduce the following definition.

**Definition 3.8.2.** A **butterfly coproduct category** is a category with finite coproducts $(+, 0)$ and zero morphisms $0_{AB} \colon A \to B$ that satisfies any of the equivalent 'jointly monic partial projections' conditions in Lemma 3.8.1. The word 'butterfly' comes from the following commutative diagram.[3]

$$\begin{array}{ccccc} A & \overset{\kappa_1}{\searrow} & & \overset{\kappa_2}{\swarrow} & B \\ \| & & A + B & & \| \\ A & \underset{\rhd_1}{\swarrow} & & \underset{\rhd_2}{\searrow} & B \end{array}$$

[3] Binary coproducts here form 'butterfly product' in the sense of [99, § 2.1.7]. The term 'butterfly coproduct' (requiring the joint monicity condition) is due to Tull (private communication).

Clearly, every finPAC is a butterfly coproduct category. Recall from Definition 3.1.7 that we introduced partial tuples $\langle\!\langle f_j \rangle\!\rangle_j$ in a finPAC. For the definition to make sense we only need the joint monicity of partial projections $\rhd_j$. Thus we can use partial tuples $\langle\!\langle f_j \rangle\!\rangle_j$ more generally in a butterfly coproduct category. For convenience, we recall the definition here. Let $(f_j : A \to B_j)_{j \in J}$ be a finite family of morphisms in a butterfly product category. Then we write $\langle\!\langle f_j \rangle\!\rangle_j : A \to \coprod_j B_j$ for a morphism such that $\rhd_j \circ \langle\!\langle f_j \rangle\!\rangle_j = f_j$ for all $j \in J$. The morphism $\langle\!\langle f_j \rangle\!\rangle_j$ may not exist, but if it does, then it is uniquely determined by the joint monicity of $\rhd_j$.

Additionally, we will use the following notation: for two morphisms $f : A \to B$ and $g : A \to C$, we write $f \perp g$ if $\langle\!\langle f, g \rangle\!\rangle$ is defined, i.e. if $f$ and $g$ are compatible. Then we have the following basic calculation rules for partial tuples, which are pretty much similar to those for ordinary tuples $\langle f, g \rangle : A \to B \times C$ for products.

**Lemma 3.8.3.** *Let* $\mathbf{C}$ *be a butterfly coproduct category. Let* $f : A \to B$ *and* $g : A \to C$ *be morphisms with* $f \perp g$.

(i) *For each* $h : A' \to A$, *we have* $f \circ h \perp g \circ h$ *and* $\langle\!\langle f \circ h, g \circ h \rangle\!\rangle = \langle\!\langle f, g \rangle\!\rangle \circ h$.

(ii) *For each* $k : B \to B'$ *and* $l : C \to C'$, *we have* $k \circ f \perp l \circ g$ *and* $\langle\!\langle k \circ f, l \circ g \rangle\!\rangle = (k + l) \circ \langle\!\langle f, g \rangle\!\rangle$.

*Proof.* Both points follow from the following commutative diagram.

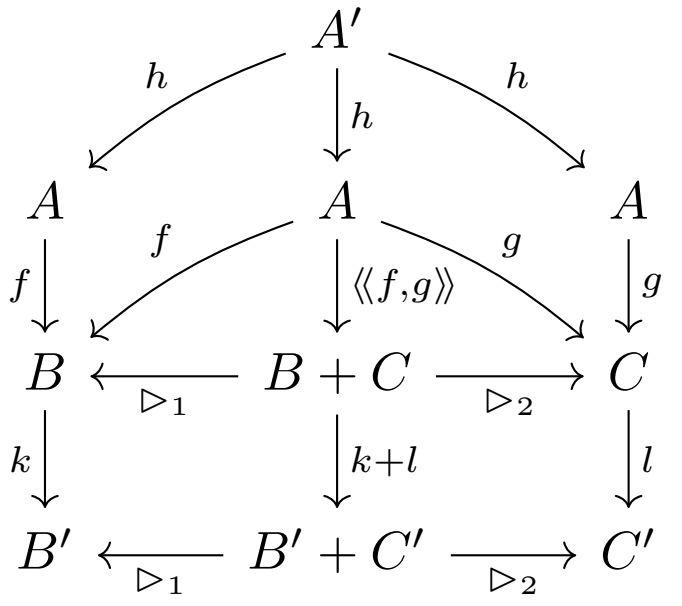

■

The partial tuple operation in a butterfly coproduct category naturally induces a partial sum operation $\ovee$ on parallel morphisms $f, g : A \to B$ as follows:

$$\begin{aligned} &f \ovee g \text{ is defined iff } \langle\!\langle f, g \rangle\!\rangle : A \to B + B \text{ is defined (i.e. } f \perp g) \\ &f \ovee g \coloneqq \big(A \xrightarrow{\langle\!\langle f, g \rangle\!\rangle} B + B \xrightarrow{\nabla} B\big) \end{aligned} \tag{3.8}$$

The definition is consistent with the addition in a finPAC by Proposition 3.1.8. We now see that a butterfly coproduct category is almost a finPAC.

**Lemma 3.8.4.** *Let* $\mathbf{C}$ *be a butterfly coproduct category. For each* $f, g : A \to B$, *the sum* $\ovee$ *satisfies:*

(i) *Commutativity: if* $f \perp g$, *then* $g \perp f$ *and* $f \ovee g = g \ovee f$.

(ii) *Unit law:* $f \ovee 0_{XY} = f$.

*Moreover, the composition is a suitable 'bihomomorphism', namely, for each* $h : A' \to A$ *and* $k : B \to B'$ *we have:*

(iii) *If* $f \perp g$, *then* $(f \ovee g) \circ h = f \circ h \ovee g \circ h$ *and* $k \circ (f \ovee g) = k \circ f \ovee k \circ g$.

(iv) $0_{XY} \circ h = 0_{X'Y}$ *and* $k \circ 0_{XY} = 0_{XY'}$.

*Finally, the following conditions hold.*

(v) *Compatible sum axiom: If* $f, g \colon A \to B$ *are compatible, then they are summable.*

(vi) *Untying axiom: If* $f, g \colon A \to B$ *are summable, then* $\kappa_1 \circ f, \kappa_2 \circ g \colon A \to B + B$ *is summable too.*

*Proof.*

(i) If $f \perp g$, then $g \perp f$ via $\langle\!\langle g, f \rangle\!\rangle = [\kappa_2, \kappa_1] \circ \langle\!\langle f, g \rangle\!\rangle$. Moreover, $g \ovee f = \nabla \circ [\kappa_2, \kappa_1] \circ \langle\!\langle f, g \rangle\!\rangle = \nabla \circ \langle\!\langle f, g \rangle\!\rangle = f \ovee g$.

(ii) One has $f \perp 0_{XY}$ via $\langle\!\langle 0, f \rangle\!\rangle = \kappa_2 \circ f$, and $f \ovee 0_{XY} = \nabla_B \circ \kappa_2 \circ f = f$.

(iii) If $f \perp g$, by Lemma 3.8.3 we have $f \circ h \perp g \circ h$ and $k \circ f \perp k \circ g$, with $\langle\!\langle f \circ h, g \circ h \rangle\!\rangle = \langle\!\langle f, g \rangle\!\rangle \circ h$ and $\langle\!\langle k \circ f, k \circ g \rangle\!\rangle = (k + k) \circ \langle\!\langle f, g \rangle\!\rangle$. Then $f \circ h \ovee g \circ h = \nabla_B \circ \langle\!\langle f, g \rangle\!\rangle \circ h = (f \ovee g) \circ h$ and $k \circ f \ovee k \circ f = \nabla_{B'} \circ (k + k) \circ \langle\!\langle f, g \rangle\!\rangle = k \circ \nabla_B \circ \langle\!\langle f, g \rangle\!\rangle = k \circ (f \ovee g)$.

(iv) Immediate by definition.

(v) Immediate by definition.

(vi) if $f, g \colon A \to B$ are summable, i.e. $\langle\!\langle f, g \rangle\!\rangle$ exists, then $\kappa_1 \circ f \perp \kappa_2 \circ g$ via $\langle\!\langle \kappa_1 \circ f, \kappa_2 \circ g \rangle\!\rangle = (\kappa_1 + \kappa_2) \circ \langle\!\langle f, g \rangle\!\rangle$. ■

This leads to the following characterization of finPACs.

**Theorem 3.8.5** (cf. [7, § 5.3]). *Let* **C** *be a butterfly coproduct category. The following are equivalent.*

(i) **C** *is a finPAC.*

(ii) *For each* $A, B \in \mathbf{C}$, *the operation* $\ovee$ *on* $\mathbf{C}(A, B)$ *defined by* (3.8) *satisfies associativity (see Definition* 2.2.1*).*

(iii) *For each* $A \in \mathbf{C}$ *the following square is a pullback.*

$$\begin{array}{ccc} (A + A) + A & \xrightarrow{\nabla + \mathrm{id}} & A + A \\ \downarrow{\scriptstyle \rhd_1} & & \downarrow{\scriptstyle \rhd_1} \\ A + A & \xrightarrow{\quad\nabla\quad} & A \end{array}$$

*Proof.* Implication (ii) $\Longrightarrow$ (i) is clear by Lemma 3.8.4.

(i) $\Longrightarrow$ (iii): To prove the pullback condition, let $f, g \colon B \to A + A$ be morphisms with $\nabla \circ f = \rhd_1 \circ g$. Let $f_i = \rhd_i \circ f$ and $g_i = \rhd_i \circ g$ $(i = 1, 2)$. Then $f_1 \perp f_2$, $g_1 \perp g_2$, and $f_1 \ovee f_2 = \nabla \circ f = \rhd_1 \circ g = g_1$, so that $f_1, f_2, g_2$ are summable. By (ternary) untying, $\kappa_1 \circ f_1, \kappa_2 \circ f_2, \kappa_3 \circ g_2 \colon B \to A + A + A$ are summable. Writing $\alpha \colon A + A + A \to (A + A) + A$ for the associativity isomorphism, we define $h \colon B \to (A + A) + A$ by

$$\begin{aligned} h &= \alpha \circ (\kappa_1 \circ f_1 \ovee \kappa_2 \circ f_2 \ovee \kappa_3 \circ g_2) \\ &= \kappa_1 \circ \kappa_1 \circ f_1 \ovee \kappa_1 \circ \kappa_2 \circ f_2 \ovee \kappa_2 \circ g_2 \\ &= \kappa_1 \circ f \ovee \kappa_2 \circ g_2 \qquad (\text{by } f = \kappa_1 \circ f_1 \ovee \kappa_2 \circ f_2) \,. \end{aligned}$$

Then we have $\rhd_1 \circ h = f$ immediately, and

$$(\nabla + \mathrm{id}) \circ h = \kappa_1 \circ \nabla \circ f \ovee \kappa_2 \circ g_2 = \kappa_1 \circ g_1 \ovee \kappa_2 \circ g_2 = g\,.$$

Hence $h$ is a desired mediating map. To see the uniqueness, let $k, k' \colon B \to (A+A)+A$ be morphisms with $\rhd_1 \circ k = f = \rhd_1 \circ k$ and $(\nabla + \mathrm{id}) \circ k = g = (\nabla + \mathrm{id}) \circ k'$. Then $\rhd_2 \circ k = \rhd_2 \circ (\nabla + \mathrm{id}) \circ k = \rhd_2 \circ (\nabla + \mathrm{id}) \circ k' = \rhd_2 \circ k'$. Thus $k = k'$ by the joint monicity of partial projections.

(iii) $\Longrightarrow$ (ii): Let $f, g, h \in \mathbf{C}(A, B)$ be morphisms with $f \perp g$ and $f \ovee g \perp h$. By definition, $\nabla \circ \langle\!\langle f, g \rangle\!\rangle = f \ovee g = \rhd_1 \circ \langle\!\langle f \ovee g, h \rangle\!\rangle$, so that we obtain a mediating map $k$ as in the diagram:

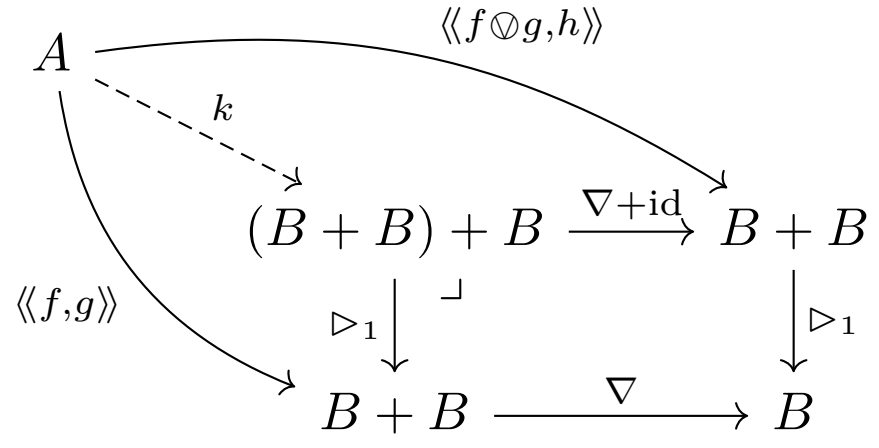

It is straightforward to check that

$$\langle\!\langle g, h \rangle\!\rangle = \big(A \xrightarrow{k} (B+B)+B \xrightarrow{\rhd_2 + \mathrm{id}} B+B\big)$$
$$\langle\!\langle f, g \ovee h \rangle\!\rangle = \big(A \xrightarrow{k} (B+B)+B \xrightarrow{[\mathrm{id}, \kappa_2]} B+B\big)\,,$$

and hence $g \perp h$ and $f \perp g \ovee h$. Moreover we have

$$\begin{aligned} f \ovee (g \ovee h) &= \nabla \circ [\mathrm{id}_{B+B}, \kappa_2] \circ k \\ &= [\nabla, \mathrm{id}_B] \circ k \\ &= \nabla \circ (\nabla + \mathrm{id}_B) \circ k \\ &= \nabla \circ \langle\!\langle f \ovee g, h \rangle\!\rangle \\ &= (f \ovee g) \ovee h\,. \end{aligned}$$

■

### 3.8.2 A characterization of effectuses

We give a characterization of an effectus based on the characterization of finPACs in §3.8.1. There we have shown that a finPAC is a butterfly coproduct category with certain additional requirements, see Theorem 3.8.5. It turns out that the additional requirements become redundant when we have other axioms for an effectus.

**Proposition 3.8.6.** *Let* $\mathbf{C}$ *be a butterfly coproduct category. Then* $\mathbf{C}$ *with an object* $I \in \mathbf{C}$ *and a family of 'truth' maps* $\mathbb{1}_A \colon A \to I$ *is an effectus if and only if all the following hold.*

(E′1) $\mathbb{1}_{A+B} = [\mathbb{1}_A, \mathbb{1}_B] \colon A + B \to I$ *for all* $A, B$.

(E′2) $\mathbb{1}_B \circ f = 0_{AI}$ *implies* $f = 0_{AB}$ *for all* $f \colon A \to B$.

(E′3) $\mathbb{1}_B \circ f \perp \mathbb{1}_B \circ g$ *implies* $f \perp g$ *for all* $f, g \colon A \to B$.

(E′4) *For each* $p\colon A \to I$*, there exists a unique* $p^\perp\colon A \to I$ *such that* $p \perp p^\perp$ *and* $p \ovee p^\perp \equiv \nabla_I \circ \langle\!\langle p, p^\perp \rangle\!\rangle = \mathbb{1}_A$*.*

*Proof.* For the 'only if' part, (E′1) follows from Lemma 3.2.4(iv), and the other conditions are a part of the definition of an effectus. We are going to prove the 'if' part.

First note that any codiagonal map $\nabla_A = [\mathrm{id}_A, \mathrm{id}_A]\colon A + A \to A$ is total in the sense that $\mathbb{1}_A \circ \nabla_A = \mathbb{1}_{A+A}$, because, for $i = 1, 2$:

$$\mathbb{1}_A \circ \nabla_A \circ \kappa_i = \mathbb{1}_A \circ \mathrm{id}_A = \mathbb{1}_A \overset{(\mathrm{E}'1)}{=} \mathbb{1}_{A+A} \circ \kappa_i \,.$$

By Theorem 3.8.5, to prove that $\mathbf{C}$ is a finPAC it suffices to show that $\ovee$ is associative. Assume $f \perp g$ and $f \ovee g \perp h$ for morphisms $f, g, h\colon A \to B$. Note that $\mathbb{1}_B \circ (f \ovee g) \perp \mathbb{1}_B \circ h$ by Lemma 3.8.3. By condition (E′3) and

$$\begin{aligned} &\mathbb{1}_{B+B} \circ \langle\!\langle f, g \rangle\!\rangle = \mathbb{1}_B \circ \nabla_B \circ \langle\!\langle f, g \rangle\!\rangle = \mathbb{1}_B \circ (f \ovee g) \\ &\mathbb{1}_{B+B} \circ \kappa_2 \circ h = \mathbb{1}_B \circ h \end{aligned}$$

we obtain $\langle\!\langle f, g \rangle\!\rangle \perp \kappa_2 \circ h$. It follows that $g \perp h$, by Lemma 3.8.3 again as

$$(\rhd_2 + \rhd_2) \circ \big\langle\!\big\langle \langle\!\langle f, g \rangle\!\rangle, \kappa_2 \circ h \big\rangle\!\big\rangle = \big\langle\!\big\langle \rhd_2 \circ \langle\!\langle f, g \rangle\!\rangle, \rhd_2 \circ \kappa_2 \circ h \big\rangle\!\big\rangle = \langle\!\langle g, h \rangle\!\rangle \,.$$

Note that the sum $\langle\!\langle f, g \rangle\!\rangle \ovee \kappa_2 \circ h\colon A \to B + B$ exists and

$$\begin{aligned} \rhd_1 \circ (\langle\!\langle f, g \rangle\!\rangle \ovee \kappa_2 \circ h) &= f \ovee 0_{XY} = f \\ \rhd_2 \circ (\langle\!\langle f, g \rangle\!\rangle \ovee \kappa_2 \circ h) &= g \ovee h \,. \end{aligned}$$

Hence $f \perp g \ovee h$ with $\langle\!\langle f, g \rangle\!\rangle \ovee \kappa_2 \circ h = \langle\!\langle f, g \ovee h \rangle\!\rangle$. Then

$$f \ovee (g \ovee h) = \nabla_B \circ \langle\!\langle f, g \ovee h \rangle\!\rangle = \nabla_B \circ (\langle\!\langle f, g \rangle\!\rangle \ovee \kappa_2 \circ h) = (f \ovee g) \ovee h \,,$$

as desired. Therefore $\mathbf{C}$ is a finPAC.

It only remains to prove that the homset $\mathbf{C}(A, I)$ is an effect algebra for each $A$. Note first that $\mathbf{C}(A, I)$ is positive, namely: $p \ovee q = 0_{AI}$ implies $p = q = 0_{AI}$. Indeed, if $p \ovee q = 0_{AI}$ then

$$0_{XI} = \mathbb{1}_I \circ 0_{AI} = \mathbb{1}_I \circ \nabla_I \circ \langle\!\langle p, q \rangle\!\rangle = \mathbb{1}_{I+I} \circ \langle\!\langle p, q \rangle\!\rangle \,,$$

so that $\langle\!\langle p, q \rangle\!\rangle = 0_{A,I+I}$ by (E′2). Thus $p = \rhd_1 \circ \langle\!\langle p, q \rangle\!\rangle = 0_{XI}$, and similarly $q = 0_{XI}$. Finally, assume $p \perp \mathbb{1}_A$ and let $q = p \ovee \mathbb{1}_A$. We have $p \ovee \mathbb{1}_A \ovee q^\perp = \mathbb{1}_A$ by (E′4). But then $p \ovee q^\perp = 0_{AI}$ by the uniqueness of orthosupplements. By positivity $p = 0_{AI}$ and we conclude that $\mathbf{C}(A, I)$ is an effect algebra. ■

It is now easy to see when a subcategory of an effectus is a 'sub-effectus'.

**Proposition 3.8.7.** *Let* $(\mathbf{C}, I)$ *be an effectus. Then a subcategory* $\mathbf{D} \subseteq \mathbf{C}$ *is an effectus if the following conditions hold.*

(i) $I \in \mathbf{D}$*,* $0 \in \mathbf{D}$*, and* $A + B \in \mathbf{D}$ *for each* $A, B \in \mathbf{D}$*.*

(ii) *For each $A \in \mathbf{D}$ the truth map $\mathbb{1}_A \colon A \to I$ belongs to $\mathbf{D}$.*

(iii) *For each $A, B \in \mathbf{D}$ the zero morphism $0_{AB} \colon A \to B$ belongs to $\mathbf{D}$.*

(iv) *For each $A, B \in \mathbf{D}$, the coprojections $\kappa_1 \colon A \to A + B$ and $\kappa_2 \colon B \to A + B$ belong to $\mathbf{D}$.*

(v) *For each $f \colon A \to C$ and $g \colon B \to C$ in $\mathbf{D}$, the cotuple $[f, g] \colon A + B \to C$ belongs to $\mathbf{D}$.*

(vi) *For each $h \colon A \to B$ and $k \colon A \to C$ in $\mathbf{D}$, the tuple $\langle\!\langle h, k \rangle\!\rangle \colon A \to B + C$ belongs to $\mathbf{D}$ whenever $\langle\!\langle h, k \rangle\!\rangle$ exists in $\mathbf{C}$.*

(vii) *For each $p \colon A \to I$ in $\mathbf{D}$, $p^{\perp}$ belongs to $\mathbf{D}$.*

*Proof.* The conditions (i), (iii), (iv) and (v) ensure that the subcategory $\mathbf{D}$ inherits finite coproducts and zero morphisms from $\mathbf{C}$. Therefore $\mathbf{D}$ also inherits partial projections such as $\rhd_1 = [\mathrm{id}_A, 0_{BA}] \colon A + B \to A$. Clearly partial projections are jointly monic in $\mathbf{D}$ too, and thus $\mathbf{D}$ is a butterfly coproduct category.

By (ii), $\mathbf{D}$ also inherits the truth maps. To prove that $\mathbf{D}$ is an effectus, we apply Proposition 3.8.6. It is obvious that (E′1) and (E′2) in Proposition 3.8.6 hold in $\mathbf{D}$. Note that by (vi) morphisms $h \colon A \to B$ and $k \colon A \to C$ are compatible in $\mathbf{D}$ if and only if $h$ and $k$ are compatible in $\mathbf{C}$. Therefore (E′3) holds in $\mathbf{D}$. Finally, condition (vii) guarantees that (E′4) holds in $\mathbf{D}$. ■

Note in particular the following special case.

**Corollary 3.8.8.** *Let $(\mathbf{C}, I)$ be an effectus. If $\mathbf{D}$ is a full subcategory of $\mathbf{C}$ such that $I \in \mathbf{D}$, $0 \in \mathbf{D}$, and $A + B \in \mathbf{D}$ for each $A, B \in \mathbf{D}$, then $(\mathbf{D}, I)$ is an effectus.* ■

By the characterization of Proposition 3.8.6, the following result can easily be verified.

**Proposition 3.8.9.** *Let $(\mathbf{C}, I_{\mathbf{C}})$ and $(\mathbf{D}, I_{\mathbf{D}})$ be effectuses. Then the product category $\mathbf{C} \times \mathbf{D}$ with unit $I = (I_{\mathbf{C}}, I_{\mathbf{D}})$ and truth maps $\mathbb{1}_{(A,B)} = (\mathbb{1}_A, \mathbb{1}_B)$ is an effectus.* ■

### 3.8.3 Deferred proofs

*Proof of Proposition* 3.4.10. Let $E$ be an effect module over $M$. The partial projections $\rhd_1, \rhd_2 \colon E + E \to E$ in $M$-$\mathbf{EMod}_{\le}^{\mathrm{op}}$, i.e. subunital module maps $\rhd_1, \rhd_2 \colon E \to E \times E$, are given by $\rhd_1(x) = (x, 0)$ and $\rhd_2(x) = (0, x)$, respectively. To prove that they are jointly epic in $M$-$\mathbf{EMod}_{\le}$, let $f, g \colon E \times E \to D$ be morphisms with $f \circ \rhd_j = g \circ \rhd_j$ for $j = 1, 2$. Then for any $x, y \in E$

$$\begin{aligned} f(x, y) &= f((x, 0) \ovee (0, y)) \\ &= f(x, 0) \ovee f(0, y) \\ &= (f \circ \rhd_1)(x) \ovee (f \circ \rhd_2)(y) \\ &= (g \circ \rhd_1)(x) \ovee (g \circ \rhd_2)(y) = \cdots = g(x, y)\,, \end{aligned}$$

so that $f = g$. Thus $\rhd_1, \rhd_2$ are jointly monic in $M$-$\mathbf{EMod}_{\le}^{\mathrm{op}}$, and hence $M$-$\mathbf{EMod}_{\le}^{\mathrm{op}}$ is a butterfly coproduct category.

Recall that we choose $M$ as a unit object and define truth maps $\mathbb{1}_E\colon M \to E$ (in $M$-$\mathbf{EMod}_{\leq}$) by $\mathbb{1}_E(s) = s \cdot 1$. We prove that $M$-$\mathbf{EMod}_{\leq}^{\mathrm{op}}$ satisfies the conditions in Proposition 3.8.6, one by one.

(E′1) For effect modules $E, D$ we have

$$\begin{aligned}\mathbb{1}_{E\times D}(s) &= s \cdot (1_E, 1_D)\\ &= (s\cdot 1_E, s\cdot 1_D)\\ &= (\mathbb{1}_E(s), \mathbb{1}_D(s))\\ &= \langle \mathbb{1}_E, \mathbb{1}_D\rangle(s)\,.\end{aligned}$$

Thus $\mathbb{1}_{E\times D} = \langle \mathbb{1}_E, \mathbb{1}_D\rangle$, i.e. $\mathbb{1}_{E\times D} = [\mathbb{1}_E, \mathbb{1}_D]$ in the opposite category.

(E′2) Let $f\colon E \to D$ be a morphism with $f \circ \mathbb{1}_E = 0$ in $M$-$\mathbf{EMod}_{\leq}$. We have

$$f(1) = (f \circ \mathbb{1}_E)(1) = 0(1) = 0\,.$$

Then for any $x \in E$ we have $0 = f(0) \leq f(x) \leq f(1) = 0$, so $f(x) = 0$. Thus $f = 0$.

(E′3) Two morphisms $f, g\colon D \to E$ in $M$-$\mathbf{EMod}_{\leq}^{\mathrm{op}}$, i.e. $f, g\colon E \to D$ in $M$-$\mathbf{EMod}_{\leq}$, are compatible, $f \perp g$, if and only if there is $h\colon D \to E \times E$ in $M$-$\mathbf{EMod}_{\leq}$ with $h \circ \rhd_1 = f$ and $h \circ \rhd_2 = g$, i.e. $h(x, 0) = f(x)$ and $h(0, x) = g(x)$ for all $x \in E$. We claim that $f \perp g$ iff $f(1) \perp g(1)$. Indeed, if $f \perp g$ then $f(1) = h(1, 0)$ and $g(1) = h(0, 1)$ are summable. Conversely, if $f(1) \perp g(1)$, then $f(x) \perp g(x)$ for any $x \in E$ and we can define a mapping $\langle\!\langle f, g\rangle\!\rangle\colon E \times E \to D$ by $\langle\!\langle f, g\rangle\!\rangle(x, y) = f(x) \ovee g(y)$. This is a subunital module maps and satisfies $\langle\!\langle f, g\rangle\!\rangle \circ \rhd_1 = f$ and $\langle\!\langle f, g\rangle\!\rangle \circ \rhd_2 = g$, so that $f \perp g$. Now the condition in question is almost obvious: if $f \circ \mathbb{1}_E \perp g \circ \mathbb{1}_E$, then $f(1) = (f \circ \mathbb{1}_E)(1) \perp (g \circ \mathbb{1}_E)(1) = g(1)$ and thus $f \perp g$.

(E′4) This holds since $M$-$\mathbf{EMod}_{\leq}^{\mathrm{op}}(E, M) = M$-$\mathbf{EMod}_{\leq}(M, E) \cong E$, with a bijection that respects $0, 1, \ovee$. ■

*Proof of Proposition* 3.5.9. Let $X$ be a weight module over $M$. The partial projections $\rhd_1, \rhd_2\colon X + X \to X$ in $M$-$\mathbf{WMod}_{\leq}$ are defined by $\rhd_1(x, x') = x$ and $\rhd_2(x, x') = x'$. Let $f, g\colon X' \to X + X$ be morphisms in $M$-$\mathbf{WMod}_{\leq}$ such that $\rhd_j \circ f = \rhd_j \circ g$ for $j = 1, 2$. Since for any $(x, x') \in X + X$,

$$(x, x') = (x, 0) \ovee (0, x') = (\rhd_1(x, x'), 0) \ovee (0, \rhd_2(x, x'))\,,$$

for all $y \in Y$ we have

$$f(y) = (\rhd_1(f(y)), 0) \ovee (0, \rhd_2(f(y))) = (\rhd_1(g(y)), 0) \ovee (0, \rhd_2(g(y))) = g(y)\,.$$

Therefore the partial projections are jointly monic, and $M$-$\mathbf{WMod}_{\leq}$ is a butterfly coproduct category.

We prove that $M$-$\mathbf{WMod}_{\leq}$ satisfies the conditions in Proposition 3.8.6, one by one.

(E′1) For any weight module $X, Y$,

$$\begin{aligned}\mathbb{1}_{X+Y}(x,y) &= |(x,y)| \\ &= |x| \ovee |y| \\ &= \mathbb{1}_X(x) \ovee \mathbb{1}_Y(y) \\ &= [\mathbb{1}_X, \mathbb{1}_Y](x,y)\,.\end{aligned}$$

Hence $\mathbb{1}_{X+Y} = [\mathbb{1}_X, \mathbb{1}_Y]$.

(E′2) Let $f\colon X \to Y$ satisfy $\mathbb{1}_Y \circ f = 0$. For any $x \in X$ we have $|f(x)| = \mathbb{1}_Y(f(x)) = 0$, so that $f(x) = 0$. Hence $f = 0$.

(E′3) Let $f, g\colon X \to Y$ be morphisms in $M$-**WMod**. We claim that $f \perp g$ in $M$-**WMod** (i.e. $\langle\!\langle f, g\rangle\!\rangle$ exists) if and only if $|f(x)| \ovee |g(x)| \le |x|$ for all $x \in X$. Suppose that there exists a partial tuple $\langle\!\langle f, g\rangle\!\rangle\colon X \to Y + Y$ satisfying $\rhd_1 \circ \langle\!\langle f, g\rangle\!\rangle = f$ and $\rhd_2 \circ \langle\!\langle f, g\rangle\!\rangle = g$. Fix $x \in X$ and let $(y_1, y_2) = \langle\!\langle f, g\rangle\!\rangle(x)$. Then

$$f(x) = \rhd_1(\langle\!\langle f, g\rangle\!\rangle(x)) = \rhd_1(y_1, y_2) = y_1$$

and similarly $g(x) = y_2$, so that $\langle\!\langle f, g\rangle\!\rangle(x) = (f(x), g(x))$. Since $\langle\!\langle f, g\rangle\!\rangle$ is weight-decreasing,

$$|x| \ge |\langle\!\langle f, g\rangle\!\rangle(x)| = |(f(x), g(x))| = |f(x)| \ovee |g(x)|$$

as desired. Conversely, assume $|f(x)| \ovee |g(x)| \le |x|$ for all $x \in X$. Then in particular $|f(x)| \perp |g(x)|$ and hence $(f(x), g(x)) \in Y + Y$ for each $x \in X$, so we can define a function $\langle\!\langle f, g\rangle\!\rangle\colon X \to Y + Y$ by

$$\langle\!\langle f, g\rangle\!\rangle(x) = (f(x), g(x))\,.$$

The assumption guarantees that $\langle\!\langle f, g\rangle\!\rangle$ is weight-decreasing: $|\langle\!\langle f, g\rangle\!\rangle(x)| = |f(x)| \ovee |g(x)| \le |x|$. It preserves the sum:

$$\begin{aligned}\langle\!\langle f, g\rangle\!\rangle(x \ovee y) &= (f(x \ovee y), g(x \ovee y)) \\ &= (f(x) \ovee f(y), g(x) \ovee g(y)) \\ &= (f(x), g(x)) \ovee (f(y), g(y)) \\ &= \langle\!\langle f, g\rangle\!\rangle(x) \ovee \langle\!\langle f, g\rangle\!\rangle(x)(y)\,.\end{aligned}$$

Similarly it preserves 0 and the scalar multiplication, showing that $\langle\!\langle f, g\rangle\!\rangle$ is a morphism in $M$-**WMod**. Clearly it satisfies $\rhd_1 \circ \langle\!\langle f, g\rangle\!\rangle = f$ and $\rhd_2 \circ \langle\!\langle f, g\rangle\!\rangle = g$. Therefore $f \perp g$.

(E′4) Let $p\colon X \to M$ be a morphism in $M$-**WMod**. Then $p(x) = |p(x)| \le |x|$, so we define $p^\perp\colon X \to M$ by $p^\perp(x) = |x| \ominus p(x)$. Clearly $p^\perp$ is weight-decreasing. It preserves the sum:

$$\begin{aligned}p^\perp(x \ovee y) &= |x \ovee y| \ominus p(x \ovee y) \\ &= (|x| \ovee |y|) \ominus (p(x) \ovee p(y)) \\ &= (|x| \ominus p(x)) \ovee (|y| \ominus p(y)) \\ &= p^\perp(x) \ovee p^\perp(y)\,.\end{aligned}$$

Similarly it preserves 0 and the scalar multiplication, so that it is a weight-decreasing module map. We see that $p \perp p^\perp$ by the characterization of compatibility given above. Note that $f \ovee g = \nabla \circ \langle\!\langle f, g \rangle\!\rangle$ is given pointwise: $(f \ovee g)(x) = f(x) \ovee g(x)$. Thus we have

$$(p \ovee p^\perp)(x) = p(x) \ovee (|x| \ominus p(x)) = |x| = \mathbb{1}_X(x)\,,$$

so $p \ovee p^\perp = \mathbb{1}_X$. It is clear that such a morphism $p^\perp$ is unique. ■

# Chapter 4

# Total Morphisms in Effectuses

In this chapter we focus on total morphisms in an effectus. We introduce *effectuses in total form* — the original formulation of effectus given by Jacobs [140] — which characterize the subcategories $\mathrm{Tot}(\mathbf{C})$ of effectuses $\mathbf{C}$ determined by total morphisms. Because any morphism $f\colon A \to B$ in an effectus $\mathbf{C}$ can be represented by a total morphism of type $A \to B + I$, the effectus $\mathbf{C}$ can be recovered from the subcategory $\mathrm{Tot}(\mathbf{C})$ as the Kleisli category of the lift monad $(-) + 1$. Therefore effectuses in total form are equivalent to our formulation of effectuses 'in partial form'. In Section 4.2 we will make the 'equivalence' more precise, formulating it as a 2-equivalence of the 2-categories of effectuses in partial form and in total form.

The chapter then continues the study of states and their convex structure, relating them to the weight module structure of substates. To do so, we introduce an additional assumption of *division* on effect monoids, which is discussed in Section 4.3. In Section 4.4 we study convex sets and weight modules over a division effect monoid. Assuming that the scalars admit division, we show that the category of weight modules with the *normalization property* is equivalent to the category of convex sets. These results will be applied to *effectuses with the normalization property* in Section 4.5. The two state-and-effect triangles over an effectus, with substates and states, are shown to be related via the 2-equivalence of effectuses in total and partial form.

## 4.1 Effectuses in total form

In this section we focus on the subcategory $\mathrm{Tot}(\mathbf{C})$, consisting of total morphisms, of an effectus $\mathbf{C}$. This leads to the notion of *effectuses in total form*, see Definition 4.1.6. As the name suggests, it gives an alternative formulation of effectuses.

We start with a few basic properties of $\mathrm{Tot}(\mathbf{C})$.

**Lemma 4.1.1.** *Let* $\mathbf{C}$ *be an effectus. The category* $\mathrm{Tot}(\mathbf{C})$ *inherits all coproducts that exist in* $\mathbf{C}$*. In particular,* $\mathrm{Tot}(\mathbf{C})$ *has all finite coproducts.*

*Proof.* Let $\coprod_j A_j$ be a coproduct in $\mathbf{C}$. Since the coprojections $\kappa_j\colon \coprod_j A_j \to A_j$ are total, the coproduct diagram lies in $\mathrm{Tot}(\mathbf{C})$. Let $(f_j\colon A_j \to B)_j$ be a family of total morphisms. Then the cotuple $[f_j]_j\colon \coprod_j A_j \to B$ satisfies

$$\mathbb{1} \circ [f_j]_j \circ \kappa_j = \mathbb{1} \circ f_j = \mathbb{1}_{A_j} = \mathbb{1} \circ \kappa_j$$

for all $j$. Thus $\mathbb{1} \circ [f_j]_j = \mathbb{1}$, i.e. the cotuple $[f_j]_j$ is total. The mediating map $[f_j]_j$ is unique in $\mathbf{C}$ and hence in $\mathrm{Tot}(\mathbf{C})$, so $\coprod_j A_j$ is a coproduct in $\mathrm{Tot}(\mathbf{C})$. ∎

**Lemma 4.1.2.** *Let* $\mathbf{C}$ *be an effectus. The unit object* $I$ *of* $\mathbf{C}$ *is final in* $\mathrm{Tot}(\mathbf{C})$.

*Proof.* For each object $A$, the truth map $\mathbb{1}\colon A \to I$ is a unique total map of this type. ∎

The following observation relates any 'partial' morphisms to total ones.

**Lemma 4.1.3.** *In an effectus* $\mathbf{C}$, *there is the following bijective correspondence:*

$$\frac{\text{a morphism } f\colon A \longrightarrow B}{\text{a total morphism } g\colon A \longrightarrow B+I}$$

*given by* $g = \langle\!\langle f, (\mathbb{1}f)^{\perp}\rangle\!\rangle$ *and* $f = \rhd_1 \circ g$.

*Proof.* First note that for any $f\colon A \to B$,

$$\langle\!\langle f, (\mathbb{1}f)^{\perp}\rangle\!\rangle = \kappa_1 \circ f \ovee \kappa_2 \circ (\mathbb{1}f)^{\perp}$$

is defined and total by Lemma 3.2.5, since $\mathbb{1} = \mathbb{1}f \ovee (\mathbb{1}f)^{\perp} = \mathbb{1} \circ f \ovee \mathbb{1} \circ (\mathbb{1}f)^{\perp}$. We check that the correspondence between $f$ and $g$ is bijective. It is clear that $\rhd_1 \circ \langle\!\langle f, (\mathbb{1}f)^{\perp}\rangle\!\rangle = f$. For the other way,

$$\begin{aligned}
\langle\!\langle \rhd_1 \circ g, (\mathbb{1} \circ \rhd_1 \circ g)^{\perp}\rangle\!\rangle &= \kappa_1 \circ \rhd_1 \circ g \ovee \kappa_2 \circ (\mathbb{1} \circ \rhd_1 \circ g)^{\perp} \\
&= \kappa_1 \circ \rhd_1 \circ g \ovee \kappa_2 \circ \rhd_2 \circ g \\
&= g\,.
\end{aligned}$$

Here we used $(\mathbb{1} \circ \rhd_1 \circ g)^{\perp} = \rhd_2 \circ g$, which holds because

$$\begin{aligned}
\mathbb{1} = \mathbb{1}g &= \mathbb{1} \circ (\kappa_1 \circ \rhd_1 \circ g \ovee \kappa_2 \circ \rhd_2 \circ g) \\
&= \mathbb{1} \circ \rhd_1 \circ g \ovee \mathbb{1}_I \circ \rhd_2 \circ g \\
&= \mathbb{1} \circ \rhd_1 \circ g \ovee \rhd_2 \circ g\,.
\end{aligned}$$

We have shown the desired bijective correspondence. ∎

By Lemmas 4.1.1 and 4.1.2, $\mathrm{Tot}(\mathbf{C})$ has finite coproducts and the final object $I$. We introduce a few definitions for such categories.

**Definition 4.1.4.** Let $\mathbf{B}$ be a category with finite coproducts $(0,+)$ and a final object 1.

(i) The **lift monad** on $\mathbf{B}$ is a monad defined by $A \mapsto A+1$. It maps a morphism $f\colon A \to B$ to $f+1\colon A+1 \to B+1$. The unit and multiplication are given respectively by the following morphisms.

$$\kappa_1\colon A \to A+1 \qquad\qquad [\mathrm{id}_{A+1}, \kappa_2]\colon (A+1)+1 \to A+1$$

(ii) We denote by $\mathrm{Par}(\mathbf{B})$ the Kleisli category of the lift monad on $\mathbf{B}$. We call morphisms in $\mathrm{Par}(\mathbf{B})$ **partial maps** and write them as $f\colon A \rightharpoonup B$. Explicitly, $\mathrm{Par}(\mathbf{B})$ has the same objects as $\mathbf{B}$. Morphisms $f\colon A \rightharpoonup B$ in $\mathrm{Par}(\mathbf{B})$ are morphisms $f\colon A \to B+1$ in $\mathbf{B}$. The identities in $\mathrm{Par}(\mathbf{B})$ are the unit $\kappa_1\colon A \to A+1$ of the monad. Composition is defined and denoted by $g \circledcirc f \coloneqq [g, \kappa_2] \circ f$ for $f\colon A \rightharpoonup B$ and $g\colon B \rightharpoonup C$.

Now we can rephrase Lemma 4.1.3 as follows.

**Proposition 4.1.5.** *For any effectus* $\mathbf{C}$*, the bijective correspondence of Lemma* 4.1.3 *defines an isomorphism of categories* $\mathbf{C} \cong \mathrm{Par}(\mathrm{Tot}(\mathbf{C}))$*. The map is identity on objects.*

*Proof.* It suffices to show that the mapping $\mathrm{Par}(\mathrm{Tot}(\mathbf{C})) \to \mathbf{C}$ given by

$$g\colon A \to B + I \ \longmapsto\ \rhd_1 \circ g\colon A \to B$$

is functorial.

(Identity) The identity in $\mathrm{Par}(\mathrm{Tot}(\mathbf{C}))$ is the coprojection $\kappa_1\colon A \to A + 1$ in $\mathrm{Tot}(\mathbf{C})$ (and $\mathbf{C}$). Then indeed, $\rhd_1 \circ \kappa_1 = \mathrm{id}_A$.

(Composition) Let $f\colon A \rightarrow\!\!\!\!\circ\ B$ and $g\colon B \rightarrow\!\!\!\!\circ\ C$ be morphisms in $\mathrm{Par}(\mathrm{Tot}(\mathbf{C}))$, i.e. total morphisms $f\colon A \to B + I$ and $g\colon B \to C + I$. Then

$$\begin{aligned}
\rhd_1 \circ (g \circledcirc f) &= \rhd_1 \circ [g, \kappa_2] \circ f \\
&= [\rhd_1 \circ g, \rhd_1 \circ \kappa_2] \circ f \\
&= [\rhd_1 \circ g, 0_{II}] \circ f \\
&= [\rhd_1 \circ g, \rhd_1 \circ g \circ 0_{IB}] \circ f \\
&= \rhd_1 \circ g \circ [\mathrm{id}_B, 0_{IB}] \circ f \\
&= (\rhd_1 \circ g) \circ (\rhd_1 \circ f)\,.
\end{aligned}$$

■

The proposition says that the 'partial map' construction $\mathrm{Par}(\mathrm{Tot}(\mathbf{C}))$ recovers the effectus $\mathbf{C}$ from its 'total' part $\mathrm{Tot}(\mathbf{C})$. In particular, any effectus appears as the category $\mathrm{Par}(\mathbf{B})$ of partial maps for some $\mathbf{B}$.

Then the following question arises: can we characterize or axiomatize the categories $\mathrm{Tot}(\mathbf{C})$ of total morphisms for effectuses $\mathbf{C}$, in the way that from such categories $\mathbf{B}$ we can obtain effectuses $\mathrm{Par}(\mathbf{B})$? To answer this question, below we define *effectuses in total form.* They have the name 'effectus' since they turn out to be equivalent, in a suitable categorical sense, to the effectuses defined in Definition 3.2.1. To distinguish the two notions of effectuses, an effectus in the sense of Definition 3.2.1 is also called an **effectus in partial form**.

**Definition 4.1.6.** An **effectus in total form** is a category with finite coproducts $(+, 0)$ and a final object 1 satisfying the following three conditions.

(T1) Diagrams of the form on the left below are pullbacks.

$$\begin{array}{ccccccc}
A + B & \xrightarrow{\mathrm{id}+g} & A + D & \qquad & A & \xrightarrow{\kappa_1} & A + B \\
{\scriptstyle f+\mathrm{id}}\downarrow & (\mathrm{T1}) & \downarrow{\scriptstyle f+\mathrm{id}} & & {\scriptstyle f}\downarrow & (\mathrm{T2}) & \downarrow{\scriptstyle f+g} \\
C + B & \xrightarrow[\mathrm{id}+g]{} & C + D & & C & \xrightarrow[\kappa_1]{} & C + D
\end{array}$$

(T2) Diagrams of the form on the right above are pullbacks.

(T3) The two morphisms below are jointly monic.

$$1 + 1 + 1 \ \underset{[\kappa_2, \kappa_1, \kappa_2]}{\overset{[\kappa_1, \kappa_2, \kappa_2]}{\rightrightarrows}}\ 1 + 1$$

The pullback of (T2) mentions only the first projections $\kappa_1$. Nevertheless, similar pullbacks for $\kappa_2$ can be obtained from (T2) via symmetry of coproducts $A+B \cong B+A$ as follows.

$$
\begin{array}{ccccc}
 & & \overset{\kappa_2}{\frown\!\!\!\longrightarrow} & & \\
B & \xrightarrow{\kappa_1} & B+A & \xrightarrow{\cong} & A+B \\
{\scriptstyle g}\downarrow & \lrcorner & \downarrow{\scriptstyle g+f} & & \downarrow{\scriptstyle f+g} \\
D & \xrightarrow{\kappa_1} & D+C & \xrightarrow{\cong} & C+D \\
 & & \underset{\kappa_2}{\smile\!\!\!\longrightarrow} & &
\end{array}
$$

We will refer to pullbacks of this form also by (T2).

In the remainder of this section, we prove that effectuses in total form are equivalent to effectuses in partial form in the following sense.

1. For any effectus in partial form $\mathbf{C}$, we prove that $\mathrm{Tot}(\mathbf{C})$ is an effectus in total form. Moreover $\mathrm{Par}(\mathrm{Tot}(\mathbf{C})) \cong \mathbf{C}$. (Theorem 4.1.11 and Proposition 4.1.5)

2. For any effectus in total form $\mathbf{B}$, we prove that $\mathrm{Par}(\mathbf{B})$ is an effectus in partial form. Moreover $\mathrm{Tot}(\mathrm{Par}(\mathbf{B})) \cong \mathbf{B}$. (Theorem 4.1.24)

These two also show that effectuses in total form are exactly the class of the categories that appear as $\mathrm{Tot}(\mathbf{C})$, for some effectus in partial form $\mathbf{C}$.

**Remark 4.1.7.** Be warned that the definition of an effectus (in total form) used in [150] is strictly stronger than Definition 4.1.6. An example that separates the two definitions is the category of convex sets (over $[0,1]$), which is an effectus in total form in our sense, but not in the sense of [150]. The definition used in this thesis agrees with the definitions used in the other publications (e.g. [36, 40, 140, 144, 248, 256]). We note that if $\mathbf{C}$ is an effectus in our sense (in partial form) such that each hom-PCM $\mathbf{C}(A,B)$ is cancellative (see § 2.3.1), then $\mathrm{Tot}(\mathbf{C})$ is an effectus in the sense of [150]. It is an open question whether the converse holds.

## 4.1.1 From partial to total form

Let $\mathbf{C}$ be an effectus in partial form. Recall from Lemmas 4.1.1 and 4.1.2 that the category $\mathrm{Tot}(\mathbf{C})$ inherits coproducts form $\mathbf{C}$ and has the final object $1 = I$.

**Lemma 4.1.8** (Condition (T1))**.** *For any total morphisms $f\colon A \to C$ and $g\colon B \to D$, the following diagram is a pullback in* $\mathrm{Tot}(\mathbf{C})$*.*

$$
\begin{array}{ccc}
A+B & \xrightarrow{\mathrm{id}+g} & A+D \\
{\scriptstyle f+\mathrm{id}}\downarrow & & \downarrow{\scriptstyle f+\mathrm{id}} \\
C+B & \xrightarrow{\mathrm{id}+g} & C+D
\end{array}
$$

*Proof.* Assume that total morphisms $h$ and $k$ are given as in the diagram, making the outer diagram commute.

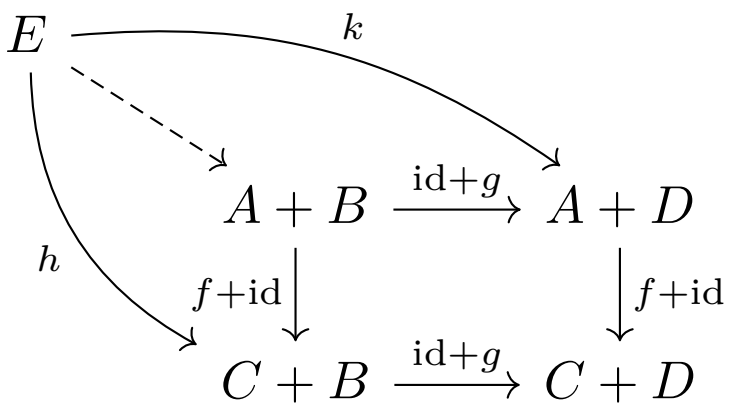

The morphism $h$ can be decomposed as $h = \langle\!\langle h_1, h_2 \rangle\!\rangle$ for $h_1 = \rhd_1 \circ h \colon E \to C$ and $h_2 = \rhd_2 \circ h \colon C \to B$. Similarly $k = \langle\!\langle k_1, k_2 \rangle\!\rangle$ with $k_i = \rhd_i \circ k$. Then

$$\begin{aligned}(\mathrm{id}_C + g) \circ h &= (\mathrm{id}_C + g) \circ \langle\!\langle h_1, h_2 \rangle\!\rangle = \langle\!\langle h_1, g \circ h_2 \rangle\!\rangle \\ (f + \mathrm{id}_D) \circ k &= (f + \mathrm{id}_D) \circ \langle\!\langle k_1, k_2 \rangle\!\rangle = \langle\!\langle f \circ k_1, k_2 \rangle\!\rangle .\end{aligned}$$

We obtain $\langle\!\langle h_1, g \circ h_2 \rangle\!\rangle = \langle\!\langle f \circ k_1, k_2 \rangle\!\rangle$, and hence $h_1 = f \circ k_1$ and $g \circ h_2 = k_2$. Note that

$$\mathbb{1}_C \circ h_1 = \mathbb{1}_C \circ f \circ k_1 = \mathbb{1}_A \circ k_1 .$$

Since $\mathbb{1} h_1 \ovee \mathbb{1} h_2 = \mathbb{1}_E$ we have $\mathbb{1} k_1 \ovee \mathbb{1} h_2 = \mathbb{1}_E$, so we may define the tuple $\langle\!\langle k_1, h_2 \rangle\!\rangle \colon C \to A + B$ that is total. We claim that $\langle\!\langle k_1, h_2 \rangle\!\rangle$ is a desired mediating map, i.e. the 'dashed' map in the diagram above. Indeed it makes the diagram commutes, since

$$(f + \mathrm{id}_B) \circ \langle\!\langle k_1, h_2 \rangle\!\rangle = \langle\!\langle f \circ k_1, h_2 \rangle\!\rangle = \langle\!\langle h_1, h_2 \rangle\!\rangle = h$$

and similarly $(\mathrm{id}_A + g) \circ \langle\!\langle k_1, h_2 \rangle\!\rangle = k$. To see the uniqueness, let $l \colon E \to A + B$ be such that $(f + \mathrm{id}_B) \circ l = h$ and $(\mathrm{id}_A + g) \circ l = k$. Then

$$\begin{aligned}\rhd_1 \circ l &= \rhd_1 \circ (\mathrm{id}_A + g) \circ l = \rhd_1 \circ k = k_1 \\ \rhd_2 \circ l &= \rhd_2 \circ (f + \mathrm{id}_B) \circ l = \rhd_2 \circ h = h_2 ,\end{aligned}$$

so that $l = \langle\!\langle k_1, h_2 \rangle\!\rangle$. ■

**Lemma 4.1.9** (Condition (T2))**.** *For any total morphism $f \colon A \to B$, the following diagram is a pullback in* $\mathrm{Tot}(\mathbf{C})$.

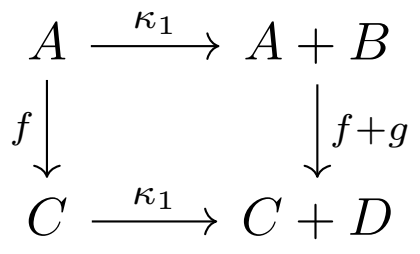

*Proof.* Assume that total morphisms $h$ and $k$ are given as in the diagram, with the outer diagram commutative.

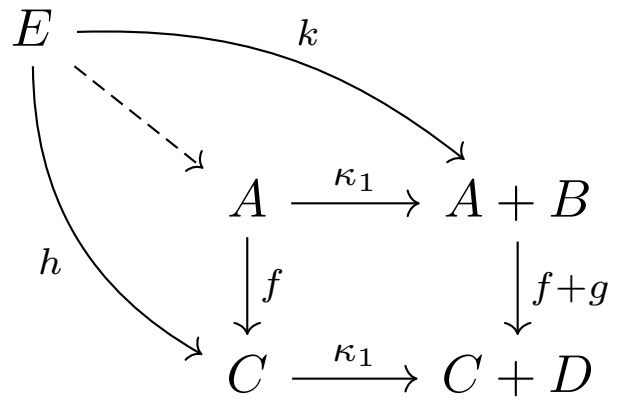

Let $k$ be decomposed as $k = \langle\langle k_1, k_2 \rangle\rangle$ for $k_i = \rhd_i \circ k$. Then

$$(f+g) \circ k = (f+g) \circ \langle\langle k_1, k_2 \rangle\rangle = \langle\langle f \circ k_1, g \circ k_2 \rangle\rangle$$
$$\kappa_1 \circ h = \kappa_1 \circ h \ovee \kappa_2 \circ 0_{ED} = \langle\langle h, 0_{ED} \rangle\rangle .$$

By assumption $\langle\langle f \circ k_1, g \circ k_2 \rangle\rangle = \langle\langle h, 0_{ED} \rangle\rangle$ and hence $f \circ k_1 = h$ and $g \circ k_2 = 0_{ED}$. But then

$$\mathbb{1}_B \circ k_2 = \mathbb{1}_D \circ g \circ k_2 = \mathbb{0}_D ,$$

so that $k_2 = 0_{EB}$. We thus obtain

$$k = \langle\langle k_1, k_2 \rangle\rangle = \kappa_1 \circ k_1 \ovee \kappa_2 \circ 0_{ZY} = \kappa_1 \circ k_1 .$$

Together with $f \circ k_1 = h$, we have shown that $k_1$ is a mediating map for the pullback. The mediating map is unique: if $l \colon E \to A$ satisfies $\kappa_1 \circ l = k$, then $k_1 = \rhd_1 \circ k = \rhd_1 \circ \kappa_1 \circ l = l$. ■

**Lemma 4.1.10** (Condition (T3)). *In the category* Tot(**C**)*, the following maps are jointly monic.*

$$I + I + I \underset{[\kappa_2,\kappa_1,\kappa_2]}{\overset{[\kappa_1,\kappa_2,\kappa_2]}{\rightrightarrows}} I + I$$

*Proof.* Assume that $p, q \colon A \to I + I + I$ in Tot(**C**) satisfies

$$[\kappa_1, \kappa_2, \kappa_2] \circ p = [\kappa_1, \kappa_2, \kappa_2] \circ q \tag{4.1}$$
$$[\kappa_2, \kappa_1, \kappa_2] \circ p = [\kappa_2, \kappa_1, \kappa_2] \circ q \tag{4.2}$$

We decompose $p$ and $q$ as $p = \langle\langle p_1, p_2, p_3 \rangle\rangle$ and $q = \langle\langle q_1, q_2, q_3 \rangle\rangle$. Since $p$ and $q$ are total we have

$$p_1 \ovee p_2 \ovee p_3 = \mathbb{1}_A = q_1 \ovee q_2 \ovee q_3 .$$

Note that equation (4.1) amounts to

$$\kappa_1 \circ p_1 \ovee \kappa_2 \circ p_2 \ovee \kappa_2 \circ p_3 = \kappa_1 \circ q_1 \ovee \kappa_2 \circ q_2 \ovee \kappa_2 \circ q_3 .$$

By composing the partial projection $\rhd_1 \colon I + I \to I$, we obtain $p_1 = q_1$. By a similar reasoning for equation (4.2), we obtain $p_2 = q_2$. Then we have

$$p_3 = (p_1 \ovee p_2)^\perp = (q_1 \ovee q_2)^\perp = q_3 ,$$

concluding that $p = q$. ■

**Theorem 4.1.11.** *For any effectus in partial form* **C***, the subcategory* Tot(**C**) *of total morphisms is an effectus in total form.*

*Proof.* By Lemmas 4.1.1, 4.1.2, 4.1.8, 4.1.9 and 4.1.10. ■

### 4.1.2 From total to partial form

Let **B** be an effectus in total form. Let us start with a property of coproducts in **B**.

**Lemma 4.1.12.** *Any coproduct $A+B$ in* **B** *is disjoint. This means that the coproduct satisfies the following two conditions.*

(i) *The coprojections $\kappa_1\colon A \to A+B$ and $\kappa_2\colon B \to A+B$ are monic.*

(ii) *The following diagram is a pullback.*

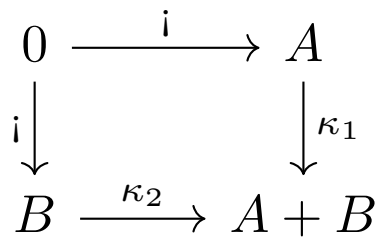

*In other words, the intersection of the coprojections is the initial object.*

*Proof.* The outer diagram below is a pullback, via a pullback (T2) on the left.

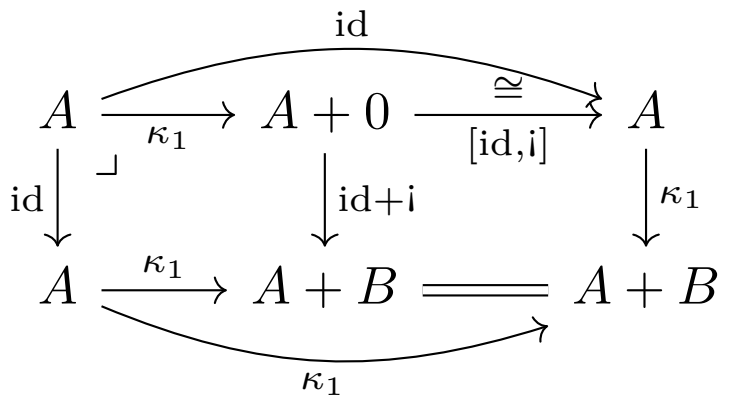

This implies (in fact, is equivalent to saying) that $\kappa_1\colon A \to A+B$ is monic. Similarly the second projection is monic. Point (ii) also follows via a pullback (T2):

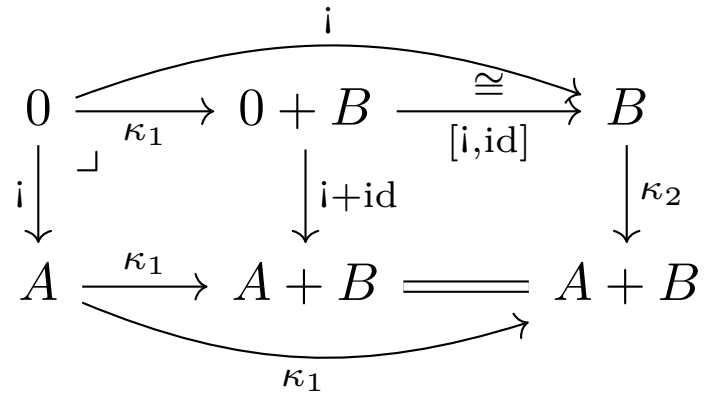

■

Recall from Definition 4.1.4 that Par(**B**) is the Kleisli category of the lift monad $(-)+1$ on **B**. We introduce some notations.

**Definition 4.1.13.** As Par(**B**) is the Kleisli category, there is an identity-on-objects functor $\mathbf{B} \to \mathrm{Par}(\mathbf{B})$ that sends

$$A \xrightarrow{f} B \qquad \text{to} \qquad A \xrightarrow{f} B \xrightarrow{\kappa_1} B+1$$

We denote this functor by $\langle - \rangle$. Explicitly, $\langle A \rangle := A$ and $\langle f \rangle := \kappa_1 \circ f$. By the functoriality, $\langle g \circ f \rangle = \langle g \rangle \circledcirc \langle f \rangle$. The identities in Par(**B**) are $\langle \mathrm{id}_A \rangle = \kappa_1\colon A \to A+1$. When no confusion is likely to arise, we simply write $\mathrm{id}_A$ for the identities in Par(**B**).

**Lemma 4.1.14.** *The functor $\langle - \rangle\colon \mathbf{B} \to \mathrm{Par}(\mathbf{B})$ is faithful.*

Therefore the functor $\langle - \rangle$ embeds $\mathbf{B}$ into $\mathrm{Par}(\mathbf{B})$, so that $\mathbf{B}$ may be seen as a subcategory of $\mathrm{Par}(\mathbf{B})$.

*Proof.* The mapping $f \mapsto \langle f \rangle \equiv \kappa_1 \circ f$ is injective since the coprojections $\kappa_1$ are monic by Lemma 4.1.12. ∎

The following is a general fact that holds for any Kleisli category.

**Lemma 4.1.15.** *The functor* $\langle - \rangle \colon \mathbf{B} \to \mathrm{Par}(\mathbf{B})$ *preserves all coproducts that exist in* $\mathbf{B}$*. In particular,* $\mathrm{Par}(\mathbf{B})$ *has all finite coproducts, since so does* $\mathbf{B}$*.*

*Proof.* Straightforward. ∎

The coprojections $A_i \mathrel{\circ\!\!\!\rightarrow} \coprod_i A_i$ in $\mathrm{Par}(\mathbf{B})$ are thus given by $\langle \kappa_i \rangle = \kappa_1 \circ \kappa_i$, where $\kappa_i \colon A_i \to \coprod_i A_i$ is the coprojection in $\mathbf{B}$. When the context is clear, we simply write $\kappa_i \colon A_i \mathrel{\circ\!\!\!\rightarrow} \coprod_i A_i$ for the coprojections in $\mathrm{Par}(\mathbf{B})$.

**Lemma 4.1.16.** *The category* $\mathrm{Par}(\mathbf{B})$ *has zero morphisms* $0_{AB} \colon A \mathrel{\circ\!\!\!\rightarrow} B$ *given by*

$$A \xrightarrow{!} 1 \xrightarrow{\kappa_2} B + 1 \quad \textit{in } \mathbf{B}.$$

*Proof.* For any $f \colon A \mathrel{\circ\!\!\!\rightarrow} B$ and $g \colon C \mathrel{\circ\!\!\!\rightarrow} D$ in $\mathrm{Par}(\mathbf{B})$,

$$\begin{aligned}
0_{BD} \circledcirc f &= (\kappa_2 \circ !_B) \circledcirc f \\
&= [\kappa_2 \circ !_B, \kappa_2] \circ f \\
&= \kappa_2 \circ [!_B, \mathrm{id}_1] \circ f \\
&= \kappa_2 \circ !_{B+1} \circ f \\
&= \kappa_2 \circ !_A \; (\equiv 0_{AD}) \\
&= [g, \kappa_2] \circ \kappa_2 \circ !_A \\
&= g \circledcirc (\kappa_2 \circ !_A) = g \circledcirc 0_{AC} \, .
\end{aligned}$$

∎

**Lemma 4.1.17.** *For any morphisms* $f \colon A \to C$ *and* $g \colon B \to D$ *in* $\mathbf{B}$*, the following diagrams are pullbacks in* $\mathrm{Par}(\mathbf{B})$*.*

$$\begin{array}{ccc}
A + B & \xrightarrow{\mathrm{id} \hat{+} \langle g \rangle} & A + D \\
{\scriptstyle \langle f \rangle \hat{+} \mathrm{id}} \downarrow & & \downarrow {\scriptstyle \langle f \rangle \hat{+} \mathrm{id}} \\
C + B & \xrightarrow{\mathrm{id} \hat{+} \langle g \rangle} & C + D
\end{array}
\qquad\qquad
\begin{array}{ccc}
A + B & \xrightarrow{\rhd_1} & A \\
{\scriptstyle \langle f \rangle \hat{+} \mathrm{id}} \downarrow & & \downarrow {\scriptstyle \langle f \rangle} \\
C + B & \xrightarrow{\rhd_1} & C
\end{array}$$

*Proof.* Note that the first diagram is a pullback in $\mathrm{Par}(\mathbf{B})$ if and only if the following diagram is a pullback in $\mathbf{B}$.

$$\begin{array}{ccc}
(A + B) + 1 & \xrightarrow{[\mathrm{id} \hat{+} \langle g \rangle, \kappa_2]} & (A + D) + 1 \\
{\scriptstyle [\langle f \rangle \hat{+} \mathrm{id}, \kappa_2]} \downarrow & & \downarrow {\scriptstyle [\langle f \rangle \hat{+} \mathrm{id}, \kappa_2]} \\
(C + B) + 1 & \xrightarrow{[\mathrm{id} \hat{+} \langle g \rangle, \kappa_2]} & (C + D) + 1
\end{array}$$

This diagram can be recognized as a pullback of (T1), via the associativity of coproducts, as below.

$$\begin{array}{ccccccc}
(A+B)+1 & \xrightarrow{\cong} & A+(B+1) & \xrightarrow{\mathrm{id}+(g+\mathrm{id})} & A+(D+1) & \xrightarrow{\cong} & (A+D)+1 \\
\Big\downarrow{\scriptstyle [\langle f\rangle \diamondplus \mathrm{id},\kappa_2]} & & \Big\downarrow{\scriptstyle f+\mathrm{id}} \quad\lrcorner & & \Big\downarrow{\scriptstyle f+\mathrm{id}} & & \Big\downarrow{\scriptstyle [\langle f\rangle \diamondplus \mathrm{id},\kappa_2]} \\
(C+B)+1 & \xrightarrow{\cong} & C+(B+1) & \xrightarrow{\mathrm{id}+(g+\mathrm{id})} & C+(D+1) & \xrightarrow{\cong} & (C+D)+1
\end{array}$$

(with the top composite $(A+B)+1 \xrightarrow{[\mathrm{id}\diamondplus\langle g\rangle,\kappa_2]} (A+D)+1$ and the bottom composite $(C+B)+1 \xrightarrow{[\mathrm{id}\diamondplus\langle g\rangle,\kappa_2]} (C+D)+1$)

Similarly, the second diagram is a pullback in Par(**B**) if and only if the diagram on the left below is a pullback in **B**.

$$\begin{array}{ccc}
(A+B)+1 & \xrightarrow{[\rhd_1,\kappa_2]} & A+1 \\
\Big\downarrow{\scriptstyle [\langle f\rangle \diamondplus \mathrm{id},\kappa_2]} & & \Big\downarrow{\scriptstyle [\langle f\rangle,\kappa_2]} \\
(C+B)+1 & \xrightarrow{[\rhd_1,\kappa_2]} & C+1
\end{array}
\qquad\qquad
\begin{array}{ccc}
A+(B+1) & \xrightarrow{\mathrm{id}+!} & A+1 \\
\Big\downarrow{\scriptstyle f+\mathrm{id}} \quad\lrcorner & & \Big\downarrow{\scriptstyle f+\mathrm{id}} \\
C+(B+1) & \xrightarrow{\mathrm{id}+!} & C+1
\end{array}$$

Up to the associativity isomorphism, the left-hand diagram is the same as a pullback on the right. ■

We use the following general lemma on jointly monic morphisms.

**Lemma 4.1.18.** *In a category, suppose that we have the following commutative diagram. (Bullets • denote arbitrary objects.)*

$$\begin{array}{ccccc}
\bullet & \xrightarrow{a} & \bullet & \xrightarrow{b} & \bullet \\
\Big\downarrow{\scriptstyle c} & & \Big\downarrow{\scriptstyle d} & & \Big\downarrow{\scriptstyle e} \\
\bullet & \xrightarrow{f} & \bullet & \xrightarrow{g} & \bullet \\
\Big\downarrow{\scriptstyle h} & & \Big\downarrow{\scriptstyle i} & & \\
\bullet & \xrightarrow{j} & \bullet & &
\end{array}$$

*If each of the four pairs $(a,c),(b,d),(f,h),(g,i)$ is jointly monic, then the composites $b\circ a$ and $h\circ c$ are jointly monic too. In particular, $b\circ a$ and $h\circ c$ are jointly monic whenever the three inner squares are pullbacks and the pair $(g,i)$ is jointly monic.*

*Proof.* In the proof we write simply $gf$ for composite $g\circ f$. Let $k$ and $l$ be parallel morphisms such that

$$bak = bal \qquad \text{and} \qquad hck = hcl\,.$$

Let $m \coloneqq da \equiv fc$. Then

$$gmk = gdak = ebak = ebal = gdal = gml$$

and similarly $imk = iml$. By the joint monicity of $g$ and $i$, we obtain $mk = ml$. Since we have $bak = bal$ and $dak = dal$, by the joint monicity of $b$ and $d$ we obtain $ak = al$. Similarly, by the joint monicity of $f$ and $h$ we obtain $ck = cl$. We conclude that $k = l$ by the joint monicity of $a$ and $c$. ■

**Lemma 4.1.19.** *Partial projections are jointly monic in* Par($\mathbf{B}$).

*Proof.* Consider the following diagram in Par($\mathbf{B}$).

$$\begin{array}{ccccc}
A+B & \xrightarrow{\mathrm{id}\mathbin{\hat{+}}\mathbb{1}} & A+1 & \xrightarrow{\rhd_1} & A \\
\downarrow{\scriptstyle \mathbb{1}\hat{+}\mathrm{id}} \; \lrcorner & & \downarrow{\scriptstyle \mathbb{1}\hat{+}\mathrm{id}} \; \lrcorner & & \downarrow{\scriptstyle \mathbb{1}} \\
1+B & \xrightarrow{\mathrm{id}\mathbin{\hat{+}}\mathbb{1}} & 1+1 & \xrightarrow{\rhd_1} & 1 \\
\downarrow{\scriptstyle \rhd_2} \; \lrcorner & & \downarrow{\scriptstyle \rhd_2} & & \\
B & \xrightarrow{\quad\mathbb{1}\quad} & 1 & &
\end{array}$$

It is commutative, and moreover, the three inner squares are pullbacks by Lemma 4.1.17, since $\mathbb{1} = \langle ! \rangle$. Therefore by Lemma 4.1.18, it suffices to prove that the partial projections $\rhd_1, \rhd_2 \colon 1+1 \mathrel{\circ\!\!\rightarrow} 1$ on $1+1$ are jointly monic. They are jointly monic in Par($\mathbf{B}$) if and only if the maps

$$[\rhd_1, \kappa_2], [\rhd_2, \kappa_2] \colon (1+1)+1 \to 1+1$$

are jointly monic in $\mathbf{B}$. The latter holds since $\rhd_1 = [\kappa_1, \kappa_2]$ and $\rhd_2 = [\kappa_2, \kappa_1]$ in $\mathbf{B}$, and the maps

$$[\kappa_1, \kappa_2, \kappa_2], [\kappa_2, \kappa_1, \kappa_2] \colon 1+1+1 \to 1+1$$

are jointly monic by the condition (T3). ■

**Proposition 4.1.20.** *The category* Par($\mathbf{B}$) *is a finPAC.*

*Proof.* We have shown that the category Par($\mathbf{B}$) has finite coproducts and a zero object, and that partial projections are jointly monic, in Lemmas 4.1.15, 4.1.16 and 4.1.19. By applying Theorem 3.8.5, it suffices to prove that the following square is a pullback in Par($\mathbf{B}$).

$$\begin{array}{ccc}
(A+A)+A & \xrightarrow{\nabla' \hat{+} \mathrm{id}} & A+A \\
{\scriptstyle \rhd_1}\downarrow & & \downarrow{\scriptstyle \rhd_1} \\
A+A & \xrightarrow{\quad\nabla'\quad} & A
\end{array} \tag{4.3}$$

Here we write $\nabla'_A \colon A+A \mathrel{\circ\!\!\rightarrow} A$ for the codiagonal in Par($\mathbf{B}$), in order to distinguish it from the codiagonal $\nabla_A$ in $\mathbf{B}$. Since the identity in Par($\mathbf{B}$) is the coprojection $\kappa_1 \colon A \to A+1$, we have

$$\nabla'_A = [\kappa_1, \kappa_1] = \kappa_1 \circ [\mathrm{id}_A, \mathrm{id}_A] = \kappa_1 \circ \nabla_A \equiv \langle \nabla_A \rangle .$$

Thus by Lemma 4.1.17 the square (4.3) is a pullback. ■

To prove that the category Par($\mathbf{B}$) is an effectus in partial form, we need to choose a unit object in Par($\mathbf{B}$). There is a canonical choice — namely, the final object 1 in $\mathbf{B}$. Then the set of predicates on $A$ will be $\mathrm{Par}(\mathbf{B})(A,1) = \mathbf{B}(A,1+1)$. We define truth predicates $\mathbb{1}_A \colon A \mathrel{\circ\!\!\rightarrow} 1$ by $\mathbb{1}_A := \langle !_A \rangle = \kappa_1 \circ !_A$. Moreover, there is an obvious 'negation' of predicates, via the swap isomorphism: for $p \colon A \to 1+1$, we define

$$p^{\perp} := \big( A \xrightarrow{p} 1+1 \xrightarrow{[\kappa_2, \kappa_1]} 1+1 \big) .$$

This satisfies:

$$(p^\perp)^\perp = p \qquad \mathbb{1}_A^\perp = \mathbb{0}_A \qquad \mathbb{0}_A^\perp = \mathbb{1}_A$$

where the falsity predicate is given by $\mathbb{0}_A := 0_{A1} = \kappa_2 \circ !_A$.

Before proving that these structures indeed define effect algebras, we show a lemma that identifies the total morphisms in $\mathrm{Par}(\mathbf{B})$.

**Lemma 4.1.21.** *For any partial map $f\colon A \rightharpoonup B$, we have $\mathbb{1}_B \circledcirc f = \mathbb{1}_A$ if and only if there exists $g\colon A \to B$ in $\mathbf{B}$ such that $f = \langle g \rangle$.*

*Proof.* The 'if' part is easy: if $f = \langle g \rangle$ then

$$\mathbb{1}_B \circledcirc f = \langle !_B \rangle \circledcirc \langle g \rangle = \langle !_B \circ g \rangle = \langle !_A \rangle = \mathbb{1}_A \,.$$

Conversely, assume $\mathbb{1}_B \circledcirc f = \mathbb{1}_A$, that is,

$$\kappa_1 \circ !_A = \mathbb{1}_A = \mathbb{1}_B \circledcirc f = (!_B + \mathrm{id}_1) \circ f \,.$$

We then use a pullback (T2), as below,

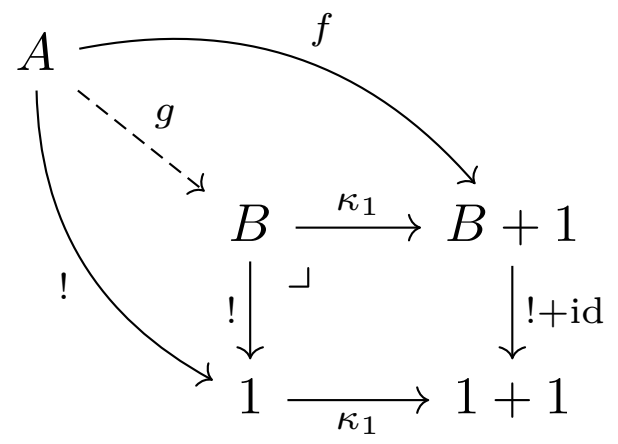

and obtain $g\colon A \to B$ with $f = \kappa_1 \circ g \equiv \langle g \rangle$, as desired. ∎

The following result that $\mathbf{B}(A, 1+1)$ forms an effect algebra is due to Jacobs [140, Proposition 4.4]. For the sake of completeness we include the proof.

**Proposition 4.1.22.** *For each $A \in \mathbf{B}$, the hom-PCM $\mathrm{Par}(\mathbf{B})(A,1) = \mathbf{B}(A, 1+1)$ is an effect algebra, Its top element is*

$$\mathbb{1}_A := \big(A \xrightarrow{!} 1 \xrightarrow{\kappa_1} 1+1\big) \quad \text{in } \mathbf{B},$$

*and the orthosupplement of $p\colon A \to 1+1$ is given by*

$$p^\perp := \big(A \xrightarrow{p} 1+1 \xrightarrow{[\kappa_2,\kappa_1]} 1+1\big)$$

*Proof.* Let $p\colon A \rightharpoonup 1$ be a predicate. Note that the following diagram in $\mathbf{B}$ commutes.

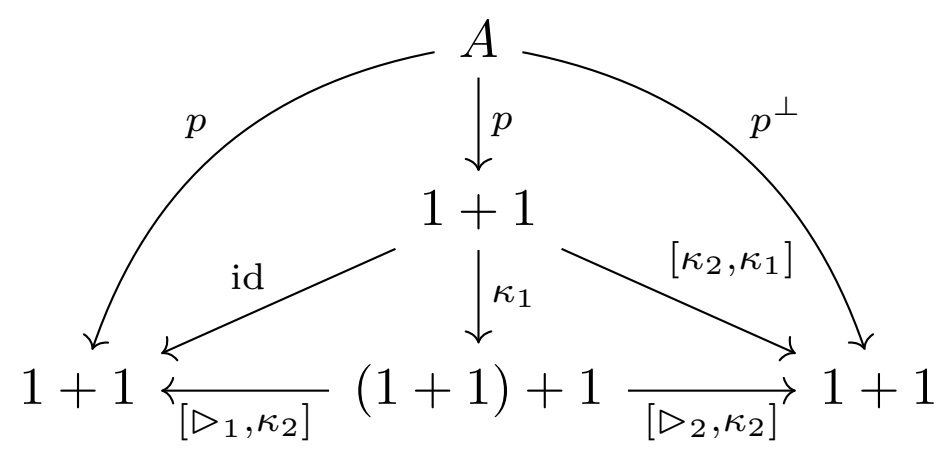

Thus $p$ and $p^\perp$ are compatible via $p \circ \kappa_1 = \langle p \rangle \colon A \multimap 1+1$. The sum satisfies:

$$p \ovee p^\perp \equiv \langle \nabla_1 \rangle \circledcirc \langle p \rangle = \langle \nabla_1 \circ p \rangle = \langle !_A \rangle \equiv \mathbb{1}_A \,,$$

as desired. To show that the $p^\perp$ is a unique map with $p \ovee p^\perp = \mathbb{1}_A$, let us assume that $q \colon A \multimap 1$ satisfies $p \ovee q = \mathbb{1}_A$. This means that there is a map $b \colon A \multimap 1+1$ such that the following diagrams in $\mathrm{Par}(\mathbf{B})$ commute.

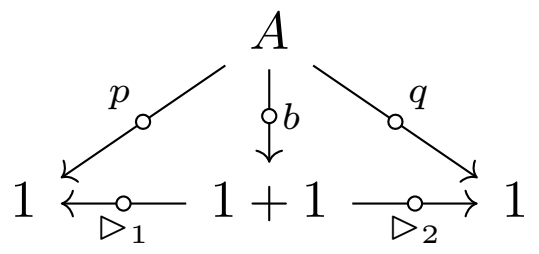

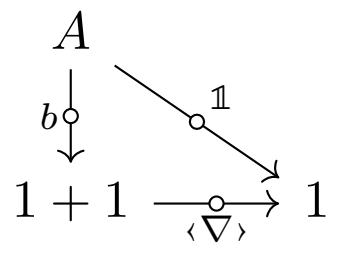

Since $\langle \nabla_1 \rangle = \langle !_{1+1} \rangle \equiv \mathbb{1}_{1+1}$ in the diagram on the right, by Lemma 4.1.21 there exists $c \colon A \to 1+1$ in $\mathbf{B}$ such that $b = \langle c \rangle$. Then

$$\begin{aligned} p &= \rhd_1 \circledcirc \langle c \rangle = \rhd_1 \circ c = c \\ q &= \rhd_2 \circledcirc \langle c \rangle = \rhd_2 \circ c = [\kappa_2, \kappa_1] \circ c \,, \end{aligned}$$

showing that $q = [\kappa_2, \kappa_1] \circ p = p^\perp$. To see that $\mathrm{Par}(\mathbf{B})(A, 1)$ is an effect algebra, it only remains to prove that $\mathbb{1}_A \perp p$ implies $p = \mathbb{0}_A$. So assume that we have a map $b \colon A \multimap 1+1$ such that $\mathbb{1}_A = \rhd_1 \circledcirc b$ and $p = \rhd_2 \circledcirc b$. We use a pullback (T2) as follows,

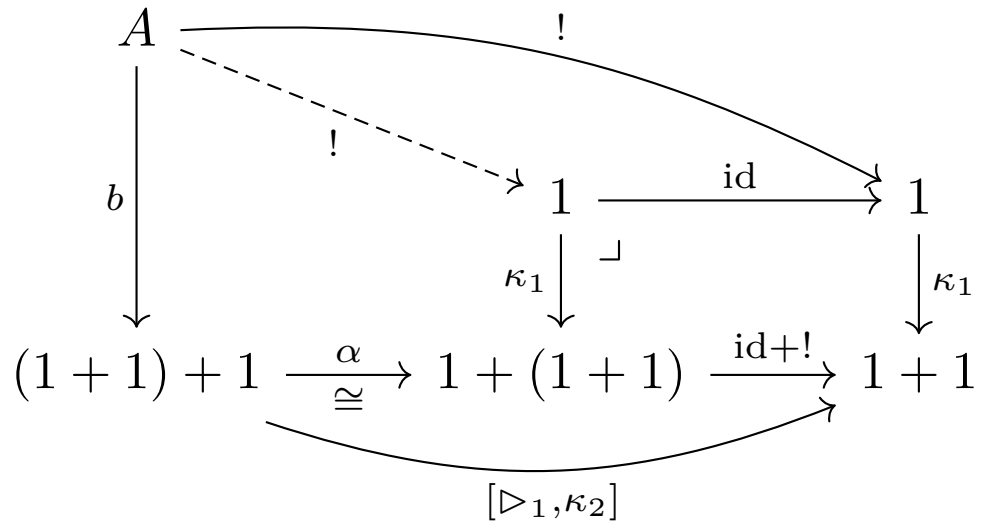

where $\alpha \colon (1+1)+1 \to 1+(1+1)$ is the associativity isomorphism. Therefore $b = \alpha^{-1} \circ \kappa_1 \circ !_A = \kappa_1 \circ \kappa_1 \circ !_A = \langle \kappa_1 \circ !_A \rangle$, and hence

$$p = \rhd_2 \circledcirc \langle \kappa_1 \circ !_A \rangle = \rhd_2 \circ \kappa_1 \circ !_A = \kappa_2 \circ !_A \equiv \mathbb{0}_A \,. \qquad \blacksquare$$

**Lemma 4.1.23.** *Let* $\mathbf{B}$ *be an effectus in total form. Let* $f, g \colon A \multimap B$ *be morphisms in* $\mathrm{Par}(\mathbf{B})$.

(i) $\mathbb{1}_B \circledcirc f = \mathbb{0}_A$ *implies* $f = 0_{AB}$.

(ii) $\mathbb{1}_B \circledcirc f \perp \mathbb{1}_B \circledcirc g$ *implies* $f \perp g$.

*Proof.*

(i) Assume $\mathbb{1}_B \circledcirc f = \mathbb{0}_A$. Then

$$\kappa_2 \circ !_A = \mathbb{0}_A = \mathbb{1}_B \circledcirc f = (!_B + \mathrm{id}_1) \circ f \,,$$

so we can use a pullback (T2):

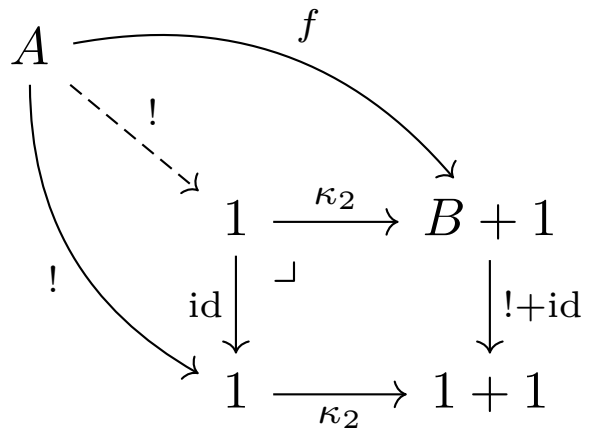

Thus $f = 0_{AB}$ by definition of zero morphisms in Par(**B**) (see Lemma 4.1.16).

(ii) Assume $\mathbb{1}_B \circledcirc f \perp \mathbb{1}_B \circledcirc g$. Then the two morphisms are compatible, i.e. there is a morphism $b\colon A \multimap 1 + 1$ such that $\rhd_1 \circ b = \mathbb{1}_B \circledcirc f$ and $\rhd_2 \circ b = \mathbb{1}_B \circledcirc g$. Since $\mathbb{1}_B = \langle !_B \rangle$, we can use a pullback on the right in Lemma 4.1.17, obtaining $c\colon A \multimap B + 1$ as in the diagram:

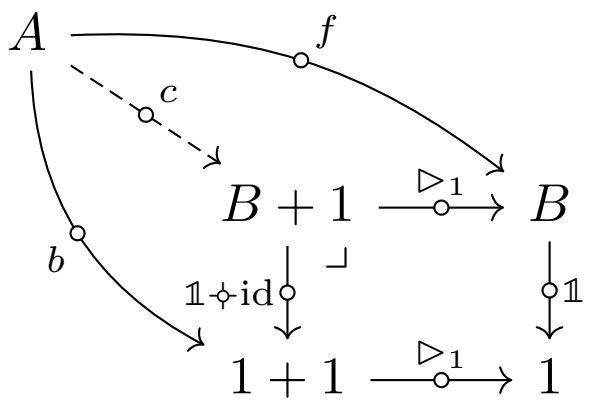

(4.4)

This $c$ satisfies:

$$\mathbb{1}_B \circledcirc g = \rhd_2 \circledcirc b = \rhd_2 \circledcirc (\mathbb{1}_B \mathbin{\hat{+}} \mathrm{id}_1) \circledcirc c = \rhd_2 \circledcirc c$$

we obtain $d\colon A \multimap B + B$, using a similar pullback (that exists via symmetry) as in the diagram:

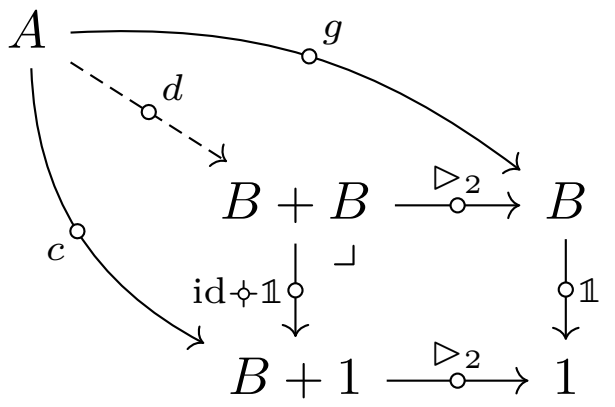

(4.5)

It follows that $f$ and $g$ are compatible via this $d$, as

$$\begin{aligned} \rhd_1 \circledcirc d &= \rhd_1 \circledcirc (\mathrm{id} \mathbin{\hat{+}} \mathbb{1}) \circledcirc d = \rhd_1 \circledcirc c = f \\ \rhd_2 \circledcirc d &= g \end{aligned}$$

by the commutativity of diagrams (4.4) and (4.5). ■

**Theorem 4.1.24.** *The category* Par(**B**)*, with* 1 *as a unit object, is an effectus in partial form. Moreover we have an isomorphism of categories* $\mathbf{B} \cong \mathrm{Tot}(\mathrm{Par}(\mathbf{B}))$*, which is obtained by restricting codomain of the functor* $\langle - \rangle\colon \mathbf{B} \to \mathrm{Par}(\mathbf{B})$.

*Proof.* The first claim follows by Propositions 4.1.20 and 4.1.22 and Lemma 4.1.23. The second claim follows by Lemmas 4.1.14 and 4.1.21. ■

## 4.2 Categorical equivalence of effectuses in partial and total form

Recall that in the previous section we have proven effectuses in partial and total form are equivalent in the following sense.

1. For any effectus in partial form $\mathbf{C}$, we prove that $\mathrm{Tot}(\mathbf{C})$ is an effectus in total form. Moreover $\mathrm{Par}(\mathrm{Tot}(\mathbf{C})) \cong \mathbf{C}$. (Theorem 4.1.11 and Proposition 4.1.5)
2. For any effectus in total form $\mathbf{B}$, we prove that $\mathrm{Par}(\mathbf{B})$ is an effectus in partial form. Moreover $\mathrm{Tot}(\mathrm{Par}(\mathbf{B})) \cong \mathbf{B}$. (Theorem 4.1.24)

Note that the word 'equivalent' here is used in an informal sense, rather than a mathematically rigorous sense. We can make it rigorous by using the categorical language — namely as a equivalence of 2-categories.

We first define 2-categories of effectuses. In this thesis, by 2-categories we mean *strict* 2-categories.

**Definition 4.2.1.** The 2-category $\mathbf{Ef}$ of effectuses in partial form is defined as follows.

- An object (0-cell) is an effectus in partial form $(\mathbf{C}, I)$.
- A morphism (1-cell) of type $(\mathbf{C}, I_{\mathbf{C}}) \to (\mathbf{D}, I_{\mathbf{D}})$ is a functor $F\colon \mathbf{C} \to \mathbf{D}$ that preserves finite coproducts, together with an isomorphism $u\colon I_{\mathbf{D}} \to FI_{\mathbf{C}}$ in $\mathbf{D}$ such that $F\mathbb{1}_A = u \circ \mathbb{1}_{FA}$ for each $A \in \mathbf{C}$.
- A 2-cell of type $(F, u) \Rightarrow (G, v)\colon (\mathbf{C}, I_{\mathbf{C}}) \to (\mathbf{D}, I_{\mathbf{D}})$ is a natural transformation $\alpha\colon F \Rightarrow G$ such that $\alpha_{I_{\mathbf{C}}} \circ u = v$.

The 2-category $\mathbf{Eft}$ of effectuses in total form is defined as follows.

- An object is an effectus in total form $\mathbf{B}$.
- A morphism $F\colon \mathbf{A} \to \mathbf{B}$ is a functor that preserves finite coproducts and the final object.
- A 2-cell $\alpha\colon F \Rightarrow G$ is a natural transformation.

It is straightforward to check that $\mathbf{Ef}$ and $\mathbf{Eft}$ are indeed 2-categories, in a similar manner to the 2-category of categories.

The definition of $\mathbf{Ef}$ is slightly complicated, compared to $\mathbf{Eft}$. So we first give some equivalent conditions for morphisms and 2-cells in $\mathbf{Ef}$. Note that a morphism in $\mathbf{Ef}$ involves an extra structure $u\colon I_{\mathbf{D}} \to FI_{\mathbf{C}}$, similarly to the structure of a (strong) monoidal functor. It turns out that the structure $u$ is uniquely determined, and thus being a morphism of $\mathbf{Ef}$ can be considered as a property of a functor.

**Lemma 4.2.2.** *Let $F\colon \mathbf{C} \to \mathbf{D}$ be a finite-coproduct-preserving functor between effectuses in partial form.*

(i) *If $u\colon I_{\mathbf{D}} \to FI_{\mathbf{C}}$ is an isomorphism, then $u^{-1} = \mathbb{1}_{FI_{\mathbf{C}}}\colon FI_{\mathbf{C}} \to I_{\mathbf{D}}$.*

(ii) *Let $u\colon I_{\mathbf{D}} \to FI_{\mathbf{C}}$ be an isomorphism. The following are equivalent.*

  (a) *The pair $(F, u)$ is a morphism in $\mathbf{Ef}$.*

(b) *$F\mathbb{1}_A\colon FA \to FI_{\mathbf{C}}$ is total for each $A \in \mathbf{C}$.*

(c) *$F$ preserves total morphisms.*

*Therefore, an isomorphism $FI_{\mathbf{C}} \cong I_{\mathbf{D}}$ is unique if exists. The morphisms $(\mathbf{C}, I_{\mathbf{C}}) \to (\mathbf{D}, I_{\mathbf{D}})$ in $\mathbf{Ef}$ are identified with finite-coproduct-preserving functors $F\colon \mathbf{C} \to \mathbf{D}$ satisfying $FI_{\mathbf{C}} \cong I_{\mathbf{D}}$ and either (hence both) of conditions* (b) *and* (c)*.*

*Proof.*

(i) The map $u^{-1}\colon FI_{\mathbf{C}} \to I_{\mathbf{D}}$ is an isomorphism and hence total by Lemma 3.2.4(iii). Then $u^{-1} = \mathbb{1}_{FI_{\mathbf{C}}}$, since $\mathbb{1}_{FI_{\mathbf{C}}}\colon FI_{\mathbf{C}} \to I_{\mathbf{D}}$ is the unique total map of this type.

(ii) By (i), one has $F\mathbb{1}_A = u \circ \mathbb{1}_{FA}$ if and only if $\mathbb{1}_{FI_{\mathbf{C}}} \circ F\mathbb{1}_A = \mathbb{1}_{FA}$, i.e. $F\mathbb{1}_A$ is total. Hence (a) $\iff$ (b). Assume (b), and let $f\colon A \to B$ be a total morphism in $\mathbf{C}$. Then

$$\mathbb{1}_{FB} \circ Ff = \mathbb{1}_{FI} \circ F\mathbb{1}_B \circ Ff = \mathbb{1}_{FI} \circ F(\mathbb{1}_B \circ f) = \mathbb{1}_{FI} \circ F(\mathbb{1}_A) = \mathbb{1}_{FA}$$

and hence $Ff$ is total. We are done since (c) $\implies$ (b) is trivial. ∎

Therefore we will simply say that $F\colon \mathbf{C} \to \mathbf{D}$ is a morphism in $\mathbf{Ef}$, leaving $u$ implicit. We can then characterize 2-cells in $\mathbf{Ef}$ too.

We also note that for functors between FinPACs, preservation of finite coproducts can be characterized via the PCM structure. This is analogous to a well-known result for biproduct categories, see Lemma 7.1.30.

**Lemma 4.2.3.** *Let $F\colon \mathbf{C} \to \mathbf{D}$ be a functor between finPACs. Then $F$ preserves finite coproducts if and only if both of the following hold.*

(i) *$F$ is enriched over PCMs, i.e. for each $A, B \in \mathbf{C}$, the map $F\colon \mathbf{C}(A, B) \to \mathbf{D}(FA, FB)$ is a PCM morphism.*

(ii) *$F$ preserves compatibility of two morphisms: if $f\colon A \to B$ and $g\colon A \to C$ are compatible, so are $Ff$ and $Fg$.*

*Proof.* It is easy to see that $F$ preserves the initial object if and only if $F$ preserves zero morphisms. Now assume that $F$ preserves finite coproducts (so $F$ preserves zero morphisms). Let $f\colon A \to B$ and $g\colon A \to C$ be compatible morphisms in $\mathbf{C}$. Then $Ff$ and $Fg$ are compatible via

$$\langle\!\langle Ff, Fg \rangle\!\rangle = \Big( FA \xrightarrow{F\langle\!\langle f,g \rangle\!\rangle} F(B + C) \xrightarrow{\cong} FB + FC \Big),$$

where the isomorphism is the inverse of $[F\kappa_1, F\kappa_2]$. If $B = C$, we have $F(f \ovee g) = Ff \ovee Fg$ since the following diagram commutes.

$$\begin{array}{ccccc} FA & \xrightarrow{F\langle\!\langle f,g \rangle\!\rangle} & F(B+B) & & \\ & \searrow^{\langle\!\langle Ff,Fg \rangle\!\rangle} & \downarrow{\cong} & \searrow^{F\nabla} & \\ & & FB + FB & \xrightarrow[\nabla]{} & FB \end{array}$$

Conversely, assume that $F$ satisfies the latter two conditions. Let $A, B \in \mathbf{C}$. Since $\rhd_1\colon A + B \to A$ and $\rhd_2\colon A + B \to B$ are compatible, so are $F\rhd_1$ and $F\rhd_2$ by assumption. We claim that the tuple $\langle\!\langle F\rhd_1, F\rhd_2\rangle\!\rangle$ is the inverse of $[F\kappa_1, F\kappa_2]\colon FA + FB \to F(A+B)$. Indeed, we have

$$\begin{aligned}
\mathrm{id}_{F(A+B)} &= F\mathrm{id}_{A+B} \\
&= F(\kappa_1 \circ \rhd_1 \ovee \kappa_2 \circ \rhd_2) \\
&= F\kappa_1 \circ F\rhd_1 \ovee F\kappa_2 \circ F\rhd_2 && F \text{ is enriched over PCMs} \\
&= [F\kappa_1, F\kappa_2] \circ \langle\!\langle F\rhd_1, F\rhd_2\rangle\!\rangle .
\end{aligned}$$

We also have $\langle\!\langle F\rhd_1, F\rhd_2\rangle\!\rangle \circ [F\kappa_1, F\kappa_2] = \mathrm{id}_{FA+FB}$ since

$$\rhd_j \circ \langle\!\langle F\rhd_1, F\rhd_2\rangle\!\rangle \circ [F\kappa_1, F\kappa_2] \circ \kappa_k = \rhd_j \circ \kappa_k$$

for each $j, k \in \{1, 2\}$. ■

**Lemma 4.2.4.** *Let $F, G\colon \mathbf{C} \to \mathbf{D}$ be morphisms in* **Ef***, and $\alpha\colon F \Rightarrow G$ a natural transformation. The following are equivalent.*

(i) *$\alpha$ is a 2-cell in* **Ef***.*

(ii) *For each $A \in \mathbf{C}$ the component $\alpha_A\colon FA \to GA$ is total in* **D***.*

(iii) *The component $\alpha_{I_\mathbf{C}}\colon FI_\mathbf{C} \to GI_\mathbf{C}$ is total in* **D***.*

*Proof.* The equivalence (i) $\Longleftrightarrow$ (ii) follows from Lemma 4.2.2(i). It is trivial that (ii) $\Longrightarrow$ (iii). Assume (iii), i.e. that $\alpha_{I_\mathbf{C}}\colon FI_\mathbf{C} \to GI_\mathbf{C}$ is total. Then for each $A \in \mathbf{C}$,

$$\begin{aligned}
\mathbb{1}_{GX} \circ \alpha_A &= \mathbb{1}_{GI_\mathbf{C}} \circ G\mathbb{1}_A \circ \alpha_A && G\mathbb{1}_A \text{ is total} \\
&= \mathbb{1}_{GI_\mathbf{C}} \circ \alpha_{I_\mathbf{C}} \circ F\mathbb{1}_A && \text{naturality of } \alpha \\
&= \mathbb{1}_{FI_\mathbf{C}} \circ F\mathbb{1}_A && \alpha_{I_\mathbf{C}} \text{ is total} \\
&= \mathbb{1}_{FA} && F\mathbb{1}_A \text{ is total.}
\end{aligned}$$

Hence (ii) holds. ■

The goal of this section is to prove the following theorem.

**Theorem 4.2.5.** *The mappings $\mathbf{C} \mapsto \mathrm{Tot}(\mathbf{C})$ and $\mathbf{B} \mapsto \mathrm{Par}(\mathbf{B})$ extend to 2-functors* $\mathrm{Tot}\colon \mathbf{Ef} \to \mathbf{Eft}$ *and* $\mathrm{Par}\colon \mathbf{Eft} \to \mathbf{Ef}$ *respectively. Moreover, they form a 2-equivalence of 2-categories* $\mathbf{Ef} \simeq \mathbf{Eft}$*.*

**Lemma 4.2.6.** *The mapping $\mathbf{C} \mapsto \mathrm{Tot}(\mathbf{C})$ extends to a 2-functor* $\mathrm{Tot}\colon \mathbf{Ef} \to \mathbf{Eft}$*.*

*Proof.* Let $F\colon \mathbf{C} \to \mathbf{D}$ be a morphism in **Ef**. We define $\mathrm{Tot}(F)\colon \mathrm{Tot}(\mathbf{C}) \to \mathrm{Tot}(\mathbf{D})$ simply to be the restriction of the functor $F$. The restriction is well-defined: for any total morphism $f\colon A \to B$ in $\mathbf{C}$, $Ff\colon FA \to FB$ is total, since

$$\mathbb{1}_{FB} \circ Ff = \mathbb{1}_{FI} \circ F\mathbb{1}_B \circ Ff = \mathbb{1}_{FI} \circ F\mathbb{1}_A = \mathbb{1}_{FA}$$

using the fact that $F\mathbb{1}_A$ and $F\mathbb{1}_B$ are total in $\mathbf{D}$. The functor $\mathrm{Tot}(F)$ preserves finite coproducts, since so does $F$. It preserves the final object since $FI_\mathbf{C} \cong I_\mathbf{D}$. Hence $\mathrm{Tot}(F)$ is a morphism in **Eft**.

For a 2-cell $\alpha\colon F \Rightarrow G$ in **Ef**, similarly we define $\mathrm{Tot}(\alpha)\colon \mathrm{Tot}(F) \Rightarrow \mathrm{Tot}(G)$ simply by $\mathrm{Tot}(\alpha)_A \coloneqq \alpha_A$. This works since $\mathrm{Tot}(F)A = FA$, and each component $\alpha_A$ is total, so it sits in $\mathrm{Tot}(\mathbf{D})$. It is clear that $\mathrm{Tot}(\alpha)_A$ is natural in $A$. Thus $\mathrm{Tot}(\alpha)$ is a 2-cell in **Eft**.

To say that Tot is a 2-functor means that the following equations hold.

(1) $\mathrm{Tot}(G \circ F) = \mathrm{Tot}(G) \circ \mathrm{Tot}(F)$ and $\mathrm{Tot}(\mathrm{id}_{\mathbf{C}}) = \mathrm{id}_{\mathrm{Tot}(\mathbf{C})}$.

(2) $\mathrm{Tot}(\gamma \circ \alpha) = \mathrm{Tot}(\gamma) \circ \mathrm{Tot}(\alpha)$ and $\mathrm{Tot}(\mathrm{id}_F) = \mathrm{id}_{\mathrm{Tot}(F)}$.

(3) $\mathrm{Tot}(G \circ \alpha) = \mathrm{Tot}(G) \circ \mathrm{Tot}(\alpha)$ and $\mathrm{Tot}(\beta \circ F) = \mathrm{Tot}(\beta) \circ \mathrm{Tot}(F)$.

All the conditions are straightforward to check. ■

We need more work for the other direction. We start with a definition of morphisms $\varphi_{F,A}\colon F(A+1) \to FA+1$ that 'distributes' [141, § 5.2] the lift monads.

**Definition 4.2.7.** Let $F\colon \mathbf{A} \to \mathbf{B}$ be a morphism in **Eft**. For each $A \in \mathbf{A}$, the mediating map below exists and is an isomorphism, since $F$ preserves the final object and finite coproducts.

$$\begin{array}{ccccc} FA & \xrightarrow{\kappa_1} & FA+1 & \xleftarrow{\kappa_2} & 1 \\ & \searrow^{F\kappa_1} & \Big\downarrow{\cong} & & \Big\downarrow{\cong} \\ & & F(A+1) & \xleftarrow[F\kappa_2]{} & F1 \end{array}$$

We denote the inverse of the isomorphism by $\varphi_{F,A}\colon F(A+1) \to FA+1$, or simply by $\varphi_A$ when the context is clear. By definition it satisfies the following equations.

$$\varphi_{F,A} \circ F\kappa_1 = \kappa_1 \tag{4.6}$$

$$\varphi_{F,A} \circ F\kappa_2 = \kappa_2 \circ !_{F1} \tag{4.7}$$

We need a few commutative diagrams involving the maps $\varphi_{F,A}$.

**Lemma 4.2.8.** *Let $F, H\colon \mathbf{A} \to \mathbf{B}$, and $G\colon \mathbf{B} \to \mathbf{C}$ be morphisms in* **Eft**. *The 'distributive law' maps $\varphi$ defined above satisfies the following properties.*

(i) *For each morphism $f\colon A \to B+1$ in $\mathbf{A}$, the following diagram commutes.*

$$\begin{array}{ccc} F(A+1) & \xrightarrow{\varphi_A} & FA+1 \\ {\scriptstyle F[f,\kappa_2]}\Big\downarrow & & \Big\downarrow{\scriptstyle [\varphi_B \circ Ff,\kappa_2]} \\ F(B+1) & \xrightarrow{\varphi_B} & FB+1 \end{array}$$

(ii) *For each natural transformation $\alpha\colon F \Rightarrow H$ and each $A \in \mathbf{A}$, the following diagram commutes.*

$$\begin{array}{ccc} F(A+1) & \xrightarrow{\varphi_{F,A}} & FA+1 \\ {\scriptstyle \alpha_{A+1}}\Big\downarrow & & \Big\downarrow{\scriptstyle \alpha_A+\mathrm{id}} \\ H(A+1) & \xrightarrow{\varphi_{H,A}} & HA+1 \end{array}$$

(iii) *For each $A \in \mathbf{A}$, the following diagram commutes.*

$$\begin{array}{ccc} GF(A+1) & & \\ {\scriptstyle G\varphi_{F,A}}\downarrow & \searrow^{\varphi_{GF,A}} & \\ G(FA+1) & \xrightarrow[\varphi_{G,FA}]{} & GFA+1 \end{array}$$

*Proof.* (i) Note that the diagram below is a coproduct

$$FA \xrightarrow{F\kappa_1} F(A+1) \xleftarrow{F\kappa_2} F1\,,$$

since $F$ preserves finite coproducts. Thus we prove $\varphi_B \circ F[f,\kappa_2] = [\varphi_B \circ Ff, \kappa_2] \circ \varphi_A$ by the following reasoning.

$$\begin{aligned} \varphi_B \circ F[f,\kappa_2] \circ F\kappa_1 &= \varphi_B \circ Ff \\ &= [\varphi_B \circ Ff, \kappa_2] \circ \kappa_1 \\ &= [\varphi_B \circ Ff, \kappa_2] \circ \varphi_A \circ F\kappa_1 && \text{by (4.6)} \\ \varphi_B \circ F[f,\kappa_2] \circ F\kappa_2 &= \varphi_B \circ F\kappa_2 \\ &= \kappa_2 \circ !_{F1} && \text{by (4.7)} \\ &= [\varphi_B \circ Ff, \kappa_2] \circ \kappa_2 \circ !_{F1} \\ &= [\varphi_B \circ Ff, \kappa_2] \circ \varphi_A \circ F\kappa_2 && \text{by (4.7)} \end{aligned}$$

(ii) Similarly we prove $(\alpha_A + \mathrm{id}) \circ \varphi_{F,A} = \varphi_{G,A} \circ \alpha_{A+1}$ by:

$$\begin{aligned} (\alpha_A + \mathrm{id}) \circ \varphi_{F,A} \circ F\kappa_1 &= (\alpha_A + \mathrm{id}) \circ \kappa_1 && \text{by (4.6)} \\ &= \kappa_1 \circ \alpha_A \\ &= \varphi_{G,A} \circ G\kappa_1 \circ \alpha_A && \text{by (4.6)} \\ &= \varphi_{G,A} \circ \alpha_{A+1} \circ F\kappa_1 && \text{naturality of } \alpha \\ (\alpha_A + \mathrm{id}) \circ \varphi_{F,A} \circ F\kappa_2 &= (\alpha_A + \mathrm{id}) \circ \kappa_2 \circ !_{F1} && \text{by (4.7)} \\ &= \kappa_2 \circ !_{F1} \\ &= \kappa_2 \circ !_{G1} \circ \alpha_1 \\ &= \varphi_{G,A} \circ G\kappa_2 \circ \alpha_1 && \text{by (4.7)} \\ &= \varphi_{G,A} \circ \alpha_{A+1} \circ F\kappa_2 && \text{naturality of } \alpha. \end{aligned}$$

(iii) We use the coproduct $GFA \xrightarrow{GF\kappa_1} GF(A+1) \xleftarrow{GF\kappa_2} GF1$ and show $\varphi_{G,FA} \circ G\varphi_{F,A} = \varphi_{GF,A}$ by:

$$\begin{aligned} \varphi_{G,FA} \circ G\varphi_{F,A} \circ GF\kappa_1 &= \varphi_{G,FA} \circ G\kappa_1 && \text{by (4.6) for } F \\ &= \kappa_1 && \text{by (4.6) for } G \\ &= \varphi_{GF,A} \circ GF\kappa_1 && \text{by (4.6) for } GF \\ \varphi_{G,FA} \circ G\varphi_{F,A} \circ GF\kappa_2 &= \varphi_{G,FA} \circ G\kappa_2 \circ G!_{F1} && \text{by (4.7) for } F \\ &= \kappa_2 \circ !_{G1} \circ G!_{F1} && \text{by (4.7) for } G \\ &= \kappa_2 \circ !_{GF1} \\ &= \varphi_{GF,A} \circ GF\kappa_2 && \text{by (4.7) for } GF. \end{aligned}$$

■

**Lemma 4.2.9.** *The mapping* $\mathbf{B} \mapsto \mathrm{Par}(\mathbf{B})$ *extends to a 2-functor* $\mathrm{Par}\colon \mathbf{Eft} \to \mathbf{Ef}$.

*Proof.* For a morphism $F\colon \mathbf{A} \to \mathbf{B}$ in $\mathbf{Eft}$, we define $\mathrm{Par}(F)\colon \mathrm{Par}(\mathbf{A}) \to \mathrm{Par}(\mathbf{B})$ by $\mathrm{Par}(F)A := FA$ and

$$\mathrm{Par}(F)(A \xrightarrow{f} B+1) := \big(FA \xrightarrow{Ff} F(B+1) \xrightarrow{\varphi_B} FB+1\big)\,.$$

Note that $\mathrm{Par}(F)$ is a 'lifting' of $F$ in the sense that the following diagram commutes.

$$\begin{array}{ccc} \mathrm{Par}(\mathbf{A}) & \xrightarrow{\mathrm{Par}(F)} & \mathrm{Par}(\mathbf{B}) \\ {\scriptstyle \langle - \rangle}\uparrow & & \uparrow{\scriptstyle \langle - \rangle} \\ \mathbf{A} & \xrightarrow{\quad F \quad} & \mathbf{B} \end{array}$$

Indeed, for $f\colon A \to B$ in $\mathbf{A}$,

$$\mathrm{Par}(F)\langle h \rangle = \mathrm{Par}(F)(\kappa_1 \circ h) = \varphi_B \circ F\kappa_1 \circ Fh = \kappa_1 \circ Fh = \langle Fh \rangle\,.$$

In particular, $\mathrm{Par}(F)$ preserves the identities in $\mathrm{Par}(\mathbf{A})$ as $\mathrm{Par}(F)\langle \mathrm{id}_A \rangle = \langle F\mathrm{id}_A \rangle = \langle \mathrm{id}_{FA} \rangle$. For $f\colon A \rightharpoondown B$ and $g\colon B \rightharpoondown C$ in $\mathrm{Par}(\mathbf{A})$,

$$\begin{aligned} \mathrm{Par}(F)(g \circledcirc f) &= \varphi_C \circ F[g, \kappa_2] \circ Ff \\ &= [\varphi_C \circ Fg, \kappa_2] \circ \varphi_B \circ Ff && \text{by Lemma 4.2.8(i)} \\ &= \mathrm{Par}(F)g \circledcirc \mathrm{Par}(F)f\,, \end{aligned}$$

showing that $\mathrm{Par}(F)$ is a functor. It preserves finite coproducts since $\mathrm{Par}(F)$ is a lifting of $F$, and $\mathrm{Par}(\mathbf{A})$ inherits finite coproducts from $\mathbf{A}$. We have $\mathrm{Par}(F)1 \equiv F1 \cong 1$ since $F$ preserves the final object, and $\mathrm{Par}(F)$ preserves total morphisms as it is a lifting of $F$. Therefore $\mathrm{Par}(F)$ is a morphism in $\mathbf{Ef}$.

Let $\alpha\colon F \Rightarrow H\colon \mathbf{A} \to \mathbf{B}$ be a 2-cell in $\mathbf{Eft}$. We define $\mathrm{Par}(\alpha)\colon \mathrm{Par}(F) \Rightarrow \mathrm{Par}(H)\colon \mathrm{Par}(\mathbf{A}) \to \mathrm{Par}(\mathbf{B})$ by $\mathrm{Par}(\alpha)_A := \langle \alpha_A \rangle\colon \mathrm{Par}(F)A \to \mathrm{Par}(H)A$. We prove that $\mathrm{Par}(\alpha)$ is natural: for $f\colon A \rightharpoondown B$ in $\mathrm{Par}(\mathbf{A})$,

$$\begin{aligned} \mathrm{Par}(\alpha_B) \circledcirc \mathrm{Par}(F)f &= \langle \alpha_B \rangle \circledcirc (\varphi_{F,A} \circ Ff) \\ &= (\alpha_B + \mathrm{id}_1) \circ \varphi_{F,B} \circ Ff \\ &= \varphi_{H,B} \circ \alpha_{B+1} \circ Ff && \text{by Lemma 4.2.8(ii)} \\ &= \varphi_{H,B} \circ Hf \circ \alpha_A && \text{naturality of } \alpha \\ &= \mathrm{Par}(H)f \circledcirc \langle \alpha_A \rangle \\ &= \mathrm{Par}(H)f \circledcirc \mathrm{Par}(\alpha_A)\,. \end{aligned}$$

Since each component of $\mathrm{Par}(\alpha)$ is total, it is a 2-cell in $\mathbf{Ef}$.

We are going to verify that Par is a 2-functor. Let $F\colon \mathbf{A} \to \mathbf{B}$ and $G\colon \mathbf{B} \to \mathbf{C}$ be morphisms in $\mathbf{Eft}$. Then

$$\begin{aligned} \mathrm{Par}(GF)f &= \varphi_{GF,B} \circ GFf \\ &= \varphi_{G,FB} \circ G\varphi_{F,B} \circ GFf && \text{by Lemma 4.2.8(iii)} \\ &= \mathrm{Par}(G)(\varphi_{F,B} \circ Ff) \\ &= \mathrm{Par}(G)\mathrm{Par}(F)f\,. \end{aligned}$$

For an identity functor $\mathrm{id}_{\mathbf{A}} \colon \mathbf{A} \to \mathbf{A}$, we have $\mathrm{Par}(\mathrm{id}_{\mathbf{A}})f = \varphi_{\mathrm{id}_{\mathbf{A}},A} \circ f$. It is easy to see that $\varphi_{\mathrm{id}_{\mathbf{A}},A} = \mathrm{id}_{A+1}$, so that $\mathrm{Par}(\mathrm{id}_{\mathbf{A}}) = \mathrm{id}_{\mathrm{Par}(\mathbf{A})}$. The rest of the verification of 2-functoriality is straightforward. ■

**Theorem 4.2.10.** *The 2-functors* $\mathrm{Par} \colon \mathbf{Eft} \to \mathbf{Ef}$ *and* $\mathrm{Tot} \colon \mathbf{Ef} \to \mathbf{Eft}$ *form a 2-equivalence of 2-categories* $\mathbf{Eft} \simeq \mathbf{Ef}$*. In other words, there are 2-natural isomorphisms* $\mathrm{id}_{\mathbf{Eft}} \cong \mathrm{Tot} \circ \mathrm{Par}$ *and* $\mathrm{id}_{\mathbf{Ef}} \cong \mathrm{Par} \circ \mathrm{Tot}$*.*

*Proof.* Let $\Phi_{\mathbf{B}} \colon \mathbf{B} \to \mathrm{Tot}(\mathrm{Par}(\mathbf{B}))$ be the isomorphism of categories in Theorem 4.1.24, given by $\Phi_{\mathbf{B}}A = A$ and $\Phi_{\mathbf{B}}f = \langle f \rangle$. It preserves finite coproducts and the final object, so that $\Phi_{\mathbf{B}}$ is a morphism in $\mathbf{Eft}$. Let $F \colon \mathbf{A} \to \mathbf{B}$ be a morphism in $\mathbf{Eft}$. Because $\mathrm{Par}(F)$ is a lifting of $F$, and $\mathrm{Tot}(\mathrm{Par}(F))$ is a restriction of $\mathrm{Par}(F)$, the following diagram commutes.

$$\begin{array}{ccc} \mathbf{A} & \xrightarrow{\quad F \quad} & \mathbf{B} \\ {\scriptstyle \Phi_{\mathbf{A}}}\big\downarrow{\scriptstyle \cong} & & {\scriptstyle \cong}\big\downarrow{\scriptstyle \Phi_{\mathbf{B}}} \\ \mathrm{Tot}(\mathrm{Par}(\mathbf{A})) & \xrightarrow{\mathrm{Tot}(\mathrm{Par}(F))} & \mathrm{Tot}(\mathrm{Par}(\mathbf{B})) \end{array}$$

Let $\alpha \colon F \Rightarrow G$ be a 2-cell in $\mathbf{Eft}$. Then

$$(\mathrm{Tot}(\mathrm{Par}(\alpha))\Phi_{\mathbf{A}})_A = (\mathrm{Tot}(\mathrm{Par}(\alpha)))_{\Phi_{\mathbf{A}}A} = (\mathrm{Par}(\alpha))_A = \langle \alpha \rangle_A = \Phi_{\mathbf{B}}\alpha_A = (\Phi_{\mathbf{B}}\alpha)_A \,,$$

so that $\mathrm{Tot}(\mathrm{Par}(\alpha))\Phi_{\mathbf{A}} = \Phi_{\mathbf{B}}\alpha$. Hence $\Phi$ defines a 2-natural isomorphism $\mathrm{id}_{\mathbf{Eft}} \Rightarrow \mathrm{Tot} \circ \mathrm{Par}$.

Next let $\Psi_{\mathbf{C}} \colon \mathrm{Par}(\mathrm{Tot}(\mathbf{C})) \to \mathbf{C}$ be the isomorphism of categories in Proposition 4.1.5, defined by $\Psi_{\mathbf{C}}A = A$ and $\Psi_{\mathbf{C}}f = \rhd_1 \circ f$. It preserves finite coproducts, the unit object $I$, and total morphisms. Hence $\Psi_{\mathbf{C}}$ is a morphism in $\mathbf{Ef}$. Let $F \colon \mathbf{C} \to \mathbf{D}$ be an arrow in $\mathbf{Ef}$. Note that the diagram below commutes,

$$\begin{array}{ccc} F(A + I_{\mathbf{C}}) & & \\ {\scriptstyle \varphi_{\mathrm{Tot}(F),A}}\big\downarrow & \searrow^{F\rhd_1} & \\ FA + I_{\mathbf{D}} & \xrightarrow[\rhd_1]{} & FA \end{array} \tag{4.8}$$

since (using the coproduct $FA \xrightarrow{GF\kappa_1} F(A + I_{\mathbf{C}}) \xleftarrow{GF\kappa_2} FI_{\mathbf{C}}$)

$$\begin{aligned} \rhd_1 \circ \varphi_{\mathrm{Tot}(F),A} \circ F\kappa_1 &= \rhd_1 \circ \kappa_1 && \text{by (4.6)} \\ &= \mathrm{id}_{FA} \\ &= F\mathrm{id}_A \\ &= F\rhd_1 \circ F\kappa_1 \\ \rhd_1 \circ \varphi_{\mathrm{Tot}(F),A} \circ F\kappa_2 &= \rhd_1 \circ \kappa_2 \circ 1_{FI_{\mathbf{C}}} && \text{by (4.7)} \\ &= 0_{FI_{\mathbf{C}},FA} \\ &= F0_{I_{\mathbf{C}},A} \\ &= F\rhd_1 \circ F\kappa_2 \,. \end{aligned}$$

Here $F0_{I_{\mathbf{C}},A} = 0_{FI_{\mathbf{C}},FA}$ because $F$ preserves the zero object and hence zero morphisms. Then for $f\colon A \rightharpoonup B$ in $\mathrm{Par}(\mathrm{Tot}(\mathbf{C}))$,

$$\begin{aligned}
\Psi_{\mathbf{D}}\mathrm{Par}(\mathrm{Tot}(F))f &= \rhd_1 \circ \varphi_{\mathrm{Tot}(F),B} \circ Ff \\
&= F\rhd_1 \circ Ff && \text{by (4.8)} \\
&= F(\rhd_1 \circ f) \\
&= F\Psi_{\mathbf{C}}f\,,
\end{aligned}$$

and hence $\Psi_{\mathbf{D}}\mathrm{Par}(\mathrm{Tot}(F)) = F\Psi_{\mathbf{C}}$. Let $\alpha\colon F \Rightarrow G$ be a 2-cell in $\mathbf{Ef}$. Then

$$\begin{aligned}
(\Psi_{\mathbf{D}}\mathrm{Par}(\mathrm{Tot}(\alpha)))_A &= \Psi_{\mathbf{D}}\mathrm{Par}(\mathrm{Tot}(\alpha))_A \\
&= \rhd_1 \circ \langle \alpha_A \rangle \\
&= \alpha_A \\
&= \alpha_{\Psi_{\mathbf{C}}A} \\
&= (\alpha\Psi_{\mathbf{C}})_A\,,
\end{aligned}$$

so that $\Psi_{\mathbf{D}}\mathrm{Par}(\mathrm{Tot}(\alpha)) = \alpha\Psi_{\mathbf{C}}$. Therefore $\Psi$ defines a 2-natural isomorphism $\mathrm{Par} \circ \mathrm{Tot} \Rightarrow \mathrm{id}_{\mathbf{Ef}}$. ■

### 4.2.1 State-and-effect triangles revisited

Now we have the definitions of 2-categories **Ef** and **Eft** of effectuses in partial and total form, and the equivalence $\mathbf{Ef} \simeq \mathbf{Eft}$ between them. Now we can describe state-and-effect triangles (Section 3.7) in a slightly better way, as diagrams in the 2-categories.

**Lemma 4.2.11.** *Let* $\mathbf{C}$ *be an effectus in partial form. Then both predicate and substate functors* $\mathrm{Pred}\colon \mathbf{C} \to \mathcal{S}\text{-}\mathbf{EMod}_{\leq}^{\mathrm{op}}$ *and* $\mathrm{St}_{\leq}\colon \mathbf{C} \to \mathcal{S}^{\mathrm{op}}\text{-}\mathbf{WMod}_{\leq}$ *are morphisms in* $\mathbf{Ef}$*.*

*Proof.* We already know that both functors preserve finite coproducts, see Propositions 3.4.13 and 3.5.11. Moreover both functors preserve (strictly) the unit object since

$$\mathcal{S} = \mathbf{C}(I, I) = \mathrm{Pred}(I) = \mathrm{St}_{\leq}(I)\,.$$

By verifying that $\mathrm{Pred}(\mathbb{1}_A) = \mathbb{1}_{\mathrm{Pred}(A)}$ and $\mathrm{St}_{\leq}(\mathbb{1}_A) = \mathbb{1}_{\mathrm{St}_{\leq}(A)}$, we conclude that they are morphisms in $\mathbf{Ef}$. (Alternatively, we know that both functors preserve total morphisms, so we can apply Lemma 4.2.2.) ■

**Lemma 4.2.12.** *For any effect monoid* $M$*, the adjunction*

$$M\text{-}\mathbf{EMod}_{\leq}^{\mathrm{op}} \underset{\mathrm{Hom}(-,M)}{\overset{\mathrm{Hom}(-,M)}{\underset{\longleftarrow}{\overset{\longrightarrow}{\top}}}} M^{\mathrm{op}}\text{-}\mathbf{WMod}_{\leq}$$

*from Proposition* 3.7.1 *is an adjunction in the 2-category* $\mathbf{Ef}$*.*

*Proof.* The two functors $\mathrm{Hom}(-,M)$ are respectively the substate and predicate functors of the effectuses $M\text{-}\mathbf{EMod}_{\le}^{\mathrm{op}}$ and $M^{\mathrm{op}}\text{-}\mathbf{WMod}_{\le}$. Therefore by Lemma 4.2.11, both functors are morphisms in $\mathbf{Ef}$. Thus it suffices to prove that the unit and counit of the adjunction are 2-cells in $\mathbf{Ef}$. By Lemma 4.2.4 it is equivalent to saying that each component of the unit and counit is total. The latter is the case by Proposition 3.7.4. ■

**Corollary 4.2.13.** *For each effectus in partial form* $\mathbf{C}$*, the state-and-effect triangle below, from Corollary* 3.7.2*, sits in the 2-category* $\mathbf{Ef}$ *of effectuses in partial form.*

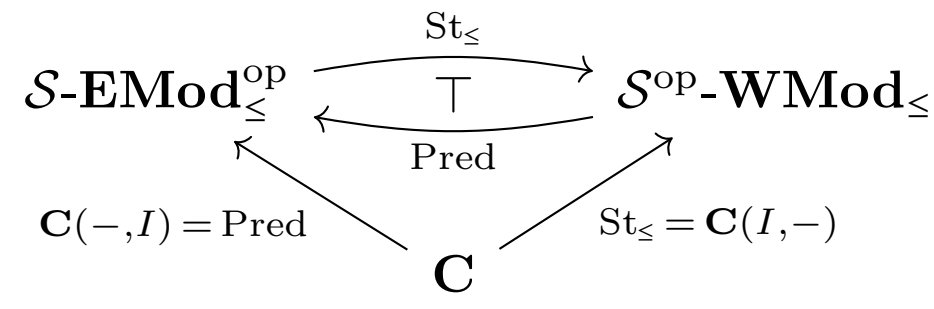

(4.9)

*Specifically, the categories, functors, and adjunction are objects, morphisms, and an adjunction in the 2-category* $\mathbf{Ef}$*, respectively.* ■

**Corollary 4.2.14.** *The 2-functor* $\mathrm{Tot}\colon \mathbf{Ef} \to \mathbf{Eft}$ *sends the state-and-effect triangle* (4.9) *in* $\mathbf{Ef}$ *to the following one in* $\mathbf{Eft}$.

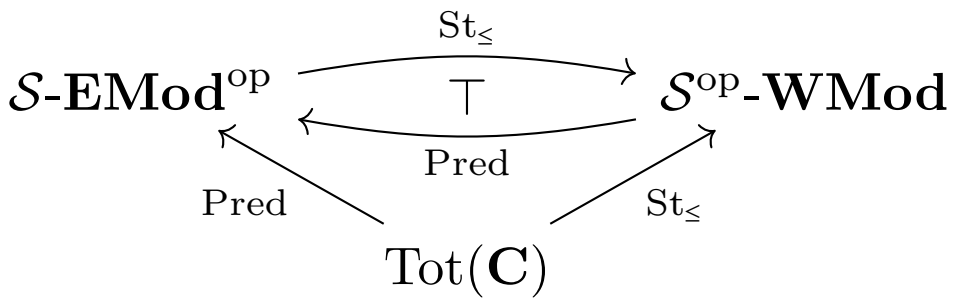

(4.10)

*Since* $\mathrm{Tot}\colon \mathbf{Ef} \to \mathbf{Eft}$ *is a part of the 2-equivalence, the two triangles* (4.9) *and* (4.10) *are related in the 2-equivalence* $\mathbf{Ef} \simeq \mathbf{Eft}$. ■

Note that we can also start with an effectus in total form $\mathbf{B}$. Then we define the predicate functor $\mathrm{Pred}\colon \mathbf{B} \to \mathcal{S}\text{-}\mathbf{EMod}^{\mathrm{op}}$ by $\mathrm{Pred}(X) = \mathrm{Par}(\mathbf{B})(X,1) = \mathbf{B}(X,1+1)$, and the substate functor $\mathrm{St}_{\le}\colon \mathbf{B} \to \mathcal{S}^{\mathrm{op}}\text{-}\mathbf{WMod}$ by $\mathrm{St}_{\le}(X) = \mathrm{Par}(\mathbf{B})(1,X) = \mathbf{B}(1,X+1)$, with scalars $\mathcal{S} = \mathrm{Par}(\mathbf{B})(1,1) = \mathbf{B}(1,1+1)$. Then we have the state-and-effect triangle on the left below, which is sent by $\mathrm{Par}\colon \mathbf{Eft} \to \mathbf{Ef}$ to the one on the right, up to isomorphism.

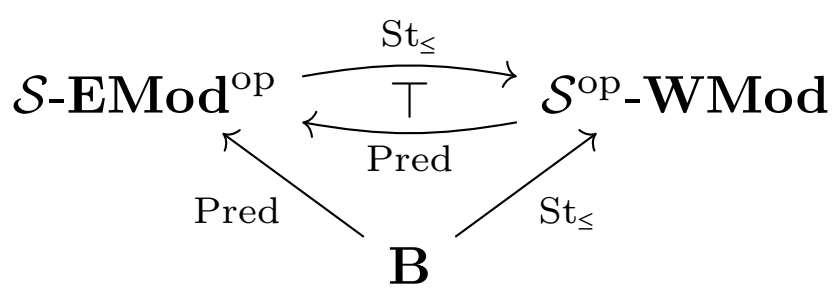

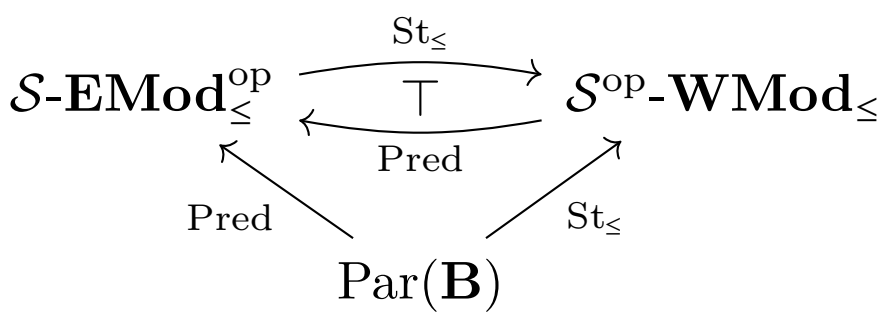

(Here it is 'up to isomorphism' since the functor $\mathrm{Par}(\mathrm{Pred})\colon \mathrm{Par}(\mathbf{B}) \to \mathrm{Par}(\mathcal{S}\text{-}\mathbf{EMod}^{\mathrm{op}})$ is not equal but isomorphic to $\mathrm{Pred}\colon \mathrm{Par}(\mathbf{B}) \to \mathcal{S}\text{-}\mathbf{EMod}_{\le}^{\mathrm{op}}$.)

In general, the equivalence $\mathbf{Ef} \simeq \mathbf{Eft}$ tells us that whether we start with an effectus in partial form or in total form, we get 'equivalent' results. It is thus a matter of choice to start with which one. On the one hand, the partial form is more convenient to work with, because we have the partially additive structure and we can avoid distractions in dealing with the lift monad. On the other hand, definitions are sometimes simpler in the total form; for example, compare the definitions of effectuses (Definitions 3.2.1 and 4.1.6; see also Proposition 3.8.6), and the definitions of the 2-categories (Definition 4.2.1).

## 4.3 Division effect monoids

Effect monoids with division operation $t\backslash s$ — called *division effect monoids* — are a convenient and well-behaved class of effect monoids. In the next section we will prove several additional results on convex sets over $M$, assuming that $M$ is a division effect monoid. They also naturally arise from effectuses with the normalization property, see Section 4.5. In this section we briefly study division effect monoids.

**Definition 4.3.1.** An effect monoid $M$ it said to admit **division** if for all $s, t \in M$ with $s \leq t$ and $t \neq 0$, there exists a unique $q$ such that $t \cdot q = s$. The $q$ is called the quotient and denoted by $t\backslash s$. An effect monoid that admits division is called a **division effect monoid**.

Since an effect monoid $M$ may be noncommutative, $M$ admitting division may differ from $M^{\mathrm{op}}$ admitting division. In the latter case, one has 'right' quotients $s/t$ satisfying $(s/t) \cdot t = s$ in $M$. When $M^{\mathrm{op}}$ admits division, we call $M$ a **right-division effect monoid**.

The prime example of an division effect monoid is the unit interval $[0, 1]$. We note that the two-element effect monoid $\{0, 1\}$ also admits division in a trivial way.

**Lemma 4.3.2.** *Let $M$ be a division effect monoid. For $r, s, t, u \in M$ with $r \leq s$, $t \cdot s \leq u$, $s \neq 0$ and $u \neq 0$, one has $(u\backslash ts) \cdot (s\backslash r) = u\backslash tr$. Setting $t = 1$, we have $(u\backslash s) \cdot (s\backslash r) = u\backslash r$ for all $r \leq s \leq u$ with $s \neq 0$.*

*Proof.* By $u \cdot (u\backslash ts) \cdot (s\backslash r) = t \cdot s \cdot (s\backslash r) = t \cdot r$. ■

**Lemma 4.3.3.** *A division effect monoid has no nontrivial zero divisors: if $r \cdot s = 0$, then $r = 0$ or $s = 0$.*

*Proof.* We prove that $r \cdot s = 0$ and $r \neq 0$ imply $s = 0$. Note that the quotient $r\backslash rs$ exists under the assumption. Then by $r \cdot s = 0 = r \cdot 0$ we have $s = r\backslash rs = 0$. ■

For an effect module $E$ over an effect monoid $M$ and for $a \in E$, let

$$\downarrow(a) = \{b \in E \mid b \leq a\}$$

be the (principal) downset. Then $\downarrow(a)$ is an effect module over $M$ with the top $a$, where the sum $b \ovee c$ is defined iff $b \ovee c$ is defined in $E$ and $b \ovee c \leq a$. In particular, for an effect monoid $M$ and for $t \in M$, the downset $\downarrow(t)$ is an effect module over $M^{\mathrm{op}}$, since $M$ itself is an effect module over $M^{\mathrm{op}}$ via $s \cdot r = r \cdot s$.

**Lemma 4.3.4.** *Let $M$ be a division effect monoid. For each nonzero $t \in M$, the 'multiplication by $t$' map $t \cdot (-) \colon M \to \downarrow(t)$ is a unital $M^{\mathrm{op}}$-module isomorphism, with the inverse $t\backslash(-) \colon \downarrow(t) \to M$.*

*Proof.* It is easy to verify that $(-) \cdot t \colon M \to {\downarrow}(t)$ is a unital module map. Moreover, $M$ admitting division implies that $t \cdot (-) \colon M \to {\downarrow}(t)$ is bijective. To prove the claim, therefore, it suffices to show that the map reflects the summability, namely that $t \cdot r \perp t \cdot s$ and $t \cdot r \ovee t \cdot s \le t$ imply $r \perp s$. If $r = 0$, the condition is trivially satisfied. Assume $r \neq 0$. Then $t \cdot r$ is nonzero by Lemma 4.3.3. Note that $t \cdot s^{\perp} = t \ominus t \cdot s \ge t \cdot r$ and hence $t \cdot s^{\perp}$ is nonzero as well. Using Lemma 4.3.2, we have

$$r = t\backslash tr = (t\backslash(t \cdot s^{\perp})) \cdot \big((t \cdot s^{\perp})\backslash tr\big) = s^{\perp} \cdot \big((t \cdot s^{\perp})\backslash tr\big) \le s^{\perp}\,,$$

that is, $r \perp s$. ∎

**Corollary 4.3.5.** *Let $M$ be a division effect monoid. For any nonzero $t \in M$ the following hold.*

(i) $t\backslash 0 = 0$.

(ii) $t\backslash t = 1$.

(iii) $t\backslash(r \ovee s) = t\backslash r \ovee t\backslash s$ *for all* $r, s \in M$ *such that* $r \ovee s \le t$

(iv) $t\backslash(r \cdot s) = (t\backslash r) \cdot s$ *for all* $r, s \in M$ *such that* $r \cdot s \le t$.

*Proof.* Because $t\backslash(-) \colon {\downarrow}(t) \to M$ is a unital $M^{\mathrm{op}}$-module map. ∎

Note that both $\{0,1\}$ and $\{0\}$ are division effect monoids, in a trivial way. In fact, these are the only examples of finite division effect monoids.

**Corollary 4.3.6.** *If a division effect monoid $M$ is finite, then $M = \{0,1\}$ or $M = \{0\}$.*

*Proof.* Assume, towards a contradiction, that $t \in M$ is an element that is neither 0 or 1. By Lemma 4.3.4 we have $M \cong {\downarrow}(t)$. But this is not possible if $M$ is finite, since the cardinalities of $M$ and ${\downarrow}(t)$ are different. Thus $M = \{0,1\}$ or $M = \{0\}$. ∎

One might wonder how many division effect monoids exist, other than obvious examples such as $[0,1]$, $\{0,1\}$, and $\{0\}$. Below we will show that a partially ordered division ring induces a division effect monoid under certain mild conditions. Based on this result, we give examples of noncommutative division effect monoids. From this it follows that there are examples of effectuses whose scalars form noncommutative division effect monoids, since $M$-$\mathbf{EMod}_{\le}^{\mathrm{op}}$ and $M$-$\mathbf{WMod}_{\le}$ are effectuses for any effect monoid $M$.

Recall that a **division ring** is a ring $R$ such that for each nonzero element $a \in R$ there is a 'multiplicative inverse' $a^{-1}$ satisfying $a^{-1} \cdot a = 1 = a \cdot a^{-1}$. (For each $a$ a multiplicative inverse $a^{-1}$ is unique.) A **partially ordered division ring** is a division ring that is at the same time a partially ordered ring. A partially ordered division ring is said to be **division-closed** if $a > 0$ implies $a^{-1} > 0$.

**Lemma 4.3.7.** *Let $R$ be a division-closed partially ordered division ring with* $0 \le 1$. *Then the unit interval* $[0,1]_R$ *is a division effect monoid. (It is also a right-division effect monoid.)*

*Proof.* It is straightforward to see that the unit interval $[0,1]_R$ of a partially ordered ring with $0 \leq 1$ is an effect monoid. To prove that $[0,1]_R$ admits division, let $s,t \in [0,1]_R$ be such that $s \leq t$ and $t \neq 0$. We claim $t \backslash s = t^{-1} \cdot s$. Since $t^{-1} > 0$ we have

$$0 = 0 \cdot s \leq t^{-1} \cdot s \leq t^{-1} \cdot t = 1\,,$$

so that $t^{-1} \cdot s \in [0,1]_R$. Clearly $t \cdot (t^{-1} \cdot s) = s$. If $q \in [0,1]_R$ satisfies $t \cdot q = s$, then $q = t^{-1} \cdot t \cdot q = t^{-1} \cdot s$, so that the quotient is unique. (Similarly one proves $s/t = s \cdot t^{-1}$.) ∎

**Lemma 4.3.8.** *Any totally ordered division ring $R$ satisfies the assumption of Lemma* 4.3.7*. Thus $[0,1]_R$ is division effect monoid.*

*Proof.* If $1 < 0$, then $1 < 0 \leq 1 \cdot 1 = 1$, a contradiction. Since $R$ is totally ordered, $0 \leq 1$ holds. To see that $R$ is division-closed, let $a \in R$ satisfy $a > 0$. If $a^{-1} < 0$, then $1 = a \cdot a^{-1} < 0$, which contradicts $0 \leq 1$. Since $a^{-1} \neq 0$, we have $a^{-1} > 0$. ∎

**Example 4.3.9.** We can now give examples of noncommutative division effect monoids, using the 'skew Laurent series ring' construction of totally ordered division ring. We here sketch the construction and refer to [184] for more details. Let $R$ be a ring. A formal Laurent series in an indeterminate $x$ over $R$ is a formal expression of the form $\sum_{i\in\mathbb{Z}} a_i x^i$ where $a_i \in R$ for each $i \in \mathbb{Z}$ such that $a_i = 0$ for all but finitely many negative indices $i < 0$. We fix an automorphism $\sigma$ on $R$. Then formal Laurent series in $x$ over $R$ form a ring, denoted by $R(\!(x;\sigma)\!)$, with the obvious addition and the 'skewed' multiplication:

$$\Big(\sum_{i\in\mathbb{Z}} a_i x^i\Big) \cdot \Big(\sum_{j\in\mathbb{Z}} b_j x^j\Big) = \sum_{i,j\in\mathbb{Z}} a_i \sigma^i(b_j) x^{i+j}\,.$$

This means that $a \in R$ and the indeterminate $x$ commute as $xa = \sigma(a)x$ via the automorphism $\sigma$. It can be shown that if $R$ is a totally ordered division ring and $\sigma$ is an order-preserving automorphism, then $R(\!(x;\sigma)\!)$ is a totally ordered division ring too [184, Proposition 18.5]. (Concrete examples of $R$ and $\sigma$ can be found in [184], see the paragraphs after Corollary 18.6.) The positive cone of $R(\!(x;\sigma)\!)$ is given by:

$$\{0\} \cup \left\{ \sum_{i=n}^{\infty} a_i x^i \;\middle|\; n \in \mathbb{Z},\ (a_i)_{i=n}^{\infty} \text{ in } R, \text{ and } a_n > 0 \right\}.$$

By Lemma 4.3.8, the unit interval $[0,1]_{R(\!(x;\sigma)\!)}$ is a division effect monoid. Explicitly, $[0,1]_{R(\!(x;\sigma)\!)}$ is:

$$\{0\} \cup \left\{ \sum_{i=0}^{\infty} a_i x^i \;\middle|\; (a_i)_{i=0}^{\infty} \text{ in } R \text{ and } 0 < a_0 < 1 \right\}$$
$$\cup \left\{ 1 + \sum_{i=n}^{\infty} a_i x^i \;\middle|\; n \in \mathbb{Z} \text{ with } n > 0,\ (a_i)_{i=n}^{\infty} \text{ in } R, \text{ and } a_n < 0 \right\}.$$

If the automorphism $\sigma$ is nontrivial, then $[0,1]_{R(\!(x;\sigma)\!)}$ is noncommutative. To see this, let $a \in R$ be such that $a > 0$ and $\sigma(a) \neq a$. Then both $1 - x$ and $1 - ax$ are in the unit interval $[0,1]_{R(\!(x;\sigma)\!)}$, but

$$(1 - ax) \cdot (1 - x) = 1 - (a+1)x + ax^2$$
$$(1 - x) \cdot (1 - ax) = 1 - (a+1)x + \sigma(a)x^2 \,.$$

Hence $(1 - ax) \cdot (1 - x) \neq (1 - x) \cdot (1 - ax)$.

## 4.4 From convex sets to weight modules

In Section 3.6 we defined a functor $\mathrm{B}\colon M\text{-}\mathbf{WMod} \to M\text{-}\mathbf{Conv}$ by taking the bases:

$$\mathrm{B}(X) = \big\{x \in X \;\big|\; |x| = 1\big\}\,.$$

In this section we give a construction in the other direction $M\text{-}\mathbf{Conv} \to M\text{-}\mathbf{WMod}$, under the assumption that $M$ is a division effect monoid.

We use the 'lifting' construction for convex sets from [150, Definition 3], with a generalization of the scalars $[0,1]$ to an arbitrary division effect monoid.

**Definition 4.4.1.** Let $K$ be a convex set over a division effect monoid $M$. We define the **lifting** $\mathcal{L}(K)$ of $K$ to be the quotient

$$\mathcal{L}(K) = (M \times K)/\!\sim\,,$$

where $\sim$ is the equivalence relation generated by $(0,x) \sim (0,y)$ for $x,y \in K$. We write $(0,*)$ for the equivalence class consisting of $(0,x)$. We may describe $\mathcal{L}(K)$ concretely as:

$$\mathcal{L}(K) \cong \big((M \setminus \{0\}) \times K\big) \uplus \{(0,*)\}\,.$$

We equip $\mathcal{L}(K)$ with the structure of a weight module as follows. For $(r,x) \in \mathcal{L}(K)$ we define the weight by $|(r,x)| = r$. This means that summability is defined as follows:

$$(r,x) \perp (s,y) \iff |(r,x)| \perp |(s,y)| \iff r \perp s\,.$$

We define the sum by

$$(r,x) \ovee (s,y) = \begin{cases} (t, [\![t\backslash r|x\rangle + t\backslash s|y\rangle]\!]) & \text{if } t \neq 0 \\ (0,*) & \text{if } t = 0\,, \end{cases}$$

where $t := r \ovee s$. Note that $t\backslash r \ovee t\backslash s = 1$ whenever $t \neq 0$. The zero in $\mathcal{L}(K)$ is $(0,*)$, and the scalar multiplication is defined by $s \cdot (r,x) = (s \cdot r, x)$.

In [150] the lifting $\mathcal{L}(K)$ is shown to be a convex set. Here we prove that $\mathcal{L}(K)$ in fact forms a weight module. This then implies that $\mathcal{L}(K)$ is a convex set by Proposition 3.6.9.

**Lemma 4.4.2.** *The lifting $\mathcal{L}(K)$ is a weight module over $M$.*

*Proof.* We first show that $\mathcal{L}(K)$ is a PCM. Commutativity $(r,x)\ovee(s,y) = (s,y)\ovee(r,x)$ and zero law $(r,x) \ovee (0,*) = (r,x)$ are easy. To see associativity, assume that $(r,x) \perp (s,y)$ and $(r,x) \ovee (s,y) \perp (t,z)$. Then $(s,y) \perp (t,z)$ and $(r,x) \perp (s,y) \ovee (t,z)$ since the summability is determined by the associated scalars. We need to show that:

$$((r,x) \ovee (s,y)) \ovee (t,z) = (r,x) \ovee ((s,y) \ovee (t,z)) \,. \tag{4.11}$$

If at least one of $r,s,t$ is zero, this follows from the zero law. So assume all of them are nonzero. Let $u = r \ovee s$ and $v = u \ovee t = r \ovee s \ovee t$. Then

$$\begin{aligned}((r,x) \ovee (s,y)) \ovee (t,z) &= (u, [\![u\backslash r|x\rangle + u\backslash s|y\rangle]\!]) \ovee (z,t) \\ &= \big(v, [\![v\backslash u\big|[\![u\backslash r|x\rangle + u\backslash s|y\rangle]\!]\big\rangle + v\backslash t|z\rangle]\!]\big)\,.\end{aligned}$$

By the axioms of convex sets,

$$\begin{aligned}&[\![v\backslash u\big|[\![u\backslash r|x\rangle + u\backslash s|y\rangle]\!]\big\rangle + v\backslash t|z\rangle]\!] \\ &= [\![(v\backslash u)\cdot(u\backslash r)|x\rangle + (v\backslash u)\cdot(u\backslash s)|y\rangle + v\backslash t|z\rangle]\!] \\ &= [\![v\backslash r|x\rangle + v\backslash s|y\rangle + v\backslash t|z\rangle]\!]\,.\end{aligned}$$

Hence

$$((r,x) \ovee (s,y)) \ovee (t,z) = (v, [\![v\backslash r|x\rangle + v\backslash s|y\rangle + v\backslash t|z\rangle]\!])\,. \tag{4.12}$$

Similarly one has

$$(r,x) \ovee ((s,y) \ovee (t,z)) = (v, [\![v\backslash r|x\rangle + v\backslash s|y\rangle + v\backslash t|z\rangle]\!])\,,$$

proving the identity (4.11).

Next we show that $\mathcal{L}(K)$ is a partial module over $M$. Clearly $t\cdot(0,*) = (0,*)$, and

$$\begin{aligned}t\cdot((r,x) \ovee (s,y)) &= t\cdot(u, [\![u\backslash r|x\rangle + u\backslash s|y\rangle]\!]) && \text{where } u := r \ovee s \\ &= (tu, [\![u\backslash r|x\rangle + u\backslash s|y\rangle]\!]) \\ &= (tu, [\![tu\backslash tr|x\rangle + tu\backslash ts|y\rangle]\!]) \\ &= (tr,x) \ovee (ts,y) && \text{since } tu = tr \ovee ts \\ &= t\cdot(r,x) \ovee t\cdot(s,y)\,.\end{aligned}$$

We also have $0\cdot(r,x) = (0,*)$, and

$$\begin{aligned}(s \ovee t)\cdot(r,x) &= ((s \ovee t)\cdot r, x) \\ &= (sr \ovee tr, x) \\ &= (u, [\![1|x\rangle]\!]) && \text{where } u := sr \ovee tr \\ &= (u, [\![u\backslash sr|x\rangle \ovee u\backslash tr|x\rangle]\!]) \\ &= (sr,x) \ovee (tr,x) \\ &= s\cdot(r,x) \ovee t\cdot(r,x)\,.\end{aligned}$$

Hence the scalar multiplication $\mathcal{L}(K) \times M \to \mathcal{L}(K)$ is a PCM bimorphism. We can easily verify that $1\cdot(r,x) = (r,x)$ and $(s\cdot t)\cdot(r,x) = s\cdot(t\cdot(r,x))$, showing that $\mathcal{L}(K)$ is a partial module.

Essentially by construction, the weight $|(r,x)| = r$ satisfies the axioms of weight modules. Thus $\mathcal{L}(K)$ is a weighed partial module over $M$. ∎

In the proof we obtained the identity (4.12) about three summable elements in $\mathcal{L}(K)$. This can be generalized to $n$-ary sum.

**Lemma 4.4.3.** *Let $(r_1, x_1), \dots, (r_n, x_n)$ be summable elements in $\mathcal{L}(K)$. Then*

$$\textstyle\bigovee_i (r_i, x_i) = \big(t, [\![\sum_i t\backslash r_i | x_i\rangle]\!]\big) \qquad \text{where} \quad t = \bigovee_i r_i\,.$$

*This holds also for $t = 0$ if the right-hand side is interpreted as $(0, *)$. In particular, if $\bigovee_i r_i = 1$, we have*

$$\textstyle\bigovee_i (r_i, x_i) = \big(1, [\![\sum_i r_i | x_i\rangle]\!]\big)\,.$$

*Proof.* By induction on $n$. ■

**Lemma 4.4.4.** *The map $\eta_K \colon K \to \mathrm{B}(\mathcal{L}(K))$ defined by $\eta_K(x) = (1, x)$ is an isomorphism of convex sets.*

*Proof.* Clearly $\eta_K$ is a bijection, with the inverse $\mathrm{B}(\mathcal{L}(K)) \to K$ given by the second projection. By abstract nonsense (see [227, Lemma 5.6.1]), the forgetful functor

$$M\text{-}\mathbf{Conv} = \mathcal{K}\ell(\mathcal{D}_M) \longrightarrow \mathbf{Set}$$

reflects isomorphisms. Hence it suffices to prove that $\eta_K$ is affine. Let $\sum_i r_i|x_i\rangle \in \mathcal{D}_M(K)$ be a formal convex sum. Then

$$\begin{aligned}
\eta_K([\![\textstyle\sum_i r_i|x_i\rangle]\!]) &= (1, [\![\textstyle\sum_i r_i|x_i\rangle]\!]) \\
&= \textstyle\bigovee_i (r_i, x_i) && \text{by Lemma 4.4.3} \\
&= \textstyle\bigovee_i r_i \cdot (1, x_i) \\
&= \textstyle\bigovee_i r_i \cdot \eta_K(x_i)\,.
\end{aligned}$$

The last expression is the convex sum $[\![\sum_i r_i|\eta_K(x_i)\rangle]\!]$ in $\mathrm{B}(\mathcal{L}(K))$. Thus $\eta_K$ is affine. ■

**Proposition 4.4.5.** *The mapping $K \mapsto \mathcal{L}(K)$, with the unit $\eta_K \colon K \to \mathrm{B}(\mathcal{L}(K))$ from Lemma 4.4.4, forms a left adjoint to the functor $\mathrm{B} \colon M\text{-}\mathbf{WMod} \to M\text{-}\mathbf{Conv}$,*

*Proof.* Let $X \in M\text{-}\mathbf{WMod}$, and $f \colon K \to B(X)$ be a morphism in $M\text{-}\mathbf{Conv}$. We define a map $\overline{f} \colon \mathcal{L}(K) \to X$ by $\overline{f}(r, x) = r \cdot f(x)$. Then $\overline{f}$ is a weight-preserving module map. For example, it preserves the sum as follows.

$$\begin{aligned}
\overline{f}((r, x) \ovee (s, x)) &= \overline{f}(t, [\![t\backslash r|x\rangle + t\backslash s|y\rangle]\!]) && \text{where } t = r \ovee s \\
&= t \cdot f([\![t\backslash r|x\rangle + t\backslash s|y\rangle]\!]) \\
&= t \cdot [\![t\backslash r|f(x)\rangle + t\backslash s|f(y)\rangle]\!] \\
&= t \cdot ((t\backslash r) \cdot f(x) \ovee (t\backslash s) \cdot f(y)) \\
&= (t \cdot (t\backslash r)) \cdot f(x) \ovee (t \cdot (t\backslash s)) \cdot f(y) \\
&= r \cdot f(x) \ovee s \cdot f(y) \\
&= \overline{f}(r, x) \ovee \overline{f}(s, y)
\end{aligned}$$

It makes the following diagram in $M$-**Conv** commute.

$$\begin{array}{ccc} \mathrm{B}(\mathcal{L}(K)) & \xrightarrow{\mathrm{B}(\overline{f})} & \mathrm{B}(X) \\ {\scriptstyle \eta_K}\uparrow & \nearrow_{f} & \\ K & & \end{array} \tag{4.13}$$

Indeed,

$$(\mathrm{B}(\overline{f}) \circ \eta_K)(x) = \mathrm{B}(\overline{f})(\eta_K(x)) = \mathrm{B}(\overline{f})(1,x) = \overline{f}(1,x) = 1 \cdot f(x) = f(x)\,.$$

The map $\overline{f}$ is unique with this property. If $g\colon \mathcal{L}(K) \to X$ makes the diagram (4.13) commute, then

$$g(r,x) = g(r \cdot (1,x)) = r \cdot g(1,x) = r \cdot g(\eta_K(x)) = r \cdot f(x) = \overline{f}(r,x)\,. \qquad \blacksquare$$

We thus have an adjunction:

$$M\text{-}\mathbf{WMod} \underset{\mathcal{L}}{\overset{\mathrm{B}}{\rightleftarrows}} M\text{-}\mathbf{Conv} \quad (\top) \tag{4.14}$$

In particular, the assignment $K \mapsto \mathcal{L}(K)$ extends to a functor $\mathcal{L}\colon M\text{-}\mathbf{Conv} \to M\text{-}\mathbf{WMod}$. Explicitly, for an affine map $f\colon K \to L$ between convex sets, the map $\mathcal{L}(f)\colon \mathcal{L}(K) \to \mathcal{L}(L)$ in $M$-**WMod** is given by

$$\mathcal{L}(f)(r,x) = r \cdot (1, f(x)) = (r, f(x))\,. \tag{4.15}$$

Moreover, by Lemma 4.4.4, the unit $\eta$ of the adjunction is an isomorphism. We obtain the following result.

**Proposition 4.4.6.** *The adjunction* (4.14) *is a coreflection (see Definition* 2.1.6*). In particular, the left adjoint functor* $\mathcal{L}\colon M\text{-}\mathbf{Conv} \to M\text{-}\mathbf{WMod}$ *is full and faithful.* ■

Thus the category $M$-**Conv** may be seen, up to equivalence, as a coreflective subcategory of $M$-**WMod**. Any convex set can be identified with the base $\mathrm{B}(X)$ of some weight module $X$.

Next we aim to restrict the adjunction (4.14) to an equivalence. Note that the counit $\varepsilon_X\colon \mathcal{L}(\mathrm{B}(X)) \to X$ of (4.14) is given by $\varepsilon_X(r,x) = r \cdot x$. By Proposition 2.1.8, we only need to characterize weight modules $X$ such that $\varepsilon_X\colon \mathcal{L}(\mathrm{B}(X)) \to X$ is an isomorphism.

**Definition 4.4.7.** We say that a weight module $X$ satisfies the **normalization property** if for each nonzero $x \in X$, there is a unique $y \in \mathrm{B}(X)$ with $x = |x| \cdot y$.

**Lemma 4.4.8.** *For a weight module $X$, the component of the counit $\varepsilon_X\colon \mathcal{L}(\mathrm{B}(X)) \to X$ is an isomorphism if and only if $X$ satisfies the normalization property.*

*Proof.* Since the counit $\varepsilon_X$ is weight-preserving, $x \perp y$ if and only if $\varepsilon_X(x) \perp \varepsilon_X(y)$, i.e. $\varepsilon_X$ reflects summability. From this it follows that $\varepsilon_X$ is an isomorphism in $M$-**WMod** if and only if $\varepsilon_X$ is a bijection. The latter means that for each $x \in X$, there is a unique $(r, y) \in \mathcal{L}(\mathrm{B}(X))$ such that $r \cdot y = x$. Since $|x| = |r \cdot y| = r$, when $x \neq 0$ it follows that there is a unique $y \in Y$ with $|x| \cdot y = x$. Conversely, suppose that $X$ satisfies the normalization property. Let $x \in X$ be arbitrary. Define a function $k\colon X \to \mathcal{L}(\mathrm{B}(X))$ by $k(0) = (0, *)$ and $k(x) = (|x|, \overline{x})$ for nonzero $x \in X$, where $\overline{x}$ is the unique element satisfying $x = |x| \cdot \overline{x}$. Then $k$ is the inverse of $\varepsilon_X$. ■

We denote the full subcategories consisting of weight modules with the normalization property as follows.

$$M\text{-}\mathbf{WModn} \hookrightarrow M\text{-}\mathbf{WMod} \qquad M\text{-}\mathbf{WModn}_{\le} \hookrightarrow M\text{-}\mathbf{WMod}_{\le}$$

**Corollary 4.4.9.** *For any division effect monoid $M$, the adjunction $M$-**WMod** $\rightleftarrows$ $M$-**Conv** restricts to an adjoint equivalence:*

$$M\text{-}\mathbf{WModn} \underset{\mathcal{L}}{\overset{\mathrm{B}}{\underset{\longleftarrow}{\overset{\longrightarrow}{\simeq}}}} M\text{-}\mathbf{Conv}$$ ■

*Proof.* By Lemma 4.4.8, objects $X \in M$-**WModn** are precisely $X \in M$-**WMod** such that $\varepsilon_X\colon \mathcal{L}(\mathrm{B}(X)) \to X$ is an isomorphism. Thus we obtain the equivalence by Proposition 2.1.8. ■

The following useful result is an adaptation of the result of Tull about normalization in an effectus, see Proposition 4.5.4.

**Proposition 4.4.10.** *Any weight module over $[0,1]$ satisfies the normalization property. As a consequence,* $[0,1]$-**WMod** $= [0,1]$-**WModn** $\simeq [0,1]$-**Conv**.

As is clear from the proof below, the proposition also holds for suitably well-behaved effect monoids, such as the unit interval $[0,1]_{\mathbb{Q}}$ of rational numbers.

*Proof.* Let $X$ be a weight module over $[0,1]$. Let $x \in X$ be a nonzero element. Note that we can always find a natural number $n$ and $r \in [0,1]$ such that $\bigovee_{j=1}^{n} r \cdot |x| = 1$. For instance, take $n = \lceil 1/|x| \rceil$ and $r = 1/(n|x|)$. Then $\bigovee_{j=1}^{n} r \cdot |x| = \bigovee_{j=1}^{n} |r \cdot x|$ is defined in $[0,1]$, so that $\bigovee_{j=1}^{n} r \cdot x \in X$ is defined $X$. We claim that $\bigovee_{j=1}^{n} r \cdot x$ is a normalization of $x$. Indeed, $\bigovee_{j=1}^{n} r \cdot x \in \mathrm{B}(X)$ since

$$\Big| \bigovee_{j=1}^{n} r \cdot x \Big| = \bigovee_{j=1}^{n} r \cdot |x| = 1\,,$$

and moreover,

$$|x| \cdot \Big( \bigovee_{j=1}^{n} r \cdot x \Big) = \bigovee_{j=1}^{n} (|x| \cdot r) \cdot x = \Big( \bigovee_{j=1}^{n} r \cdot |x| \Big) \cdot x = 1 \cdot x = x\,.$$

The normalization is unique: if $x = |x| \cdot y$ for $y \in \mathrm{B}(X)$, then

$$y = \Big( \bigovee_{j=1}^{n} r \cdot |x| \Big) \cdot y = \bigovee_{j=1}^{n} r \cdot (|x| \cdot y) = \bigovee_{j=1}^{n} r \cdot x\,.$$ ■

**Lemma 4.4.11.** *Let $M$ be a division effect monoid. Let $X$ be a weight module over $M$, with the normalization property. For $x, y \in X$ and $r \in M$, if $r \cdot x = r \cdot y$ and $r \neq 0$, then $x = y$.*

*Proof.* Under the given condition, it follows that $x = 0 \iff y = 0$, via Lemma 4.3.3. Hence we assume that $x \neq 0$ and $y \neq 0$. Let

$$s := |r \cdot x| = |r \cdot y| \,.$$

Then $r \cdot |x| = s = r \cdot |y|$, so that $|x| = r \backslash s = |y|$. There are $\overline{x}, \overline{y}, z \in \mathrm{B}(X)$ such that $x = |x| \cdot \overline{x}$, $y = |y| \cdot \overline{y}$, and

$$r \cdot x = s \cdot z = r \cdot y \,.$$

Now

$$\begin{aligned}(s \cdot |x|) \cdot \overline{x} &= s \cdot (|x| \cdot \overline{x}) \\ &= s \cdot x = s \cdot y \\ &= s \cdot (|y| \cdot \overline{y}) \\ &= (s \cdot |y|) \cdot \overline{y} = (s \cdot |x|) \cdot \overline{y} \,.\end{aligned}$$

Since $(s \cdot |x|) \cdot \overline{x} = (s \cdot |x|) \cdot \overline{y}$ is nonzero, $\overline{x} = \overline{y}$ by the uniqueness of the base condition. Therefore $x = |x| \cdot \overline{x} = |y| \cdot \overline{y} = y$. ■

**Lemma 4.4.12.** *If $M$ is a division effect monoid, the subcategory $M$-$\mathbf{WModn}_{\leq} \hookrightarrow M$-$\mathbf{WMod}_{\leq}$ is closed under binary coproducts: $X + Y \in M$-$\mathbf{WModn}_{\leq}$ for each $X, Y \in M$-$\mathbf{WModn}_{\leq}$.*

*Proof.* Let $X, Y \in M$-$\mathbf{WModn}_{\leq}$. Recall that the binary coproduct in $M$-$\mathbf{WMod}_{\leq}$ is given by:

$$X + Y = \{(x, y) \in X \times Y \mid |x| \perp |y|\}$$

with $|(x, y)| = |x| \ovee |y|$. Suppose that $(x, y) \in X + Y$ is nonzero, i.e. $x \neq 0$ or $y \neq 0$. By symmetry we may assume $x \neq 0$. Let $\overline{x} \in X$ be the unique element with $x = |x| \cdot \overline{x}$.

*Case 1:* $y = 0$. The element $(\overline{x}, 0) \in X + Y$ satisfies $(\overline{x}, 0) \in \mathrm{B}(X + Y)$ and $(x, 0) = |(x, 0)| \cdot (\overline{x}, 0)$.

*Case 2:* $y \neq 0$. Let $\overline{y} \in X$ be the unique element with $y = |y| \cdot \overline{y}$. Let $r = |(x, y)| = |x| \ovee |y|$. Then we have

$$((r \backslash |x|) \cdot \overline{x}, (r \backslash |y|) \cdot \overline{y}) \in \mathrm{B}(X + Y)$$

and

$$(x, y) = |(x, y)| \cdot ((r \backslash |x|) \cdot \overline{x}, (r \backslash |y|) \cdot \overline{y}) \,.$$

We proved the existence part of normalization. To see the uniqueness, suppose that

$$|(x, y)| \cdot (x', y') = (x, y) = |(x, y)| \cdot (x'', y'')$$

for $(x', y'), (x'', y'') \in \mathrm{B}(X + Y)$. We have $|(x, y)| \cdot x' = x = |(x, y)| \cdot x''$. Since $|(x, y)|$ is nonzero, by Lemma 4.4.11 we obtain $x' = x''$. Similarly $y' = y''$, so that $(x', y') = (x'', y'')$. ■

Note that we assume that $M$ is a division effect monoid. in Lemma 4.4.12. It turns out that division is necessary.

**Lemma 4.4.13.** *Let $M$ be an effect monoid, and let $M + M$ be a coproduct in $M$-$\mathbf{WMod}_{\le}$. If $M + M$ satisfies the normalization property, then $M$ admits division.*

*Proof.* Let $s, t \in M$ be such that $s \le t$ and $t \ne 0$. Then $(s, t \ominus s) \in M + M$, and hence there is a unique $(q_1, q_2) \in \mathrm{B}(M + M)$ such that $|(s, t \ominus s)| \cdot (q_1, q_2) = (s, t \ominus s)$. Note that

$$|(s, t \ominus s)| = |s| \ovee |t \ominus s| = s \ovee (t \ominus s) = t\,.$$

that is, $t \cdot q_1 = s$ and $t \cdot q_2 = t \ominus s$. Thus a quotient $q_1 = t\backslash s$ exists. To prove that the quotient is unique, assume that $q_1'$ satisfies $t \cdot q_1' = s$. Then we have

$$t \cdot (q_1')^{\perp} = t \ominus (t \cdot q_1') = t \ominus s\,,$$

so that $|(s, t \ominus s)| \cdot (q_1', (q_1')^{\perp}) = (s, t \ominus s)$, with $(q_1', (q_1')^{\perp}) \in \mathrm{B}(M + M)$. By the uniqueness of normalization, $(q_1', (q_1')^{\perp}) = (q_1, q_2)$, hence $q_1' = q_1$. ∎

Therefore the subcategory $M$-$\mathbf{WModn}_{\le} \hookrightarrow M$-$\mathbf{WMod}_{\le}$ is closed under binary coproducts if and only if $M$ admits division.

**Theorem 4.4.14.** *If $M$ is a division effect monoid, the category $M$-$\mathbf{WModn}_{\le}$ is an effectus in partial form.*

*Proof.* The category $M$-$\mathbf{WModn}_{\le}$ is a full subcategory of the effectus $M$-$\mathbf{WMod}_{\le}$. Clearly $1 = \{0\}$ and $M$ are in $M$-$\mathbf{WModn}_{\le}$, i.e. they satisfy the normalization property, and the subcategory is closed under binary coproducts. Therefore $M$-$\mathbf{WModn}_{\le}$ is an effectus by Corollary 3.8.8. ∎

**Corollary 4.4.15.** *If $M$ is a division effect monoid, the category $M$-$\mathbf{Conv}$ is an effectus in total form.*

*Proof.* The total subcategory $M$-$\mathbf{WModn} = \mathrm{Tot}(M\text{-}\mathbf{WModn}_{\le})$ of the effectus $M$-$\mathbf{WModn}_{\le}$ is an effectus in total form (Theorem 4.1.11). By Corollary 4.4.9, the category $M$-$\mathbf{Conv}$ is equivalent to $M$-$\mathbf{WModn}$, and hence an effectus in total form too. ∎

In fact, the lifting construction $\mathcal{L}(K)$ was introduced in [150] in order to obtain a concrete description of finite coproducts in $M$-$\mathbf{Conv}$. It is now obtained as a corollary.

**Corollary 4.4.16.** *Let $M$ be a division effect monoid. The coproduct of a finite family $(K_j)_j$ in $M$-$\mathbf{Conv}$ can be obtained as follows.*

$$\coprod_j K_j = \mathrm{B}\Big(\coprod_j \mathcal{L}(K_j)\Big) = \Big\{(x_j)_j \in \prod_j \mathcal{L}(K_j) \;\Big|\; \bigovee_j |x_j| = 1\Big\}$$

*Proof.* Note that we have the equivalence $M$-$\mathbf{WModn} \simeq M$-$\mathbf{Conv}$ by Corollary 4.4.9, and that the subcategories $M$-$\mathbf{WModn} \hookrightarrow M$-$\mathbf{WModn}_{\le} \hookrightarrow M$-$\mathbf{WMod}_{\le}$ are closed under finite coproducts, see Lemma 4.4.12. ∎

In particular, we have the following result, which explains the term 'lifting'. Note that the final object in $M$-**Conv** is the singleton 1.

**Corollary 4.4.17.** *Let $M$ be a division effect monoid, and $K \in M\text{-}\mathbf{Conv}$. Then the coproduct of $K$ and the final object* 1 *in $M$-**Conv** is the lifting $\mathcal{L}(K)$, seen as a convex set by Proposition* 3.6.9*.*

*Proof.* Because $\mathcal{L}(1) \cong M$, by Corollary 4.4.16 we have

$$K + 1 = \{(x, r) \in \mathcal{L}(K) \times M \mid |x| \ovee r = 1\} \cong \mathcal{L}(K)\,. \qquad \blacksquare$$

Hence partial maps in the effectus $M$-**Conv** are affine maps $f\colon K \to \mathcal{L}(L)$. Since $M\text{-}\mathbf{WModn} = \mathrm{Tot}(M\text{-}\mathbf{WModn}_{\leq})$, we have equivalences

$$\mathrm{Par}(M\text{-}\mathbf{Conv}) \simeq \mathrm{Par}(\mathrm{Tot}(M\text{-}\mathbf{WModn}_{\leq})) \cong M\text{-}\mathbf{WModn}_{\leq}\,.$$

Thus we have the 'Kleisli' adjunction

$$M\text{-}\mathbf{Conv} \rightleftarrows \mathrm{Par}(M\text{-}\mathbf{Conv}) \simeq M\text{-}\mathbf{WModn}_{\leq}$$

between convex sets and weight modules with weight-decreasing maps, which induces the lift monad on $M$-**Conv**. For a later reference, we note that this adjunction can be explicitly described as follows.

**Lemma 4.4.18.** *There is an adjunction:*

$$M\text{-}\mathbf{Conv} \underset{\text{forget}}{\overset{\mathcal{L} \cong (-)+1}{\underset{\longleftarrow}{\overset{\longrightarrow}{\bot}}}} M\text{-}\mathbf{WModn}_{\leq}$$

*Proof.* It is straightforward to see that for each $X \in M\text{-}\mathbf{WModn}_{\leq}$, any affine map $f\colon K \to X$ extends uniquely to a morphism $\overline{f}\colon \mathcal{L}(K) \to X$ in $M\text{-}\mathbf{WModn}_{\leq}$ by $\overline{f}(r, x) = r \cdot f(x)$. This means that $\mathcal{L}$ is left adjoint to the forgetful functor $M\text{-}\mathbf{WModn}_{\leq} \to M\text{-}\mathbf{Conv}$. $\blacksquare$

We note that morphisms in $M\text{-}\mathbf{WModn}_{\leq}$ are merely module maps, so the forgetful functor $M\text{-}\mathbf{WModn}_{\leq} \to M\text{-}\mathbf{PMod}$ is full and faithful. This is parallel with the fact that any module map between effect modules is subunital.

**Proposition 4.4.19.** *A module map $f\colon X \to Y$ between weight modules is weight-decreasing if $X$ satisfies the normalization property.*

*Proof.* Let $x \in X$ be fixed. By normalization there is $\bar{x} \in \mathrm{B}(X)$ with $x = |x| \cdot \bar{x}$. Then $|f(x)| = |x| \cdot |f(\bar{x})| \leq |x|$. $\blacksquare$

We end the section with some remarks.

**Remark 4.4.20.** Bas Westerbaan proved that $M$-**Conv** is an effectus in total form under the assumption that the effect monoid $M$ is a so-called *effect divisoid*, which is a weaker assumption than being a division effect monoid; see [256]. It is an open question whether $M$-**Conv** is an effectus in total form for arbitrary effect monoids $M$. In the general case, the singleton convex set 1 is still a final object in $M$-**Conv**. Since $M$-**Conv** $= \mathcal{EM}(\mathcal{D}_M)$ is the Eilenberg-Moore category of a monad on **Set**, the category $M$-**Conv** also has coproducts as a consequence of a general result [20, Theorem 4.3.5]. However, the verification of the axioms of effectus is not easy, because for a general effect monoid $M$, we do not have a simple representation of coproducts in $M$-**Conv** like in Corollary 4.4.16.

**Remark 4.4.21.** In this thesis, a convex set $K$ over an effect monoid $M$ is by definition an Eilenberg-Moore algebra $[\![-]\!]\colon \mathcal{D}_M(K) \to K$ for the distribution monad $\mathcal{D}_M$ over $M$. There is an alternative, more traditional definition of a convex set $K$ via an $M$-indexed family of binary operations, i.e. $\langle -; -, - \rangle\colon M \times K \times K \to K$. The supposed interpretation of the operation is a binary convex sum: $\langle r; x, y\rangle = r\cdot x + r^{\perp}\cdot y$. The operation $\langle -; -, - \rangle$ is required to satisfy the following conditions:

(a) $\langle r; x, y\rangle = \langle r^{\perp}; y, x\rangle$.

(b) $\langle 0; x, y\rangle = y$.

(c) $\langle r; x, x\rangle = x$.

(d) $\big\langle r; x, \langle s; y, z\rangle\big\rangle = \big\langle t; \langle t\backslash r; x, y\rangle, z\big\rangle$ whenever $t = r \ovee r^{\perp}\cdot s \neq 0$.

Since the last condition involves division, the effect monoid $M$ is required to admit division. This style of definition of convex sets over $M = [0, 1]$ can be found in e.g. [106, 108, 134]. (The axiomatization of convex sets goes back to Stone [244], which uses a slightly different formulation where scalars are given by an ordered division ring.)

Let $M$ be a division effect monoid. Temporarily, let us mean by a *binary convex set* over $M$ a set $K$ with an operation $\langle -; -, - \rangle\colon M \times K \times K \to K$ satisfying the four requirements above. In [134, Theorem 4] it is shown that binary convex sets over $[0, 1]$ are the same as convex sets over $[0, 1]$, i.e. Eilenberg-Moore algebras for the distribution monad $\mathcal{D}$. In fact, this holds generally for division effect monoids $M$. If $K$ is a convex set, then it is straightforward to verify that $K$ is a binary convex set via $\langle r; x, y\rangle = [\![ r|x\rangle + r^{\perp}|y\rangle ]\!]$. The other direction is more laborious since it involves defining a $\mathcal{D}_M$-algebra structure $\mathcal{D}_M(K) \to K$. In [134, Theorem 4] this is done by induction. We here sketch an alternative proof via weight modules. Given a binary convex set $K$, define the lifting $\mathcal{L}(K)$ of $K$ and its weight module structure in the same way as Definition 4.4.1, except that the sum is given by:

$$(r, x) \ovee (s, y) = \begin{cases} (r \ovee s, \langle (r \ovee s)\backslash r; x, y\rangle) & \text{if } r \ovee s \neq 0 \\ (0, *) & \text{if } r \ovee s = 0\,, \end{cases}$$

Then one can verify that $\mathcal{L}(K)$ is a weight module satisfying $\mathrm{B}(\mathcal{L}(K)) \cong K$. Since $\mathrm{B}(\mathcal{L}(K))$ is a convex set, $K$ is also equipped with the structure of a convex set.

## 4.5 Effectuses with the normalization property

In general, the weight modules $\mathrm{St}_{\leq}(A)$ of substates in an effectus do not satisfy the normalization property in the sense of Definition 4.4.7. Normalization is a convenient and also fairly reasonable condition to assume. In this section we study such effectuses with the normalization property.

**Definition 4.5.1.** An effectus $\mathbf{C}$ is said to satisfy the **normalization property** if for each object $A \in \mathbf{C}$, the weight module $\mathrm{St}_{\le}(A)$ of substates satisfies the normalization property. Explicitly, $\mathbf{C}$ satisfies the normalization property iff for each substate $\omega\colon I \to A$ with $\omega \neq 0$, there exists a unique *state* $\overline{\omega}\colon I \to A$ such that $\omega = |\omega| \cdot \overline{\omega} \equiv \overline{\omega} \circ \mathbb{1} \circ \omega$, see the diagram below:

$$\begin{array}{ccc} I & \xrightarrow{\ \omega\ } & A \\ {\scriptstyle\omega}\big\downarrow & & \big\downarrow{\scriptstyle\mathbb{1}} \\ A & \xleftarrow{\ \overline{\omega}\ } & I \end{array}$$

We say that an effectus in total form $\mathbf{B}$ satisfies the normalization property if $\mathrm{Par}(\mathbf{B})$ satisfies the condition above.

The normalization property implies that the scalars admit division.

**Proposition 4.5.2.** *If an effectus* $\mathbf{C}$ *satisfies the normalization property, then the scalars* $\mathcal{S} = \mathbf{C}(I, I)$ *admit right-division, that is,* $\mathcal{S}^{\mathrm{op}}$ *is a division effect monoid.*

*Proof.* Since the functor $\mathrm{St}_{\le}\colon \mathbf{C} \to \mathcal{S}^{\mathrm{op}}\text{-}\mathbf{WMod}_{\le}$ preserves finite coproducts (Proposition 3.5.11), we have $\mathrm{St}_{\le}(I + I) \cong \mathrm{St}_{\le}(I) + \mathrm{St}_{\le}(I) = \mathcal{S}^{\mathrm{op}} + \mathcal{S}^{\mathrm{op}}$ in $\mathcal{S}^{\mathrm{op}}\text{-}\mathbf{WMod}_{\le}$. By assumption, $\mathrm{St}_{\le}(I + I)$ satisfies the normalization property, and hence so does $\mathcal{S}^{\mathrm{op}} + \mathcal{S}^{\mathrm{op}}$. Then $\mathcal{S}^{\mathrm{op}}$ admits division by Lemma 4.4.13. ∎

**Example 4.5.3.** There are effectuses that do not satisfy the normalization property. For example, the product categories $\mathbf{Set} \times \mathbf{Set}$ and $\mathcal{K}\ell(\mathcal{D}) \times \mathcal{K}\ell(\mathcal{D})$ are effectuses by Proposition 3.8.9, but either of them does not satisfy the normalization property. Indeed, their scalars are respectively $\{0, 1\} \times \{0, 1\}$ and $[0, 1] \times [0, 1]$, which are not division effect monoids. By Proposition 4.5.2, they cannot satisfy the normalization property.

The following useful result was originally shown by Tull (private communication).

**Proposition 4.5.4.** *Any real effectus satisfies the normalization property.*

*Proof.* Immediate from Proposition 4.4.10. ∎

Next we describe state-and-effect triangles over effectuses with the normalization property.

**Proposition 4.5.5.** *Let* $\mathbf{C}$ *be an effectus with the normalization property. Then the substate functor* $\mathrm{St}_{\le}\colon \mathbf{C} \to \mathcal{S}^{\mathrm{op}}\text{-}\mathbf{WMod}_{\le}$ *restricts to* $\mathrm{St}_{\le}\colon \mathbf{C} \to \mathcal{S}^{\mathrm{op}}\text{-}\mathbf{WModn}_{\le}$. *Moreover the latter is a morphism of effectuses in partial form.*

*Proof.* By definition, the functor restricts to $\mathrm{St}_{\le}\colon \mathbf{C} \to \mathcal{S}^{\mathrm{op}}\text{-}\mathbf{WModn}_{\le}$. By Proposition 4.5.2 the effect monoid $\mathcal{S}^{\mathrm{op}}$ admits division, and thus $\mathcal{S}^{\mathrm{op}}\text{-}\mathbf{WModn}_{\le}$ is an effectus by Theorem 4.4.14. The restriction $\mathrm{St}_{\le}\colon \mathbf{C} \to \mathcal{S}^{\mathrm{op}}\text{-}\mathbf{WModn}_{\le}$ is a morphism of effectuses since $\mathcal{S}^{\mathrm{op}}\text{-}\mathbf{WModn}_{\le}$ inherits the effectus structure from $\mathcal{S}^{\mathrm{op}}\text{-}\mathbf{WMod}_{\le}$. ∎

**Lemma 4.5.6.** *Let* $M$ *be an effect monoid. The effectus* $M\text{-}\mathbf{EMod}_{\le}^{\mathrm{op}}$ *satisfies the normalization property if and only if* $M$ *is a right-division effect monoid.*

*Proof.* The 'only if' part follows by Proposition 4.5.2, since

$$\mathcal{S} = M\text{-}\mathbf{EMod}^{\mathrm{op}}_{\leq}(M, M) \cong M\,,$$

see Proposition 3.4.11. Conversely, assume that $M$ is a right-division effect monoid. A substate on $E$ in $M$-$\mathbf{EMod}^{\mathrm{op}}_{\leq}$ is a (subunital) $M$-module map $\omega\colon E \to M$. Note that $\omega = 0$ if and only if $\omega(1) = 0$. Suppose that $\omega \neq 0$. Then we define a state $\overline{\omega}\colon E \to M$ by $\overline{\omega}(a) = \omega(a)/\omega(1)$. By Corollary 4.3.5 we can verify that $\overline{\omega}$ is indeed a unital module map. Then we have

$$(|\omega| \cdot \overline{\omega})(a) = \overline{\omega}(a) \cdot |\omega| = (\omega(a)/\omega(1)) \cdot \omega(1) = \omega(a)\,,$$

hence $|\omega| \cdot \overline{\omega} = \omega$. (Note that the states form an $M^{\mathrm{op}}$-module, so the scalars act from right.) To check the uniqueness, let $\omega'$ be a state such that $|\omega| \cdot \omega' = \omega$. For each $a \in E$, we have $\omega'(a) \cdot \omega(1) = \omega(a)$, so that $\omega'(a) = \omega(a)/\omega(1) = \overline{\omega}(a)$. Therefore $\omega' = \overline{\omega}$. ∎

**Lemma 4.5.7.** *Let $M$ be a division effect monoid. Then the effectus $M$-$\mathbf{WModn}_{\leq}$ satisfies the normalization property.*

*Proof.* By Proposition 3.5.10, for each $X \in M$-$\mathbf{WModn}_{\leq}$ the substates $\mathrm{St}_{\leq}(X)$ are isomorphic to $X$ as weight modules. Since $X$ satisfies the normalization property, so does $\mathrm{St}_{\leq}(X)$. ∎

**Proposition 4.5.8.** *Let $M$ be a right-division effect monoid. Then the adjunction $M$-$\mathbf{EMod}^{\mathrm{op}}_{\leq} \rightleftarrows M^{\mathrm{op}}$-$\mathbf{WMod}_{\leq}$ from Proposition* 3.7.1 *restricts to an adjunction $M$-$\mathbf{EMod}^{\mathrm{op}}_{\leq} \rightleftarrows M^{\mathrm{op}}$-$\mathbf{WModn}_{\leq}$.*

*Proof.* Since $M^{\mathrm{op}}$-$\mathbf{WModn}_{\leq} \hookrightarrow M^{\mathrm{op}}$-$\mathbf{WMod}_{\leq}$ is a full subcategory, it suffices to show that the functor $\mathrm{St}_{\leq}\colon M$-$\mathbf{EMod}^{\mathrm{op}}_{\leq} \to M^{\mathrm{op}}$-$\mathbf{WMod}_{\leq}$ restricts to $\mathrm{St}_{\leq}\colon M$-$\mathbf{EMod}^{\mathrm{op}}_{\leq} \to M^{\mathrm{op}}$-$\mathbf{WModn}_{\leq}$. Indeed, this is the case by Lemma 4.5.6. ∎

**Corollary 4.5.9.** *Let $\mathbf{C}$ be an effectus with the normalization property. Then we have the following state-and-effect triangle, which sits in the 2-category $\mathbf{Ef}$ of effectuses.*

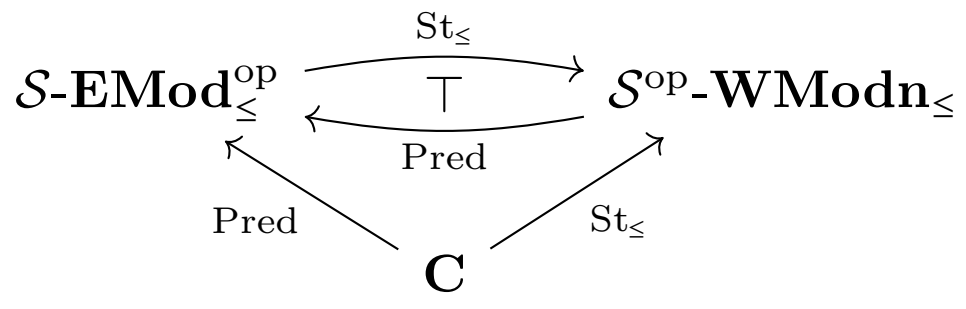

(4.16)

∎

We note that the triangle is filled by the two 'validity' natural transformations as in Proposition 3.7.3.

Applying the 2-functor $\mathrm{Tot}\colon \mathbf{Ef} \to \mathbf{Eft}$ to the diagram (4.16), we get the following diagram in $\mathbf{Eft}$.

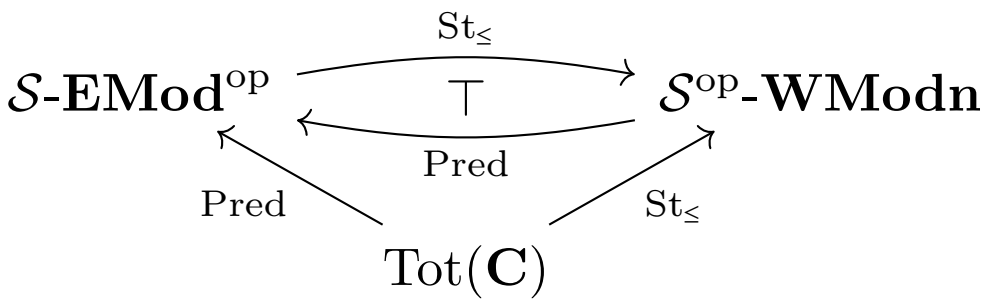

Via the equivalence $\mathcal{S}^{\mathrm{op}}\text{-}\mathbf{WModn} \simeq \mathcal{S}^{\mathrm{op}}\text{-}\mathbf{Conv}$ that commutes with the substate and state functors as below,

$$\begin{array}{ccc} \mathrm{Tot}(\mathbf{C}) & \xrightarrow{\mathrm{St}_{\leq}} & M\text{-}\mathbf{WModn} \\ & \searrow^{\mathrm{St}} & \simeq \downarrow \mathrm{B} \\ & & M\text{-}\mathbf{Conv} \end{array}$$

(see Corollaries 3.6.5 and 4.4.9) we obtain the following result.

**Corollary 4.5.10.** *The functor* $\mathrm{Tot}\colon \mathbf{Ef} \to \mathbf{Eft}$ *sends the triangle* (4.16) *to the following one in* **Eft**, *up to equivalence.*

$$\begin{array}{ccc} \mathcal{S}\text{-}\mathbf{EMod}^{\mathrm{op}} & \underset{\mathrm{Pred}}{\overset{\mathrm{St}}{\underset{\top}{\rightleftarrows}}} & \mathcal{S}^{\mathrm{op}}\text{-}\mathbf{Conv} \\ & {}_{\mathrm{Pred}}\nwarrow \quad \nearrow_{\mathrm{St}} & \\ & \mathrm{Tot}(\mathbf{C}) & \end{array} \tag{4.17}$$

*The two triangles* (4.16) *and* (4.17) *are related in the 2-equivalence* $\mathbf{Ef} \simeq \mathbf{Eft}$. ■

Thus, if $\mathbf{C}$ is an effectus with the normalization property, the state functor

$$\mathrm{St}\colon \mathrm{Tot}(\mathbf{C}) \longrightarrow \mathcal{S}^{\mathrm{op}}\text{-}\mathbf{Conv}$$

is a morphism of an effectus, and in particular, preserves binary coproducts. In fact, the converse is true if we assume that the scalars admit division.

**Proposition 4.5.11.** *Let* $\mathbf{C}$ *be an effectus such that the scalars* $\mathcal{S} = \mathbf{C}(I, I)$ *admit right-division, that is,* $\mathcal{S}^{\mathrm{op}}$ *is a division effect monoid. Then the following are equivalent.*

(i) $\mathbf{C}$ *satisfies the normalization property.*

(ii) *The state functor* $\mathrm{St}\colon \mathrm{Tot}(\mathbf{C}) \to \mathcal{S}^{\mathrm{op}}\text{-}\mathbf{Conv}$ *is a morphism of effectuses in total form.*

(iii) *The state functor* $\mathrm{St}\colon \mathrm{Tot}(\mathbf{C}) \to \mathcal{S}^{\mathrm{op}}\text{-}\mathbf{Conv}$ *preserves binary coproducts.*

*Proof.* (i) $\Longrightarrow$ (ii) is shown by the discussions above, and (ii) $\Longrightarrow$ (iii) is trivial. We will prove (iii) $\Longrightarrow$ (i).

Assume that $\mathrm{St}\colon \mathrm{Tot}(\mathbf{C}) \to \mathcal{S}^{\mathrm{op}}\text{-}\mathbf{Conv}$ preserves binary coproducts. Let $A \in \mathbf{C}$. Then there exist canonical isomorphisms:

$$\mathcal{L}(\mathrm{St}(A)) \cong \mathrm{St}(A) + 1 \cong \mathrm{St}(A) + \mathrm{St}(I) \cong \mathrm{St}(A + I) \cong \mathrm{St}_{\leq}(A)$$

where the first isomorphism exists by Corollary 4.4.17, and the last one by Lemma 4.1.3. Let $\alpha\colon \mathcal{L}(\mathrm{St}(A)) \to \mathrm{St}_{\leq}(A)$ be the isomorphism, which is affine. It is not hard to see that $\alpha$ satisfies $\alpha(1, \omega) = \omega$ and $\alpha(0, *) = 0$. Then

$$\begin{aligned} \alpha(r, \omega) &= \alpha(r(1, \omega) \ovee r^{\perp}(0, *)) \\ &= r \cdot \alpha(1, \omega) \ovee r^{\perp} \cdot \alpha(0, *) \\ &= r \cdot \omega \ovee r^{\perp} \cdot 0 \\ &= r \cdot \omega \,. \end{aligned}$$

Since $\mathrm{St}(A) = \mathrm{B}(\mathrm{St}_{\leq}(A))$, the map $\alpha$ equals the component $\varepsilon\colon \mathcal{L}(\mathrm{B}(\mathrm{St}_{\leq}(A))) \to \mathrm{St}_{\leq}(A)$ of the counit of the adjunction $\mathcal{S}^{\mathrm{op}}$-**WMod** $\rightleftarrows$ $\mathcal{S}^{\mathrm{op}}$-**Conv**. By Lemma 4.4.8, $\mathrm{St}_{\leq}(A)$ satisfies the normalization property. ■

In particular, we obtain the following result about an effectus in total form.

**Corollary 4.5.12.** *Let* $\mathbf{B}$ *be an effectus in total form such that the scalars* $\mathcal{S} = \mathbf{B}(1, 1+1)$ *admit right-division. Then* $\mathbf{B}$ *satisfies the normalization property if and only if the state functor* $\mathrm{St} = \mathbf{B}(1, -)\colon \mathbf{B} \to \mathcal{S}^{\mathrm{op}}\text{-}\mathbf{Conv}$ *is a morphism of effectuses in total form.* ■

# Chapter 5

# Logical Structures in Effectuses

In this chapter we study effectuses from a logical point of view, systematically using the language of indexed categories/posets and fibrations, a standard tool in categorical logic [133, 163, 174, 189, 190, 221]. Specifically, for each effectus $\mathbf{C}$ we construct a suitable category $\mathrm{Pred}_{\Box}(\mathbf{C})$ of predicates $p \in \mathrm{Pred}(A)$. It carries a 'forgetful' functor $\mathrm{Pred}_{\Box}(\mathbf{C}) \to \mathbf{C}$, forming a fibration. The notions of *kernel* and *image* in an effectus can be then introduced as certain universal (i.e. cartesian/opcartesian) liftings along the fibration $\mathrm{Pred}_{\Box}(\mathbf{C}) \to \mathbf{C}$. We also introduce the notions of *comprehension* $\{p \,|\, A\}$ and *quotient* $A/p$ for a predicates $p$ on an object $A$ in an effectus. They are nicely captured categorically as right adjoint to truth and left adjoint to falsity, forming a chain of adjunctions:

$$\text{quotient} \dashv \text{falsity} \dashv \text{fibration} \dashv \text{truth} \dashv \text{comprehension}$$

Our leading examples, $\mathbf{Pfn}$, $\mathcal{K}\ell(\mathcal{D}_{\leq})$, and $\mathbf{Wstar}_{\leq}^{\mathrm{op}}$, all admit comprehension and quotients, hence the chains of adjunctions.

After we establish these notions, in Section 5.5 we study *sharp* predicates — which capture *projections* among predicates/effects in the effectus $\mathbf{Wstar}_{\leq}^{\mathrm{op}}$ — using images and comprehension. We show under a mild assumption that sharp predicates form orthomodular lattices. Additionally, we show that sharp predicates yields a bifibration, and that the subcategory of an effectus determined by *sharp morphisms* forms an effectus again.

Finally in Section 5.6, we make a comparison to Janelidze and Weighill's theory of non-abelian algebras, where a similar chain of adjunctions appears.

## 5.1 Fibrational setup

In this section we review basics of fibrations, and describe fibrations of predicates in an effectus. Then we look at the fibrations of predicates in our main examples of effectuses.

### 5.1.1 The Grothendieck construction and fibrations

In Chapter 3 we saw that for each effectus $\mathbf{C}$ there is a predicate functor $\mathrm{Pred}\colon \mathbf{C}^{\mathrm{op}} \to \mathcal{S}\text{-}\mathbf{EMod}_{\leq}$. Since every effect algebra/module is partially ordered, the predicate functor is an instance of an **indexed poset**, that is, a contravariant poset-valued functor $\Phi\colon \mathbf{C}^{\mathrm{op}} \to \mathbf{Poset}$. Here $\mathbf{Poset}$ denotes the category of posets and monotone functions. Indexed posets can be seen as models of predicate logic in a very general sense [133, 221]: the category $\mathbf{C}$ interprets types and terms, and the functor $\Phi$ assigns to each object/type $A \in \mathbf{C}$ a poset $\Phi A$ of predicates on $A$, and to each morphism/term $f\colon A \to B$ a substitution map $\Phi f\colon \Phi B \to \Phi A$. In the context of program semantics — here morphisms in $\mathbf{C}$ are thought of as programs — the reindexing maps $\Phi f\colon \Phi B \to \Phi A$ may be seen as weakest precondition operators.

The following is a version of the so-called Grothendieck construction applied to indexed posets (rather than more general *indexed categories*).

**Definition 5.1.1** (**Grothendieck construction**)**.** Let $\Phi\colon \mathbf{C}^{\mathrm{op}} \to \mathbf{Poset}$ be a functor (indexed poset). We define a category $\int_{\mathbf{C}} \Phi$ as follows.

- Objects are pairs $(A, a)$ of $A \in \mathbf{C}$ and $a \in \Phi A$.
- Morphisms $(A, a) \to (B, b)$ are $f\colon A \to B$ in $\mathbf{C}$ satisfying $a \le \Phi f(b)$.

There is an obvious forgetful functor $\varphi\colon \int_{\mathbf{C}} \Phi \to \mathbf{C}$, given by $\varphi(A, a) = A$.

The construction pieces the posets $\Phi A$ together into one category $\int_{\mathbf{C}} \Phi$. Intuitively, if one sees $f$ as programs and thus reindexing $\Phi f\colon \Phi B \to \Phi A$ as weakest precondition operators, then one may think of the morphisms $f\colon (A, a) \to (B, b)$ in $\int_{\mathbf{C}} \Phi$ as Hoare triples $\{a\}\, f\, \{b\}$ — if predicate $a$ holds before the execution of program $f$, then $b$ holds afterwards. The resulting functor $\varphi\colon \int_{\mathbf{C}} \Phi \to \mathbf{C}$ has a structure of so-called *fibrations*, which we briefly recall below:

**Definition 5.1.2.** With respect to a given functor $\varphi\colon \mathbf{E} \to \mathbf{C}$, we use the following notation and terminology.

(i) We say that an object $X \in \mathbf{E}$ is **above** $A \in \mathbf{C}$ if $\varphi X = A$, and a morphism $h\colon X \to Y$ in $\mathbf{E}$ is **above** $f\colon A \to B$ in $\mathbf{C}$ if $\varphi h = f$.

(ii) For an object $A \in \mathbf{C}$, the **fibre** $\mathbf{E}_A$ over $A$ is the subcategory $\mathbf{E}_A \subseteq \mathbf{E}$ consisting of objects above $A$ and morphisms above $\mathrm{id}_A$. In short: $\mathbf{E}_A = \varphi^{-1}(A)$.

(iii) Let $f\colon A \to B$ be a morphism in $\mathbf{C}$ and $Y \in \mathbf{E}_B$ be an object above $B$. A **cartesian lifting** of $f$ to $Y$ (along $\varphi$) is a morphism $l\colon X \to Y$ above $f$ (hence $\varphi X = A$) such that for any $k\colon Z \to Y$ in $\mathbf{E}$ and $g\colon \varphi Z \to A$ in $\mathbf{C}$ satisfying $\varphi k = f \circ g$, there exists a unique map $m\colon Z \to X$ in $\mathbf{E}$ above $g$ such that $k = l \circ m$. Pictorially:

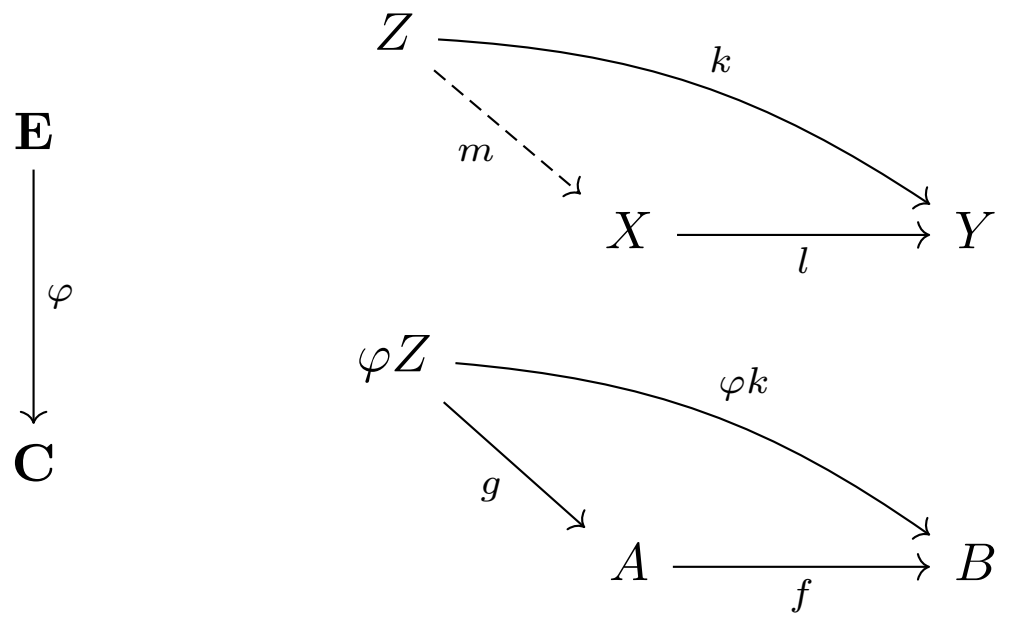

**Definition 5.1.3.** A **fibration** is a functor $\varphi\colon \mathbf{E} \to \mathbf{C}$ such that for every $f\colon A \to B$ in $\mathbf{C}$ and $Y \in \mathbf{E}_B$, there exists a cartesian lifting of $f$ to $Y$. A **poset fibration** is a fibration such that every fibre $\mathbf{E}_A$ is a poset.

Cartesian liftings are unique up to isomorphism in general, and unique 'on the nose' for a poset fibration. We write $\overline{f}(Y)\colon f^*(Y) \to Y$ for the cartesian lifting of $f\colon A \to B$ to $Y$. A fundamental result about fibrations is that they are equivalent to indexed categories, via the Grothendieck construction. The following is a version of the result for poset fibrations.

**Proposition 5.1.4.**

(i) *Let $\Phi\colon \mathbf{C}^{\mathrm{op}} \to \mathbf{Poset}$ be an indexed poset. The functor $P\colon \int_{\mathbf{C}} \Phi \to \mathbf{C}$ given by the Grothendieck construction is a poset fibration. The cartesian lifting of $f\colon A \to B$ in $\mathbf{C}$ to $(B,y) \in \int_{\mathbf{C}} \Phi$ is given by $f^*(B,y) = (A, \Phi f(y))$ and $\overline{f}(B,y) = f$.*

(ii) *Let $\varphi\colon \mathbf{E} \to \mathbf{C}$ be a poset fibration. For each $f\colon A \to B$ in $\mathbf{C}$, the mapping $Y \mapsto f^*(Y)$ defines a monotone map $f^*\colon \mathbf{E}_B \to \mathbf{E}_A$. The assignments $A \mapsto \mathbf{E}_A$ and $f \mapsto f^*$ yield an indexed poset $\mathbf{C}^{\mathrm{op}} \to \mathbf{Poset}$.*

*The two constructions between poset fibrations and functors $\mathbf{C}^{\mathrm{op}} \to \mathbf{Poset}$ are inverse of each other in a suitable sense.*

*Proof.* It is straightforward to verify (i) and (ii). For the precise meaning of 'inverse of each other in a suitable sense', we refer to [133, § 1.10] (where a result about general fibrations is presented). ■

### 5.1.2 Fibrations of predicates in an effectus

Now let $\mathbf{C}$ be an effectus. We first focus on total morphisms and hence on the subcategory $\mathrm{Tot}(\mathbf{C})$. Then we have a predicate functor $\mathrm{Pred}\colon \mathrm{Tot}(\mathbf{C})^{\mathrm{op}} \to \mathbf{EA}$. Since effect algebras are posets and there is inclusion $\mathbf{EA} \hookrightarrow \mathbf{Poset}$, we can apply the Grothendieck construction.

**Definition 5.1.5.** For an effectus $\mathbf{C}$, we write

$$\mathrm{Pred}(\mathrm{Tot}(\mathbf{C})) = \textstyle\int_{\mathrm{Tot}(\mathbf{C})} \mathrm{Pred}$$

for the category obtained by applying the Grothendieck construction to the predicate functor $\mathrm{Pred}\colon \mathrm{Tot}(\mathbf{C})^{\mathrm{op}} \to \mathbf{EA}$. Explicitly, objects in $\mathrm{Pred}(\mathrm{Tot}(\mathbf{C}))$ are pairs $(A,p)$ of $A \in \mathbf{C}$ and $p \in \mathrm{Pred}(A)$, and morphisms $f\colon (A,p) \to (B,q)$ are total morphisms $f\colon A \to B$ in $\mathbf{C}$ such that $p \le f^*(q)$.

In this setting we can see falsity and truth predicates as functors $\mathbb{0}, \mathbb{1}\colon \mathrm{Tot}(\mathbf{C}) \to \mathrm{Pred}(\mathrm{Tot}(\mathbf{C}))$ by defining $\mathbb{0}(A) = (A, \mathbb{0}_A)$ and $\mathbb{1}(A) = (A, \mathbb{1}_A)$. In fact they are adjoint functors:

**Lemma 5.1.6.** *The falsity* $\mathbb{0}$ *and truth* $\mathbb{1}$ *form left and right adjoints to the fibration as below.*

$$\begin{array}{c} \mathrm{Pred}(\mathrm{Tot}(\mathbf{C})) \\ \mathbb{0}\Big(\dashv \Big\downarrow \dashv\Big)\mathbb{1} \\ \mathrm{Tot}(\mathbf{C}) \end{array}$$

*Proof.* Let $f\colon A \to B$ be a morphism in $\mathrm{Tot}(\mathbf{C})$. Then $f\colon (A, \mathbb{0}) \to (B, \mathbb{0})$ is indeed a morphism in $\mathrm{Pred}(\mathrm{Tot}(\mathbf{C}))$ since $\mathbb{0} \le f^*(\mathbb{0})$, and so is $f\colon (A, \mathbb{1}) \to (B, \mathbb{1})$ since $\mathbb{1} \le \mathbb{1} = f^*(\mathbb{1})$. Thus they indeed form functors. Moreover, for each $q \in \mathrm{Pred}(B)$ and $p \in \mathrm{Pred}(A)$ there are the obvious bijections:

$$\frac{(A, \mathbb{0}) \to (B, q) \text{ in } \mathrm{Pred}(\mathrm{Tot}(\mathbf{C}))}{A \to B \text{ in } \mathrm{Tot}(\mathbf{C})} \qquad \frac{(A, p) \to (B, \mathbb{1}) \text{ in } \mathrm{Pred}(\mathrm{Tot}(\mathbf{C}))}{A \to B \text{ in } \mathrm{Tot}(\mathbf{C})}$$

This shows that they are left and right adjoints. ■

We would like to have a similar result for general 'partial' maps. A naive attempt would be to apply the Grothendieck construction to the predicate functor $\mathrm{Pred}\colon \mathbf{C}^{\mathrm{op}} \to \mathbf{EA}_{\le}$ and obtain a fibration $\int_{\mathbf{C}} \mathrm{Pred} \to \mathbf{C}$. A morphism $f\colon (A, p) \to (B, q)$ in $\int_{\mathbf{C}} \mathrm{Pred}$ is $f\colon A \to B$ in $\mathbf{C}$ such that $p \le f^*(q)$. This requirement for morphisms in $\int_{\mathbf{C}} \mathrm{Pred}$ is however too restrictive: the obvious mapping $A \mapsto (A, \mathbb{1}_A)$ does not form a functor $\mathbf{C} \to \int_{\mathbf{C}} \mathrm{Pred}$, because $\mathbb{1} \le f^*(\mathbb{1})$ does not hold in general. Therefore we introduce a more relaxed notion of predicate transformers, which is obtained via the De Morgan dual of $f^*$.

**Definition 5.1.7.** Let $f\colon A \to B$ be a morphism. We write $f^\square(p) = f^*(p^\perp)^\perp$ for $p \in \mathrm{Pred}(B)$ and call the mapping $f^\square\colon \mathrm{Pred}(B) \to \mathrm{Pred}(A)$ the **liberal predicate transformer** for $f$.

Here the word 'liberal' comes from connection with weakest liberal preconditions, and the notation $f^\square$ comes from similarity to the box modality $\Box$ in modal logic; see examples in § 5.1.3. By the De Morgan duality, $f^*$ may be associated with the diamond duality $\Diamond$, and indeed the notation $f^\Diamond = f^*$ was used in [40]. We will not use the notation $f^\Diamond$ in this thesis.

We show basic properties of $f^\square$.

**Lemma 5.1.8.** *Let* $f\colon A \to B$ *and* $g\colon B \to C$.

(i) $f^\square$ *is monotone:* $p \le q$ *implies* $f^\square(p) \le f^\square(q)$.

(ii) *the mappings* $(-)^\square$ *are (contravariantly) functorial:* $\mathrm{id}^\square = \mathrm{id}$ *and* $(g \circ f)^\square = f^\square \circ g^\square$.

(iii) $f^\square(\mathbb{1}) = \mathbb{1}$.

(iv) $f^\square(p) = \mathbb{1} \iff p^\perp \circ f = \mathbb{0} \iff \mathbb{1} \circ f = p \circ f$ *for all* $p \in \mathrm{Pred}(B)$.

(v) $f^\square(p) = (\mathbb{1}f)^\perp \ovee f^*(p)$ *for all* $p \in \mathrm{Pred}(B)$. *In particular,* $f^*(p) \le f^\square(p)$.

(vi) *If* $f$ *is total,* $f^\square(p) = f^*(p)$ *for all* $p \in \mathrm{Pred}(B)$.

The points (i) and (ii) show that the assignments $A \mapsto \mathrm{Pred}(A)$ and $f \mapsto f^\square$ define a functor / indexed poset $\mathbf{C}^{\mathrm{op}} \to \mathbf{Poset}$. We denote it as $\mathrm{Pred}_\square\colon \mathbf{C}^{\mathrm{op}} \to \mathbf{Poset}$.

*Proof.*

(i) If $p \leq q$, then $q^\perp \leq p^\perp$. Thus $f^*(q^\perp) \leq f^*(p^\perp)$ and so $f^\square(p) = f^*(p^\perp)^\perp \leq f^*(q^\perp)^\perp = f^\square(q)$.

(ii) We have $\mathrm{id}^\square = \mathrm{id}$ by $\mathrm{id}^\square(p) = \mathrm{id}^*(p^\perp)^\perp = p^{\perp\perp} = p$. We prove $(g \circ f)^\square = f^\square \circ g^\square$ as follows:

$$\begin{aligned}(g \circ f)^\square(p) = (g \circ f)^*(p^\perp)^\perp &= f^*(g^*(p^\perp))^\perp \\ &= f^*(g^*(p^\perp)^{\perp\perp})^\perp = f^\square(g^\square(p)) = (f^\square \circ g^\square)(p)\,.\end{aligned}$$

(iii) $f^\square(\mathbb{1}) = f^*(\mathbb{0})^\perp = \mathbb{0}^\perp = \mathbb{1}$.

(iv) The first equivalence holds as follows.

$$f^\square(p) = \mathbb{1} \iff (p^\perp \circ f)^\perp = \mathbb{1} \iff p^\perp \circ f = \mathbb{0}\,.$$

The second one holds via $\mathbb{1} \circ f = p \circ f \ovee p^\perp \circ f$.

(v) We have:

$$\begin{aligned}f^\square(p) = f^*(p^\perp)^\perp &= \mathbb{1} \ominus f^*(\mathbb{1} \ominus p) \\ &= \mathbb{1} \ominus (f^*(\mathbb{1}) \ominus f^*(p)) \\ &= (\mathbb{1} \ominus f^*(\mathbb{1})) \ovee f^*(p) \\ &= (\mathbb{1}f)^\perp \ovee f^*(p)\,.\end{aligned}$$

(vi) We have $(\mathbb{1}f)^\perp = \mathbb{0}$ if $f$ is total. Thus $f^\square(p) = f^*(p) \ovee (\mathbb{1}f)^\perp = f^*(p)$. ■

Note that $f^\square \colon \mathrm{Pred}(B) \to \mathrm{Pred}(A)$ no longer preserves the structure of effect modules nor algebras. In particular, $f^\square$ need not preserve the falsity $\mathbb{0}$, but instead preserves $\mathbb{1}$.

**Definition 5.1.9.** Let $\mathbf{C}$ be an effectus. We write $\mathrm{Pred}_\square(\mathbf{C}) = \int_{\mathbf{C}} \mathrm{Pred}_\square$ for the category obtained by applying the Grothendieck construction to $\mathrm{Pred}_\square \colon \mathbf{C}^{\mathrm{op}} \to \mathbf{Poset}$. Explicitly, the objects in $\mathrm{Pred}_\square(\mathbf{C})$ are pairs $(A, p)$ where $A \in \mathbf{C}$ and $p \in \mathrm{Pred}(A)$. The morphisms $(A, p) \to (B, q)$ are morphisms $f \colon A \to B$ satisfying $p \leq f^\square(q)$.

The fibration $\mathrm{Pred}_\square(\mathbf{C}) \to \mathbf{C}$ nicely extends the fibration over total morphisms defined in Definition 5.1.5, and indeed a result similar to Lemma 5.1.6 holds.

**Proposition 5.1.10.** *The category* $\mathrm{Pred}(\mathrm{Tot}(\mathbf{C}))$ *from Definition* 5.1.5 *is a subcategory of* $\mathrm{Pred}_\square(\mathbf{C})$*, and the following diagram commutes.*

$$\begin{array}{ccc}\mathrm{Pred}(\mathrm{Tot}(\mathbf{C})) & \hookrightarrow & \mathrm{Pred}_\square(\mathbf{C}) \\ \downarrow & & \downarrow \\ \mathrm{Tot}(\mathbf{C}) & \hookrightarrow & \mathbf{C}\end{array}$$

*Moreover, the inclusion functors preserves cartesian liftings — thus the inclusions are a 'morphism of fibrations'.*

*Proof.* Note that $\mathrm{Pred}(\mathrm{Tot}(\mathbf{C}))$ and $\mathrm{Pred}_{\Box}(\mathbf{C})$ have the same objects. Every morphism $f\colon (A,p) \to (B,q)$ in $\mathrm{Pred}(\mathrm{Tot}(\mathbf{C}))$ is a morphism in $\mathrm{Pred}_{\Box}(\mathbf{C})$ too, because $f^{\Box}(q) = f^{*}(q)$ for a total map $f$. This also shows that inclusion preserves cartesian liftings. Commutativity of the diagram is obvious. ∎

**Proposition 5.1.11.** *The assignments $A \mapsto (A, \mathbb{0})$ and $A \mapsto (A, \mathbb{1})$ form functors $\mathbf{C} \to \mathrm{Pred}_{\Box}(\mathbf{C})$ that are respectively a left and a right adjoint to the fibration $\mathrm{Pred}_{\Box}(\mathbf{C}) \to \mathbf{C}$. Moreover, the obvious squares in the diagram below commute,*

$$\begin{array}{ccc} \mathrm{Pred}(\mathrm{Tot}(\mathbf{C})) & \hookrightarrow & \mathrm{Pred}_{\Box}(\mathbf{C}) \\ \mathbb{0}\left(\dashv\downarrow\dashv\right)\mathbb{1} & & \mathbb{0}\left(\dashv\downarrow\dashv\right)\mathbb{1} \\ \mathrm{Tot}(\mathbf{C}) & \hookrightarrow & \mathbf{C} \end{array}$$

*where the adjunctions on the left are the ones from Lemma* 5.1.6.

*Proof.* The proof of the adjunctions is basically the same as that of Lemma 5.1.6, because reindexing $f^{\Box}\colon \mathrm{Pred}(B) \to \mathrm{Pred}(A)$ preserves the top $\mathbb{1}$. ∎

The fibration $\mathrm{Pred}_{\Box}(\mathbf{C}) \to \mathbf{C}$ will play an important role in the subsequent sections. We will see that notions of images, comprehension, and quotients in an effectus can be formulated neatly using the fibration $\mathrm{Pred}_{\Box}(\mathbf{C}) \to \mathbf{C}$. In particular, comprehension and quotients yield a chain of adjunctions, extending the adjunctions given by falsity and truth functors.

The following lemma shows that liberal predicate transformers $f^{\Box}$ interact nicely with coproducts.

**Lemma 5.1.12.** *Suppose that $\coprod_j A_j$ is a coproduct of (possibly infinitely many) objects $A_j$ in an effectus.*

(i) *For any morphisms $f_j\colon A_j \to B$ and predicate $p \in \mathrm{Pred}(B)$,*

$$[f_j]_j^{\Box}(q) = [f_j^{\Box}(q)]_j \,.$$

(ii) *Suppose also that a coproduct $\coprod_j B_j$ exists. Then for any morphisms $g_j\colon A_j \to B_j$ and predicates $q_j \in \mathrm{Pred}(B_j)$,*

$$\big(\textstyle\coprod_j g_j\big)^{\Box}([q_j]_j) = [g_j^{\Box}(q_j)]_j \,.$$

*Proof.* Point (i) follows by:

$$[f_j]_j^{\Box}(q) \equiv (q^{\perp} \circ [f_j]_j)^{\perp} = [q^{\perp} \circ f_j]_j^{\perp} \overset{\star}{=} [(q^{\perp} \circ f_j)^{\perp}]_j = [f_j^{\Box}(q)]_j \,.$$

The marked equality $\overset{\star}{=}$ holds by Proposition 3.4.13. To show (ii), note that $\coprod_j g_j = [\kappa_j \circ g_j]_j$. Thus we can apply (i) and obtain:

$$\begin{aligned} \big(\textstyle\coprod_j g_j\big)^{\Box}([q_j]_j) &= \big[(\kappa_j \circ g_j)^{\Box}([q_j]_j)\big]_j \\ &= \big[([q_j^{\perp}]_j \circ \kappa_j \circ g_j)^{\perp}\big]_j \\ &= [(q_j^{\perp} \circ g_j)^{\perp}]_j = [g_j^{\Box}(q_j)]_j \,. \end{aligned}$$ ∎

### 5.1.3 Examples

We describe the fibrations $\mathrm{Pred}_{\Box}(\mathbf{C}) \to \mathbf{C}$ for our main examples of effectuses, for deterministic, probabilistic, and quantum settings. It boils down to the description of liberal predicate transformers $f^{\Box}\colon \mathrm{Pred}(B) \to \mathrm{Pred}(A)$. We will see that in each example, $f^{\Box}$ correspond to weakest liberal preconditions, while $f^*$ correspond to weakest ('conservative') preconditions, capturing the well-known notion of weakest (liberal) preconditions for deterministic programs [67], and also the suitable generalizations for probabilistic programs [180, 204] and for quantum programs [62, 78].

**Example 5.1.13** (Deterministic)**.** Consider the effectus **Pfn**. The morphisms, partial functions $f\colon X \rightharpoonup Y$, may be viewed as models of deterministic programs. A value $f(x)$ being undefined means that the program $f$ does not terminate for the input/state $x$. Recall that predicates $X \rightharpoonup 1$ are identified with subsets $P \subseteq X$. Let $f\colon X \rightharpoonup Y$ be a morphism and $Q \subseteq Y$ a predicate.

The predicate transformation $f^*(Q) \subseteq X$ is given by:

$$x \in f^*(Q) \iff f(x) \text{ defined and } f(x) \in Q\,. \tag{5.1}$$

This $f^*(Q)$ coincides with weakest ('conservative') precondition $\mathrm{wp}(f)(Q)$ in the deterministic setting.

The *liberal* predicate transformation $f^{\Box}(Q) = f^*(Q^{\perp})^{\perp}$ is:

$$x \in f^{\Box}(Q) \iff f(x) \text{ defined implies } f(x) \in Q\,. \tag{5.2}$$

We see that $f^{\Box}(Q)$ is the weakest liberal precondition $\mathrm{wlp}(f)(Q)$. Therefore $f\colon X \rightharpoonup Y$ is a morphism $(X,P) \to (Y,Q)$ in $\mathrm{Pred}_{\Box}(\mathbf{Pfn})$ if and only if for each $x \in P$, $f(x) \in Q$ whenever $f(x)$ is defined. This coincides with the usual notion of a Hoare triple $\{P\}\, f\, \{Q\}$ for partial correctness.

**Example 5.1.14** (Probabilistic)**.** Next we consider the effectus $\mathcal{K}\ell(\mathcal{D}_{\le})$. Let $f\colon X \to \mathcal{D}_{\le}(Y)$ be a morphism in $\mathcal{K}\ell(\mathcal{D}_{\le})$. The predicate transformer $f^*\colon [0,1]^Y \to [0,1]^X$ is calculated as:

$$f^*(q)(x) = \sum_{y\in Y} q(y) \cdot f(x)(y)\,.$$

To understand the meaning of the operation $f^*$, let us view the map $f\colon X \to \mathcal{D}_{\le}(Y)$ as a probabilistic program: for each input $x \in X$, program $f$ returns an output $y \in Y$ with probability $f(x)(y)$, but with some probability $f$ may return no output (i.e. may diverge). We can express this in the usual notation of probability theory:

$$\mathrm{P}(f{\downarrow}, \mathbf{y} = y \mid \mathbf{x} = x) = f(x)(y)$$

where $\mathbf{x}$ and $\mathbf{y}$ are random variables for the input and output respectively, and $f{\downarrow}$ denotes the event that $f$ terminates (returns some output). Following § 3.3.2 and Example 3.4.6(ii), we introduce an event $\mathbf{q}$ corresponding to a predicate $q \in [0,1]^Y$ such that $\mathrm{P}(\mathbf{q} \mid \mathbf{y} = y) = q(y)$. We assume that $\mathbf{q}$ is conditionally independent of $\mathbf{x}$ and $f{\downarrow}$ given $\mathbf{y}$, so that

$$\mathrm{P}(\mathbf{q} \mid f{\downarrow}, \mathbf{y} = y, \mathbf{x} = x) = \mathrm{P}(\mathbf{q} \mid \mathbf{y} = y)\,.$$

Then we can reason with the standard rules of probability as follows:

$$\begin{aligned}
f^*(q)(x) &= \sum_y q(y) f(x)(y) \\
&= \sum_y \mathrm{P}(\mathbf{q} \mid \mathbf{y} = y)\, \mathrm{P}(f{\downarrow}, \mathbf{y} = y \mid \mathbf{x} = x) \\
&= \sum_y \mathrm{P}(\mathbf{q} \mid f{\downarrow}, \mathbf{y} = y, \mathbf{x} = x)\, \mathrm{P}(f{\downarrow}, \mathbf{y} = y \mid \mathbf{x} = x) \\
&= \sum_y \mathrm{P}(f{\downarrow}, \mathbf{q}, \mathbf{y} = y \mid \mathbf{x} = x) \\
&= \mathrm{P}(f{\downarrow}, \mathbf{q} \mid \mathbf{x} = x)\,.
\end{aligned}$$

Thus we can interpret the predicate $f^*(q)$ as the event '$f{\downarrow}$ and $\mathbf{q}$' i.e. $f$ terminates and predicate $q$ holds. Therefore $f^*(q)$ can be understood as the weakest (conservative) precondition in the probabilistic setting (cf. (5.1)).

For the liberal predicate transformer $f^\square \colon [0,1]^Y \to [0,1]^X$, we have

$$\begin{aligned}
f^\square(q)(x) &= 1 - (\mathbb{1}f)(x) + f^*(q)(x) && \text{by Lemma 5.1.8(v)} \\
&= 1 - \mathrm{P}(f{\downarrow} \mid \mathbf{x} = x) + \mathrm{P}(f{\downarrow} \text{ and } \mathbf{q} \mid \mathbf{x} = x) \\
&= \mathrm{P}(\text{not } f{\downarrow} \mid \mathbf{x} = x) + \mathrm{P}(f{\downarrow} \text{ and } \mathbf{q} \mid \mathbf{x} = x) \\
&= \mathrm{P}((\text{not } f{\downarrow}) \text{ or } (f{\downarrow} \text{ and } \mathbf{q}) \mid \mathbf{x} = x) \\
&= \mathrm{P}((\text{not } f{\downarrow}) \text{ or } \mathbf{q} \mid \mathbf{x} = x) \\
&= \mathrm{P}(f{\downarrow} \text{ implies } \mathbf{q} \mid \mathbf{x} = x)\,.
\end{aligned}$$

Thus we can interpret $f^\square(q)$ as the event '$f{\downarrow}$ implies $\mathbf{q}$' and hence the weakest liberal precondition.

Morphisms in $\mathrm{Pred}_\square(\mathcal{K}\ell(\mathcal{D}_\le))$ can be understood in terms of weakest liberal preconditions. Concretely, a morphism in $\mathrm{Pred}_\square(\mathcal{K}\ell(\mathcal{D}_\le))$ from $(X, p \in [0,1]^X)$ to $(Y, q \in [0,1]^Y)$ is $f \colon X \to \mathcal{D}_\le(Y)$ with $p \le f^\square(q)$. The condition $p \le f^\square(q)$ can be read as 'for each given input $x \in X$, the probability that $p$ holds is less than or equal to the probability that $q$ holds whenever $f$ terminates'. It is a probabilistic analogue of a Hoare triple for partial correctness.

**Example 5.1.15** (Quantum)**.** Consider the effectus $\mathbf{Wstar}_\le^{\mathrm{op}}$ of $W^*$-algebras for quantum processes. Let $f \colon \mathscr{A} \to \mathscr{B}$ be a morphism in $\mathbf{Wstar}_\le^{\mathrm{op}}$, i.e. a normal subunital CP map $f \colon \mathscr{B} \to \mathscr{A}$. Predicates in $\mathbf{Wstar}_\le^{\mathrm{op}}$ are identified with effects $a \in [0,1]_\mathscr{A}$, and the predicate transformer $f^*$ is simply the restriction $f \colon [0,1]_\mathscr{B} \to [0,1]_\mathscr{A}$. Generalizing the definition in [62, § 3.1] to $W^*$-algebras, we say that $p \in [0,1]_\mathscr{A}$ is a precondition of $q \in [0,1]_\mathscr{B}$ for $f$ if

$$\omega \vDash p \;\le\; f_*(\omega) \vDash q \quad \text{for every state } \omega \in \mathrm{St}(\mathscr{A})\,. \tag{5.3}$$

Since

$$\omega \vDash p \;\le\; f_*(\omega) \vDash q \iff \omega \vDash p \;\le\; \omega \vDash f^*(q)$$

and normal states separate points of a $W^*$-algebra, $p$ is a precondition of $q$ if and only if $p \le f^*(q)$. We thus see that $f^*(q)$ is the weakest (i.e. largest) precondition of $q$.

We can explain the condition (5.3) as follows. The validity $f_*(\omega) \vDash q$ can be rewritten as:

$$f_*(\omega) \vDash q \;=\; |f_*(\omega)| \cdot \Big(\frac{f_*(\omega)}{|f_*(\omega)|} \vDash q\Big)\,.$$

Here $|f_*(\omega)| = \omega(f(1))$ is the probability that 'program' $f$ converges/terminates, when executed in state $\omega$. If $f$ converges, the state after the execution is given by normalization $f_*(\omega)/|f_*(\omega)|$. Thus the product $|f_*(\omega)| \cdot (f_*(\omega)/|f_*(\omega)| \vDash q)$ is the probability that $f$ converges *and* $q$ holds afterwords. Therefore the condition (5.3) can be read as 'for every initial state, the probability that $p$ holds is less than or equal to the probability that $f$ terminates and $q$ holds afterwords'.

By Lemma 5.1.8(v), the liberal predicate transformer $f^\square \colon [0,1]_{\mathscr{B}} \to [0,1]_{\mathscr{A}}$ is given by

$$f^\square(q) = 1 - f^*(1) + f^*(q)\,.$$

The validity of $f^\square(q)$ in a state $\omega \in \mathrm{St}(\mathscr{A})$ is the probability that $f$ diverges *or* $f^*(q)$ holds:

$$\omega \vDash f^\square(q) \;=\; (1 - |f_*(\omega)|) + (\omega \vDash f^*(q))\,,$$

that is, the probability that $q$ holds whenever $f$ converges. Thus we can view $f^\square(q)$ as a quantum analogue of weakest liberal precondition. Morphisms in $\mathrm{Pred}_\square(\mathbf{Wstar}_{\leq}^{\mathrm{op}})$ can be understood as a quantum analogue of Hoare triples for partial correctness.

**Remark 5.1.16.** Weakest preconditions have been studied in general categorical settings [118, 128], which cover various examples from program semantics and logics. So far our effectus theoretic framework does not cover as many examples from program semantics/logics as [118, 128], since the current examples of effectuses are motivated by theories of physics, or come from extensive categories (see Section 6.6). For instance, it is not obvious to accommodate nondeterministic computation in effectus theory: the Kleisli category $\mathcal{K}\ell(\mathcal{P})$ of the powerset monad $\mathcal{P}$, the standard model of nondeterministic computation, is not an effectus (in partial or total form). It will be interesting future work to find more examples of effectuses motivated by the programming perspectives.

## 5.2 Kernels and images

Throughout the section we consider an effectus $\mathbf{C}$, and the fibration $\mathrm{Pred}_\square(\mathbf{C}) \to \mathbf{C}$ defined in the previous section. We will introduce and study notions of kernels $\ker(f)$ and images $\mathrm{im}(f)$ for morphisms $f \colon A \to B$ in an effectus. We define kernels as certain predicates $\ker(f) \in \mathrm{Pred}(A)$ on the domain of $f$, and images as predicates $\mathrm{im}(f) \in \mathrm{Pred}(B)$ on the codomain, using the language of fibrations. Note that in a traditional, more common sense, kernels and images in a category are certain monos/subobjects (see e.g. [20]). One may see from Examples 5.2.2 and 5.2.6 that the fibrational definitions suitably generalize the traditional ones.

### 5.2.1 Kernels

We define kernels in terms of fibrations.

**Definition 5.2.1.** The **kernel** of a morphism $f\colon A \to B$ is the predicate $\ker(f) \in \mathrm{Pred}(A)$ obtained by reindexing the falsity with respect to fibration $\mathrm{Pred}_\square(\mathbf{C}) \to \mathbf{C}$, that is, $\ker(f) := f^\square(\mathbb{0})$. In other words, it is the predicate $\ker(f) \in \mathrm{Pred}(A)$ such that $f\colon (A, \ker(f)) \to (B, \mathbb{0})$ is a cartesian lifting of $f\colon A \to B$ to $(B, \mathbb{0})$.

By definition of $f^\square$, we have $f^\square(\mathbb{0}) = f^*(\mathbb{0}^\perp)^\perp = (\mathbb{1}f)^\perp$, that is, the kernel is simply the orthosupplement of the 'domain' predicate $\mathbb{1}f$. Thus we can intuitively see $\ker(f) = (\mathbb{1}f)^\perp$ as the predicate that '$f$ is undefined'.

**Example 5.2.2.** The following example is *not* an effectus, but it illustrates how Definition 5.2.1 captures fibrationally the notion of kernels. We write **Grp** for the category of groups and group homomorphisms. There is a fibration $\mathrm{Sub}(\mathbf{Grp}) \to \mathbf{Grp}$ of subgroups over groups (it is an instance of *subobject fibrations* [133]): objects of $\mathrm{Sub}(\mathbf{Grp})$ are subgroups $S \subseteq G$, and morphisms $(S \subseteq G) \to (T \subseteq H)$ are group homomorphisms $f\colon G \to H$ that send elements of $S$ to $T$. The functor $\mathrm{Sub}(\mathbf{Grp}) \to \mathbf{Grp}$ sends $S \subseteq G$ to $G$. A cartesian lifting of $f\colon G \to H$ to a subgroup $T \subseteq H$ is given by $(f^*(T) \subseteq G) \to (T \subseteq H)$ where $f^*(T) = \{x \in G \mid f(x) \in T\}$ is the inverse image. In particular, the cartesian lifting of $f$ to the smallest subgroup $\{1\} \subseteq H$ (here $1 \in H$ is the unit) is given by

$$f^*(\{1\}) = \{x \in G \mid f(x) = 1\} = \ker(f)\,,$$

that is, the kernel of a group homomorphism in the usual sense.

**Example 5.2.3.** We briefly review kernels (= orthosupplements of $\mathbb{1}f$) in our main examples of effectuses.

(i) Let $f\colon X \rightharpoonup Y$ be a partial function, i.e. a morphism in the effectus **Pfn**. Since the domain predicate $\mathbb{1}f \in \mathrm{Pred}(X) = \mathcal{P}(X)$ is precisely the domain of definition, the kernel is the subset $\ker(f) = \{x \in X \mid f(x) \text{ undefined}\}$.

(ii) In the effectus $\mathcal{K}\ell(\mathcal{D}_\le)$, the kernel $\ker(f) \in [0,1]^X$ of a morphism $f\colon X \to \mathcal{D}_\le(Y)$ is given by:

$$\ker(f)(x) = 1 - \sum_{y \in Y} f(x)(y)\,.$$

This is the probability that $f(x)$ is undefined.

(iii) In the effectus $\mathbf{Wstar}_\le^{\mathrm{op}}$ the kernel of $f\colon \mathscr{A} \to \mathscr{B}$, i.e. a morphism $f\colon \mathscr{B} \to \mathscr{A}$ in **Wstar**, is $\ker(f) = 1 - f(1) \in [0,1]_{\mathscr{A}}$. This can be seen as the effect that says 'quantum process $f$ does not occur'.

### 5.2.2 Images

We define images fibrationally, as a suitable dual of kernels. We thus need a notion that is dual to cartesian liftings.

**Definition 5.2.4.** Let $\varphi\colon \mathbf{E} \to \mathbf{C}$ be a functor. Let $f\colon A \to B$ be a morphism in $\mathbf{C}$, and $X \in \mathbf{E}_A$ be an object above $A$. An **opcartesian lifting** of $f$ to $A$ along $\varphi$ is a cartesian lifting of $f$ to $A$ along the opposite functor $\varphi^{\mathrm{op}}\colon \mathbf{E}^{\mathrm{op}} \to \mathbf{C}^{\mathrm{op}}$. Explicitly in terms of $\varphi\colon \mathbf{E} \to \mathbf{C}$, it is a morphism $l\colon X \to Y$ above $f$ (hence $\varphi Y = B$) such that for any $k\colon X \to Z$ in $\mathbf{E}$ and $g\colon B \to \varphi Z$ in $\mathbf{C}$ satisfying $\varphi k = g \circ f$, there exists a unique map $m\colon Y \to Z$ in $\mathbf{E}$ above $g$ such that $k = m \circ l$. See the diagram:

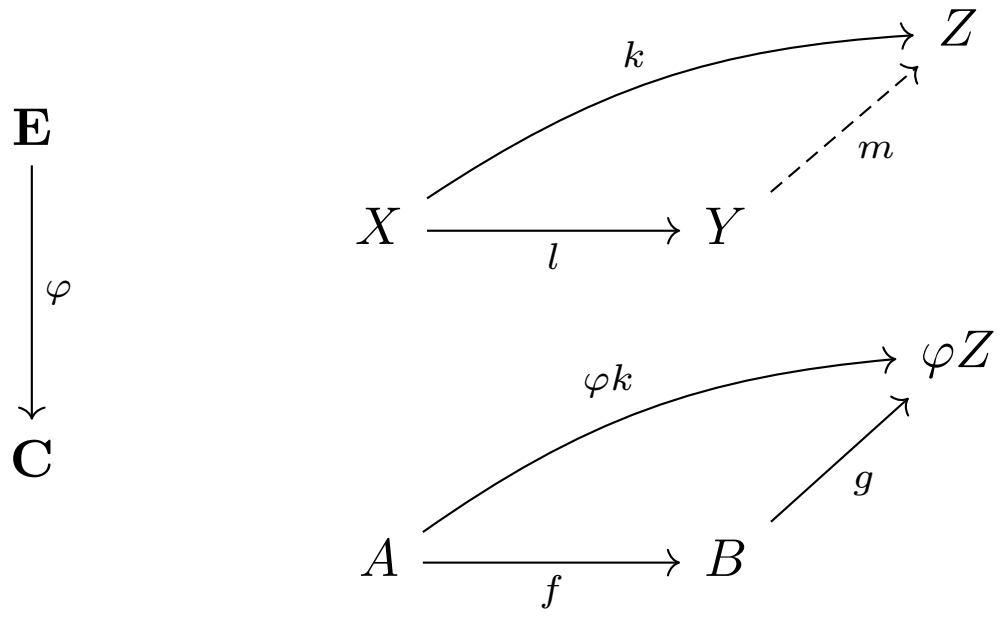

A functor $\varphi\colon \mathbf{E} \to \mathbf{C}$ is an **opfibration** if the opposite functor $\varphi^{\mathrm{op}}\colon \mathbf{E}^{\mathrm{op}} \to \mathbf{C}^{\mathrm{op}}$ is a fibration — i.e. if every morphism $f\colon A \to B$ in $\mathbf{C}$ has a opcartesian lifting to each object above $A$. A functor that is both a fibration and an opfibration is called a **bifibration**.

**Definition 5.2.5.** An **image** of a morphism $f\colon A \to B$ in an effectus is a predicate $\mathrm{im}(f) \in \mathrm{Pred}(B)$ such that $f\colon (A, \mathbb{1}) \to (B, \mathrm{im}(f))$ is an opcartesian lifting of $f\colon A \to B$ to $(A, \mathbb{1})$ along $\mathrm{Pred}_\Box(\mathbf{C}) \to \mathbf{C}$. Note that images need not exist. We say that an effectus **has images** if image $\mathrm{im}(f)$ exists for every morphism $f\colon A \to B$.

For illustration we give an example of groups.

**Example 5.2.6.** Continuing Example 5.2.2, we illustrate the fibrational definition of images by the fibration of subgroups $\mathrm{Sub}(\mathbf{Grp}) \to \mathbf{Grp}$. (Note again that $\mathbf{Grp}$ is not an effectus.) The fibration is in fact a bifibration, i.e. there exists an opcartesian lifting of any homomorphism $f\colon G \to H$ to any subgroup $S \subseteq G$. The lifting is given as $(S, G) \to (f[S], H)$ where $f[S] = \{f(g) \mid g \in S\}$ is the direct image. In particular the opcartesian lifting of a homomorphism $f\colon G \to H$ to the largest subgroup $G \subseteq G$ is precisely the image $\mathrm{im}(f) = f[G]$ of the homomorphism.

We note that the usual notion of images in categories (i.e. the smallest subobject of the codomain of $f$ through which $f$ factors) coincides with images in the sense of Definition 5.2.5 applied to subobject fibrations $\mathrm{Sub}(\mathbf{C}) \to \mathbf{C}$; see e.g. [133, § 4.4].

The following proposition gives a more convenient, direct definition of images:

**Proposition 5.2.7.** *Let $f\colon A \to B$ be a morphism in an effectus. Then $q \in \mathrm{Pred}(B)$ is an image of $f$ if and only if $q$ is a least predicate such that $f^\Box(q) = \mathbb{1}_A$.*

*Proof.* Suppose that $q = \mathrm{im}(f)$, i.e. $(A, \mathbb{1}) \to (B, \mathrm{im}(f))$ is an opcartesian lifting of $f$ to $(A, \mathbb{1})$. Then $\mathbb{1} \le f^\Box(\mathrm{im}(f))$ and thus $f^\Box(\mathrm{im}(f)) = \mathbb{1}$. Assume that $q \in \mathrm{Pred}(B)$ satisfies $f^\Box(q) = \mathbb{1}$. Then $f\colon (A, \mathbb{1}) \to (B, q)$ is a morphism in $\mathrm{Pred}_\Box(\mathbf{C})$. By the universality of a cartesian lifting, $\mathrm{id}_B\colon B \to B$ in $\mathbf{C}$ must lift to $\mathrm{id}_B\colon (B, \mathrm{im}(f)) \to (B, q)$ in $\mathrm{Pred}_\Box(\mathbf{C})$. This implies $\mathrm{im}(f) \le q$. Therefore $\mathrm{im}(f)$ is a least predicate such that $f^\Box(\mathrm{im}(f)) = \mathbb{1}_A$.

Conversely, let $q \in \mathrm{Pred}(B)$ be a least predicate with $f^\Box(q) = \mathbb{1}_A$. Then $f\colon (A, \mathbb{1}) \to (B, q)$ is a morphism in $\mathrm{Pred}_\Box(\mathbf{C})$. To prove that it is a opcartesian lifting, let $h\colon (A, \mathbb{1}) \to (C, r)$ be a morphism in $\mathrm{Pred}_\Box(\mathbf{C})$ and $g\colon B \to C$ in $\mathbf{C}$ such that $h = g \circ f$. Note that

$$\mathbb{1} \le h^\Box(r) = (g \circ f)^\Box(r) = f^\Box(g^\Box(r)) \,.$$

Then by assumption we have $q \le g^\Box(r)$, so that $g$ is a morphism $(B, q) \to (C, r)$ in $\mathrm{Pred}_\Box(\mathbf{C})$. We conclude that $f\colon (A, \mathbb{1}) \to (B, q)$ is an opcartesian lifting. ■

An important fact about images, due to Abraham Westerbaan, is that they are always ortho-sharp.

**Proposition 5.2.8.** *Let* $\mathrm{im}(f) \in \mathrm{Pred}(B)$ *be an image of* $f\colon A \to B$*. Then* $\mathrm{im}(f)$ *is ortho-sharp in* $\mathrm{Pred}(B)$*, i.e.* $\mathrm{im}(f) \wedge \mathrm{im}(f)^\perp = \mathbb{0}$.

*Proof.* Suppose that $p \le \mathrm{im}(f)$ and $p \le \mathrm{im}(f)^\perp$. Then $p \perp \mathrm{im}(f)^\perp$. Using $\mathrm{im}(f)^\perp \circ f = \mathbb{0}$, we have

$$(p \ovee \mathrm{im}(f)^\perp) \circ f = p \circ f \ovee \mathrm{im}(f)^\perp \circ f = p \circ f \le \mathrm{im}(f)^\perp \circ f = \mathbb{0} \,,$$

so $\mathrm{im}(f) \le (p \ovee \mathrm{im}(f)^\perp)^\perp$, i.e. $p \ovee \mathrm{im}(f)^\perp \le \mathrm{im}(f)^\perp$. Then $p \ovee \mathrm{im}(f)^\perp = \mathrm{im}(f)^\perp$ and hence $p = \mathbb{0}$. ■

This motivates the definition of sharp predicates in an effectus using images, see Section 5.5.

**Example 5.2.9.** Our main examples of effectuses have images.

(i) In the effectus **Pfn**, the image of a partial function $f\colon X \to Y$ is given by $\mathrm{im}(f) = \{y \in Y \mid \exists x \in X.\, y = f(x)\}$. Indeed, it is easy to see via (5.2) that $\mathrm{im}(f)$ is the least subset $Q \subseteq Y$ such that $f^\Box(Q) = X$.

(ii) Let $f\colon X \to \mathcal{D}_\le(Y)$ be a morphism in $\mathcal{K}\ell(\mathcal{D}_\le)$. We claim that

$$\mathrm{im}(f)(y) = \begin{cases} 1 & \text{if there exists } x \in X \text{ with } f(x)(y) > 0 \\ 0 & \text{otherwise.} \end{cases}$$

With this definition of $\mathrm{im}(f)$, indeed we have:

$$\begin{aligned} f^\Box(q) = \mathbb{1} &\iff q^\perp \odot f = \mathbb{0} \\ &\iff q^\perp(y) \cdot f(x)(y) = 0 \;\text{ for all } x \in X \text{ and } y \in Y \\ &\iff f(x)(y) > 0 \text{ implies } q(y) = 1 \;\text{ for all } x \in X \text{ and } y \in Y \\ &\iff \mathrm{im}(f)(y) \le q(y) \;\text{ for all } y \in Y \,. \end{aligned}$$

(iii) Let $f\colon \mathscr{A} \to \mathscr{B}$ be a morphism in $\mathbf{Wstar}_\le^{\mathrm{op}}$, i.e. a normal subunital CP map $f\colon \mathscr{B} \to \mathscr{A}$. By definition, the image of $f$ is the least effect $q \in \mathcal{P}r(\mathscr{B})$ such that $f(q^\perp) = 0$, if such an effect exists. By Proposition 5.2.8, the image $\mathrm{im}(f)$ is ortho-sharp and hence a projection (see Proposition 2.6.11). Thus $\mathrm{im}(f)$ is is the least *projection* $\mathfrak{q} \in \mathcal{P}r(\mathscr{B})$ such that $f(\mathfrak{q}^\perp) = 0$. Conversely, we claim that such a projection, if it exists, is the image of $f$. To verify this, let $\mathrm{im}'(f)$ be the least projection $\mathfrak{q} \in \mathcal{P}r(\mathscr{B})$ with $f(\mathfrak{q}^\perp) = 0$. Assume $f(q^\perp) = 0$ for an effect $q \in [0,1]_{\mathscr{B}}$. By [254, Proposition 45], we have $f(\lfloor q \rfloor^{\perp}) = f(\lceil q^{\perp} \rceil) = 0$, where $\lfloor p \rfloor$ and $\lceil p \rceil$ denote respectively the greatest projection below $p$ and the least projection above $p$ (see Definition 2.6.15). Then $\mathrm{im}'(f) \le \lfloor q \rfloor \le q$, showing that $\mathrm{im}'(f)$ is the image of $f$.

What we have shown is that the image of a map $f$ in $\mathbf{Wstar}_{\le}^{\mathrm{op}}$ is precisely the *carrier* of $f$ in the sense of Westerbaan [253, 63 I]. The image/carrier of a normal subunital CP map $f\colon \mathscr{B} \to \mathscr{A}$ exists and can be obtained by

$$\mathrm{im}(f) = \bigwedge \{ \mathfrak{q} \in \mathcal{Pr}(\mathscr{B}) \mid f(\mathfrak{q}^{\perp}) = 0 \},$$

where the meet $\bigwedge$ is taken in the complete lattice $\mathcal{Pr}(\mathscr{B})$ of projections; see Proposition 2.6.13. We refer to [253] for further details.

**Remark 5.2.10.** Images in the effectus $\mathcal{K}\ell(\mathcal{G}_{\le})$ are not very well-behaved. Indeed, even the Lebesgue measure $\mu$ on $[0,1]$, seen as a state $\mu\colon 1 \to \mathcal{G}_{\le}([0,1])$, does not have an image. By definition, the image of the Lebesgue measure $\mu \in \mathcal{G}_{\le}([0,1])$ is a least measurable function $p\colon [0,1] \to [0,1]$ such that $\int p^{\perp}\, \mathrm{d}\mu = 0$. Note that for each $x \in [0,1]$, the obvious Dirac function $\delta_x\colon [0,1] \to [0,1]$ is measurable and $\int \delta_x \, \mathrm{d}\mu = 0$. Thus, if the image $\mathrm{im}(\mu)$ exists, one has $\mathrm{im}(\mu) \le (\delta_x)^{\perp}$ for all $x \in [0,1]$, which implies $\mathrm{im}(\mu)(x) = 0$ for all $x \in X$. But then $\int \mathrm{im}(\mu)^{\perp}\, \mathrm{d}\mu = 1$, which contradicts to the assumption that $\mathrm{im}(\mu)$ is the image. Therefore the Lebesgue measure/state $\mu$ does not have an image in the effectus $\mathcal{K}\ell(\mathcal{G}_{\le})$.

We note that in measure theory, there is a notion of *support* of a measure, which is similar to images of states in $\mathcal{K}\ell(\mathcal{G}_{\le})$. However, the definition assumes a *topology*: the **support** of a measure $\mu$ is a least $\mu$-conegligible *closed* set. For example, the support of the Lebesgue measure on $[0,1]$ is the whole space $[0,1]$. It seems impossible to capture the notion of support in terms of the category $\mathcal{K}\ell(\mathcal{G}_{\le})$, since $\mathcal{K}\ell(\mathcal{G}_{\le})$ is defined in terms of measurable spaces only, and not topological spaces.

Note that a morphism $f\colon A \to B$ is total iff $\mathbb{1}f = \mathbb{1}$ iff $\ker(f) = \mathbb{0}$. For this reason, total morphisms are also called *internal monos* in [40]. A dual notion (*internal epis*) is defined as follows.

**Definition 5.2.11.** A morphism $f\colon A \to B$ is said to be **faithful** if ($\mathrm{im}(f)$ exists and) $\mathrm{im}(f) = \mathbb{1}$.

See Example 5.2.13(iii) for the reason for the terminology ‘faithful’.

The following explicit characterization of faithful morphisms is convenient.

**Lemma 5.2.12.** *Let $f\colon A \to B$ be a morphism. The following are equivalent:*

(i) *$f$ is faithful, i.e.* $\mathrm{im}(f) = \mathbb{1}$.

(ii) $f^{\square}(q) = \mathbb{1}$ *implies* $q = \mathbb{1}$ *for each* $q \in \mathrm{Pred}(B)$.

(iii) $q \circ f = \mathbb{0}$ *implies* $q = \mathbb{0}$ *for each* $q \in \mathrm{Pred}(B)$.

*Proof.* If $f$ is faithful, then $f^{\square}(q) = \mathbb{1}$ implies $\mathbb{1} = \mathrm{im}(f) \le q$, i.e. $q = \mathbb{1}$. Thus (ii) holds. Conversely, if (ii) holds, then clearly $f^{\square}(q) = \mathbb{1}$ implies $\mathbb{1} \le q$. Therefore (i) $\iff$ (ii). The equivalence (ii) $\iff$ (iii) is straightforward. ∎

**Example 5.2.13.**

(i) Faithful maps in **Pfn** are precisely surjective partial functions.

(ii) In the effectus $\mathcal{K}\ell(\mathcal{D}_{\leq})$, a map $f\colon X \to \mathcal{D}_{\leq}(Y)$ is faithful if and only if for each $y \in Y$ there exists $x \in X$ such that $f(x)(y) > 0$. In particular, a substate $\omega \in \mathcal{D}_{\leq}(X)$ is faithful if and only if $\omega$ has full support.

(iii) By Lemma 5.2.12(iii), a morphism $f\colon \mathscr{A} \to \mathscr{B}$ in $\mathbf{Wstar}_{\leq}^{\mathrm{op}}$, i.e. $f\colon \mathscr{B} \to \mathscr{A}$ in $\mathbf{Wstar}_{\leq}$, is faithful iff $f(q) = 0$ implies $q = 0$ for every $q \in [0,1]_{\mathscr{B}}$. In particular, a state is faithful precisely when it is a *faithful* state $\omega\colon \mathscr{A} \to \mathbb{C}$ in the usual sense, see e.g. [246, Definition 9.4]. This is where the terminology comes from.

Note that 'external' epis are 'internal' epis, i.e. faithful maps.

**Proposition 5.2.14.** *Any epi $f\colon A \to B$ is faithful.*

*Proof.* If $q \circ f = \mathbb{0}$, then $q \circ f = \mathbb{0} \circ f$, so $q = \mathbb{0}$ since $f$ is epic. ∎

In particular, all identities $\mathrm{id}\colon A \to A$ and partial projections $\rhd_i\colon A_1 + A_2 \to A_i$ are faithful, since they are (split) epis.

In the rest of the subsection, we investigate how images interact with the constructions in effectuses such as composition, tupling, addition, and cotupling.

**Lemma 5.2.15.** *Let $f\colon A \to B$ and $g\colon B \to C$ be morphisms.*

(i) $\mathrm{im}(g \circ f) \leq \mathrm{im}(g)$ *(if both sides exist).*

(ii) *If $f$ is faithful,* $\mathrm{im}(g \circ f) = \mathrm{im}(g)$, *where* $\mathrm{im}(g \circ f)$ *exists if and only if so does* $\mathrm{im}(g)$.

*Proof.*

(i) This follows from $(g \circ f)^{\square}(\mathrm{im}(g)) = f^{\square}(g^{\square}(\mathrm{im}(g))) = f^{\square}(\mathbb{1}) = \mathbb{1}$.

(ii) For any $p \in \mathrm{Pred}(C)$,

$$(g \circ f)^{\square}(p) = \mathbb{1} \iff f^{\square}(g^{\square}(p)) = \mathbb{1} \iff g^{\square}(p) = \mathbb{1}\,,$$

by the faithfulness of $f$. This proves the claim. ∎

**Lemma 5.2.16.** *Let $f\colon A \to B$ and $g\colon A \to C$ be compatible morphisms (i.e. $\langle\!\langle f, g\rangle\!\rangle\colon A \to B + C$ exists). For each predicate $p \in \mathrm{Pred}(B)$ and $q \in \mathrm{Pred}(C)$,*

$$\langle\!\langle f, g\rangle\!\rangle^{\square}([p,q]) = \mathbb{1} \iff f^{\square}(p) = g^{\square}(q) = \mathbb{1}\,.$$

*Proof.* By the following equivalence:

$$\begin{aligned}
\langle\!\langle f, g\rangle\!\rangle^{\square}([p,q]) = \mathbb{1} &\iff [p,q]^{\perp} \circ \langle\!\langle f, g\rangle\!\rangle = \mathbb{0} \\
&\iff p^{\perp} \circ f \ovee q^{\perp} \circ g = \mathbb{0} \\
&\iff p^{\perp} \circ f = q^{\perp} \circ g = \mathbb{0} \\
&\iff f^{\square}(p) = g^{\square}(q) = \mathbb{1}\,.
\end{aligned}$$ ∎

The following result is useful, allowing us to calculate images componentwise.

**Proposition 5.2.17.** *In the setting of Lemma* 5.2.16*,* $\mathrm{im}\langle\!\langle f, g\rangle\!\rangle = [\mathrm{im}(f), \mathrm{im}(g)]$*. Here* $\mathrm{im}\langle\!\langle f, g\rangle\!\rangle$ *exists if and only if both* $\mathrm{im}(f)$ *and* $\mathrm{im}(g)$ *exist.*

*Proof.* Now assume that both $\mathrm{im}(f)$ and $\mathrm{im}(g)$ exist. By Lemma 5.2.16, for any $[p, q] \in \mathrm{Pred}(B + C)$ we have

$$\begin{aligned} \langle\!\langle f, g\rangle\!\rangle^{\square}([p, q]) = \mathbb{1} &\iff f^{\square}(p) = g^{\square}(q) = \mathbb{1} \\ &\iff \mathrm{im}(f) \le p \text{ and } \mathrm{im}(g) \le q \\ &\iff [\mathrm{im}(f), \mathrm{im}(g)] \le [p, q]\,, \end{aligned}$$

proving that $\mathrm{im}\langle\!\langle f, g\rangle\!\rangle = [\mathrm{im}(f), \mathrm{im}(g)]$.

Conversely, assume that $\mathrm{im}\langle\!\langle f, g\rangle\!\rangle$ exists. Then for any $p \in \mathrm{Pred}(B)$,

$$\begin{aligned} f^{\square}(p) = \mathbb{1} &\iff f^{\square}(p) = g^{\square}(\mathbb{1}) = \mathbb{1} \\ &\iff \langle\!\langle f, g\rangle\!\rangle^{\square}([p, \mathbb{1}]) = \mathbb{1} \\ &\iff \mathrm{im}\langle\!\langle f, g\rangle\!\rangle \le [p, \mathbb{1}] \\ &\iff \kappa_1 \circ \mathrm{im}\langle\!\langle f, g\rangle\!\rangle \le p\,, \end{aligned}$$

proving $\mathrm{im}(f) = \kappa_1 \circ \mathrm{im}\langle\!\langle f, g\rangle\!\rangle$. Similarly we obtain $\mathrm{im}(g) = \kappa_2 \circ \mathrm{im}\langle\!\langle f, g\rangle\!\rangle$ and hence $\mathrm{im}\langle\!\langle f, g\rangle\!\rangle = [\mathrm{im}(f), \mathrm{im}(g)]$. ■

From this several interesting consequences follow.

**Corollary 5.2.18.** *Let* $f : A \to B$ *and* $g : A \to C$ *be faithful morphisms that are compatible. Then* $\langle\!\langle f, g\rangle\!\rangle : A \to B + C$ *is faithful.* ■

**Corollary 5.2.19.** *Suppose that images of total morphisms exist in an effectus* $\mathbf{C}$*. Then images of arbitrary morphisms exist in* $\mathbf{C}$*.*

*Proof.* Let $f : A \to B$ be an arbitrary morphism. Then $\langle\!\langle f, (\mathbb{1}f)^{\perp}\rangle\!\rangle : A \to B + I$ is a total morphism, so the image $\mathrm{im}\langle\!\langle f, (\mathbb{1}f)^{\perp}\rangle\!\rangle$ exist. By Proposition 5.2.17, $\mathrm{im}(f)$ exists too. ■

**Corollary 5.2.20.** *Let* $f : A \to B$ *and* $h : C \to D$ *be morphisms. Then* $\mathrm{im}(f + h) = [\mathrm{im}(f), \mathrm{im}(h)]$*. Here* $\mathrm{im}(f + h)$ *exists if and only if both* $\mathrm{im}(f)$ *and* $\mathrm{im}(h)$ *exist.*

*Proof.* Since $f + h = \langle\!\langle f \circ \rhd_1, h \circ \rhd_2\rangle\!\rangle$, the claim follows from Proposition 5.2.17 and Lemma 5.2.15(ii). ■

**Proposition 5.2.21.**

(i) $\mathrm{im}(0) = \mathbb{0}$ *for zero morphisms* $0 : A \to B$*.*

(ii) *Let* $f, g : A \to B$ *be summable morphisms. If images* $\mathrm{im}(f)$ *and* $\mathrm{im}(g)$ *exist, then*

$$\mathrm{im}(f \ovee g) = \mathrm{im}(f) \vee \mathrm{im}(g)\,.$$

*Here the image* $\mathrm{im}(f \ovee g)$ *exists if and only if the join* $\mathrm{im}(f) \vee \mathrm{im}(g)$ *(in* $\mathrm{Pred}(B)$*) exists.*

*Proof.*

(i) we have $0^{\square}(p) = \mathbb{1} \iff p^{\perp} \circ 0 = \mathbb{0}$, but the latter holds for any $p$ and hence is equivalent to $\mathbb{0} \leq p$. Thus $\mathrm{im}(0) = \mathbb{0}$.

(ii) For every $p \in \mathrm{Pred}(B)$,

$$\begin{aligned}(f \ovee g)^{\square}(p) = \mathbb{1} &\iff p^{\perp} \circ (f \ovee g) = \mathbb{0} \\ &\iff p^{\perp} \circ f \ovee p^{\perp} \circ g = \mathbb{0} \\ &\iff p^{\perp} \circ f = \mathbb{0} \;\text{ and }\; p^{\perp} \circ g = \mathbb{0} \\ &\iff f^{\square}(p) = \mathbb{1} \;\text{ and }\; g^{\square}(p) = \mathbb{1} \\ &\iff \mathrm{im}(f) \leq p \;\text{ and }\; \mathrm{im}(g) \leq p\,.\end{aligned}$$

The equivalence proves the claim, since both $\mathrm{im}(f \ovee g)$ and $\mathrm{im}(f) \vee \mathrm{im}(g)$ are a least predicate $p$ satisfying the above equivalent conditions. ∎

**Corollary 5.2.22.** *One has* $\mathrm{im}(\kappa_1) = [\mathbb{1}, \mathbb{0}]$ *and* $\mathrm{im}(\kappa_2) = [\mathbb{0}, \mathbb{1}]$ *for coprojections* $\kappa_1 \colon A \to A + B$ *and* $\kappa_2 \colon B \to A + B$.

*Proof.* Since $\kappa_1 = \langle\!\langle \mathrm{id}, 0 \rangle\!\rangle$, by Proposition 5.2.17 we have $\mathrm{im}(\kappa_1) = [\mathrm{im}(\mathrm{id}), \mathrm{im}(0)] = [\mathbb{1}, \mathbb{0}]$. Similarly $\mathrm{im}(\kappa_2) = [\mathbb{0}, \mathbb{1}]$. ∎

**Lemma 5.2.23.** *Let* $(f_j \colon A_j \to B)_j$ *be a (possibly infinite) family of morphisms with a common codomain. Assume that a coproduct* $\coprod_j A_j$ *and images* $\mathrm{im}(f_j)$ *exist for all* $j$*. Then*

$$\mathrm{im}([f_j]_j) = \bigvee\nolimits_j \mathrm{im}(f_j)\,.$$

*Here the image* $\mathrm{im}([f_j]_j)$ *of cotuple* $[f_j]_j \colon \coprod_j A_j \to B$ *exists if and only if the join* $\bigvee_j \mathrm{im}(f_j)$ *exists.*

*Proof.* We reason similarly to the proof of Proposition 5.2.21(ii) using the following equivalence for each $p \in \mathrm{Pred}(B)$:

$$\begin{aligned}[f_j]_j^{\square}(p) = \mathbb{1} &\iff [f_j^{\square}(p)]_j = [\mathbb{1}]_j && \text{by Lemma 5.1.12} \\ &\iff f_j^{\square}(p) = \mathbb{1} \;\text{ for all } j \\ &\iff \mathrm{im}(f_j) \leq p \;\text{ for all } j\,.\end{aligned}$$

∎

## 5.3 Comprehension

*Comprehension* is an operation that turns a predicate $p$ on $A$ into a type/object $\{A \mid p\}$ which, intuitively, contains elements of $A$ satisfying $p$. It carries an embedding $\{A \mid p\} \to A$. The prototypical example is *set comprehension* that turns a predicate $P$ into the subset $\{x \in A \mid P(x)\} \hookrightarrow A$. Comprehension has been well studied in the context of categorical logic and type theory, using the language of fibrations [76, 132, 133, 189]. Let us briefly describe a fibrational formulation of comprehension that is known as a *D-category* [76], a *comprehension category with unit* [132, 133], or a *fibration with subset types* [133]. Let $\varphi \colon \mathbf{E} \to \mathbf{C}$ be a poset fibration with a *fibred final* *object* — which means that each fibre $\mathbf{E}_A$ has a greatest element $\mathbb{1}_A$ ('truth') and each reindexing $f^*\colon \mathbf{E}_B \to \mathbf{E}_A$ preserves $\mathbb{1}$. The assignment $A \mapsto \mathbb{1}_A$ then yields a right adjoint $\mathbb{1}\colon \mathbf{C} \to \mathbf{E}$ to $\varphi$. In this setting one defines comprehension as a right adjoint to the truth functor $\mathbb{1}\colon \mathbf{C} \to \mathbf{E}$, as in:

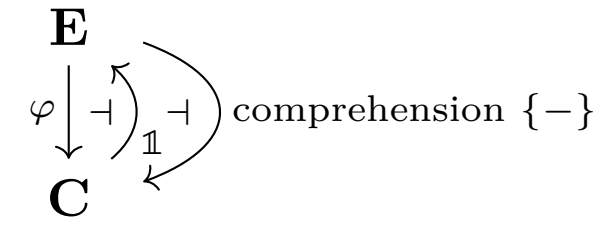

The bijective correspondence for the adjunction

$$\frac{\mathbb{1}_A \longrightarrow Y \quad \text{in } \mathbf{E}}{A \longrightarrow \{Y\} \quad \text{in } \mathbf{C}}$$

captures introduction and elimination rules for comprehension/subset types [133, § 4.6].

The discussion in § 5.1.2 shows that for each effectus $\mathbf{C}$, the fibration of predicates $\mathrm{Pred}_\Box(\mathbf{C}) \to \mathbf{C}$ has a fibred final object. Therefore we can apply the above formulation of comprehension to effectuses.

**Definition 5.3.1.** Let $\mathbf{C}$ be an effectus. A **comprehension** of a predicate $p \in \mathrm{Pred}(A)$ is a universal morphism from the truth functor $\mathbb{1}\colon \mathbf{C} \to \mathrm{Pred}_\Box(\mathbf{C})$ to $(A,p) \in \mathrm{Pred}_\Box(\mathbf{C})$. We write $\pi_p\colon (\{A\,|\,p\},\mathbb{1}) \to (A,p)$ for the comprehension of $p$. Explicitly, it is an object $\{A\,|\,p\} \in \mathbf{C}$ with a morphism $\pi_p\colon (\{A\,|\,p\},\mathbb{1}) \to (A,p)$ in $\mathrm{Pred}_\Box(\mathbf{C})$ such that for each $f\colon (B,\mathbb{1}) \to (A,p)$ in $\mathrm{Pred}_\Box(\mathbf{C})$, there exists a unique morphism $\overline{f}\colon B \to \{A\,|\,p\}$ in $\mathbf{C}$ with $f = \pi_p \circ \overline{f}$.

We often write simply $\pi_p\colon \{A\,|\,p\} \to A$ for comprehension maps, since morphisms in $\mathrm{Pred}_\Box(\mathbf{C})$ are morphisms in $\mathbf{C}$ satisfying extra condition. We say that an effectus **has comprehension** if comprehensions $\pi_p\colon \{A\,|\,p\} \to A$ exist for all $A \in \mathbf{C}$ and $p \in \mathrm{Pred}(A)$. By the standard characterization of adjunction in terms of universal morphisms [199], we obtain the following result.

**Proposition 5.3.2.** *An effectus* $\mathbf{C}$ *has comprehension if and only if there is a right adjoint* $\{-\,|\,-\}\colon \mathrm{Pred}_\Box(\mathbf{C}) \to \mathbf{C}$ *to truth* $\mathbb{1}\colon \mathbf{C} \to \mathrm{Pred}_\Box(\mathbf{C})$. ∎

We give explicit characterizations of comprehension in an effectus.

**Lemma 5.3.3.** *Let $p$ be a predicate on $A$ and $\pi\colon C \to A$ a morphism. The following are equivalent.*

(i) *$\pi\colon C \to A$ is a comprehension of $p$.*

(ii) *$\pi^\Box(p) = \mathbb{1}$, and for each $f\colon B \to A$ in $\mathbf{C}$ satisfying $f^\Box(p) = \mathbb{1}$, there exists a unique morphism $\overline{f}\colon B \to C$ with $f = \pi_p \circ \overline{f}$.*

(iii) *The following diagram is an equalizer in $\mathbf{C}$.*

$$C \xrightarrow{\ \pi\ } A \underset{p}{\overset{\mathbb{1}}{\rightrightarrows}} I$$

*Proof.* We obtain (i) $\Longleftrightarrow$ (ii) by unfolding the definition. (ii) $\Longleftrightarrow$ (iii) follows from the fact that $f^{\square}(p) = \mathbb{1} \iff \mathbb{1} \circ f = p \circ f$ for any $f\colon B \to A$ (see Lemma 5.1.8(iv)). ■

**Example 5.3.4.** Our main examples of effectuses have comprehension.

(i) In the effectus **Pfn**, predicates are subsets $P \subseteq X$. A comprehension $\{X \mid P\}$ is $P$ itself, with inclusion $P \hookrightarrow X$. The universality amounts to the following bijection:

$$\frac{(Y \subseteq Y) \xrightarrow{f} (P \subseteq X) \quad \text{in } \mathrm{Pred}_{\square}(\mathbf{Pfn})}{\overline{Y \xrightarrow{g} \{X \,|\, P\} = P \quad \text{in } \mathbf{Pfn}}}$$

Here $f$ is a morphism in $\mathrm{Pred}_{\square}(\mathbf{Pfn})$ iff $f(y) \in P$ whenever $f(y)$ is defined, for each $y \in Y$. It is then obvious that $f$ and $g$ correspond one-to-one.

(ii) Next we consider the effectus $\mathcal{K}\ell(\mathcal{D}_{\leq})$ for probability. Let $p \in [0,1]^X$ be a 'fuzzy' predicate on a set $X$. We claim that comprehension is the set

$$\{X \,|\, p\} = \{x \in X \mid p(x) = 1\}\,,$$

with the obvious embedding $\{X \,|\, p\} \hookrightarrow X \to \mathcal{D}_{\leq}(X)$. We need to verify the following correspondence:

$$\frac{(\mathbb{1} \in [0,1]^Y) \xrightarrow{f} (p \in [0,1]^X) \quad \text{in } \mathrm{Pred}_{\square}(\mathcal{K}\ell(\mathcal{D}_{\leq}))}{\overline{Y \xrightarrow{g} \{X \,|\, p\} \quad \text{in } \mathcal{K}\ell(\mathcal{D}_{\leq})}}$$

In the top row $f\colon Y \to \mathcal{D}_{\leq}(X)$ satisfies for each $y \in Y$,

$$1 \leq f^{\square}(p)(x) \equiv \sum_{x \in X} p(x) f(y)(x) + 1 - \sum_{x \in X} f(y)(x)\,,$$

that is, $\sum_{x \in X} p(x) f(y)(x) = \sum_{x \in X} f(y)(x)$. It holds if and only if $f(y)(x) > 0$ implies $p(x) = 1$ for all $y \in Y$. From this it is straightforward to see the bijective correspondence, via:

$$\begin{aligned} f(y)(x) &= g(y)(x) && \text{for } x \in \{X \mid p\} \\ f(y)(x) &= 0 && \text{for } x \in X \setminus \{X \mid p\}\,. \end{aligned}$$

We note that the effectus $\mathcal{K}\ell(\mathcal{G}_{\leq})$ also has comprehension, which can be shown by much the same argument [41].

(iii) Consider the effectus $\mathbf{Wstar}_{\leq}^{\mathrm{op}}$ of $W^*$-algebras. Let $p \in [0,1]_{\mathscr{A}}$ be a predicate on $\mathscr{A}$. Then the comprehension of $p$ is:

$$\{\mathscr{A} \,|\, p\} := \lfloor p \rfloor \mathscr{A} \lfloor p \rfloor = \{\lfloor p \rfloor a \lfloor p \rfloor \mid a \in \mathscr{A}\}$$

with the map $\pi_p\colon \mathscr{A} \to \{\mathscr{A} \,|\, p\}$ in $\mathbf{Wstar}_{\leq}$ given by $\pi_p(a) = \lfloor p \rfloor a \lfloor p \rfloor$. Note that $\{\mathscr{A} \,|\, p\} = \lfloor p \rfloor \mathscr{A} \lfloor p \rfloor$ is a (nonunital) $*$-subalgebra of $\mathscr{A}$ that is weak* closed [232, Lemma 1.7.6]. It follows that $\{\mathscr{A} \,|\, p\}$ is a $W^*$-algebra (with unit $\lfloor p \rfloor$). One can also show that $\pi_p\colon \mathscr{A} \to \{\mathscr{A} \,|\, p\}$ is a morphism in $\mathbf{Wstar}_{\leq}$. The universal property of $\pi_p$ as comprehension amounts to: $\pi_p(p) = \pi_p(1)$, and for any $f\colon \mathscr{A} \to \mathscr{B}$ in $\mathbf{Wstar}_{\leq}$ satisfying $f(p) = f(1)$ there exists a unique $\overline{f}\colon \{\mathscr{A}\,|\,p\} \to \mathscr{B}$ in $\mathbf{Wstar}_{\leq}$ such that $\overline{f} \circ \pi_p = f$. This property of $\pi_p$ was proved by Abraham and Bas Westerbaan [254]; see also [253, 256]. Here we do not reproduce the proof, which requires substantial knowledge of operator algebras.

We give a few basic properties of comprehension.

**Proposition 5.3.5.** *Let* $\mathbf{C}$ *be an effectus.*

(i) *The identity map* $\mathrm{id}\colon A \to A$ *is a comprehension of truth* $\mathbb{1}_A$.

(ii) *Every comprehension map* $\pi_p\colon \{A\,|\,p\} \to A$ *is monic in* $\mathbf{C}$.

(iii) *For a morphism* $f\colon A \to B$ *and a predicate* $q \in \mathrm{Pred}(B)$, *if comprehensions for* $q$ *and* $f^{\square}(q)$ *exist, then the following square is a pullback in* $\mathbf{C}$.

$$\begin{array}{ccc} \{A\,|\,f^{\square}(q)\} & \overset{f'}{\dashrightarrow} & \{B\,|\,q\} \\ {\scriptstyle \pi_{f^{\square}(q)}}\big\downarrow & \lrcorner & \big\downarrow{\scriptstyle \pi_q} \\ A & \xrightarrow{\quad f \quad} & B \end{array}$$

*Here the dashed arrow* $f'$ *is the canonical mediating map.*

We note that these properties hold generally for poset fibrations with comprehension (subset types), see [133, Lemma 4.6.2].

*Proof.*

(i) Straightforward.

(ii) By Lemma 5.3.3(iii), every comprehension $\pi_p\colon \{A\,|\,p\} \to A$ is an equalizer and hence a (regular) mono in $\mathbf{C}$.

(iii) First note that the dashed map $f'$ indeed exists since

$$(f \circ \pi_{f^{\square}(q)})^{\square}(q) = \pi_{f^{\square}(q)}^{\square}(f^{\square}(q)) \geq \mathbb{1}\,.$$

Assume that $\alpha\colon C \to A$ and $\beta\colon C \to \{B\,|\,q\}$ satisfy $f \circ \alpha = \pi_q \circ \beta$. Then

$$\alpha^{\square}(f^{\square}(q)) = (f \circ \alpha)^{\square}(q) = (\pi_q \circ \beta)^{\square}(q) = \beta^{\square}(\pi_q^{\square}(q)) \geq \beta^{\square}(\mathbb{1}) = \mathbb{1}$$

Hence there is a unique $\overline{\alpha}\colon C \to \{A\,|\,f^{\square}(q)\}$ with $\alpha = \pi_p \circ \overline{\alpha}$. From the fact that $\pi_q$ is monic it follows that $f' \circ \overline{\alpha} = \beta$. Therefore the square is a pullback. ∎

**Corollary 5.3.6.** *Let* $\mathbf{C}$ *be an effectus with comprehension. Then comprehensions are closed under pullbacks along arbitrary morphisms. In particular, pullbacks/intersections of comprehensions exist as below:*

$$\begin{array}{ccc} \{A\,|\,p\} \cap \{A\,|\,q\} & \rightarrowtail & \{A\,|\,q\} \\ \big\downarrow & \lrcorner & \big\downarrow{\scriptstyle \pi_q} \\ \{A\,|\,p\} & \overset{\pi_p}{\rightarrowtail} & A \end{array}$$

*where*

$$\{A\,|\,p\} \cap \{A\,|\,q\} := \{\{A\,|\,p\}\,|\,\pi_p^\square(q)\} \cong \{\{A\,|\,q\}\,|\,\pi_q^\square(p)\}\,. \qquad \blacksquare$$

Comprehension of falsity $\mathbb{0}$ is what you would expect (cf. Proposition 5.3.5(i)).

**Proposition 5.3.7.** *The unique morphism* $¡_A\colon 0 \to A$ *from the zero object is a comprehension of falsity* $\mathbb{0}_A \in \mathrm{Pred}(A)$.

*Proof.* We prove this via characterization (iii) in Lemma 5.3.3. By initiality we have $\mathbb{1}_A \circ ¡_A = \mathbb{0}_A \circ ¡_A$. Suppose that $f\colon B \to A$ satisfies $\mathbb{1}_A \circ f = \mathbb{0}_A \circ f$. Then $\mathbb{1}_A \circ f = \mathbb{0}_B$, so $f = 0_{BA}$. Thus $f$ factors through $¡_A\colon 0 \to A$ uniquely, as desired. ∎

Notice that in Example 5.3.4 all the comprehension maps $\pi_p\colon \{A\,|\,p\} \to A$ are total. We will now focus on comprehension with the extra property.

**Definition 5.3.8.** A **total comprehension** of $p \in \mathrm{Pred}(A)$ is a comprehension $\pi_p\colon \{A\,|\,p\} \to A$ of $p$ that is total: $\mathbb{1} \circ \pi_p = \mathbb{1}$.

It is unclear whether there exists a non-total comprehension. In Proposition 5.4.10 we will show that in an effectus with quotients (introduced in § 5.4), every comprehension is total.

A nice fact about total comprehensions is that they are compatible with coproducts:

**Proposition 5.3.9.** *Let* $\pi_p\colon \{A\,|\,p\} \to X$ *and* $\pi_q\colon \{B\,|\,q\} \to Y$ *be total comprehensions of* $p \in \mathrm{Pred}(X)$ *and* $q \in \mathrm{Pred}(Y)$. *Then the coproduct*

$$\{A\,|\,p\} + \{B\,|\,q\} \xrightarrow{\pi_p+\pi_q} A + B$$

*is a total comprehension of* $[p,q] \in \mathrm{Pred}(A+B)$.

*Proof.* Clearly $\pi_p+\pi_q$ is total. We prove the universality of $\pi_p+\pi_q$. Let $f\colon C \to A+B$ be a morphism in $\mathbf{C}$ with $f^\square([p,q]) = \mathbb{1}$. Let $f = \langle\!\langle f_1, f_2\rangle\!\rangle$ be decomposed with $f_1\colon C \to A$ and $f_2\colon C \to B$. Then, via Lemma 5.1.8(iv),

$$\mathbb{0} = [p,q]^\perp \circ f = [p^\perp, q^\perp] \circ \langle\!\langle f_1, f_2\rangle\!\rangle = p^\perp \circ f_1 \owedge q^\perp \circ f_2\,.$$

By positivity of an effect algebra, $p^\perp \circ f_1 = \mathbb{0} = q^\perp \circ f_2$. Then by the universality of $\pi_p$ and $\pi_q$, we obtain $\overline{f_1}\colon C \to \{A\,|\,p\}$ and $\overline{f_2}\colon C \to \{B\,|\,q\}$ with $f_1 = \pi_p \circ \overline{f_1}$ and $f_2 = \pi_q \circ \overline{f_2}$, respectively. Since $\pi_p$ and $\pi_q$ are total, $\mathbb{1}f_1 = \mathbb{1}\overline{f_1}$ and $\mathbb{1}f_2 = \mathbb{1}\overline{f_2}$, so the tuple $\langle\!\langle \overline{f_1}, \overline{f_2}\rangle\!\rangle\colon C \to \{A\,|\,p\} + \{B\,|\,q\}$ exists. Then (by Lemma 3.8.3)

$$(\pi_p + \pi_q) \circ \langle\!\langle \overline{f_1}, \overline{f_2}\rangle\!\rangle = \langle\!\langle \pi_p \circ \overline{f_1}, \pi_q \circ \overline{f_2}\rangle\!\rangle = \langle\!\langle f_1, f_2\rangle\!\rangle = f\,.$$

To prove the uniqueness, assume that $(\pi_p+\pi_q)\circ g = f$ for some $g\colon C \to \{A\,|\,p\}+\{B\,|\,q\}$. Then

$$f_1 = \rhd_1 \circ f = \rhd_1 \circ (\pi_p + \pi_q) \circ g = \pi_p \circ \rhd_1 \circ g\,,$$

so that $\rhd_1 \circ g = \overline{f_1}$ by the universality of $\pi_p$. Similarly $\rhd_2 \circ g = \overline{f_2}$. Therefore $g = \langle\!\langle \overline{f_1}, \overline{f_2}\rangle\!\rangle$. ∎

**Corollary 5.3.10.** *Let* $\mathbf{C}$ *be an effectus with total comprehension. Then the comprehension functor* $\{-\,|\,-\}\colon \mathrm{Pred}_\square(\mathbf{C}) \to \mathbf{C}$ *preserves finite coproducts.*

*Proof.* This follows from Proposition 5.3.7 and Proposition 5.3.9, since $(A, \mathbb{0}_A)$ is initial in $\mathrm{Pred}_{\Box}(\mathbf{C})$, and $(A + B, [p, q])$ is a coproduct of $(A, p)$ and $(B, q)$ in $\mathrm{Pred}_{\Box}(\mathbf{C})$. ■

Let $\mathbf{C}$ be an effectus with total comprehension. Then we can show (see Lemma 5.3.12 below; also recall Lemma 5.1.6) that the defining adjunction for comprehension can be restricted to total morphisms as in:

$$\begin{array}{ccc} \mathrm{Pred}(\mathrm{Tot}(\mathbf{C})) & \hookrightarrow & \mathrm{Pred}_{\Box}(\mathbf{C}) \\ \downarrow \dashv \uparrow \mathbb{1} \dashv \downarrow \{-\,|\,-\} & & \downarrow \dashv \uparrow \mathbb{1} \dashv \downarrow \{-\,|\,-\} \\ \mathrm{Tot}(\mathbf{C}) & \hookrightarrow & \mathbf{C} \end{array}$$

This means that the fibration $\mathrm{Pred}(\mathrm{Tot}(\mathbf{C})) \to \mathrm{Tot}(\mathbf{C})$ has comprehension in the sense of a right adjoint to truth. Then a natural question is whether the converse holds: if fibration $\mathrm{Pred}(\mathrm{Tot}(\mathbf{C})) \to \mathrm{Tot}(\mathbf{C})$ has comprehension, does $\mathrm{Pred}_{\Box}(\mathbf{C}) \to \mathbf{C}$ have comprehension too? We give an affirmative answer to this, Theorem 5.3.14 below, under a reasonable assumption that comprehension commutes with coproducts. The theorem characterizes total comprehension purely in terms of total morphisms, and thus is convenient when we start with effectuses in total form.

For convenience we introduce some terminology.

**Definition 5.3.11.** Let $\mathbf{C}$ be an effectus. Let $p \in \mathrm{Pred}(A)$ be a predicate. A **T-comprehension** of $p$ is a universal morphism from the truth functor $\mathbb{1} : \mathrm{Tot}(\mathbf{C}) \to \mathrm{Pred}(\mathrm{Tot}(\mathbf{C}))$ to $(A, p) \in \mathrm{Pred}(B)$. For T-comprehension we use the same notation $\pi_p : (\{A\,|\,p\}, \mathbb{1}) \to (A, p)$ as comprehension.

Then an effectus $\mathbf{C}$ has T-comprehension (i.e. T-comprehensions $\pi_p : \{A\,|\,p\} \to A$ exist for all $p \in \mathrm{Pred}(A)$) if and only if there is a right adjoint $\{-\,|\,-\} : \mathrm{Pred}(\mathrm{Tot}(\mathbf{C})) \to \mathrm{Tot}(\mathbf{C})$ to truth $\mathbb{1} : \mathrm{Tot}(\mathbf{C}) \to \mathrm{Pred}(\mathrm{Tot}(\mathbf{C}))$.

Note that T-comprehensions $\pi_p : \{A\,|\,p\} \to A$ are total by definition, but need not be comprehensions in the sense of Definition 5.3.1. Total comprehensions are T-comprehensions.

**Lemma 5.3.12.** *If $\pi_p : \{A\,|\,p\} \to A$ is a total comprehension of a predicate $p \in \mathrm{Pred}(A)$, then $\pi_p$ is a T-comprehension of $p$.*

*Proof.* Let $f : B \to A$ be a morphism in $\mathrm{Tot}(\mathbf{C})$ such that $f^*(p) = \mathbb{1}$. Then there exists a unique morphism $\overline{f} : B \to \{A\,|\,p\}$ with $\pi_p \circ \overline{f} = f$. The morphism $\overline{f}$ is total, since

$$\mathbb{1} \circ \overline{f} = \mathbb{1} \circ \pi_p \circ \overline{f} = \mathbb{1} \circ f = \mathbb{1}\,.$$

Thus $\pi_p$ is a T-comprehension. ■

The following is a key lemma to prove Theorem 5.3.14.

**Lemma 5.3.13.** *Let $\mathbf{C}$ be an effectus. Let $\pi_p : \{A\,|\,p\} \to A$ be a T-comprehension of $p \in \mathrm{Pred}(A)$. Then $\pi_p$ is a (total) comprehension of $p$ in $\mathbf{C}$ if and only if $\pi_p + \mathrm{id}_I : \{A\,|\,p\} + I \to X + I$ is a T-comprehension of $[p, \mathbb{1}_I] \in \mathrm{Pred}(A + I)$.*

*Proof.* The 'only if' part follows by Proposition 5.3.9 and Lemma 5.3.12, since $\mathrm{id}\colon I \to I$ is a total comprehension of $\mathbb{1}_I$. To prove the converse, let $f\colon B \to A$ be a morphism in $\mathbf{C}$ with $f^\square(p) = \mathbb{1}$. Let $g \coloneqq \langle\!\langle f, \ker(f)\rangle\!\rangle\colon B \to A + I$. Then $g$ is total and satisfies:

$$g^*([p, \mathbb{1}]) = [p, \mathbb{1}] \circ \langle\!\langle f, \ker(f)\rangle\!\rangle = p \circ f \owedge \ker(f) = f^\square(p) = \mathbb{1}\,.$$

Hence by the universality of $\pi_{[p,\mathbb{1}]} = \pi_p + \mathrm{id}_I$, there exists a unique map $\overline{g}\colon B \to \{A \,|\, p\} + I$ with $(\pi_p + \mathrm{id}_I) \circ \overline{g} = g$. Then

$$f = \rhd_1 \circ g = \rhd_1 \circ (\pi_p + \mathrm{id}_I) \circ \overline{g} = \pi_p \circ \rhd_1 \circ \overline{g}\,.$$

Hence $\rhd_1 \circ \overline{g}\colon B \to \{A \,|\, p\}$ is a mediating map for $f$. To see the uniqueness, suppose that $f = \pi_p \circ h$ for some $h\colon B \to \{A \,|\, p\}$. Then the tuple $\langle\!\langle h, \ker(h)\rangle\!\rangle\colon B \to \{A \,|\, p\} + I$ satisfies

$$(\pi_p + \mathrm{id}_I) \circ \langle\!\langle h, \ker(h)\rangle\!\rangle = \langle\!\langle \pi_p \circ h, \ker(h)\rangle\!\rangle = \langle\!\langle f, \ker(h)\rangle\!\rangle = \langle\!\langle f, \ker(f)\rangle\!\rangle = g\,.$$

Thus $\langle\!\langle h, \ker(h)\rangle\!\rangle = \overline{g}$, so that $h = \rhd_1 \circ \overline{g}$. ∎

**Theorem 5.3.14.** *For each effectus* $\mathbf{C}$*, the following are equivalent.*

(i) $\mathbf{C}$ *has total comprehension — that is, the truth functor* $\mathbb{1}\colon \mathbf{C} \to \mathrm{Pred}_\square(\mathbf{C})$ *has a right adjoint* $\{- \,|\, -\}\colon \mathrm{Pred}_\square(\mathbf{C}) \to \mathbf{C}$*, and each component of the counit* $\pi_p\colon (\{A \,|\, p\}, \mathbb{1}) \to (A, p)$ *is total.*

(ii) $\mathbf{C}$ *has T-comprehension that commutes with finite coproducts — that is, the truth functor* $\mathbb{1}\colon \mathrm{Tot}(\mathbf{C}) \to \mathrm{Pred}(\mathrm{Tot}(\mathbf{C}))$ *for total maps has a right adjoint* $\{- \,|\, -\}\colon \mathrm{Pred}(\mathrm{Tot}(\mathbf{C})) \to \mathrm{Tot}(\mathbf{C})$ *that preserves finite coproducts.*

We note that (ii) is equivalent to a weaker condition that $\mathbf{C}$ has T-comprehension which commutes with the lift $(-) + I$. This is clear from the proof below.

*Proof.* That (i) $\Longrightarrow$ (ii) follows by Corollary 5.3.10 and Lemma 5.3.12. Conversely, if we assume (ii), then for each T-comprehension $\pi_p\colon \{A \,|\, p\} \to A$ the coproduct $\pi_p + \mathrm{id}_I\colon \{A \,|\, p\} + I \to A + I$ is a T-comprehension of $[p, \mathbb{1}_I]$ since $\mathrm{id}_I\colon I \to I$ is a T-comprehension of $\mathbb{1}_I$. Thus (i) follows by Lemma 5.3.13. ∎

## 5.4 Quotients

We here study a notion of quotients by predicates. It is an operation that sends a predicate $p \in \mathrm{Pred}(A)$ to the object $A/p$ where, intuitively, elements that satisfy $p$ are identified with zero/unit. This is thus reminiscent of quotient algebras such as quotient groups by subgroups and quotient rings by ideals. These quotients can be described via fibrations as a suitable dual of comprehension. We first illustrate this by the example of groups, continuing Examples 5.2.2 and 5.2.6.

**Example 5.4.1.** Recall from Example 5.2.2 the fibration of subgroups over groups $\mathrm{Sub}(\mathbf{Grp}) \to \mathbf{Grp}$, $(S \subseteq G) \mapsto G$. The functor admits the following chain of

adjunctions:

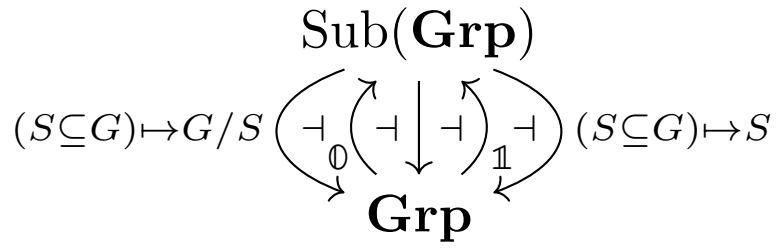

The first left and right adjoints to $\mathrm{Sub}(\mathbf{Grp}) \to \mathbf{Grp}$ are 'falsity' and 'truth' functors — like in Proposition 5.1.11 — respectively, they send a group $G$ to the smallest subgroup $\mathbb{0}(G) = (\{1\} \subseteq G)$ and the largest subgroup $\mathbb{1}(G) = (G \subseteq G)$. There is 'comprehension' of subgroups, i.e. a right adjoint to truth $\mathbb{1}$. It simply sends $S \subseteq G$ to $S$. Dually, there exists a left adjoint to falsity $\mathbb{0}$. The left adjoint sends $S \subseteq G$ to the quotient group $G/S := G/{\sim_S}$ where $\sim_S$ is the congruence relation generated by $x \sim_S y \iff x^{-1}y \in S$. If $S$ is a normal subgroup, the quotient is the set of cosets, i.e. $G/S = \{xS \mid x \in X\}$. Otherwise, $G/S$ is equal to the quotient $G/\overline{S}$ by the normal closure $\overline{S}$. To see that quotients are indeed left adjoint to falsity $\mathbb{0}$, we need to establish the following bijections.

$$\frac{(S \subseteq G) \xrightarrow{f} \mathbb{0}(H) = (\{1\} \subseteq H) \quad \text{in } \mathrm{Sub}(\mathbf{Grp})}{G/S \xrightarrow{g} H \qquad\qquad \text{in } \mathbf{Grp}} \tag{5.4}$$

We here assume that $S \subseteq G$ is a normal subgroup and leave the general case to the interested reader. A morphism on the top of (5.4) is a group homomorphism $f\colon G \to H$ such that $S \subseteq f^{-1}(1)$, i.e. $f(s) = 1$ for all $s \in S$. Given such an $f$ we define $g\colon G/S \to H$ by $g(xS) = f(x)$. The map $g$ is well-defined: if $xS = yS$, then $x^{-1}y \in S$ and hence

$$f(y) = f(xx^{-1}y) = f(x)f(x^{-1}y) = f(x) \cdot 1 = f(x) \ .$$

Conversely, given $g$ on the bottom of (5.4), we define $f\colon G \to H$ by $f(x) = g(xS)$. Then $f(s) = g(sS) = g(1S) = 1$ for all $s \in S$, so it is indeed a map $(S \subseteq G) \to \mathbb{0}(H)$ in $\mathrm{Sub}(\mathbf{Grp})$. This establishes the desired bijection (5.4).

Thus we define quotients in an effectus as follows.

**Definition 5.4.2.** A **quotient** for a predicate $p \in \mathrm{Pred}(A)$ is a universal morphism from $(A, p) \in \mathrm{Pred}_\Box(\mathbf{C})$ to the falsity functor $\mathbb{0}\colon \mathbf{C} \to \mathrm{Pred}_\Box(\mathbf{C})$. We denote a quotient for $p$ as $\xi_p\colon (A, p) \to (A/p, \mathbb{0})$. Explicitly, it is a morphism $\xi_p\colon (A, p) \to (A/p, \mathbb{0})$ in $\mathrm{Pred}_\Box(\mathbf{C})$ such that for each $f\colon (A, p) \to (B, \mathbb{0})$ in $\mathrm{Pred}_\Box(\mathbf{C})$ there exists a unique morphism $\overline{f}\colon A/q \to B$ with $\overline{f} \circ \xi_p = f$.

We say that an effectus $\mathbf{C}$ **has quotients** if quotients $\xi_p\colon A \to A/p$ exist for all $A \in \mathbf{C}$ and $p \in \mathrm{Pred}(A)$. The following proposition is immediate.

**Proposition 5.4.3.** *An effectus* $\mathbf{C}$ *has quotients if and only if there is a left adjoint* $(-)/(-)\colon \mathrm{Pred}_\Box(\mathbf{C}) \to \mathbf{C}$ *to the falsity* $\mathbb{0}\colon \mathbf{C} \to \mathrm{Pred}_\Box(\mathbf{C})$. ■

Recall that total comprehension commutes with finite coproducts. A similar claim for quotients is immediate, since the quotient functor is a left adjoint.

**Corollary 5.4.4.** *Let* $\mathbf{C}$ *be an effectus with quotients. Then the quotient functor* $(-)/(-)\colon \mathrm{Pred}_\Box(\mathbf{C}) \to \mathbf{C}$ *preserves finite coproducts:* $0/\mathbb{0} \cong 0$ *and* $(A+B)/[p,q] \cong A/p + B/q$. ■

**Example 5.4.5.** All of our main examples of effectuses have quotients.

(i) Let $P \subseteq X$ be a predicate in the effectus **Pfn**. Then the quotient $X/P$ is the complement:

$$X/P := \{x \in X \mid x \notin P\} = P^\perp .$$

The universality amounts to the following bijective correspondence:

$$\frac{(P \subseteq X) \xrightarrow{f} (\varnothing \subseteq Y) \quad \text{in } \mathrm{Pred}_\Box(\mathbf{Pfn})}{P^\perp = X/P \overset{g}{\rightharpoonup} Y \quad \text{in } \mathbf{Pfn}}$$

Here on the top is a partial function $f\colon X \rightharpoonup Y$ such that

$$P \subseteq f^\Box(\varnothing) = \{x \in X \mid f(x) \text{ is undefined}\} ,$$

that is, $f(x)$ is undefined for every $x \in P$. It is easy to see that those partial functions are in bijection with partial functions $g\colon P^\perp \rightharpoonup Y$. The quotient map $\xi_P\colon X \rightharpoonup X/P$ is the partial function with $\xi_P(x) = x$ if $x \in X/P = P^\perp$, and undefined otherwise.

(ii) Next consider a 'fuzzy' predicate $p \in [0,1]^X$ in the effectus $\mathcal{K}\ell(\mathcal{D}_\le)$. The quotient for $p$ is given as

$$X/p := \{x \in X \mid p(x) < 1\} .$$

We verify the universality, that is, the following bijections:

$$\frac{(p \in [0,1]^X) \xrightarrow{f} (0 \in [0,1]^Y) \quad \text{in } \mathrm{Pred}_\Box(\mathcal{K}\ell(\mathcal{D}_\le))}{X/p \xrightarrow{g} Y \quad \text{in } \mathcal{K}\ell(\mathcal{D}_\le)}$$

The morphism $f$ on the top is a function $f\colon X \to \mathcal{D}_\le(Y)$ such that $p \le f^\Box(\mathbb{0})$, i.e. $\sum_{y\in Y} f(x)(y) \le p^\perp(x)$ for each $x \in X$. Given $f$ we define $\overline{f}\colon X/p \to \mathcal{D}_\le(Y)$ by $\overline{f}(x)(y) = f(x)(y)/p^\perp(x)$ for $x \in X/p$ and $y \in Y$. Note that $p^\perp(x) \ne 0$ for $x \in X/p$. The value $\overline{f}(x)$ is a subdistribution since $\sum_{y\in Y} f(x)(y) \le p^\perp(x)$.

From the bottom to top, given $g\colon X/p \to \mathcal{D}_\le(Y)$ we define $\overline{g}\colon X \to \mathcal{D}_\le(Y)$ by $\overline{g}(x)(y) = g(x)(y) \cdot p^\perp(x)$. Then for each $x \in X$,

$$\sum_{y\in Y} \overline{g}(x)(y) = \sum_{y\in Y} g(x)(y) \cdot p^\perp(x) \le p^\perp(x) ,$$

showing that $\overline{g}$ is a morphism $(p \in [0,1]^X) \to (\mathbb{0} \in [0,1]^Y)$ in $\mathrm{Pred}_\Box(\mathcal{K}\ell(\mathcal{D}_\le))$.

Clearly the mappings $f \mapsto \overline{f}$ and $g \mapsto \overline{g}$ are inverses of each other. The quotient map $\xi_p\colon X \to \mathcal{D}_\le(X/p)$ is given by $\xi_p(x) = p^\perp(x)|x\rangle$.

One can similarly prove that the effectus $\mathcal{K}\ell(\mathcal{G}_\le)$ has quotients [41].

(iii) Let $p \in [0,1]_{\mathscr{A}}$ be a predicate/effect in the effectus $\mathbf{Wstar}_{\leq}^{\mathrm{op}}$ of $W^*$-algebras, for quantum processes. The quotient for $p$ is then:

$$\mathscr{A}/p := \lceil p^\perp \rceil \mathscr{A} \lceil p^\perp \rceil = \{\lceil p^\perp \rceil a \lceil p^\perp \rceil \mid a \in \mathscr{A}\},$$

where $\lceil q \rceil$ denotes the smallest projection above $q$. The universal morphism $\xi_p \colon \mathscr{A} \to \mathscr{A}/p$ in $\mathbf{Wstar}_{\leq}^{\mathrm{op}}$, i.e.$\xi_p \colon \mathscr{A}/p \to \mathscr{A}$ in $\mathbf{Wstar}_{\leq}$, is given by $\xi_p(a) = \sqrt{p^\perp} \cdot a \cdot \sqrt{p^\perp}$. Here for positive $x \in \mathscr{A}$, we denote by $\sqrt{x}$ the unique positive element satisfying $\sqrt{x} \cdot \sqrt{x} = x$. The universality of $\xi_p$ amounts to the following statement: $\xi_p(1) \leq p^\perp$, and for any $f \colon \mathscr{B} \to \mathscr{A}$ in $\mathbf{Wstar}_{\leq}$ satisfying $f(1) \leq p^\perp$, there exists a unique morphism $\overline{f} \colon \mathscr{B} \to \mathscr{A}/p$ in $\mathbf{Wstar}_{\leq}$ such that $\xi_p \circ \overline{f} = f$. The proof of this statement is highly nontrivial. See Abraham and Bas Westerbaan's paper [254] or their theses [253, 256] for details.

**Lemma 5.4.6.** *Let $p$ be a predicate on $A$. Let $\xi \colon A \to Q$ be a morphism. The following are equivalent.*

(i) *$\xi$ is a quotient for $p$, i.e. it lifts to a morphism $\xi \colon (A,p) \to (Q,\mathbb{0})$ in* $\mathrm{Pred}_\Box(\mathbf{C})$ *that is universal.*

(ii) $p \leq \ker(\xi)$, *and for each morphism $f \colon A \to B$ satisfying* $p \leq \ker(f)$, *there exists a unique morphism $\overline{f} \colon Q \to B$ such that $\overline{f} \circ \xi = f$.*

(iii) $\ker(\xi) = p$, *and for each morphism $f \colon A \to B$ satisfying* $\ker(f) = p$, *there exists a unique morphism $\overline{f} \colon Q \to B$ such that $\overline{f} \circ \xi = f$.*

*Proof.* The equivalence (i) $\iff$ (ii) is immediate by unfolding the definitions. We prove (ii) $\implies$ (iii). Since $\ker(p^\perp) = p$, there exists $\overline{p^\perp} \colon Q \to I$ such that $\overline{p^\perp} \circ \xi = p^\perp$. Then $p^\perp = \overline{p^\perp} \circ \xi_p \leq \mathbb{1} \circ \xi$, i.e. $\ker(\xi) \leq p$. Thus $\ker(\xi) = p$. The latter part is trivial.

Now we prove the converse (iii) $\implies$ (ii). It is trivial that $p \leq \ker(\xi)$. Let $f \colon A \to B$ be a morphism with $p \leq \ker(f)$. Let

$$q := \ker(f) \ominus p = p^\perp \ominus \mathbb{1}f \qquad \text{and} \qquad g := \langle\langle f, q \rangle\rangle \colon A \to B + I\,.$$

Then $\mathbb{1}g = \mathbb{1}f \ovee q = p^\perp$, i.e. $\ker(g) = p$. Therefore there exists $\overline{g} \colon Q \to B + I$ such that $g = \overline{g} \circ \xi$. Then $f = \rhd_1 \circ g = \rhd_1 \circ \overline{g} \circ \xi$, so $\rhd_1 \circ \overline{g} \colon Q \to B$ is a mediating map for $f$. To check the uniqueness, assume that $h \colon Q \to B$ satisfies $f = h \circ \xi$. Then

$$\ker(h) \circ \xi = \mathbb{1} \circ \xi \ominus \mathbb{1} \circ h \circ \xi = p^\perp \ominus \mathbb{1} \circ f = q\,.$$

Thus the tuple $\langle\langle h, \ker(h) \rangle\rangle \colon A \to B + I$ satisfies $\langle\langle h, \ker(h) \rangle\rangle \circ \xi = \langle\langle f, q \rangle\rangle = g$. It follows that $\langle\langle h, \ker(h) \rangle\rangle = \overline{g}$ and hence $h = \rhd_1 \circ \overline{g}$. ∎

**Lemma 5.4.7.** *Let $\xi_p \colon A \to A/p$ be a quotient for $p \in$* $\mathrm{Pred}(A)$.

(i) *$\xi_p$ is an epi.*

(ii) *If $f \colon A \to B$ satisfies* $\ker(f) = p$, *then the mediating map $\overline{f} \colon A/p \to B$ (such that $f = \overline{f} \circ \xi_p$) is total.*

*Proof.*

(i) Let $h, k\colon A/p \to B$ be morphisms satisfying $h \circ \xi_p = k \circ \xi_p =: f$. Then

$$\mathbb{1} \circ f = \mathbb{1} \circ h \circ \xi \leq \mathbb{1} \circ \xi_p = p^\perp ,$$

i.e. $p \leq \ker(f)$. Therefore $h = k$ by the universal property of $\xi_p$.

(ii) Note that

$$\mathbb{1} \circ \overline{f} \circ \xi_p = \mathbb{1} \circ f = p^\perp = \mathbb{1} \circ \xi_p .$$

Since $\xi_p$ is an epi, we obtain $\mathbb{1} \circ \overline{f} = \mathbb{1}$. ∎

Let $p \in \mathrm{Pred}(A)$ be a predicate such that quotients $\xi_p\colon A/p \to A$ and $\xi_{p^\perp}\colon A/p^\perp \to A$ exist. Since $\mathbb{1}\xi_{p^\perp} = p$ and $\mathbb{1}\xi_p = p^\perp$, we can tuple the quotient maps as

$$A \xrightarrow{\langle\!\langle \xi_{p^\perp}, \xi_p \rangle\!\rangle} A/p^\perp + A/p .$$

The tuple is total: $\mathbb{1} \circ \langle\!\langle \xi_{p^\perp}, \xi_p \rangle\!\rangle = p \ovee p^\perp = \mathbb{1}$. We write $\mathrm{dc}_p = \langle\!\langle \xi_{p^\perp}, \xi_p \rangle\!\rangle$, and call it the **decomposition map** for $p$. In fact, the decomposition map $\mathrm{dc}_p$ can be characterized by a suitable universal property as follows.

**Proposition 5.4.8.** *Let* $\mathbf{C}$ *be an effectus. Let* $p \in \mathrm{Pred}(A)$ *be a predicate, and* $d = \langle\!\langle d_1, d_2 \rangle\!\rangle\colon A \to A_1 + A_2$ *be a total morphism. The following are equivalent.*

(i) $d\colon A \to A_1 + A_2$ *is (isomorphic to) the decomposition map* $\mathrm{dc}_p$ *for* $p$ *— that is,* $d_1\colon A \to A_1$ *and* $d_2\colon A \to A_2$ *are quotients for* $p^\perp$ *and* $p$*, respectively.*

(ii) *We have*

$$(\mathbb{1}_{A_1} + \mathbb{1}_{A_2}) \circ d = \langle\!\langle p, p^\perp \rangle\!\rangle\colon A \to I + I .$$

*Moreover,* $d\colon A \to A_1 + A_2$ *is universal among such morphisms: for each total morphism* $f\colon A \to B + C$ *satisfying* $(\mathbb{1}_B + \mathbb{1}_C) \circ f = \langle\!\langle p, p^\perp \rangle\!\rangle$*, there exist unique total morphisms* $g_1\colon A_1 \to B$ *and* $g_2\colon A_2 \to C$ *such that* $(g_1 + g_2) \circ d = f$*.*

*Proof.* (i) $\Longrightarrow$ (ii): We have

$$(\mathbb{1}_{A_1} + \mathbb{1}_{A_2}) \circ d = (\mathbb{1}_{A_1} + \mathbb{1}_{A_2}) \circ \langle\!\langle d_1, d_2 \rangle\!\rangle = \langle\!\langle \mathbb{1}d_1, \mathbb{1}d_2 \rangle\!\rangle = \langle\!\langle p, p^\perp \rangle\!\rangle .$$

If $f\colon A \to B + C$ satisfies $(\mathbb{1}_B + \mathbb{1}_C) \circ f = \langle\!\langle p, p^\perp \rangle\!\rangle$, then $\mathbb{1}f_1 = p$ and $\mathbb{1}f_2 = p^\perp$ where $f_1 \coloneqq \rhd_1 \circ f$ and $f_2 \coloneqq \rhd_2 \circ f$. Since $d_1$ and $d_2$ are quotients, by Lemma 5.4.7(ii), we obtain total morphisms $g_1\colon A_1 \to B$ and $g_2\colon A_2 \to C$ such that $g_1 \circ d_1 = f_1$ and $g_2 \circ d_2 = f_2$, respectively. Then clearly $(g_1 + g_2) \circ d = f$. To see the uniqueness suppose that some maps $h_1, h_2$ satisfy $(h_1 + h_2) \circ d = f$. It follows that $f_1 = h_1 \circ d_1$ and $f_2 = h_2 \circ d_2$. Therefore $g_1 = h_1$ and $g_2 = h_2$ By the universality of $d_1, d_2$.

(ii) $\Longrightarrow$ (i): Clearly $\mathbb{1}d_1 = p$ and $\mathbb{1}d_2 = p^\perp$. We prove that $d_1\colon A \to A_1$ is a quotient for $p^\perp$. Let $f\colon A \to B$ be a morphism with $\ker(f) = p^\perp$. Then the tuple $\langle\!\langle f, \ker(f) \rangle\!\rangle\colon A \to B + I$ satisfies $(\mathbb{1} + \mathbb{1}) \circ \langle\!\langle f, \ker(f) \rangle\!\rangle = \langle\!\langle p, p^\perp \rangle\!\rangle$. Thus there exist morphisms $g_1\colon A_1 \to B$ and $g_2\colon A_2 \to I$ such that $(g_1 + g_2) \circ d = \langle\!\langle f, \ker(f) \rangle\!\rangle$. It follows that $g_1 \circ d_1 = f$. To check the uniqueness let $h\colon A_1 \to B$ be a morphism with $h \circ d_1 = f$. Since $\mathbb{1} \circ d_2 = p^\perp = \ker(f)$, we have $(h + \mathbb{1}) \circ d = \langle\!\langle f, \ker(f) \rangle\!\rangle$. By the universality of the decomposition $d$, we obtain $g_1 = h$ (and $g_2 = \mathbb{1}$). Therefore $d_1$ is a quotient for $p^\perp$, by Lemma 5.4.6. We can similarly prove that $d_2$ is a quotient for $p$. ∎

Note that condition (ii) involves only total morphisms. Therefore we obtain a characterization of effectuses with quotients purely in terms of total morphisms.

**Corollary 5.4.9.** *An effectus* $\mathbf{C}$ *has quotients if and only if for each predicate* $p \in \mathrm{Pred}(A)$*, there exists a universal decomposition map* $d\colon A \to A_1 + A_2$ *in* $\mathrm{Tot}(\mathbf{C})$ *with respect to* $\langle\!\langle p, p^\perp \rangle\!\rangle\colon A \to I + I$*, in the sense of Proposition* 5.4.8(ii). ■

We end the section with a useful observation, due to Bas Westerbaan, that in presence of quotients, comprehension is always total.

**Proposition 5.4.10.** *In an effectus with quotients, every comprehension* $\pi_p\colon \{A \,|\, p\} \to A$ *is total.*

*Proof.* We will prove $\ker(\pi_p) \equiv (\mathbb{1}\pi_p)^\perp = \mathbb{0}$. Let $q = \ker(\pi_p)$. By the universality of the quotient $\xi_q\colon \{A \,|\, p\} \to \{A \,|\, p\}/q$, there is $f\colon \{A \,|\, p\}/q \to A$ such that $\pi_p = f \circ \xi_q$. Then

$$\mathbb{1} \circ f \circ \xi_q = \mathbb{1} \circ \pi_p = p \circ \pi_p = p \circ f \circ \xi_q \,,$$

so that $\mathbb{1} \circ f = p \circ f$, since $\xi_q$ is epic. Therefore $f$ in turn factors through comprehension $\pi_p$ via $\overline{f}\colon \{A \,|\, p\}/q \to \{A \,|\, p\}$ as $f = \pi_p \circ \overline{f}$. From $\pi_p = f \circ \xi_q = \pi_p \circ \overline{f} \circ \xi_q$ we obtain $\overline{f} \circ \xi_q = \mathrm{id}_{\{A|p\}}$ because $\pi_p$ is monic. Thus $\xi_q$ is a split mono and hence total. Therefore $\ker(\pi_p) = q = \ker(\xi_q) = \mathbb{0}$. ■

# 5.5 Sharp predicates

We study sharp predicates in an effectus, using the notions of images and comprehension defined above. We will use some results on Galois connections, which we first recap briefly.

## 5.5.1 Recap on Galois connections

Recall that we can think of any poset as a category by viewing relation $x \leq y$ as a (unique) morphism from $x$ to $y$. Functors $f\colon P \to Q$ between posets are precisely monotone maps. Thus there are a notion of adjunctions $P \leftrightarrows Q$ between posets, and notions of monads and comonads on posets. Adjunctions and (co)monads on posets are quite special: they are always *idempotent* in the following sense.

**Lemma 5.5.1.** *Let* $P, Q$ *be posets. Let* $f\colon P \to Q$ *and* $g\colon Q \to P$ *be monotone maps in adjunction* $f \dashv g$*, Let* $h = g \circ f\colon P \to P$ *and* $k = f \circ g\colon Q \to Q$ *be the monad and comonad induced by the adjunction. Then the following hold for all* $x \in P$ *and* $y \in Q$*.*

(i) $f(g(f(x))) = f(x)$.

(ii) $g(f(g(y))) = g(y)$.

(iii) $h(h(x)) = h(x)$.

(iv) $k(k(y)) = k(y)$.

*Proof.* We only prove the first assertion. The rest is similar. From $g(f(x)) \leq g(f(x))$ we obtain $f(g(f(x))) \leq f(x)$. From $f(x) \leq f(x)$ we obtain $x \leq g(f(x))$, so that $f(x) \leq f(g(f(x)))$ by applying $f$. Thus $f(g(f(x))) = f(x)$. ■

An adjunction between posets is commonly called a **Galois connection**. A monad on a poset is called a **closure operator**, and a comonad is called a **co-closure** (or **kernel**, or **interior**) **operator**. For a function $h\colon P \to P$ we denote the set of $h$-fixed points by

$$\mathrm{FP}(h) = \{x \in P \mid h(x) = x\}\,.$$

Note that if $h$ is a (co)monad / (co-)closure operator, $\mathrm{FP}(h)$ is the Eilenberg-Moore category of $h$. Fixed points of a (co-)closure operator are often called **(co-)closed elements**. They are of great importance in the theory of Galois connections.

**Proposition 5.5.2.** *We continue in the setting of Lemma* 5.5.1*.*

(i) *An element $x \in P$ is an $h$-fixed point if and only if there is $y \in Q$ such that $x = g(y)$. In other words:* $\mathrm{FP}(h) = g[Q]$*. Dually, we have* $\mathrm{FP}(k) = f[P]$*.*

(ii) *For each $x \in P$, $h(x)$ is a least $h$-fixed point above $x$. Dually: for each $y \in Q$, $k(y)$ is a greatest $k$-fixed point below $y$.*

(iii) *The inclusion* $\mathrm{FP}(h) \hookrightarrow P$ *has a left adjoint* $h\colon P \to \mathrm{FP}(h)$ *given by co-restricting* $h\colon P \to P$*. Dually:* $\mathrm{FP}(k) \hookrightarrow Q$ *has a right adjoint* $k\colon Q \to \mathrm{FP}(k)$*.*

(iv) *The restrictions* $f\colon \mathrm{FP}(h) \to Q$ *and* $g\colon \mathrm{FP}(k) \to P$ *are order-embeddings:* $x \le x' \iff f(x) \le f(x')$ *for each* $x, x' \in \mathrm{FP}(h)$*, and* $y \le y' \iff g(y) \le g(y')$ *for each* $y, y' \in \mathrm{FP}(k)$*.*

(v) *The Galois connection* $f\colon P \rightleftarrows Q\colon g$ *restricts to a poset isomorphism* $\mathrm{FP}(h) \cong \mathrm{FP}(k)$ *between the fixed points.*

*Proof.* For (i)–(iv) we prove only the first one of the two dual claims.

(i) If $x = h(x)$ then $x = g(f(x))$, so $x \in g[Q]$. If $x = g(y)$ for some $y \in Q$, then $g(y) = g(f(g(y))) = h(g(y))$ by Lemma 5.5.1.

(ii) Since $h$ is a closure operator / monad, $h(x)$ is a $h$-fixed point above $x$, i.e. $h(h(x)) = h(x)$ and $x \le h(x)$ hold. Let $y \in \mathrm{FP}(h)$ be such that $x \le y$. Then $h(x) \le h(y) = y$. Thus $h(x)$ is least among those.

(iii) This just rephrases (ii).

(iv) If $f(x) \le f(x')$ for $x, x' \in \mathrm{FP}(h)$, then

$$x = h(x) = g(f(x)) \le g(f(x')) = h(x') = x'\,.$$

The converse is trivial.

(v) The restriction $f\colon \mathrm{FP}(h) \to \mathrm{FP}(k)$ is well-defined by (i), and injective by (iv). It is surjective since for each $y \in \mathrm{FP}(k)$, we have $y = k(y) = f(g(y))$ with $g(y) \in \mathrm{FP}(h)$. ∎

## 5.5.2 Sharp predicates via a Galois connection

Henceforth in this section, we work in an effectus with images and total comprehension. Let us give a name to such effectuses.

**Definition 5.5.3.** An effectus is **pre-comprehensive** if it has images and total comprehension.

As the name suggests, we later, in Definition 5.5.23, define a *comprehensive effectus* as a pre-comprehensive effectus satisfying a certain additional condition.

Our main examples of effectuses **Pfn**, $\mathcal{K}\ell(\mathcal{D}_{\leq})$, and $\mathbf{Wstar}_{\leq}^{\mathrm{op}}$ are pre-comprehensive. We note that the requirement of total comprehension is rather mild, but that of images is relatively strong. Indeed, in Remark 5.2.10 we observed that the effectus $\mathcal{K}\ell(\mathcal{G}_{\leq})$ does not have all images. In Example 6.6.11, we will also find an example of an effectus that is extensive (in total form) but does not have images.

We recall the notion of subobjects.

**Definition 5.5.4.** A **subobject** of $A \in \mathbf{C}$ is an equivalence class of monos $m\colon U \rightarrowtail A$, where two monos $m\colon U \rightarrowtail A$ and $n\colon V \rightarrowtail A$ are equivalent if there is an isomorphism $k\colon U \xrightarrow{\cong} V$ such that $n \circ k = m$. We denote by $\mathrm{Sub}(A)$ the set of subobjects of $A$. The set $\mathrm{Sub}(A)$ is partially ordered: $(U \overset{m}{\rightarrowtail} A) \leq (V \overset{n}{\rightarrowtail} A)$ iff the dashed map below exists.

$$\begin{array}{ccc} U & \dashrightarrow & V \\ & {\scriptstyle m}\searrow \quad \swarrow{\scriptstyle n} & \\ & A & \end{array}$$

Subobjects $U \overset{m}{\rightarrowtail} A$ are denoted simply by $U$ when no confusion is likely to arise.

For each predicate $p \in \mathrm{Pred}(A)$, the comprehension map $\pi_p\colon \{A\,|\,p\} \to A$ is a mono, so $(\{A\,|\,p\} \overset{\pi_p}{\rightarrowtail} A) \in \mathrm{Sub}(A)$ is a subobject. If $p \leq q$, then there is a dashed map in the diagram below

$$\begin{array}{ccc} \{A\,|\,p\} & \dashrightarrow & \{A\,|\,q\} \\ & {\scriptstyle \pi_p}\searrow \quad \swarrow{\scriptstyle \pi_q} & \\ & A & \end{array}$$

by the universality of $\pi_q$, since $\pi_p^{\square}(q) \geq \pi_p^{\square}(p) = \mathbb{1}$. Therefore $\{A\,|\,p\} \leq \{A\,|\,q\}$ in $\mathrm{Sub}(A)$. Thus we have shown the following.

**Lemma 5.5.5.** *The mapping* $p \mapsto (\{A\,|\,p\} \overset{\pi_p}{\rightarrowtail} A)$ *defines a monotone map* $\{A\,|\,-\}\colon \mathrm{Pred}(A) \to \mathrm{Sub}(A)$. ■

Now we claim that the mapping has a left adjoint given by images.

**Proposition 5.5.6.** *The mapping* $(U \overset{m}{\rightarrowtail} A) \mapsto \mathrm{im}(m)$ *defines a left adjoint to* $\{A\,|\,-\}\colon \mathrm{Pred}(A) \to \mathrm{Sub}(A)$. *We thus obtain the following Galois connection.*

$$\mathrm{Pred}(A) \underset{\mathrm{im}}{\overset{\{A|-\}}{\rightleftarrows}} \mathrm{Sub}(A) \qquad (\top)$$

*Proof.* We need to verify the following two points.

(1) $(U \overset{m}{\rightarrowtail} A) \leq (\{A\,|\,\mathrm{im}(m)\} \rightarrowtail A)$ in $\mathrm{Sub}(A)$ for all $(U \overset{m}{\rightarrowtail} A) \in \mathrm{Sub}(A)$.

(2) $(U \overset{m}{\rightarrowtail} A) \leq (\{A\,|\,p\} \rightarrowtail A)$ in $\mathrm{Sub}(A)$ implies $\mathrm{im}(m) \leq p$ in $\mathrm{Pred}(A)$ for all $(U \overset{m}{\rightarrowtail} A) \in \mathrm{Sub}(A)$ and $p \in \mathrm{Pred}(A)$.

For (1), note that $m^{\square}(\mathrm{im}(m)) = \mathbb{1}$ by the definition of image, and hence there is a dashed map in:

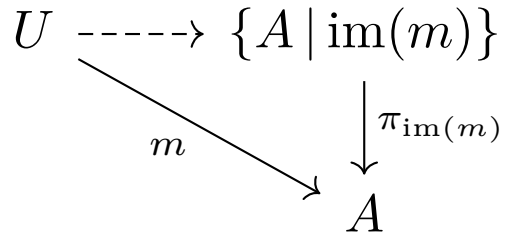

Thus $U \leq \{A \,|\, \mathrm{im}(m)\}$ in $\mathrm{Sub}(A)$. For (2), assume that $U \leq \{A \,|\, p\}$ in $\mathrm{Sub}(A)$, i.e. there is an morphism $k$ in the commutative diagram:

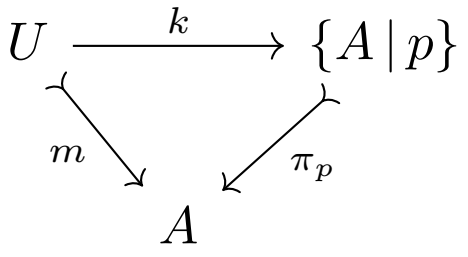

Then $\mathrm{im}(m) = \mathrm{im}(\pi_p \circ k) \leq \mathrm{im}(\pi_p) \leq p$ by Lemma 5.2.15. ■

**Definition 5.5.7.** In a pre-comprehensive effectus, we define the **floor** operation $\lfloor - \rfloor \colon \mathrm{Pred}(A) \to \mathrm{Pred}(A)$ on predicates by

$$\lfloor p \rfloor := \mathrm{im}\big(\{A \,|\, p\} \overset{\pi_p}{\rightarrowtail} A\big)\,.$$

We say that a predicate $p \in \mathrm{Pred}(A)$ is **sharp** if $\lfloor p \rfloor = p$. We write $\mathrm{ShPred}(A) \subseteq \mathrm{Pred}(A)$ for the subset of sharp predicates. We use Fraktur symbols $\mathfrak{p}, \mathfrak{q}, \ldots$ to denote sharp predicates.

In Example 5.5.10 below, we will see that the floor notation $\lfloor p \rfloor$ is consistent with Definition 2.6.15, that is: for a predicate $p$ in a $W^*$-algebra, $\lfloor p \rfloor = \mathrm{im}(\pi_p)$ is the greatest projection below $p$.

The basic fact that motivates the definition of sharp predicates is the following.

**Proposition 5.5.8.** *Every sharp predicate* $\mathfrak{p} \in \mathrm{ShPred}(A)$ *is ortho-sharp, that is,* $\mathfrak{p} \wedge \mathfrak{p}^{\perp} = \mathbb{0}$ *in* $\mathrm{Pred}(A)$.

*Proof.* By Proposition 5.2.8. ■

Note that the operation $\lfloor - \rfloor \colon \mathrm{Pred}(A) \to \mathrm{Pred}(A)$ is the co-closure operator induced by the Galois connection $\mathrm{im} \dashv \{A \,|\, -\}$ and sharp predicates are $\lfloor - \rfloor$-fixed points. Quite a few desirable properties of floor $\lfloor - \rfloor$ and sharp predicates follow immediately from general results on Galois connections.

**Proposition 5.5.9.** *The following hold for all* $p, q \in \mathrm{Pred}(A)$.

(i) $p \leq q \implies \lfloor p \rfloor \leq \lfloor q \rfloor$

(ii) $\lfloor p \rfloor \leq p$

(iii) $\lfloor \lfloor p \rfloor \rfloor = \lfloor p \rfloor$

(iv) $\{A \,|\, p\} = \{A \,|\, \lfloor p \rfloor\}$ *in* $\mathrm{Sub}(A)$ *for each* $p \in \mathrm{Pred}(A)$.

(v) $p$ *is sharp if and only if* $p = \mathrm{im}(m)$ *for some* $(U \overset{m}{\rightarrowtail} A) \in \mathrm{Sub}(A)$.

(vi) $\lfloor p \rfloor$ *is a greatest sharp predicate below* $p$.

(vii) *The co-restricted floor map* $\lfloor - \rfloor \colon \mathrm{Pred}(A) \to \mathrm{ShPred}(A)$ *is right adjoint to the inclusion* $\mathrm{ShPred}(A) \hookrightarrow \mathrm{Pred}(A)$. *In consequence:*

(a) *The inclusion* $\mathrm{ShPred}(A) \hookrightarrow \mathrm{Pred}(A)$ *preserves joins* $\bigvee$.

(b) *The floor map* $\lfloor - \rfloor \colon \mathrm{Pred}(A) \to \mathrm{ShPred}(A)$ *preserves meets* $\bigwedge$.

(viii) *The restricted map* $\pi_{(-)} \colon \mathrm{ShPred}(A) \to \mathrm{Sub}(A)$ *is an order-embedding, i.e. for each sharp predicates* $\mathfrak{p}, \mathfrak{q} \in \mathrm{ShPred}(A)$,

$$\mathfrak{p} \le \mathfrak{q} \iff \{A \,|\, \mathfrak{p}\} \le \{A \,|\, \mathfrak{q}\} \text{ in } \mathrm{Sub}(A)\,.$$

*Proof.* (i) and (ii) are immediate since $\lfloor - \rfloor$ is a comonad. (iii) and (iv) hold by Lemma 5.5.1. (v)–(viii) hold by Proposition 5.5.2. ∎

**Example 5.5.10.**

(i) Recall that for a predicate $P \subseteq X$ in the effectus **Pfn**, we have comprehension $\pi_P \colon \{X \,|\, P\} = P \hookrightarrow X$. Then

$$\lfloor P \rfloor = \mathrm{im}(\pi_P) = \pi_P[P] = P\,.$$

Therefore all predicates in **Pfn** are sharp.

(ii) In $\mathcal{K}\ell(\mathcal{D})$, the comprehension of a predicate $p \in [0,1]^X$ is given by

$$\{X \,|\, p\} = \{x \in X \mid p(x) = 1\}$$

with $\pi_p \colon \{X \,|\, p\} \to \mathcal{D}_{\le}(X)$, $\pi_p(x) = 1|x\rangle$. Thus $\lfloor p \rfloor \in [0,1]^X$ is given as:

$$\lfloor p \rfloor(x) = \mathrm{im}(\pi_p)(x) = \begin{cases} 1 & \text{if } p(x) = 1 \\ 0 & \text{otherwise.} \end{cases}$$

It is easy to see that $\lfloor p \rfloor = p$ if and only if $p$ is Boolean-valued: $p(x) \in \{0,1\}$ for all $x \in X$. Hence sharp predicates $p \colon X \to \{0,1\} \subseteq [0,1]$ can be identified with subsets $P \subseteq X$.

(iii) In the effectus $\mathbf{Wstar}_{\le}^{\mathrm{op}}$ of $W^*$-algebras, we claim that a predicate/effect $p \in [0,1]_{\mathscr{A}}$ is sharp if and only if $p$ is a projection. The 'only if' follows from Proposition 5.5.8, since ortho-sharp elements in $[0,1]_{\mathscr{A}}$ are projections, see [254, Lemma 31]. To prove the converse, let $\mathfrak{p} \in \mathscr{A}$ be a projection. Then

$$\mathscr{A} \xrightarrow{\pi_{\mathfrak{p}}} \{\mathscr{A} \,|\, \mathfrak{p}\} = \mathfrak{p}\mathscr{A}\mathfrak{p}$$

where $\pi_{\mathfrak{p}}(a) = \mathfrak{p}a\mathfrak{p}$. Using the formula in Example 5.2.9,

$$\mathrm{im}(\pi_{\mathfrak{p}}) = \bigwedge \{\mathfrak{q} \in \mathscr{A} \mid \mathfrak{q} \text{ is a projection such that } \mathfrak{p}\mathfrak{q}\mathfrak{p} = \mathfrak{p}\}$$

We have $\mathfrak{p}\mathfrak{q}\mathfrak{p} = \mathfrak{p} \iff \mathfrak{p} \le \mathfrak{q}$ by Lemma 2.6.12, so it follows that $\mathrm{im}(\pi_{\mathfrak{p}}) = \mathfrak{p}$, and hence $\mathfrak{p}$ is sharp. Therefore sharp predicates in $[0,1]_{\mathscr{A}}$ are precisely projections. By Proposition 5.5.9(vi), $\lfloor p \rfloor = \mathrm{im}(\pi_p)$ is the greatest projection below $p$, and hence the notation $\lfloor p \rfloor$ is consistent with Definition 2.6.15.

Let $\mathrm{Comp}(A) \subseteq \mathrm{Sub}(A)$ denotes the set of comprehensions of predicates on $A$, modulo equivalence. In other words, $\mathrm{Comp}(A)$ is the image of the function

$$\{A \mid -\} \colon \mathrm{Pred}(A) \longrightarrow \mathrm{Sub}(A)\,.$$

By Proposition 5.5.2(i), $\mathrm{Comp}(A)$ is the set of fixed points of the closure operator $\mathrm{Sub}(A) \to \mathrm{Sub}(A)$ induced by the Galois connection $\mathrm{Pred}(A) \rightleftarrows \mathrm{Sub}(A)$. We obtain the following result by Proposition 5.5.2(v).

**Proposition 5.5.11.** *The Galois connection on the left below restricts to the poset isomorphism on the right.*

$$\mathrm{Pred}(X) \underset{\mathrm{im}}{\overset{\{A|-\}}{\underset{\longleftarrow}{\overset{\longrightarrow}{\top}}}} \mathrm{Sub}(X) \qquad\qquad \mathrm{ShPred}(X) \underset{\mathrm{im}}{\overset{\{A|-\}}{\underset{\longleftarrow}{\overset{\longrightarrow}{\cong}}}} \mathrm{Comp}(X)$$ ■

Proposition 5.5.9(v) means that images of monos are always sharp. In fact, this is the case for arbitrary maps.

**Lemma 5.5.12.** *For any predicate $p \in \mathrm{Pred}(A)$ and any morphism $f\colon B \to A$, we have $f^\square(p) = \mathbb{1} \iff f^\square(\lfloor p \rfloor) = \mathbb{1}$*

*Proof.* The implication $\Longleftarrow$ is clear because $\lfloor p \rfloor \le p$ and $f^\square$ is monotone. To see $\Longrightarrow$, assume $f^\square(p) = \mathbb{1}$. There is then $\overline{f}\colon B \to \{A \mid p\}$ with $f = \pi_p \circ \overline{f}$. Since $\pi_p^\square(\mathrm{im}(\pi_p)) = \mathbb{1}$, we have $\lfloor p \rfloor \circ \pi_p = \mathrm{im}(\pi_p) \circ \pi_p = \mathbb{1} \circ \pi_p$ and hence

$$\lfloor p \rfloor \circ f = \lfloor p \rfloor \circ \pi_p \circ \overline{f} = \mathbb{1} \circ \pi_p \circ \overline{f} = \mathbb{1} \circ f\,.$$

Thus $f^\square(\lfloor p \rfloor) = \mathbb{1}$. ■

**Proposition 5.5.13.** *For any map $f\colon B \to A$, the image $\mathrm{im}(f) \in \mathrm{Pred}(A)$ is sharp.*

*Proof.* We have $f^\square(\mathrm{im}(f)) = \mathbb{1}$ and hence $f^\square(\lfloor \mathrm{im}(f) \rfloor) = \mathbb{1}$ by Lemma 5.5.12. This means $\mathrm{im}(f) \le \lfloor \mathrm{im}(f) \rfloor$. We are done since $\lfloor \mathrm{im}(f) \rfloor \le \mathrm{im}(f)$. ■

We can thus characterize sharp predicates without mentioning comprehension.

**Corollary 5.5.14.** *A predicate $p \in \mathrm{Pred}(A)$ is sharp if and only if $p = \mathrm{im}(f)$ for some morphism $f\colon B \to A$.* ■

**Remark 5.5.15.** In a pre-comprehensive effectus, there too exists an adjunction $\mathrm{Pred}(A) \rightleftarrows \mathbf{C}/A$ given by comprehension and images, between the poset of predicates and the slice category over $A$. In fact, this is how Lawvere [189] originally described comprehension categorically; see also [132, Example 4.18]. The induced comonad coincides with the floor operator $\lfloor - \rfloor$. Corollary 5.5.14 may also be shown via this fact.

**Proposition 5.5.16.**

(i) *For each $A \in \mathbf{C}$, the falsity $\mathbb{0} \in \mathrm{Pred}(A)$ is sharp.*

(ii) *For each $A \in \mathbf{C}$, the truth $\mathbb{1} \in \mathrm{Pred}(A)$ is sharp.*

(iii) *For each predicate* $p \in \mathrm{Pred}(A)$ *and* $q \in \mathrm{Pred}(B)$*, i.e.* $[p, q] \in \mathrm{Pred}(A + B)$*, we have*
$$\lfloor [p, q] \rfloor = [\lfloor p \rfloor, \lfloor q \rfloor] \,.$$

(iv) *Predicates* $p \in \mathrm{Pred}(A)$ *and* $q \in \mathrm{Pred}(B)$ *are both sharp if and only if the cotuple* $[p, q] \in \mathrm{Pred}(A + B)$ *is sharp.*

*Proof.*

(i) We have $\lfloor \mathbb{0} \rfloor \leq \mathbb{0}$, so $\lfloor \mathbb{0} \rfloor = \mathbb{0}$.

(ii) Truth $\mathbb{1} \in \mathrm{Pred}(A)$ is the image of identity $\mathrm{id}\colon A \to A$.

(iii) By Proposition 5.3.9 and Corollary 5.2.20,
$$\lfloor [p, q] \rfloor = \mathrm{im}(\pi_{[p,q]}) = \mathrm{im}(\pi_p + \pi_q) = [\mathrm{im}(\pi_p), \mathrm{im}(\pi_q)] = [\lfloor p \rfloor, \lfloor q \rfloor] \,.$$

(iv) This follows easily from the previous point. ■

### 5.5.3 Lattice structure in sharp predicates

Here we prove that sharp predicates $\mathrm{ShPred}(A)$ form a lattice, i.e. that they admit finite joins and meets.

**Proposition 5.5.17.** *For any sharp predicates* $\mathfrak{p}, \mathfrak{q} \in \mathrm{ShPred}(A)$ *there exists a join* $\mathfrak{p} \vee \mathfrak{q}$ *in* $\mathrm{Pred}(A)$*. The join is given by* $\mathfrak{p} \vee \mathfrak{q} = \mathrm{im}([\pi_{\mathfrak{p}}, \pi_{\mathfrak{q}}])$ *and it is sharp. Therefore* $\mathfrak{p} \vee \mathfrak{q}$ *is also a join in* $\mathrm{ShPred}(A)$*.*

*Proof.* By Lemma 5.2.23,
$$\mathrm{im}([\pi_{\mathfrak{p}}, \pi_{\mathfrak{q}}]) = \mathrm{im}(\pi_{\mathfrak{p}}) \vee \mathrm{im}(\pi_{\mathfrak{q}}) = \lfloor \mathfrak{p} \rfloor \vee \lfloor \mathfrak{q} \rfloor = \mathfrak{p} \vee \mathfrak{q} \,.$$
The join is sharp by Proposition 5.5.13. ■

An important consequence is that sharp predicates are closed under partial addition $\ovee$.

**Proposition 5.5.18.** *Let* $\mathfrak{p}, \mathfrak{q} \in \mathrm{ShPred}(A)$ *be sharp predicates with* $\mathfrak{p} \perp \mathfrak{q}$*. Then the sum* $\mathfrak{p} \ovee \mathfrak{q}$ *is sharp, and moreover* $\mathfrak{p} \ovee \mathfrak{q} = \mathfrak{p} \vee \mathfrak{q}$*.*

*Proof.* By Corollary 2.3.14, one has $\mathfrak{p} \ovee \mathfrak{q} = \mathfrak{p} \vee \mathfrak{q}$ since join $\mathfrak{p} \vee \mathfrak{q}$ exists by Proposition 5.5.17 and the predicates $\mathfrak{p}$ and $\mathfrak{q}$ are ortho-sharp (Proposition 5.5.8) and hence disjoint. Then $\mathfrak{p} \ovee \mathfrak{q} = \mathfrak{p} \vee \mathfrak{q}$ is sharp by Proposition 5.5.17. ■

**Lemma 5.5.19.** *Let* $(\mathfrak{p}_j)_j$ *be a family of sharp predicates on* $A$*. Suppose that an intersection* $\bigcap_j \{A \,|\, \mathfrak{p}_j\}$ *of the subobjects (i.e. a meet in* $\mathrm{Sub}(A)$*) exists. Then the image*
$$\mathrm{im}\Big(\bigcap\nolimits_j \{A \,|\, \mathfrak{p}_j\} \rightarrowtail A\Big)$$
*is a meet of* $(\mathfrak{p}_j)_j$ *in* $\mathrm{ShPred}(A)$*.*

*Proof.* Let $m$ denote the subobject $\bigcap_j \{A \,|\, \mathfrak{p}_j\} \rightarrowtail A$. We have $\mathrm{im}(m) \leq \mathrm{im}(\pi_{\mathfrak{p}_j}) = \mathfrak{p}_j$ since $\bigcap_j \{A \,|\, \mathfrak{p}_j\} \leq \{A \,|\, \mathfrak{p}_j\}$ in $\mathrm{Sub}(A)$. Now suppose that $\mathfrak{q} \in \mathrm{ShPred}(A)$ satisfies $\mathfrak{q} \leq \mathfrak{p}_j$ for all $j$. Then $\{A \,|\, \mathfrak{q}\} \leq \{A \,|\, \mathfrak{p}_j\}$ for all $j$, and hence $\{A \,|\, \mathfrak{q}\} \leq \bigcap_j \{A \,|\, \mathfrak{p}_j\}$. Therefore $\mathfrak{q} = \mathrm{im}(\pi_{\mathfrak{q}}) \leq \mathrm{im}(m)$. ■

We note that the lemma above is an instance of the general fact that a right adjoint of a reflection (here $\{A \,|\, -\}\colon \mathrm{ShPred}(A) \to \mathrm{Sub}(A)$) creates limits.

**Proposition 5.5.20.** *For each* $\mathfrak{p}, \mathfrak{q} \in \mathrm{ShPred}(A)$*, there exists a meet* $\mathfrak{p} \wedge \mathfrak{q}$ *in* $\mathrm{ShPred}(A)$*. Concretely it is given as the image of a composite of comprehensions (in two ways):*

$$\begin{aligned} \mathfrak{p} \wedge \mathfrak{q} &= \mathrm{im}\big(\{\{A \,|\, \mathfrak{p}\} \,|\, \pi_{\mathfrak{p}}^*(\mathfrak{q})\} \rightarrowtail \{A \,|\, \mathfrak{p}\} \rightarrowtail A\big) \\ &= \mathrm{im}\big(\{\{A \,|\, \mathfrak{q}\} \,|\, \pi_{\mathfrak{q}}^*(\mathfrak{p})\} \rightarrowtail \{A \,|\, \mathfrak{q}\} \rightarrowtail A\big) \end{aligned}$$

*Proof.* This follows by Corollary 5.3.6 and Lemma 5.5.19. ■

**Corollary 5.5.21.** *In a pre-comprehensive effectus, sharp predicates* $\mathrm{ShPred}(A)$ *form a lattice.*

*Proof.* By Propositions 5.5.16, 5.5.17 and 5.5.20. ■

Finally we note that sharp predicates even form a complete lattice when the effectus has suitable limits or colimits (and satisfies a certain smallness condition).

**Proposition 5.5.22.** *Let* $\mathbf{C}$ *be a pre-comprehensive effectus. Let* $A \in \mathbf{C}$ *be an object. Assume that* $\mathrm{ShPred}(A)$ *is a small set (the condition is met when* $\mathbf{C}$ *is locally small, or when* $\mathbf{C}$ *is well-powered). Then sharp predicates* $\mathrm{ShPred}(A)$ *form a complete lattice if at least one of the following conditions hold.*

(i) $\mathbf{C}$ *has all small coproducts.*

(ii) $\mathbf{C}$ *has all small wide pullbacks of comprehensions. (This, of course, holds when* $\mathbf{C}$ *is complete).*

*Proof.* Note that a lattice is complete if and only if it has either all joins or all meets.

Assume (i). Let $(\mathfrak{p}_j)_j$ be a (small) family of sharp predicates on $A$. By assumption, there exists a coproduct $\coprod_j \{A \,|\, \mathfrak{p}_j\}$, and by the same reasoning as Proposition 5.5.17, via Lemma 5.2.23, we have a join $\bigvee_j \mathfrak{p}_j = \mathrm{im}([\pi_{\mathfrak{p}_j}]_j)$ in $\mathrm{ShPred}(A)$.

Assume (ii). Let $(\mathfrak{p}_j)_j$ be a (small) family of sharp predicates on $A$. By assumption, there exists an intersection $\bigcap \{A \,|\, \mathfrak{p}_j\}$ of comprehensions. By Lemma 5.5.19, we have a meet $\bigwedge_j \mathfrak{p}_j = \mathrm{im}(\bigcap \{A \,|\, \mathfrak{p}_j\} \rightarrowtail A)$ in $\mathrm{ShPred}(A)$. ■

All of our main examples of effectuses $\mathbf{Pfn}$, $\mathcal{K}\ell(\mathcal{D}_{\leq})$, and $\mathbf{Wstar}_{\leq}^{\mathrm{op}}$ are locally small and have small coproducts. Therefore sharp predicates in the effectuses form complete lattices.

### 5.5.4 Comprehensive effectuses

So far we have worked in a pre-comprehensive effectus, i.e. an effectus with images and total comprehension. Under this rather natural assumption, we have already proved quite a few properties of sharp predicates — for example, sharp predicates are closed under addition and form a lattice. Unfortunately, however, it is unclear whether sharp predicates are closed under orthosupplements. Since such a property is highly desirable, we take it as a definition of *comprehensive effectus*.

**Definition 5.5.23.** A **comprehensive effectus** is a pre-comprehensive effectus in which the orthosupplement $\mathfrak{p}^\perp$ is sharp for each sharp predicate $\mathfrak{p}$.

By Example 5.5.10 it is easy to see that **Pfn**, $\mathcal{K}\ell(\mathcal{D}_\leq)$, and $\mathbf{Wstar}_\leq^{\mathrm{op}}$ are comprehensive effectuses.

By what we have already shown, we immediately obtain the following results.

**Proposition 5.5.24.** *For each object $A$ in a comprehensive effectus, sharp predicates* $\mathrm{ShPred}(A)$ *form an effect subalgebra of* $\mathrm{Pred}(A)$*.*

*Proof.* This follows from Propositions 5.5.16 and 5.5.18, and the definition of a comprehensive effectus. ■

**Theorem 5.5.25.** *For each object $A$ in a comprehensive effectus, sharp predicates* $\mathrm{ShPred}(A)$ *form an orthomodular lattice.*

*Proof.* By Corollary 5.5.21 and Proposition 5.5.24, sharp predicates $\mathrm{ShPred}(A)$ form a lattice effect algebra. By Proposition 5.5.8 every element $\mathfrak{p} \in \mathrm{ShPred}(A)$ is ortho-sharp. By Proposition 2.3.17, $\mathrm{ShPred}(A)$ is an orthomodular lattice. ■

We introduce the De Morgan dual of $\lfloor - \rfloor$.

**Definition 5.5.26.** In a comprehensive effectus, we define the **ceiling** operation $\lceil - \rceil$ on predicates $p \in \mathrm{Pred}(A)$ by $\lceil p \rceil = \lfloor p^\perp \rfloor^\perp$.By the assumption of a comprehensive effectus, $\lceil p \rceil$ is sharp.

**Proposition 5.5.27.** *For any predicate $p$ in a comprehensive effectus, the predicate $\lceil p \rceil$ is the least sharp predicate above $p$. In other words, $\lceil - \rceil \colon \mathrm{Pred}(X) \to \mathrm{ShPred}(X)$ is a left adjoint to the inclusion $\mathrm{ShPred}(X) \hookrightarrow \mathrm{Pred}(X)$.*

*Proof.* We have $p \leq \lceil p \rceil$ since $\lfloor p^\perp \rfloor \leq p^\perp$ and hence $p = p^{\perp\perp} \leq \lfloor p^\perp \rfloor^\perp = \lceil p \rceil$. Now suppose that $p \leq \mathfrak{q}$ for a sharp predicate $\mathfrak{q}$. Then $\mathfrak{q}^\perp \leq p^\perp$ and so $\mathfrak{q}^\perp \leq \lfloor p^\perp \rfloor$, since $\mathfrak{q}^\perp$ is sharp too. Hence $\lceil p \rceil = \lfloor p^\perp \rfloor^\perp \leq \mathfrak{q}^{\perp\perp} = \mathfrak{q}$. ■

**Corollary 5.5.28.** *The following hold in a comprehensive effectus.*

(i) *The inclusion* $\mathrm{ShPred}(A) \hookrightarrow \mathrm{Pred}(A)$ *preserves joins and meets.*

(ii) *The floor* $\lfloor - \rfloor \colon \mathrm{Pred}(A) \to \mathrm{Pred}(A)$ *preserves meets.*

(iii) *The ceiling* $\lceil - \rceil \colon \mathrm{Pred}(A) \to \mathrm{Pred}(A)$ *preserves joins.*

*Proof.* Claim (i) holds since the inclusion $\mathrm{ShPred}(A) \hookrightarrow \mathrm{Pred}(A)$ has both left adjoint $\lceil - \rceil$ and right adjoint $\lfloor - \rfloor$. Then (ii) follows from (i) since the right adjoint $\lfloor - \rfloor \colon \mathrm{Pred}(A) \to \mathrm{ShPred}(A)$ preserves meets. Similarly (iii) holds. ■

The definition of comprehensive effectuses is rather minimalistic. The following one imposes an alternative stronger condition, which holds in **Pfn**, $\mathcal{K}\ell(\mathcal{D}_{\leq})$, and $\mathbf{Wstar}_{\leq}^{\mathrm{op}}$, and might be more natural.

**Definition 5.5.29.** A **strongly comprehensive effectus** is a pre-comprehensive effectus where every ortho-sharp predicate is sharp.

Strongly comprehensive effectuses are indeed comprehensive.

**Proposition 5.5.30.** *Every strongly comprehensive effectus is comprehensive.*

*Proof.* In a strongly comprehensive effectus, a predicate is sharp if and only if it is ortho-sharp by Proposition 5.5.8. Since $p^{\perp}$ is ortho-sharp for each ortho-sharp predicate $p$, $\mathbf{C}$ is comprehensive. ■

This stronger variant is sometimes more convenient than comprehensive effectuses, see e.g. Corollary 5.5.49.

## 5.5.5 The bifibration of sharp predicates

Throughout the rest of the section, $\mathbf{C}$ is a comprehensive effectus. In this setting we show that sharp predicates form a bifibration, that is, both a fibration and an opfibration.

**Definition 5.5.31.** Let $f\colon A \to B$ be a morphism. We define:

$$\begin{aligned} f_{\blacklozenge}&\colon \mathrm{Pred}(A) \to \mathrm{Pred}(B) \quad \text{by } f_{\blacklozenge}(p) = \mathrm{im}(f \circ \pi_p) \\ f^{\blacksquare}&\colon \mathrm{Pred}(B) \to \mathrm{Pred}(A) \quad \text{by } f^{\blacksquare}(q) = \lfloor f^{\square}(q) \rfloor \,. \end{aligned}$$

Clearly these maps can restrict to sharp predicates, yielding $f_{\blacklozenge}\colon \mathrm{ShPred}(A) \to \mathrm{ShPred}(B)$ and $f^{\blacksquare}\colon \mathrm{ShPred}(B) \to \mathrm{ShPred}(A)$. It is nevertheless convenient to allow non-sharp predicates in their domains.

The notations $f_{\blacklozenge}$ and $f^{\blacksquare}$ are taken from Bas Westerbaan's notations $f_{\diamond}$ and $f^{\square}$ in his thesis [256]. To avoid a clash of notation we use the closed diamond and box instead.

**Lemma 5.5.32.** *Let $f\colon A \to B$ be a morphism. For each predicate $p \in \mathrm{Pred}(A)$ and $q \in \mathrm{Pred}(B)$, we have*

$$f_{\blacklozenge}(p) \leq q \iff \lfloor p \rfloor \leq f^{\square}(q) \iff \lfloor p \rfloor \leq f^{\blacksquare}(q) \,.$$

*Proof.* The latter equivalence holds since $f^{\blacksquare}(q) \equiv \lfloor f^{\square}(q) \rfloor$ is the largest sharp predicate below $f^{\square}(q)$. The first equivalence is shown as follows.

$$\begin{aligned} f_{\blacklozenge}(p) \equiv \mathrm{im}(f \circ \pi_p) \leq q &\iff (f \circ \pi_p)^{\square}(q) = \mathbb{1} \\ &\iff \pi_p^{\square}(f^{\square}(q)) = \mathbb{1} \\ &\iff \lfloor p \rfloor \equiv \mathrm{im}(\pi_p) \leq f^{\square}(q) \end{aligned}$$ ■

**Proposition 5.5.33.** *For each morphism $f\colon A \to B$, $f_\blacklozenge\colon \mathrm{ShPred}(A) \to \mathrm{ShPred}(B)$ and $f^\blacksquare\colon \mathrm{ShPred}(B) \to \mathrm{ShPred}(A)$ are monotone maps in a Galois connection:*

$$\mathrm{ShPred}(A) \underset{f^\blacksquare}{\overset{f_\blacklozenge}{\underset{\longleftarrow}{\overset{\longrightarrow}{\bot}}}} \mathrm{ShPred}(B)$$

*Proof.* The map $f^\blacksquare$ monotone since it is a composite of $f^\square$ and $\lfloor - \rfloor$. By Lemma 5.5.32 we have $f_\blacklozenge(\mathfrak{p}) \le \mathfrak{q} \iff \mathfrak{p} \le f^\blacksquare(\mathfrak{q})$ for sharp predicates $\mathfrak{p}$ and $\mathfrak{q}$, so $f_\blacklozenge$ is left adjoint to $f^\blacksquare$. It follows that $f_\blacklozenge$ is monotone too. ■

**Lemma 5.5.34.** *For each morphism $f\colon A \to B$ and $q \in \mathrm{Pred}(B)$, we have $\lfloor f^\square(q) \rfloor = \lfloor f^\square(\lfloor q \rfloor) \rfloor$, i.e. $f^\blacksquare(q) = f^\blacksquare(\lfloor q \rfloor)$.*

*Proof.* By $\lfloor f^\square(q) \rfloor \le f^\square(q)$ and

$$\begin{aligned}
\lfloor f^\square(q) \rfloor \le f^\square(q) &\iff f_\blacklozenge(f^\square(q)) \le q && \text{Lemma 5.5.32}\\
&\iff f_\blacklozenge(f^\square(q)) \le \lfloor q \rfloor && f_\blacklozenge(f^\square(q)) \text{ is sharp}\\
&\iff \lfloor f^\square(q) \rfloor \le f^\square(\lfloor q \rfloor) && \text{Lemma 5.5.32}\\
&\iff \lfloor f^\square(q) \rfloor \le \lfloor f^\square(\lfloor q \rfloor) \rfloor\,,
\end{aligned}$$

we have $\lfloor f^\square(q) \rfloor \le \lfloor f^\square(\lfloor q \rfloor) \rfloor$. On the other hand, $\lfloor f^\square(\lfloor q \rfloor) \rfloor \le \lfloor f^\square(q) \rfloor$ holds since $f^\square(\lfloor q \rfloor) \le f^\square(q)$. We conclude that $\lfloor f^\square(q) \rfloor = \lfloor f^\square(\lfloor q \rfloor) \rfloor$. ■

**Proposition 5.5.35.** *For a comprehensive effectus $\mathbf{C}$, the mappings $A \mapsto \mathrm{ShPred}(A)$ and*

$$(A \xrightarrow{f} B) \mapsto (\mathrm{ShPred}(B) \xrightarrow{f^\blacksquare} \mathrm{ShPred}(A))$$

*define a functor $\mathrm{ShPred}_\blacksquare\colon \mathbf{C}^{\mathrm{op}} \to \mathbf{Poset}$.*

*Proof.* It preserves identities: for each sharp predicate $\mathfrak{p}$,

$$\mathrm{id}^\blacksquare(\mathfrak{p}) = \lfloor \mathrm{id}^\square(\mathfrak{p}) \rfloor = \lfloor \mathfrak{p} \rfloor = \mathfrak{p} = \mathrm{id}(\mathfrak{p})\,.$$

We now check that it preserves composition: for each $f\colon A \to B$ and $g\colon B \to C$, for each $\mathfrak{p} \in \mathrm{ShPred}(C)$,

$$\begin{aligned}
(g \circ f)^\blacksquare(\mathfrak{p}) &= \lfloor (g \circ f)^\square(\mathfrak{p}) \rfloor\\
&= \lfloor (f^\square(g^\square(\mathfrak{p})) \rfloor\\
&= \lfloor (f^\square(\lfloor g^\square(\mathfrak{p}) \rfloor) \rfloor && \text{by Lemma 5.5.34}\\
&= f^\blacksquare(g^\blacksquare(\mathfrak{p}))\,.
\end{aligned}$$

Hence $(g \circ f)^\blacksquare = f^\blacksquare \circ g^\blacksquare$. ■

We can thus obtain a fibration by the Grothendieck construction.

**Definition 5.5.36.** Let $\mathrm{ShPred}_\blacksquare(\mathbf{C})$ denote the category obtained by applying the Grothendieck construction to $\mathrm{ShPred}_\blacksquare\colon \mathbf{C}^{\mathrm{op}} \to \mathbf{Poset}$. Explicitly, the objects of $\mathrm{ShPred}_\blacksquare(\mathbf{C})$ are pairs $(A, u)$ of $A \in \mathbf{C}$ and $u \in \mathrm{ShPred}(A)$. The morphisms $(A, u) \to (B, v)$ are morphisms $f\colon A \to B$ in $\mathbf{C}$ satisfying $u \le f^\blacksquare(v)$.

We note that $\mathrm{ShPred}_{\blacksquare}(\mathbf{C})$ also arises as a subcategory of $\mathrm{Pred}_{\square}(\mathbf{C})$.

**Proposition 5.5.37.** *The category* $\mathrm{ShPred}_{\blacksquare}(\mathbf{C})$ *equals the full subcategory of* $\mathrm{Pred}_{\square}(\mathbf{C})$ *consisting of pairs* $(A, p)$ *such that* $p$ *is sharp. The inclusion commutes with the forgetful functors, i.e. the following diagram commutes.*

$$\begin{array}{ccc} \mathrm{ShPred}_{\blacksquare}(\mathbf{C}) & \xhookrightarrow{\ \mathit{full}\ } & \mathrm{Pred}_{\square}(\mathbf{C}) \\ & \searrow \quad \swarrow & \\ & \mathbf{C} & \end{array}$$

*Proof.* For objects the claim is obvious. Let $(A, u)$ and $(B, v)$ be objects in $\mathrm{ShPred}_{\blacksquare}(\mathbf{C})$. Then $u \le f^{\blacksquare}(v) \iff u \le f^{\square}(v)$ for each $f\colon A \to B$ in $\mathbf{C}$. Therefore $f$ is a morphism $(A, u) \to (A, v)$ in $\mathrm{ShPred}_{\blacksquare}(\mathbf{C})$ if and only if $f$ is a morphism $(A, u) \to (B, v)$ in $\mathrm{Pred}_{\square}(\mathbf{C})$. ∎

Note, however, that the inclusion need not be a morphism of fibrations, since it need not preserve the cartesian liftings — usually $f^{\blacksquare}(v) \neq f^{\square}(v)$.

There is a convenient result that characterizes bifibrations.

**Proposition 5.5.38** ([133, Lemma 9.1.2])**.** *A fibration* $\varphi\colon \mathbf{E} \to \mathbf{C}$ *is a bifibration if and only if for each* $f\colon A \to B$ *in* $\mathbf{C}$*, the reindexing* $f^*\colon \mathbf{E}_B \to \mathbf{E}_A$ *has a left adjoint.* ∎

As an easy corollary we obtain:

**Corollary 5.5.39.** *The fibration* $\mathrm{ShPred}_{\blacksquare}(\mathbf{C}) \to \mathbf{C}$ *is a bifibration.*

*Proof.* By Proposition 5.5.33, every reindexing map $f^{\blacksquare}\colon \mathrm{ShPred}(B) \to \mathrm{ShPred}(A)$ has a left adjoint $f_{\blacklozenge}$. ∎

Concretely, the opcartesian lifting of $f\colon A \to B$ in $\mathbf{C}$ to $(A, \mathfrak{p})$ in $\mathrm{ShPred}_{\blacksquare}(\mathbf{C})$ is given by $(A, \mathfrak{p}) \to (B, f_{\blacklozenge}(\mathfrak{p}))$.

**Remark 5.5.40.** The indexed poset $\mathrm{ShPred}_{\blacksquare}\colon \mathbf{C}^{\mathrm{op}} \to \mathbf{Poset}$ sends objects $A \in \mathbf{C}$ to orthomodular lattices $\mathrm{ShPred}(A)$. However the reindexing maps $f^{\blacksquare}\colon \mathrm{ShPred}(B) \to \mathrm{ShPred}(A)$ are in general not homomorphisms of orthomodular lattices. In particular, it need not preserve orthocomplements $\mathfrak{p}^{\perp}$, even if $f\colon A \to B$ is a total morphism. Indeed, assuming that $f$ is total we have $f^{\blacksquare}(\mathfrak{p}^{\perp}) = \lfloor f^*(\mathfrak{p})^{\perp} \rfloor$ and $f^{\blacksquare}(\mathfrak{p})^{\perp} = \lceil f^*(p)^{\perp} \rceil$. Thus $f^{\blacksquare}(\mathfrak{p}^{\perp}) = f^{\blacksquare}(\mathfrak{p})^{\perp}$ implies that $f^*(\mathfrak{p})$ is sharp — which is not the case in general. A special class of morphisms that preserves sharpness will be studied in the next section.

We can, nevertheless, view the mappings $A \mapsto \mathrm{ShPred}(A)$ as a functor $\mathbf{C}^{\mathrm{op}} \to \mathbf{OMLGal}$. Here **OMLGal** is the category of orthomodular lattices and *antitone Galois connections* between them. Explicitly, a morphism $X \to Y$ in **OMLGal** is a pair $(f_*, f^*)$ of monotones maps $f_*\colon X^{\mathrm{op}} \to Y$ and $f^*\colon Y \to X^{\mathrm{op}}$ in adjunction $f^* \dashv f_*$. The category **OMLGal** was first studied by Crown [60], and later by Jacobs [136] more systematically in terms of *dagger kernel categories* [123].

For a comprehensive effectus $\mathbf{C}$, the functor $F\colon \mathbf{C}^{\mathrm{op}} \to \mathbf{OMLGal}$ is defined as follows. For objects $A \in \mathbf{C}$ we define $FA = \mathrm{ShPred}(A)$. For a morphism $f\colon A \to B$ in $\mathbf{C}$, we define $Ff = ((Ff)_*, (Ff)^*)\colon \mathrm{ShPred}(B) \to \mathrm{ShPred}(A)$ in **OMLGal** by $(Ff)_*(\mathfrak{q}) = f_\blacklozenge(\mathfrak{q})^\perp$ and $(Ff)^*(\mathfrak{p}) = f^\blacksquare(\mathfrak{p}^\perp)$. The well-definedness of the functor follows easily from Propositions 5.5.33 and 5.5.35.

### 5.5.6 Sharp morphisms

Continuing in the setting of a comprehensive effectus $\mathbf{C}$, here we study morphisms $f\colon A \to B$ that are compatible with sharp predicates. There are several equivalent ways to express such compatibility.

**Lemma 5.5.41.** *For a morphism $f\colon A \to B$, the following are equivalent.*

(i) $f^\square\colon \mathrm{Pred}(B) \to \mathrm{Pred}(A)$ *preserves sharp predicates:* $f^\square(\mathfrak{q}) \in \mathrm{ShPred}(A)$ *for all* $\mathfrak{q} \in \mathrm{ShPred}(B)$.

(ii) $f^*\colon \mathrm{Pred}(B) \to \mathrm{Pred}(A)$ *preserves sharp predicates:* $f^*(\mathfrak{q}) \in \mathrm{ShPred}(A)$ *for all* $\mathfrak{q} \in \mathrm{ShPred}(B)$.

(iii) $f^\square(\mathfrak{q}) = f^\blacksquare(\mathfrak{q})$ $(\equiv \lfloor f^\square(\mathfrak{q})\rfloor)$ *for all* $\mathfrak{q} \in \mathrm{ShPred}(B)$.

(iv) $f^\square(\lfloor q\rfloor) = \lfloor f^\square(q)\rfloor$ *for all* $q \in \mathrm{Pred}(B)$.

*Proof.* It is obvious that (i) $\Longleftrightarrow$ (iii) and (iv) $\Longrightarrow$ (i) hold. Equivalence (i) $\Longleftrightarrow$ (ii) follows easily via $f^\square(\mathfrak{q}) = f^*(\mathfrak{q}^\perp)^\perp$. Finally, (iii) $\Longrightarrow$ (iv) follows by Lemma 5.5.34 as $f^\square(\lfloor q\rfloor) = \lfloor f^\square(\lfloor q\rfloor)\rfloor = \lfloor f^\square(q)\rfloor$. ■

**Definition 5.5.42.** A morphism $f\colon A \to B$ is said to be **sharp** if it satisfies one, and hence all, of the equivalent conditions of Lemma 5.5.41. It is clear that sharp morphisms (with all the objects) form a subcategory of $\mathbf{C}$. The (wide) subcategory is denoted by $\mathrm{Sharp}(\mathbf{C}) \subseteq \mathbf{C}$. We also write $\mathrm{Sharp}(\mathrm{Tot}(\mathbf{C})) \subseteq \mathrm{Tot}(\mathbf{C})$ for the wide subcategory of sharp total morphisms.

We first address the issue mentioned in Remark 5.5.40.

**Proposition 5.5.43.** *Let $f\colon A \to B$ be a sharp total morphism. Then reindexing $f^\blacksquare\colon \mathrm{ShPred}(B) \to \mathrm{ShPred}(A)$ is a homomorphism of orthomodular lattices. Therefore, the restriction of the functor $\mathrm{ShPred}_\blacksquare\colon \mathbf{C}^{\mathrm{op}} \to \mathbf{Poset}$ to sharp total morphisms yields an 'indexed orthomodular lattice', i.e. a functor $\mathrm{Sharp}(\mathrm{Tot}(\mathbf{C}))^{\mathrm{op}} \to \mathbf{OML}$.*

*Proof.* If $f\colon A \to B$ is sharp and total, then we have $f^\blacksquare(\mathfrak{q}) = f^\square(\mathfrak{q}) = f^*(\mathfrak{q})$. The map $f^*$ is a homomorphism of effect algebras, and thus preserves $\mathbb{0}$, $\mathbb{1}$, and $(-)^\perp$. Since $f^\blacksquare\colon \mathrm{ShPred}(B) \to \mathrm{ShPred}(A)$ has a left adjoint $f_\blacklozenge$ (Proposition 5.5.33), $f^\blacksquare$ preserves (arbitrary) meets $\wedge$. Then $f^\blacksquare$ also preserves joins $\vee$, because $\mathfrak{p} \vee \mathfrak{q} = (\mathfrak{p}^\perp \wedge \mathfrak{q}^\perp)^\perp$. Therefore $f^\blacksquare$ is a homomorphism of orthomodular lattices. ■

Next we list basic properties of sharp morphisms.

**Lemma 5.5.44.** *Let $\mathbf{C}$ be a comprehensive effectus.*

(i) *All coprojections $\kappa_1\colon A \to A+B$ and $\kappa_2\colon B \to A+B$ are sharp.*

(ii) *All zero morphisms $0_{AB}\colon A \to B$ are sharp.*

(iii) *If* $f\colon A \to C$ *and* $g\colon B \to C$ *are sharp morphisms, then the cotuple* $[f,g]\colon A+B \to C$ *is sharp too. In particular, partial projections* $\rhd_1 = [\mathrm{id}, 0]$ *and* $\rhd_2 = [0, \mathrm{id}]$ *are sharp.*

(iv) *If* $h\colon A \to B$ *and* $k\colon A \to C$ *are sharp morphisms that are compatible, then the tuple* $\langle\!\langle h,k\rangle\!\rangle\colon A \to B+C$ *is sharp too.*

*Proof.*

(i) Let $\mathfrak{p} \in \mathrm{ShPred}(A+B)$. By Proposition 5.5.16(iv), $\mathfrak{p} = [\mathfrak{p}_1, \mathfrak{q}_2]$ for $\mathfrak{p}_1 \in \mathrm{ShPred}(A)$ and $\mathfrak{p}_2 \in \mathrm{ShPred}(B)$. Therefore $\kappa_1^*(\mathfrak{p}) = \mathfrak{p}_1$ and $\kappa_2^*(\mathfrak{p}) = \mathfrak{p}_2$ are sharp predicates.

(ii) For any $\mathfrak{p} \in \mathrm{ShPred}(B)$ we have $0_{AB}^{\square}(\mathfrak{p}) = \mathbb{1}_A \in \mathrm{ShPred}(A)$.

(iii) For each $\mathfrak{p} \in \mathrm{ShPred}(C)$, we have $[f,g]^{\square}(\mathfrak{p}) = [f^{\square}(\mathfrak{p}), g^{\square}(\mathfrak{p})]$. Therefore by Proposition 5.5.16(iv), if $f$ and $g$ are sharp morphisms, so is $[f,g]$.

(iv) Let $\mathfrak{p} \in \mathrm{ShPred}(B+C)$, i.e. $\mathfrak{p} = [\mathfrak{p}_1, \mathfrak{p}_2]$ for $\mathfrak{p}_1 \in \mathrm{ShPred}(B)$ and $\mathfrak{p}_2 \in \mathrm{ShPred}(C)$, by Proposition 5.5.16(iv). Then

$$\langle\!\langle h,k\rangle\!\rangle^*([\mathfrak{p}_1, \mathfrak{p}_2]) = [\mathfrak{p}_1, \mathfrak{p}_2] \circ \langle\!\langle h,k\rangle\!\rangle = \mathfrak{p}_1 \circ h \varovee \mathfrak{p}_2 \circ k\,,$$

which is sharp since so are both $\mathfrak{p}_1 \circ h$ and $\mathfrak{p}_2 \circ k$. ∎

The lemma above shows that the subcategory $\mathrm{Sharp}(\mathbf{C}) \subseteq \mathbf{C}$ is closed under most of the constructions in an effectus. As a result we obtain:

**Theorem 5.5.45.** *Let* $\mathbf{C}$ *be a comprehensive effectus such that all truth maps* $\mathbb{1}_A\colon A \to I$ *are sharp morphisms. Then the subcategory* $\mathrm{Sharp}(\mathbf{C})$ *is an effectus.*

*Proof.* We apply Proposition 3.8.7. By Lemma 5.5.44, it is clear that all the conditions in Proposition 3.8.7 but (vii) hold. Thus we will prove that if $p\colon A \to I$ is a sharp morphism, then so is $p^{\perp}\colon A \to I$. Let $p\colon A \to I$ be a sharp morphism and let $\mathfrak{s} \in \mathrm{ShPred}(I)$. Then

$$(p^{\perp})^*(\mathfrak{s}) = \mathfrak{s} \circ p^{\perp} = \mathfrak{s} \circ \mathbb{1} \ominus \mathfrak{s} \circ p\,.$$

This is a sharp predicate since both $\mathfrak{s} \circ \mathbb{1}$ and $\mathfrak{s} \circ p$ are sharp, and sharp predicates are closed under $\ominus$. ∎

The theorem above is rather unsatisfactory, not only because we have an additional assumption that $\mathbb{1}_A\colon A \to I$ is sharp, but also because predicates in $\mathrm{Sharp}(\mathbf{C})$ do not necessarily coincide with sharp predicates in $\mathbf{C}$. Clearly we have $\mathrm{Sharp}(\mathbf{C})(A,I) \subseteq \mathrm{ShPred}(A)$, since if $p\colon A \to I$ is a sharp morphism, then $p = \mathbb{1}_I \circ p = p^*(\mathbb{1}_I)$ must be a sharp predicate. However, it is not clear whether the converse is the case.

It turns out that such issues can be solved in a strongly comprehensive effectus.

**Lemma 5.5.46.** *For each sharp scalar* $\mathfrak{s} \in \mathrm{ShPred}(I)$ *we have* $\mathfrak{s} \circ \mathfrak{s}^{\perp} = 0 = \mathfrak{s}^{\perp} \circ \mathfrak{s}$, $\mathfrak{s} \circ \mathfrak{s} = \mathfrak{s}$, *and* $\mathfrak{s}^{\perp} \circ \mathfrak{s}^{\perp} = \mathfrak{s}^{\perp}$.

*Proof.* We have $\mathfrak{s} \circ \mathfrak{s}^{\perp} \le \mathfrak{s} \circ 1 \le \mathfrak{s}$ and similarly $\mathfrak{s} \circ \mathfrak{s}^{\perp} \le \mathfrak{s}^{\perp}$. Thus by ortho-sharpness $\mathfrak{s} \circ \mathfrak{s}^{\perp} = 0$. Similarly we prove $\mathfrak{s}^{\perp} \circ \mathfrak{s} = 0$. Then $\mathfrak{s} = \mathfrak{s} \circ (\mathfrak{s} \varovee \mathfrak{s}^{\perp}) = \mathfrak{s} \circ \mathfrak{s} \varovee 0 = \mathfrak{s} \circ \mathfrak{s}$, and similarly $\mathfrak{s}^{\perp} \circ \mathfrak{s}^{\perp} = \mathfrak{s}^{\perp}$. ∎

**Lemma 5.5.47.** *For each sharp predicate* $\mathfrak{p} \in \mathrm{ShPred}(A)$ *and sharp scalar* $\mathfrak{s} \in \mathrm{ShPred}(I)$*, the predicate* $\mathfrak{s} \circ \mathfrak{p}$ *is ortho-sharp.*

*Proof.* Let $q \leq \mathfrak{s} \circ \mathfrak{p}$ and $q \leq (\mathfrak{s} \circ \mathfrak{p})^{\perp}$. Since $q \leq \mathfrak{s} \circ \mathfrak{p} \leq \mathfrak{p}$ and $\mathfrak{p}$ is ortho-sharp, to prove $q = \mathbb{0}$ it suffices to show $q \leq \mathfrak{p}^{\perp}$. By $\mathfrak{s}^{\perp} \circ q \leq \mathfrak{s}^{\perp} \circ \mathfrak{s} \circ \mathfrak{p} = 0$, i.e. $\mathfrak{s}^{\perp} \circ q = 0$, we have

$$q = \mathfrak{s} \circ q \ovee \mathfrak{s}^{\perp} \circ q = \mathfrak{s} \circ q\,.$$

We also have $\mathfrak{s} \circ (\mathfrak{s} \circ \mathfrak{p})^{\perp} = \mathfrak{s} \circ \mathfrak{p}^{\perp}$ because $(\mathfrak{s} \circ \mathfrak{p})^{\perp} = \mathfrak{s} \circ \mathfrak{p}^{\perp} \ovee \mathfrak{s}^{\perp} \circ \mathbb{1}$. Then

$$q = \mathfrak{s} \circ q \leq \mathfrak{s} \circ (\mathfrak{s} \circ \mathfrak{p})^{\perp} = \mathfrak{s} \circ \mathfrak{p}^{\perp} \leq \mathfrak{p}^{\perp}\,.$$ ■

**Theorem 5.5.48.** *In a strongly comprehensive effectus, a predicate* $p\colon A \to I$ *is a sharp morphism if and only if* $p$ *is a sharp predicate.*

*Proof.* If $p$ is a sharp morphism, then $p = \mathbb{1}_I \circ p = p^*(\mathbb{1}_I)$ is a sharp predicate. Suppose that $p$ is a sharp predicate. Then for each $\mathfrak{s} \in \mathrm{ShPred}(I)$, $p^*(\mathfrak{s}) = \mathfrak{s} \circ p$ is ortho-sharp by Lemma 5.5.47. In a strongly comprehensive effectus, $p^*(\mathfrak{s})$ is a sharp predicate. Therefore $p$ is a sharp morphism. ■

**Corollary 5.5.49.** *Let* $\mathbf{C}$ *be a strongly comprehensive effectus. Then the subcategory* $\mathrm{Sharp}(\mathbf{C})$ *is an effectus. Moreover we have:*

(i) *Predicates in* $\mathrm{Sharp}(\mathbf{C})$ *are precisely sharp predicates in* $\mathbf{C}$.

(ii) *For each object* $A$*, predicates on* $A$ *in* $\mathrm{Sharp}(\mathbf{C})$ *form an orthomodular lattice.*

(iii) *For each total morphism* $f\colon A \to B$ *in* $\mathrm{Sharp}(\mathbf{C})$*, reindexing* $f^*\colon \mathrm{Pred}(B) \to \mathrm{Pred}(A)$ *is a homomorphism of orthomodular lattice. Therefore the predicate functor restricted on total morphisms is:*

$$\mathrm{Pred}\colon \mathrm{Tot}(\mathrm{Sharp}(\mathbf{C}))^{\mathrm{op}} \to \mathbf{OML}\,,$$

*where* $\mathbf{OML}$ *is the category of orthomodular lattices.*

*Proof.* By Theorems 5.5.45 and 5.5.48 and Proposition 5.5.43. ■

We conclude the section with examples of sharp morphisms.

**Example 5.5.50.**

(i) In the effectus $\mathbf{Pfn}$, all morphisms are sharp since all predicates are sharp. Therefore $\mathrm{Sharp}(\mathbf{Pfn}) = \mathbf{Pfn}$.

(ii) In the effectus $\mathcal{K}\ell(\mathcal{D}_{\leq})$, a morphism $f\colon X \to \mathcal{D}_{\leq}(Y)$ is sharp if and only if $f(x)(y) \in \{0,1\}$ for all $x \in X$ and $y \in Y$. These sharp morphisms can be identified with with partial functions $\tilde{f}\colon X \rightharpoonup Y$ via $\tilde{f}(x) = y$ iff $f(x)(y) = 1$. Thus $\mathrm{Sharp}(\mathcal{K}\ell(\mathcal{D}_{\leq})) \cong \mathbf{Pfn}$.

We note that in $\mathcal{K}\ell(\mathcal{G}_{\leq})$ — for general measure-theoretic probability — sharp morphisms $f\colon X \to \mathcal{G}_{\leq}(Y)$ can be characterized by condition $f(x)(V) \in \{0,1\}$ for all $x \in X$ and $V \in \Sigma_Y$. But these morphisms cannot be identified with partial measurable functions $f\colon X \rightharpoonup Y$, since there can exist $\{0,1\}$-valued measures that are neither zero or Dirac measures.

(iii) Let $f\colon \mathscr{A} \to \mathscr{B}$ be a morphism in the effectus $\mathbf{Wstar}_{\leq}^{\mathrm{op}}$ of $W^*$-algebras, i.e. a normal subuntial CP map $f\colon \mathscr{B} \to \mathscr{A}$. Then $f$ is sharp if and only if $f(\mathfrak{q})$ is a projection for each projection $\mathfrak{q}$ in $\mathscr{B}$. If $f$ is multiplicative — thus a $*$-homomorphism — then clearly $f$ is sharp, as $f(\mathfrak{q}) \cdot f(\mathfrak{q}) = f(\mathfrak{q} \cdot \mathfrak{q}) = f(\mathfrak{q})$. In fact, the converse is true: if $f$ is sharp, then $f$ is multiplicative [254, Proposition 47][1]. Therefore the subcategory $\mathrm{Sharp}(\mathbf{Wstar}_{\leq}^{\mathrm{op}})$ precisely consists of normal $*$-homomorphisms. (Note that any $*$-homomorphism is subunital.)

## 5.6 Comparison to Janelidze and Weighill's theory of forms

We have described comprehension and quotients in an effectus $\mathbf{C}$ as the following chain of adjunctions.

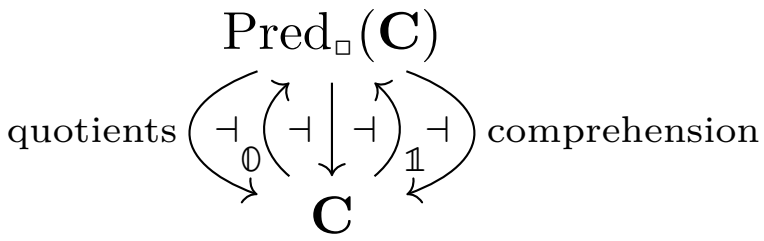

These comprehension and quotients were mainly inspired by the fibrational perspectives in categorical logic [133]. Here we mention another related work where similar chains of adjunctions appeared, namely a recent categorical study of non-abelian algebras by (mainly) Janelidze and Weighill [97, 156–159, 252]. One can indeed find chains of adjunction depicted in [252, § 3.5] and in [97, § 5]. In fact, the (bi)fibration of subgroups over groups $\mathrm{Sub}(\mathbf{Grp}) \to \mathbf{Grp}$, presented in Examples 5.2.2, 5.2.6 and 5.4.1, is a prototypical example in Janelidze and Weighill's theory. We briefly describe basic notions and terminology in their theory.

(1) Their theory is based on a notion of *forms*, which are functors $\varphi\colon \mathbf{E} \to \mathbf{C}$ that are both faithful and amnestic. Recall that a functor $\varphi\colon \mathbf{E} \to \mathbf{C}$ is *amnestic* if every isomorphism $f$ in $\mathbf{E}$ is an identity whenever $\varphi f$ is an identity in $\mathbf{C}$. If $\varphi\colon \mathbf{E} \to \mathbf{C}$ is a form, every fibre $\mathbf{E}_A$ is a poset. Conversely, it is not hard to see that any poset (op)fibration $\varphi\colon \mathbf{E} \to \mathbf{C}$ is a form.

(2) Let $\varphi\colon \mathbf{E} \to \mathbf{C}$ be a form and $X \in \mathbf{E}$ be an object. For a morphism $f\colon A \to \varphi X$ in $\mathbf{C}$ with codomain $\varphi X$, consider the following property:

(LU) For each $Y \in \mathbf{E}_A$, there exists a morphism $f'\colon Y \to X$ in $\mathbf{E}$ such that $\varphi f' = f$.

A *left universalizer* of an object $X \in \mathbf{E}$ is a universal (terminal) one among morphisms $f\colon A \to FX$ satisfying the property (LU). A *right universalizer* is defined in the dual manner: it is a left universalizer with respect to $\varphi^{\mathrm{op}}\colon \mathbf{E}^{\mathrm{op}} \to \mathbf{C}^{\mathrm{op}}$.

(3) A form $\varphi\colon \mathbf{E} \to \mathbf{C}$ is *locally bounded* if every fibre $\mathbf{E}_A$ is bounded as poset, i.e. it has a least element $\mathbb{0}$ and a greatest element $\mathbb{1}$.

[1] The reference proves the result for unital maps, from which one can obtain a similar result for subunital maps (e.g. using Lemma 5.5.44).

(4) A form $\varphi\colon \mathbf{E} \to \mathbf{C}$ is *bounded* if it is locally bounded and for each $f\colon A \to B$ in $\mathbf{C}$, there exist a cartesian lifting of $f$ to $\mathbb{0} \in \mathbf{E}_B$ and an opcartesian lifting of $f$ to $\mathbb{1} \in \mathbf{E}_A$ — which are denoted by $f^*(\mathbb{0}) \to \mathbb{0}$ and $\mathbb{1} \to f_!(\mathbb{1})$, respectively.

(5) For a morphism $f\colon A \to B$ in $\mathbf{C}$, the object $f^*(\mathbb{0}) \in \mathbf{E}_A$ is called the *right norm* of $f$, and $f_!(\mathbb{1}) \in \mathbf{E}_B$ is the *left norm* of $f$.

(6) Objects in $\mathbf{E}$ of the form $f^*(\mathbb{0})$ for some $f$ are said to be *normal* (or *right normal*), and objects of the form $f_!(\mathbb{1})$ are *conormal* (or *left normal*).

(7) A morphism $f\colon A \to B$ in $\mathbf{C}$ is called *thin* if $f^*(\mathbb{0}) = \mathbb{0}$; and *thick* if $f_!(\mathbb{1}) = \mathbb{1}$.

(8) (From [252, § 3.3]) Let $\varphi\colon \mathbf{E} \to \mathbf{C}$ be a bounded form. Then the mapping $A \in \mathbf{C} \mapsto \mathbb{1} \in \mathbf{E}_A$ canonically extends to a functor $\mathbb{1}\colon \mathbf{C} \to \mathbf{E}$ that is a right adjoint to $\varphi$. Moreover, a left universalizer of $X$ is precisely universal morphism from $\mathbb{1}\colon \mathbf{C} \to \mathbf{E}$ to $X$. Therefore $\varphi$ has all left universalizers if and only if $\mathbb{1}\colon \mathbf{C} \to \mathbf{E}$ has a right adjoint. The dual statement holds for $\mathbb{0}$ and right universalizers.

(9) To summarize, if $\varphi\colon \mathbf{E} \to \mathbf{C}$ is a bounded form that has both left and right universalizers, then we have the following chain of adjunctions:

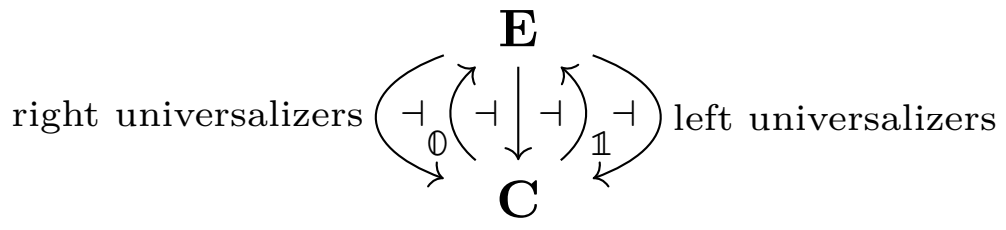

Now we consider an effectus $\mathbf{C}$ and the induced fibration $\mathrm{Pred}_\Box(\mathbf{C}) \to \mathbf{C}$. Then $\mathrm{Pred}_\Box(\mathbf{C}) \to \mathbf{C}$ is a locally bounded form. We can thus describe some basic notions of effectus theory in the terminology of the theory of forms:

(1) Let $f\colon A \to B$ be a morphism in the effectus $\mathbf{C}$. A right norm of $f$ is the kernel $\ker(f) \in \mathrm{Pred}(A)$, and a left norm of $f$ is an image $\mathrm{im}(f) \in \mathrm{Pred}(A)$. Therefore $\mathrm{Pred}_\Box(\mathbf{C}) \to \mathbf{C}$ is a bounded form if and only if $\mathbf{C}$ has images.

(2) A left universalizer of $p \in \mathrm{Pred}(A)$ is precisely a comprehension $\pi_p\colon \{A \,|\, p\} \to A$. A right universalizer of $p$ is precisely a quotient $\xi_p\colon A \to A/p$.

(3) Thin morphisms are precisely total ones, and thick morphisms are precisely faithful ones.

(4) Conormal objects in $\mathrm{Pred}_\Box(\mathbf{C})$ are precisely sharp predicates, see Corollary 5.5.14.

We note that the form $\mathrm{Pred}_\Box(\mathbf{C}) \to \mathbf{C}$ over an effectus $\mathbf{C}$ is rather ill-behaved from the viewpoint of Janelidze and Weighill's theory. As their motivating example is the form $\mathrm{Sub}(\mathbf{Grp}) \to \mathbf{Grp}$ of groups, their theory is often concerned with stronger properties than effectus theory. For example, in their theory the form $\varphi\colon \mathbf{E} \to \mathbf{C}$ is often assumed to be a bifibration, but the fibration $\mathrm{Pred}_\Box(\mathbf{C}) \to \mathbf{C}$ over an effectus $\mathbf{C}$ is in general not an bifibration — in particular, it is not the case for our leading examples $\mathcal{K}\ell(\mathcal{D}_\leq)$ and $\mathbf{Wstar}_\leq^{\mathrm{op}}$.

Another property that usually fails for the form $\mathrm{Pred}_\Box(\mathbf{C}) \to \mathbf{C}$ over an effectus $\mathbf{C}$ is *exactness*, or the 'first isomorphism theorem'. Note that in an effectus we have the following factorization.

**Proposition 5.6.1.** *Let* $\mathbf{C}$ *be an effectus with images, comprehension and quotients. Then any morphism* $f\colon A \to B$ *in* $\mathbf{C}$ *can be factorized as follows, for some unique map* $\theta_f$*.*

$$\begin{array}{ccccc} A & & \xrightarrow{\quad f \quad} & & B \\ & {\scriptstyle \xi_{\ker(f)}} \searrow & & \nearrow {\scriptstyle \pi_{\mathrm{im}(f)}} & \\ & A/\ker(f) & \xrightarrow{\ \theta_f\ } & \{A \,|\, \mathrm{im}(f)\} & \end{array}$$

*Proof.* By the universality of quotient $\xi_{\ker(f)}\colon A \to A/\ker(f)$, there is $\overline{f}\colon A/\ker(f) \to B$ with $f = \overline{f} \circ \xi_{\ker(f)}$ as in:

$$\begin{array}{ccc} A & & \\ {\scriptstyle \xi_{\ker(f)}} \downarrow & \searrow^{f} & \\ A/\ker(f) & \underset{\overline{f}}{\dashrightarrow} & B \end{array}$$

Now

$$\mathrm{im}(f)^{\perp} \circ \overline{f} \circ \xi_{\ker(f)} = \mathrm{im}(f)^{\perp} \circ f = \mathbb{0} = \mathbb{0} \circ \xi_{\ker(f)}$$

and because $\xi_{\ker(f)}$ is an epi, we have $\mathrm{im}(f)^{\perp} \circ \overline{f} = \mathbb{0}$, i.e. $\overline{f}^{\Box}(\mathrm{im}(f)) = \mathbb{1}$. Therefore $\overline{f}$ factors through comprehension $\{B \,|\, \mathrm{im}(f)\}$ as in:

$$\begin{array}{ccc} A/\ker(f) & \xrightarrow{\ \overline{f}\ } & B \\ {\scriptstyle \theta_f} \downarrow & \nearrow_{\pi_{\mathrm{im}(f)}} & \\ \{B \,|\, \mathrm{im}(f)\} & & \end{array}$$

We thus obtain a desired factorization $f = \pi_{\mathrm{im}(f)} \circ \theta_f \circ \xi_{\ker(f)}$. The map $\theta_f$ is unique since $\pi_{\mathrm{im}(f)}$ is monic and $\xi_{\ker(f)}$ is epic. ■

In the terminology of Janelidze and Weighill, the proposition shows that the form $\mathrm{Pred}_{\Box}(\mathbf{C}) \to \mathbf{C}$ is *orthogonal*, see [158, Definition 3.2 and Theorem 3.4]. An orthogonal form is *exact* (see [158, Definition 3.2 and Theorem 3.6]) if $\theta_f$ defined as above is an isomorphism for every morphism $f\colon A \to B$, i.e. if the 'first isomorphism theorem' holds.

The form $\mathrm{Pred}_{\Box}(\mathbf{C}) \to \mathbf{C}$ over an effectus is rarely exact: indeed, the exactness fails for all our leading examples $\mathbf{Pfn}$, $\mathcal{K}\ell(\mathcal{D}_{\leq})$, and $\mathbf{Wstar}_{\leq}^{\mathrm{op}}$. In Bas Westerbaan's thesis [256], a morphism $f\colon A \to B$ in an effectus such that $\theta_f$ is an isomorphism is said to be *pure*, and such morphisms are studied in detail. Pure morphisms are a very special class of morphisms: for example, a morphism $f\colon \mathcal{B}(\mathscr{H}) \to \mathcal{B}(\mathscr{K})$ in $\mathbf{Wstar}_{\leq}^{\mathrm{op}}$ is pure if and only if it is given by $f(A) = T^*AT$ for some bounded operator $T\colon \mathscr{H} \to \mathscr{K}$.

Finally, let us mention Grandis' work on categorical homological algebras [98, 99], which generalizes abelian categories. It is closely related to Janelidze and Weighill's theory. A comparison of effectus theory to Grandis' work can be found in Bas Westerbaan's thesis [256].

# Chapter 6

# Measurements in Effectuses

The notion of *measurement* is at the heart of quantum theory [27, 28, 120, 210]. In this chapter we study measurements in terms of an effectus. Recall that for a state $\omega\colon I \to A$ and a predicate $p\colon A \to I$ in an effectus, one has the validity $(\omega \vDash p) = p \circ \omega$, which is understood as an *abstract Born rule*: $\omega \vDash p$ is the probability of predicate $p$ in state $\omega$. This captures an aspect of quantum measurements, but is not sufficient because:

- it concerns only 'yes-no' measurement; and
- it gives no information about the state of the system after a measurement.

To discuss measurements in an effectus in a sufficiently general way, we use a notion of 'test', which comes from the operational probabilistic framework of Chiribella et al. [33–35, 61]. The close connection of effectuses to the operational probabilistic framework was discovered and studied by Tull [248–250]. We can think of an effectus as a certain type of an operational probabilistic theory, and hence can use the language of the operational probabilistic framework for an effectus.

Let $X$ be a finite set, which represents the set of outcomes of a measurement. In an effectus, a *test* with outcome set $X$ from object $A$ to object $B$ can be defined as a total morphism of the form:

$$f\colon A \longrightarrow X \cdot B\,,$$

where $X \cdot B = \coprod_{x \in X} B$ is the copower of $B$ by $X$. It describes a measuring process on a system of type $A$ that yields an outcome $x \in X$ and leaves the system in type $B$ when the process ends.

In particular, we are interested in tests of the form $f\colon A \to X \cdot A$, called *instruments*. An instrument represents a measurement that is 'non-destructive' in the sense that after the measurement, the system is still present in the same type (but usually in a different state; cf. side-effect-freeness). We mainly discuss three properties of instruments: repeatability, side-effect-freeness, and Booleanness. Here Booleanness is a property of being both repeatable and side-effect-free. In our abstract setting we discuss repeatable instruments, relating them to sharp predicates. Side-effect-freeness is shown to be related to compatibility/commutativity of observables. As the name suggests, Boolean instruments are related to Boolean algebras.

We say that an effectus is *Boolean* if every observable can be measured by a Boolean instrument. We will give a characterization of Boolean effectuses, under some assumptions (comprehension or quotients), by *extensivity* of coproducts — a well-established notion of 'nice' coproducts. This can also be seen as a new characterization of extensive categories (with a final object) as Boolean effectuses.

## 6.1 The operational probabilistic framework

The operational probabilistic framework was introduced by Chiribella et al. [33–35, 61] to describe experiments on (physical) systems and predictions about them (i.e. probabilities of outcomes from experiments). The framework is based on the notion of *operational probabilistic theories* (*OPTs*), categories equipped with certain structures. Tull first pointed out the close connection of effectuses to the operational probabilistic framework, and he proved that effectuses may be identified with a certain kind of OPTs [248].

In this thesis, we use the operational probabilistic framework as a language for the theory of measurements in effectuses. For this purpose, we will introduce a suitably modified version of OPTs (which we call *abstract operational probabilistic theories*) based on Tull's work [248]. For example, the original definition of OPTs uses a monoidal category, but our modified version will be just a category, focusing on sequential composition. For other differences, see Remark 6.1.8.

**Definition 6.1.1.** A (tensor-free) **operational structure** is a category $\mathbf{C}$ with a specified 'unit' object $I \in \mathbf{C}$ and for each $A, B \in \mathbf{C}$ and each finite set $X$, a specified set $\mathrm{Test}_X(A,B)$ of families $(f_x : A \to B)_{x\in X}$ of morphisms, i.e. $\mathrm{Test}_X(A,B) \subseteq \mathbf{C}(A,B)^X$, satisfying the following conditions.

(i) If $(f_x)_{x\in X} \in \mathrm{Test}_X(A,B)$ and $(g_y)_{y\in Y} \in \mathrm{Test}_Y(B,C)$, then $(g_y \circ f_x)_{(x,y)\in X\times Y} \in \mathrm{Test}_{X\times Y}(A,C)$.

(ii) $(\mathrm{id}_A)_{*\in 1} \in \mathrm{Test}_1(A,A)$ for any singleton $1 = \{*\}$ and $A \in \mathbf{C}$.

(iii) If $(f_x)_{x\in X} \in \mathrm{Test}_X(A,B)$ and $\sigma : Y \overset{\cong}{\to} X$ is a bijection, then $(f_{\sigma(y)})_{y\in Y} \in \mathrm{Test}_Y(A,B)$.

Let $\mathbf{C}$ be an operational structure. The objects in $\mathbf{C}$ are called **types** of systems. The special unit type $I$ represents the trivial system, i.e. 'nothing' or a system with no information. Morphisms in $\mathbf{C}$ are called **transformations** or **events**. Families $(f_x)_{x\in X}$ in $\mathrm{Test}_X(A,B)$ are called **tests** (with **outcome set** $X$), and elements $x \in X$ are called **outcomes**. We interpret a test $(f_x : A \to B)_{x\in X}$ as an operation on a system of type $A$ that turns the system to type $B$ and yields an outcome $x \in X$. An occurrence of outcome $x \in X$ corresponds to that of the event/transformation $f_x : A \to B$, which determines how the system transforms. The condition (i) above asserts that we can (sequentially) compose two tests: for tests $f = (f_x : A \to B)_{x\in X}$ and $g = (g_y : B \to C)_{y\in Y}$, the composed test is defined and denoted by

$$f; g \coloneqq (g_y \circ f_x : A \to C)_{(x,y)\in X\times Y}\,.$$

It has the cartesian product $X \times Y$ as outcome set.

A test of type $(\omega_x : I \to A)_{x\in X}$ is called a **preparation test**: it is a test starting in 'nothing' $I$ and resulting in a system $A$. Dually, a test of type $(p_x : A \to I)_{x\in X}$ is called a **observation test** or **observable**. An observation test turns a system to 'nothing' $I$ and hence discards the system, yielding only an outcome $x \in X$.

We say that a test is **closed** if it is of type $(s_x : I \to I)_{x\in X}$. An **experiment** is a composable sequence of tests $f^{(1)}, f^{(2)}, \dots, f^{(n)}$ such that the composed test $f^{(1)}; f^{(2)}; \cdots; f^{(n)}$ is closed — that is: $f^{(1)}$ is a preparation test and $f^{(n)}$ is an observation test. We call transformations $s\colon I \to I$ **scalars**, and see them as 'abstract probabilities'. Then a closed test $(s_x\colon I \to I)_{x\in X}$ may be viewed as an '(abstract) probability distribution' on $X$. Thus each experiment induces a probability distribution of outcomes — this is the basic intuition in the operational probabilistic framework.

We will make two assumptions on an operational structure: *causality* and *coarse-graining*. Let us introduce some more terminology. Tests $(f\colon A \to B)_{*\in 1}$ that have the singleton outcome set $1 = \{*\}$ are called **deterministic**, in the sense that when one performs such a test, the transformation $f$ always occurs with the unique outcome $* \in 1$. We call a transformation $f\colon A \to B$ a **channel** if the singleton family $(f\colon A \to B)_{*\in 1}$ is a test, and identify channels with deterministic tests. A channel of type $\omega\colon I \to A$ — a deterministic preparation test — is called a **state**. Dually, a channel of type $A \to I$ may be understood as the operation of discarding a system of type $A$ without any observation. The first assumption, called *causality*, is that for each type $A$ there exists a unique such discarding operation.

**Definition 6.1.2.** An operational structure $\mathbf{C}$ satisfies **causality** if for each $A \in \mathbf{C}$, there is a unique channel of type $A \to I$. We denote the unique 'discarding' channel by $\mathbb{1}_A\colon A \to I$.

Later we will explain the causality property as the principle of 'no signalling from the future', see Proposition 6.1.7. See also [50, 55] where causality in (monoidal) categories is more systematically studied.

Next we introduce the *coarse-graining* operation, which yields a 'coarse-grained' test by identifying some outcomes of a test. For example, consider a test $(f_{x_1}, f_{x_2}, f_{x_3})$ with outcomes $\{x_1, x_2, x_3\}$. Suppose that we identify outcomes $x_1$ and $x_2$, say as $x_{12}$. Then we obtain a 'coarse-grained' test $(f_{x_{12}}, f_{x_3})$ with outcomes $\{x_{12}, x_3\}$ together with a 'coarse-grained' transformation $f_{x_{12}} = f_{x_1} \ovee f_{x_2}$. Coarse-graining can be defined axiomatically as the PCM structure compatible with the operational structure.

**Definition 6.1.3.** An operational structure $(\mathbf{C}, I, \mathrm{Test})$ admits **coarse-graining** if

(a) $\mathbf{C}$ is enriched over PCMs;

(b) every test $(f_x\colon A \to B)_{x\in X}$ is a summable family in the hom-PCM $\mathbf{C}(A, B)$;

(c) for each test $(f_x)_{x\in X}$ and each function $\alpha\colon X \to Y$ (where $Y$ is a finite set), the family $(g_y)_{y\in Y}$ defined by $g_y = \bigovee_{x\in\alpha^{-1}(y)} f_x$ is a test.

Now we define the notion of 'abstract' operational probabilistic theory.

**Definition 6.1.4.** An **abstract operational probabilistic theory** (**AOPT** for short) is an operational structure that satisfies causality and admits coarse-graining.

In an AOPT, every test $f = (f_x\colon A \to B)_{x\in X}$ induces an observable on $A$ by composing the discarding channel $\mathbb{1}\colon B \to I$, namely, the observable $f; \mathbb{1} = (\mathbb{1} f_x\colon A \to I)_{x\in X}$. For a test $f = (f_x\colon A \to B)_{x\in X}$ and an observable $p = (p_x\colon A \to I)_{x\in X}$, we say that $f$ is $p$-**compatible** or $f$ **measures** $p$ if $f; \mathbb{1} = p$ holds.

Since the discarding channel on $I$ must be the identity, each observation test $p = (p_x\colon A \to I)_{x\in X}$ is itself a $p$-compatible test. By performing the test $p$ we obtain an outcome $x \in X$ but discard the system at the same time. Such measurements are sometimes called 'destructive' measurements. In contrast, tests of type $(f_x \colon A \to A)_{x \in X}$, may be understood as 'non-destructive' measurements, which keep the system in the same type. We call such a test $(f_x \colon A \to A)_{x \in X}$ an **instrument**.

Although the scalars $\mathbf{C}(I, I)$ in an AOPT need not form an effect monoid, they have a similar probability-like structure. We can multiply scalars $s \cdot t = s \circ t$ via composition and add scalars $s \ovee t$ via the PCM structure. By causality, the unique discarding channel $\mathbb{1}_I \colon I \to I$ is equal to the identity $\mathrm{id}_I \colon I \to I$ and hence is the unit of multiplication. The scalar $\mathbb{1}_I = \mathrm{id}_I$ is seen as 'probability one'.

**Proposition 6.1.5.** *In an AOPT, every observable $(p_x \colon A \to I)_{x \in X}$ satisfies*

$$\bigovee_{x \in X} p_x = \mathbb{1}_A \,.$$

*Proof.* By coarse-graining along the unique function $X \to 1 = \{*\}$, we obtain the deterministic test $(\bigovee_x p_x \colon A \to I)_{* \in 1}$. By causality, $\bigovee_x p_x = \mathbb{1}_A$. ■

**Corollary 6.1.6.** *In an AOPT, every closed test $(s_x \colon I \to I)_{x \in X}$ satisfies $\bigovee_{x \in X} s_x = \mathbb{1}$. Thus every closed test is a 'probability distribution' on $X$.* ■

Let $f^{(1)}, \ldots, f^{(n)}$ be an experiment. Suppose that each test $f^{(j)}$ has outcome set $X^{(j)}$. Then the composite $f^{(1)}; \cdots; f^{(n)}$ is a closed test with outcome set $X^{(1)} \times \cdots \times X^{(n)}$—i.e. a 'joint probability distribution' on $X^{(1)}, \ldots, X^{(n)}$. To reason about such joint probability distributions, it is convenient to employ the common notation from probability theory. We write $\mathbf{o}_j$ for the random variable denoting the outcome from the $j$th test in the experiment. Then we denote the probability (scalar) that outcomes $x^{(1)}, \ldots, x^{(n)}$ occur from the experiment by

$$\mathrm{P}_{f^{(1)}, \ldots, f^{(n)}}(\mathbf{o}_1 = x^{(1)}, \ldots, \mathbf{o}_n = x^{(n)}) = f^{(n)}_{x^{(n)}} \circ \cdots \circ f^{(1)}_{x^{(1)}} \quad \colon I \to I \,.$$

Following the usual convention, when no confusion is likely to arise, we omit random variables $\mathbf{o}_j$ and simply write:

$$\mathrm{P}_{f^{(1)}, \ldots, f^{(n)}}(x^{(1)}, \ldots, x^{(n)}) = \mathrm{P}_{f^{(1)}, \ldots, f^{(n)}}(\mathbf{o}_1 = x^{(1)}, \ldots, \mathbf{o}_n = x^{(n)}) \,.$$

**Marginal distributions** can be defined in the usual manner via sum $\ovee$. For instance, consider the experiment consisting of tests $\omega = (\omega_y \colon I \to A)_{y \in Y}$, $f = (f_x \colon A \to B)_{x \in X}$, and $p = (p_z \colon B \to I)_{z \in Z}$. Then the joint probability distribution is:

$$\mathrm{P}_{\omega,f,p}(y, x, z) \equiv \mathrm{P}_{\omega,f,p}(\mathbf{o}_1 = y, \mathbf{o}_2 = x, \mathbf{o}_3 = z) = p_z \circ f_x \circ \omega_y \,.$$

The marginal distribution, say on the variables $\mathbf{o}_1$ and $\mathbf{o}_3$, is defined by summing over $\mathbf{o}_2$:

$$\mathrm{P}_{\omega,f,p}(y, z) = \bigovee_{x \in X} \mathrm{P}_{\omega,f,p}(y, x, z) = \bigovee_{x \in X} p_z \circ f_x \circ \omega_y \,.$$

In particular, if $f$ is a deterministic test (channel), say with $X = \{x_1\}$, then $\mathrm{P}_{\omega,f,p}(y, x_1, z) = \mathrm{P}_{\omega,f,p}(y, z)$, so the unique outcome $x_1$ may be omitted in the notation.

It is often the case that the scalars $\mathbf{C}(I, I)$ form a commutative division effect monoid. In that case, we can also define **conditional probability** in the usual manner via division, for example:

$$\mathrm{P}_{\omega,f,p}(x \mid y, z) = \mathrm{P}_{\omega,f,p}(y, x, z) / \mathrm{P}_{\omega,f,p}(y, z)\,.$$

It is defined when $\mathrm{P}_{\omega,f,p}(y, z)$ is nonzero. Clearly, the usual calculation rule such as $\mathrm{P}_{\omega,f,p}(x \mid y, z) \cdot \mathrm{P}_{\omega,f,p}(y, z) = \mathrm{P}_{\omega,f,p}(y, x, z)$ is valid.

With this notation in hand, we now describe causality as the principle of 'no signalling from the future'.

**Proposition 6.1.7.** *Let $f^{(1)}, \ldots, f^{(n)}$ be an experiment. For each number $k$ such that $1 \leq k < n$, we have*

$$\mathrm{P}_{f^{(1)},\ldots,f^{(n)}}(x^{(1)}, \ldots, x^{(k)}) = \mathrm{P}_{f^{(1)},\ldots,f^{(k)},\mathbb{1}}(x^{(1)}, \ldots, x^{(k)})\,.$$

*In words, the (marginal) probability of the outcomes from the first $k$ tests does not depends on the subsequent tests, and in fact it is equal to the probability in the experiment where we discard the system immediately after the $k$th test.*

*Proof.* Calculate as follows.

$$\begin{aligned}
&\mathrm{P}_{f^{(1)},\ldots,f^{(n)}}(x^{(1)}, \ldots, x^{(k)}) \\
&= \bigovee_{x^{(k+1)},\ldots,x^{(n)}} \mathrm{P}_{f^{(1)},\ldots,f^{(n)}}(x^{(1)}, \ldots, x^{(n)}) \\
&= \bigovee_{x^{(k+1)},\ldots,x^{(n)}} f^{(n)}_{x^{(n)}} \circ \cdots \circ f^{(1)}_{x^{(1)}} \\
&= \Big( \bigovee_{x^{(k+1)},\ldots,x^{(n)}} f^{(n)}_{x^{(n)}} \circ \cdots \circ f^{(k+1)}_{x^{(k+1)}} \Big) \circ f^{(k)}_{x^{(k)}} \circ \cdots \circ f^{(1)}_{x^{(1)}} \\
&= \mathbb{1} \circ f^{(k)}_{x^{(k)}} \circ \cdots \circ f^{(1)}_{x^{(1)}} \qquad \text{by Proposition 6.1.5} \\
&= \mathrm{P}_{f^{(1)},\ldots,f^{(k)},\mathbb{1}}(x^{(1)}, \ldots, x^{(k)})
\end{aligned}$$

■

In the light of the above result, we can generalize the notion of experiments as follows: an **experiment** is a composable sequence of tests $f^{(1)}, \ldots, f^{(n)}$ where the first test $f^{(1)}$ is an preparation test (but $f^{(n)}$ need not be an observation test). Then the probability of outcomes from the experiment is calculated by adding the discarding channel $\mathbb{1}$ at the end, that is:

$$\begin{aligned}
\mathrm{P}_{f^{(1)},\ldots,f^{(n)}}(x^{(1)}, \ldots, x^{(n)}) &:= \mathrm{P}_{f^{(1)},\ldots,f^{(n)},\mathbb{1}}(x^{(1)}, \ldots, x^{(n)}) \\
&= \mathbb{1} \circ f^{(n)}_{x^{(n)}} \circ \cdots \circ f^{(1)}_{x^{(1)}}\,.
\end{aligned}$$

**Remark 6.1.8.** Tull introduced a suitable reformulation of OPTs of Chiribella et al. called *operational theories with control* [248, Definition 1]. Our AOPTs can be seen as a simplification of Tull's operational theories with control; for example we omitted the *control* structure.

For comparison, let us describe the original definition of OPTs by Chiribella et al. [33, 35, 61] stripped of its monoidal structure:

- A (tensor-free version of) *operational probabilistic theory* is an operational structure $(\mathbf{C}, I, \mathrm{Test})$ equipped with a monoid morphism $[\![-]\!]\colon \mathbf{C}(I,I) \to [0,1]$ such that $\sum_{x\in X} [\![s_x]\!] = 1$ for every closed test $(s_x\colon I \to I)_{x\in X}$.

Here one has the interpretation $[\![s]\!] \in [0,1]$ of scalars in concrete probabilities, so that every closed test induces an ordinary probability distribution. Although causality and coarse-graining are not included in the above minimal definition of OPTs, both of them are basic additional assumptions used in the operational probabilistic framework, see [33, 35, 61]. We note that coarse-graining there is introduced via representation of transformations in vector spaces, while Definition 6.1.3 is more axiomatic.

## 6.2 Effectuses as operational probabilistic theories

In this section we put effectuses in the context of the operational probabilistic framework. To do so, we define tests in an effectus, and then show that every effectus can be seen as an AOPT.

**Definition 6.2.1.** A **test** in an effectus $\mathbf{C}$ is a family $(f_x\colon A \to B)_{x\in X}$ of morphisms indexed by a finite set $X$ such that $\bigovee_{x\in X} \mathbb{1} f_x = \mathbb{1}$.

In other words, by Lemma 3.2.5, a test in an effectus is a summable family $(f_x)_x$ of morphisms such that the sum $\bigovee_x f_x$ is a total morphism. Clearly, a channel is precisely a total morphism.

**Proposition 6.2.2.** *An effectus* $\mathbf{C}$ *with tests defined above and the unit* $I \in \mathbf{C}$ *form an AOPT.*

*Proof.* Let $(f_x\colon A \to B)_{x\in X}$ and $(g_y\colon B \to C)_{y\in Y}$ be tests. Then

$$\bigovee_{(x,y)\in X\times Y} \mathbb{1} \circ g_y \circ f_x = \bigovee_{x\in X} \Big(\bigovee_{y\in Y} \mathbb{1} \circ g_y\Big) \circ f_x = \bigovee_{x\in X} \mathbb{1} \circ f_x = \mathbb{1}\,,$$

whence $(g_y \circ f_x)_{(x,y)\in X\times Y}$ is a test. The rest of axioms of an operational structure are obvious. The operational structure $\mathbf{C}$ satisfies causality since for each $A \in \mathbf{C}$, the truth predicate $\mathbb{1}\colon A \to I$ is the unique channel (= total morphism) of this type. The effectus $\mathbf{C}$ is enriched over PCMs, and every test $(f_x\colon A \to B)_{x\in X}$ is a summable family since the domain predicates $(\mathbb{1} f_x)_{x\in X}$ are summable. Let $(f_x)_{x\in X}$ be a test and a $\alpha\colon X \to Y$ be a function. Let $g_y = \bigovee_{x\in\alpha^{-1}(y)} f_x$. Then

$$\bigovee_{y\in Y} \mathbb{1} g_y = \bigovee_{y\in Y} \bigovee_{x\in\alpha^{-1}(y)} \mathbb{1} f_x = \bigovee_{x\in X} \mathbb{1} f_x = \mathbb{1}\,,$$

so that the family $(g_y)_{y\in Y}$ is a test. Therefore $\mathbf{C}$ admits coarse-graining, proving that $\mathbf{C}$ is an AOPT. ∎

From now on, effectuses will be viewed as AOPTs in the manner above.

**Proposition 6.2.3.** *An observation test* $(p_x\colon A \to I)_{x\in X}$ *in an effectus is precisely a family of predicates* $p_x \in \mathrm{Pred}(A)$ *such that* $\bigovee_{x\in X} p_x = \mathbb{1}$. ∎

For an effect algebra $E$, an $n$-tuple $(a_1, \dots, a_n)$ of elements $a_j \in E$ satisfying $a_1 \ovee \cdots \ovee a_n = 1$ is called an **$n$-test** in $E$, see e.g. [229, 230, 243]. Therefore the proposition above says that an observation test $(p_j \colon A \to I)_{j\in[n]}$ with $n$ outcomes $[n] = \{1, \dots, n\}$ is precisely an $n$-test in the effect algebra $\mathrm{Pred}(A)$.

There is an alternative concise description of tests in an effectus, via copowers.

**Definition 6.2.4.** Let $A \in \mathbf{C}$ be an object in a category. For a set $X$, the **copower** $X \cdot A$ of $A$ by $X$ is the $X$-fold coproduct of $A$ in $\mathbf{C}$, namely:

$$X \cdot A := \coprod_{x\in X} A \,.$$

In particular when $X = [n] \equiv \{1, \dots, n\}$ for $n \in \mathbb{N}$, we write:

$$n \cdot A := [n] \cdot A = \overbrace{A + \cdots + A}^{n \text{ times}} \,.$$

Moreover, for a morphism $f \colon A \to B$ we will write $X \cdot f \colon X \cdot A \to X \cdot B$ for the morphism given as $X \cdot f = \coprod_{x\in X} f$.

**Proposition 6.2.5.** *In an effectus, tests $(f_x \colon A \to B)_{x\in X}$ are in bijective correspondence with total morphisms of the form $f \colon A \to X \cdot B$, where $X \cdot B$ is the copower of $B$ by $X$. They corresponds via $f_x = \rhd_x \circ f$ and $f = \langle\!\langle f_x \rangle\!\rangle_{x\in X}$.*

*Proof.* This is an immediate consequence of Lemma 3.2.5. ■

We will henceforth identify tests $(f_x \colon A \to B)_{x\in X}$ in an effectus with total morphisms of the form $f = \langle\!\langle f_x \rangle\!\rangle_x \colon A \to X \cdot B$. In particular, two-outcome observables are identified with total morphisms of type $A \to I + I$, which are precisely predicates in the effectus in total form $\mathrm{Tot}(\mathbf{C})$.

Given tests $f \colon A \to X \cdot B$ and $g \colon B \to Y \cdot C$, we can describe the composite $f; g \colon A \to (X \times Y) \cdot C$ concretely as follow. For this the obvious isomorphism $X \cdot (Y \cdot C) \cong (X \times Y) \cdot C$ will be used.

**Proposition 6.2.6.** *In the situation described above, the following diagram commutes.*

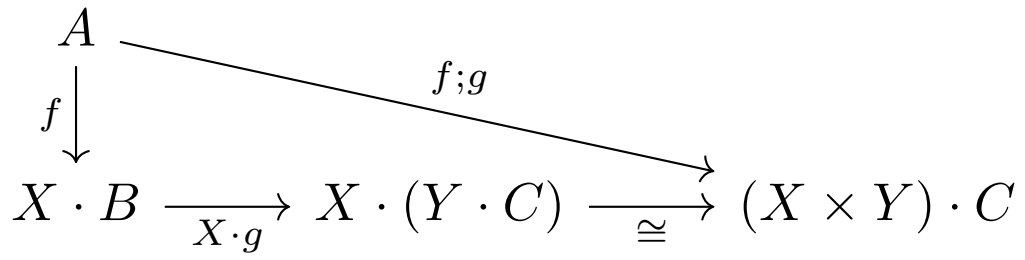

*Proof.* The two morphisms are equal when composed with the partial projections $\rhd_{x,y} \colon (X \times Y) \cdot C \to C$, as shown in the following commutative diagram:

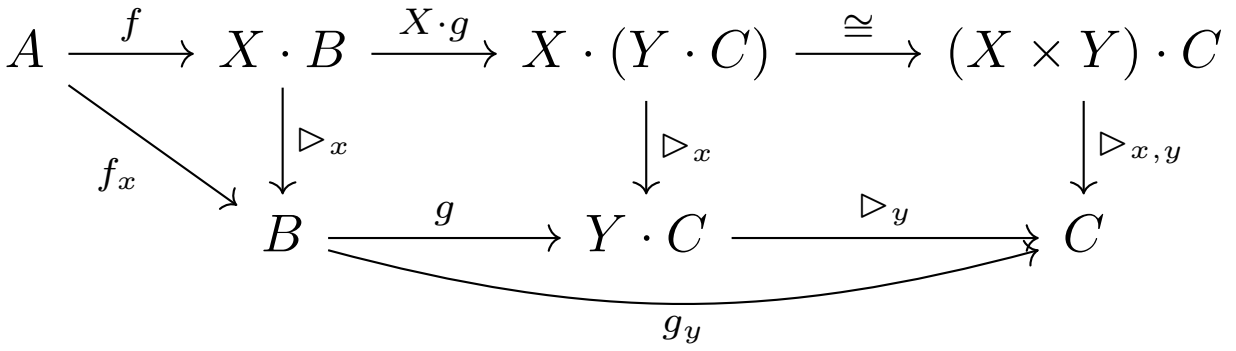

Therefore the diagram in question commutes by the joint monicity of partial projections. ■

As a special case we have:

**Corollary 6.2.7.** *Let $f\colon A \to X \cdot B$ be a test and $p\colon A \to X \cdot I$ an observable. The test $f$ is $p$-compatible, i.e. $f;\mathbb{1} = p$, if and only if the following diagram commutes.*

$$\begin{array}{ccc} A & \xrightarrow{f} & X \cdot B \\ & {\scriptstyle p}\searrow & \downarrow {\scriptstyle X\cdot\mathbb{1}} \\ & & X \cdot I \end{array}$$

∎

To discuss effectuses in the operational probabilistic framework, it is convenient to introduce 'operationally well-behaved' effectuses by imposing some additional assumptions. The first assumption is the normalization property from Section 4.5 — substates can be normalized into proper states. It is a fairly reasonable assumption and indeed, any real effectus satisfies the normalization property. Chiribella et al. also use normalization in their operational probabilistic framework, as an operation that is possible under other operationally reasonable assumptions, see [61, § 5.4.1] and [35, § 4.1.4].

The second assumption is the following separation property.

**Definition 6.2.8.** We say that an effectus satisfies the **separation property** if for each pair of morphisms $f, g\colon A \to B$, one has $f = g$ whenever $p \circ f \circ \omega = q \circ f \circ \omega$ for all $\omega \in \mathrm{St}(A)$ and $p \in \mathrm{Pred}(A)$.

Two morphisms $f, g\colon A \to B$ may be considered to be 'statistically equivalent' if $p \circ f \circ \omega = q \circ f \circ \omega$ for all $\omega \in \mathrm{St}(A)$ and $p \in \mathrm{Pred}(A)$, because the scalar $p \circ f \circ \omega$ represent the probability of the predicate $p$ holds after the transformation $f$ in the initial state $\omega$. The separation property thus asserts that any statistically equivalent morphisms are equal. In the operational probabilistic framework of Chiribella et al., they use the OPT quotiented by such statistical equivalence, and therefore the separation property always holds; see [61, § 3.2] or [35, § 2.2] for details.

We note that possible variations of the separation properties are equivalent to the one above, when normalization is assumed.

**Lemma 6.2.9.** *The following are equivalent in an effectus with the normalization property.*

(i) *The separation property in the sense of Definition 6.2.8 holds.*

(ii) *Total morphisms are separated by states and predicates.*

(iii) *Morphisms are separated by substates and predicates.*

(iv) *Total morphisms are separated by substates and predicates.*

(v) *Tests are separated by preparation and observation tests: for each pair of tests $f, g\colon A \to X \cdot B$, if $\omega; f; p = \omega; g; p$ for all tests $\omega\colon I \to Y \cdot A$ and $p\colon B \to Z \cdot I$, then $f = g$.*

(vi) *The category $\mathbf{C}$ is well-pointed and well-copointed with respect to the object $I$. Here well-pointedness w.r.t. $I$ is the property that for each $f, g\colon A \to B$ in $\mathbf{C}$, if $f \circ \omega = g \circ \omega$ for all $\omega\colon I \to A$, then $f = g$. Well-copointedness is the dual property.*

*Proof.* Implications (i) $\implies$ (ii) $\implies$ (iv) and (i) $\implies$ (iii) $\implies$ (iv) are obvious. Below we will first show (iv) $\implies$ (ii) $\implies$ (i), which implies that (i)–(iv) are all equivalent. Then we will prove the rest of the equivalence.

(iv) $\implies$ (ii): Assume that total morphisms $f, g\colon A \to B$ satisfies $p \circ f \circ \omega = p \circ g \circ \omega$ for all $\omega \in \mathrm{St}(A)$ and $p \in \mathrm{Pred}(B)$. Let $\omega \in \mathrm{St}_{\leq}(A)$ and $p \in \mathrm{Pred}(B)$ be an arbitrary substate and predicate. If $\omega = 0$ then $p \circ f \circ \omega = 0 = p \circ g \circ \omega$. Thus assume $\omega \neq 0$ and let $\overline{\omega}$ be the normalization of $\omega$ such that $\omega = \overline{\omega} \circ |\omega|$. Then

$$p \circ f \circ \omega = p \circ f \circ \overline{\omega} \circ |\omega| = p \circ g \circ \overline{\omega} \circ |\omega| = p \circ g \circ \omega$$

using the assumption. By (iv) we conclude that $f = g$.

(ii) $\implies$ (i): Let $f, g\colon A \to B$ be possibly non-total morphisms such that $p \circ f \circ \omega = p \circ g \circ \omega$ for all $\omega \in \mathrm{St}(A)$ and $p \in \mathrm{Pred}(B)$. Then for each $\omega \in \mathrm{St}(A)$ we have

$$(\mathbb{1}f)^{\perp} \circ \omega = (\mathbb{1} \circ f \circ \omega)^{\perp} = (\mathbb{1} \circ g \circ \omega)^{\perp} = (\mathbb{1}g)^{\perp} \circ \omega\,.$$

Therefore for any $\omega \in \mathrm{St}(A)$ and $p = [p_1, p_2] \in \mathrm{Pred}(B + I)$,

$$\begin{aligned} p \circ \langle\!\langle f, (\mathbb{1}f)^{\perp} \rangle\!\rangle \circ \omega &= p_1 \circ f \circ \omega \ovee p_2 \circ (\mathbb{1}f)^{\perp} \circ \omega \\ &= p_1 \circ g \circ \omega \ovee p_2 \circ (\mathbb{1}g)^{\perp} \circ \omega \\ &= p \circ \langle\!\langle g, (\mathbb{1}g)^{\perp} \rangle\!\rangle \circ \omega\,. \end{aligned}$$

Since $\langle\!\langle f, (\mathbb{1}f)^{\perp} \rangle\!\rangle$ and $\langle\!\langle g, (\mathbb{1}g)^{\perp} \rangle\!\rangle$ are total morphisms (of type $A \to B + I$), we obtain $\langle\!\langle f, (\mathbb{1}f)^{\perp} \rangle\!\rangle = \langle\!\langle g, (\mathbb{1}g)^{\perp} \rangle\!\rangle$ and hence $f = g$.

(i) $\implies$ (v): Suppose that two tests $f, g\colon A \to X \cdot B$ satisfies $\omega; f; p = \omega; g; p$ for all tests $\omega\colon I \to Y \cdot A$ and $p\colon B \to Z \cdot I$. Then in particular for any state $\omega \in \mathrm{St}(A)$ and predicate $p \in \mathrm{Pred}(A)$ we have $\omega; f; \langle\!\langle p, p^{\perp} \rangle\!\rangle = \omega; g; \langle\!\langle p, p^{\perp} \rangle\!\rangle$, so that $p \circ f_x \circ \omega = p \circ g_x \circ \omega$ for each $x \in X$. Hence by (i) we obtain $f_x = g_x$ for each $x \in X$, that is, $f = g$.

(v) $\implies$ (iv): Assume that two total morphisms $f, g\colon A \to B$ satisfies $p \circ f \circ \omega = p \circ g \circ \omega$ for all $\omega \in \mathrm{St}_{\leq}(A)$ and $p \in \mathrm{Pred}(B)$. We will consider $f, g$ as tests with singleton outcome set, i.e. channels. Let $\omega = \langle\!\langle \omega_y \rangle\!\rangle_y \colon I \to Y \cdot A$ and $p = \langle\!\langle p_z \rangle\!\rangle_z \colon B \to Z \cdot I$ be arbitrary preparation and observation tests. By assumption, for any $y \in Y$ and $z \in Z$ we have $p_z \circ f \circ \omega_y = p_z \circ g \circ \omega_y$. Therefore $\omega; f; p = \omega; g; p$. By (v) we obtain $f = g$.

(iii) $\implies$ (vi): Suppose that $f \circ \omega = g \circ \omega$ for all $\omega\colon I \to A$. Then $p \circ f \circ \omega = p \circ g \circ \omega$ for all $\omega \in \mathrm{St}(A)$ and $p \in \mathrm{Pred}(A)$, so that $f = g$ by (iii). Therefore $\mathbf{C}$ is well-pointed. Well-copointedness is shown similarly.

(vi) $\implies$ (iii): Suppose that $p \circ f \circ \omega = p \circ g \circ \omega$ for all $\omega \in \mathrm{St}_{\leq}(A)$ and $p \in \mathrm{Pred}(B)$. By well-copointedness, for each $\omega \in \mathrm{St}_{\leq}(A)$ we have $f \circ \omega = g \circ \omega$. Then $f = g$ by well-pointedness. ∎

Finally we assume that scalars are commutative. To summarize, we introduce the following definition:

**Definition 6.2.10.** We say that an effectus $\mathbf{C}$ is **operationally well-behaved** if it satisfies the following conditions.

(i) $\mathbf{C}$ satisfies the normalization property (see Definition 4.5.1).

(ii) $\mathbf{C}$ satisfies the separation property.

(iii) The scalars are commutative: $s \cdot t = t \cdot s$ for all $s, t \colon I \to I$.

By Proposition 4.5.2 the normalization property implies that scalars $\mathcal{S} = \mathbf{C}(I, I)$ admit division. Therefore in an operationally well-behaved effectus one can define *conditional probability*: $\mathrm{P}(x \mid y) = \mathrm{P}(x, y) / \mathrm{P}(y)$, see §6.1.

Real effectuses are of great importance, since experiments $f^{(1)}, \ldots, f^{(n)}$ yield probability distributions $\mathrm{P}_{f^{(1)},\ldots,f^{(n)}}$ in the usual sense. The following proposition is obvious but worth noting.

**Proposition 6.2.11.** *Any real effectus with the separation property is operationally well-behaved.*

*Proof.* Any real effectus satisfies the normalization property (Proposition 4.4.10) and has commutative scalars. ■

We end the section with examples.

**Example 6.2.12.** We describe the notions in the operational probabilistic framework in our main examples of effectuses.

(i) In the effectus **Pfn** of sets and partial function, tests with outcome set $X$ are families of partial functions $(f_x \colon A \rightharpoonup B)_{x \in X}$ satisfying the condition that for each $a \in A$, there exists a unique $x \in X$ such that $f_x(a)$ is defined. By Proposition 6.2.5 they are equivalently total functions $f \colon A \to X \cdot B$. As one has $X \cdot B = \coprod_{x \in X} B \cong X \times B$, tests send states $a \in A$ to states $b \in B$ together with outcomes $x \in X$. In particular, observables are total functions of the form $p \colon A \to X$. They are identified with partition $(p^{-1}(x))_{x \in X}$ of the set $A$. Closed tests $s \colon 1 \to X$ correspond elements $s \in X$, which may also be seen as Boolean-valued distributions $X \to \{0, 1\}$ in the obvious way. Thus each experiment $f^{(0)}, \ldots, f^{(n)}$ yields a 'deterministic' outcome $(x^{(0)}, \ldots, x^{(n)})$.

(ii) In the effectus $\mathcal{K}\ell(\mathcal{D}_{\leq})$ of sets and subprobabilistic maps, tests with outcome set $X$ are families of functions $(f_x \colon A \to \mathcal{D}_{\leq}(B))_{x \in X}$ such that for each $a \in A$ one has $\sum_{x \in X} \sum_{b \in B} f_x(a)(b) = 1$. Equivalently, they are functions $f \colon A \to \mathcal{D}(X \cdot B) \cong \mathcal{D}(X \times B)$. Observables are functions $p \colon A \to \mathcal{D}(X)$, i.e. ones that map elements $a \in A$ to probability distributions $p(a) \in \mathcal{D}(X)$ on the outcome set $X$. Closed tests $1 \to \mathcal{D}(X)$ are exactly probability distributions on $X$. Thus each experiment $f^{(0)}, \ldots, f^{(n)}$ with outcome sets $X^{(0)}, \ldots, X^{(n)}$ induces a 'joint' probability distribution $\mathrm{P}_{f^{(0)},\ldots,f^{(n)}}$ on the product $X^{(0)} \times \cdots \times X^{(n)}$.

(iii) In the effectus $\mathbf{Wstar}_{\leq}^{\mathrm{op}}$ of $W^*$-algebras, tests (from $\mathscr{A}$ to $\mathscr{B}$) with outcome set $X$ are families of normal subunital CP maps $(f_x \colon \mathscr{B} \to \mathscr{A})_{x \in X}$ such that for each $b \in B$ one has $\sum_{x \in X} f_x(b) = 1$, or equivalently the sum $\bigovee_{x \in X} f_x$ is a unital map. By Proposition 6.2.5 tests can also be described as normal *unital* CP maps $f \colon X \cdot \mathscr{B} \to \mathscr{A}$. Observables are families $(p_x \in [0, 1]_{\mathscr{A}})_{x \in X}$ of predicates/effects with $\sum_{x \in X} p_x = 1$. When $\mathscr{A} = \mathcal{B}(\mathscr{H})$ for a Hilbert space $\mathscr{H}$, the observables are commonly known as *positive-operator valued measures* (*POV measures* or *POVMs*) [27, 65, 181]. POVMs are a generalization of 'sharp' observables, i.e. self-adjoint operators on $\mathscr{H}$, and thus also called *unsharp observables* or simply *observables*. Note that general POVMs may have measurable spaces $(X, \Sigma_X)$ as the spaces of outcomes, while in our setting outcome spaces $X$ are restricted to finite discrete ones. Similarly if $\mathscr{A} = \mathscr{B} = \mathcal{B}(\mathscr{H})$, tests/instruments $(f_x : \mathcal{B}(\mathscr{H}) \to \mathcal{B}(\mathscr{H}))_{x\in X}$ are precisely *instruments* in the sense of Ozawa [213], with finite outcome spaces $X$ (in general, the outcome spaces may be measurable spaces). See also [65, 66, 120] for the notion of quantum instruments. Finally, closed tests $(s_x : \mathbb{C} \to \mathbb{C})_{x\in X}$ in $\mathbf{Wstar}_{\lesssim}^{\mathrm{op}}$ correspond to probability distributions on $X$, via the identification of scalars $s_x : \mathbb{C} \to \mathbb{C}$ with probabilities $s_x(1) \in [0, 1]$.

# 6.3 Repeatable measurements and sharp observables

Repeatability of measurement refers to a property that if a quantity is measured twice consecutively, then the two measurements yield the same outcomes. Von Neumann [210, § III.3 and § IV.3] introduced the property, sometimes called the *repeatability hypothesis*. From the hypothesis he deduced the well-known *projection postulate* (or *collapse postulate*) that determines the state after a measurement by projection. His discussion was then refined by Lüders [194].

Later Davies and Lewis [66] initiated a modern quantum measurement theory based on the notion of instruments, where repeatability is a property of instruments that is not necessarily satisfied. In the modern framework repeatable measurements/instruments have been studied in relation to sharp observables and measurements of von Neumann and Lüders, see e.g. [25, 66, 183].

In this section we will discuss repeatability and related concepts in our abstract setting of effectuses.

## 6.3.1 Repeatable and idempotent instruments

We start with the definition of repeatability and a stronger property of idempotency.

**Definition 6.3.1.** We say that an instrument $f = \langle\!\langle f_x \rangle\!\rangle_x : A \to X \cdot A$ is

(i) **repeatable** if $f_{x'} \circ f_x = 0_{AA}$ for each $x, x' \in X$ with $x \neq x'$;

(ii) **idempotent** if each $f_x$ is idempotent, i.e. $f_x \circ f_x = f_x$, for each $x \in X$.

We give a few characterization of repeatable instruments.

**Proposition 6.3.2.**

(i) *An instrument $f = \langle\!\langle f_x \rangle\!\rangle_x : A \to X \cdot A$ is repeatable if and only if $\mathbb{1} \circ f_x \circ f_x = \mathbb{1} \circ f_x$ for each $x \in X$.*

(ii) *Suppose that the effectus is operationally well-behaved. Then an instrument $f$ is repeatable if and only if*

$$\mathrm{P}_{\omega,f,f}(\mathbf{o}_1 = y, \mathbf{o}_2 = x, \mathbf{o}_3 = x') = \begin{cases} \mathrm{P}_{\omega,f}(\mathbf{o}_1 = y, \mathbf{o}_2 = x) & \text{if } x = x' \\ 0 & \text{if } x \neq x' \end{cases}$$

*for any preparation test $\omega : I \to Y \cdot A$.*

*Proof.*

(i) Assume that $f$ is repeatable. Then for each $x' \in X$, one has

$$\mathbb{1} \circ f_{x'} = \mathbb{1} \circ \Big( \bigovee_{x\in X} f_x \Big) \circ f_{x'} = \mathbb{1} \circ \Big( \bigovee_{x\in X} f_x \circ f_{x'} \Big) = \mathbb{1} \circ f_{x'} \circ f_{x'} \,.$$

Conversely, assume that $\mathbb{1} \circ f_x \circ f_x = \mathbb{1} \circ f_x$ for each $x \in X$. Then for each $x' \in X$ we have

$$\mathbb{1} \circ f_{x'} = \mathbb{1} \circ \Big( \bigovee_{x\in X} f_x \Big) \circ f_{x'} = \bigovee_{x\in X} \mathbb{1} \circ f_x \circ f_{x'} = \mathbb{1} \circ f_{x'} \ovee \bigovee_{x\neq x'} \mathbb{1} \circ f_x \circ f_{x'} \,.$$

By cancellation and positivity, for each $x \in X$ with $x \neq x'$ we have $\mathbb{1} \circ f_x \circ f_{x'} = \mathbb{0}$ and hence $f_x \circ f_{x'} = 0_{AA}$.

(ii) If $f$ is repeatable then for any preparation $\omega\colon I \to Y \cdot A$

$$\mathrm{P}_{\omega,f,f}(\mathbf{o}_2 = x, \mathbf{o}_3 = x) = \mathbb{1} \circ f_x \circ f_x \circ \omega = \mathbb{1} \circ f_x \circ \omega = \mathrm{P}_{\omega,f}(\mathbf{o}_2 = x)$$

and if $x \neq x'$

$$\mathrm{P}_{\omega,f,f}(\mathbf{o}_2 = x, \mathbf{o}_3 = x') = \mathbb{1} \circ f_{x'} \circ f_x \circ \omega = 0 \,.$$

The converse follows by separation. ■

Proposition 6.3.2(ii) expresses the condition that repeating measurement by the instrument yields the same outcome. We obtain the following corollary as a consequence of Proposition 6.3.2(i).

**Corollary 6.3.3.** *Any idempotent instrument is repeatable.* ■

We give a few characterization of idempotent instruments.

**Proposition 6.3.4.** *An instrument $f\colon A \to X \cdot A$ is idempotent if and only if the following diagram commutes.*

$$\begin{array}{ccc} A & \xrightarrow{f} & X\cdot A \\ & {\scriptstyle f;f}\searrow & \downarrow{\scriptstyle \Delta\cdot A} \\ & & (X\times X)\cdot A \end{array} \qquad \textit{that is:} \qquad \begin{array}{ccc} A & \xrightarrow{\qquad f \qquad} & X\cdot A \\ {\scriptstyle f}\downarrow & & \downarrow{\scriptstyle \Delta\cdot A} \\ X\cdot A \xrightarrow{X\cdot f} X\cdot(X\cdot A) & \xrightarrow{\cong} & (X\times X)\cdot A \end{array}$$

*where $\Delta\colon X \to X \times X$ is the diagonal.*

*Proof.* It is easy to verify that $\rhd_{x,x'} \circ (f;f) = \rhd_{x,x'} \circ (\Delta \cdot A) \circ f$ for each $x, x' \in X$, using

$$\Big( X \cdot A \xrightarrow{\Delta\cdot A} (X\times X)\cdot A \xrightarrow{\rhd_{x,x'}} A \Big) = \begin{cases} \rhd_x & \text{if } x = x' \\ 0 & \text{if } x \neq x' \,. \end{cases}$$

Thus $f;f = (\Delta \cdot A) \circ f$ by the joint monicity of partial projections $\rhd_{x,x'}$. ■

**Proposition 6.3.5.** *In an operationally well-behaved effectus, an instrument* $f\colon A \to X \cdot A$ *is idempotent if and only if for any preparation test* $\omega\colon I \to Y \cdot A$ *and observable* $p\colon A \to Z \cdot I$*, one has*

$$\begin{aligned} &\mathrm{P}_{\omega,f,f,p}(\mathbf{o}_1 = y, \mathbf{o}_2 = x, \mathbf{o}_3 = x', \mathbf{o}_4 = z) \\ &= \begin{cases} \mathrm{P}_{\omega,f,p}(\mathbf{o}_1 = y, \mathbf{o}_2 = x, \mathbf{o}_3 = z) & \text{if } x = x' \\ 0 & \text{if } x \neq x' . \end{cases} \end{aligned}$$

*Proof.* Similar to the proof of Proposition 6.3.2(ii). ∎

Therefore from an operational point of view, measuring twice with an idempotent instrument is exactly the same as measuring only once. Note that, in contrast, measuring twice with a repeatable instrument may introduce additional disturbance of the state, compared to measuring once.

**Definition 6.3.6.** An observable $p = (p_x)_x\colon A \to X \cdot I$ in an comprehensive effectus is **sharp** if for each $x \in X$ the predicate $p_x\colon A \to I$ is sharp.

**Example 6.3.7.** We give example of repeatable (in fact, idempotent) instruments in $\mathbf{Wstar}_{\leq}^{\mathrm{op}}$. Let $\mathscr{A}$ be a $W^*$-algebra, and $(\mathfrak{p}_x)_{x\in X}$ be a sharp observable, that is, a family $(\mathfrak{p}_x)_{x\in X}$ of projections $\mathfrak{p}_x \in \mathscr{A}$ such that $\sum_{x\in X} \mathfrak{p}_x = 1$. Then we define an instrument $f\colon \mathscr{A} \to X \cdot \mathscr{A}$ in $\mathbf{Wstar}_{\leq}^{\mathrm{op}}$, i.e. $f\colon \mathscr{A}^X \to \mathscr{A}$ in $\mathbf{Wstar}_{\leq}$, by $f((a_x)_x) = \sum_{x\in X} \mathfrak{p}_x a_x \mathfrak{p}_x$. In other words, $f$ is the partial tuple of $f_x\colon A \to A$ where $f_x(a) = \mathfrak{p}_x a \mathfrak{p}_x$. Since $f_x(1) = \mathfrak{p}_x$, $f$ is a $(\mathfrak{p}_x)_x$-compatible instrument. For each $x \in X$ we have $f_x(f_x(a)) = \mathfrak{p}_x\mathfrak{p}_x a \mathfrak{p}_x\mathfrak{p}_x = \mathfrak{p}_x a \mathfrak{p}_x = f_x(a)$. Therefore $f_x \circ f_x = f_x$ and the instrument $f$ is idempotent, hence repeatable. The instrument $f$ is called the Lüders instrument [27, 28, 120].

Note that this works only when the observable is sharp. For a general 'unsharp' observable $(p_x)_{x\in X}$, one can still construct a $(p_x)_x$-compatible instrument $f\colon \mathscr{A}^X \to \mathscr{A}$ by $f_x(a) = \sqrt{p_x}a\sqrt{p_x}$ using square roots. The instrument is sometimes called the *generalized Lüders instrument* (e.g. in [26]), or simply the *Lüders instrument* (e.g. in [120]). It is not repeatable in general, since $f_x(f_x(1)) = p_x^2 \neq p_x = f_x(1)$. Clearly it is repeatable if and only if the observable $(p_x)_x$ is sharp.

As one can see from the example above, there is a certain relationship between repeatability of instruments and sharpness of observables. The examples below show, however, that the relationship is not so easy as one might expect. This leads us to consider additional conditions such as ideality on instruments in the following subsections.

**Example 6.3.8.** The observable measured by a repeatable (or idempotent) instrument is not necessarily sharp. We give a counterexample in $\mathbf{Wstar}_{\leq}^{\mathrm{op}}$. Consider the $W^*$-algebra $\mathcal{M}_3 = \mathbb{C}^{3\times 3}$ of $3 \times 3$-matrices. Let

$$A_0 = |0\rangle\langle 0| + \frac{1}{2}|2\rangle\langle 2| \qquad\qquad A_1 = |1\rangle\langle 1| + \frac{1}{2}|2\rangle\langle 2|$$

be matrices, and define maps $f_0, f_1\colon \mathcal{M}_3 \to \mathcal{M}_3$ by

$$f_0(B) = \langle 0|B|0\rangle A_0 \qquad\qquad f_1(B) = \langle 1|B|1\rangle A_1 .$$

It is easy to verify that $\langle\!\langle f_0, f_1\rangle\!\rangle\colon \mathcal{M}_3 \to \mathcal{M}_3 + \mathcal{M}_3$ is an instrument in $\mathbf{Wstar}_{\leq}^{\mathrm{op}}$, which measures the observable given by $f_0(1) = A_0$ and $f_1(1) = A_1$. Since $\langle j|A_k|j\rangle = \delta_{jk}$, we have $f_j \circ f_j = f_j$ and $f_j \circ f_k = 0_{AA}$ for $k \neq j$. Therefore the instrument $\langle\!\langle f_0, f_1\rangle\!\rangle$ is idempotent and hence repeatable, but the measured observable $(A_0, A_1)$ is not sharp.

Note that the example involves only diagonal entries of matrices. Therefore we can construct a similar counterexample as an instrument on the commutative algebra $\mathbb{C}^3$, and also an counterexample in $\mathcal{K}\ell(\mathcal{D}_{\leq})$.

**Example 6.3.9.** Conversely, not all instruments that measure sharp observables are repeatable either. Consider the $2 \times 2$-matrix algebra $\mathcal{M}_2$ in the effectus $\mathbf{Wstar}_{\leq}^{\mathrm{op}}$. Let $P_0 = |0\rangle\langle 0|$ and $P_1 = |1\rangle\langle 1|$ be projections, and $H$ be the Hadamard unitary matrix, that is, $H = \frac{1}{\sqrt{2}}\left[\begin{smallmatrix} 1 & 1 \\ 1 & -1 \end{smallmatrix}\right]$. Then define maps $g_1, g_2\colon \mathcal{M}_2 \to \mathcal{M}_2$ by

$$g_0(B) = P_0 H B H P_0 \qquad g_1(B) = P_1 B P_1$$

They form an instrument $\langle\!\langle f_0, f_1\rangle\!\rangle\colon \mathcal{M}_2 \to \mathcal{M}_2 + \mathcal{M}_2$ in $\mathbf{Wstar}_{\leq}^{\mathrm{op}}$ that measures projections $P_0, P_1$ (since $g_0(1) = P_0$ and $g_1(1) = P_1$). But the instrument is not repeatable, because

$$g_1(g_0(1)) = P_0 H P_1 H P_0 = \frac{1}{2} P_0 \neq 0\,,$$

that is, $\mathbb{1} \circ g_0 \circ g_1 \neq 0$ in $\mathbf{Wstar}_{\leq}^{\mathrm{op}}$. The instrument $f$ corresponds to the procedure where we first perform the Lüders measurement for $P_0$ and $P_1$, and if we obtain an outcome 0 (associated to $P_0$), we then apply the Hadamard gate. Thus if we see the outcome 0, the state after the measurement is $H|0\rangle = (|0\rangle + |1\rangle)/\sqrt{2}$. Therefore, if we perform a measurement by $f$ twice in a row, it is possible to get different outcomes.

*Note.* Davies and Lewis [66] use 'repeatable' for the property we call 'idempotent'; and 'weakly repeatable' for what we call 'repeatable'. We follow Busch and others [25, 27, 28, 183], who use 'repeatable' for the weaker property. Note that [66] uses 'strongly repeatable' for the property even stronger than 'idempotent' (see Theorem 6.3.28 and the following paragraph), while 'strongly repeatable' in [183] means 'idempotent' in our sense.

## 6.3.2 C- and Q-idempotents

This subsection contains preliminary results about idempotents in effectuses. There are two special kinds of idempotents in effectuses: *C-idempotents* and *Q-idempotents*. Here C stands for comprehension and Q for quotients, and they respectively related to comprehension and quotients in a certain way.

Recall that an *idempotent* is an endomap $f\colon A \to A$ such that $f \circ f = f$. An idempotent $f\colon A \to A$ *splits* if there exists morphisms $m\colon A' \to A$ and $e\colon A \to A'$ such that $e \circ m = \mathrm{id}_{A'}$ and $m \circ e = f$. Note that $m$ and $e$ are respectively a split mono and epi, and hence in an effectus, $m$ is total and $e$ is faithful.

**Lemma 6.3.10.** *Let $f\colon A \to A$ be an idempotent that splits as $A \xrightarrow{e} A' \xrightarrow{m} A$. Let $p = \mathbb{1} f$. The following are equivalent.*

(i) *$m\colon A' \to A$ is a (total) comprehension of $p$.*

(ii) $h^{\square}(p) = \mathbb{1}$ *implies* $f \circ h = h$ *for any* $h\colon B \to A$.

*Proof.* (i) $\Longrightarrow$ (ii): Suppose that $h\colon B \to A$ satisfies $h^{\square}(p) = \mathbb{1}$. By the universality of comprehension $m\colon A' \to A$ we obtain $\overline{h}\colon B \to A'$ such that $h = m \circ \overline{h}$. Then

$$f \circ h = m \circ e \circ m \circ \overline{h} = m \circ \overline{h} = h\,.$$

(ii) $\Longrightarrow$ (i): To prove that $m\colon A' \to A$ is a comprehension of $p$, suppose that $h\colon B \to A$ satisfies $h^{\square}(p) = \mathbb{1}$. Then $f \circ h = h$, which shows that $\overline{h} \coloneqq e \circ h\colon B \to A'$ is a desired mediating map, as:

$$m \circ \overline{h} = m \circ e \circ h = f \circ h = h\,.$$

Moreover the mediating map is unique since $m$ is monic. ■

**Lemma 6.3.11.** *Let* $f\colon A \to A$ *be an idempotent that splits as* $A \xrightarrow{e} A' \xrightarrow{m} A$. *Let* $p = \mathbb{1}f$. *The following are equivalent.*

(i) $e\colon A \to A'$ *is a quotient for* $p^{\perp}$.

(ii) $\mathbb{1}g \leq p$ *implies* $g \circ f = g$ *for any* $g\colon A \to B$.

*Proof.* (i) $\Longrightarrow$ (ii): If $g\colon A \to B$ satisfies $\mathbb{1}g \leq p$, then $g$ factors through the quotient $e\colon A \to A'$ via $\overline{g}\colon A' \to B$ as $g = \overline{g} \circ e$. Then

$$g \circ f = \overline{g} \circ e \circ m \circ e = \overline{g} \circ e = g\,.$$

(ii) $\Longrightarrow$ (i): Let $g\colon A \to B$ satisfy $\mathbb{1}g \leq p$ (i.e. $p^{\perp} \leq \ker(g)$). Then $g \circ f = g$, and $\overline{g} \coloneqq g \circ m\colon A' \to B$ is a desired mediating map, since

$$\overline{g} \circ e = g \circ m \circ e = g \circ f = g\,.$$

The mediating map is unique since $e$ is epic. ■

We define C- and Q-idempotents based on the observations above.

**Definition 6.3.12.** Let $f\colon A \to A$ be an endomorphism and $p \coloneqq \mathbb{1}f$. We say that $f$ is

(i) **C-idempotent** if $f$ is idempotent such that $h^{\square}(p) = \mathbb{1}$ implies $f \circ h = h$ for any $h\colon B \to A$.

(ii) **Q-idempotent** if $f$ is idempotent such that $\mathbb{1}g \leq p$ implies $g \circ f = g$ for any $g\colon A \to B$.

(iii) **CQ-idempotent** if $f$ is both $C$-idempotent and $Q$-idempotent.

It is well known that an idempotent $f\colon A \to A$ splits if and only if there exists an equalizer or a coequalizer of $f$ and $\mathrm{id}_A\colon A \to A$. Similar statements for C- and Q-idempotents hold, but with comprehension and quotients instead of (co)equalizers.

**Proposition 6.3.13.** *Let $f\colon A \to A$ be a C-idempotent. Let $p = \mathbb{1}f$. Then the idempotent $f$ splits if and only if there exists a comprehension $\pi_p\colon \{A\,|\,p\} \to A$ of $p$. In that case, the splitting of $f$ is given by the universality of the comprehension as below.*

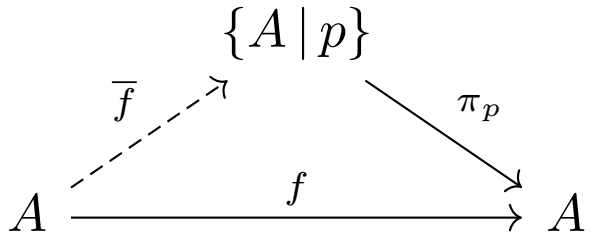

*Proof.* If $f$ splits, say as $A \xrightarrow{r} A' \xrightarrow{s} A$, then $s\colon A' \to A$ is a comprehension of $p$ by Lemma 6.3.10. Conversely, assume that there exists a comprehension $\pi_p\colon \{A\,|\,p\} \to A$. Since $p \circ f = \mathbb{1} \circ f \circ f = \mathbb{1} \circ f$ and so $f^{\square}(p) = \mathbb{1}$, there exists $\overline{f}\colon A \to \{A\,|\,p\}$ such that $f = \pi_p \circ \overline{f}$. Note that $\pi_p^{\square}(p) = \mathbb{1}$ and hence $f \circ \pi_p = \pi_p$ by C-idempotency. Then

$$\pi_p \circ \mathrm{id} = \pi_p = f \circ \pi_p = \pi_p \circ \overline{f} \circ \pi_p \,.$$

Since $\pi_p$ is monic, $\overline{f} \circ \pi_p = \mathrm{id}$. Therefore $f$ splits. ■

**Proposition 6.3.14.** *Let $f\colon A \to A$ be a Q-idempotent. Let $p = \mathbb{1}f$. Then the idempotent $f$ splits if and only if there exists a quotient $\xi_{p^\perp}\colon A \to A/p^\perp$ for $p^\perp$. In that case, the splitting of $f$ is given by the universality of the quotient as below.*

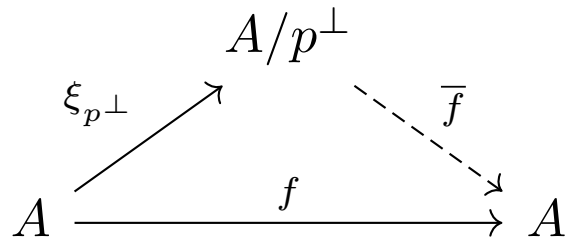

*Proof.* If $f$ splits, say as $A \xrightarrow{r} A' \xrightarrow{s} A$, then $r\colon A \to A'$ is a quotient for $p^\perp$ by Lemma 6.3.11. Conversely assume that there is a quotient $\xi_{p^\perp}\colon A \to A/p^\perp$. Because $\mathbb{1}f = p$, i.e. $\ker(f) = p^\perp$, we obtain the mediating map $\overline{f}\colon A/p^\perp \to A$ such that $f = \overline{f} \circ \xi_{p^\perp}$. Since $\mathbb{1}\xi_{p^\perp} = p$ we have $\xi_{p^\perp} \circ f = \xi_{p^\perp}$ by Q-idempotency. Then

$$\mathrm{id} \circ \xi_{p^\perp} = \xi_{p^\perp} = \xi_{p^\perp} \circ f = \xi_{p^\perp} \circ \overline{f} \circ \xi_{p^\perp} = \xi_{p^\perp} \circ \overline{f} \circ \xi_{p^\perp} \,.$$

Then $\xi_{p^\perp} \circ \overline{f} = \mathrm{id}$ as $\xi_{p^\perp}$ is epi. Therefore $f$ splits. ■

C- and Q-idempotent instruments will mean what you would expect:

**Definition 6.3.15.** An instrument $f = \langle\!\langle f_x \rangle\!\rangle_x\colon A \to X \cdot A$ is **C-idempotent** (resp. **Q-** and **CQ-idempotent**) if $f_x$ is C-idempotent (resp. Q- and CQ-idempotent) for each $x \in X$.

## 6.3.3 C-ideal and Q-ideal instruments

We define two kinds of ideality of instruments, which will turn out to be related with comprehension and quotients, respectively.

**Definition 6.3.16.** An instrument $f\colon A \to X \cdot A$ is said to be:

(i) **C-ideal** provided that for any morphism $h\colon B \to A$ and $x' \in X$, if $f_x \circ h = 0$ for all $x \in X \setminus \{x'\}$, then $f_{x'} \circ h = h$;

(ii) **Q-ideal** provided that for any morphism $g\colon A \to B$ and $x' \in X$, if $g \circ f_x = 0$ for all $x \in X \setminus \{x'\}$, then $g \circ f_{x'} = g$;

(iii) **CQ-ideal** if it is both C-ideal and Q-ideal.

We first give characterizations of C-ideality in an operationally well-behaved effectus. Intuitively, C-ideality is a property that the measurement does not disturb the state of a system whenever some outcome is certain (deterministic). In fact, C-ideality is equivalent to the property known as *d-ideality* [25, 27, 183], when interpreted in the effectus $\mathbf{Wstar}_{\leq}^{\mathrm{op}}$ of $W^*$-algebras.

**Proposition 6.3.17.** *Let $f\colon A \to X \cdot A$ be an instrument in an operationally well-behaved effectus. The following are equivalent.*

(i) *$f$ is C-ideal.*

(ii) *For each state $\omega \in \mathrm{St}(A)$ and $x' \in X$, if $f_x \circ \omega = 0$ for all $x \in X \setminus \{x'\}$, then $f_{x'} \circ \omega = \omega$.*

(iii) *For each preparation test $\omega\colon I \to Y \cdot A$ and $y \in Y$, if there exists $x' \in X$ such that ($\mathrm{P}_{\omega,f}(y) \neq 0$ and) $\mathrm{P}_{\omega,f}(x' \mid y) = \mathbb{1}$, then $\mathrm{P}_{\omega,f,p}(y,z) = \mathrm{P}_{\omega,p}(y,z)$ for any observable $p\colon A \to Z \cdot I$ and $z \in Z$.*

(iv) *(d-ideality) For each $\omega \in \mathrm{St}(A)$ and $x' \in X$, if $\mathrm{P}_{\omega,f}(x') = \mathbb{1}$, then $f_{x'} \circ \omega = \omega$.*

*Proof.* (i) $\iff$ (ii): The direction $\Longrightarrow$ is trivial. We prove $\Longleftarrow$. Suppose that $h\colon B \to A$ and $x' \in X$ satisfy $f_x \circ h = 0$ for all $x \in X \setminus \{x'\}$. By the separation property, to conclude that $f_{x'} \circ h = h$ it suffices to prove that $f_{x'} \circ h \circ \omega = h \circ \omega$ for all $\omega \in \mathrm{St}(B)$. Take an arbitrary state $\omega \in \mathrm{St}(B)$. If $h \circ \omega = 0$ the desired equation clearly holds. Thus assume $h \circ \omega \neq 0$ and let $\sigma \in \mathrm{St}(A)$ be the normalization of $h \circ \omega$. Then for all $x \in X \setminus \{x'\}$,

$$f_x \circ \sigma \circ |h \circ \omega| = f_x \circ h \circ \omega = 0\,.$$

From $|h \circ \omega| \neq 0$ and $|f_x \circ \sigma| \cdot |h \circ \omega| = 0$, we obtain $f_x \circ \sigma = 0$. By (ii), $f_{x'} \circ \sigma = \sigma$. Therefore

$$f_{x'} \circ h \circ \omega = f_{x'} \circ \sigma \circ |h \circ \omega| = \sigma \circ |h \circ \omega| = h \circ \omega\,.$$

(i) $\Longrightarrow$ (iii): Suppose that $x' \in X$ satisfies $\mathrm{P}_{\omega,f}(x' \mid y) = \mathbb{1}$. Then

$$\mathrm{P}_{\omega,f}(y,x') = \mathrm{P}_{\omega,f}(x' \mid y)\,\mathrm{P}_{\omega,f}(y) = \mathrm{P}_{\omega,f}(y) \equiv \bigovee_{x \in X} \mathrm{P}_{\omega,f}(y,x)\,.$$

By cancellation, $\bigovee_{x \in X \setminus \{x'\}} \mathrm{P}_{\omega,f}(y,x) = 0$. Then for all $x \in X \setminus \{x'\}$,

$$0 = \mathrm{P}_{\omega,f}(y,x) \equiv \mathbb{1} \circ f_x \circ \omega_y\,,$$

so that $f_x \circ \omega_y = 0$. Hence $f_{x'} \circ \omega_y = \omega_y$ by C-ideality. Therefore for any observable $p\colon A \to Z \cdot I$,

$$\mathrm{P}_{\omega,f,p}(y,z) = \bigovee_{x \in X} p_z \circ f_x \circ \omega_y = p_z \circ f_{x'} \circ \omega_y = p_z \circ \omega_y = \mathrm{P}_{\omega,p}(y,z)\,.$$

(iii) $\Longrightarrow$ (iv): Suppose that $\mathrm{P}_{\omega,f}(x') = \mathbb{1}$ for a state $\omega\colon I \to X$ and $x' \in X$. Let us view $\omega$ as a test with outcome set $\{y\}$. Then

$$\mathrm{P}_{\omega,f}(y, x') = \mathrm{P}_{\omega,f}(x') = \mathbb{1}\,,$$

and by causality

$$\mathrm{P}_{\omega,f}(y) = \mathrm{P}_{\omega}(y) = \mathbb{1} \circ \omega = \mathbb{1}\,.$$

Therefore

$$\mathrm{P}_{\omega,f}(x' \mid y) = \mathrm{P}_{\omega,f}(y, x') / \mathrm{P}_{\omega,f}(y) = \mathbb{1}\,.$$

Reasoning similarly to the proof of (i) $\Longrightarrow$ (iii), we obtain $f_x \circ \omega = 0$ for all $x \in X \setminus \{x'\}$. Now by (iii), for any observable $p\colon A \to Z \cdot I$ and $z \in Z$,

$$p_z \circ f_{x'} \circ \omega = \bigovee_{x \in X} p_z \circ f_x \circ \omega \equiv \mathrm{P}_{\omega,f,p}(y, z) = \mathrm{P}_{\omega,p}(y, z) \equiv p_z \circ \omega\,.$$

By separation, $f_{x'} \circ \omega = \omega$.

(iv) $\Longrightarrow$ (ii): Suppose that $\omega \in \mathrm{St}(A)$ and $x' \in X$ satisfies $f_x \circ \omega = 0$ for all $x \in X \setminus \{x'\}$. Then

$$\mathrm{P}_{\omega,f}(x') = \mathbb{1} \circ f_{x'} \circ \omega = \bigovee_{x \in X} \mathbb{1} \circ f_x \circ \omega = \mathbb{1}\,.$$

By (iv) we obtain $f_{x'} \circ \omega = \omega$. ■

Next we give characterizations of Q-ideality. Formally Q-ideality is dual to C-ideality, but its operational meaning is slightly more complicated. It roughly means the following: if a later observation $z \in Z$ makes some outcome from the measurement by $f$ certain, then the measurement by $f$ does not change the probability that we observe the outcome $z$.

**Proposition 6.3.18.** *Let $f\colon A \to X \cdot A$ be an instrument in an operationally well-behaved effectus. The following are equivalent.*

(i) *$f$ is Q-ideal.*

(ii) *For any $p \in \mathrm{Pred}(A)$ and $x' \in X$, if $p \circ f_x = 0$ for all $x \in X \setminus \{x'\}$, then $p \circ f_{x'} = p$.*

(iii) *For each observable $p\colon A \to Z \cdot I$ and $z \in Z$, if there exists $x' \in X$ such that $\mathrm{P}_{\omega,f,p}(x' \mid z) = \mathbb{1}$ (whenever $\mathrm{P}_{\omega,f,p}(z) \neq 0$) for all $\omega \in \mathrm{St}(A)$, then $\mathrm{P}_{\omega,f,p}(y, z) = \mathrm{P}_{\omega,p}(y, z)$ for any preparation test $\omega\colon I \to Y \cdot A$ and $y \in Y$.*

*Proof.* (i) $\Longrightarrow$ (iii): Let $p\colon A \to Z \cdot A$ be an observable and $z \in Z$. Suppose that $x' \in X$ satisfies $\mathrm{P}_{\omega,f,p}(x' \mid z) = \mathbb{1}$ for any $\omega \in \mathrm{St}(A)$ with $\mathrm{P}_{\omega,f,p}(z) \neq 0$. Take an arbitrary state $\omega \in \mathrm{St}(A)$. If $\mathrm{P}_{\omega,f,p}(z) \neq 0$, then

$$\mathrm{P}_{\omega,f,p}(x', z) = \mathrm{P}_{\omega,f,p}(x' \mid z)\, \mathrm{P}_{\omega,f,p}(z) = \mathrm{P}_{\omega,f,p}(z) = \bigovee_{x \in X} \mathrm{P}_{\omega,f,p}(x, z)\,.$$

By cancellation $\mathrm{P}_{\omega,f,p}(x, z) = 0$ for all $x \in X \backslash x'$. If $\mathrm{P}_{\omega,f,p}(z) = 0$, then $\mathrm{P}_{\omega,f,p}(x, z) = 0$ for any $x \in X$. Thus in either case, we have $p_x \circ f_x \circ \omega = \mathrm{P}_{\omega,f,p}(x, z) = 0$ for all $x \in X \backslash \{x'\}$. Since $\omega$ was arbitrary, by separation we have $f_x \circ p_z = 0$ for all $x \in X \backslash \{x'\}$. Therefore $f_{x'} \circ p_z = p_z$ by (ii). Then for any preparation test $\omega\colon I \to Y \cdot A$,

$$\mathrm{P}_{\omega,f,p}(y,z) \equiv \bigovee_{x \in X} \mathrm{P}_{\omega,f,p}(y,x,z) = \bigovee_{x \in X} p_z \circ f_x \circ \omega_y = p_z \circ \omega_y \equiv \mathrm{P}_{\omega,p}(y,z)$$

(iii) $\Longrightarrow$ (ii): Suppose that $p \in \mathrm{Pred}(A)$ and $x \in X$ satisfy $p \circ f_x = 0$ for all $x \in X \setminus \{x'\}$. Let $\hat{p}\colon A \to Z \cdot I$ be an observable given by $Z = \{z_1, z_2\}$, $\hat{p}_{z_1} = p$ and $\hat{p}_{z_2} = p^{\perp}$. For any state $\omega \in \mathrm{St}(A)$, by assumption we have

$$\mathrm{P}_{\omega,f,\hat{p}}(z_1) = \bigovee_{x \in X} \mathrm{P}_{\omega,f,\hat{p}}(x, z_1) = \bigovee_{x \in X} p \circ f_x \circ \omega = p \circ f_{x'} \circ \omega \equiv \mathrm{P}_{\omega,f,\hat{p}}(x', z_1)\,. \quad (6.1)$$

It follows that $\mathrm{P}_{\omega,f,p}(x' \mid z) = \mathbb{1}$ whenever $\mathrm{P}_{\omega,f,p}(z) \neq 0$. By (iii), for any state $\omega \in \mathrm{St}(A)$ we have $\mathrm{P}_{\omega,f,\hat{p}}(z_1) = \mathrm{P}_{\omega,\hat{p}}(z_1)$, and hence

$$p \circ \omega \equiv \mathrm{P}_{\omega,\hat{p}}(z_1) = \mathrm{P}_{\omega,f,\hat{p}}(z_1) \overset{(6.1)}{=} \mathrm{P}_{\omega,f,\hat{p}}(x', z_1) \equiv p \circ f_{x'} \circ \omega\,.$$

Therefore $p = p \circ f_{x'}$ by separation.

(ii) $\Longrightarrow$ (i): Suppose that $g\colon A \to B$ and $x' \in X$ satisfy $g \circ f_x = 0$ for all $x \in X \backslash \{x'\}$. Let $p \in \mathrm{Pred}(B)$ be an arbitrary predicate. Then $p \circ g \circ f_x = 0$ for all $x \in X \setminus \{x'\}$. By (ii), we obtain $p \circ g \circ f_{x'} = p \circ g$. Since $p$ was arbitrary, $g \circ f_{x'} = g$ by separation. ■

We will study repeatable instruments that are C- or Q-ideal. First of all, such instruments are always idempotent:

**Proposition 6.3.19.** *Let $f\colon A \to X \cdot A$ is a repeatable instrument. If $f$ is C-ideal or Q-ideal, then $f$ is idempotent.*

*Proof.* Suppose that $f$ is C-ideal. Let $x \in X$ be fixed. Since $f_{x'} \circ f_x = 0$ for each $x' \in X \setminus \{x\}$, we obtain $f_x \circ f_x = f_x$ by the C-ideality. Therefore $f$ is idempotent. The proof is similar when $f$ is Q-ideal. ■

**Theorem 6.3.20.** *Let $f\colon A \to X \cdot A$ be an instrument. The following are equivalent.*

(i) *$f$ is repeatable and C-ideal.*

(ii) *$f$ is C-idempotent.*

*Proof.* We write $f = \langle\!\langle f_x \rangle\!\rangle_x$ and $p_x = \mathbb{1} f_x$.

(i) $\Longrightarrow$ (ii): By Proposition 6.3.19, $f$ is idempotent. Fix $x' \in X$. Suppose that $h\colon B \to A$ satisfies $h^{\square}(p_{x'}) = \mathbb{1}$. Then $p_{x'}^{\perp} \circ h = \mathbb{0}$. Since $p_{x'}^{\perp} = \bigovee_{x \neq x'} \mathbb{1} f_x$, we have $\bigovee_{x \neq x'} \mathbb{1} \circ f_x \circ h = \mathbb{0}$. Hence by positivity $\mathbb{1} \circ f_x \circ h = \mathbb{0}$ and so $f_x \circ h = 0$ for each $x \in X \setminus \{x'\}$. By C-ideality we obtain $f_{x'} \circ h = h$. Therefore $f_{x'}$ is a C-idempotent.

(ii)$\Longrightarrow$ (i): By Corollary 6.3.3, $f$ is repeatable. Suppose that $h\colon B \to A$ and $x' \in X$ satisfy $f_x \circ h = 0$ for all $x \in X \setminus \{x'\}$. Then

$$p_{x'}^{\perp} \circ h = \Big( \bigovee_{x \neq x'} \mathbb{1} f_x \Big) \circ h = \bigovee_{x \neq x'} \mathbb{1} \circ f_x \circ h = \mathbb{0}\,.$$

Therefore $h^{\square}(p_{x'}) = \mathbb{1}$, and $f_{x'} \circ h = h$ by the C-idempotency of $f_{x'}$. ■

**Lemma 6.3.21.** *Let $f = \langle\!\langle f_x \rangle\!\rangle_x \colon A \to X \cdot A$ be a Q-ideal repeatable instrument. Then for each $x' \in X$,*

$$\operatorname{im}\Big( \bigovee_{x \neq x'} f_x \Big) = \bigovee_{x \neq x'} \mathbb{1} f_x \equiv (\mathbb{1} f_{x'})^{\perp} .$$

*Proof.* Let $g = \bigovee_{x \neq x'} f_x$. Then

$$\mathbb{1} f_{x'} \circ g = \bigovee_{x \neq x'} \mathbb{1} \circ f_{x'} \circ f_x = \mathbb{0} ,$$

that is, $g^{\square}((\mathbb{1} f_{x'})^{\perp}) = \mathbb{1}$. Now assume that $p \in \mathrm{Pred}(A)$ satisfies $g^{\square}(p) = \mathbb{1}$, i.e. $p^{\perp} \circ g = \mathbb{0}$. Then

$$\bigovee_{x \neq x'} p^{\perp} \circ f_x = \mathbb{0} ,$$

and by positivity $p^{\perp} \circ f_x = \mathbb{0}$ for all $x \in X \setminus \{x'\}$. By Q-ideality, $p^{\perp} \circ f_{x'} = p^{\perp}$. Then

$$p^{\perp} = p^{\perp} \circ f_{x'} \leq \mathbb{1} \circ f_{x'} \quad ,$$

so that $(\mathbb{1} f_{x'})^{\perp} \leq p$. ■

**Lemma 6.3.22.** *Assume that the effectus is comprehensive. Let $f = \langle\!\langle f_x \rangle\!\rangle_x \colon A \to X \cdot A$ be a Q-idempotent instrument. Then $\operatorname{im}(\bigovee_{x \neq x'} f_x) = \bigovee_{x \neq x'} \mathbb{1} f_x$ for each $x' \in X$ if and only if $\operatorname{im}(f_x) = \mathbb{1} f_x$ for each $x \in X$.*

*Proof.* Let $f = \langle\!\langle f_x \rangle\!\rangle_x \colon A \to X \cdot A$ be a Q-idempotent instrument in a comprehensive effectus. Since $f$ is repeatable, for each $x' \in X$,

$$(\mathbb{1} f_{x'})^{\perp} \circ f_{x'} = \Big( \bigovee_{x \neq x'} \mathbb{1} f_x \Big) \circ f_{x'} = \bigovee_{x \neq x'} \mathbb{1} \circ f_x \circ f_{x'} = 0 ,$$

so that $\operatorname{im}(f_{x'}) \leq \mathbb{1} f_{x'}$. In particular, the images $\operatorname{im}(f_x)$ are summable. Hence by Propositions 5.2.21 and 5.5.18, for each $x' \in X$ we have

$$\operatorname{im}\Big( \bigovee_{x \neq x'} f_x \Big) = \bigvee_{x \neq x'} \operatorname{im}(f_x) = \bigovee_{x \neq x'} \operatorname{im}(f_x) .$$

Therefore if $\operatorname{im}(f_x) = \mathbb{1} f_x$ holds for each $x \in X$, clearly $\operatorname{im}(\bigovee_{x \neq x'} f_x) = \bigovee_{x \neq x'} \mathbb{1} f_x$ holds for each $x' \in X$.

Conversely, if $\operatorname{im}(\bigovee_{x \neq x'} f_x) = \bigovee_{x \neq x'} \mathbb{1} f_x$ for each $x' \in X$, then $\bigovee_{x \neq x'} \operatorname{im}(f_x) = \bigovee_{x \neq x'} \mathbb{1} f_x$. Since $\operatorname{im}(f_x) \leq \mathbb{1} f_x$, it follows that $\operatorname{im}(f_x) = \mathbb{1} f_x$ for all $x \in X \setminus \{x'\}$. Thus we have $\operatorname{im}(f_x) = \mathbb{1} f_x$ for all $x \in X$ if $X$ contains at least two elements. If $X$ is empty, the equivalence in question is trivial since both conditions are vacuous. Finally suppose that $X$ is a singleton, say $X = \{x\}$. Then $f_x \colon A \to A$ is a Q-idempotent with $\mathbb{1} f_x = \mathbb{1}$. By Q-idempotency, $f_x = \mathrm{id}_A \circ f_x = \mathrm{id}_A$. Therefore $\operatorname{im}(f_x) = \mathbb{1} = \mathbb{1} f_x$. ■

**Theorem 6.3.23.** *For any instrument $f = \langle\!\langle f_x \rangle\!\rangle_x \colon A \to X \cdot A$ in an effectus, the conditions* (i) *and* (ii) *below are equivalent. Moreover, in a comprehensive effectus, all conditions* (i)–(iii) *are equivalent.*

(i) *$f$ is repeatable and Q-ideal.*

(ii) *$f$ is Q-idempotent and* $\mathrm{im}(\bigovee_{x\neq x'} f_x) = \bigovee_{x\neq x'} \mathbb{1} f_x$ *for each $x' \in X$.*

(iii) *$f$ is Q-idempotent and* $\mathrm{im}(f_x) = \mathbb{1} f_x$ *for each $x \in X$.*

*Proof.* Equivalence (ii) $\iff$ (iii) in a comprehensive effectus holds by Lemma 6.3.22. Thus we focus on (i) $\iff$ (ii).

(i) $\Longrightarrow$ (ii): Lemma 6.3.21 proves $\mathrm{im}(\bigovee_{x\neq x'} f_x) = \bigovee_{x\neq x'} \mathbb{1} f_x$ for each $x' \in X$. By Proposition 6.3.19, $f$ is idempotent. Fix $x' \in X$. Suppose that $g\colon A \to B$ satisfies $\mathbb{1} g \le \mathbb{1} f_{x'}$. Then for each $x \in X \setminus \{x'\}$,

$$\mathbb{1} \circ g \circ f_x \le \mathbb{1} \circ f_{x'} \circ f_x = \mathbb{0}\,,$$

so that $g \circ f_x = 0$. Then $g \circ f_{x'} = g$ by Q-ideality. Hence $f_{x'}$ is a Q-idempotent.

(ii) $\Longrightarrow$ (i): Idempotency implies repeatability (Corollary 6.3.3). To see $Q$-ideality, suppose that $g\colon A \to B$ and $x' \in X$ satisfy $g \circ f_x = 0$ for all $x \in X \setminus \{x'\}$. Then

$$\mathbb{1} g \circ \Big(\bigovee_{x\neq x'} f_x\Big) = \bigovee_{x\neq x'} \mathbb{1} \circ g \circ f_x = \mathbb{0}\,.$$

By $\mathrm{im}(\bigovee_{x\neq x'} f_x) = \bigovee_{x\neq x'} \mathbb{1} f_x = (\mathbb{1} f_{x'})^\perp$, we obtain $(\mathbb{1} f_{x'})^\perp \le (\mathbb{1} g)^\perp$, i.e. $\mathbb{1} g \le \mathbb{1} f_{x'}$. Hence $g \circ f_{x'} = g$ by the Q-idempotency of $f_{x'}$. ∎

In a comprehensive effectus, images $\mathrm{im}(f_x)$ are always sharp predicates; see Proposition 5.5.13. Therefore we obtain the following corollary.

**Corollary 6.3.24.** *If $f\colon A \to X \cdot A$ is a Q-ideal repeatable instrument in a comprehensive effectus, then the measured observable $f; \mathbb{1}\colon A \to X \cdot I$ is sharp.* ∎

The corollary answers in a way the questions of the relationship between repeatable instruments and sharp observables. Note that a similar statement for C-ideality does not hold.

**Example 6.3.25.** In Example 6.3.8 we saw an example of a repeatable instrument that measures a non-sharp observable. In fact, the instrument given there is C-ideal. By Theorem 6.3.20, it suffices to show that the instrument $f = \langle\!\langle f_0, f_1 \rangle\!\rangle$ is C-idempotent. Recall that $f_0\colon \mathcal{M}_3 \to \mathcal{M}_3$ is given by $f_0(B) = \langle 0|B|0\rangle A_0$ using some matrix $A_0$. The map $f$ is an idempotent that splits as:

$$\mathcal{M}_3 \xrightarrow{\langle 0|-|0\rangle} \mathbb{C} \xrightarrow{(-)\cdot A_0} \mathcal{M}_3\,.$$

Here $\langle 0|-|0\rangle\colon \mathcal{M}_3 \to \mathbb{C}$ is a comprehension of the projection $|0\rangle\langle 0|$, as

$$|0\rangle\langle 0|\mathcal{M}_3|0\rangle\langle 0| = \big\{ |0\rangle\langle 0|B|0\rangle\langle 0| \;\big|\; B \in \mathcal{M}_3 \big\} \cong \mathbb{C}\,.$$

This proves that $f_0\colon \mathcal{M}_3 \to \mathcal{M}_3$ is a C-idempotent by Lemma 6.3.10. Therefore the observable measured by a C-ideal repeatable instrument is not necessarily sharp, unlike Q-ideal repeatable instruments.

Under an additional assumption on instruments, however, we can prove a version of Corollary 6.3.24 for C-ideality. The following definition comes from [66, 183].

**Definition 6.3.26.** We say that an instrument $f = \langle\!\langle f_x \rangle\!\rangle_x \colon A \to X \cdot A$ is **nondegenerate** provided that for each predicate $p \in \mathrm{Pred}(A)$, if $p \circ f_x = \mathbb{0}$ for all $x \in X$, then $p = \mathbb{0}$.

**Lemma 6.3.27.** *An instrument $f = \langle\!\langle f_x \rangle\!\rangle_x \colon A \to X \cdot A$ is nondegenerate if and only if the sum $\bigovee_{x \in X} f_x \colon A \to A$ is faithful.*

*Proof.* This follows from the fact that for each predicate $p \in \mathrm{Pred}(A)$, one has $p \circ (\bigovee_{x \in X} f_x) = \mathbb{0}$ if and only if $\bigovee_{x \in X} p \circ f_x = \mathbb{0}$ if and only if $p \circ f_x = \mathbb{0}$ for all $x \in X$. ∎

**Theorem 6.3.28.** *If $f \colon A \to X \cdot A$ is a C-ideal, repeatable, and nondegenerate instrument in a comprehensive effectus, then the measured observable $f ; \mathbb{1}$ is sharp.*

*Proof.* Let $f = \langle\!\langle f_x \rangle\!\rangle_x$ be as in the hypothesis. We write $p_x = \mathbb{1} f_x$. Then $f$ is C-idempotent by Theorem 6.3.20. By assumption, the effectus has comprehension, so by Proposition 6.3.13 every idempotent $f_x$ splits through comprehension as follows.

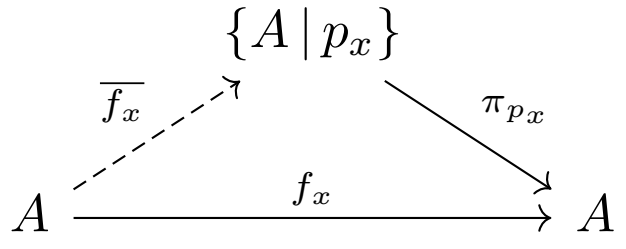

Note here that $\overline{f_x}$ is faithful since it is a (split) epi. By assumption $f$ is nondegenerate and thus $\bigovee_{x \in X} f_x$ is faithful by Lemma 6.3.27. So we have

$$
\begin{aligned}
\mathbb{1} = \mathrm{im}\Big(\bigovee_{x \in X} f_x\Big) &= \bigvee_{x \in X} \mathrm{im}(f_x) && \text{by Proposition 5.2.21} \\
&= \bigvee_{x \in X} \mathrm{im}(\pi_{p_x} \circ \overline{f_x}) \\
&= \bigvee_{x \in X} \mathrm{im}(\pi_{p_x}) && \text{by Lemma 5.2.15} \\
&= \bigvee_{x \in X} \lfloor p_x \rfloor \\
&= \bigovee_{x \in X} \lfloor p_x \rfloor && \text{by Proposition 5.5.18}\,.
\end{aligned}
$$

Since $\bigovee_{x \in X} p_x = \mathbb{1}$ and $\lfloor p_x \rfloor \le p_x$ it follows that $\lfloor p_x \rfloor = p_x$ for all $x \in X$, that is, every $p_x = \mathbb{1} f_x$ is sharp. ∎

Davies and Lewis [66] called C-ideal (= d-ideal), repeatable, and nondegenerate instruments *strongly repeatable*. The theorem above captures a part of [66, Theorem 10] in our abstract setting. In the next subsection we prove a statement that corresponds to [66, Theorem 10] for finite outcome sets, under an additional assumption.

### 6.3.4 Lüders instruments

In Example 6.3.7 we presented a 'Lüders' repeatable instrument for each sharp observable in $\mathbf{Wstar}_{\lesssim}^{\mathrm{op}}$. We now define Lüders instruments in the abstract setting of effectuses. They are defined in terms of comprehension and quotients.

**Definition 6.3.29.** Let $\mathbf{C}$ be a comprehensive effectus. We say that **C has Lüders instruments** if for each sharp predicate $\mathfrak{p} \in \mathrm{ShPred}(A)$, there exists a morphism $\zeta_{\mathfrak{p}} \colon A \to \{A \,|\, \mathfrak{p}\}$ such that

- $\big(\{A \,|\, \mathfrak{p}\} \xrightarrow{\pi_{\mathfrak{p}}} A \xrightarrow{\zeta_{\mathfrak{p}}} \{A \,|\, \mathfrak{p}\}\big) = \mathrm{id}_{\{A|\mathfrak{p}\}}$, so that $\{A \,|\, \mathfrak{p}\}$ is a retract of $A$;
- the map $\zeta_{\mathfrak{p}} \colon A \to \{A \,|\, \mathfrak{p}\}$ is a quotient for $\mathfrak{p}^{\perp}$.

Then for each $\mathfrak{p} \in \mathrm{ShPred}(A)$, the retraction $\zeta_{\mathfrak{p}}$ yields a split idempotent as below.

$$\mathrm{asrt}_{\mathfrak{p}} \coloneqq \big(A \xrightarrow{\zeta_{\mathfrak{p}}} \{A \,|\, \mathfrak{p}\} \xrightarrow{\pi_{\mathfrak{p}}} A\big)$$

It is called the **assert map** for $\mathfrak{p}$. It satisfies $\mathbb{1} \circ \mathrm{asrt}_{\mathfrak{p}} = \mathfrak{p}$, see Lemma 6.3.30 below. Therefore, for each sharp observable $\mathfrak{p} = \langle\!\langle \mathfrak{p}_x \rangle\!\rangle_x \colon A \to X \cdot I$ one can define a $\mathfrak{p}$-compatible instrument by

$$\mathrm{instr}_{\mathfrak{p}} \coloneqq \langle\!\langle \mathrm{asrt}_{\mathfrak{p}_x} \rangle\!\rangle_x \colon A \longrightarrow X \cdot A\,.$$

It is called the **Lüders instrument** for observable $\mathfrak{p}$.

**Lemma 6.3.30.** *For each* $\mathfrak{p} \in \mathrm{ShPred}(A)$*, the following equations hold.*

(i) $\mathbb{1} \circ \mathrm{asrt}_{\mathfrak{p}} = \mathfrak{p}$.

(ii) $\mathrm{im}(\mathrm{asrt}_{\mathfrak{p}}) = \mathfrak{p}$.

(iii) $\mathfrak{p} \circ \mathrm{asrt}_{\mathfrak{p}} = \mathfrak{p}$.

(iv) $\mathfrak{p}^{\perp} \circ \mathrm{asrt}_{\mathfrak{p}} = \mathbb{0}$.

*Proof.*

(i) Since $\pi_{\mathfrak{p}}$ is total and $\zeta_{\mathfrak{p}}$ is a quotient for $p^{\perp}$,

$$\mathbb{1} \circ \mathrm{asrt}_{\mathfrak{p}} = \mathbb{1} \circ \pi_{\mathfrak{p}} \circ \zeta_{\mathfrak{p}} = \mathbb{1} \circ \zeta_{\mathfrak{p}} = \mathfrak{p}\,.$$

(ii) Note that $\zeta_{\mathfrak{p}_x}$ is a split epi and hence faithful. Thus by Lemma 5.2.15,

$$\mathrm{im}(\mathrm{asrt}_{\mathfrak{p}}) = \mathrm{im}(\pi_{\mathfrak{p}} \circ \zeta_{\mathfrak{p}}) = \mathrm{im}(\pi_{\mathfrak{p}}) = \lfloor \mathfrak{p} \rfloor = \mathfrak{p}\,.$$

(iii) By (i) and idempotency of $\mathrm{asrt}_{\mathfrak{p}}$,

$$\mathfrak{p} \circ \mathrm{asrt}_{\mathfrak{p}} = \mathbb{1} \circ \mathrm{asrt}_{\mathfrak{p}} \circ \mathrm{asrt}_{\mathfrak{p}} = \mathbb{1} \circ \mathrm{asrt}_{\mathfrak{p}} = \mathfrak{p}\,.$$

(iv) By (i) and (iii), we have $\mathfrak{p} \circ \mathrm{asrt}_{\mathfrak{p}} = \mathfrak{p} = \mathbb{1} \circ \mathrm{asrt}_{\mathfrak{p}}$, which implies $\mathfrak{p}^{\perp} \circ \mathrm{asrt}_{\mathfrak{p}} = \mathbb{0}$. ∎

**Example 6.3.31.** Our main examples of effectuses have Lüders instruments. Recall comprehensions and quotients in the effectuses from Example 5.3.4 and Example 5.4.5, respectively.

(i) Let $P \subseteq A$ be a predicate/subset on a set $A$ in **Pfn** (all predicates are sharp in **Pfn**). We have a quotient and a comprehension:

$$A \xrightarrow{\xi_{P^\perp} = \zeta_P} A/P^\perp = P = \{A \,|\, P\} \xrightarrow{\pi_P} A\,.$$

Here $\xi_{P^\perp} = \zeta_P$ is the obvious partial function defined on $P$, and $\pi_P$ is the (total) inclusion. Clearly, $\zeta_P \circ \pi_P = \mathrm{id}$. The assert map $\mathrm{asrt}_P \equiv \pi_P \circ \zeta_P \colon A \to A$ is:

$$\mathrm{asrt}_P(a) = \begin{cases} a & \text{if } a \in P \\ \text{undefined} & \text{if } a \notin P\,. \end{cases}$$

Now let $p\colon A \to X \cdot 1$ be an observable with outcome set $X$. We identify it with a partition $(P_x)_{x\in X}$ of the set $A$. Then the Lüders instrument $\mathrm{instr}_P \equiv \langle\!\langle \mathrm{asrt}_{P_x} \rangle\!\rangle_x \colon A \to X \cdot A$ is given by for each $a \in P_x$,

$$\mathrm{instr}_P(a) = \kappa_x(a)\,.$$

(ii) Let $\mathfrak{p}$ be a sharp predicate on $A$ in $\mathcal{K}\ell(\mathcal{D}_\le)$. It is a predicate $\mathfrak{p} \in [0,1]^A$ such that $p(a) \in \{0,1\}$ for all $a \in A$. Then the object parts of the quotient for $\mathfrak{p}^\perp$ and comprehension of $\mathfrak{p}$ agree:

$$\begin{aligned} A/\mathfrak{p}^\perp &= \{a \in A \mid \mathfrak{p}^\perp(a) < 1\} \\ &= \{a \in A \mid \mathfrak{p}(a) = 1\} \\ &= \{a \in A \mid \mathfrak{p}(a) > 0\} = \{A \,|\, \mathfrak{p}\}\,. \end{aligned}$$

Since we focus on sharp predicates $\mathfrak{p}$ — which are identified with subsets $P \subseteq A$ — assert maps and instruments in $\mathcal{K}\ell(\mathcal{D}_\le)$ are essentially the same as those in **Pfn**. For example, for a sharp observable $\mathfrak{p} = \langle\!\langle \mathfrak{p}_x \rangle\!\rangle_{x\in X}$, the Lüders instrument $\mathrm{instr}_\mathfrak{p} \colon A \to \mathcal{D}(X \cdot A)$ is given by

$$\mathrm{instr}_\mathfrak{p}(a) = 1|\kappa_x(a)\rangle$$

for each $a \in A$ with $\mathfrak{p}_x(a) = 1$.

(iii) In $\mathbf{Wstar}_\le^{\mathrm{op}}$, sharp predicates on a $W^*$-algebra $\mathscr{A}$ are projections i.e. elements $\mathfrak{p} \in \mathscr{A}$ with $\mathfrak{p}^* = \mathfrak{p} = \mathfrak{p}^2$. For such a sharp predicate $\mathfrak{p}$, recall from Examples 5.3.4 and 5.4.5 that

$$\mathscr{A}/\mathfrak{p}^\perp = \lceil \mathfrak{p} \rceil \mathscr{A} \lceil \mathfrak{p} \rceil = \mathfrak{p}\mathscr{A}\mathfrak{p} = \lfloor \mathfrak{p} \rfloor \mathscr{A} \lfloor \mathfrak{p} \rfloor = \{\mathscr{A} \,|\, \mathfrak{p}\}\,.$$

The quotient map $\xi_{\mathfrak{p}^\perp} = \zeta_\mathfrak{p} \colon \mathfrak{p}\mathscr{A}\mathfrak{p} \to \mathscr{A}$ and comprehension map $\pi_\mathfrak{p} \colon \mathscr{A} \to \mathfrak{p}\mathscr{A}\mathfrak{p}$ are given by $\zeta_\mathfrak{p}(a') = \mathfrak{p}a'\mathfrak{p}$ and $\pi_\mathfrak{p}(a) = \mathfrak{p}a\mathfrak{p}$. By definition, elements in $\mathfrak{p}\mathscr{A}\mathfrak{p}$ are of the form $\mathfrak{p}a\mathfrak{p}$ for $a \in \mathscr{A}$, and thus by

$$\pi_\mathfrak{p}(\zeta_\mathfrak{p}(\mathfrak{p}a\mathfrak{p})) = \mathfrak{p}\mathfrak{p}\mathfrak{p}a\mathfrak{p}\mathfrak{p}\mathfrak{p} = \mathfrak{p}a\mathfrak{p}$$

we obtain $\pi_\mathfrak{p} \circ \zeta_\mathfrak{p} = \mathrm{id}$. Therefore $\zeta_\mathfrak{p} \circ \pi_\mathfrak{p} = \mathrm{id}$ in the opposite $\mathbf{Wstar}_\le^{\mathrm{op}}$, as desired. Then the assert map $\mathrm{asrt}_\mathfrak{p} \colon \mathscr{A} \to \mathscr{A}$ is given by $\mathrm{asrt}_\mathfrak{p}(a) = \mathfrak{p}a\mathfrak{p}$. For a sharp observable $\mathfrak{p} = (\mathfrak{p}_x)_x \in \mathscr{A}^X$, the Lüders instrument $\mathrm{instr}_{\mathfrak{p}} = \langle\!\langle \mathrm{asrt}_{\mathfrak{p}_x} \rangle\!\rangle_x$ is precisely what is called the 'Lüders instrument' in quantum theory, see e.g. [120]. As a morphism $\mathscr{A}^X \to \mathscr{A}$ it is given by:

$$\mathrm{instr}_{\mathfrak{p}}((a_x)_x) = \sum_{x \in X} \mathfrak{p}_x a_x \mathfrak{p}_x \,.$$

Below in Proposition 6.3.33, we characterize Lüders instruments as unique CQ-ideal repeatable instruments. For this we first observe the uniqueness of CQ-ideal repeatable instruments.

**Lemma 6.3.32.** *Let $f = \langle\!\langle f_x \rangle\!\rangle_x \colon A \to X \cdot A$ and $g = \langle\!\langle g_x \rangle\!\rangle_x \colon A \to X \cdot A$ be instruments that measure a common observable, i.e. $\mathbb{1} f_x = \mathbb{1} g_x$ for all $x \in X$. If*

- *$f$ is repeatable and Q-ideal; and*
- *$g$ is C-ideal,*

*then $f = g$. In particular, CQ-ideal repeatable instruments that measure a common observable are unique.*

*Proof.* Let us fix an arbitrary $x \in X$. Then for each $x' \in X \setminus \{x\}$,

$$\mathbb{1} \circ g_x \circ f_{x'} = \mathbb{1} \circ f_x \circ f_{x'} = \mathbb{0}$$

and hence $g_x \circ f_{x'} = 0$. By the Q-ideality of $f$, we obtain $g_x \circ f_x = g_x$. On the other hand, for each $x' \in X \setminus \{x\}$,

$$\mathbb{1} \circ g_{x'} \circ f_x = \mathbb{1} \circ f_{x'} \circ f_x = \mathbb{0}$$

and thus $g_{x'} \circ f_x = 0$. By the C-ideality of $g$, we obtain $g_x \circ f_x = f_x$. Therefore $f_x = g_x \circ f_x = g_x$. ∎

**Proposition 6.3.33.** *Let $\mathbf{C}$ be a comprehensive effectus. The following are equivalent.*

(i) *$\mathbf{C}$ has Lüders instruments.*

(ii) *For each sharp predicate $\mathfrak{p} \in \mathrm{ShPred}(A)$, there exists a quotient $\xi_{\mathfrak{p}^\perp} \colon A \to A/p^\perp$ for $\mathfrak{p}^\perp$, and the composite*

$$\{A \,|\, \mathfrak{p}\} \xrightarrow{\pi_{\mathfrak{p}}} A \xrightarrow{\xi_{\mathfrak{p}^\perp}} A/\mathfrak{p}^\perp$$

*is an isomorphism.*

(iii) *For each sharp observable $\mathfrak{p} \colon A \to X \cdot I$, there exists a CQ-ideal repeatable instrument $f \colon A \to X \cdot A$ that measures $\mathfrak{p}$.*

(iv) *For each sharp predicate $\mathfrak{p} \in \mathrm{ShPred}(A)$, there exists a CQ-ideal repeatable instrument $f \colon A \to A + A$ that measures $\langle\!\langle \mathfrak{p}, \mathfrak{p}^\perp \rangle\!\rangle$.*

*By Lemma* 6.3.32*, if CQ-ideal repeatable instruments exist, they are unique. When the equivalent conditions above hold, the Lüders instrument* $\mathrm{instr}_{\mathfrak{p}} \colon A \to X \cdot A$ *for a sharp observable* $\mathfrak{p} \colon A \to X \cdot I$ *is precisely the unique CQ-ideal repeatable instrument.*

*Proof.* To see (ii) $\Longrightarrow$ (i), let $\theta$ be the inverse of the isomorphism $\xi_{\mathfrak{p}^\perp} \circ \pi_{\mathfrak{p}}$. Then the composite

$$A \xrightarrow{\xi_{\mathfrak{p}^\perp}} A/\mathfrak{p}^\perp \underset{\cong}{\xrightarrow{\theta}} \{A \,|\, p\}$$

is a quotient for $\mathfrak{p}^\perp$ and is a retraction of $\pi_{\mathfrak{p}}$, i.e. (i) holds. Since the converse is obvious, we have (i) $\Longleftrightarrow$ (ii).

Implication (iii) $\Longrightarrow$ (iv) is trivial. We next prove (iv) $\Longrightarrow$ (i). Let $\mathfrak{p} \in \mathrm{ShPred}(A)$ be an arbitrary sharp predicate, and let $f = \langle\!\langle f_1, f_2 \rangle\!\rangle \colon A \to A + A$ be a CQ-ideal repeatable instrument that measures $\langle\!\langle \mathfrak{p}, \mathfrak{p}^\perp \rangle\!\rangle$. Then $\mathbb{1} f_1 = \mathfrak{p}$. By Theorems 6.3.20 and 6.3.23, the map $f_1 \colon A \to A$ is a CQ-idempotent. Since $f_1$ is a C-idempotent and the effectus has comprehension, by Proposition 6.3.13, the idempotent $f_1$ splits via comprehension $\{A \,|\, \mathfrak{p}\}$ as follows.

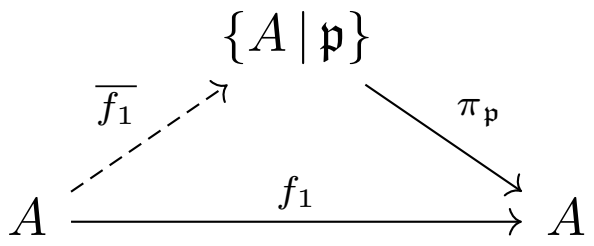

Then $f_1$ is a split Q-idempotent, so that $\overline{f_1} \colon A \to \{A \,|\, \mathfrak{p}\}$ is a quotient for $\mathfrak{p}^\perp$ by Lemma 6.3.11. Therefore condition (i) is met.

Finally we prove (i) $\Longrightarrow$ (iii) and the last assertion. Let $\mathfrak{p} \colon A \to X \cdot I$ be a sharp observable. We claim that the Lüders instrument $\mathrm{instr}_{\mathfrak{p}} = \langle\!\langle \mathrm{asrt}_{\mathfrak{p}_x} \rangle\!\rangle_x \colon A \to X \cdot A$ from Definition 6.3.29 is CQ-ideal and repeatable. By construction, each $\mathrm{asrt}_{\mathfrak{p}_x}$ is a split CQ-idempotent. Thus $\mathrm{instr}_{\mathfrak{p}}$ is repeatable and C-ideal by Theorem 6.3.20. By Lemma 6.3.30 we have $\mathrm{im}(\mathrm{asrt}_{\mathfrak{p}_x}) = \mathbb{1} \circ \mathrm{asrt}_{\mathfrak{p}_x}$, so that $\mathrm{instr}_{\mathfrak{p}}$ is Q-ideal by Theorem 6.3.23. ■

Existence of Lüders instruments is a powerful assumption. For example, Corollary 6.3.24 and Theorem 6.3.28 can be strengthened as follows.

**Theorem 6.3.34.** *Let $f \colon A \to X \cdot A$ be an instrument in a comprehensive effectus with Lüders instruments. Write $p = f; \mathbb{1}$ for the observable measured by $f$. Then the following are equivalent.*

(i) *$p$ is sharp and $f$ is equal to the Lüders instrument* $\mathrm{instr}_p \colon A \to X \cdot A$.

(ii) *$p$ is sharp and $f$ is C-ideal.*

(iii) *$f$ is repeatable and Q-ideal.*

(iv) *$f$ is repeatable, C-ideal, and nondegenerate.*

*Proof.* By Proposition 6.3.33, Lüders instruments are CQ-ideal and repeatable. Thus (i) $\Longrightarrow$ (ii) and (i) $\Longrightarrow$ (iii) follow immediately. Also the Lüders instruments $\mathrm{instr}_p =$

$\langle\!\langle \mathrm{asrt}_{p_x} \rangle\!\rangle_x$ are nondegenerate by Lemma 6.3.27 and

$$\begin{aligned}
\mathrm{im}\Big(\bigovee_{x\in X} \mathrm{asrt}_{p_x}\Big) &= \bigvee_{x\in X} \mathrm{im}(\mathrm{asrt}_{p_x}) && \text{by Proposition 5.2.21} \\
&= \bigvee_{x\in X} p_x && \text{by Lemma 6.3.30} \\
&= \bigovee_{x\in X} p_x && \text{by Proposition 5.5.18} \\
&= \mathbb{1}\,.
\end{aligned}$$

Therefore (i) $\Longrightarrow$ (iv) holds.

Now assume (ii). Since $p$ is sharp there is the Lüders instrument $\mathrm{instr}_p$ for $p$. But then $f = \mathrm{instr}_p$ follows by Lemma 6.3.32. We proved (ii) $\Longrightarrow$ (i). By a similar reasoning one can prove (iii) $\Longrightarrow$ (i), as $p$ is sharp by Corollary 6.3.24, and also (iv) $\Longrightarrow$ (i) by Theorem 6.3.28. ∎

Instantiating (i) $\Longleftrightarrow$ (iv) in $\mathbf{Wstar}_{\leq}^{\mathrm{op}}$, we obtain [66, Theorem 10], for finite outcome sets. Similarly from (i) $\Longleftrightarrow$ (ii) we obtain [183, Corollary 3] or [27, Corollary 4.7.3]. As a corollary, sharp predicates can be characterized via repeatable instruments.

**Corollary 6.3.35.** *Let $p \in \mathrm{Pred}(A)$ be a predicate in a comprehensive effectus with Lüders instruments. Then the following are equivalent.*

(i) *$p$ is sharp.*

(ii) *There exists a (unique) repeatable Q-ideal instrument that measures $\langle\!\langle p, p^{\perp} \rangle\!\rangle$.*

(iii) *There exists a (unique) repeatable, C-ideal, nondegenerate instrument that measures $\langle\!\langle p, p^{\perp} \rangle\!\rangle$.* ∎

We present another application of Lüders instruments, which shows that for instruments that measures sharp observables, repeatability is equivalent to the property called *first-kindness* and *value reproducibility*. The result was originally proved by Busch et al. for usual quantum instruments [25, Theorem D].

**Theorem 6.3.36.** *Assume that the effectus is comprehensive, operationally well-behaved, and has Lüders instruments. Let $f\colon A \to X \cdot A$ be a instrument that measures a sharp observable, i.e. the observable $f;\mathbb{1}$ is sharp. Then the following are equivalent.*

(i) *$f$ is repeatable.*

(ii) *(first-kindness) $\mathrm{P}_{\omega,f}(\mathbf{o}_2 = x) = \mathrm{P}_{\omega,f,f}(\mathbf{o}_3 = x)$ for each state $\omega \in \mathrm{St}(A)$ and $x \in X$.*

(iii) *(value reproducibility) $\mathrm{P}_{\omega,f}(\mathbf{o}_2 = x) = \mathbb{1}$ implies $\mathrm{P}_{\omega,f,f}(\mathbf{o}_3 = x) = \mathbb{1}$ for each state $\omega \in \mathrm{St}(A)$ and $x \in X$.*

*Proof.* (i) $\Longrightarrow$ (ii) is easy:

$$\begin{aligned}
\mathrm{P}_{\omega,f}(\mathbf{o}_2 = x) &= \mathbb{1} \circ f_x \circ \omega \\
&= \bigovee_{x' \in X} \mathbb{1} \circ f_x \circ f_{x'} \circ \omega \\
&= \bigovee_{x' \in X} \mathrm{P}_{\omega,f,f}(\mathbf{o}_2 = x', \mathbf{o}_3 = x) \\
&= \mathrm{P}_{\omega,f,f}(\mathbf{o}_3 = x)\,,
\end{aligned}$$

and (ii) $\Longrightarrow$ (iii) is trivial. We will prove (iii) $\Longrightarrow$ (i). We write $\mathfrak{p} = \langle\!\langle \mathfrak{p}_x \rangle\!\rangle_x = \langle\!\langle \mathbb{1} f_x \rangle\!\rangle_x$ for the measured observable, which is sharp by assumption. We fix $x \in X$ towards proving $\mathfrak{p}_x \circ f_x = \mathfrak{p}_x$. Let $x \in X$ be an arbitrary outcome and $\omega \in \mathrm{St}(A)$ an arbitrary state. We assume $|\mathrm{asrt}_{\mathfrak{p}_x} \circ \omega| \equiv \mathfrak{p}_x \circ \omega \neq 0$ and let $\sigma$ be the normalization of $\mathrm{asrt}_{\mathfrak{p}_x} \circ \omega$. Then $\mathfrak{p}_x \circ \sigma = \mathbb{1}$, since

$$\mathfrak{p}_x \circ \sigma \circ (\mathfrak{p}_x \circ \omega) = \mathfrak{p}_x \circ \mathrm{asrt}_{\mathfrak{p}_x} \circ \omega = \mathfrak{p}_x \circ \omega = \mathbb{1} \circ (\mathfrak{p}_x \circ \omega)\,.$$

Therefore

$$\mathrm{P}_{\sigma,f}(\mathbf{o}_2 = x) = \mathbb{1} \circ f_x \circ \sigma = \mathfrak{p}_x \circ \sigma = \mathbb{1}\,.$$

By assumption, we have $\mathrm{P}_{\sigma,f,f}(\mathbf{o}_3 = x) = 1$, i.e. $\bigovee_{x' \in X} \mathbb{1} \circ f_x \circ f_{x'} \circ \sigma = \mathbb{1}$. Multiplying $\mathfrak{p}_x \circ \omega$ to both sides, we obtain

$$\begin{aligned}
\mathfrak{p}_x \circ \omega &= \bigovee_{x' \in X} \mathbb{1} \circ f_x \circ f_{x'} \circ \sigma \circ (\mathfrak{p}_x \circ \omega) \\
&= \bigovee_{x' \in X} \mathfrak{p}_x \circ f_{x'} \circ \mathrm{asrt}_{\mathfrak{p}_x} \circ \omega \\
&= \mathfrak{p}_x \circ f_x \circ \mathrm{asrt}_{\mathfrak{p}_x} \circ \omega\,.
\end{aligned}$$

Here the last equality holds since for each $x' \neq x$, by Lemma 6.3.30(iv),

$$\mathbb{0} \leq \mathfrak{p}_x \circ f_{x'} \circ \mathrm{asrt}_{\mathfrak{p}_x} \circ \omega \leq \mathfrak{p}_{x'} \circ \mathrm{asrt}_{\mathfrak{p}_x} \circ \omega. \leq \mathfrak{p}_x^{\perp} \circ \mathrm{asrt}_{\mathfrak{p}_x} \circ \omega = \mathbb{0}\,.$$

If $\mathfrak{p}_x \circ \omega = 0$, then $\mathfrak{p}_x \circ f_x \circ \mathrm{asrt}_{\mathfrak{p}_x} \circ \omega = 0 = \mathfrak{p}_x \circ \omega$. Therefore $\mathfrak{p}_x \circ f_x \circ \mathrm{asrt}_{\mathfrak{p}_x} = \mathfrak{p}_x$ by separation. Since $\mathfrak{p}_x = \mathbb{1} \circ \mathrm{asrt}_{\mathfrak{p}_x}$, it follows that $(\mathrm{asrt}_{\mathfrak{p}_x})^{\square}(\mathfrak{p}_x \circ f_x) = \mathbb{1}$ and hence $\mathfrak{p}_x = \mathrm{im}(\mathrm{asrt}_{\mathfrak{p}_x}) \leq \mathfrak{p}_x \circ f_x$. From $\mathfrak{p}_x \circ f_x \leq \mathbb{1} \circ f_x = \mathfrak{p}_x$ we obtain $\mathfrak{p}_x \circ f_x = \mathfrak{p}_x$, that is, $\mathbb{1} \circ f_x \circ f_x = \mathbb{1} \circ f_x$. By Proposition 6.3.2, $f$ is repeatable. ■

**Remark 6.3.37.** We defined Lüders instruments only for sharp observables, following the standard usage of 'Lüders instrument' in quantum theory [26–28]. Nevertheless, as mentioned in Example 6.3.7, in the effectus $\mathbf{Wstar}_{\leq}^{\mathrm{op}}$ we can define the *generalized Lüders instrument* $f\colon \mathscr{A} \to X \cdot \mathscr{A}$ for general 'unsharp' observable $(p_x)_{x \in X}$ by $f_x(a) = \sqrt{p_x} a \sqrt{p_x}$, via square root. Bas Westerbaan in his thesis [256] worked on a categorical axiomatization of generalized Lüders instruments in an effectus. Note however that the axiomatization requires more assumptions on an effectus.

## 6.4 Side-effect-free measurements

In the previous section we discussed (C- and Q-) ideality of instruments. Roughly, ideality is a condition of 'minimal' disturbance: it asserts that a measurement does not disturb the state of a system when some condition is satisfied. Here, in contrast, we discuss *side-effect-free* instruments — measurements by such instruments cause no disturbance at all.

It is well known that measurements on quantum systems necessarily cause disturbance (cf. Heisenberg's uncertainty principle), and thus side-effect-free measurements are in general impossible. Below, indeed, we will show that in the effectus $\mathbf{Wstar}_{\leq}^{\mathrm{op}}$ of $W^*$-algebras, if all observables on a $W^*$-algebra $\mathscr{A}$ are side-effect-freely measurable, then $\mathscr{A}$ must be a commutative $W^*$-algebra — which means the system represented by $\mathscr{A}$ is 'classical'.

**Definition 6.4.1.** An instrument $f = \langle\!\langle f_x \rangle\!\rangle_x \colon A \to X \cdot A$ is **side-effect-free** if

$$\bigovee_{x\in X} f_x = \mathrm{id}_A \,.$$

In words, side-effect-freeness is a property that if we perform a measurement by the instrument and forget the outcome, then it is the same as doing nothing. We give characterizations of side-effect-freeness as a commutative diagram and in the operational language.

**Proposition 6.4.2.**

(i) *An instrument $f \colon A \to X \cdot A$ is side-effect-free if and only if the following diagram commutes.*

$$\begin{array}{ccc} A & \xrightarrow{\;f\;} & X \cdot A \\ & \searrow^{=} & \downarrow{\scriptstyle\nabla} \\ & & A \end{array}$$

(ii) *Suppose that the effectus is operationally well-behaved. An instrument $f \colon A \to X \cdot A$ is side-effect-free if and only if for all preparation and observation tests $\omega \colon I \to Y \cdot A$ and $\omega \colon A \to Z \cdot I$,*

$$\forall y \in Y.\, \forall z \in Z. \quad \mathrm{P}_{\omega,f,p}(y,z) = \mathrm{P}_{\omega,p}(y,z)\,.$$

*Proof.*

(i) Immediate from

$$\nabla \circ f = \nabla \circ \langle\!\langle f_x \rangle\!\rangle_{x\in X} = \bigovee_{x\in X} f_x \,,$$

see Proposition 3.1.8.

(ii) Suppose that $f$ is side-effect-free. Then for any preparation and observation tests $\omega \colon I \to Y \cdot A$ and $\omega \colon A \to Z \cdot I$,

$$\mathrm{P}_{\omega,f,p}(y,z) = \bigovee_{x\in X} p_z \circ f_x \circ \omega_y = p_z \circ \Big(\bigovee_{x\in X} f_x\Big) \circ \omega_y = p_z \circ \omega_y = \mathrm{P}_{\omega,p}(y,z)\,.$$

The converse follows by separation. ∎

As is clear form the intuition, side-effect-freeness implies C- and Q-ideality.

**Proposition 6.4.3.** *Any side-effect-free instrument is CQ-ideal.*

*Proof.* Let $f = \langle\!\langle f_x \rangle\!\rangle_x \colon A \to X \cdot A$ be a side-effect-free instrument. Suppose that $h \colon B \to A$ and $x' \in X$ satisfies $f_x \circ h = 0$ for all $x \in X \setminus \{x'\}$. Then

$$h = \mathrm{id} \circ h = \Big( \bigovee_{x \in X} f_x \Big) \circ h = \bigovee_{x \in X} f_x \circ h = f_{x'} \circ h \,.$$

Similarly, if $g \colon A \to B$ and $x' \in X$ satisfies $g \circ f_x = 0$ for all $x \in X \setminus \{x'\}$, then

$$g = g \circ \mathrm{id} = g \circ \Big( \bigovee_{x \in X} f_x \Big) = \bigovee_{x \in X} g \circ f_x = g \circ f_{x'} \,. \qquad \blacksquare$$

**Definition 6.4.4.**

(i) An observable $p \colon A \to X \cdot I$ is **side-effect-free** if there exists a side-effect-free instrument that measures $p$.

(ii) An object $A$ is **side-effect-free** if all observables $p \colon A \to X \cdot I$ on $A$ are side-effect-free.

**Example 6.4.5.**

(i) All objects in the effectus **Pfn** are side-effect-free. Indeed, for each observable on a set $A$, i.e. a partition $P = (P_x)_{x \in X}$ of $A$, the standard Lüders instrument $\mathrm{instr}_P \colon A \to X \cdot A$ from Example 6.3.31(i) is side-effect-free: it is easy to verify $\nabla \circ \mathrm{instr}_P = \mathrm{id}$.

(ii) All objects in the effectus $\mathcal{K}\ell(\mathcal{D})$ are side-effect-free, too. Let $p = \langle\!\langle p_x \rangle\!\rangle_x \colon A \to \mathcal{D}(X \cdot 1)$ be a observable in $\mathcal{K}\ell(\mathcal{D}_{\leq})$. We define an instrument $f \colon A \to \mathcal{D}(X \cdot A)$ by $f(a) = \sum_{x \in X} p_x(a) | \kappa_x(a) \rangle$. Then $f$ is side-effect-free, as:

$$(\nabla \circledcirc f)(a) = \nabla\Big( \sum_{x \in X} p_x(a) | \kappa_x(a) \rangle \Big) = \sum_{x \in X} p_x(a) | a \rangle = 1 | a \rangle \,.$$

Unlike the examples above, observables in $\mathbf{Wstar}_{\leq}^{\mathrm{op}}$ are in general not side-effect-free (as expected!). We investigate side-effect-free instruments/observables in $\mathbf{Wstar}_{\leq}^{\mathrm{op}}$ in some detail below.

**Proposition 6.4.6.** *Let $p = \langle\!\langle p_x \rangle\!\rangle_x \colon \mathscr{A} \to X \cdot \mathbb{C}$ be an observable in the effectus $\mathbf{Wstar}_{\leq}^{\mathrm{op}}$ of $W^*$-algebras — that is, a family $(p_x)_{x \in X}$ of effects $p_x \in \mathscr{A}$ such that $\bigovee_{x \in X} p_x = 1$. Assume that $p_x$ is central for each $x \in X$. Then maps $f_x \colon \mathscr{A} \to \mathscr{A}$ given by $f_x(a) = p_x \cdot a$ form a $p$-compatible side-effect-free instrument $f = \langle\!\langle f_x \rangle\!\rangle_x \colon \mathscr{A} \to X \cdot \mathscr{A}$.*

*Proof.* As $\sqrt{p_x}$ can be defined by functional calculus for $p_x$, $\sqrt{p_x}$ belongs to the unital $C^*$-subalgebra of $\mathscr{A}$ generated by $p_x$ (see e.g. [246, § I.4]). Since $p_x$ is central, it follows that $\sqrt{p_x}$ is central too. Thus $f_x(a) = p_x \cdot a = \sqrt{p_x} a \sqrt{p_x}$ and hence $f_x$ is a normal subunital CP map. Clearly $\langle\!\langle f_x \rangle\!\rangle_x$ is a $p$-compatible instrument, which is side-effect-free as:

$$\Big( \bigovee_{x \in X} f_x \Big)(a) = \sum_{x \in X} f_x(a) = \sum_{x \in X} p_x a = \Big( \sum_{x \in X} p_x \Big) a = a \,. \qquad \blacksquare$$

**Proposition 6.4.7.** *Let $f = \langle\!\langle f_x \rangle\!\rangle_x \colon \mathscr{A} \to X \cdot \mathscr{A}$ be a side-effect-free instrument in $\mathbf{Wstar}_{\leq}^{\mathrm{op}}$, and let $(p_x)_x = (f_x(1))_x$ be the observable measured by $f$. Then $p_x$ is central for each $x \in X$. Moreover, the instrument $f = \langle\!\langle f_x \rangle\!\rangle_x$ satisfies $f_x(a) = p_x \cdot a$ for each $x \in X$, or equivalently, $f((a_x)_{x\in X}) = \sum_{x\in X} p_x \cdot a_x$ as a map $\mathscr{A}^X \to \mathscr{A}$.*

*Proof.* The side-effect-free instrument is a normal unital CP map $f\colon \mathscr{A}^X \to \mathscr{A}$ such that $f \circ \Delta = \mathrm{id}_{\mathscr{A}}$. Here $\Delta\colon \mathscr{A} \to \mathscr{A}^X$ is the diagonal map, given by $\Delta(a) = (a)_{x\in X}$, which is a unital $*$-homomorphism. By Tomiyama's theorem [247, Theorem 1] (see also [91, Lemma 5]), we have

$$a \cdot f(b) = f(\Delta(a) \cdot b) \quad \text{and} \quad f(b) \cdot a = f(b \cdot \Delta(a)) \tag{6.2}$$

for each $a \in \mathscr{A}$ and $b \in \mathscr{A}^X$. Write $\delta_x \in \mathscr{A}^X$ for the tuple having 1 at $x$th coordinate and 0 elsewhere, that is, $\delta_x = \rhd_x(1)$ using the partial (co)projection $\rhd_x\colon \mathscr{A} \to \mathscr{A}^X$. Then for each $(a_x)_x \in \mathscr{A}^X$ one has $(a_x)_x = \sum_x \delta_x \cdot \Delta(a_x)$ and hence

$$\begin{aligned} f((a_x)_{x\in X}) &= \sum_{x\in X} f(\delta_x \cdot \Delta(a_x)) \\ &= \sum_{x\in X} f(\delta_x) \cdot a_x && \text{by (6.2)} \\ &= \sum_{x\in X} f_x(1) \cdot a_x = \sum_{x\in X} p_x \cdot a_x\,, \end{aligned}$$

as desired. Similarly one can also prove $f((a_x)_{x\in X}) = \sum_{x\in X} a_x \cdot p_x$. For each $x \in X$, this implies that $p_x \cdot a = f_x(a) = a \cdot p_x$, showing that $p_x$ is central. ■

**Corollary 6.4.8.** *An observable $p = (p_x)_{x\in X}$ on a $W^*$-algebra $\mathscr{A}$ can be measured by a side-effect-free instrument if and only if $p_x$ is central for all $x \in X$.* ■

As a consequence, we obtain a neat characterization of side-effect-free objects in $\mathbf{Wstar}_{\leq}^{\mathrm{op}}$.

**Corollary 6.4.9.** *A $W^*$-algebra $\mathscr{A}$ is side-effect-free if and only if $\mathscr{A}$ is commutative.*

*Proof.* If $\mathscr{A}$ is side-effect-free, then for each effect/predicate $p \in \mathscr{A}$, the yes-no observable $\langle\!\langle p, p^{\perp} \rangle\!\rangle$ is side-effect-free. By Corollary 6.4.8 it follows that all effects in $\mathscr{A}$ are central. This implies that $\mathscr{A}$ is commutative, since any element of $\mathscr{A}$ is written as a linear combination of effects. The converse is obvious by Corollary 6.4.8. ■

**Remark 6.4.10.** The results above for $W^*$-algebras — Propositions 6.4.6 and 6.4.7 and Corollaries 6.4.8 and 6.4.9 — hold more generally for the effectus $\mathbf{Cstar}_{\leq}^{\mathrm{op}}$ of $C^*$-algebras, with the same proofs. (Note in particular that Tomiyama's theorem [247, Theorem 1] holds for $C^*$-algebras.) Our proofs are a rather obvious adaptation of Jacobs's similar results in [140, § 7], see in particular [140, Corollary 8.7]. Note, however, that our and Jacobs' settings are different. For an observable $p\colon A \to X \cdot I$, we call *any* morphism $f\colon A \to X \cdot A$ satisfying $f ; \mathbb{1} = p$ an instrument for $p$. In [140, § 7], on the other hand, *instruments* are defined as a fixed family of morphisms $\mathrm{instr}_p\colon A \to n \cdot A$ for $n$-outcome observables $p\colon A \to n \cdot I$ satisfying certain conditions [140, Assumption 2].

We turn to general effectuses, showing that side-effect-free predicates satisfy a certain effect-algebraic property, and can be related to MV-algebras.

**Definition 6.4.11.** A pair of elements $a, b \in E$ in an effect algebra is said to be **Mackey compatible** [74] (or **coexistent** [87]), written $a \leftrightarrow b$, if there exist summable elements $a', b', c \in E$ (i.e. $a' \ovee b' \ovee c$ exists) such that $a = a' \ovee c$ and $b = b' \ovee c$.

Following [74] we say that an effect algebra $E$ has the **Mackey property** if every pair of elements in $E$ is Mackey compatible, that is, $a \leftrightarrow b$ for all $a, b \in E$.

Mackey compatibility $a \leftrightarrow b$ intuitively means that it is possible to jointly measure $a$ and $b$ at once. More concretely, consider 2-outcome observables $\langle\!\langle p, p^\perp \rangle\!\rangle \colon A \to I + I$ and $\langle\!\langle q, q^\perp \rangle\!\rangle \colon A \to I + I$. Then $p$ and $q$ are Mackey compatible in $\mathrm{Pred}(A)$ if and only if there exists a 4-outcome observable $\langle\!\langle r_1, r_2, r_3, r_4 \rangle\!\rangle \colon A \to I + I + I + I$ such that $r_1 \ovee r_2 = p$ and $r_1 \ovee r_3 = q$. In that case, instead of measuring $\langle\!\langle p, p^\perp \rangle\!\rangle$ and $\langle\!\langle q, q^\perp \rangle\!\rangle$ separately, one can measure $\langle\!\langle r_1, r_2, r_3, r_4 \rangle\!\rangle$, getting an outcome in $\{1, 2, 3, 4\}$. Then outcome 1 is interpreted as the result 'both $p$ and $q$ are true', and outcome 2 as '$p$ is true and $q$ is false', and so on.

**Proposition 6.4.12.** *Let $p \in \mathrm{Pred}(A)$ be a predicate such that $\langle\!\langle p, p^\perp \rangle\!\rangle$ is side-effect-free. Then $p$ is Mackey compatible with any predicate $q \in \mathrm{Pred}(A)$.*

*Proof.* Let $\langle\!\langle f_1, f_2 \rangle\!\rangle \colon A \to A + A$ measure $\langle\!\langle p, p^\perp \rangle\!\rangle$. Then for any $q \in \mathrm{Pred}(A)$, we have

$$\begin{aligned} p &= \mathbb{1} \circ f_1 = (q \ovee q^\perp) \circ f_1 = q \circ f_1 \ovee q^\perp \circ f_1 \\ q &= q \circ \mathrm{id}_A = q \circ (f_1 \ovee f_2) = q \circ f_1 \ovee q \circ f_2 \,, \end{aligned}$$

and sum $q \circ f_1 \ovee q^\perp \circ f_1 \ovee q \circ f_2$ exists. Hence $p \leftrightarrow q$. ■

**Corollary 6.4.13.** *If an object $A$ is side-effect-free, then the effect algebra* $\mathrm{Pred}(A)$ *of predicates has the Mackey property.* ■

We now recall a standard result that relates the Mackey property of effect algebras to MV-algebras.

**Definition 6.4.14.** An **MV-algebra** is a commutative monoid $(E, +, 0)$ with a unary operation $(-)^\perp \colon E \to E$ satisfying the three equations below, for each $a, b \in E$:

$$a^{\perp\perp} = a \qquad a + 0^\perp = 0^\perp \qquad (a^\perp + b)^\perp + b = (b^\perp + a)^\perp + a \,.$$

The element $0^\perp$ is denoted by 1.

MV-algebras are introduced as algebraic models of the infinitely-many-valued logic of Łukasiewicz [32, 46], in a way analogous to Boolean algebras being models of classical propositional logic. Chovanec and Kôpka [44, 45] showed that MV-algebras can be identified with effect algebras satisfying extra properties.

**Theorem 6.4.15.** *MV-algebras and lattice effect algebras with the Mackey property are the same structures. More specifically, there is a one-to-one correspondence between the two structures, given as follows.*

- *Given an MV-algebra* $(E, +, 0, (-)^\perp)$*, define* $a \ovee b$ *if* $a^\perp + b^\perp = 1$*, and in that case* $a \ovee b = a + b$*. Then* $(E, \ovee, 0, 1)$ *is a lattice effect algebra with the Mackey property.*
- *Given a lattice effect algebra* $(E, \ovee, 0, 1)$ *with the Mackey property, define* $a + b = a \ovee (a^\perp \wedge b)$*. Then* $(E, +, 0, (-)^\perp)$ *is an MV-algebra.*

*Proof.* See Theorems 1.3.4, 1.8.12, and 1.10.6 of [74]. ■

**Corollary 6.4.16.** *Let* $A$ *be a side-effect-free object in an effectus. If* $\mathrm{Pred}(A)$ *has binary joins* $p \vee q$ *or meets* $p \wedge q$*, then* $\mathrm{Pred}(A)$ *forms an MV-algebra.*

*Proof.* Clearly if $\mathrm{Pred}(A)$ has binary joins or meets, then it has all finite joins and meets, i.e. $\mathrm{Pred}(A)$ is a lattice effect algebra. Then the assertion follows by Corollary 6.4.13 and Theorem 6.4.15. ■

We end the section by showing that in our main examples of effectuses, side-effect-free objects do satisfy the assumption of Corollary 6.4.16 — existence of joins/meets — and hence the predicates form an MV-algebra. It is an open problem to find a more natural condition on an effectus that implies the assumption of Corollary 6.4.16.

**Example 6.4.17.**

(i) In **Pfn**, predicates $\mathrm{Pred}(A) \cong \mathcal{P}(A)$ admit joins and meets given by unions and intersections. Hence predicates $\mathcal{P}(A)$ form an MV-algebra. (This is however immediate from the fact that any Boolean algebra is an MV-algebra.)

(ii) In $\mathcal{K}\ell(\mathcal{D}_{\leq})$, predicates $\mathrm{Pred}(A) \cong [0,1]^A$ admit joins/meets, calculated pointwise. Therefore predicates $[0,1]^A$ form an MV-algebra too.

(iii) Let $\mathscr{A}$ be a commutative $W^*$-algebra. Then by the Gelfand duality theorem, $\mathscr{A}$ is isomorphic to the algebra $\mathrm{C}(K)$ of $\mathbb{C}$-valued continuous functions on some compact Hausdorff space $K$. Predicates/effects $\mathrm{Pred}(\mathscr{A}) \cong [0,1]_{\mathrm{C}(K)}$ are then $[0,1]$-valued continuous functions, so that finite joins/meets exist, calculated pointwise. Therefore in $\mathbf{Wstar}_{\leq}^{\mathrm{op}}$, predicates on a side-effect-free object (= commutative $W^*$-algebra) form an MV-algebra.

## 6.5 Boolean measurements

In Sections 6.3 and 6.4 we have studied repeatability and side-effect-freeness of measurements. In this section we will study the combination of the two properties — which we call Booleanness. We will relate Boolean measurements with Boolean algebra structure of predicates.

**Definition 6.5.1.** We say that an instrument $f\colon A \to X \cdot A$ is **Boolean** if it is both repeatable and side-effect-free.

**Lemma 6.5.2.** *In any effectus:*

(i) *Boolean instruments are CQ-ideal and CQ-idempotent.*

(ii) *For each observable* $p\colon A \to X \cdot I$*, a* $p$*-compatible Boolean instrument is unique if it exists.*

*Proof.* Since Boolean instruments are side-effect-free by definition, they are CQ-ideal by Proposition 6.4.3. Then they are also CQ-idempotent by Theorems 6.3.20 and 6.3.23. The uniqueness holds since CQ-ideal repeatable instruments are unique, see Lemma 6.3.32. ■

**Definition 6.5.3.** We say that an endomorphism $f\colon A \to A$ is a **Boolean idempotent** if $f$ belongs to some Boolean instrument $g = \langle\!\langle g_x \rangle\!\rangle_x \colon A \to X \cdot A$, i.e. $f = g_x$ for some $x \in X$. We write $\mathrm{BIdem}(A)$ for the set of Boolean idempotents on $A$.

By Lemma 6.5.2(i) we immediately obtain:

**Corollary 6.5.4.** *Boolean idempotents are CQ-idempotents.* ■

Boolean idempotents admit the following characterization.

**Proposition 6.5.5.** *A morphism $f\colon A \to A$ is a Boolean idempotent if and only if there exists a morphism $g\colon A \to A$ satisfying*

- $\mathbb{1} g = (\mathbb{1} f)^{\perp}$;
- $f \ovee g = \mathrm{id}_A$;
- $f \circ g = 0_{AA} = g \circ f$.

*In that case, the partial tuple $\langle\!\langle f, g \rangle\!\rangle\colon A \to A + A$ is a Boolean instrument.*

*Proof.* If $g\colon A \to A$ is a morphism satisfying the required conditions, then clearly $\langle\!\langle f, g \rangle\!\rangle\colon A \to A + A$ is a Boolean instrument. Conversely assume that $f$ is a Boolean idempotent, i.e. there is a Boolean instrument $h = \langle\!\langle h_x \rangle\!\rangle_x \colon A \to X \cdot A$ and $f = h_{x'}$. Then it is easy to verify that the sum $g = \bigovee_{x \neq x'} h_x \colon A \to A$ satisfies the required conditions. ■

**Definition 6.5.6.** Let $f\colon A \to A$ be a Boolean idempotent. Then the morphism $g\colon A \to A$ given in Proposition 6.5.5 is unique by Lemma 6.5.2(ii). We write $f^{\perp} = g$ for this morphism and call it the **complement** of $f$.

It turns out that Boolean idempotents are commutative:

**Lemma 6.5.7.** *For any Boolean idempotents $f, g\colon A \to A$, one has $f \circ g = g \circ f$.*

*Proof.* Note that

$$0 = f \circ f^{\perp} = f \circ (g \ovee g^{\perp}) \circ f^{\perp} = f \circ g \circ f^{\perp} \ovee f \circ g^{\perp} \circ f^{\perp} \;.$$

Hence $f \circ g \circ f^{\perp} = 0$ by positivity (see Lemma 3.2.6). Similarly we have $f^{\perp} \circ g \circ f = 0$. Then

$$\begin{aligned} f \circ g &= f \circ g \circ (f \ovee f^{\perp}) \\ &= f \circ g \circ f \ovee f \circ g \circ f^{\perp} \\ &= f \circ g \circ f \\ &= f \circ g \circ f \ovee f^{\perp} \circ g \circ f \\ &= (f \ovee f^{\perp}) \circ g \circ f \\ &= g \circ f \,. \end{aligned}$$ ■

**Proposition 6.5.8.** *For each $A \in \mathbf{C}$ the following hold.*

(i) $\mathrm{id}_A, 0_{AA} \in \mathrm{BIdem}(A)$.

(ii) *Boolean idempotents $f, g \in \mathrm{BIdem}(A)$ are summable if and only if $f \circ g = 0$. In that case, $f \ovee g \in \mathrm{BIdem}(A)$.*

(iii) $\mathrm{BIdem}(A)$ *forms an effect algebra with the PCM structure of $\mathbf{C}(A, A)$, the complement $(-)^\perp$, and the top $\mathrm{id}_A$.*

*Proof.*

(i) Clearly $\langle\!\langle \mathrm{id}_A, 0_{AA} \rangle\!\rangle \colon A \to A + A$ forms a Boolean instrument.

(ii) Assume $f \perp g$. Then $\mathbb{1}f \perp \mathbb{1}g$, so that $\mathbb{1}f \leq (\mathbb{1}g)^\perp = \mathbb{1}g^\perp$. Therefore $\mathbb{1} \circ f \circ g \leq \mathbb{1} \circ g^\perp \circ g = 0$ and so $f \circ g = 0$. Conversely, assume $f \circ g = 0$. Then

$$\begin{aligned} \mathbb{1}f = \mathbb{1} \circ f \circ (g \ovee g^\perp) &= \mathbb{1} \circ f \circ g \ovee \mathbb{1} \circ f \circ g^\perp \\ &= \mathbb{1} \circ f \circ g^\perp \\ &\leq \mathbb{1} \circ g^\perp = (\mathbb{1}g)^\perp . \end{aligned}$$

Hence $\mathbb{1}f \perp \mathbb{1}g$ and thus $f \perp g$. Now we assume $f \perp g$ and prove that $f \ovee g \in \mathrm{BIdem}(A)$. To do so we show that $\langle\!\langle f \ovee g, f^\perp \circ g^\perp \rangle\!\rangle$ forms a Boolean instrument. By $f \circ g = 0$, we have

$$f = f \circ (g \ovee g^\perp) = f \circ g \ovee f \circ g^\perp = f \circ g^\perp$$

and similarly $g = f^\perp \circ g$. Then

$$\begin{aligned} \mathrm{id} &= (f \ovee f^\perp) \circ (g \ovee g^\perp) \\ &= f \circ g \ovee f \circ g^\perp \ovee f^\perp \circ g \ovee f^\perp \circ g^\perp \\ &= f \ovee g \ovee f^\perp \circ g^\perp . \end{aligned}$$

By the commutativity of Boolean idempotent, it is clear that

$$(f \ovee g) \circ (f^\perp \circ g^\perp) = 0 = (f^\perp \circ g^\perp) \circ (f \ovee g) .$$

Therefore $f \ovee g$ is a Boolean idempotent with $(f \ovee g)^\perp = f^\perp \circ g^\perp$.

(iii) The previous points show that $\mathrm{BIdem}(A)$ is a sub-PCM of $\mathbf{C}(A, A)$. By the definition of complements $f^\perp$, we have $f \ovee f^\perp = \mathrm{id}$. An element $g \in \mathrm{BIdem}(A)$ satisfying $f \ovee g = \mathrm{id}$ is unique: indeed, $f \perp g$ implies $g \circ f = 0$, since $\mathbb{1} \circ g \circ f = \mathbb{1} \circ f^\perp \circ f = \mathbb{0}$. Similarly $f \circ g = 0$ and hence $g = f^\perp$. Now assume $f \perp \mathrm{id}$. Then $\mathbb{1}f \perp \mathbb{1}$ in $\mathrm{Pred}(A)$, so that $\mathbb{1}f = \mathbb{0}$ and thus $f = 0$. We have proved that $\mathrm{BIdem}(A)$ is an effect algebra. ■

Since $\mathrm{BIdem}(A, A)$ is an effect algebra, there is a canonical partial order associated to it. This order turns out to coincide with the one of $\mathbf{C}(A, A)$.

**Lemma 6.5.9.** *Let $f, g \colon A \to A$ be Boolean idempotents. The following are equivalent.*

(i) *$f \leq g$ in $\mathrm{BIdem}(A, A)$, i.e. there exists $h \in \mathrm{BIdem}(A, A)$ such that $f \ovee h = g$.*

(ii) $f \leq g$ *in* $\mathbf{C}(A,A)$*, i.e. there exists* $h \in \mathbf{C}(A,A)$ *such that* $f \ovee h = g$*.*

(iii) $f \circ g^{\perp} = 0$.

(iv) $f \circ g = f$.

*Proof.* As is the case for any effect algebras, one has $f \leq g$ in $\mathrm{BIdem}(A,A)$ if and only if $f \perp g^{\perp}$. The latter is equivalent to $f \circ g = 0$ by Proposition 6.5.8(ii). Therefore (i) $\iff$ (iii). Now we prove the other conditions are equivalent.

(ii) $\Longrightarrow$ (iii): If $f \ovee h = g$, then $0 = g \circ g^{\perp} = (f \ovee h) \circ g^{\perp} = f \circ g^{\perp} \ovee h \circ g^{\perp}$. By positivity, $f \circ g^{\perp} = 0$.

(iii) $\Longrightarrow$ (iv): $f = f \circ (g \ovee g^{\perp}) = f \circ g \ovee f \circ g^{\perp} = f \circ g$.

(iv) $\Longrightarrow$ (ii): $g = (f \ovee f^{\perp}) \circ g = f \circ g \ovee f^{\perp} \circ g = f \ovee f^{\perp} \circ g$. Thus $f \leq g$ in $\mathbf{C}(A,A)$. ∎

We continue to study the order structure of Boolean idempotents, showing that $\mathrm{BIdem}(A)$ is a Boolean algebra.

**Proposition 6.5.10.** *For each* $A \in \mathbf{C}$ *the following hold.*

(i) $f \circ g \in \mathrm{BIdem}(A)$ *for each* $f, g \in \mathrm{BIdem}(A)$*, with*

$$(f \circ g)^{\perp} = f \circ g^{\perp} \ovee f^{\perp} \circ g \ovee f^{\perp} \circ g^{\perp} .$$

*Moreover* $f \circ g$ *is a meet of* $f$ *and* $g$ *in* $\mathrm{BIdem}(A)$*.*

(ii) $\mathrm{BIdem}(A)$ *is a Boolean algebra.*

(iii) *Moreover* $\mathrm{BIdem}(A)$ *is a Boolean effect algebra with the effect algebra structure from Proposition* 6.5.8 *— that is,* $f \perp g$ *iff* $f \wedge g = 0$*, and in that case* $f \ovee g = f \vee g$*.*

*Proof.*

(i) Let

$$h = f \circ g^{\perp} \ovee f^{\perp} \circ g \ovee f^{\perp} \circ g^{\perp} .$$

By the commutativity of Boolean idempotents, it is easy to verify that $\langle\!\langle f \circ g, h \rangle\!\rangle$ forms a Boolean instrument. We prove that $f \circ g$ is a meet of $f$ and $g$. Clearly $f \circ g$ is a lower bound of $f$ and $g$. Now assume $h \leq f$ and $h \leq g$. Then $h \circ f = h = h \circ g$ by Lemma 6.5.9. Therefore $h \circ (f \circ g) = h \circ g = h$, so that $h \leq f \circ g$.

(ii) Since meets in $\mathrm{BIdem}(A)$ are given by $f \wedge g = f \circ g$ and $(-)^{\perp}$ is an order-reversing involution on $\mathrm{BIdem}(A)$, joins are given via $(-)^{\perp}$ by

$$\begin{aligned} f \vee g = (f^{\perp} \wedge g^{\perp})^{\perp} = (f^{\perp} \circ g^{\perp})^{\perp} &= f \circ g \ovee f \circ g^{\perp} \ovee f^{\perp} \circ g \\ &= f \ovee f^{\perp} \circ g . \end{aligned}$$

Note then that $f^{\perp}$ is an order-theoretic complement:

$$\begin{aligned} f \wedge f^{\perp} &= f \circ f^{\perp} = 0 ; \\ f \vee f^{\perp} &= (f^{\perp} \wedge f)^{\perp} = 0^{\perp} = \mathrm{id} . \end{aligned}$$

Therefore BIdem$(A)$ is a complemented lattice. Finally we prove that BIdem$(A)$ is distributive.

$$\begin{aligned}(f \wedge g) \vee (f \wedge h) &= (f \circ g) \vee (f \circ h) \\ &= f \circ g \ovee (f \circ g)^{\perp} \circ (f \circ h) \\ &= f \circ g \ovee (f \circ g^{\perp} \ovee f^{\perp} \circ g \ovee f^{\perp} \circ g^{\perp}) \circ (f \circ h) \\ &= f \circ g \ovee f \circ g^{\perp} \circ h \\ &= f \circ (g \ovee g^{\perp} \circ h) \\ &= f \wedge (g \vee h)\,.\end{aligned}$$

(iii) By Proposition 6.5.8(ii) we have $f \perp g$ if and only if $f \wedge g = f \circ g = 0$. In that case, we have $g = (f \ovee f^{\perp}) \circ g = f^{\perp} \circ g$, and hence

$$f \vee g = f \ovee f^{\perp} \circ g = f \ovee g\,. \qquad \blacksquare$$

**Proposition 6.5.11.** *For each $A \in \mathbf{C}$, the mapping*

$$\mathrm{BIdem}(A) \longrightarrow \mathrm{Pred}(A)\,, \qquad f \longmapsto \mathbb{1}f$$

*that sends Boolean idempotents to their domain predicates is a homomorphism of effect algebras that reflects summability.*

*Proof.* Clearly $\mathbb{1}(-)\colon \mathrm{BIdem}(A) \to \mathrm{Pred}(A)$ is a PCM morphism. It sends the top $\mathrm{id}_A$ to $\mathbb{1}_A$, so $\mathbb{1}(-)$ is a homomorphism of effect algebras. It reflects the summability: if $\mathbb{1}f \perp \mathbb{1}g$, then $f \perp g$. ∎

**Definition 6.5.12.** An observable $p = \langle\!\langle p_x \rangle\!\rangle_x \colon A \to X \cdot I$ is **Boolean** if there exists a Boolean instrument $f\colon A \to X \cdot A$ that measures $p$. A predicate $p \in \mathrm{Pred}(A)$ is **Boolean** if the 2-outcome observable $\langle\!\langle p, p^{\perp} \rangle\!\rangle$ is Boolean. We write $\mathrm{BPred}(A) \subseteq \mathrm{Pred}(A)$ for the set of Boolean predicates.

We will use the following standard facts on effect algebras.

**Lemma 6.5.13.** *Let $f\colon A \to B$ be a homomorphism of effect algebras that reflects summability. Then*

(i) *The image $f[A] \subseteq B$ is an effect subalgebra of $B$.*

(ii) *$f$ is injective.*

*Proof.* It is straightforward to verify (i). For (ii), see e.g. [230, Proposition 2.2.2]. ∎

**Theorem 6.5.14.** *For each object $A \in \mathbf{C}$ in an effectus, the following hold.*

(i) *A predicate $p \in \mathrm{Pred}(A)$ is Boolean if and only if there exists a Boolean idempotent $f\colon A \to A$ such that $p = \mathbb{1}f$.*

(ii) $\mathrm{BPred}(A)$ *is an effect subalgebra of* $\mathrm{Pred}(A)$.

(iii) *The mapping $f \mapsto \mathbb{1}f$ defines an isomorphism* $\mathrm{BIdem}(A) \cong \mathrm{BPred}(A)$ *of effect algebras. As a consequence,* $\mathrm{BPred}(A)$ *is a Boolean effect algebra.*

*Proof.* Assertion (i) holds by Proposition 6.5.5. But then BPred($A$) is the image of the summability-reflecting effect algebra homomorphism $\mathbb{1}(-)\colon \mathrm{BIdem}(A) \to \mathrm{Pred}(A)$. Therefore (ii) follows by Lemma 6.5.13(i). Now by Lemma 6.5.13(ii), the map $\mathbb{1}(-)$ is injective. Thus the co-restriction $\mathbb{1}(-)\colon \mathrm{BIdem}(A) \to \mathrm{BPred}(A)$ is a bijective homomorphism of effect algebras that reflects summability, which is an isomorphism. We proved (iii). ∎

**Proposition 6.5.15.** *An instrument $f = \langle\!\langle f_x \rangle\!\rangle_x \colon A \to X \cdot A$ is Boolean if and only if $f_x \colon A \to A$ is a Boolean idempotent for each $x \in X$.*

*Proof.* The 'only if' holds by definition. Conversely assume that $f_x \colon A \to A$ is a Boolean idempotent for each $x \in X$. For each $x \neq x'$, by Proposition 6.5.8(ii) we have $f_x \circ f_{x'} = 0$. Again by Proposition 6.5.8(ii) the sum $\bigovee_{x\in X} f_x$ is a Boolean idempotent. Then $\bigovee_{x\in X} f_x = \mathrm{id}_A$ follows from

$$\mathbb{1} \circ \Big( \bigovee_{x\in X} f_x \Big) = \mathbb{1} = \mathbb{1} \circ \mathrm{id}_A \,.$$

and the injectivity of $\mathbb{1}(-)\colon \mathrm{BIdem}(A) \to \mathrm{Pred}(A)$. ∎

**Corollary 6.5.16.** *An observable $p = \langle\!\langle p_x \rangle\!\rangle_x \colon A \to X \cdot I$ is Boolean if and only if the predicate $p_x$ is Boolean for all $x \in X$.* ∎

**Definition 6.5.17.** We say that an effectus $\mathbf{C}$ is **Boolean** if all observables in $\mathbf{C}$ are Boolean, or equivalently (by Corollary 6.5.16), if all predicates in $\mathbf{C}$ are Boolean.

**Lemma 6.5.18.** *Let $E, D$ be Boolean effect algebras. If $f\colon E \to D$ is a homomorphism of effect algebras, then $f$ is a homomorphism of Boolean algebras, that is, $f$ preserves $\wedge$ and $\vee$.*

*Proof.* Since $f$ preserves orthosupplements/complements $(-)^\perp$, it suffices to prove that $f$ preserves $\vee$. Let $a, b \in E$ be arbitrary elements. Then we have $a = (a \wedge b) \ovee (a \wedge b^\perp)$ and $b = (a \wedge b) \ovee (a^\perp \wedge b)$, and hence

$$a \vee b = (a \wedge b) \vee (a \wedge b^\perp) \vee (a^\perp \wedge b) = (a \wedge b) \ovee (a \wedge b^\perp) \ovee (a^\perp \wedge b)\,.$$

Therefore

$$\begin{aligned} f(a) \vee f(b) &= (f(a \wedge b) \ovee f(a \wedge b^\perp)) \vee (f(a \wedge b) \ovee f(a^\perp \wedge b)) \\ &= f(a \wedge b) \vee f(a \wedge b^\perp) \vee f(a^\perp \wedge b) \\ &= f(a \wedge b) \ovee f(a \wedge b^\perp) \ovee f(a^\perp \wedge b) \\ &= f(a \vee b)\,. \end{aligned}$$ ∎

**Theorem 6.5.19.** *Let $\mathbf{C}$ be a Boolean effectus. Then the predicate functor on total morphisms is defined into the category* $\mathbf{BA}$ *of Boolean algebras, that is,* $\mathrm{Pred}\colon \mathrm{Tot}(\mathbf{C}) \to \mathbf{BA}^{\mathrm{op}}$.

*Proof.* For each $A \in \mathbf{C}$, $\mathrm{Pred}(A) = \mathrm{BPred}(A)$ is a Boolean effect algebra by Theorem 6.5.14. For each total map $f\colon A \to B$, the predicate transformer $f^*\colon \mathrm{Pred}(B) \to \mathrm{Pred}(A)$ is a homomorphism of effect algebras, and hence a homomorphism of Boolean algebras by Lemma 6.5.18. ∎

**Example 6.5.20.** We describe Boolean predicates in our main examples of effectuses.

(i) In the effectus **Pfn** all predicates are Boolean, i.e. **Pfn** is a Boolean effectus. Indeed, in Examples 6.3.31 and 6.4.5 we saw that all observables can be measured by repeatable and side-effect-free instruments.

(ii) By Corollary 6.3.24, Boolean predicates in a comprehensive effectus must be sharp. But then by what we observed in Examples 6.3.31 and 6.4.5, all sharp predicate in $\mathcal{K}\ell(\mathcal{D}_{\leq})$ are Boolean. Thus Boolean predicates in $\mathcal{K}\ell(\mathcal{D}_{\leq})$ are precisely sharp predicates, i.e. two-valued functions $p\colon A \to \{0,1\} \subseteq [0,1]$, which are identified with subsets $P \subseteq A$.

(iii) By Proposition 6.4.7, it is easy to see that Boolean predicates in $\mathbf{Wstar}_{\leq}^{\mathrm{op}}$ are precisely central projections $\mathfrak{p} \in \mathscr{A}$, that is, projections $\mathfrak{p}$ such that $\mathfrak{p}a = a\mathfrak{p}$ for all $a \in \mathscr{A}$.

**Remark 6.5.21.** Manes and Arbib introduced a notion of *guards* in a partially (countably) additive category, and proved that guards form a Boolean algebra [202, § 3.3]. Our notion of Boolean idempotents is basically the same as Manes and Arbib's guards (see Proposition 6.5.5), and so is our proof that Boolean idempotents form a Boolean algebra. In Manes and Arbib's proof, countable additivity is not necessary, but we do use the property that sums $\ovee$ are positive (i.e. $f \ovee g = 0$ implies $f = g = 0$), which follows from countable additivity.

**Remark 6.5.22.** If we assume that the effectus is comprehensive, the Booleanness of the effectus implies that all predicates are sharp by Corollary 6.3.24, and thus they form orthomodular lattices. By Corollary 6.4.16 predicates moreover form MV-algebras. From this it follows that predicates form Boolean algebras, since effect algebras that are both orthomodular lattices and MV-algebras are Boolean. In this section, however, we have shown a stronger result that Boolean predicates always form Boolean algebras, without the assumption that the effectus is comprehensive.

## 6.6 Extensive categories in effectus theory

*Extensivity* is a notion that captures 'well-behaved' coproducts. The notion is well-established and appears in a broad context, see e.g. [29–31, 47–49, 182, 219, 240, 242]. Examples of *extensive categories* — categories with extensive coproducts — include: any toposes, the category of topological spaces, and the opposite of the category of commutative rings.

Jacobs [140] observed relevance of extensivity in effectus theory, proving that every extensive category with a final object is an effectus (in total form). In this section, we strengthen Jacobs' observation by showing that extensive categories with a final object can be characterized as Boolean effectuses (Definition 6.5.17), in total form, satisfying certain additional properties. The result may be interpreted negatively: extensive categories (in particular, any toposes) viewed as effectuses are 'degenerate' in the sense that all predicates are Boolean.

### 6.6.1 Extensive categories

Here we briefly review basics of extensive categories. Extensive coproducts can be defined in a few equivalent ways.

**Proposition 6.6.1.** *Let* $\mathbf{B}$ *be a category with finite coproducts. Let* $A+B$ *be a binary coproduct in* $\mathbf{B}$*. The following are equivalent.*

(i) *The canonical functor given as below*

$$\begin{gathered} \mathbf{B}/A \times \mathbf{B}/B \xrightarrow{+} \mathbf{B}/(A+B) \\ (C \xrightarrow{f} A, D \xrightarrow{g} B) \longmapsto (C+D \xrightarrow{f+g} A+B) \end{gathered} \tag{6.3}$$

*is an equivalence of categories. Here* $\mathbf{B}/A$ *denotes the slice category over* $A$.

(ii) *The following two conditions hold.*

(a) *For any morphism* $h\colon E \to A+B$*, there exist objects* $E_1, E_2 \in \mathbf{B}$*, an isomorphism* $\theta\colon E_1 + E_2 \xrightarrow{\cong} E$*, and morphisms* $f_1\colon E_1 \to A$*,* $f_2\colon E_2 \to B$ *such that* $f \circ \theta = f_1 + f_2$.

(b) *For any morphisms* $f\colon C \to A$ *and* $g\colon D \to B$*, the following squares are pullbacks.*

$$\begin{array}{ccccc} C & \xrightarrow{\kappa_1} & C+D & \xleftarrow{\kappa_2} & D \\ {\scriptstyle f}\downarrow & \lrcorner & \downarrow{\scriptstyle f+g} & \llcorner & \downarrow{\scriptstyle g} \\ A & \xrightarrow{\kappa_1} & A+B & \xleftarrow{\kappa_2} & B \end{array}$$

(iii) *The following two conditions hold.*

(a) *There exist pullbacks of any morphism* $h\colon E \to A+B$ *along the coprojections into* $A+B$*, as in:*

$$\begin{array}{ccccc} E_1 & \longrightarrow & E & \longleftarrow & E_2 \\ \downarrow & \lrcorner & \downarrow{\scriptstyle h} & \llcorner & \downarrow \\ A & \xrightarrow{\kappa_1} & A+B & \xleftarrow{\kappa_2} & B \end{array}$$

(b) *Whenever we have a commutative diagram of the following form,*

$$\begin{array}{ccccc} E_1 & \longrightarrow & E & \longleftarrow & E_2 \\ \downarrow & & \downarrow & & \downarrow \\ A & \xrightarrow{\kappa_1} & A+B & \xleftarrow{\kappa_2} & B \end{array}$$

*the two inner squares are pullbacks if and only if the top row* $(E_1 \to E \leftarrow E_2)$ *is a coproduct.*

*Proof.* It is not hard to see that (ii)(a) holds iff the functor (6.3) is essentially surjective; and (ii)(b) holds iff (6.3) is full and faithful (see [30, Lemma 1] for details). Therefore (i) $\iff$ (ii) holds.

To see (ii) $\iff$ (iii), note first that the 'if' part of (iii)(b) is equivalent to (ii)(b). In the presence of (iii)(b), the top row of the pullbacks in (iii)(a) is a coproduct, yielding a canonical isomorphism $E \cong E_1 + E_2$ and proving (ii)(a). Thus (iii) $\Longrightarrow$ (ii).

Conversely, in the presence of (ii)(b), the decomposition $E \cong E_1 + E_2$ in (ii)(a) yields pullbacks in (iii)(a). Since pullbacks are unique up to isomorphism, this also implies that whenever we have pullbacks as in (iii)(a), the top row is a coproduct, i.e. the 'only if' part of (iii)(a) holds. Therefore (ii) $\Longrightarrow$ (iii). ■

**Definition 6.6.2.** Let **B** be a category with finite coproducts. A binary coproduct $A + B$ in **B** is **extensive** if any (and hence all) of the equivalent conditions in Proposition 6.6.1 holds. A category **B** is **extensive** if it has finite coproducts and all binary coproducts are extensive.

Below we will collect useful results on extensive categories. First we show that any finite coproduct $A_1 + \cdots + A_n$ in an extensive category is 'extensive', generalizing Proposition 6.6.1(iii).

**Proposition 6.6.3.** *Let $A_1 + \ldots + A_n$ be a coproduct in an extensive category.*

(i) *There exist pullbacks of any morphism $f\colon B \to A_1 + \cdots + A_n$ along any coprojection $\kappa_j\colon A_j \to A_1 + \cdots + A_n$, as in:*

$$\begin{array}{ccc} \bullet & \longrightarrow & B \\ \big\downarrow & \lrcorner & \big\downarrow{\scriptstyle f} \\ A_j & \xrightarrow{\kappa_j} & A_1 + \cdots + A_n \end{array}$$

(ii) *Suppose that we have a family of commutative diagrams below, for $j = 1, \ldots, n$.*

$$\begin{array}{ccc} B_j & \xrightarrow{h_j} & B \\ {\scriptstyle g_j}\big\downarrow & & \big\downarrow{\scriptstyle f} \\ A_j & \xrightarrow{\kappa_j} & A_1 + \cdots + A_n \end{array}$$

*Then every diagram is a pullback if and only if the family $(h_j\colon B_j \to B)_j$ forms an (n-ary) coproduct.*

*Proof.* Condition (i) and the 'if' part of (ii) hold since $n$-ary coprojections $\kappa_j\colon A_j \to \coprod_j A_j$ may be seen as binary coprojections $\kappa_1\colon A_j \to A_j + \coprod_{j' \neq j} A'_j \cong \coprod_j A_j$. To see the 'only if' part of (ii), let a family of pullbacks be given. Using Proposition 6.6.1(ii)(b) repeatedly, we can decompose $f\colon B \to \coprod_j A_j$ as $\coprod_j f_j\colon \coprod_j B_j \to \coprod_j A_j$ via some isomorphism $B \cong \coprod_j B_j$. We thus have the following pullback for each $j$:

$$\begin{array}{ccccc} B_j & \xrightarrow{\kappa_j} & B_1 + \cdots + B_n & \xrightarrow{\cong} & B \\ {\scriptstyle f_j}\big\downarrow & \lrcorner & \big\downarrow{\scriptstyle f_1 + \cdots + f_n} & \swarrow{\scriptstyle f} & \\ A_j & \xrightarrow{\kappa_j} & A_1 + \cdots + A_n & & \end{array}$$

By uniqueness of pullbacks, it follows that the family $(h_j\colon B_j \to B)$ is a coproduct. ■

**Lemma 6.6.4.** *In an extensive category* $\mathbf{B}$*, pullbacks are closed under taking coproducts: if the first two squares below are pullbacks, so is the third one.*

$$
\begin{array}{ccc}
A_1 & \xrightarrow{f_1} & B_1 \\
{\scriptstyle g_1}\downarrow & & \downarrow{\scriptstyle h_1} \\
C_1 & \xrightarrow{k_1} & D_1
\end{array}
\qquad
\begin{array}{ccc}
A_2 & \xrightarrow{f_2} & B_2 \\
{\scriptstyle g_2}\downarrow & & \downarrow{\scriptstyle h_2} \\
C_2 & \xrightarrow{k_2} & D_2
\end{array}
\qquad
\begin{array}{ccc}
A_1 + A_2 & \xrightarrow{f_1+f_2} & B_1 + B_2 \\
{\scriptstyle g_1+g_2}\downarrow & & \downarrow{\scriptstyle h_1+h_2} \\
C_1 + C_2 & \xrightarrow{k_1+k_2} & D_1 + D_2
\end{array}
$$

*Proof.* Note that pullbacks are products in a slice category. Thus if the first two squares above are pullbacks, they are products in $\mathbf{B}/D_1$ and $\mathbf{B}/D_2$, respectively. Then they together form a product in the product category $\mathbf{B}/D_1 \times \mathbf{B}/D_2$. By extensivity, the canonical functor $+\colon \mathbf{B}/D_1 \times \mathbf{B}/D_2 \to \mathbf{B}/(D_1 + D_2)$ is an equivalence, and hence preserves products. It follows that the third diagram above is a product in $\mathbf{B}/(D_1 + D_2)$, that is, a pullback in $\mathbf{B}$. ■

**Lemma 6.6.5.** *Let* $f, g\colon A \to B_1 + \cdots + B_n$ *be morphisms in an extensive category. Then* $f = g$ *if and only if for each* $j = 1, \ldots, n$*, there exists a common pullback of* $f$ *and* $g$ *along the coprojection* $\kappa_j\colon B_j \to B_1 + \cdots + B_n$*, i.e. there exist* $h_j\colon C_j \to A$ *and* $k_j\colon C_j \to B_j$ *such that the square*

$$
\begin{array}{ccc}
C_j & \xrightarrow{\quad h_j \quad} & A \\
{\scriptstyle k_j}\downarrow & & \downarrow{\scriptstyle ?} \\
B_j & \xrightarrow{\kappa_j} & B_1 + \cdots + B_n
\end{array}
$$

*is a pullback for both* $? = f, g$*.*

*Proof.* The 'only if' is trivial. Conversely, assume the latter condition. Then we have

$$f \circ [h_1, \ldots, h_n] = k_1 + \cdots + k_n = g \circ [h_1, \ldots, h_n]\,.$$

By extensivity, $(h_j\colon C_j \to A)_j$ is a coproduct, so the cotuple

$$[h_1, \ldots, h_n]\colon C_1 + \cdots + C_n \longrightarrow A$$

is an isomorphism. Therefore $f = g$. ■

**Lemma 6.6.6.** *Let* $\mathbf{B}$ *be a category with finite coproducts and a final object* $1$*. Then* $\mathbf{B}$ *is extensive if and only if the coproduct* $1 + 1$ *is extensive.*

*Proof.* See [31, Proposition 4.1]. ■

### 6.6.2 Extensive categories and Boolean effectuses

Notice that condition (ii)(b) in Proposition 6.6.1 for extensive categories are the same as condition (T2) for effectuses in total form. This suggests some relationship between extensive categories and effectuses. Indeed, the following result was shown by Jacobs.

**Theorem 6.6.7** (Jacobs)**.** *Every extensive category with a final object is an effectus in total form.*

The result appeared in [140, Example 4.7] without a proof. Later a proof is presented in a preprint [40, Proposition 88], but still some details on the joint monicity condition are omitted there. For the sake of completeness, we will give a full proof here.

*Proof.* Let **B** be an extensive category with a final object 1. We verify that **B** satisfies the three conditions (T1), (T2) and, (T3) for effectuses in total form.

Condition (T1): For any morphisms $f\colon A \to C$ and $g\colon B \to D$, the first two squares below are trivially pullbacks.

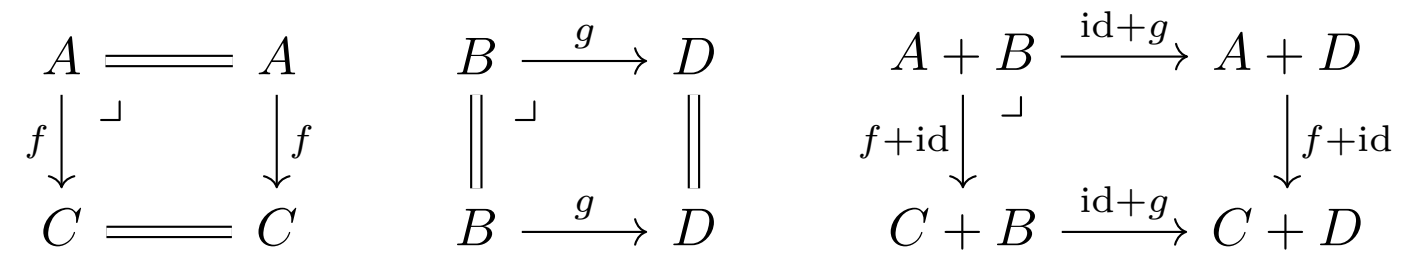

By Lemma 6.6.4, the third square above is a pullback too.

Condition (T2): It is precisely Proposition 6.6.1(ii)(b).

Condition (T3): We need to prove that the morphisms

$$\mathbb{W} := [\kappa_1, \kappa_2, \kappa_2]\colon 1+1+1 \to 1+1$$
$$\rotatebox[origin=c]{180}{$\mathbb{W}$} := [\kappa_2, \kappa_1, \kappa_2]\colon 1+1+1 \to 1+1$$

are jointly monic. To this end, let $f, g\colon A \to 1+1+1$ be morphisms such that

$$\mathbb{W} \circ f = \mathbb{W} \circ g =: h_1 \qquad \text{and} \qquad \rotatebox[origin=c]{180}{$\mathbb{W}$} \circ f = \rotatebox[origin=c]{180}{$\mathbb{W}$} \circ g =: h_2\,.$$

We invoke Lemma 6.6.5 to prove that $f = g$. We construct a common pullback of $f$ and $g$ along the first projection $\kappa_1\colon 1 \to 1+1+1$ as follows, by taking a pullback $h_1^*(\kappa_1)\colon A_1 \to A$ of $h_1$ along $\kappa_1$ as in the outer rectangle below.

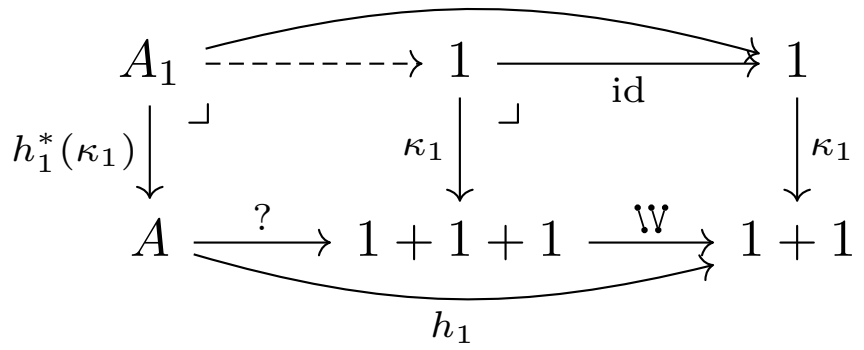

Here ? denotes either $f$ or $g$. The right-hand inner square is a pullback by extensivity. By the 'pullback lemma' [10, Lemma 5.8], the left-hand square is a pullback, which is common for $f$ and $g$, since the dashed morphism $A_1 \to 1$ does not depend on them. In a similar manner, we obtain a common pullback of $f$ and $g$ along $\kappa_2\colon 1 \to 1+1+1$ as in the following diagram.

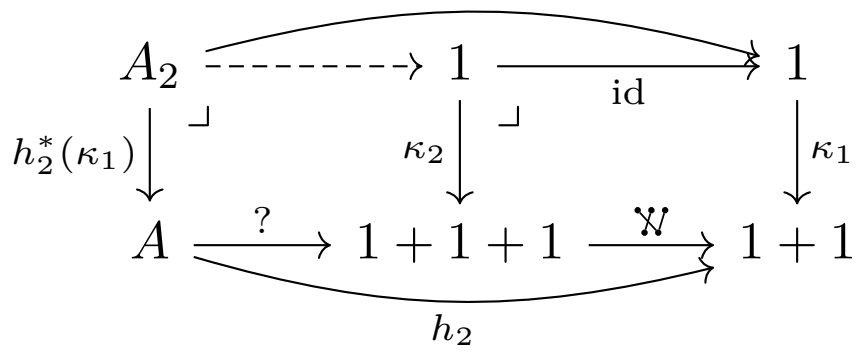

It requires more work to construct a common pullback along the third coprojection $\kappa_3 \colon 1 \to 1+1+1$. First note that we have the following pullbacks similarly to the above.

$$
\begin{array}{ccccc}
A_{23} & \overset{k_1^{(?)}}{\dashrightarrow} & 1+1 & \overset{!}{\longrightarrow} & 1 \\
{\scriptstyle h_1^*(\kappa_2)}\big\downarrow \; \lrcorner & & {\scriptstyle [\kappa_2,\kappa_3]}\big\downarrow \; \lrcorner & & \big\downarrow{\scriptstyle \kappa_2} \\
A & \overset{?}{\longrightarrow} & 1+1+1 & \overset{\rotatebox{180}{W}}{\longrightarrow} & 1+1
\end{array}
\qquad \text{(outer arrows: } A_{23} \to 1,\ h_1 \colon A \to 1+1)
\tag{6.4}
$$

$$
\begin{array}{ccccc}
A_{13} & \overset{k_2^{(?)}}{\dashrightarrow} & 1+1 & \overset{!}{\longrightarrow} & 1 \\
{\scriptstyle h_2^*(\kappa_2)}\big\downarrow \; \lrcorner & & {\scriptstyle [\kappa_1,\kappa_3]}\big\downarrow \; \lrcorner & & \big\downarrow{\scriptstyle \kappa_2} \\
A & \overset{?}{\longrightarrow} & 1+1+1 & \overset{\rotatebox{180}{W}}{\longrightarrow} & 1+1
\end{array}
\qquad \text{(outer arrows: } A_{13} \to 1,\ h_2 \colon A \to 1+1)
\tag{6.5}
$$

Here $k_1^{(?)} \colon A_{23} \to 1+1$ and $k_2^{(?)} \colon A_{13} \to 1+1$ might depend on $? = f, g$, but it turns out that they do not:

$$
\begin{aligned}
k_1^{(f)} &= \rotatebox{180}{W} \circ [\kappa_2, \kappa_3] \circ k_1^{(f)} \\
&= \rotatebox{180}{W} \circ f \circ h_1^*(\kappa_2) \\
&= \rotatebox{180}{W} \circ g \circ h_1^*(\kappa_2) \\
&= \rotatebox{180}{W} \circ [\kappa_2, \kappa_3] \circ k_1^{(g)} = k_1^{(g)}
\end{aligned}
$$

and similarly $k_2^{(f)} = k_2^{(g)}$. Therefore we will simply write them as $k_1$ and $k_2$. Since $h_1^*(\kappa_2)$ and $h_2^*(\kappa_2)$ are (isomorphic to) coprojections, we have a pullback:

$$
\begin{array}{ccc}
A_3 & \overset{l_2}{\longrightarrow} & A_{13} \\
{\scriptstyle l_1}\big\downarrow \; \lrcorner & & \big\downarrow{\scriptstyle h_2^*(\kappa_2)} \\
A_{23} & \overset{h_1^*(\kappa_2)}{\longrightarrow} & A
\end{array}
\tag{6.6}
$$

The maps $l_1$ and $l_2$ make the outer diagram below commute, so that the diagrams involving the unique map $A_3 \to 1$ commute by a pullback.

$$
\begin{array}{ccccc}
A_3 & \overset{l_2}{\longrightarrow} & A_{13} & & \\
{\scriptstyle l_1}\big\downarrow & \overset{!}{\searrow} & & \overset{k_2}{\searrow} & \\
A_{23} & & 1 & \overset{\kappa_2}{\longrightarrow} & 1+1 \\
& \overset{k_1}{\searrow} & {\scriptstyle \kappa_2}\big\downarrow \; \lrcorner & & \big\downarrow{\scriptstyle [\kappa_1,\kappa_3]} \\
& & 1+1 & \overset{[\kappa_2,\kappa_3]}{\longrightarrow} & 1+1+1
\end{array}
\tag{6.7}
$$

Write $m = h_1^*(\kappa_2) \circ l_1 = h_2^*(\kappa_2) \circ l_2$. We now claim that the following square is a pullback for both $? = f, g$.

$$
\begin{array}{ccc}
A_3 & \overset{!}{\longrightarrow} & 1 \\
{\scriptstyle m}\big\downarrow & & \big\downarrow{\scriptstyle \kappa_3} \\
A & \overset{?}{\longrightarrow} & 1+1+1
\end{array}
\tag{6.8}
$$

Let us prove that the square is a pullback for $? = f$. Its commutativity follows by commutativity of (6.7). Let $\alpha\colon Z \to A$ satisfy $f \circ \alpha = \kappa_3 \circ !_Z$. Then $f \circ \alpha = [\kappa_2, \kappa_3] \circ \kappa_2 \circ !_Z$. By the left-hand pullback in (6.4), we obtain $\alpha_{23}\colon Z \to A_{23}$ such that $\alpha = h_1^*(\kappa_2) \circ \alpha_{23}$ (and $\kappa_2 \circ !_Z = k_1 \circ \alpha_{23}$). Similarly by $f \circ \alpha = [\kappa_1, \kappa_3] \circ \kappa_2 \circ !_Z$, via the left-hand pullback in (6.5), we obtain $\alpha_{13}\colon Z \to A_{13}$ such that $\alpha = h_2^*(\kappa_2) \circ \alpha_{13}$. But then $h_1^*(\kappa_2) \circ \alpha_{23} = \alpha = h_2^*(\kappa_2) \circ \alpha_{13}$, so by the pullback (6.6), we obtain $\alpha_1\colon A \to A_3$ such that $\alpha_{23} = l_1 \circ \alpha_1$ and $\alpha_{13} = l_2 \circ \alpha_1$. The $\alpha_1$ is a desired mediating map:

$$m \circ \alpha_1 = h_1^*(\kappa_2) \circ l_1 \circ \alpha_1 = h_1^*(\kappa_2) \circ \alpha_{23} = \alpha\,.$$

The uniqueness of a mediating map follows automatically since $m$ is monic. Similarly the square (6.8) is a pullback for $? = g$, and we are done. ■

Below we will give necessary and sufficient conditions for an effectus (in total form) to be extensive — which include Booleanness, in particular. We thus obtain a new characterization of extensive categories (with a final object) as Boolean effectuses in total form with an additional property.

**Lemma 6.6.8.** *Let $p \in \mathrm{Pred}(A)$ be a predicate in an effectus $\mathbf{C}$. The following are equivalent.*

(i) *$\langle\!\langle p, p^\perp \rangle\!\rangle$ can be measured by a Boolean instrument $\langle\!\langle f_1, f_2 \rangle\!\rangle\colon A \to A + A$ such that both $f_1$ and $f_2$ are split idempotents.*

(ii) *There exist $A_1, A_2$ and an isomorphism $\chi\colon A \xrightarrow{\cong} A_1 + A_2$ such that $(\mathbb{1} + \mathbb{1}) \circ \chi = \langle\!\langle p, p^\perp \rangle\!\rangle$.*

(iii) *Comprehensions of $p$ and $p^\perp$ exist; and the cotuple*

$$[\pi_p, \pi_{p^\perp}]\colon \{A \,|\, p\} + \{A \,|\, p^\perp\} \longrightarrow A$$

*is an isomorphism. (Thus both $\pi_p$ and $\pi_{p^\perp}$ must be total comprehensions.)*

(iv) *T-comprehensions of $p$ and $p^\perp$ exist; and the cotuple*

$$[\pi_p, \pi_{p^\perp}]\colon \{A \,|\, p\} + \{A \,|\, p^\perp\} \longrightarrow A$$

*is an isomorphism.*

(v) *Quotients for $p$ and $p^\perp$ exists; and the tuple of quotients (i.e. the decomposition map $\mathrm{dc}_p$)*

$$\mathrm{dc}_p = \langle\!\langle \xi_{p^\perp}, \xi_p \rangle\!\rangle\colon A \longrightarrow A/p^\perp + A/p$$

*is an isomorphism.*

*Proof.* (i) $\Longrightarrow$ (iii) and (v): Note that Boolean instruments are CQ-ideal and hence CQ-idempotent (Theorems 6.3.20 and 6.3.23). Therefore $f_1\colon A \to A$ is a C-idempotent, and by Proposition 6.3.13 the splitting $f_1$ is given via a comprehension of $p = \mathbb{1} f_1$:

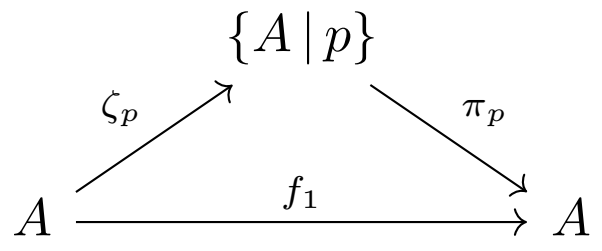

At the same time $f_1\colon A \to A$ is a Q-idempotent too, therefore by Proposition 6.3.14, the map $\zeta_p\colon A \to \{A\,|\,p\}$ is a quotient for $p^\perp$. Similarly the splitting of $f_2$ is given by $f_2 = \pi_{p^\perp} \circ \zeta_{p^\perp}$ where $\pi_{p^\perp}$ is a comprehension of $p^\perp$ and $\zeta_{p^\perp}$ is a quotient for $p$. Then the tuple $\langle\!\langle \zeta_p, \zeta_{p^\perp} \rangle\!\rangle\colon A \to \{A\,|\,p\} + \{A\,|\,p^\perp\}$ exists, and we have

$$[\pi_p, \pi_{p^\perp}] \circ \langle\!\langle \zeta_p, \zeta_{p^\perp} \rangle\!\rangle = \pi_p \circ \zeta_p \ovee \pi_{p^\perp} \circ \zeta_{p^\perp} = f_1 \ovee f_2 = \mathrm{id}_A\,.$$

We claim that $\langle\!\langle \zeta_p, \zeta_{p^\perp} \rangle\!\rangle \circ [\pi_p, \pi_{p^\perp}] = \mathrm{id}$ holds too. To prove this, it suffices to show that the following equation holds for each $j,k \in \{1,2\}$,

$$\rhd_k \circ \langle\!\langle \zeta_p, \zeta_{p^\perp} \rangle\!\rangle \circ [\pi_p, \pi_{p^\perp}] \circ \kappa_j = \rhd_k \circ \mathrm{id} \circ \kappa_j\,,$$

namely, that the following four equations hold.

$$\zeta_p \circ \pi_p = \mathrm{id} \qquad\qquad \zeta_{p^\perp} \circ \pi_{p^\perp} = \mathrm{id}$$
$$\zeta_p \circ \pi_{p^\perp} = 0 \qquad\qquad \zeta_{p^\perp} \circ \pi_p = 0\,.$$

The upper two equations hold by construction. We have $\zeta_p \circ \pi_{p^\perp} = 0$ since $\mathbb{1} \circ \zeta_p \circ \pi_{p^\perp} = p \circ \pi_{p^\perp} = \mathbb{0}$. Similarly $\zeta_{p^\perp} \circ \pi_p = 0$ holds. Therefore both maps $[\pi_p, \pi_{p^\perp}]\colon \{A\,|\,p\} + \{A\,|\,p^\perp\} \to A$ and $\langle\!\langle \zeta_p, \zeta_{p^\perp} \rangle\!\rangle\colon A \to \{A\,|\,p\} + \{A\,|\,p^\perp\}$ are isomorphisms, where the latter map is a universal decomposition for $p$.

(ii) $\Longrightarrow$ (i): Let $\psi = [\psi_1, \psi_2]\colon A_1 + A_2 \to A$ be the inverse of $\chi$. Let $\chi = \langle\!\langle \chi_1, \chi_2 \rangle\!\rangle\colon A \to A_1 + A_2$ decompose. Let $f_1 = \psi_1 \circ \chi_1$ and $f_2 = \psi_2 \circ \chi_2$. By $(\mathbb{1} + \mathbb{1}) \circ \chi = \langle\!\langle p, p^\perp \rangle\!\rangle$ we have $\mathbb{1}\chi_1 = p$ and $\mathbb{1}\chi_2 = p^\perp$. Since $\psi$ is an isomorphism and hence total, both $\psi_1$ and $\psi_2$ are total too. Then $\mathbb{1} f_1 = \mathbb{1} \circ \psi \circ \chi_1 = \mathbb{1}\chi_1 = p$ and similarly $\mathbb{1} f_2 = p^\perp$. Now we have

$$\mathrm{id}_A = [\psi_1, \psi_2] \circ \langle\!\langle \chi_1, \chi_2 \rangle\!\rangle = \psi_1 \circ \chi_1 \ovee \psi_2 \circ \chi_2 = f_1 \ovee f_2\,.$$

Moreover, both $f_1$ and $f_2$ are split idempotents, since

$$\chi_1 \circ \psi_1 = \rhd_1 \circ \chi \circ \psi \circ \kappa_1 = \rhd_1 \circ \kappa_1 = \mathrm{id}$$

and similarly $\chi_2 \circ \psi_2 = \mathrm{id}$. Therefore $\langle\!\langle f_1, f_2 \rangle\!\rangle\colon A \to A + A$ forms an instrument for $p$ that is side-effect-free and idempotent, and hence Boolean.

(iii) $\Longrightarrow$ (iv) is obvious, since total comprehensions are T-comprehensions.

(iv) $\Longrightarrow$ (ii): It suffices to prove that $\langle\!\langle p, p^\perp \rangle\!\rangle \circ [\pi_p, \pi_{p^\perp}] = \mathbb{1} + \mathbb{1}$, i.e. $\langle\!\langle p, p^\perp \rangle\!\rangle \circ \pi_p = \kappa_1 \circ \mathbb{1}$ and $\langle\!\langle p, p^\perp \rangle\!\rangle \circ \pi_{p^\perp} = \kappa_2 \circ \mathbb{1}$. By definition of T-comprehension we have $p \circ \pi_p = \mathbb{1}$ and $p^\perp \circ \pi_p = \mathbb{0}$. Therefore

$$\kappa_1 \circ \mathbb{1} = \kappa_1 \circ p \circ \pi_p \ovee \kappa_2 \circ p^\perp \circ \pi_p = \langle\!\langle p \circ \pi_p, p^\perp \circ \pi_p \rangle\!\rangle = \langle\!\langle p, p^\perp \rangle\!\rangle \circ \pi_p\,.$$

Similarly we prove $\kappa_2 \circ \mathbb{1} = \langle\!\langle p, p^\perp \rangle\!\rangle \circ \pi_p$.

(v) $\Longrightarrow$ (ii) is immediate since $(\mathbb{1} + \mathbb{1}) \circ \mathrm{dc}_p = \langle\!\langle p, p^\perp \rangle\!\rangle$. ∎

**Theorem 6.6.9.** *For an effectus* $\mathbf{C}$*, the following are equivalent.*

(i) *The subcategory* $\mathrm{Tot}(\mathbf{C})$ *of total morphisms is extensive.*

(ii) $\mathbf{C}$ *is Boolean and all Boolean idempotents* $f\colon A \to A$ *split.*

(iii) **C** *is Boolean and has comprehension.*

(iv) **C** *is Boolean and has quotients.*

(v) **C** *has comprehension or T-comprehension, and for each predicate* $p \in \text{Pred}(A)$*, the cotuple*

$$[\pi_p, \pi_{p^\perp}] \colon \{A \,|\, p\} + \{A \,|\, p^\perp\} \longrightarrow A$$

*is an isomorphism.*

(vi) **C** *has quotients, and for each predicate* $p \in \text{Pred}(A)$*, the tuple (decomposition map)*

$$\text{dc}_p = \langle\!\langle \xi_{p^\perp}, \xi_p \rangle\!\rangle \colon A \longrightarrow A/p^\perp + A/p$$

*is an isomorphism.*

*Moreover, when any of these conditions hold,* **C** *has total comprehension.*

*Proof.* The category Tot(**C**) is an effectus in total form and hence satisfies the pullback condition (T2) of Definition 4.1.6, so that all binary coproducts satisfy (ii)(b) of Proposition 6.6.1. Since $I$ is a final object in Tot(**C**), by Lemma 6.6.6, Tot(**C**) is extensive if and only if the coproduct $I + I$ satisfies (ii)(a) of Proposition 6.6.1. This is equivalent to saying that condition (ii) of Lemma 6.6.8 holds for all predicates $p$ in **C**. It follows that Tot(**C**) is extensive if and only if all predicates in **C** satisfies any of the equivalent conditions of Lemma 6.6.8. This proves that (i), (ii), (v), and (vi) are equivalent; and that if any of the conditions holds, **C** has total comprehension. Now note that Boolean idempotents are QC-idempotents. Hence by Propositions 6.3.13 and 6.3.14 a Boolean idempotent $f \colon A \to A$ splits if and only if there exists a comprehension of $\mathbb{1}f$ if and only if there exists a quotient for $(\mathbb{1}f)^\perp$. Therefore (ii), (iii), and (iv) are equivalent. ■

We say that an effectus in total form **B** is Boolean if the effectus Par(**B**) is Boolean, see Definition 6.5.17. Note that an instrument is Boolean if and only if it is both idempotent and side-effect-free, and that idempotency can be characterized by Proposition 6.3.4, in terms of total morphisms. Thus Booleanness of **B** can be described directly in **B**. We now characterize extensive categories as a special kind of effectuses in total form.

**Corollary 6.6.10.** *Let* **B** *be a category with finite coproducts and a final object. Then* **B** *is extensive if and only if* **B** *is an effectus with total form and it satisfies any (and hence all) of the following equivalent conditions.*

(i) **B** *is Boolean and all Boolean idempotents split in* Par(**B**)*.*

(ii) **B** *is Boolean and has total comprehension.*

(iii) **B** *is Boolean and has quotients.*

(iv) **B** *has total comprehension (or T-comprehension) and for each predicate* $p \colon A \to 1 + 1$ *in* **B***, the cotuple*

$$[\pi_p, \pi_{p^\perp}] \colon \{A \,|\, p\} + \{A \,|\, p^\perp\} \longrightarrow A$$

*is an isomorphism.*

(v) **B** *has quotients and for each predicate* $p\colon A \to 1+1$ *in* **B***, the decomposition map*

$$\mathrm{dc}_p = \langle\!\langle \xi_{p^\perp}, \xi_p \rangle\!\rangle \colon A \longrightarrow A/p^\perp + A/p$$

*is an isomorphism.*

*Proof.* By Theorems 6.6.7 and 6.6.9, and the equivalence of effectuses in partial and total form, see Sections 4.1 and 4.2. ■

Hence, any extensive category with a final object is a Boolean effectus in total form and moreover has total comprehension and quotients. In particular, predicates (i.e. maps $A \to 1+1$) in an extensive category with a final object form Boolean algebras. This was also shown in [166, Theorem 4.11] as a corollary of more general results about extensive restriction categories.

The example below shows that extensive categories, viewed as effectuses, may not have images.

**Example 6.6.11.** The opposite $\mathbf{BA}^{\mathrm{op}}$ of the category of Boolean algebras is an extensive category with a final object, and hence an effectus in total form. This follows from the standard fact that the opposite $\mathbf{CRing}^{\mathrm{op}}$ of the category of commutative rings is extensive, together with the fact that $\mathbf{BA}^{\mathrm{op}}$ is identified with the full subcategory of $\mathbf{CRing}^{\mathrm{op}}$ consisting of Boolean rings, which is closed under finite coproducts and pullbacks along coprojections. The 2-element Boolean algebra 2 is initial in **BA** and hence final in $\mathbf{BA}^{\mathrm{op}}$. Thus states in $\mathbf{BA}^{\mathrm{op}}$ are Boolean algebra homomorphisms $\omega\colon A \to 2$, which correspond bijectively to ultrafilters $U \subseteq A$ via $a \in U \iff \omega(a) = 1$. Since predicates on $A$ are simply elements $a \in A$, an image of a state/ultrafilter $U \subseteq A$ is precisely a least element in $U$. Therefore there exists an image of an ultrafilter $U \subseteq A$ if and only if $U$ is a principal filter. However, it is well known that nonprincipal ultrafilters exist (assuming the Axiom of Choice). We conclude that the effectus $\mathbf{BA}^{\mathrm{op}}$ does not have images.

Therefore, extensive categories with a final object are not necessarily (pre-)comprehensive effectuses (Definitions 5.5.3 and 5.5.23). They are, nevertheless, a reasonably well-behaved class of effectuses. Indeed, the predicates $p\colon A \to 1+1$ there form a Boolean algebra (by Theorem 6.5.19), and hence an orthomodular lattice. This suggests that it might be possible to improve the definition of comprehensive effectuses so that it includes all extensive categories as examples. We leave it to future work.

# Chapter 7

# Miscellaneous Topics in Effectus Theory

This chapter presents miscellaneous topics in effectus theory.

Section 7.1 is concerned with totalization of effectuses. Totalization is a construction that turns partial algebraic structures into total ones. Since the homsets of an effectus (in partial form) are PCMs, one can apply totalization to effectuses. Totalization of an effectus yields a biproduct (a.k.a. semiadditive) category, and moreover a 2-coreflection between the 2-categories of effectuses and 'grounded' biproduct categories. The section is concluded with an observation which reveals some connection between effectus theory and categorical quantum mechanics.

In Section 7.2 we reveal a relationship between effectuses and the convex operational approach. The latter is a well-established framework for quantum theory or generalized probability theory. There, a system is modelled by a dual pair of a base-norm and an order-unit space, which is called a *convex operational model.* The connection between effectuses and convex operational models is made in two ways. First, we show that the category of convex operational models forms an effectus. Second, we show that an effectus that satisfies certain conditions (i.e. real and has an order-separation property) can be embedded into the category of convex operational models, via a faithful functor that preserves the structure of effectuses.

In Section 7.3 we study partially $\sigma$-additive (i.e. countably additive) structure in relation to effectus theory. Such partially $\sigma$-additive structures were studied by Arbib and Manes [7, 202] in the context of program semantics, but countable structures are also relevant in quantum foundations (e.g. [200]). We introduce *$\sigma$-effectuses*, an extension of effectuses with the structure of partially $\sigma$-additive categories, and present basic results about them.

## 7.1 Totalization and grounded biproduct categories

*Totalization* is a construction that turns partial algebraic structures into total ones. It was studied in [146] from a categorical perspective for PCMs and effect algebras. For example, PCMs can be totalized into commutative monoids, forming a left adjoint to the forgetful functor. Moreover, equipping commutative monoids with a suitable structure (i.e. downset), one obtains a *coreflection* (Definition 2.1.6). It means that PCMs can be nicely embedded into such 'total' structures.

In this section, we study the totalization of effectuses. Biproduct categories (a.k.a.

semiadditive categories) will arise via totalization of effectuses. Furthermore, equipping the 'ground' structure $\ground$ with biproduct categories, we obtain a coreflection for effectuses. Biproducts and ground structure are also used in *categorical quantum mechanics* [2, 54, 127] of the Oxford school, and an example of grounded biproduct categories naturally appears there. Thus the totalization of effectuses may be considered as a first step in making connection between effectus theory and categorical quantum mechanics. This direction has been further developed by Tull [250].

This section is based on (unpublished) joint work with Tull. See his thesis [250, Chapter 3] for his account of this topic and further related results.

In § 7.1.1 we recall basic definitions and results about totalization of PCMs and effect algebras. Then in § 7.1.2 we apply totalization to effectuses, and introduce grounded biproduct categories. In § 7.1.3 we show a coreflection between effectuses and grounded biproduct categories.

## 7.1.1 Totalization of PCMs

This subsection reviews basic definitions and results from Jacobs and Mandemaker's article [146] on totalization of PCMs and effect algebras. We also include results that are useful to identify a concrete presentation of totalization, Propositions 7.1.11 and 7.1.13; these are new.

**Definition 7.1.1.** A **finite multiset** on a set $X$ is a function $\varphi\colon X \to \mathbb{N}$ with finite support, i.e. such that $\mathrm{supp}(\varphi) = \{x \mid \varphi(x) \neq 0\}$ is finite. We will refer to finite multisets simply as **multisets** in this thesis, since we use only finite ones. Multisets are 'sets' in which multiple instances of an element can occur: the number $\varphi(x) \in \mathbb{N}$ is the multiplicity of an element $x$. We denote by $\mathcal{M}(X)$ the set of multisets on $X$.

Similarly to distributions $\sum_j r_j|x_j\rangle \in \mathcal{D}(X)$, we represent multisets as formal weighted sums $\sum_j n_j|x_j\rangle$, where $n_j$ is the multiplicity of $x_j \in X$. For example, $1|x\rangle$ denotes the multiset containing a single element $x$. We write 0 for empty multisets.

The set $\mathcal{M}(X)$ of multisets has the structure of commutative monoid in a pointwise manner: for $\varphi, \psi \in \mathcal{M}(X)$ the sum is defined by $(\varphi + \psi)(x) = \varphi(x) + \psi(x)$, and the empty multiset 0, which sends every element to zero, is the neutral element. It is straightforward to see that $\mathcal{M}(X)$ is a free commutative monoid over $X$, that is:

**Lemma 7.1.2.** *The assignment $X \mapsto \mathcal{M}(X)$, with the unit $\eta_X\colon X \to \mathcal{M}(X)$ given by $\eta_X(x) = 1|x\rangle$, forms a left adjoint to the forgetful functor* **CMon** $\to$ **Set***. Here* **CMon** *denotes the category of commutative monoids and monoid morphisms.* ■

Hence we obtain a monad $\mathcal{M}\colon \mathbf{Set} \to \mathbf{Set}$. The monad structure of $\mathcal{M}$ is similar to the distribution monad $\mathcal{D}$, see Definition 2.4.1.

We now introduce totalization of partial commutative monoids (PCMs). Let $M$ be a PCM. Then we have a commutative monoid $\mathcal{M}(M)$ (its structure is not related to the PCM structure of $M$). We denote by $\sim$ the smallest congruence on $\mathcal{M}(M)$ satisfying

- $1|x \ovee y\rangle \sim 1|x\rangle + 1|y\rangle$ for all summable $x, y \in M$;
- $1|0\rangle \sim 0$.

**Definition 7.1.3.** The **totalization** of a PCM $M$ is the quotient $\mathcal{T}(M) := \mathcal{M}(M)/\sim$ of the monoid $\mathcal{M}(M)$ by the congruence $\sim$ defined above.

Totalization yields free commutative monoids over PCMs:

**Proposition 7.1.4.** *Totalization $M \mapsto \mathcal{T}(M)$ defines a left adjoint to the obvious forgetful functor* **CMon** $\to$ **PCM**. *The unit $\eta_M \colon M \to \mathcal{T}(M)$ is the map given by $\eta_M(x) = 1|x\rangle$.*

In the proof below and elsewhere, we write $n \cdot a = a + \cdots + a$ for the $n$-fold sum in a commutative monoid.

*Proof.* Let $f \colon M \to A$ be a PCM morphism between $M \in \mathbf{PCM}$ and $A \in \mathbf{CMon}$. Since $\mathcal{M}(M)$ is a free commutative monoid over a set $M$, we have a monoid morphism $f' \colon \mathcal{M}(M) \to A$ given by $f'(\sum_j n_j|x_j\rangle) = \sum_j n_j \cdot f(x_j)$. Now we claim that $\varphi \sim \psi$ implies $f'(\varphi) = f'(\psi)$ for all $\varphi, \psi \in \mathcal{M}(M)$. To this end, let

$$R = \{(\varphi, \psi) \mid f'(\psi) = f'(\varphi)\} \subseteq \mathcal{M}(M) \times \mathcal{M}(M)\,.$$

Then $R$ is a congruence on $\mathcal{M}(M)$ that moreover satisfies

$$(1|x \ovee y\rangle, 1|x\rangle + 1|y\rangle) \in R \qquad \text{and} \qquad (1|0\rangle, 0) \in R$$

for all summable $x, y \in M$. Since $\sim$ is the smallest one among such congruences, we have $\sim \subseteq R$, that is, $\varphi \sim \psi$ implies $f'(\varphi) = f'(\psi)$. Therefore we obtain a monoid homomorphism $f'' \colon \mathcal{M}(M)/\sim = \mathcal{T}(M) \to A$ by

$$f''\big(\textstyle\sum_j n_j|x_j\rangle\big) = f'\big(\textstyle\sum_j n_j|x_j\rangle\big) = \textstyle\sum_j n_j \cdot f(x_j)\,.$$

Clearly $\eta \circ f'' = f$. If $g \colon \mathcal{T}(M) \to A$ is a monoid morphism such that $g \circ \eta = f$, then $g(1|x\rangle) = f(x)$ for all $x \in M$. This implies that

$$g\big(\textstyle\sum_j n_j|x_j\rangle\big) = \textstyle\sum_j n_j \cdot f(x_j) = f''\big(\textstyle\sum_j n_j|x_j\rangle\big)\,.$$

Therefore $f''$ is a unique monoid morphism satisfying $\eta \circ f'' = f$, and we are done. ■

**Corollary 7.1.5.** *Totalization defines a functor $\mathcal{T} \colon$ **PCM** $\to$ **CMon**: for a PCM morphism $f \colon M \to N$, the monoid morphism $\mathcal{T}(f) \colon \mathcal{T}(M) \to \mathcal{T}(N)$ is given by $\mathcal{T}(f)(\sum_j n_j|x_j\rangle) = \sum_j n_j|f(x_j)\rangle$.* ■

**Example 7.1.6.** Here are some examples of totalization. These can be easily verified by Propositions 7.1.11 and 7.1.13 shown below.

(i) $\mathcal{T}(2) \cong \mathbb{N}$, where $2 = \{0, 1\}$ is the 2-element effect algebra.

(ii) $\mathcal{T}([0, 1]) \cong \mathbb{R}_+$, where $[0, 1] \subseteq \mathbb{R}$ is the real unit interval.

(iii) For each set $X$, we can see $X + \{0\}$ as a PCM where 0 is a zero element, and any pair of $x, y \in X$ is not summable. Then we have:

$$\mathcal{T}(X + \{0\}) \cong \mathcal{M}(X)$$

(iv) The powerset $\mathcal{P}(X)$ of a set $X$ forms an effect algebra, and hence a PCM, with disjoint union as sum. Then

$$\mathcal{T}(\mathcal{P}(X)) \cong \{\varphi\colon X \to \mathbb{N} \mid \varphi \text{ is bounded}\}\,.$$

Here a function $\varphi\colon X \to \mathbb{N}$ is **bounded** if $\sup_{x\in X} \varphi(x) < \infty$.

(v) The set $[0,1]^X$ of fuzzy subsets of a set $X$ forms an effect algebra. Its totalization is:

$$\mathcal{T}([0,1]^X) \cong \{\varphi\colon X \to \mathbb{R}_+ \mid \varphi \text{ is bounded}\}\,.$$

(vi) For a $C^*$-algebra $\mathscr{A}$, consider the effect algebra $[0,1]_{\mathscr{A}}$ of effects. Then

$$\mathcal{T}([0,1]_{\mathscr{A}}) \cong \{x \in \mathscr{A} \mid x \geq 0\}.$$

For later use, we prove that totalization $\mathcal{T}\colon \mathbf{PCM} \to \mathbf{CMon}$ sends PCM bimorphisms to monoid bimorphisms. Categorically this means that the functor $\mathcal{T}\colon \mathbf{PCM} \to \mathbf{CMon}$ is lax monoidal with respect to the tensor products, which is (indirectly) shown in [146]. Here we give a more direct proof.

**Lemma 7.1.7.** *Let $M, N, L$ be PCMs. Let $f\colon M \times N \to L$ be a PCM bimorphism. Then the map $\hat{f}\colon \mathcal{T}(M) \times \mathcal{T}(N) \to \mathcal{T}(L)$ given by*

$$\hat{f}(\textstyle\sum_j n_j|x_j\rangle, \sum_k m_k|y_k\rangle) = \sum_{jk} n_j m_k|f(x_j, y_k)\rangle$$

*is well-defined, and moreover it is a monoid bimorphism.*

*Proof.* For each $x \in M$, we have a PCM morphism $f(x,-)\colon N \to L$, and hence a monoid morphism $\mathcal{T}(f(x,-))\colon \mathcal{T}(N) \to \mathcal{T}(L)$ given by $\mathcal{T}(f(x,-))(\sum_k m_k|y_k\rangle) = \sum_k m_k|f(x,y_k)\rangle$. Now let $\sum_k m_k|y_k\rangle \in \mathcal{T}(N)$ be fixed, and let $g\colon M \to \mathcal{T}(L)$ be a map given by $g(x) = \mathcal{T}(f(x,-))(\sum_k m_k|y_k\rangle)$. It is easy to see that $g$ is a PCM morphism (when $\mathcal{T}(L)$ is viewed as a PCM). By freeness of totalization, $g$ extends to a monoid morphism $\overline{g}\colon \mathcal{T}(M) \to \mathcal{T}(L)$, which is given by

$$\begin{aligned}\overline{g}(\textstyle\sum_j n_j|x_j\rangle) = \sum_j n_j \cdot g(x_j) &= \textstyle\sum_j n_j \cdot (\sum_k m_k|f(x_j, y_k)\rangle\rangle) \\ &= \textstyle\sum_{jk} n_j m_k|f(x_j, y_k)\rangle\rangle) \\ &= \hat{f}(\textstyle\sum_j n_j|x_j\rangle, \sum_k m_k|y_k\rangle)\,.\end{aligned}$$

This proves that $\hat{f}$ is well-defined, and it is a monoid morphism in the first argument. By a symmetric argument, $\hat{f}$ is a monoid morphism in the second argument. ∎

The lemma below about 'Kleene equality' is convenient. To describe it we need some definitions. Let $\sum_j n_j|x_j\rangle \in \mathcal{M}(M)$ be a multiset over a PCM $M$. Then we define an **interpretation** of $\sum_j n_j|x_j\rangle$ in $M$ as:

$$[\![\textstyle\sum_j n_j|x_j\rangle]\!] = \bigovee_j n_j \cdot x_j \quad \in M\,,$$

where $n \cdot x$ denotes the $n$-fold sum: $x \ovee \cdots \ovee x$. The interpretation $[\![-]\!]\colon \mathcal{M}(M) \rightharpoonup M$ is a partial function, since $\ovee$ is a partial operation. We say that two multisets $\varphi, \psi \in \mathcal{M}(M)$ are **Kleene-equal**, and write $\varphi \simeq \psi$, if both

- $[\![\varphi]\!]$ is defined $\iff$ $[\![\psi]\!]$ is defined;
- $[\![\varphi]\!] = [\![\psi]\!]$ whenever either (hence both) side is defined.

**Lemma 7.1.8** ([146, Lemma 6(i)]). *Let $M$ be a PCM and $\varphi, \psi \in \mathcal{M}(M)$ multisets. If $\varphi = \psi$ in $\mathcal{T}(M)$ (i.e. $\varphi \sim \psi$ for the congruence $\sim$ defining $\mathcal{T}(M)$), then $\varphi$ and $\psi$ are Kleene-equal.*

*Proof.* This holds since the Kleene equality $\simeq$ is a congruence on $\mathcal{M}(M)$ satisfying $1|x\rangle + 1|y\rangle \simeq 1|x \ovee y\rangle$ for all summable $x, y \in M$ and $1|0\rangle \simeq 0$. ■

Now we describe a coreflection given by totalization of PCMs. Let $A$ be a commutative monoid. A **downset** in $A$ is a nonempty subset $U \subseteq A$ that is downward closed: if $x \leq y$ and $y \in U$, then $x \in U$. Here $\leq$ is the algebraic preorder on $A$: $x \leq y$ iff $x + z = y$ for some $z \in A$. Clearly every downset $U$ in $A$ contains $0 \in A$, and forms a PCM via $x \perp y \iff x + y \in U$ and $x \ovee y = x + y$.

We write **DCM** for the category of downsets in commutative monoids. The objects are pairs $(A, U)$ of $A \in \mathbf{CMon}$ and a downset $U \subseteq A$. The morphisms $f\colon (A, U) \to (B, V)$ are morphisms $f\colon A \to B$ that sends $x \in U$ to $f(x) \in V$. Then there is a functor $\mathrm{Down}\colon \mathbf{DCM} \to \mathbf{PCM}$ given by $\mathrm{Down}(A, U) = U$. For a morphism $f\colon (A, U) \to (B, V)$, $\mathrm{Down}(f)\colon U \to V$ is the restriction of $f$.

Note that the totalization $\mathcal{T}(M)$ of a PCM $M$ carries a canonical downset:

$$U = \{1|x\rangle \mid x \in M\} \subseteq \mathcal{T}(M)\,,$$

and thus the functor $\mathcal{T}\colon \mathbf{PCM} \to \mathbf{CMon}$ lifts to $\mathcal{T}\colon \mathbf{PCM} \to \mathbf{DCM}$.

**Theorem 7.1.9.** *The functors $\mathcal{T}$ and* Down *form a coreflection:*

$$\mathbf{PCM} \underset{\mathrm{Down}}{\overset{\mathcal{T}}{\underset{\longleftarrow}{\overset{\longrightarrow}{\bot}}}} \mathbf{DCM} \qquad\qquad \mathrm{Down} \circ \mathcal{T} \cong \mathrm{id}\,.$$

*Proof.* See [146, Theorem 6]. ■

It is often possible to find a concrete presentation of the totalization $\mathcal{T}(M)$. We give two useful results (Propositions 7.1.11 and 7.1.13) for that purpose.

**Definition 7.1.10.** A commutative monoid $A$ has the **refinement property** [68] (see also [110] and [141, §4.2]) if whenever $a_1 + a_2 = b_1 + b_2$ in $A$, there exist $c_{11}, c_{12}, c_{21}, c_{22} \in A$ such that

$$a_1 = c_{11} + c_{12} \qquad\qquad a_2 = c_{21} + c_{22}$$
$$b_1 = c_{11} + c_{21} \qquad\qquad b_2 = c_{12} + c_{22}\,.$$

The equations can be described in the following table:

$$\begin{array}{cc|c} c_{11} & c_{12} & a_1 \\ c_{21} & c_{22} & a_2 \\ \hline b_1 & b_2 & \end{array}$$

**Proposition 7.1.11.** *Let $A$ be a commutative monoid with the refinement property, and $U \subseteq A$ be a downset. Let $\overline{U} \subseteq A$ be the commutative monoid generated by $U$. Then $\overline{U}$ is the totalization of the PCM $U$.*

*Proof.* We will prove that the inclusion $\overline{U} \hookrightarrow U$ has the universal property of totalization, that is: for any $B \in \mathbf{CMon}$ and $f: U \to B$ in $\mathbf{PCM}$, there is a unique monoid homomorphism $\overline{f}: \overline{U} \to B$ that extends $f$. Given such an $f$, we define $\overline{f}: \overline{U} \to B$ by

$$\overline{f}\Big(\sum\nolimits_i a_i\Big) \coloneqq \sum\nolimits_i f(a_i)$$

for $a_1, \dots, a_n \in U$. We need to show that the function is well-defined. Let

$$a_1, \dots, a_n, b_1, \dots, b_m \in U \quad \text{satisfy} \quad \sum\nolimits_i a_i = \sum\nolimits_j b_j \quad \text{in } A\,.$$

We claim that $\sum_i f(a_i) = \sum_j f(b_j)$. If $n < 2$ or $m < 2$, the claim is obvious since the sum $\sum_i a_i = \sum_j b_j$ is defined in $U$. We assume $n, m \geq 2$. By the refinement property, there exists a matrix $(c_{ij})_{ij}$ of elements in $A$ such that

$$a_i = \sum\nolimits_j c_{ij} \qquad\qquad b_j = \sum\nolimits_i c_{ij}\,,$$

see [110, Proposition 5.10]. We have $c_{ij} \in U$ since $U$ is a downset. Then

$$\sum\nolimits_i f(a_i) = \sum\nolimits_i f(\sum\nolimits_j c_{ij}) = \sum\nolimits_i \sum\nolimits_j f(c_{ij}) = \sum\nolimits_j f(\sum\nolimits_i c_{ij}) = \sum\nolimits_j f(b_j)\,.$$

Therefore the function $\overline{f}: \overline{U} \to B$ is well-defined. It is then easy to see that $\overline{f}$ is a unique monoid homomorphism that extends $f$. We conclude that $\overline{U} \cong \mathcal{T}(U)$. ■

**Definition 7.1.12.**

(i) A **rig** ('ring without negatives' [80], also known as a *semiring*) is a set $R$ equipped with a commutative monoid structure $(+, 0)$ and a monoid structure $(\cdot, 1)$ such that the multiplication $\cdot$ distributes over $(+, 0)$, i.e. $\cdot : R \times R \to R$ is a monoid bihomomorphism with respect to $(R, +, 0)$.

(ii) Let $R$ be a rig. A **module over the rig** $R$ (or simply, an $R$-module) is a commutative monoid $(A, +, 0)$ with an $R$-action $\cdot : R \times A \to A$ that is a monoid bihomomorphism (w.r.t. $(R, +, 0)$) and satisfies $r \cdot (s \cdot a) = (r \cdot s) \cdot a$ for all $r, s \in R$ and $a \in A$.

Put categorically, a rig is a monoid in the monoidal category **CMon** (with the monoidal structure given by tensor product), and a module over a rig is a module over a monoid in **CMon**.

The following proposition is based on the result of Tull [250, § 3.2.5].

**Proposition 7.1.13.** *Let $A$ be a commutative monoid and $U \subseteq A$ be a downset. We assume that $A$ is a module over the rig $\mathbb{R}_+$ of nonnegative reals, and that for each $a \in A$, there exists an $n \in \mathbb{N}_{>0}$ such that $(1/n) \cdot a \in U$. Then $A$ is the totalization of the PCM $U$.*

We note that $\mathbb{R}_+$ in the statement may be generalized to an arbitrary rig $R$ with a certain property used in the proof below. For example, the proposition holds true for the rig $\mathbb{Q}_+$ of nonnegative rational numbers, with exactly the same proof.

*Proof.* We prove that the inclusion $U \hookrightarrow A$ satisfies the universal property of totalization. Let $B \in \mathbf{CMon}$ and $f\colon U \to B$ in $\mathbf{PCM}$ be given. We define $\overline{f}\colon A \to B$ by

$$\overline{f}(a) = n \cdot f\big(\tfrac{1}{n} \cdot a\big) \equiv \overbrace{f\big(\tfrac{1}{n} \cdot a\big) + \cdots + f\big(\tfrac{1}{n} \cdot a\big)}^{n \text{ times}} ,$$

where $n \in \mathbb{N}_{>0}$ is a number with $(1/n) \cdot a \in U$, which exists by definition. We need to show that $\overline{f}(a)$ does not depend on the choice of $n$. So suppose that $n, m \in \mathbb{N}_{>0}$ satisfy $(1/n) \cdot a \in U$ and $(1/m) \cdot a \in U$. Then $(1/nm) \cdot a \in U$ because $U$ is a downset and

$$n \cdot (\tfrac{1}{nm} \cdot a) = \tfrac{1}{nm} \cdot a + \cdots + \tfrac{1}{nm} \cdot a = \tfrac{1}{m} \cdot a .$$

Thus

$$\begin{aligned} n \cdot f\big(\tfrac{1}{n} \cdot a\big) = n \cdot f\big(m \cdot \tfrac{1}{nm} \cdot a\big) &= nm \cdot f\big(\tfrac{1}{nm} \cdot a\big) \\ &= m \cdot f\big(n \cdot \tfrac{1}{nm} \cdot a\big) = m \cdot f\big(\tfrac{1}{m} \cdot a\big) , \end{aligned}$$

showing that the function $\overline{f}\colon A \to B$ is well-defined. It is easy to see that $\overline{f}$ is a unique monoid morphism that extends $f$. ■

Finally, we briefly review a coreflection given by totalization of effect algebras. A **barred commutative monoid** is a commutative monoid $A$ with a distinguished 'unit' element $u \in A$ that satisfies:

- positivity: $a + b = 0$ implies $a = b = 0$;
- barred cancellativity: $a + b = a + c = u$ implies $b = c$.

We write **BCM** for the category of barred commutative monoids. The morphisms are monoid homomorphisms that preserve the units. For any barred commutative monoid $(A, u)$, there is a canonical (principal) downset $\downarrow(u) = \{a \in A \mid a \leq u\}$. In this way we can think of **BCM** as a (non-full) subcategory of **DCM**. Note that **EA** can also be viewed as a subcategory of **PCM**.

**Proposition 7.1.14.** *The coreflection* $\mathbf{PCM} \rightleftarrows \mathbf{DCM}$ *of Theorem* 7.1.9 *restricts to a coreflection of the subcategories:*

$$\mathbf{EA} \underset{\text{Down}}{\overset{\mathcal{T}}{\underset{\longleftarrow}{\overset{\longrightarrow}{\bot}}}} \mathbf{BCM} \qquad\qquad \text{Down} \circ \mathcal{T} \cong \text{id} .$$

*Proof.* See [146, Proposition 3]. ■

Explicitly, for each effect algebra $E$, the totalization $\mathcal{T}(E)$ (as PCM) together with $1|1\rangle \in \mathcal{T}(E)$ forms a barred commutative monoid. For each barred commutative monoid $(A, u)$, the downset $\downarrow(u)$ forms an effect algebra.

### 7.1.2 Totalization of effectuses

We now apply the totalization construction to an effectus. More specifically, totalization is applied to the hom-PCMs of an effectus, yielding a new category. (Recall that 'effectus' in this thesis means 'effectus in partial form'.) Let $\mathbf{C}$ be an effectus. We define a category $\mathcal{T}(\mathbf{C})$, called the **totalization** of $\mathbf{C}$, as follows. The category $\mathcal{T}(\mathbf{C})$ has the same objects as $\mathbf{C}$. For each $A, B \in \mathcal{T}(\mathbf{C})$, the homset is given by $\mathcal{T}(\mathbf{C})(A,B) = \mathcal{T}(\mathbf{C}(A,B))$. The identity on $A$ is $1|\mathrm{id}_A\rangle \in \mathcal{T}(\mathbf{C})(A,A)$, where $\mathrm{id}_A$ is the identity in $\mathbf{C}$. For morphisms $\sum_j n_j|f_j\rangle \in \mathcal{T}(\mathbf{C})(A,B)$ and $\sum_k m_k|g_k\rangle \in \mathcal{T}(\mathbf{C})(B,C)$, the composition is given as follows.

$$(\textstyle\sum_k m_k|g_k\rangle) \circ (\sum_j n_j|f_j\rangle) = \sum_{kj} m_k n_j|g_k \circ f_j\rangle \quad \in \mathcal{T}(\mathbf{C})(A,C)$$

**Proposition 7.1.15.** *For each effectus $\mathbf{C}$, the totalization $\mathcal{T}(\mathbf{C})$ is a category enriched over commutative monoids. Moreover there is an identity-on-objects faithful functor $\mathbf{C} \to \mathcal{T}(\mathbf{C})$ given by $f \mapsto 1|f\rangle$.*

Here a category $\mathbf{D}$ is **enriched over commutative monoids** if each homset $\mathbf{D}(A,B)$ is a commutative monoid and the composition is a monoid bihomomorphism.

*Proof.* The composition is well-defined by Lemma 7.1.7, and it is easy to verify the unit law and associativity, so that $\mathcal{T}(\mathbf{C})$ is a category. By construction, every homset $\mathcal{T}(\mathbf{C})(A,B) = \mathcal{T}(\mathbf{C}(A,B))$ is a commutative monoid, and by Lemma 7.1.7, composition in $\mathcal{T}(\mathbf{C})$ is a monoid bimorphism. Therefore $\mathcal{T}(\mathbf{C})$ is enriched over commutative monoids. Clearly the mapping $f \mapsto 1|f\rangle$ defines an identity-on-objects functor $\mathbf{C} \to \mathcal{T}(\mathbf{C})$. Since the totalization of PCMs yields injections $\mathbf{C}(A,B) \to \mathcal{T}(\mathbf{C}(A,B))$, the functor $\mathbf{C} \to \mathcal{T}(\mathbf{C})$ is faithful. ■

We will study the structures/properties of categories $\mathcal{T}(\mathbf{C})$ arising from effectuses via totalization. Most importantly, the categories $\mathcal{T}(\mathbf{C})$ have finite *biproducts*.

**Definition 7.1.16.** Let $\mathbf{E}$ be a category with zero morphisms. Let $(A_j)_{j\in J}$ be a family of objects. A **biproduct** of $(A_j)_{j\in J}$ is an object $\bigoplus_j A_j$ that is both a product and a coproduct of $(A_j)_{j\in J}$ such that the projections $\pi_j \colon \bigoplus_j A_j \to A_j$ and coprojections $\kappa_j \colon A_j \to \bigoplus_j A_j$ satisfy

$$\pi_j \circ \kappa_k = \begin{cases} \mathrm{id} & \text{if } j = k \\ 0 & \text{if } j \neq k \end{cases}$$

for all $j, k \in J$.

**Definition 7.1.17.** A **biproduct category** (also called a **semiadditive category**) is a category with zero morphisms and finite biproducts. Equivalently, it is a category with a zero object 0 and binary biproducts $A \oplus B$. (Note that the zero object is a nullary biproduct.)

For morphisms $f \colon A \to B$ and $g \colon C \to D$ in a biproduct category, a morphism $A \oplus C \to B \oplus D$ can be formed in two ways, as a product $f \times g$ and as a coproduct $f + g$. The two morphisms turns out to be equal, and thus the morphism $f \times g = f + g$ is written as $f \oplus g$.

We recall characterizations of biproduct categories.

**Lemma 7.1.18.** *Let* $\mathbf{E}$ *be a category. The following are equivalent.*

(i) $\mathbf{E}$ *is a biproduct category.*

(ii) $\mathbf{E}$ *has finite coproducts, and is enriched over commutative monoids.*

(iii) $\mathbf{E}$ *is enriched over commutative monoids, and for each* $A, B \in \mathbf{E}$*, there is an object* $A \oplus B$ *and morphisms:*

$$\begin{array}{ccccc} A & \xrightarrow{\kappa_1} & A \oplus B & \xleftarrow{\kappa_2} & B \\ A & \xleftarrow{\pi_1} & A \oplus B & \xrightarrow{\pi_2} & B \end{array}$$

*such that* $\kappa_1 \circ \pi_1 \dotplus \kappa_2 \circ \pi_2 = \mathrm{id}_{A \oplus B}$ *and*

$$\pi_j \circ \kappa_k = \begin{cases} \mathrm{id} & \text{if } j = k \\ 0 & \text{if } j \neq k. \end{cases}$$

We write $f \dotplus g$ for the sum in the commutative monoid $\mathbf{E}(A, B)$ in order to avoid confusion with the coproduct $f + g \colon A + A \to B + B$ of morphisms.

*Proof.* As the result is well known (see e.g. [20, 205]), we only sketch the proof.

(i) $\Longrightarrow$ (ii): For parallel morphisms $f, g \colon A \to B$, the sum is defined as

$$f \dotplus g = \left( A \xrightarrow{\Delta} A \oplus A \xrightarrow{f \oplus g} B \oplus B \xrightarrow{\nabla} A \right),$$

where $\Delta$ and $\nabla$ are the diagonal and codiagonal.

(ii) $\Longrightarrow$ (iii): Take $A \oplus B = A + B$, $\pi_1 = [\mathrm{id}, 0]$, and $\pi_2 = [0, \mathrm{id}]$.

(iii) $\Longrightarrow$ (i): Let $f \colon C \to A$ and $g \colon C \to B$ be morphisms. Then define $\langle f, g \rangle \colon C \to A \oplus B$ by

$$\langle f, g \rangle = \kappa_1 \circ f \dotplus \kappa_2 \circ g\,.$$

This tupling operation makes $A \oplus B$ into a product. Similarly $A \oplus B$ is a coproduct. ■

**Proposition 7.1.19.** *The totalization* $\mathcal{T}(\mathbf{C})$ *of an effectus* $\mathbf{C}$ *is a biproduct category.*

*Proof.* We show that $\mathcal{T}(\mathbf{C})$ satisfies condition (iii) of Lemma 7.1.18. By Proposition 7.1.15, $\mathcal{T}(\mathbf{C})$ is a category enriched over commutative monoids. For each $A, B \in \mathbf{C}$ we have

$$\begin{array}{ccccc} A & \xrightarrow{\kappa_1} & A + B & \xleftarrow{\kappa_2} & B \\ A & \xleftarrow{\rhd_1} & A + B & \xrightarrow{\rhd_2} & B \end{array} \qquad \text{in } \mathbf{C}\,.$$

These maps satisfy $\kappa_1 \circ \rhd_1 \ovee \kappa_2 \circ \rhd_2 = \mathrm{id}$ and

$$\rhd_j \circ \kappa_k = \begin{cases} \mathrm{id} & \text{if } j = k \\ 0 & \text{if } j \neq k. \end{cases} \tag{7.1}$$

The functor $\mathbf{C} \to \mathcal{T}(\mathbf{C})$ sends the maps $\kappa_k$ and $\rhd_j$ in $\mathcal{T}(\mathbf{C})$, preserving the equations (7.1). Moreover we have

$$1|\mathrm{id}_A\rangle = 1|\kappa_1 \circ \rhd_1 \ovee \kappa_2 \circ \rhd_2\rangle = 1|\kappa_1\rangle \circ 1|\rhd_1\rangle \dotplus 1|\kappa_2\rangle \circ 1|\rhd_2\rangle\,.$$

Therefore $A \oplus B = A + B$ forms a biproduct in $\mathcal{T}(\mathbf{C})$. ■

**Remark 7.1.20.** Totalization can be applied more generally to finPACs, and it also yields biproduct categories. Hoshino [130] studied a similar construction for *(strong) unique decomposition categories*, which are a generalization of partially $\sigma$-additive categories.

The totalization $\mathcal{T}(\mathbf{C})$ of an effectus is not only a biproduct category, but also is equipped with the 'ground' structure $\bar{\top}$. Below we will define *grounded biproduct categories*, which are biproduct categories with the ground structure satisfying certain axioms. Then it will be shown that $\mathcal{T}(\mathbf{C})$ is a grounded biproduct category, and conversely that every grounded biproduct category induces an effectus. In the next subsection, § 7.1.3, these constructions are shown to form a coreflection.

**Definition 7.1.21.** A **grounded biproduct category** is a biproduct category $(\mathbf{E}, \oplus, 0)$ with a distinguished object $I$ and a family of 'ground' maps $\bar{\top}_A \colon A \to I$ for $A \in \mathbf{E}$ satisfying the four conditions below.

(G1) $\bar{\top}_I = \mathrm{id}_I \colon I \to I$.

(G2) $\bar{\top}_{A\oplus B} = [\bar{\top}_A, \bar{\top}_B] \colon A \oplus B \to I$ for each $A, B \in \mathbf{E}$.

(G3) $\bar{\top}_B \circ f = 0$ implies $f = 0$ for each $f \colon A \to B$.

(G4) $p \dotplus q = p \dotplus r = \bar{\top}_A$ implies $p = q$ for each $p, q, r \colon A \to I$.

We say that a morphism $f \colon A \to B$ is **causal** if $\bar{\top}_B \circ f = \bar{\top}_A$. A morphism $f \colon A \to B$ is **subcausal** if $\bar{\top}_B \circ f \leq \bar{\top}_A$ in the algebraic ordering, i.e. if there is $p \colon A \to I$ such that $\bar{\top}_B \circ f \dotplus p = \bar{\top}_A$. Causal and subcausal morphisms form wide subcategories of $\mathbf{E}$, which will be denoted respectively by $\mathrm{Caus}(\mathbf{E})$ and $\mathrm{Caus}_{\leq}(\mathbf{E})$.

**Proposition 7.1.22.** *The totalization $\mathcal{T}(\mathbf{C})$ of an effectus $\mathbf{C}$ is a grounded biproduct category with the ground maps given by the truth maps $1|\mathbb{1}_A\rangle \colon A \to I$. Moreover, $\mathbf{C}$ is isomorphic to the category $\mathrm{Caus}_{\leq}(\mathcal{T}(\mathbf{C}))$ of subcausal maps (via $f \mapsto 1|f\rangle$).*

*Proof.* Conditions (G1) and (G2) are immediate since the truth maps $\mathbb{1}_A$ satisfy similar equations in $\mathbf{C}$. To show (G3), let $f = \sum_j n_j|f_j\rangle \colon A \to B$ be a morphism in $\mathcal{T}(\mathbf{C})$ such that $\bar{\top} \circ f = 0$. Then

$$1|0\rangle = 0 = 1|\mathbb{1}\rangle \circ (\textstyle\sum_j n_j|f_j\rangle) = \textstyle\sum_j n_j|\mathbb{1} \circ f_j\rangle\,.$$

By Lemma 7.1.8, it follows that the sum $\bigovee_j n_j \cdot (\mathbb{1} \circ f_j)$ is defined in $\mathbf{C}(A, I)$ and $\bigovee_j n_j \cdot (\mathbb{1} \circ f_j) = 0$. By positivity of morphisms in $\mathbf{C}$, $\mathbb{1} \circ f_j = 0$ for all $j$, so that $f_j = 0$ for all $j$. Therefore $f = \sum_j n_j|f_j\rangle = 0$. Note that (G4) follows easily from the latter assertion $\mathrm{Caus}_{\leq}(\mathcal{T}(\mathbf{C})) \cong \mathbf{C}$; hence we prove the isomorphism. Since the canonical functor $\mathbf{C} \to \mathcal{T}(\mathbf{C})$ is faithful, it suffices to prove that for each $f \colon A \to B$ in $\mathcal{T}(\mathbf{C})$, $f$ is subcausal if and only if $f = 1|g\rangle$ for some $g\colon A \to B$ in $\mathbf{C}$. The ‘if’ part is immediate since

$$\ground \circ f = 1|\mathbb{1}\rangle \circ 1|g\rangle = 1|\mathbb{1} \circ g\rangle \leq 1|\mathbb{1}\rangle = \ground\,.$$

Conversely, assume that $f\colon A \to B$ is subcausal. Then we have $\ground \circ f \dotplus p = \ground$ for some $p\colon A \to I$. Let $f = \sum_j n_j|f_j\rangle$ and $p = \sum_k m_k|p_k\rangle$ for morphisms $f_j, p_k$ in $\mathbf{C}$. Then we have

$$\textstyle\sum_j n_j|\mathbb{1} \circ f_j\rangle \dotplus \sum_k m_k|p_k\rangle = 1|\mathbb{1}\rangle\,.$$

By Lemma 7.1.8, the interpretation of the left-hand side is defined, so the sum $\bigovee_j n_j \cdot (\mathbb{1} \circ f_j)$ in $\mathbf{C}(A, I)$ is defined. Hence the sum $\bigovee_j n_j \cdot f_j$ is defined. It follows that

$$f = \textstyle\sum_j n_j|f_j\rangle = 1\big|\bigovee_j n_j \cdot f_j\big\rangle\,,$$

as desired. ■

We now describe a construction in the opposite direction: from grounded biproduct categories to effectuses.

**Lemma 7.1.23.** *In a grounded biproduct category, morphisms are positive: $f \dotplus g = 0$ implies $f = g = 0$.*

*Proof.* Let $f, g\colon A \to B$ be morphisms with $f \dotplus g = 0$. Note that the codiagonal $\nabla_B\colon B \oplus B \to B$ is causal: $\ground_B \circ \nabla_B = [\ground_B, \ground_B] = \ground_{B\oplus B}$. Thus

$$\ground_{B\oplus B} \circ \langle f, g\rangle = \ground_B \circ \nabla_B \circ \langle f, g\rangle = \ground_B \circ (f \dotplus g) = \ground_B \circ 0 = 0\,.$$

Then $\langle f, g\rangle = 0$ by (G3). It follows that $f = g = 0$. ■

**Lemma 7.1.24.** *Let $\mathbf{E}$ be a grounded biproduct category. Then the category $\mathrm{Caus}_{\leq}(\mathbf{E})$ of subcausal morphisms is a finPAC.*

*Proof.* Since the subcategory $\mathrm{Caus}_{\leq}(\mathbf{E})$ include all zero morphisms in $\mathbf{E}$, the zero object $0$ in $\mathbf{E}$ is a zero object in $\mathrm{Caus}_{\leq}(\mathbf{E})$ too. Note that coprojections $A \xrightarrow{\kappa_1} A \oplus B \xleftarrow{\kappa_2} B$ are causal by (G2). Let $f\colon A \to C$ and $g\colon B \to C$ be subcausal morphisms. Then the cotuple $[f, g]\colon A \oplus B \to C$ is causal as

$$\ground_C \circ [f, g] = [\ground_C \circ f, \ground_C \circ g] \leq [\ground_A, \ground_B] = \ground_{A\oplus B}\,.$$

Here the cotupling $[-,-]$ preserves the (pre)order since it is a monoid morphism. Therefore $A \oplus B$ is a coproduct of $A$ and $B$ in $\mathrm{Caus}_{\leq}(\mathbf{E})$. We have shown that $\mathrm{Caus}_{\leq}(\mathbf{E})$ has finite coproducts and a zero object.

For each $A, B \in \mathbf{E}$, subcausal morphisms $\mathrm{Caus}_{\leq}(\mathbf{E})(A, B)$ form a downset in the commutative monoid $\mathbf{E}(A, B)$. Therefore $\mathrm{Caus}_{\leq}(\mathbf{E})(A, B)$ is a PCM by defining sum $f \ovee g = f + g$ iff $f + g$ is subcausal. It is straightforward to see that compositions in $\mathrm{Caus}_{\leq}(\mathbf{E})$ are PCM bimorphisms. Therefore $\mathrm{Caus}_{\leq}(\mathbf{E})$ is enriched over PCMs.

Finally we check the axioms of finPACs. Note that partial projections $\rhd_j\colon A_1 \oplus A_2 \to A_j$ in $\mathrm{Caus}_{\leq}(\mathbf{E})$ are the projections $\pi_j$ of the biproduct. Clearly, subcausal morphisms $f, g \colon A \to B$ are compatible in $\mathrm{Caus}_{\leq}(\mathbf{E})$ iff the tuple $\langle f, g\rangle \colon A \to B \oplus B$ is subcausal. Moreover,

$$\bar{\top}_{B\oplus B} \circ \langle f, g\rangle = [\bar{\top}_B, \bar{\top}_B] \circ \langle f, g\rangle = \bar{\top}_B \circ f \dotplus \bar{\top}_B \circ g = \bar{\top}_B \circ (f + g)\,,$$

so that $\langle f, g\rangle$ is subcausal iff $f + g$ is subcausal. From this it is immediate that $\mathrm{Caus}_{\leq}(\mathbf{E})$ satisfies the compatible sum axiom. The untying axiom also holds since $\langle f, g\rangle = \kappa_1 \circ f \dotplus \kappa_2 \circ g$. ■

**Proposition 7.1.25.** *Let* $\mathbf{E}$ *be a grounded biproduct category. Then the category* $\mathrm{Caus}_{\leq}(\mathbf{E})$ *of subcausal morphisms is an effectus with* $I$ *as the unit object and ground maps* $\bar{\top}_A \colon A \to I$ *as truth maps.*

*Proof.* By (G4) and Lemma 7.1.23, for each $A \in \mathbf{E}$ the homset $\mathbf{E}(A, I)$ is a barred commutative monoid with $\bar{\top}_A$ as unit. The canonical downset $\downarrow(\bar{\top}_A)$ coincides with $\mathrm{Caus}_{\leq}(\mathbf{E})(A, I)$. Thus $\mathrm{Caus}_{\leq}(\mathbf{E})(A, I)$ is an effect algebra with top $\mathbb{1}_A = \bar{\top}_A$, as required by (E1) in the definition of an effectus (Definition 3.2.1). Note that (E2) is immediate from (G3). Finally we check (E3). Let $f, g \colon A \to B$ be morphisms in $\mathrm{Caus}_{\leq}(\mathbf{E})$ such that $\bar{\top} \circ f \ovee \bar{\top} \circ g$ is defined. The latter means that $\bar{\top} \circ f \dotplus \bar{\top} \circ g \leq \bar{\top}$. Thus

$$\bar{\top} \circ (f \dotplus g) = \bar{\top} \circ f \dotplus \bar{\top} \circ g \leq \bar{\top}\,,$$

so that $f \dotplus g$ is subcausal, i.e. $f \ovee g$ is defined. ■

**Corollary 7.1.26.** *If* $\mathbf{E}$ *is a grounded biproduct category, then the category* $\mathrm{Caus}(\mathbf{E})$ *is an effectus in total form.* ■

**Example 7.1.27.** To give examples of grounded biproduct categories, we generalize the multiset monad $\mathcal{M}$ (Definition 7.1.1), following [59, § 5]. Let $(R, +, 0, \cdot, 1)$ be a rig, see Definition 7.1.12. Then the multiset monad $\mathcal{M}_R \colon \mathbf{Set} \to \mathbf{Set}$ over $R$ is defined by

$$\mathcal{M}_R(X) = \{\varphi \colon X \to R \mid \mathrm{supp}(\varphi) \text{ is finite}\}$$

with the unit and the multiplication defined similarly to the distribution monads $\mathcal{D}_M$, see Definition 2.4.1. Then the monad $\mathcal{M}_R$ is *additive* in the sense that $\mathcal{M}_R(X + Y) \cong \mathcal{M}_R(Y) \times \mathcal{M}_R(Y)$ and $\mathcal{M}_R(0) \cong 1$, which implies that the Kleisli category $\mathcal{K}\ell(\mathcal{M}_R)$ is a biproduct category with $X \oplus Y = X + Y$, see [59, § 4 and § 5] for details.

We claim that the Kleisli category $\mathcal{K}\ell(\mathcal{M}_R)$ is a grounded biproduct category if the rig $R$ satisfies:

- (positivity) $s + t = 0$ implies $s = t = 0$; and
- (barred cancellativity) $s + t = s + r = 1$ implies $t = r$.

The conditions are equivalent to saying that $(R, +, 0)$ is a barred commutative monoid with unit 1. A sufficient condition is that $R$ is a cancellative positive rig.

The 'ground' structure on $\mathcal{K}\ell(\mathcal{M}_R)$ is given as follows: the unit object is the singleton $1 = \{*\}$, and the ground map $\bar{\top}_X \colon X \to \mathcal{M}_R(1)$ is defined by $\bar{\top}_X = \eta_1 \circ !_X$, i.e. $\bar{\top}_X(x) = 1|*\rangle$. We verify the axioms of grounded biproduct categories:

(G1) $\bar{\top}_1 = \eta_1 \circ !_1 = \eta_1$.

(G2) $\bar{\top}_{X\oplus Y} = \eta_1 \circ !_{X+Y} = \eta_1 \circ [!_X, !_Y] = [\eta_1 \circ !_X, \eta_1 \circ !_Y] = [\bar{\top}_X, \bar{\top}_Y]$.

(G3) If $\bar{\top}_Y \odot f = 0$ for $f\colon X \to \mathcal{M}_R(Y)$, then

$$0 = (\bar{\top}_Y \odot f)(x)(*) = \sum_{y\in Y} \bar{\top}_Y(y)(*) \cdot f(x)(y) = \sum_{y\in Y} f(x)(y)\,.$$

By positivity, $f(x)(y) = 0$. Thus $f = 0$.

(G4) Note that the additive structure on homsets $\mathcal{K}\ell(\mathcal{M}_R)(X,Y) = \mathbf{Set}(X, \mathcal{M}_R(Y))$ is the obvious pointwise addition. Thus if $p \dotplus q = p \dotplus r = \bar{\top}_X$ for $p, q, r\colon X \to \mathcal{M}_R(1)$, for each $x \in X$ one has

$$p(x)(*) + q(x)(*) = p(x)(*) + r(x)(*) = 1\,.$$

By the barred cancellation, $q(x)(*) = r(x)(*)$. Therefore $q = r$.

We now give concrete examples of rigs $R$.

(i) Take $R = \mathbb{N}$, the rig of natural numbers. Then $\mathcal{M}_{\mathbb{N}}$ is the usual (finite) multiset monad introduced in Definition 7.1.1. Clearly $\mathbb{N}$ is positive and cancellative, so that $\mathcal{K}\ell(\mathcal{M}_{\mathbb{N}})$ is a biproduct category. The causal maps $f\colon X \to \mathcal{M}_{\mathbb{N}}(Y)$ in $\mathcal{K}\ell(\mathcal{M}_{\mathbb{N}})$ are precisely (total) functions $f\colon X \to Y$, and the subcausal maps are partial functions $f\colon X \rightharpoonup Y$. Thus it yields the effectus $\mathrm{Caus}_{\leq}(\mathcal{K}\ell(\mathcal{M}_{\mathbb{N}})) \cong \mathbf{Pfn}$ for deterministic processes, and its total part $\mathrm{Caus}(\mathcal{K}\ell(\mathcal{M}_{\mathbb{N}})) \cong \mathbf{Set}$.

We note that the totalization $\mathcal{T}(\mathbf{Pfn})$ does not coincide with $\mathcal{K}\ell(\mathcal{M}_{\mathbb{N}})$. By Proposition 7.1.11 it is not hard to see that $\mathcal{T}(\mathbf{Pfn})$ is isomorphic to the wide subcategory of $\mathcal{K}\ell(\mathcal{M}_{\mathbb{N}})$ consisting of morphisms $f\colon X \to \mathcal{M}_{\mathbb{N}}(Y)$ that are *bounded* in the sense that $\sup_{x\in X, y\in Y} f(x)(y) < \infty$.

(ii) Take $R = \mathbb{R}_+$, the rig of nonnegative real numbers, which is positive and cancellative. Then we obtain a grounded biproduct category $\mathcal{K}\ell(\mathcal{M}_{\mathbb{R}_+})$. The causal maps $f\colon X \to \mathcal{M}_{\mathbb{R}_+}(Y)$ in $\mathcal{K}\ell(\mathcal{M}_{\mathbb{R}_+})$ are bijective with $\mathcal{D}$-Kleisli maps $f\colon X \to \mathcal{D}(Y)$, and the subcausal maps are $\mathcal{D}_{\leq}$-Kleisli maps $f\colon X \to \mathcal{D}_{\leq}(Y)$. Thus it yields the effectus $\mathrm{Caus}_{\leq}(\mathcal{K}\ell(\mathcal{M}_{\mathbb{R}_+})) \cong \mathcal{K}\ell(\mathcal{D}_{\leq})$ for probabilistic processes, and its total part $\mathrm{Caus}(\mathcal{K}\ell(\mathcal{M}_{\mathbb{R}_+})) \cong \mathcal{K}\ell(\mathcal{D})$.

Similarly to the previous example, the totalization $\mathcal{T}(\mathcal{K}\ell(\mathcal{D}_{\leq}))$ does not coincide with $\mathcal{K}\ell(\mathcal{M}_{\mathbb{R}_+}$, but by Proposition 7.1.11 (or by Proposition 7.1.13), $\mathcal{T}(\mathcal{K}\ell(\mathcal{D}_{\leq}))$ is isomorphic to the wide subcategory of $\mathcal{K}\ell(\mathcal{M}_{\mathbb{R}_+})$ consisting of morphisms $f\colon X \to \mathcal{M}_{\mathbb{R}_+}(Y)$ with $\sup_{x\in X, y\in Y} f(x)(y) < \infty$.

In general, if $R$ is positive and barred-cancellative, then the unit interval $[0,1]_R$ in $R$ (with respect to the algebraic ordering) is an effect monoid. Thus we can form the distribution monad $\mathcal{D}_{[0,1]_R}$ over $[0,1]_R$. (The monad $\mathcal{D}_{[0,1]_R}$ can also be described abstractly as the *affine part* [192] of the monad $\mathcal{M}_R$, see [144, Proposition 11 and Lemma 14].) Then $\mathrm{Caus}(\mathcal{K}\ell(\mathcal{M}_R)) \cong \mathcal{K}\ell(\mathcal{D}_{[0,1]_R})$. Moreover, one can describe the subcausal part $\mathrm{Caus}_{\leq}(\mathcal{K}\ell(\mathcal{M}_R))$ as the Kleisli category of the '*sub*distribution monad over $[0,1]_R$'. We leave the details as an exercise to the reader.

**Example 7.1.28.** We turn to quantum examples. We claim that the opposite $\mathbf{Wstar}_{\mathrm{CP}}^{\mathrm{op}}$ of the category of $W^*$-algebras and (arbitrary) normal CP maps is a grounded biproduct category. The category $\mathbf{Wstar}_{\mathrm{CP}}$ has finite products in the same way as $\mathbf{Wstar}_{\leq}$ does — that is, the trivial algebra $\{0\}$ is a final object and the direct sum $\mathscr{A}\oplus\mathscr{B}$ is a product; see Section 3.3.3. Clearly $\{0\}$ is also initial in $\mathbf{Wstar}_{\mathrm{CP}}$, and hence a zero object. Moreover $\mathscr{A}\oplus\mathscr{B}$ is also a coproduct. Indeed, there are coprojections given by $\kappa_1(a)=(a,0)$ and $\kappa_2(b)=(0,b)$. If $f\colon\mathscr{A}\to\mathscr{C}$ and $g\colon\mathscr{B}\to\mathscr{C}$ are normal CP maps, we can define $[f,g]\colon\mathscr{A}\oplus\mathscr{B}\to\mathscr{C}$ by $[f,g](a,b)=f(a)+g(b)$. The map $[f,g]$ is normal CP, and it is the unique mediating map. Therefore $\mathbf{Wstar}_{\mathrm{CP}}$, and hence $\mathbf{Wstar}_{\mathrm{CP}}^{\mathrm{op}}$, is a biproduct category.

The ground structure on $\mathbf{Wstar}_{\mathrm{CP}}^{\mathrm{op}}$ is as follows. The unit object is $\mathbb{C}$, and the ground map $\bar{\top}_{\mathscr{A}}\colon\mathscr{A}\to\mathbb{C}$ is the normal CP map $\bar{\top}_{\mathscr{A}}\colon\mathbb{C}\to\mathscr{A}$ defined by $\bar{\top}_{\mathscr{A}}(\lambda)=1\lambda$. We verify the axioms of grounded biproduct categories:

(G1) Clearly, $\bar{\top}_{\mathbb{C}}=\mathrm{id}_{\mathbb{C}}$.

(G2) In $\mathbf{Wstar}_{\mathrm{CP}}$ we have $\bar{\top}_{\mathscr{A}\oplus\mathscr{B}}(\lambda)=(1,1)\lambda=(1\lambda,1\lambda)=\langle\bar{\top}_{\mathscr{A}},\bar{\top}_{\mathscr{B}}\rangle(\lambda)$. Thus $\bar{\top}_{\mathscr{A}\oplus\mathscr{B}}=[\bar{\top}_{\mathscr{A}},\bar{\top}_{\mathscr{B}}]$ in the opposite.

(G3) Assume $\bar{\top}_{\mathscr{B}}\circ f=0$ for $f\colon\mathscr{A}\to\mathscr{B}$ in $\mathbf{Wstar}_{\mathrm{CP}}^{\mathrm{op}}$. Then $f(1)=0$, which implies $f=0$ (use Corollary 2.6.10).

(G4) Morphisms $p\colon\mathscr{A}\to\mathbb{C}$ in $\mathbf{Wstar}_{\mathrm{CP}}^{\mathrm{op}}$, i.e. normal CP maps $p\colon\mathbb{C}\to\mathscr{A}$, can be identified with positive elements $p(1)\in\mathscr{A}_+$. The addition $\dot{+}$ in $\mathbf{Wstar}_{\mathrm{CP}}^{\mathrm{op}}(\mathscr{A},\mathbb{C})$ corresponds to the usual addition in $\mathscr{A}_+$. Since $\mathscr{A}_+$ is cancellative, the desired property is satisfied.

Therefore $\mathbf{Wstar}_{\mathrm{CP}}^{\mathrm{op}}$ is a grounded biproduct category. The causal maps in $\mathbf{Wstar}_{\mathrm{CP}}^{\mathrm{op}}$ are unital maps, and the subcausal maps are subunital maps. Thus it yields the effectus $\mathrm{Caus}_{\leq}(\mathbf{Wstar}_{\mathrm{CP}}^{\mathrm{op}})=\mathbf{Wstar}_{\leq}^{\mathrm{op}}$ and its total part $\mathrm{Caus}(\mathbf{Wstar}_{\mathrm{CP}}^{\mathrm{op}})=\mathbf{Wstar}^{\mathrm{op}}$.

Unlike the examples we saw in Example 7.1.27, in this case we have

$$\mathcal{T}(\mathbf{Wstar}_{\leq}^{\mathrm{op}})\cong\mathbf{Wstar}_{\mathrm{CP}}^{\mathrm{op}}\,.$$

To see this, note that the subunital maps $\mathbf{Wstar}_{\leq}^{\mathrm{op}}(\mathscr{A},\mathscr{B})$ form a downset in the commutative monoid $\mathbf{Wstar}_{\mathrm{CP}}^{\mathrm{op}}(\mathscr{A},\mathscr{B})$, and that $\mathbf{Wstar}_{\mathrm{CP}}^{\mathrm{op}}(\mathscr{A},\mathscr{B})$ is a module over the rig $\mathbb{R}_+$. For each normal CP map $f\colon\mathscr{B}\to\mathscr{A}$, take $n\in\mathbb{N}_{>0}$ such that $n\geq\|f(1)\|$. Then the map $(1/n)\cdot f\colon\mathscr{B}\to\mathscr{A}$ is subunital, since $\|(1/n)\cdot f(1)\|\leq 1$ and hence $(1/n)\cdot f(1)\leq 1$ by Lemma 2.6.9. By Proposition 7.1.13, we obtain $\mathcal{T}(\mathbf{Wstar}_{\leq}^{\mathrm{op}}(\mathscr{A},\mathscr{B}))\cong\mathbf{Wstar}_{\mathrm{CP}}^{\mathrm{op}}(\mathscr{A},\mathscr{B})$, and therefore $\mathcal{T}(\mathbf{Wstar}_{\leq}^{\mathrm{op}})\cong\mathbf{Wstar}_{\mathrm{CP}}^{\mathrm{op}}$.

The above arguments work also for $C^*$-algebras. Therefore, the opposite $\mathbf{Cstar}_{\mathrm{CP}}^{\mathrm{op}}$ of the category of $C^*$-algebras and CP maps is a grounded biproduct category, with $\mathrm{Caus}_{\leq}(\mathbf{Cstar}_{\mathrm{CP}}^{\mathrm{op}})=\mathbf{Cstar}_{\leq}^{\mathrm{op}}$ and $\mathrm{Caus}(\mathbf{Cstar}_{\mathrm{CP}}^{\mathrm{op}})=\mathbf{Cstar}^{\mathrm{op}}$. Moreover one has $\mathcal{T}(\mathbf{Cstar}_{\leq}^{\mathrm{op}})\cong\mathbf{Cstar}_{\mathrm{CP}}^{\mathrm{op}}$.

### 7.1.3 A coreflection between effectuses and grounded biproduct categories

We have shown that every effectus $\mathbf{C}$ yields a grounded biproduct category $\mathcal{T}(\mathbf{C})$ via totalization, and conversely that every grounded biproduct category $\mathbf{E}$ induces an effectus $\mathrm{Caus}_{\leq}(\mathbf{C})$ via subcausal maps. These constructions do not form an equivalence, unlike the equivalence between effectuses in partial and total form (Section 4.2). Nevertheless, there is a *coreflection* between effectuses and grounded biproduct categories (Theorem 7.1.32).

**Definition 7.1.29.** We define a 2-category **GBC** of grounded biproduct categories as follows.

- An object is a grounded biproduct category $(\mathbf{E}, I)$.
- A morphism of type $(\mathbf{E}, I_{\mathbf{E}}) \to (\mathbf{F}, I_{\mathbf{F}})$ is a functor $F\colon \mathbf{E} \to \mathbf{F}$ that preserves finite biproducts, together with an isomorphism $u\colon I_{\mathbf{F}} \to F I_{\mathbf{E}}$ in $\mathbf{E}$ such that $F⏚_A = u \circ ⏚_{FA}$ for each $A \in \mathbf{E}$.
- A 2-cell of type $(F, u) \Rightarrow (G, v)\colon (\mathbf{E}, I_{\mathbf{E}}) \to (\mathbf{F}, I_{\mathbf{F}})$ is a natural transformation $\alpha\colon F \Rightarrow G$ such that $\alpha_{I_{\mathbf{C}}} \circ u = v$.

**Lemma 7.1.30.** *Let $F\colon \mathbf{E} \to \mathbf{F}$ be a functor between biproduct categories. The following are equivalent.*

(i) *$F$ preserves finite biproducts.*

(ii) *$F$ preserves finite coproducts.*

(iii) *For each $A, B \in \mathbf{E}$, the map $F\colon \mathbf{E}(A, B) \to \mathbf{F}(FA, FB)$ is a monoid morphism. (In other words, $F$ is a functor enriched over commutative monoids.)*

*Proof.* It is not hard to see that $F$ preserves the zero object iff $F$ preserves the initial object iff $F$ preserves zero morphisms. Thus we may assume that $F$ preserves the zero objects and zero morphisms. Note then that for each $A, B \in \mathbf{E}$, the following two canonical morphisms

$$F(A \oplus B) \underset{[F\kappa_1, F\kappa_2]}{\overset{\langle F\pi_1, F\pi_2\rangle}{\rightleftarrows}} FA \oplus FB \tag{7.2}$$

always satisfy $\langle F\pi_1, F\pi_2\rangle \circ [F\kappa_1, F\kappa_2] = \mathrm{id}_{FA \oplus FB}$. Then clearly (ii) $\Longrightarrow$ (i) holds, since if $[F\kappa_1, F\kappa_2]$ is invertible, then so is $\langle F\pi_1, F\pi_2\rangle$. Now we assume (iii). Then the morphisms in (7.2) satisfies

$$\begin{aligned}
[F\kappa_1, F\kappa_2] \circ \langle F\pi_1, F\pi_2\rangle &= F\kappa_1 \circ F\pi_1 \dotplus F\kappa_2 \circ F\pi_2 \\
&= F(\kappa_1 \circ \pi_1 \dotplus \kappa_2 \circ \pi_2) \\
&= F\mathrm{id}_{A \oplus B} = \mathrm{id}_{F(A \oplus B)}
\end{aligned}$$

Therefore the morphisms in (7.2) are isomorphisms, showing that (ii) (and (i)) hold. Finally we prove (iii) assuming (i). For $f, g\colon A \to B$ in $\mathbf{E}$, equation $F(f \dotplus g) = Ff \dotplus Fg$ holds by the following diagram chasing.

$$\begin{array}{ccccccc}
 & \nearrow^{F\Delta} & F(A \oplus A) & \xrightarrow{F(f \oplus g)} & F(B \oplus B) & \searrow^{F\nabla} & \\
FA & & \downarrow\cong & & \downarrow\cong & & FB \\
 & \searrow_{\Delta} & FA \oplus FA & \xrightarrow{Ff \oplus Fg} & FB \oplus FB & \nearrow_{\nabla} &
\end{array}$$

■

**Proposition 7.1.31.** *The mapping* $\mathbf{E} \mapsto \mathrm{Caus}_\le(\mathbf{E})$ *yields a functor* $\mathrm{Caus}_\le \colon \mathbf{GBC} \to \mathbf{Ef}$.

*Proof.* Let $(F,u)\colon (\mathbf{E}, I_\mathbf{E}) \to (\mathbf{F}, I_\mathbf{F})$ be a morphism in $\mathbf{GBC}$. Let $f\colon A \to B$ be a subcausal morphism in $\mathbf{E}$. Then $\ground \circ f \le \ground$, so that $F\ground \circ Ff \le F\ground$, since $F$ preserves sums and hence the algebraic preorder. By $F\ground = u \circ \ground$, we have $u \circ \ground \circ Ff \le u \circ \ground$ and hence $\ground \circ Ff \le \ground$ by composing $u^{-1}$. That is, $Ff$ is subcausal. Therefore we can restrict $F$ to a functor $\mathrm{Caus}_\le(F)\colon \mathrm{Caus}_\le(\mathbf{E}) \to \mathrm{Caus}_\le(\mathbf{F})$ between the subcategories. The functor $\mathrm{Caus}_\le(F)$ preserves finite coproducts, because finite coproducts in $\mathrm{Caus}_\le(\mathbf{E})$ are biproducts in $\mathbf{E}$. Note that

$$\mathrm{id}_{FI} = F\mathrm{id}_I = F\ground_I = u \circ \ground_{FI}\,,$$

and hence $u^{-1} = \ground_{FI}$, so $\ground_{FI} \circ u = \mathrm{id}_I = \ground_I$. Therefore $u\colon I_\mathbf{F} \to FI_\mathbf{E}$ is causal and belongs to $\mathrm{Caus}_\le(\mathbf{F})$. Because $\mathbb{1} = \ground$, the functor $\mathrm{Caus}_\le(F)$ forms a morphism of effectuses. Let $\alpha\colon (F,u) \Rightarrow (G,v)\colon (\mathbf{E}, I_\mathbf{E}) \to (\mathbf{F}, I_\mathbf{F})$ be a 2-cell in $\mathbf{GBC}$. Then for each $A \in \mathbf{E}$,

$$\begin{aligned}
\ground_{GA} \circ \alpha_A &= v^{-1} \circ G\ground_A \circ \alpha_A && \text{by } G\ground_A = v \circ \ground_{GA} \\
&= v^{-1} \circ \alpha_I \circ F\ground_A && \\
&= u^{-1} \circ F\ground_A && \text{by } \alpha_I \circ u = v \\
&= \ground_{FA} && \text{by } F\ground_A = u \circ \ground_{FA}\,.
\end{aligned}$$

Therefore $\alpha_A$ is causal, so we can define $\mathrm{Caus}_\le(\alpha)\colon \mathrm{Caus}_\le(F) \Rightarrow \mathrm{Caus}_\le(G)$ by $\mathrm{Caus}_\le(\alpha)_A = \alpha_A$. Clearly $\mathrm{Caus}_\le(\alpha)$ is natural and forms a 2-cell in $\mathbf{Ef}$. It is easy to check that $\mathrm{Caus}_\le\colon \mathbf{GBC} \to \mathbf{Ef}$ is 2-functorial. ∎

**Theorem 7.1.32.** *The totalization* $\mathbf{C} \mapsto \mathcal{T}(\mathbf{C})$ *of effectuses yields a left (strict) 2-adjoint to the 2-functor* $\mathrm{Caus}_\le\colon \mathbf{GBC} \to \mathbf{Ef}$.

$$\mathbf{Ef} \underset{\mathrm{Caus}_\le}{\overset{\mathcal{T}}{\underset{\bot}{\rightleftarrows}}} \mathbf{GBC}$$

*Moreover, it is a '2-coreflection' in the sense that the unit of the 2-adjunction* $\eta\colon \mathrm{id}_\mathbf{Ef} \Rightarrow \mathrm{Caus}_\le \circ \mathcal{T}$ *is an isomorphism.*

*Proof.* We define the unit by the isomorphisms $\eta_\mathbf{C}\colon \mathbf{C} \to \mathrm{Caus}_\le(\mathcal{T}(\mathbf{C}))$, $\eta_\mathbf{C}(f) = 1|f\rangle$, from Proposition 7.1.22. It is clear that every $\eta_\mathbf{C}$ is a morphism in $\mathbf{Ef}$. We show that $\eta_\mathbf{C}$ is universal. Let $F\colon \mathbf{C} \to \mathrm{Caus}_\le(\mathbf{E})$ be a morphism of effectuses. It consists of PCM morphisms $F_{A,B}\colon \mathbf{C}(A,B) \to \mathrm{Caus}_\le(\mathbf{E})(FA,FB)$ for $A,B \in \mathbf{C}$. Since $\mathrm{Caus}_\le(\mathbf{E})(FA,FB)$ is a downset in a monoid $\mathbf{E}(FA,FB)$, the adjunction $\mathcal{T}\colon \mathbf{PCM} \rightleftarrows \mathbf{DCM}\colon \mathrm{Down}$ yields monoid morphisms

$$\overline{F}_{A,B}\colon \mathcal{T}(\mathbf{C})(A,B) = \mathcal{T}(\mathbf{C}(A,B)) \to \mathbf{E}(A,B)\,.$$

(Explicitly, $\overline{F}(\sum_j n_j|f_j\rangle) = \sum_j n_j \cdot Ff_j$.) It is straightforward to verify that these maps $\overline{F}_{A,B}$ define a functor $\overline{F}\colon \mathcal{T}(\mathbf{C}) \to \mathbf{E}$. Since it is a $\mathbf{CMon}$-enriched functor between biproduct categories, it preserves finite biproducts by Lemma 7.1.30. Therefore $\overline{F}\colon \mathcal{T}(\mathbf{C}) \to \mathbf{E}$ (with $u\colon I_{\mathbf{E}} \overset{\cong}{\to} FI_{\mathbf{C}}$) forms a morphism in **GBC**. Clearly $F = \mathrm{Caus}_{\leq}(\overline{F}) \circ \eta_{\mathbf{C}}$ holds. To see the uniqueness, let $G\colon \mathcal{T}(\mathbf{C}) \to \mathbf{E}$ be a morphism in **GBC** such that $F = \mathrm{Caus}_{\leq}(G) \circ \eta_{\mathbf{C}}$. Then $GA = FA$ for each $A \in \mathbf{C}$, and the following diagram commutes for each $A, B \in \mathbf{C}$:

$$\begin{array}{ccc} \mathrm{Caus}_{\leq}(\mathcal{T}(\mathbf{C})(A,B)) & \xrightarrow{\mathrm{Caus}_{\leq}(G)} & \mathrm{Caus}_{\leq}(\mathbf{E})(FA,FB) \\ {\scriptstyle \eta_{\mathbf{C}}}\uparrow & \nearrow_{F} & \\ \mathbf{C}(A,B) & & \end{array}$$

Here $\mathrm{Caus}_{\leq}(G)$ is a restriction of the monoid morphism

$$G_{A,B}\colon \mathcal{T}(\mathbf{C})(A,B) \longrightarrow \mathbf{E}(FA,FB)\,.$$

Therefore it follows via adjunction $\mathbf{PCM} \rightleftarrows \mathbf{DCM}$ that $G_{A,B}$ is equal to $\overline{F}_{A,B}$ given above. Therefore $G = \overline{F}\colon \mathcal{T}(\mathbf{C}) \to \mathbf{E}$. Now we verify the universality of $\eta_{\mathbf{C}}$ with respect to 2-cells. Let $\alpha\colon F_1 \Rightarrow F_2\colon \mathbf{C} \to \mathrm{Caus}_{\leq}(\mathbf{E})$ be a 2-cell in **Ef**. We need to define $\overline{\alpha}\colon \overline{F_1} \Rightarrow \overline{F_2}\colon \mathcal{T}(\mathbf{C}) \to \mathbf{E}$, but this is easy: simply define $\overline{\alpha}_A = \alpha_A$, since $\mathbf{C}$ and $\mathcal{T}(\mathbf{C})$ has the same objects and $\mathrm{Caus}_{\leq}(\mathbf{E})$ is a subcategory of $\mathbf{E}$. It is clear that $\overline{\alpha}$ is a 2-cell in **GBC** and is a unique one such that $\alpha = \mathrm{Caus}_{\leq}(\overline{\alpha}) \circ \eta_{\mathbf{C}}$. We conclude that the 2-adjunction $\mathcal{T} \dashv \mathrm{Caus}_{\leq}$ holds, with the unit $\eta_{\mathbf{C}}$ an isomorphism. ■

**Corollary 7.1.33.** *The mapping* $\mathbf{C} \mapsto \mathcal{T}(\mathbf{C})$ *yields a 2-functor* $\mathcal{T}\colon \mathbf{Ef} \to \mathbf{GBC}$. ■

Even though **Ef** and **GBC** are not equivalent, the coreflection $\mathbf{Ef} \rightleftarrows \mathbf{GBC}$ gives us a good justification to work with grounded biproduct categories instead of effectuses. This may be seen as a first step in connecting effectus theory and *categorical quantum mechanics* [2, 54, 127] initiated by Abramsky and Coecke [1] and developed by them and many others around Oxford. Both biproducts $\oplus$ and ground structures $\bar{\top}$ are familiar concepts in categorical quantum mechanics; see e.g. [2, § 5] and [127, Chapter 2] for biproducts, and [54, § 6.6] and [127, Chapter 7] for ground structures (called there 'discarding' or 'environment structure'). Moreover, the following results show that a grounded biproduct category naturally appears in categorical quantum mechanics. The category **FdHilb** of finite-dimensional Hilbert spaces and linear maps is the archetypal example in categorical quantum mechanics. Although **FdHilb** itself is not a grounded biproduct category, applying to **FdHilb** any of the abstract categorical constructions below:

(i) the *CPM construction* together with the free biproduct completion [236];

(ii) the CPM construction together with the Karoubi envelope construction for unital dagger idempotents [125, 237] (see also [57]);

(iii) the *CP* construction* [52, 53];

one obtains the category $\mathbf{FdCstar}_{\mathrm{CP}}$ of finite-dimensional $C^*$-algebras (which are always $W^*$-algebras) and CP maps [125, Theorem 2.5 and Example 3.4]. Namely, we have the following isomorphisms:

$$\mathrm{CPM}(\mathbf{FdHilb})^{\oplus} \cong \mathrm{Split}_{\mathscr{I}}(\mathrm{CPM}(\mathbf{FdHilb})) \cong \mathrm{CP}^{*}(\mathbf{FdHilb}) \cong \mathbf{FdCstar}_{\mathrm{CP}} \quad (= \mathbf{FdWstar}_{\mathrm{CP}})\,,$$

where $(-)^{\oplus}$ denotes the biproduct completion and $\mathrm{Split}_{\mathscr{I}}$ the Karoubi envelope for unital dagger idempotents. Thus the category $\mathbf{FdCstar}_{\mathrm{CP}}$ appears naturally in categorical quantum mechanics. At the same time, the opposite $\mathbf{FdCstar}_{\mathrm{CP}}^{\mathrm{op}}$ forms a grounded biproduct category, see Example 7.1.28.[1] Therefore, $\mathbf{FdCstar}_{\mathrm{CP}}$ serves as a model for finite-dimensional quantum systems both in effectus theory and in categorical quantum mechanics. A further investigation, covering for instance logic and measurements, of the connection between effectus theory and categorical quantum mechanics will be left for future work. Closely related work here is Tull's thesis [250], which develops a categorical approach to the study of operational theories of physics using notions inspired by effectus theory.

## 7.2 Effectuses and the convex operational framework

The convex operational framework is a well-established framework for generalized probability theory, based on *base-norm* and *order-unit spaces*, which are certain kind of ordered normed spaces in a dual relationship (see § 7.2.2). The approach goes back (at least) to Ludwig [195–198], who studied an axiomatic framework where states and effects are respectively embedded in a base-norm and an order-unit space (see [198, Chapter IV], see also [197, Th. III.3.1]), and to Davies and Lewis [66], who used base-norm spaces as abstract state spaces. The term 'convex operational' referring to this approach seems to be coined by Barnum and Wilce [11–14, 258]. They also introduced *convex operational models* in [11, 258]. On the mathematical side, the notions of base-norm spaces and order-unit spaces are well-established too, see e.g. [4, 5, 9, 155, 259]. A fundamental result shown by Edward [75] and Ellis [77] is the duality between base-norm spaces and order-unit spaces — categorically they form a dual adjunction, see Theorem 7.2.40.

The notion of base-norm spaces is closely related to convex sets. Indeed, each base-norm space by definition has a *base*, which is a convex subset that generates the whole space in a suitable way. An analogous relationship between order-unit spaces and effect modules was revealed in [147, 148]. Since every effectus induces convex sets of states and effect modules of predicates, there should be some relationship between effectuses and the convex operational framework. The goal of this section is to make the relationship explicit and precise.

The work in this section is inspired by a result of van de Wetering [257, Theorem 2.9], which asserts that each *operational effect theory* $\mathbf{E}$ — a type of category inspired by effectus theory, see [257, Definitions 2.3 and 2.8] — induces a faithful functor from $\mathbf{E}$ to the category of order-unit spaces. Theorem 7.2.61 shown below can be seen as an extension of his result in the setting of effectuses, replacing order-unit spaces with *convex operational models* (see Definition 7.2.46).

[1] In fact, the finite dimensionality makes $\mathbf{FdCstar}_{\mathrm{CP}}$ self-dual, i.e. $\mathbf{FdCstar}_{\mathrm{CP}}^{\mathrm{op}} \cong \mathbf{FdCstar}_{\mathrm{CP}}$. Hence $\mathbf{FdCstar}_{\mathrm{CP}}$ itself also forms a grounded biproduct category.

In § 7.2.1 we establish equivalences of categories:

(i) (effect modules) $\simeq$ (semi-order-unit spaces)

(ii) (cancellative convex sets) $\simeq$ (cancellative weight modules) $\simeq$ (semi-base-norm spaces)

and in § 7.2.2 we restrict the above equivalences to the subcategories:

(iii) (Archimedean effect modules) $\simeq$ (order-unit spaces)

(iv) (metric convex sets) $\simeq$ (metric weight modules) $\simeq$ (pre-base-norm spaces)

These equivalences describe the precise relationship between effect modules and order-unit spaces, and between convex sets, weight modules, and base-norm spaces. Note that the equivalences (i) and (iii) are simply taken from [147, 148]. If we ignore weight modules, the equivalences of convex sets and base-norm spaces in (ii) and (iv) may be seen as a categorical presentation of known results, though the relevant results are scattered around e.g. [106–108, 134, 169, 223, 244]. In § 7.2.3 we give some more results about order-unit and base-norm spaces. In § 7.2.4 we introduce *convex operational models* as suitable dual pairs of pre-base-norm spaces and order-unit spaces. We also define *state-effect models* as certain dual pairs of convex sets and effect modules, and prove that the categories of convex operational models and state-effect models are equivalent. The equivalence may seen as a variant of Ludwig's embedding theorem [198, Chapter IV]. Finally in § 7.2.5 we make a precise explicit connection between effectuses and the convex operational framework. This is done by first showing that the categories of pre-base-norm spaces, order-unit spaces, and convex operational models are all effectuses. Then we show that for each real effectus $\mathbf{C}$ satisfying a certain separation condition, there is a faithful functor $F\colon \mathbf{C} \to \mathbf{COM}_{\leq}$ into the category $\mathbf{COM}$ of convex operational models (with 'subunital' morphisms). Moreover the functor $F\colon \mathbf{C} \to \mathbf{COM}_{\leq}$ is shown to be a morphism of effectuses.

### 7.2.1 Effect/weight modules, convex sets, and ordered vector spaces

Here we describe how effect modules, weight modules, and convex sets can be represented (or embedded) in — moreover, categorically equivalent to — certain ordered vector spaces. We call the ordered vector spaces corresponding to effect modules *semi-order-unit spaces*, and those corresponding to weight modules and convex sets *semi-base-norm spaces*. They are, respectively, relaxed 'semi-norm' versions of order-unit spaces and base-norm spaces. More specifically, it will be shown that every effect module can be represented as the *unit interval* $[0, u]_{\mathscr{A}}$ of some semi-order-unit space $(\mathscr{A}, u)$, and that every cancellative convex set and every cancellative weight module can be represented as the *base* $\mathrm{B}(\mathscr{V})$ and the *subbase* $\mathrm{B}_{\leq}(\mathscr{V})$, respectively, of some semi-base-norm space $(\mathscr{V}, \tau)$.

To discuss such representation in (real) ordered vector spaces, we focus on effect/weight modules and convex sets over the real unit interval $[0, 1] \subseteq \mathbb{R}$. Therefore, henceforth in this section, we refer to effect modules over $[0,1]$ simply by effect modules, and write simply **EMod** $=[0,1]$-**EMod**. We use similar convention for weight modules and convex sets, and so in particular we write **WMod** $=[0,1]$-**WMod** and **Conv** $=[0,1]$-**Conv**. Note that weight modules over $[0,1]$ always satisfy the normalization property by Proposition 4.4.10, and therefore weight modules and convex sets are equivalent by Corollary 4.4.9 — that is: **WMod** $=$ **WModn** $\simeq$ **Conv**.

Since we will present similar results for effect modules and weight modules, we start with a general result that applies to positive partial modules over $[0,1]$. We write **PPMod** $\hookrightarrow [0,1]$-**PMod** for the full subcategory consisting of positive partial modules over $[0,1]$.

**Definition 7.2.1.** A **cone** $C$ is a module over the rig $\mathbb{R}_+$ (see Definition 7.1.12) that satisfies *positivity*: $x+y=0$ implies $x=y=0$ for all $x,y\in C$. A cone $C$ is said to be **cancellative** if $x+y=x+z$ implies $y=z$ for all $x,y,z\in C$. Any cone $C$ carries the algebraic preorder given by $x\le y \iff \exists z.\, x+z=y$. If $C$ is cancellative, the preorder is a partial order. We write **Cone** for the category of cones and $\mathbb{R}_+$-module maps, and **CCone** $\hookrightarrow$ **Cone** for the full subcategory of cancellative cones.

**Definition 7.2.2.** We will refer to partially ordered real vector spaces simply as **ordered vector spaces**. Specifically, an ordered vector space is a real vector space $\mathcal{V}$ with a subset $\mathcal{V}_+$ satisfying:

(a) $\mathcal{V}_+$ is a $\mathbb{R}_+$-submodule of $\mathcal{V}$;

(b) $\mathcal{V}_+\cap(-\mathcal{V}_+)=\{0\}$.

The set $\mathcal{V}_+$ is called the **positive cone** of $\mathcal{V}$. Then $\mathcal{V}$ is partially ordered via $x\le y \iff y-x\in\mathcal{V}_+$. A map $f\colon\mathcal{V}\to\mathcal{W}$ between ordered vector spaces is **positive** if it sends positive elements to positive elements, which is equivalent to $f$ being monotone when $f$ is linear. It is standard (see e.g. [90, § 0.2]) that an ordered vector space $\mathcal{V}$ is *directed* (as poset) if and only if it is generated by the positive cone, i.e. $\mathcal{V}=\mathcal{V}_+-\mathcal{V}_+$. We write **DOVect** for the category of directed ordered vector spaces and positive linear maps.

In the lemma below, we will use the totalization construction from Section 7.1 and the well-known *Grothendieck group* construction for commutative monoids. We briefly recall the latter. Let $(M,+,0)$ be a commutative monoid. Consider the product monoid $M\times M$. We view elements $(x,y)\in M\times M$ as formal differences $x-y$, and define a relation $\sim$ on $M\times M$ by $(x,y)\sim(x',y')$ iff $x+y'+z=x'+y+z$ for some $z\in M$. Then $\sim$ is a monoid congruence on $M\times M$, which yields the quotient monoid $\mathcal{K}(M):=(M\times M)/{\sim}$. The monoid $\mathcal{K}(M)$ forms an (abelian) group, with inverses $-(x,y)=(y,x)$, and is called the **Grothendieck group** of $M$. The construction $M\mapsto\mathcal{K}(M)$ yields a left adjoint to the forgetful functor **Ab** $\to$ **CMon** from the category of abelian group to commutative monoids.

**Lemma 7.2.3.**

(i) *The totalization $\mathcal{T}(X)$ of a positive partial $[0,1]$-module $X$ is a cone, yielding a left adjoint functor:*

$$\mathbf{PPMod} \underset{\textit{forget}}{\overset{\mathcal{T}}{\rightleftarrows}} \mathbf{Cone} \qquad (\mathcal{T} \dashv \textit{forget})$$

*Here* **Cone** $\to$ **PPMod** *is the obvious forgetful functor.*

(ii) *The Grothendieck group* $\mathcal{K}(C)$ *of a cancellative cone* $C$ *is a directed ordered vector space, yielding an equivalence:*

$$\mathbf{CCone} \underset{(-)_+}{\overset{\mathcal{K}}{\rightleftarrows}} \mathbf{DOVect} \quad (\simeq)$$

*Here* $(-)_+ \colon \mathbf{DOVect} \to \mathbf{CCone}$ *is the functor that sends* $(V, V_+) \in \mathbf{DOVect}$ *to* $V_+$.

*Proof.*

(i) Let $\mathcal{T}(X)$ be the totalization of a partial module $X$. Since $\mathbb{R}_+ \cong \mathcal{T}([0,1])$ (Example 7.1.6), the $[0,1]$-action $\cdot \colon [0,1] \times X \to X$ induces a $\mathbb{R}_+$-action $\cdot \colon \mathbb{R}_+ \times \mathcal{T}(X) \to \mathcal{T}(X)$ by Lemma 7.1.7. It is then easy to see that $\mathcal{T}(X)$ is a $\mathbb{R}_+$-module. Positivity of $\mathcal{T}(X)$ follows easily by Lemma 7.1.8. Therefore $\mathcal{T}(X)$ is a cone. From the fact that $M \mapsto \mathcal{T}(M)$ is left adjoint to **CMon** $\to$ **PCM** (Proposition 7.1.4) it follows that $X \mapsto \mathcal{T}(X)$ is left adjoint to the forgetful functor **Cone** $\to$ **PPMod**.

(ii) It is straightforward to verify that the Grothendieck group $\mathcal{K}(C)$ of a cancellative cone $C$ is a real vector space, via the fact that $\mathcal{K}(\mathbb{R}_+) \cong \mathbb{R}$. We write $\eta_C \colon C \to \mathcal{K}(C)$ for the map given by $\eta_C(x) = (x, 0)$, which is injective because $C$ is cancellative. Then we define the positive cone of $\mathcal{K}(C)$ to be the image $\mathcal{K}(C)_+ := \eta_C[C]$. Clearly $\mathcal{K}(C)_+$ is a $\mathbb{R}_+$-submodule of $\mathcal{K}(C)$. Let $x \in \mathcal{K}(C)_+ \cap (-\mathcal{K}(C)_+)$. Then there exist $y, z \in C$ such that $(y, 0) = x = (0, z)$ in $\mathcal{K}(C)$. By definition of $\mathcal{K}(C)$, we have $y + z + w = w$ for some $w \in C$. Then $y = z = 0$ by cancellativity and positivity of $C$. Therefore $x = 0$, showing that $\mathcal{K}(C)$ is an ordered vector space. Clearly $\mathcal{K}(C)$ is generated by $\mathcal{K}(C)_+$ and hence directed. It is straightforward to see that the mapping $C \mapsto \mathcal{K}(C)$ extends to a functor **CCone** $\to$ **DOVect**. In the other direction, it is obvious that $\mathscr{V} \mapsto \mathscr{V}_+$ forms a functor **DOVect** $\to$ **CCone**.

For each $C \in$ **CCone** the map $\eta_C \colon C \to \mathcal{K}(C)$ is injective, and thus its corestriction $\eta_C \colon C \to \mathcal{K}(C)_+$ is an isomorphism in **CCone**. For each $\mathscr{V} \in$ **DOVect**, define the map $\varepsilon_{\mathscr{V}} \colon \mathcal{K}(\mathscr{V}_+) \to \mathscr{V}$ by $\varepsilon_{\mathscr{V}}(x_1, x_2) = x_1 - x_2$. Then $\varepsilon_{\mathscr{V}}$ is surjective since $\mathscr{V}$ is directed. To see the injectivity, assume $x_1 - x_2 = y_1 - y_2$ for $(x_1, x_2), (y_1, y_2) \in \mathcal{K}(\mathscr{V}_+)$. Then $x_1 + y_2 = y_1 + x_2$, which implies $(x_1, x_2) = (y_1, y_2)$ in $\mathcal{K}(\mathscr{V}_+)$ by the definition of the Grothendieck group. Therefore $\varepsilon_{\mathscr{V}}$ is bijective, and hence an isomorphism in **DOVect**. We leave the verification of naturality of $\eta$ and $\varepsilon$ to the reader. (In fact, they constitute an adjoint equivalence.) ∎

Now we describe the equivalence between effect modules and semi-order-unit spaces.

**Definition 7.2.4.** A **unit cone** is a cone $C$ equipped with an **order unit** $u \in C$ such that for any $x \in C$ there exists $n \in \mathbb{N}$ such that $x \leq n \cdot u$, where $\leq$ is the algebraic (pre)ordering. A map $f \colon (C, u_C) \to (D, u_D)$ between unit cones are **unital** if $f(u_C) = u_D$; and **subunital** if $f(u_C) \leq u_D$.

Below we use the category $\mathbf{CUCone}_{\leq} \hookrightarrow \mathbf{CCone}$ of *cancellative* unit cones and subunital $\mathbb{R}_+$-module maps, and its wide subcategory $\mathbf{CUCone} \hookrightarrow \mathbf{CUCone}_{\leq}$ determined by unital maps.

**Definition 7.2.5.**

(i) A **semi-order-unit space** is an ordered vector space $\mathscr{A}$ with a specified positive element $u \in \mathscr{A}_+$, called the **order unit** of $\mathscr{A}$, such that for all $a \in \mathscr{A}$ there exists $n \in \mathbb{N}$ with $-nu \leq a \leq nu$.

(ii) A map $f\colon (\mathscr{A}, u_{\mathscr{A}}) \to (\mathscr{B}, u_{\mathscr{B}})$ between semi-order-unit spaces is **unital** if $f(u_{\mathscr{A}}) = u_{\mathscr{B}}$; and **subunital** if $f(u_{\mathscr{A}}) \leq u_{\mathscr{B}}$.

(iii) We write $\mathbf{sOUS}_{\leq}$ for the category of semi-order-unit spaces and subunital positive linear maps, and $\mathbf{sOUS} \hookrightarrow \mathbf{sOUS}_{\leq}$ for the wide subcategory determined by *unital* maps.

Note that every semi-order-unit space is directed. Semi-order-unit spaces appeared in [147, 148] as 'partially ordered vector spaces with a strong unit'. The category denoted by **poVectu** there is equal to **sOUS**.

**Theorem 7.2.6.**

(i) *The functor* $\mathcal{T}\colon \mathbf{PPMod} \to \mathbf{Cone}$ *from Lemma* 7.2.3(i) *lifts to the functor* $\mathcal{T}\colon \mathbf{EMod}_{\leq} \to \mathbf{CUCone}_{\leq}$, *which is a part of the equivalence:*

$$\mathbf{EMod}_{\leq} \underset{[0,u]_{(-)}}{\overset{\mathcal{T}}{\rightleftarrows}} \mathbf{CUCone}_{\leq} \quad (\simeq)$$

*Here the inverse functor* $\mathbf{CUCone}_{\leq} \to \mathbf{EMod}$ *is given by sending unit cones* $C$ *to the unit intervals* $[0,u]_C = \{x \in U \mid x \leq u\}$.

(ii) *The equivalence* $\mathbf{CCone} \simeq \mathbf{DOVect}$ *from Lemma* 7.2.3(ii) *lifts to the equivalence:*

$$\mathbf{CUCone}_{\leq} \underset{(-)_+}{\overset{\mathcal{K}}{\rightleftarrows}} \mathbf{sOUS}_{\leq} \quad (\simeq)$$

*Combining the two equivalences one obtains the equivalence:*

$$\mathbf{EMod}_{\leq} \underset{[0,u]_{(-)}}{\overset{\mathcal{K}\circ\mathcal{T}}{\rightleftarrows}} \mathbf{sOUS}_{\leq} \quad (\simeq)$$

*Here the functor* $[0,u]_{(-)}\colon \mathbf{sOUS}_{\leq} \to \mathbf{EMod}_{\leq}$ *sends semi-order-unit spaces* $\mathscr{A}$ *to the unit intervals* $[0,u]_{\mathscr{A}} = \{a \in A \mid 0 \leq a \leq u\}$. *Moreover, all the equivalences above can restrict to unital maps. In particular,* $\mathbf{EMod} \simeq \mathbf{sOUS}$.

*Proof.* The result for unital maps — $\mathbf{EMod} \simeq \mathbf{CUCone} \simeq \mathbf{sOUS}$ — is shown in [148, Lemma 13 and Theorem 14], and the proof works also for subunital maps. ∎

We turn to the equivalence between weight modules and semi-base-norm spaces.

**Definition 7.2.7.** A **weight cone** is a $\mathbb{R}_+$-module $C$ equipped with a $\mathbb{R}_+$-module map $|-|: C \to \mathbb{R}_+$, called a **weight**, such that $|x| = 0$ implies $x = 0$ for all $x \in C$. The **base** and **subbase** of the weight cone is given respectively by:

$$\mathrm{B}(C) = \{x \in C \mid |x| = 1\}$$
$$\mathrm{B}_{\leq}(C) = \{x \in C \mid |x| \leq 1\}\,.$$

A map $f: C \to D$ between weight cones are **weight-preserving** if $|f(x)| = |x|$; and **weight-decreasing** if $|f(x)| \leq |x|$. We write $\mathbf{WCone}_{\leq}$ for the category of weight cones and *weight-decreasing* $\mathbb{R}_+$-module maps, and $\mathbf{WCone} \hookrightarrow \mathbf{WCone}_{\leq}$ for the wide subcategory with *weight-preserving* maps.

**Definition 7.2.8.**

(i) A **semi-base-norm space** is a directed ordered vector space $\mathscr{V}$ with a specified strictly positive[2] linear functional $\tau: \mathscr{V} \to \mathbb{R}$, which is called a **trace**. The trace defines the subsets $\mathrm{B}(\mathscr{V}), \mathrm{B}_{\leq}(\mathscr{V}) \subseteq \mathscr{V}_+$ called the **base** and **subbase**, respectively, by

$$\mathrm{B}(\mathscr{V}) = \{x \in \mathscr{V}_+ \mid \tau(x) = 1\}$$
$$\mathrm{B}_{\leq}(\mathscr{V}) = \{x \in \mathscr{V}_+ \mid \tau(x) \leq 1\}\,.$$

(ii) A map $f: (\mathscr{V}, \tau_{\mathscr{V}}) \to (\mathscr{W}, \tau_{\mathscr{W}})$ between semi-base-norm spaces is said to be **trace-preserving** if $\tau_{\mathscr{W}}(f(x)) = \tau_{\mathscr{V}}(x)$ for all $x \in \mathscr{V}$; and **trace-decreasing** if $\tau_{\mathscr{W}}(f(x)) \leq \tau_{\mathscr{V}}(x)$ for all $x \in \mathscr{V}_+$.

(iii) We write $\mathbf{sBNS}_{\leq}$ for the category of semi-order-unit spaces and *trace-decreasing* positive linear maps, and $\mathbf{sBNS} \hookrightarrow \mathbf{sBNS}_{\leq}$ for the wide subcategory determined by *trace-preserving* maps.

Semi-base-norm spaces appeared in [223] as 'base ordered linear spaces', which are defined in terms of *bases*. Our definition of semi-base-norm spaces in terms of traces is a generalization of (pre-)base-norm spaces in [90]. See the following remark for the equivalence of traces and bases.

**Remark 7.2.9.** Let $\mathscr{V}$ be a directed ordered vector space. A **base for the cone** $\mathscr{V}_+$ is a convex subset $B \subseteq \mathscr{V}_+$ such that for every nonzero $x \in \mathscr{V}_+$ there exist a unique $r \in \mathbb{R}_{>0}$ and $y \in B$ such that $x = ry$. If $(\mathscr{V}, \tau)$ is a semi-base-norm space, then $\mathrm{B}(\mathscr{V})$ is a base for $\mathscr{V}_+$ in this sense. Conversely, suppose that $B$ is a base of $\mathscr{V}_+$. For each nonzero $x \in X$, let $\tau(x) \in \mathbb{R}_{>0}$ be a unique real number such that $x = \tau(x) \cdot \overline{x}$ and $\overline{x} \in B$. Setting $\tau(0) = 0$, we obtain a $\mathbb{R}_+$-module map $\tau: \mathscr{V}_+ \to \mathbb{R}_+$, which extends uniquely to a strictly positive linear functional $\tau: \mathscr{V} \to \mathbb{R}$, that is, a trace on $\mathscr{V}$. The two constructions establish a bijective correspondence between traces $\tau: \mathscr{V} \to \mathbb{R}$ and bases $B \subseteq \mathscr{V}_+$.

We aim to obtain a result analogous to Theorem 7.2.6 for weight modules. Note that if a weight module is represented in a vector space, then it is necessarily cancellative. Thus we need to assume cancellativity. The same applies to convex sets, for which cancellativity is defined as follows.

[2] A map $\tau: \mathscr{V} \to \mathbb{R}$ is strictly positive if $x > 0$ implies $\tau(x) > 0$.

**Definition 7.2.10.** A convex set $K$ is **cancellative** if $[\![r|x\rangle + r^{\perp}|y\rangle]\!] = [\![r|x\rangle + r^{\perp}|z\rangle]\!]$ and $r \neq 1$ implies $y = z$, for every $x, y, z \in K$ and $r \in [0, 1]$.

We denote the full subcategories of *cancellative* convex sets / weight modules / weight cones as follows.

$$\mathbf{CConv} \hookrightarrow \mathbf{Conv} \qquad \mathbf{CWMod}_{\leq} \hookrightarrow \mathbf{WMod}_{\leq} \qquad \mathbf{CWMod} \hookrightarrow \mathbf{WMod}$$
$$\mathbf{CWCone}_{\leq} \hookrightarrow \mathbf{WCone}_{\leq} \qquad \mathbf{CWCone} \hookrightarrow \mathbf{WCone}$$

**Lemma 7.2.11.** *Let $C$ be a weight cone. Then $C$ is cancellative if and only if* $\mathrm{B}_{\leq}(C)$ *is cancellative (as PCM) if and only if* $\mathrm{B}(C)$ *is cancellative (as convex set).*

*Proof.* It is clear that if $C$ is cancellative, then so is $\mathrm{B}_{\leq}(C)$. Now suppose that $\mathrm{B}_{\leq}(C)$ is cancellative. Assume $[\![r|x\rangle + r^{\perp}|y\rangle]\!] = [\![r|x\rangle + r^{\perp}|z\rangle]\!]$ for $x, y, z \in \mathrm{B}(C)$ and $r \in [0, 1)$. This means

$$r \cdot x \ovee r^{\perp} \cdot y = r \cdot x \ovee r^{\perp} \cdot z \qquad \text{in } \mathrm{B}_{\leq}(C)\,,$$

so that $r^{\perp} \cdot y = r^{\perp} \cdot z$ by cancellation. Because $r^{\perp} \neq 0$, we obtain $y = z$ by Lemma 4.4.11. Hence $\mathrm{B}(C)$ is cancellative. Finally, assuming that $\mathrm{B}(C)$ is cancellative, we prove that $C$ is cancellative. Suppose that $x + y = x + z$ for $x, y, z \in C$. Since $|x| + |y| = |x| + |z|$ we have $|y| = |z|$. If at least one of $x, y, z$ is zero, then it is easy to see $y = z$. We assume that all $x, y, z$ are nonzero. Let $\bar{x} = |x|^{-1} \cdot x$, $\bar{y} = |y|^{-1} \cdot y$ and $\bar{z} = |z|^{-1} \cdot z$, so that $\bar{x}, \bar{y}, \bar{z} \in \mathrm{B}(C)$. Let $r = |x|/(|x| + |y|)$. Then $r^{\perp} = |y|/(|x| + |y|)$, and

$$\begin{aligned} [\![r|\bar{x}\rangle + r^{\perp}|\bar{y}\rangle]\!] = \frac{|x|\bar{x} + |y|\bar{y}}{|x| + |y|} &= \frac{x + y}{|x| + |y|} \\ &= \frac{x + z}{|x| + |y|} = \frac{|x|\bar{x} + |y|\bar{z}}{|x| + |y|} = [\![r|\bar{x}\rangle + r^{\perp}|\bar{z}\rangle]\!]\,, \end{aligned}$$

whence $\bar{y} = \bar{z}$ by cancellation in $\mathrm{B}(C)$. Then $y = |y|^{-1} \cdot \bar{y} = |z|^{-1} \cdot \bar{z} = z$. ■

**Theorem 7.2.12.**

(i) *The functor* $\mathcal{T}\colon \mathbf{PPMod} \to \mathbf{Cone}$ *from Lemma* 7.2.3(i) *lifts to the functor* $\mathcal{T}\colon \mathbf{WMod}_{\leq} \to \mathbf{WCone}_{\leq}$*, which is a part of the equivalence:*

$$\mathbf{WMod}_{\leq} \underset{\mathrm{B}_{\leq}}{\overset{\mathcal{T}}{\rightleftarrows}} \mathbf{WCone}_{\leq} \quad (\simeq) \tag{7.3}$$

*The inverse functor* $\mathbf{WCone}_{\leq} \to \mathbf{WMod}_{\leq}$ *is given by sending weight cones $C$ to the subbases* $\mathrm{B}_{\leq}(C)$.

(ii) *The equivalence* (7.3) *above can restrict to the subcategories* $\mathbf{CWMod}_{\leq} \simeq \mathbf{CWCone}_{\leq}$ *of cancellative weight modules/cones.*

(iii) *The equivalence* $\mathbf{CCone} \simeq \mathbf{DOVect}$ *from Lemma* 7.2.3(ii) *lifts to the equivalence:*

$$\mathbf{CWCone}_{\leq} \underset{(-)_{+}}{\overset{\mathcal{K}}{\rightleftarrows}} \mathbf{sBNS}_{\leq} \quad (\simeq)$$

*Combining the equivalences in* (ii) *and* (iii), *one obtains the equivalence:*

$$\mathbf{CWMod}_{\leq} \underset{\mathrm{B}_{\leq}}{\overset{\mathcal{K}\circ\mathcal{T}}{\rightleftarrows}} \mathbf{sBNS}_{\leq} \quad (\simeq)$$

*Moreover, all the equivalences above can restrict to weight/trace-preserving maps. In particular,* $\mathbf{CWMod} \simeq \mathbf{sBNS}$.

*Proof.* (i) Let $X \in \mathbf{WMod}_{\leq}$ and $\mathcal{T}(X)$ be the cone obtained by totalization. The weight $|{-}|\colon X \to [0,1]$ is a $[0,1]$-module map, and hence it extends to a $\mathbb{R}_+$-module map $|{-}|\colon \mathcal{T}(X) \to \mathcal{T}([0,1]) \cong \mathbb{R}$ by the functoriality of $\mathcal{T}$. Assume that $|\sum_j n_j|x_j\rangle| = 0$ for $\sum_j n_j|x_j\rangle \in \mathcal{T}(X)$. Then

$$0 = \Big|\sum\nolimits_j n_j|x_j\rangle\Big| = \sum\nolimits_j n_j \cdot |x_j|\,,$$

so $|x_j| = 0$ and hence $x_j = 0$ for each $j$. Thus $\sum_j n_j|x_j\rangle = 0$, showing that $\mathcal{T}(X)$ is a weight cone. We thus obtain a functor $\mathcal{T}\colon \mathbf{WMod}_{\leq} \to \mathbf{WCone}_{\leq}$. Clearly the functor $\mathrm{B}_{\leq}\colon \mathbf{WCone}_{\leq} \to \mathbf{WMod}_{\leq}$ is well-defined.

The totalization comes with a canonical injective map $\eta\colon X \to \mathcal{T}(X)$, which restricts to a weight-preserving $[0,1]$-module map $\eta\colon X \to \mathrm{B}_{\leq}(\mathcal{T}(X))$. To prove that $\eta\colon X \to \mathrm{B}_{\leq}(\mathcal{T}(X))$ is surjective, let $\sum_j n_j|x_j\rangle \in \mathrm{B}_{\leq}(\mathcal{T}(X))$, so that $|\sum_j n_j|x_j\rangle| \leq 1$. Then

$$1 \geq \Big|\sum\nolimits_j n_j|x_j\rangle\Big| = \sum\nolimits_j n_j \cdot |x_j|$$

which implies that the sum $\bigovee_j n_j \cdot x_j$ is defined in $X$. Then $\sum_j n_j|x_j\rangle = 1|\bigovee_j n_j \cdot x_j\rangle = \eta(\bigovee_j n_j \cdot x_j)$, showing that $\eta\colon X \to \mathrm{B}_{\leq}(\mathcal{T}(X))$ is surjective, and so bijective. It follows that $\eta\colon X \to \mathrm{B}_{\leq}(\mathcal{T}(X))$ is an isomorphism in $\mathbf{WMod}_{\leq}$.

For each weight cone $C$, define $\varepsilon\colon \mathcal{T}(\mathrm{B}_{\leq}(C)) \to C$ by $\varepsilon(\sum_j n_j|x_j\rangle) = \sum_j n_j \cdot x_j$, that is, $\varepsilon$ sends formal sums to actual sums. It is surjective: for each $x \in C$, we have $n^{-1} \cdot x \in \mathrm{B}_{\leq}(C)$ for large enough $n \in \mathbb{N}$ (such that $n \geq |x|$), and hence $n|n^{-1} \cdot x\rangle \in \mathcal{T}(\mathrm{B}_{\leq}(C))$ satisfies $\varepsilon(n|n^{-1} \cdot x\rangle) = n \cdot (n^{-1} \cdot x) = x$. To prove the injectivity, let $\sum_j n_j|x_j\rangle$ and $\sum_k m_k|y_k\rangle$ be elements of $\mathcal{T}(\mathrm{B}_{\leq}(C))$ such that

$$\sum\nolimits_j n_j \cdot x_j = \eta\Big(\sum\nolimits_j n_j|x_j\rangle\Big) = \eta\Big(\sum\nolimits_k m_k|y_k\rangle\Big) = \sum\nolimits_k m_k \cdot y_k\,.$$

Let $N = \sum_j n_j + \sum_k m_k$. Since the case when $N = 0$ is trivial, we assume that $N \geq 1$. Then $\sum_j n_j|N^{-1} \cdot x_j\rangle$ is a member of $\mathcal{T}(\mathrm{B}_{\leq}(C))$ such that

$$\sum\nolimits_j n_j \cdot |N^{-1} \cdot x_j| = \sum\nolimits_j n_j \cdot N^{-1} \cdot |x_j| \leq \sum\nolimits_j n_j \cdot N^{-1} \leq 1\,,$$

i.e. the sum $\bigovee_j n_j \cdot (N^{-1} \cdot x_j)$ is defined in $\mathrm{B}_{\leq}(C)$. This implies that

$$\sum\nolimits_j n_j|N^{-1} \cdot x_j\rangle = 1\big|\bigovee\nolimits_j n_j \cdot (N^{-1} \cdot x_j)\big\rangle$$

in $\mathcal{T}(\mathrm{B}_{\leq}(C))$. Similarly, $\sum_k m_k|N^{-1}\cdot y_k\rangle = 1|\ovee_k m_k\cdot(N^{-1}\cdot y_k)\rangle$. Since sums $\ovee$ in $\mathrm{B}_{\leq}(C)$ are $\sum$ in $C$, we have

$$\begin{aligned}\ovee_j n_j\cdot(N^{-1}\cdot x_j) &= N^{-1}\cdot\sum_j n_j\cdot x_j\\ &= N^{-1}\cdot\sum\nolimits_k m_k\cdot y_k = \ovee_k m_k\cdot(N^{-1}\cdot y_k)\,,\end{aligned}$$

whence $\sum_j n_j|N^{-1}\cdot x_j\rangle = \sum_k m_k|N^{-1}\cdot y_k\rangle$ in $\mathcal{T}(\mathrm{B}_{\leq}(C))$. Since $N\cdot(N^{-1}\cdot x) = x$ in $\mathrm{B}_{\leq}(X)$, we have $N|N^{-1}\cdot x\rangle = 1|x\rangle$ in $\mathcal{T}(\mathrm{B}_{\leq}(X))$. We conclude that

$$\sum\nolimits_j n_j|x_j\rangle = N\cdot\sum\nolimits_j n_j|N^{-1}\cdot x_j\rangle = N\cdot\sum\nolimits_k m_k|N^{-1}\cdot y_k\rangle = \sum\nolimits_k m_k|y_k\rangle\,.$$

Therefore $\varepsilon\colon\mathcal{T}(\mathrm{B}_{\leq}(C))\to C$ is an isomorphism in $\mathbf{WCone}_{\leq}$. Verifying the naturality of $\eta$ and $\varepsilon$, we prove $\mathbf{WMod}_{\leq}\simeq\mathbf{WCone}_{\leq}$.

(ii) This follows by Lemma 7.2.11.

(iii) Let $C$ be a cancellative weight cone. Then from $|{-}|\colon C\to\mathbb{R}_+$ we obtain a positive linear map $\tau\colon\mathcal{K}(C)\to\mathcal{K}(\mathbb{R}_+)\cong\mathbb{R}$ by the functoriality of $\mathcal{K}$. The map $\tau$ is strictly positive and thus makes $\mathcal{K}(C)$ into a semi-base-norm space. Conversely, if $\mathscr{V}$ is a semi-base-norm space, clearly the positive cone $\mathscr{V}_+$ is a cancellative cone. It is straightforward to see that the mappings are well-defined on morphisms and the functors lift to $\mathcal{K}\colon\mathbf{WCone}_{\leq}\to\mathbf{sBNS}_{\leq}$ and $(-)_+\colon\mathbf{sBNS}_{\leq}\to\mathbf{WCone}_{\leq}$.

For each $C\in\mathbf{CWCone}_{\leq}$ and $\mathscr{V}\in\mathbf{sBNS}_{\leq}$, the isomorphisms $\eta_C\colon C\to\mathcal{K}(C)_+$ and $\varepsilon_{\mathscr{V}}\colon\mathcal{K}(\mathscr{V}_+)\to\mathscr{V}$ from the proof of Lemma 7.2.3(ii) are weight- and trace-preserving, respectively. Therefore they are isomorphisms in $\mathbf{CWCone}_{\leq}$ and $\mathbf{sBNS}_{\leq}$, and the equivalence $\mathbf{CCone}\simeq\mathbf{DOVect}$ lifts to $\mathbf{CWCone}_{\leq}\simeq\mathbf{sBNS}_{\leq}$. ■

Combining this with the equivalence between convex sets and weight modules, we obtain:

**Corollary 7.2.13.** *The categories of cancellative convex sets, cancellative weight modules (with weight-preserving maps), and semi-base-norm spaces (with trace-preserving maps) are all equivalent:*

$$\mathbf{CConv}\ \underset{\mathrm{B}}{\overset{\mathcal{L}}{\rightleftarrows}}\!\!\!\!\!\!{\simeq}\ \ \mathbf{CWMod}\ \underset{\mathrm{B}_{\leq}}{\overset{\mathcal{K}\circ\mathcal{T}}{\rightleftarrows}}\!\!\!\!\!\!{\simeq}\ \ \mathbf{sBNS}$$

*Proof.* The equivalence on the right is Theorem 7.2.12. By Proposition 4.4.10 and Corollary 4.4.9, we have $\mathbf{WMod}\simeq\mathbf{Conv}$, which restricts to the subcategories $\mathbf{CWMod}\simeq\mathbf{CConv}$ by Lemma 7.2.11. ■

The equivalence $\mathbf{CConv}\simeq\mathbf{sBNS}$ is probably the most general version of equivalences concerning convex sets and base-norm spaces. More nontrivial, special versions of equivalences can be found in [223, Theorem 3.6] and [90, Corollary 2.9, Proposition 2.4.13, Proposition 3.3.3]; see also the next subsection and § 7.3.2. An immediate consequence of $\mathbf{CConv}\simeq\mathbf{sBNS}$ is that every cancellative convex set can be represented as a convex subset of a real vector space. This representation theorem was originally proved by Stone [244], see also [107, Theorem 4].

**Remark 7.2.14.** The Grothendieck group $\mathcal{K}(C)$ of a possibly non-cancellative weight cone $C$ also forms a semi-base-norm space, and in that case we get a reflection (an adjunction whose counit is an isomorphism) between them:

$$\mathbf{WCone} \underset{(-)_+}{\overset{\mathcal{K}}{\rightleftarrows}} \mathbf{sBNS} \quad (\bot)$$

Therefore there is also a reflection $\mathbf{Conv} \rightleftarrows \mathbf{sBNS}$ between (possibly non-cancellative) convex sets and semi-base-norm spaces.

### 7.2.2 Order-unit and (pre-)base-norm spaces

In the previous subsection, we obtained the equivalences $\mathbf{EMod} \simeq \mathbf{sOUS}$ and $\mathbf{CConv} \simeq \mathbf{CWMod} \simeq \mathbf{sBNS}$. For this we introduced the notion of semi-order-unit spaces and semi-base-norm spaces, which are respectively relaxed 'semi-norm' versions of order-unit and base-norm spaces. In this section we will introduce 'proper' order-unit and (pre-)base-norm spaces, both of which have intrinsic norms. Then based on the equivalences $\mathbf{EMod} \simeq \mathbf{sOUS}$ and $\mathbf{CConv} \simeq \mathbf{CWMod} \simeq \mathbf{sBNS}$. we will identify corresponding subcategories of effect/weight modules and convex sets.

To introduce order-unit and (pre-)base-norm spaces, we first recall the notion of Minkowski functionals. Let $\mathscr{V}$ be a real vector space and $S$ a subset of $\mathscr{V}$. Then the **Minkowski functional** of $S$ is a function $\|-\|_S \colon \mathscr{V} \to \mathbb{R}_+ \cup \{\infty\}$ defined by

$$\|x\|_S = \inf\{r > 0 \mid x \in rS\}\,.$$

If $\|-\|$ is a seminorm on $\mathscr{V}$, then one has $\|-\|_U = \|-\|$, where $U = \{x \in \mathscr{V} \mid \|x\| \le 1\}$ is the unit ball with respect to $\|-\|$. Thus any seminorm on a vector space can be obtained as the Minkowski functional. The following standard result tells us when the Minkowski functional $\|-\|_S$ defines a (semi)norm.

**Definition 7.2.15.** Let $\mathscr{V}$ be a real vector space. Then a subset $S \subseteq \mathscr{V}$ is said to be:

(i) **absolutely convex** if $0 \in S$ and $rx + sy \in S$ for all $x, y \in S$ and $r, s \in \mathbb{R}$ with $|r| + |s| \le 1$;

(ii) **absorbent** if for all $x \in \mathscr{V}$ there exists $r > 0$ such that $x \in rS$;

(iii) **radially bounded** if $\{r \in \mathbb{R} \mid rx \in S\} \subseteq \mathbb{R}$ is bounded for each nonzero $x \in \mathscr{V}$.

**Lemma 7.2.16.** *Let $S$ be an absorbent absolutely convex subset of a real vector space $\mathscr{V}$. Let $\|-\|_S \colon \mathscr{V} \to \mathbb{R}_+ \cup \{\infty\}$ be the Minkowski functional of $S$. Then:*

(i) *$\|-\|_S$ is a seminorm.*

(ii) *$\|-\|_S$ is a norm if and only if $S$ is radially bounded.*

*Proof.* See [90, § 0.1]. ■

**Definition 7.2.17.** An **order-unit space** is a semi-order-unit space $(\mathscr{A}, u)$ that is **Archimedean** in the sense that $nx \le u$ for all $n \in \mathbb{N}$ implies $x \le 0$. For an order-unit space $(\mathscr{A}, u)$, the Minkowski functional $\|-\|_{[-u,u]}$ of the interval

$$[-u, u] = \{a \in \mathscr{A} \mid -u \le a \le u\}$$

is a norm, called the **order-unit norm** on $\mathscr{A}$ and written simply as $\|-\|$. Explicitly,

$$\|a\| = \inf\{r > 0 \mid -ru \le a \le ru\}\,.$$

A **Banach order-unit space** is an order-unit space that is complete with respect to the order-unit norm.

We write $\mathbf{OUS} \hookrightarrow \mathbf{sOUS}$ and $\mathbf{OUS}_{\le} \hookrightarrow \mathbf{sOUS}_{\le}$ for the full subcategories of order-unit spaces.

**Remark 7.2.18.** The interval $[-u, u]$ in a semi-order-unit space is always absolutely convex and absorbent, but not necessarily radially bounded. Hence, in general, $\|-\|_{[-u,u]}$ is a seminorm but not a norm. It is known (see [5, Proposition 1.14] or [90, Lemma A.5.3]) that a semi-order-unit space $(\mathscr{A}, u)$ is an order-unit space if and only if both the seminorm $\|-\|_{[-u,u]}$ is a norm and the positive cone $\mathscr{A}_+$ is closed in the norm (cf. Definition 7.2.23, of (pre-)base-norm spaces).

Recall from Theorem 7.2.6 that effect modules and semi-order-unit spaces are equivalent: $\mathbf{EMod} \simeq \mathbf{sOUS}$. The Archimedean property can be translated into effect modules using a 'halving' trick.

**Definition 7.2.19.** A effect module $E$ is **Archimedean** if for any $a, b \in E$, $a \le b$ holds whenever $(1/2) \cdot a \le (1/2) \cdot b \ovee (1/2n) \cdot 1$ for all $n \in \mathbb{N}_{>0}$. The definition yields the full subcategories of Archimedean effect modules, $\mathbf{AEMod} \hookrightarrow \mathbf{EMod}$ and $\mathbf{AEMod}_{\le} \hookrightarrow \mathbf{EMod}_{\le}$.

The definition involves halves $1/2$. This is to ensure that the sum $(1/2) \cdot b \ovee (1/2n) \cdot 1$ is always defined. Then we obtain:

**Theorem 7.2.20** ([148, Proposition 15]). *The equivalence* $\mathbf{EMod}_{\le} \simeq \mathbf{sOUS}_{\le}$ *from Theorem* 7.2.6 *restricts to the subcategories* $\mathbf{AEMod}_{\le} \simeq \mathbf{OUS}_{\le}$ *of Archimedean effect modules and order-unit spaces. Similarly one has* $\mathbf{AEMod} \simeq \mathbf{OUS}$ *for unital maps.*

*Proof.* The claim boils down to showing that a semi-order-unit space $(\mathscr{A}, u)$ is Archimedean if and only if the interval $[0, u]_{\mathscr{A}}$ is an Archimedean effect module. This easily follows from the fact that $(1/2) \cdot a \le (1/2) \cdot b \ovee (1/2n) \cdot 1$ in $[0, u]_{\mathscr{A}}$ is equivalent to $n(a - b) \le u$ in $\mathscr{A}$. ■

Be warned that some authors say that an effect algebra is *Archimedean* (e.g. [74, § 1.2]) if whenever the $n$-fold sum $n \cdot x \coloneqq x \ovee \dots \ovee x$ is defined for all $n \in \mathbb{N}$, one has $x = 0$. For an effect module, the condition is equivalent to saying that $x \le (1/n) \cdot 1$ for all $n \in \mathbb{N}_{>0}$ implies $x = 0$, and is weaker than the Archimedean property of Definition 7.2.19. In [229, 230], an interval effect algebra is said to be *Archimedean* if the enveloping partially ordered group is Archimedean. This definition is compatible with Definition 7.2.19.

**Example 7.2.21.** We give examples of order-unit spaces and Archimedean effect modules. Each example consists of an order-unit space $(\mathscr{A}, u)$ and the Archimedean effect module that arises as its unit interval $[0, u]_{\mathscr{A}}$, via the equivalence $\mathbf{OUS} \simeq \mathbf{AEMod}$.

(i) The set $\mathbb{R}$ of real numbers is a Banach order-unit space with unit $1 \in \mathbb{R}$. Its unit interval is the Archimedean effect module $[0,1]$.

(ii) Let $X$ be a set. Then the real $\ell^\infty$-space $\ell^\infty_{\mathbb{R}}(X)$ (i.e. the space of all bounded functions $\varphi\colon X \to \mathbb{R}$) is a Banach order-unit space, whose unit is the constant function with value 1. The order-unit norm coincides with the sup norm. Its unit interval is the Archimedean effect module $[0,1]^X$ of $[0,1]$-valued functions on $X$.

(iii) Let $X$ be a compact Hausdorff space. Then the real continuous function space $\mathrm{C}_{\mathbb{R}}(X)$ is a Banach order-unit space, in a similarly manner to the previous example. Its unit interval is the Archimedean effect module of continuous functions $X \to [0,1]$.

(iv) Let $(X,\mu)$ be a measure space. Then the real $L^\infty$-space

$$L^\infty_{\mathbb{R}}(X,\mu) = \{\varphi\colon X \to \mathbb{R} \mid \varphi \text{ is measurable and essentially bounded}\}/{=_{\text{a.e.}}}$$

is a Banach order-unit space with the constant function with value 1 as unit. The order-unit norm is the essential sup norm: $\|\varphi\| = \operatorname{ess\,sup}|\varphi|$. Its unit interval is the Archimedean effect module of measurable functions $X \to [0,1]$ modulo $\mu$-negligible sets.

(v) The previous example works for any measure $\mu$. Thus, given a measurable set $(X,\Sigma_X)$, one may take the trivial infinite measure $\mu\colon \Sigma_X \to [0,\infty]$ defined by $\mu(\varnothing) = 0$ and $\mu(U) = \infty$ for all nonempty $U \in \Sigma_X$. Then the empty set is the only $\mu$-negligible set, so that we obtain a Banach order-unit space:

$$L^\infty_{\mathbb{R}}(X,\mu) \cong \mathcal{L}^\infty_{\mathbb{R}}(X) = \{\varphi\colon X \to \mathbb{R} \mid \varphi \text{ is measurable and bounded}\}\,.$$

The order-unit norm is the sup norm. The unit interval is the Archimedean effect module $\mathbf{Meas}(X,[0,1])$ of measurable functions $X \to [0,1]$.

(vi) Let $\mathscr{H}$ be a Hilbert space. Then the space $\mathcal{B}(\mathscr{H})_{\mathrm{sa}}$ of self-adjoint bounded operators on $\mathscr{H}$ is a Banach order-unit space, with the identity operator id as unit. The order-unit norm is the operator norm. Its unit interval is the Archimedean effect module of effects on $\mathscr{H}$.

In fact, all examples above are special cases of the following one:

(vii) Let $\mathscr{A}$ be a $C^*$-algebra. Then the space $\mathscr{A}_{\mathrm{sa}}$ of self-adjoint elements is a Banach order-unit space with unit 1. The order-unit norm coincides with the original norm of the $C^*$-algebra. This follows from Lemma 2.6.9 and the fact that $\mathscr{A}_+$ is closed [246, Theorem I.6.1]. Its unit interval is the Archimedean effect module $[0,1]_{\mathscr{A}}$ of effects in $\mathscr{A}$.

In particular, for each $W^*$-algebra $\mathscr{A}$, the space $\mathscr{A}_{\mathrm{sa}}$ is a Banach order-unit space.

Indeed, we can instantiate (i), (ii), (iii), (iv), and (vi), respectively by $C^*$-algebras $\mathbb{C}$, $\ell^\infty(X)$, $\mathrm{C}(X)$, $L^\infty(X,\mu)$, and $\mathcal{B}(\mathscr{H})$.

Below we introduce pre-base-norm spaces, which are dual to order-unit spaces. By the duality theorem (Theorem 7.2.40), the dual $\mathscr{V}^*$ of a pre-base-norm space $\mathscr{V}$ is always a Banach order-unit space.

We turn to base-norm spaces. Let $(\mathscr{V}, \tau)$ be a semi-base-norm space. The definition of (pre-)base-norm spaces involves the *absolutely convex hull* of the base $\mathrm{B}(\mathscr{V})$, which is denoted by $\mathrm{absco}(\mathrm{B}(\mathscr{V}))$. We start with an explicit description of $\mathrm{absco}(\mathrm{B}(\mathscr{V}))$.

**Lemma 7.2.22.** *If* $\mathscr{V} = \{0\}$*, then* $\mathrm{absco}(\mathrm{B}(\mathscr{V})) = \{0\}$*. If* $\mathscr{V} \neq \{0\}$*, then*

$$\mathrm{absco}(\mathrm{B}(\mathscr{V})) = \{x - y \mid x, y \in \mathscr{V}_+ \text{ and } \tau(x) + \tau(y) = 1\} .$$

*Proof.* If $\mathscr{V} = \{0\}$, clearly $\mathrm{absco}(\mathrm{B}(\mathscr{V})) = \{0\}$. As $\mathbf{CConv} \simeq \mathbf{sBNS}$, we have $\mathscr{V} = \{0\}$ if and only if $\mathrm{B}(\mathscr{V}) = \varnothing$. Thus if $\mathscr{V} \neq \{0\}$, then $\mathrm{B}(\mathscr{V}) \neq \varnothing$, so that $\mathrm{absco}(\mathrm{B}(\mathscr{V})) = \mathrm{co}(\mathrm{B}(\mathscr{V}) \cup -\mathrm{B}(\mathscr{V}))$, see [90, Lemma 0.1.1]. The rest is straightforward. ■

Let $U = \mathrm{absco}(\mathrm{B}(\mathscr{V}))$. We can easily verify that $U$ is absorbent. Therefore the Minkowski functional $\|-\|_U$ is a seminorm on $\mathscr{V}$, which we call the **base seminorm** on $\mathscr{V}$ and simply write $\|-\| = \|-\|_U$. The base seminorm is not necessarily a norm, which motivates the following definition.

**Definition 7.2.23.** A **pre-base-norm space** is a semi-base-norm space $(\mathscr{V}, \tau)$ such that the base seminorm is a norm, or equivalently, $U = \mathrm{absco}(B)$ is radially bounded. Then the base seminorm $\|-\| = \|-\|_U$ is called the **base norm** on $\mathscr{V}$. A **base-norm space** is a pre-base-norm space $(\mathscr{V}, \tau)$ where the positive cone $\mathscr{V}_+$ is closed with respect to the base norm.

A **Banach pre-base-norm** (resp. **Banach base-norm**) **space** is a pre-base-norm (resp. base-norm) space that is complete with respect to the base-norm.

In this section we will be mainly concerned with pre-base-norm spaces. We write $\mathbf{pBNS} \subseteq \mathbf{sBNS}$ and $\mathbf{pBNS}_{\leq} \subseteq \mathbf{sBNS}_{\leq}$ for the full subcategories of pre-base-norm spaces.

**Remark 7.2.24.** It should be noted that inequivalent definitions of base-norm spaces are used in the literature. Our terminology and definitions of pre-base-norm and base-norm spaces follow Furber [90, § 2.2.1]. See [90, § 2.2.3] for a detailed comparison of the inequivalent definitions. Ignoring the difference on the trivial space $\{0\}$ — our definition includes $\{0\}$ as (pre-)base-norm spaces, while some definitions do not — our pre-base-norm spaces coincide with Nagel's 'base norm space' [209]; and our base-norm spaces coincide with Asimow and Ellis's 'base norm space' [9].

Some authors (e.g. [4, 5]) use a stronger definition that requires $\mathrm{absco}(\mathrm{B}(\mathscr{V}))$ to be radially compact. Note also that some authors include completeness in the definition; for example, 'base norm space' in [186] means a Banach pre-base-norm space in our terminology.

Recall from Theorem 7.2.12 that there are equivalences $\mathbf{CConv} \simeq \mathbf{CWMod} \simeq \mathbf{sBNS}$ between cancellative convex sets, cancellative weight modules, and semi-base-norm spaces. In the rest of this subsection, we aim at identifying subclasses of convex sets and weight modules that are equivalent to pre-base-norm spaces, in an analogous way to Theorem 7.2.20.

The following presentation of the base seminorm is convenient.

**Lemma 7.2.25.** *The base seminorm $\|-\|$ on a semi-base-norm space $(\mathscr{V},\tau)$ can be calculated by the following formula:*

$$\|x\| = \inf\{\tau(x_1)+\tau(x_2) \mid x_1,x_2 \in \mathscr{V}_+ \text{ and } x = x_1 - x_2\}$$

*Proof.* The assertion is trivial if $\mathscr{V} = \{0\}$, or equivalently, $\mathrm{B}(\mathscr{V}) = \varnothing$. Thus we assume that $\mathrm{B}(\mathscr{V})$ is nonempty. The equation clearly holds for $x = 0$. We fix a nonzero $x \in \mathscr{V}$, and let $S = \{r > 0 \mid x \in rU\}$, where $U = \mathrm{absco}(\mathrm{B}(\mathscr{V}))$, and let

$$T = \{\tau(x_1)+\tau(x_2) \mid x_1,x_2 \in \mathscr{V}_+ \text{ and } x = x_1 - x_2\}\,.$$

Then it suffices to prove $S = T$. By Lemma 7.2.22, $r \in S$ iff $x = r(y - z)$ for some $y, z \in \mathscr{V}_+$ such that $\tau(y) + \tau(z) = 1$. In that case, we have

$$r = r(\tau(y)+\tau(z)) = \tau(ry)+\tau(rz) \in T\,,$$

whence $S \subseteq T$. Conversely, suppose $x = x_1 - x_2$ for some $x_1, x_2 \in \mathscr{V}_+$. Then $\tau(x_1) + \tau(x_2) \in S$, since

$$x = (\tau(x_1)+\tau(x_2))\left(\frac{x_1}{\tau(x_1)+\tau(x_2)} - \frac{x_2}{\tau(x_1)+\tau(x_2)}\right).$$

Here $\tau(x_1) + \tau(x_2) > 0$ as we assumed $x \neq 0$. Therefore $T \subseteq S$ and hence $S = T$. ■

Let $(\mathscr{V},\tau)$ be a semi-base-norm space. Then any subset $S \subseteq \mathscr{V}$ is equipped with a pseudometric induced by the base seminorm. We call it the **base pseudometric** and write $d(x, y) := \|x - y\|$.

**Proposition 7.2.26.** *A semi-base-norm space $(\mathscr{V},\tau)$ is a pre-base-norm space if and only if the base pseudometric $d(x, y) = \|x - y\|$ restricted on the subbase $\mathrm{B}_{\leq}(\mathscr{V})$ is a metric.*

*Proof.* The 'only if' is trivial. Assume that $d(x, y) = \|x - y\|$ is metric on $\mathrm{B}_{\leq}(\mathscr{V})$. Suppose that $x \in \mathscr{V}$ satisfies $\|x\| = 0$. Since $\mathscr{V}^+$ is generating, there are $x_+, x_- \in \mathscr{V}_+$ such that $x = x_+ - x_-$. If $x_+ = x_- = 0$, we are done. Otherwise, let $r = (\max\{\tau(x_+), \tau(x_-)\})^{-1}$. Then $rx_+, rx_- \in \mathrm{B}_{\leq}(\mathscr{V})$, and $d(rx_+, rx_-) = r \cdot d(x_+, x_-) = r\|x\| = 0$. By assumption, $rx_+ = rx_-$ and therefore $x = x_+ - x_- = 0$. ■

Therefore, to obtain a class of weight modules that is equivalent to pre-base-norm spaces, it suffices to characterize the base pseudometric $d(x, y)$ in terms of the weight module structure. This will be done below.

**Lemma 7.2.27.** *Let $(\mathscr{V},\tau)$ is a semi-base-norm space. Let $x, y \in \mathrm{B}_{\leq}(\mathscr{V})$ be elements in the subbase. Then*

$$d(x, y) \equiv \|x - y\| = \inf\{\tau(z) + \tau(w) \mid z, w \in \mathrm{B}_{\leq}(\mathscr{V}) \text{ and } x + z = y + w\}$$

*Proof.* Let

$$\begin{aligned} S_{x,y} &= \inf\{\tau(z) + \tau(w) \mid z, w \in \mathscr{V}_+ \text{ and } x + z = y + w\} \\ T_{x,y} &= \inf\{\tau(z) + \tau(w) \mid z, w \in \mathrm{B}_{\leq}(\mathscr{V}) \text{ and } x + z = y + w\}\,. \end{aligned}$$

Then $\|x-y\| = \inf S_{x,y}$ by Lemma 7.2.25. Clearly $S_{x,y} \supseteq T_{x,y}$ and hence $\inf S_{x,y} \leq \inf T_{x,y}$. Let $r \in S_{x,y}$, that is, $r = \tau(z) + \tau(w)$ for some $z, w \in \mathscr{V}_+$ such that $x + z = y + w$. If $z, w \in \mathrm{B}_{\leq}(\mathscr{V})$, then $r \in T_{x,y}$ and hence $\inf T_{x,y} \leq r$. Assume otherwise, and without loss of generality, $z \notin \mathrm{B}_{\leq}(\mathscr{V})$. Then $\tau(z) > 1$ and hence $\tau(z) > 1 \geq \tau(y)$. By $\tau(x) + \tau(z) = \tau(y) + \tau(w)$ it follows that $\tau(x) < \tau(w)$ and

$$\inf T_{x,y} \leq \tau(x) + \tau(y) < \tau(z) + \tau(w) = r\,.$$

Therefore $\inf T_{x,y} \leq \inf S_{x,y}$, and we conclude that $\|x-y\| = \inf T_{x,y}$. ■

**Proposition 7.2.28.** *Let $(\mathscr{V}, \tau)$ is a semi-base-norm space. The pseudometric $d(x,y) = \|x-y\|$ on the subbase $\mathrm{B}_{\leq}(\mathscr{V})$ can be calculated by the weight module structure of $\mathrm{B}_{\leq}(\mathscr{V})$ as follows.*

$$d(x,y) = \inf\bigl\{ |z| + |w| \bigm| z, w \in \mathrm{B}_{\leq}(\mathscr{V}) \text{ and } \tfrac{1}{2}x \ovee \tfrac{1}{2}z = \tfrac{1}{2}y \ovee \tfrac{1}{2}w \bigr\}\,. \qquad ■$$

*Proof.* Note that the sum $\frac{1}{2}x \ovee \frac{1}{2}z = \frac{1}{2}y \ovee \frac{1}{2}w$ is always defined in $\mathrm{B}(\mathscr{V})$, and the equality is equivalent to $x + z = y + w$ in $\mathscr{V}$. Thus the claim follows by Lemma 7.2.27. ■

This justifies the following definition:

**Definition 7.2.29.** Let $X$ be a weight module. We define the **base pseudometric** $d$ on $X$ by:

$$d(x,y) = \inf\bigl\{ |z| + |w| \bigm| z, w \in X \text{ and } \tfrac{1}{2}x \ovee \tfrac{1}{2}z = \tfrac{1}{2}y \ovee \tfrac{1}{2}w \bigr\}\,.$$

This indeed defines a pseudometric, see the lemma below. We say that a weight module is **metric** if the base pseudometric $d$ is a metric (i.e. $d(x,y) = 0$ implies $x = y$). We write $\mathbf{MWMod}_{\leq} \hookrightarrow \mathbf{WMod}_{\leq}$ and $\mathbf{MWMod} \hookrightarrow \mathbf{WMod}$ for the full subcategories of metric weight modules.

**Lemma 7.2.30.** *The base pseudometric $d$ on a weight module $X$ is indeed a pseudometric.*

*Proof.* It is clear that $d(x,x) = 0$ and $d(x,y) = d(y,x)$ hold. To prove the triangle inequality $d(x,z) \leq d(x,y) + d(y,z)$, it suffices to show that for any $a, b, c, d \in X$ such that $\frac{1}{2}x \ovee \frac{1}{2}a = \frac{1}{2}y \ovee \frac{1}{2}b$ and $\frac{1}{2}y \ovee \frac{1}{2}c = \frac{1}{2}z \ovee \frac{1}{2}d$, we have

$$d(x,z) \leq |a| + |b| + |c| + |d|\,.$$

If $|a| + |c| \leq 1$ and $|b| + |d| \leq 1$, then $\frac{1}{2}x \ovee \frac{1}{2}(a \ovee c) = \frac{1}{2}z \ovee \frac{1}{2}(b \ovee d)$, so the above inequality holds. Otherwise, without loss of generality we may assume $|a| + |c| > 1$. By $|x| + |a| + |c| = |z| + |b| + |d|$ and $|z| \leq 1 < |a| + |c|$ it follows that $|x| \leq |b| + |d|$ and hence

$$d(x,z) \leq |x| + |z| \leq |a| + |b| + |c| + |d|\,. \qquad ■$$

At this point, it is clear that the categories of metric *cancellative* weight modules pre-base-norm spaces are equivalent. It turns out that cancellativity is redundant.

**Lemma 7.2.31.** *Every metric weight module is cancellative.*

*Proof.* Let $X$ be a metric weight module. Suppose that $x \ovee y = x \ovee z$ in $X$. Take an arbitrary $n \in \mathbb{N}_{>0}$. We have $\frac{1}{n}x \ovee \frac{1}{n}y = \frac{1}{n}x \ovee \frac{1}{n}z$, and using it repeatedly,

$$\frac{1}{n}x \ovee y = \frac{1}{n}x \ovee \frac{1}{n}y \ovee \cdots \ovee \frac{1}{n}y = \frac{1}{n}x \ovee \frac{1}{n}z \ovee \cdots \ovee \frac{1}{n}z = \frac{1}{n}x \ovee z\,.$$

Since $n$ is arbitrary, we obtain $d(y,z) = 0$. Because $d$ is a metric, $y = z$. ■

**Corollary 7.2.32.** *The equivalence* $\mathbf{CWMod}_{\leq} \simeq \mathbf{sBNS}_{\leq}$ *from Theorem* 7.2.12 *restricts to the equivalence* $\mathbf{MWMod}_{\leq} \simeq \mathbf{pBNS}_{\leq}$ *of the categories of metric weight modules and pre-base-norm spaces. One also has* $\mathbf{MWMod} \simeq \mathbf{pBNS}$ *for weight/trace-preserving maps.*

*Proof.* By Propositions 7.2.26 and 7.2.28 and Lemma 7.2.31. ■

Next we characterize convex sets that correspond to metric weight modules using the pseudometric $\sigma$ on a convex set introduced by Gudder [106]. For a convex set $K$, we define:

$$\sigma(x,y) = \inf\bigl\{ r \in [0,1] \bigm| [\![ r^{\perp}|x\rangle + r|z\rangle ]\!] = [\![ r^{\perp}|y\rangle + r|w\rangle ]\!] \text{ for some } z,w \in K \bigr\}\,. \quad (7.4)$$

It satisfies $\sigma(x,y) \leq 1/2$. The following proposition relates Gudder's pseudometric $\sigma$ to the base pseudometric of weight modules. This is an adaptation of Gudder's observation in [106, p. 261] about *natural seminorm* (essentially the same thing as the base seminorm of a semi-base-norm space) to the setting of weight modules.

**Proposition 7.2.33.** *Let $X$ be a weight module. Then for each $x, y \in \mathrm{B}(X)$,*

$$d(x,y) = \frac{2\sigma(x,y)}{1 - \sigma(x,y)}\,,$$

*where $d$ is the base pseudometric on $X$ and $\sigma$ is Gudder's pseudometric on the convex set* $\mathrm{B}(X)$.

*Proof.* We fix $x, y \in \mathrm{B}(X)$ and write $d \coloneqq d(x,y)$ and $\sigma \coloneqq \sigma(x,y)$. If $x = y$, then $d = 0 = \sigma$, and the desired equation holds. We assume $x \neq y$ below. Suppose that $r \in [0, 1/2]$ satisfies $r^{\perp} \cdot x \ovee r \cdot z = r^{\perp} \cdot y \ovee r \cdot w$ for some $z, w \in \mathrm{B}(X)$. Then

$$\frac{1}{2} \cdot x \ovee \frac{1}{2} \cdot \frac{r}{1-r} \cdot z = \frac{1}{2} \cdot y \ovee \frac{1}{2} \cdot \frac{r}{1-r} \cdot w \qquad \text{in } X\,.$$

By the definition of the pseudometric $d = d(x,y)$, we have $d \leq 2r/(1-r)$, so that $d/(d+2) \leq r$. We have $\sigma = \sigma(x,y)$ as an infimum over these $r \in [0,1/2]$, see (7.4); hence $d/(d+2) \leq \sigma$, that is, $d \leq 2\sigma/(1-\sigma)$. To prove the other inequality, assume $\frac{1}{2}x \ovee \frac{1}{2}z = \frac{1}{2}y \ovee \frac{1}{2}w$ for $z, w \in X$. Since $|x| = 1 = |y|$, we have $|z| = |w| =: s$. By $x \neq y$, we have $s > 0$. Writing $\overline{z}$ and $\overline{w}$ respectively for the normalization of $z$ and $w$, we have the following equation in $\mathrm{B}(X)$:

$$\frac{1}{1+s} \cdot x \ovee \frac{s}{1+s} \cdot \overline{z} = \frac{1}{1+s} \cdot y \ovee \frac{r}{1+s} \cdot \overline{w}\,.$$

Therefore $\sigma \leq s/(1+s)$ and hence $2\sigma/(1-\sigma) \leq 2s = |z| + |w|$. By the definition of $d = d(x,y)$ we obtain $2\sigma/(1-\sigma) \leq d$, which concludes the proof. ■

**Definition 7.2.34.** A convex set $K$ is **metric** if Gudder's pseudometric $\sigma$ defined by (7.4) is a metric, or equivalently, $d(x,y) = (2\sigma(x,y))/(1-\sigma(x,y))$ is a metric. We call $d(x,y)$ the **base metric** of $K$. We write **MConv** $\hookrightarrow$ **Conv** for the full subcategory of metric convex sets.

**Remark 7.2.35.** Metric convex sets are convex sets with *canonical* metric. The notion should not be confused with similar notions such as *convex metric space* in [143, 149]. The latter is a convex set that is also equipped with a metric, which need not be the canonical one.

**Proposition 7.2.36.** *The equivalence* **Conv** $\simeq$ **WMod** *from Corollary* 4.4.9 *restricts to the full subcategories* **MConv** $\simeq$ **MWMod** *of metric convex sets and metric weight modules.*

*Proof.* It suffices to show that a weight module $X$ is metric if and only if the convex set $\mathrm{B}(X)$ is metric. The 'only if' follows from Proposition 7.2.33. To prove the converse, assume that $\mathrm{B}(X)$ is metric. Let $x, y \in X$ satisfy $d(x,y) = 0$. Take an arbitrary $\varepsilon > 0$. By $d(x,y) = 0$, there exists $z, w \in X$ such that $\frac{1}{2}x \ovee \frac{1}{2}z = \frac{1}{2}y \ovee \frac{1}{2}w$ and $|z| + |w| < \varepsilon$. Then $|x| + |z| = |y| + |w|$ and

$$\big||x| - |y|\big| = \big||z| - |w|\big| \leq |z| + |w| < \varepsilon\,.$$

As $\varepsilon > 0$ is arbitrary, $|x| = |y| =: r$. If $r = 0$ we are done. If $r > 0$, then let $\overline{x}$ and $\overline{y}$ be respectively the normalization of $x$ and $y$, such that $x = r\overline{x}$ and $y = r\overline{y}$. Then it is not hard to see $d(x,y) = d(r \cdot \overline{x}, r \cdot \overline{y}) = r \cdot d(\overline{x}, \overline{y})$. Therefore $d(\overline{x}, \overline{y}) = 0$. By the assumption that $\mathrm{B}(X)$ is metric, $\overline{x} = \overline{y}$ and hence $x = y$. We conclude that $X$ is metric. ∎

We obtain the following corollary, which was also proved in [21, Theorem 2.3] (see also [223, Proposition 2.7]) in a different way.

**Corollary 7.2.37.** *Every metric convex set is cancellative.*

*Proof.* By Lemmas 7.2.11 and 7.2.31 and Proposition 7.2.36. ∎

To summarize, we have obtained the following equivalences, restricting the equivalences of Corollary 7.2.13 to the subcategories.

**Corollary 7.2.38.** *The categories of metric convex sets, metric weight modules (with weight-preserving maps), and pre-base-norm spaces (with trace-preserving maps) are all equivalent:*

$$\mathbf{MConv} \simeq \mathbf{MWMod} \simeq \mathbf{pBNS}\,. \qquad ∎$$

**Example 7.2.39.** We give examples of (pre-)base-norm spaces, metric weight modules, and metric convex sets. Each example consists of a (pre-)base-norm space $(\mathscr{V}, \tau)$ and the metric weight module $\mathrm{B}_{\leq}(\mathscr{V})$ and the metric convex set $\mathrm{B}(\mathscr{V})$, which respectively arise as the subbase and the base corresponding via the equivalence of Corollary 7.2.38.

(i) The set $\mathbb{R}$ of real numbers forms a Banach base-norm space with the identity $\mathrm{id}\colon \mathbb{R} \to \mathbb{R}$ as trace. Its subbase is the metric weight module $[0,1]$, and the base is the metric convex set $\{1\}$.

(ii) Let $X$ be a set. Then the real $\ell^1$-space $\ell^1_{\mathbb{R}}(X)$ (i.e. the space of all absolutely summable functions $\varphi\colon X \to \mathbb{R}$) is a Banach base-norm space, with summation as trace: $\tau(\varphi) = \sum_{x\in X} \varphi(x)$. The base norm coincides with the $\ell^1$ norm. Its subbase is the metric weight module $\mathcal{D}^\infty_\leq(X)$ of infinite subprobability distributions on $X$, and the base is the metric convex set $\mathcal{D}^\infty(X)$ of infinite probability distributions on $X$.

(iii) There is a finite version of the previous example. For a set $X$, define

$$\ell^1_{c,\mathbb{R}}(X) = \{\varphi\colon X \to \mathbb{R} \mid \mathrm{supp}(\varphi) \text{ is finite}\}\,.$$

Then $\ell^1_{c,\mathbb{R}}(X)$ with summation as trace is a base-norm space. Then the subbase is the metric weight module $\mathcal{D}_\leq(X)$ of finite subprobability distributions on $X$, and the base is the metric convex set $\mathcal{D}(X)$ of finite probability distributions on $X$.

(iv) Let $(X, \Sigma_X)$ be a measurable space. Then the *ca* space (see e.g. [72, § IV.2]) of finite signed measures on $X$, i.e.

$$\mathit{ca}(X) = \{\mu\colon \Sigma_X \to \mathbb{R} \mid \mu \text{ is } \sigma\text{-additive}\}$$

is a Banach base-norm space with the trace $\tau(\mu) = \mu(X)$. The base norm coincides with the *total variation*: $\|\mu\| = |\mu|(X)$. Here $|\mu|$ is the *variation measure* of $\mu$, which may be defined as $|\mu| = \mu_+ + \mu_-$ via the Jordan decomposition $\mu = \mu_+ - \mu_-$. The subbase is the metric weight module $\mathcal{G}_\leq(X)$ of subprobability measures on $X$, and the base is the metric convex set $\mathcal{G}(X)$ of probability measures on $X$.

(v) Let $(X, \mu)$ be a measure space. Then the real $L^1$-space $L^1_{\mathbb{R}}(X, \mu)$ is a Banach base-norm space, with integration as trace: $\tau(\varphi) = \int \varphi \,\mathrm{d}\mu$. Its subbase is the metric weight module of (equivalence classes of) subprobability density functions on $X$ — integrable functions $\varphi\colon X \to \mathbb{R}_+$ with $\int \varphi \,\mathrm{d}\mu \leq 1$. Its base is the metric convex set of (equivalence classes of) probability density functions on $X$ — integrable functions $\varphi\colon X \to \mathbb{R}_+$ with $\int \varphi \,\mathrm{d}\mu = 1$.

(vi) Let $\mathscr{H}$ be a Hilbert space. Then the space $\mathcal{TC}(\mathscr{H})_{\mathrm{sa}}$ of self-adjoint trace-class operators on $\mathscr{H}$ is a Banach base-norm space, with the usual trace $\mathrm{tr}\colon \mathcal{TC}(\mathscr{H})_{\mathrm{sa}} \to \mathbb{R}$. The base norm is the trace norm. Its subbase is the metric weight module $\mathcal{Den}_\leq(\mathscr{H})$ of subnormalized density operators ($\rho$ such that $\mathrm{tr}(\rho) \leq 1$), and its base is the metric convex set $\mathcal{Den}(\mathscr{H})$ of density operators (such that $\mathrm{tr}(\rho) = 1$).

(vii) Let $\mathscr{A}$ be a $C^*$-algebra. Note that the dual space $\mathscr{A}^*$ inherits the involution operation via $\varphi^*(x) = \overline{\varphi(x^*)}$. We say that $\varphi \in \mathscr{A}^*$ is **self-adjoint** if $\varphi^* = \varphi$. Then the self-adjoint part $(\mathscr{A}^*)_{\mathrm{sa}}$ is a Banach base-norm space, with trace $\tau(\varphi) = \varphi(1)$. This is a consequence from the duality between base-norm and order-unit spaces (Theorem 7.2.40), since $(\mathscr{A}^*)_{\mathrm{sa}} = (\mathscr{A}_{\mathrm{sa}})^*$. Its subbase is the metric weight module $\mathrm{St}_\leq(\mathscr{A}) = \mathbf{Cstar}_\leq(\mathscr{A}, \mathbb{C})$ of substates on $\mathscr{A}$, and its base is the metric convex set $\mathrm{St}(\mathscr{A}) = \mathbf{Cstar}(\mathscr{A}, \mathbb{C})$ of states on $\mathscr{A}$.

(viii) Let $\mathscr{A}$ be a $W^*$-algebra. By the previous point, $(\mathscr{A}^*)_{\mathrm{sa}}$ is a Banach base-norm space. There is another option: the self-adjoint part $(\mathscr{A}_*)_{\mathrm{sa}}$ of the predual space $\mathscr{A}_*$, i.e. the space of *normal* functionals on $\mathscr{A}$, is also a Banach base-norm space. In this case, the subbase is the metric weight module $\mathrm{St}_{\leq}(\mathscr{A}) = \mathbf{Wstar}_{\leq}(\mathscr{A}, \mathbb{C})$ of *normal* substates on $\mathscr{A}$, and the base is the metric convex set $\mathrm{St}(\mathscr{A}) = \mathbf{Wstar}(\mathscr{A}, \mathbb{C})$ of *normal* states on $\mathscr{A}$.

Note that (ii) is a special case of (viii), since $\ell^\infty(X)$ is a $W^*$-algebra with predual $\ell^1(X)$. Similarly so is (v), if $(X, \mu)$ is a localizable measure space [232, § 1.18]. Furthermore, (vi) is a special case of (viii) with $\mathscr{A} = \mathcal{B}(\mathscr{H})$ and $\mathscr{A}_* = \mathcal{TC}(\mathscr{H})$.

As we mentioned above, the duality theorem (Theorem 7.2.40) tells us that the dual $\mathscr{A}^*$ of an order-unit space $\mathscr{A}$ is a Banach base-norm space.

### 7.2.3 More on order-unit and (pre-)base-norm spaces

For a normed space $\mathscr{V}$, we write $\mathscr{V}^*$ for the dual space consisting of all bounded linear functionals $\varphi\colon \mathscr{V} \to \mathbb{R}$. The space $\mathscr{V}^*$ is a normed space via the *dual norm* $\|\varphi\| = \sup\{|\varphi(x)| \mid x \in \mathscr{V}, \|x\| \leq 1\}$ (i.e. the operator norm), and in fact a Banach space (see e.g. [58, 218]). If $\mathscr{V}$ is moreover an ordered vector space, then $\mathscr{V}^*$ is also an ordered vector space with positive cone $\mathscr{V}^*_+$ the set of bounded positive functionals. Equivalently, we define the ordering on $\mathscr{V}^*$ by $\varphi \leq \psi$ iff $\varphi(x) \leq \psi(x)$ for all $x \in \mathscr{V}_+$. We now recall the fundamental duality between (pre-)base-norm and order-unit spaces.

**Theorem 7.2.40.**

(i) *If $(\mathscr{V}, \tau)$ is a pre-base-norm space, then $\mathscr{V}^*$ is a Banach order-unit space with unit $\tau \in \mathscr{V}^*$. The order-unit norm on $\mathscr{V}^*$ agrees with the dual norm.*

(ii) *If $(\mathscr{A}, u)$ is an order-unit space, then $\mathscr{A}^*$ is a Banach base-norm space with trace $\tau$ given by $\tau(\varphi) = \varphi(u)$. The base norm on $\mathscr{V}^*$ agrees with the dual norm.*

(iii) *The two constructions yield a dual adjunction between pre-base-norm and order-unit spaces:*

$$\mathbf{pBNS}_{\leq} \underset{(-)^*}{\overset{(-)^*}{\rightleftarrows}} \mathbf{OUS}_{\leq}^{\mathrm{op}} \quad (\top)$$

*The adjunction restricts to trace-preserving/unital maps:* $\mathbf{pBNS} \rightleftarrows \mathbf{OUS}^{\mathrm{op}}$.

*Proof.* See respectively Proposition 2.4.17, Theorem 2.5.1, and Theorem 2.5.4 in [90]. ∎

Below we show that Archimedean effect modules, metric convex sets, and metric weight modules can be characterized in terms of (order-)separation by suitable functionals on them. We first need lemmas.

**Lemma 7.2.41.** *Let $(\mathscr{A}, u)$ be an order-unit space.*

(i) *Morphisms in $\mathbf{OUS}_{\leq}$ are bounded with respect to the order-unit norm. In particular, $\mathbf{OUS}_{\leq}(\mathscr{A}, \mathbb{R}) \subseteq \mathscr{A}^*$.*

(ii) *For each $a \in \mathscr{A}$, if $\varphi(a) \geq 0$ for all $\varphi \in \mathbf{OUS}(\mathscr{A}, \mathbb{R})$, then $a \geq 0$.*

(iii) *$\mathbf{OUS}(\mathscr{A}, \mathbb{R}) \cong \mathbf{EMod}([0, u]_{\mathscr{A}}, [0, 1])$, where the mapping from left to right is given by the restriction to $[0, u]_{\mathscr{A}}$.*

*Proof.*

(i) See [90, Proposition 1.2.8].

(ii) See [5, Lemma 1.18].

(iii) The bijection is a mapping given by the functor $[0,u]_{(-)}\colon \mathbf{OUS} \xrightarrow{\simeq} \mathbf{AEMod}$, which is an equivalence and hence full and faithful. ■

**Lemma 7.2.42.** *Let $(\mathscr{V},\tau)$ be a pre-base-norm space.*

(i) *Morphisms in $\mathbf{pBNS}_{\leq}$ are bounded with respect to the base norm. In particular, $\mathbf{pBNS}_{\leq}(\mathscr{V},\mathbb{R}) \subseteq \mathscr{V}^*$.*

(ii) *For each $x \in \mathscr{V}$, if $\varphi(x) = 0$ for all $\varphi \in \mathbf{pBNS}_{\leq}(\mathscr{V},\mathbb{R})$, then $x = 0$.*

(iii) *$\mathscr{V}$ is a base-norm space (i.e. $\mathscr{V}_+$ is closed) if and only if for each $x \in \mathscr{V}$, one has $x \geq 0$ whenever $\varphi(x) \geq 0$ for all $\varphi \in \mathbf{pBNS}_{\leq}(\mathscr{V},\mathbb{R})$.*

(iv) *$\mathbf{pBNS}_{\leq}(\mathscr{V},\mathbb{R}) \cong \mathbf{WMod}_{\leq}(\mathrm{B}_{\leq}(\mathscr{V}),[0,1]) \cong \mathbf{Conv}(\mathrm{B}(\mathscr{V}),[0,1])$, where the mappings from left to right are given by the restriction to $\mathrm{B}_{\leq}(\mathscr{V})$ and $\mathrm{B}(\mathscr{V})$.*

*Proof.*

(i) See [90, Proposition 2.2.12].

(ii) By Theorem 7.2.40(i), $\mathscr{V}^*$ is an order-unit space with unit $\tau$, and $\mathbf{pBNS}_{\leq}(\mathscr{V},\mathbb{R})$ is precisely the unit interval $[0,\tau]$ in $\mathscr{V}^*$. Therefore $\mathbf{pBNS}_{\leq}(\mathscr{V},\mathbb{R})$ spans $\mathscr{V}^*$. Thus, if $\varphi(x) = 0$ for all $\varphi \in \mathbf{pBNS}_{\leq}(\mathscr{V},\mathbb{R})$, then $\varphi(x) = 0$ for all $\varphi \in \mathscr{V}^*$. It follows that $x = 0$.

(iii) (If) Let $(x_n)_{n\in\mathbb{N}}$ be a sequence in $\mathscr{V}_+$ that converges to $x \in \mathscr{V}$ in the base-norm. For each $\varphi \in \mathbf{BNS}_{\leq}(\mathscr{V},\mathbb{R})$ the sequence $(\varphi(x_n))_n$ converges to $\varphi(x)$ in $\mathbb{R}$, since $\varphi$ is bounded. Because $\varphi(x_n) \geq 0$ for all $n$, one obtain $\varphi(x) \geq 0$. By assumption, we conclude that $x \geq 0$.

(Only if) (cf. [5, Lemma 1.18]) Fix $x \in \mathscr{V}$ and suppose that $\varphi(x) \geq 0$ for all $\varphi \in \mathbf{pBNS}_{\leq}(\mathscr{V},\mathbb{R})$. Assume, towards a contradiction, that $x \notin \mathscr{V}_+$. By the Hahn-Banach separation theorem [58, Theorem IV.3.9] there exists $\psi \in \mathscr{V}^*$ and $\alpha \in \mathbb{R}$ such that $\psi(x) < \alpha < \psi(y)$ for all $y \in \mathscr{V}_+$. Then $\psi(x) < \alpha < \psi(0) = 0$. We claim that $\psi$ is positive. Indeed, if $\psi(y) < 0$ for some $y \in \mathscr{V}_+$, then there exists an $N \in \mathbb{N}$ such that $\psi(Ny) = N\psi(y) < \alpha$ but $Ny \in \mathscr{V}_+$, leading to a contradiction. Since $\tau$ is an order unit of $\mathscr{V}^*$, there exists an $N \in \mathbb{N}_{>0}$ such that $(1/N)\cdot\psi$ is trace-decreasing, i.e. $(1/N)\cdot\psi \in \mathbf{pBNS}_{\leq}(\mathscr{V},\mathbb{R})$. Since $(1/N)\cdot\psi(x) < 0$, one obtains a contradiction.

(iv) The first bijection is given by the equivalence $\mathrm{B}_{\leq}\colon \mathbf{pBNS}_{\leq} \xrightarrow{\simeq} \mathbf{WMod}_{\leq}$. The second bijection is given by $\mathcal{L}(\mathrm{B}(\mathscr{V})) \cong \mathrm{B}_{\leq}(\mathscr{V})$ and Lemma 4.4.18. ■

We note that the separation results above (Lemma 7.2.41(ii) and Lemma 7.2.42(ii) and (iii)) require the Hahn-Banach theorem and hence (a weak form of) the Axiom of Choice.

As far as the author is aware, the following characterization of Archimedean effect modules is new, though the 'if' direction is essentially shown by van de Wetering [257, Proposition A.1]. We note that there is an analogous result for interval effect algebras [230, Corollary 3.4.12].

**Proposition 7.2.43.** *An effect module $E$ is Archimedean if and only if it is order-separated in the sense that for each $x, y \in E$, one has $x \leq y$ whenever $\varphi(x) \leq \varphi(y)$ for all unital module maps $\varphi\colon E \to [0,1]$.*

*Proof.* Assume that $E$ is order-separated. Let $x, y \in E$ be given and suppose that $(1/2)\cdot x \leq (1/2)\cdot y \ovee (1/2n)\cdot 1$ for all $n \in \mathbb{N}_{>0}$. For any unital module map $\varphi\colon E \to [0,1]$, we have $(1/2)\cdot\varphi(x) \leq (1/2)\cdot\varphi(y) + 1/2n$, i.e. $\varphi(x) \leq \varphi(y) + 1/n$. Since this holds for any $n \geq 2$, we obtain $\varphi(x) \leq \varphi(y)$. Then $x \leq y$ by order-separation.

Conversely, assume that $E$ is Archimedean. By $\mathbf{AEMod} \simeq \mathbf{OUS}$, we may assume that $E = [0,u]_{\mathscr{A}}$ for some order-unit space. Suppose that $\varphi(x) \leq \varphi(y)$ for all unital module maps $\varphi\colon [0,u]_{\mathscr{A}} \to [0,1]$. By Lemma 7.2.41(iii), this is equivalent to saying that for all $\varphi \in \mathbf{OUS}(\mathscr{A}, [0,1])$, one has $\varphi(x) \leq \varphi(y)$, and hence $\varphi(y - x) \geq 0$. By Lemma 7.2.41(ii), $y - x \geq 0$, that is, $x \leq y$. ∎

The next is an analogous result for convex sets. The result is known, see [21, Theorem 2.3], see also [223, Proposition 2.7].

**Proposition 7.2.44.** *A convex set $K$ is metric if and only if it is separated in the sense that for each $x, y \in K$, one has $x = y$ whenever $\varphi(x) = \varphi(y)$ for all affine map $\varphi\colon K \to [0,1]$.*

*Proof.* Suppose that $K$ is separated. Let $x, y \in K$ be such that $\sigma(x,y) = 0$. Let $\varphi\colon K \to [0,1]$ be an arbitrary affine map, and $\varepsilon \in \mathbb{R}$ an arbitrary real number with $0 < \varepsilon < 1$. By $\sigma(x,y) = 0$, and the definition of the pseudometric $\sigma$ (7.4), there exists $r \in [0,1]$ $z, w \in K$ such that $[\![ r^\perp|x\rangle + r|z\rangle ]\!] = [\![ r^\perp|y\rangle + r|w\rangle ]\!]$ and $r < \varepsilon$. Applying $\varphi$, we have $r^\perp \cdot \varphi(x) + r \cdot \varphi(z) = r^\perp \cdot \varphi(y) + r \cdot \varphi(w)$. Then

$$|\varphi(x) - \varphi(y)| = \frac{r}{1-r} \cdot |\varphi(w) - \varphi(z)| \leq \frac{2r}{1-r} < \frac{2\varepsilon}{1-\varepsilon}$$

Since $\varepsilon$ (such that $0 < \varepsilon < 1$) is chosen arbitrarily, $2\varepsilon/(1-\varepsilon)$ can be arbitrarily small, and therefore $\varphi(x) = \varphi(y)$. We have shown that $\varphi(x) = \varphi(y)$ for all affine map $\varphi\colon X \to [0,1]$, which implies $x = y$ by assumption.

The converse follows from Lemma 7.2.42, by a reasoning similar to the proof of Proposition 7.2.43. ∎

Finally we give an analogous separation result for weight modules. In fact, we can also characterize weight modules that correspond to base-norm spaces.

**Proposition 7.2.45.**

(i) *A weight module $X$ is metric if and only if it is separated in the sense that for each $x, y \in X$, one has $x = y$ whenever $\varphi(x) = \varphi(y)$ for all trace-decreasing module maps $\varphi\colon X \to [0,1]$.*

(ii) *A weight module $X$ is isomorphic to the subbase $\mathrm{B}_{\leq}(\mathscr{V})$ of some base-norm space $\mathscr{V}$ if and only if it is order-separated in the sense that for each $x, y \in X$, one has $x \leq y$ whenever $\varphi(x) \leq \varphi(y)$ for all trace-decreasing module maps $\varphi\colon X \to [0,1]$.*

*Proof.*

(i) Suppose that $X$ is separated. Since $\mathbf{WMod}_{\leq}(X,[0,1]) \cong \mathbf{Conv}(\mathrm{B}(X),[0,1])$ by Lemma 4.4.18, the convex set $\mathrm{B}(X)$ is separated and hence metric by Proposition 7.2.44. Then $X$ is metric by Proposition 7.2.36. The converse follows by Lemma 7.2.42(ii) and (iv).

(ii) Suppose that $X$ is order-separated. Then $X$ is separated, and hence by the previous point, metric. Thus we may assume that $X = \mathrm{B}_{\leq}(\mathscr{V})$ for some pre-base-norm space. We invoke Lemma 7.2.42(iii) to prove that $\mathscr{V}$ is a base-norm space. Suppose that $x \in \mathscr{V}$ satisfies $\varphi(x) \geq 0$ for all $\varphi \in \mathbf{pBNS}_{\leq}(\mathscr{V},\mathbb{R})$. We can find $r \in \mathbb{R}_{>0}$ and $y, z \in \mathrm{B}_{\leq}(\mathscr{V})$ such that $x = r(y-z)$. By $\varphi(x) \geq 0$ one has $\varphi(z) \leq \varphi(y)$. By Lemma 7.2.42(iv), then one has $\varphi(z) \leq \varphi(y)$ for all $\varphi \in \mathbf{WMod}_{\leq}(\mathrm{B}_{\leq}(\mathscr{V}),[0,1])$. By the assumption of order-separation, we obtain $z \leq y$. Therefore $x = r(y-z) \geq 0$, showing that $\mathscr{V}$ is a base-norm space. The converse follows by Lemma 7.2.42(iii) and (iv). ∎

### 7.2.4 Convex operational models and state-effect models

Since pre-base-norm and order-unit spaces are in dual relationship, it makes sense to consider a dual pair $(\mathscr{V}, \mathscr{A}, \langle\,,\rangle)$ of a pre-base-norm space $\mathscr{V}$ and an order-unit space $\mathscr{A}$. We call such a suitable dual pair a *convex operational model* (*COM*), following [11, 258] (see Remark 7.2.48 below for the comparison of the definitions). A convex operational model can be viewed as a model of a system, where $\mathscr{V}$ is the space of states, $\mathscr{A}$ is the space of effects, and the pairing $\langle\,,\rangle\colon \mathscr{V} \times \mathscr{A} \to \mathbb{R}$ as the assignment of probabilities to pairs of states and effects. This view is mathematically more precisely justified by a categorical equivalence of COMs and *state-effect models* (Definition 7.2.49), which are more primitive axiomatic models defined in terms of convex sets and effect modules. The equivalence may be seen as a variant of Ludwig's embedding theorem [198, Chapter IV].

**Definition 7.2.46.** A **convex operational model** (**COM**, for short) is a triple $(\mathscr{V}, \mathscr{A}, \langle\,,\rangle)$ where $\mathscr{V}$ is a pre-base-norm space with trace $\tau$, $\mathscr{A}$ is an order-unit space with unit $u$, and $\langle\,,\rangle\colon \mathscr{V} \times \mathscr{A} \to \mathbb{R}$ is a bilinear functional satisfying:

(CO1) $\langle x, a\rangle \geq 0$ for all positive $x \in \mathscr{V}_+$ and $a \in \mathscr{A}_+$.

(CO2) $\langle x, u\rangle = \tau(x)$ for all $x \in \mathscr{V}$.

(CO3) $\mathscr{V}$ order-separates $\mathscr{A}$ in the sense that for any $a \in \mathscr{A}$, $\langle x, a\rangle \geq 0$ for all positive $x \in \mathscr{V}_+$ implies $a \geq 0$

(CO4) $\mathscr{A}$ separates $\mathscr{V}$ in the sense that for any $x \in \mathscr{V}$, $\langle x, a\rangle = 0$ for all $a \in \mathscr{A}$ implies $x = 0$.

For the ease of comparison with the definitions of convex operational models in [11, 258], we give a reformulation of the above definition.

**Lemma 7.2.47.** *A convex operational model $(\mathscr{V}, \mathscr{A}, \langle\,,\rangle)$ is the same, up to isomorphism, as a triple $(\mathscr{V}, \mathscr{V}^{\#}, \tau)$ where $(\mathscr{V}, \tau)$ is a pre-base-norm space and $\mathscr{V}^{\#} \subseteq \mathscr{V}^*$ is a subspace of the dual $\mathscr{V}^*$ such that $\tau \in \mathscr{V}^{\#}$ and $\mathscr{V}^{\#}$ separates the points of $\mathscr{V}$.*

*Proof.* Let $(\mathscr{V}, \mathscr{V}^{\#}, \tau)$ be the latter triple. From the fact that $(\mathscr{V}^*, \tau)$ is an order-unit space (Theorem 7.2.40), it follows that the subspace $\mathscr{V}^{\#} \subseteq \mathscr{V}^*$ is also an order-unit space with positive cone $\mathscr{V}^{\#}_+ := \mathscr{V}^{\#} \cap \mathscr{V}^*_+$ and unit $\tau$. Note that the choice of the positive cone means that one has $\varphi \leq \psi$ in $\mathscr{V}^{\#}$ if and only if $\varphi \leq \psi$ in $\mathscr{V}^*$. Then we define a pairing $\langle\,,\rangle\colon \mathscr{V} \times \mathscr{V}^{\#} \to \mathbb{R}$ by $\langle x, \varphi\rangle = \varphi(x)$. It is straightforward to see that $(\mathscr{V}, \mathscr{V}^{\#}, \langle\,,\rangle)$ satisfies (CO1)–(CO4) and hence forms a convex operational model.

Conversely, let $(\mathscr{V}, \mathscr{A}, \langle\,,\rangle)$ be a convex operational model. For each $a \in \mathscr{A}$ we define a linear map $\iota(a)\colon \mathscr{V} \to \mathbb{R}$ by $\iota(a)(x) = \langle x, a\rangle$. Note that for each $a \in [0, u]_{\mathscr{A}}$ the map $\iota(a)$ is bounded, since $\iota(a)$ is positive and trace-decreasing by (CO1) and (CO2). Now if $a \in \mathscr{A}$ is arbitrary, it can be written as $a = r_1 a_1 + r_2 a_2$ for $r_i \in \mathbb{R}$ and $a_i \in [0, u]_{\mathscr{A}}$, and then $\iota(a) = r_1\iota(a_1) + r_2\iota(a_2)$. Therefore $\iota(a)$ is bounded i.e. $\iota(a) \in \mathscr{V}^*$, for any $a \in \mathscr{A}$. Thus we obtain a linear map $\iota\colon \mathscr{A} \to \mathscr{V}^*$. Let $\mathscr{V}^{\#} := \iota[\mathscr{A}] \subseteq \mathscr{V}^*$ be the image of $\iota$. Then $\mathscr{V}^{\#}$ is a subspace which contains $\tau$, since $\iota(u) = \langle -, u\rangle = \tau$. By (CO4), $\mathscr{V}^{\#}$ separates the points of $\mathscr{V}$. Therefore $(\mathscr{V}, \iota[\mathscr{A}], \tau)$ forms the latter triple in the statement. Furthermore, note that the map $\iota\colon \mathscr{A} \to \mathscr{V}^*$ is an order-embedding by (CO3). Thus the co-restriction $\iota\colon \mathscr{A} \to \mathscr{V}^{\#} \equiv \iota[\mathscr{A}]$ is a linear order isomorphism, which sends $u$ to $\tau$. This proves that the two constructions are inverse up to isomorphism. ■

**Remark 7.2.48.** We compare our definition of COMs with the definitions in [11, 258]. We are mainly concerned with the definition in [11], which is closer to our definition. Then we briefly mention [258].

A COM in [11, Definition 4] is a triple $(\mathscr{V}, \mathscr{V}^{\#}, \tau)$ where $(\mathscr{V}, \tau)$ is a Banach base-norm space and a weak-* dense subspace $\mathscr{V}^{\#} \subseteq \mathscr{V}^*$ equipped with a positive cone satisfying $\mathscr{V}^{\#}_+ \subseteq \mathscr{V}^*_+$. such that $\tau \in \mathscr{V}^{\#}$. First note that the weak-* denseness of a subspace $\mathscr{V}^{\#} \subseteq \mathscr{V}^*$ is known (e.g. [218, E 2.4.4]) to be equivalent to the condition that $\mathscr{V}^{\#}$ separates $\mathscr{V}$. There are still differences in our and their definition of COMs. Their definition is stronger than ours in that $\mathscr{V}$ is a Banach base-norm space. However, their definition is also more general than us in that $\mathscr{V}^{\#}$ may be equipped with a different ordering from $\mathscr{V}^*$, that is, the positive cone $\mathscr{V}^{\#}_+$ may be strictly smaller than $\mathscr{V}^{\#} \cap \mathscr{V}^*_+$. Thus, even though $(\mathscr{V}^*, \tau)$ always forms an order-unit space, it is unclear whether $(\mathscr{V}^{\#}, \tau)$, with the positive cone $\mathscr{V}^{\#}_+$, forms an order-unit space. Slightly generalizing their terminology[3], let us call a COM $(\mathscr{V}, \mathscr{V}^{\#}, \tau)$, in their sense, *saturated* if $\mathscr{V}^{\#}_+ = \mathscr{V}^{\#} \cap \mathscr{V}^*_+$. Then a saturated COM $(\mathscr{V}, \mathscr{V}^{\#}, \tau)$ in their sense coincides with a COM in our sense (via Lemma 7.2.47) with an extra property that $\mathscr{V}$ is a Banach base-norm space.

A COM in [258, Definition 2.2] is defined in a much more general way; it is a triple $(\mathscr{V}, \mathscr{V}^{\#}, \tau)$ where $\mathscr{V}$ is an ordered vector space and $\mathscr{V}^{\#} \subseteq \mathscr{V}'$ is a subspace of the algebraic dual[4] $\mathscr{V}'$, equipped with a positive cone satisfying $\mathscr{V}^{\#}_+ \subseteq \mathscr{V}'_+$, such that $\tau$ is an order unit of $\mathscr{V}^{\#}$ and $\mathscr{V}^{\#}$ separates points of $\mathscr{V}$. In this definition one does not require $\mathscr{V}$ to be a (pre-)base-norm space, nor $\mathscr{V}^{\#}$ to be an order-unit space. Clearly, any COM in our sense (via Lemma 7.2.47) is a COM in the sense of [258], with the 'saturated' positive cone $\mathscr{V}^{\#}_+ = \mathscr{V}^{\#} \cap \mathscr{V}^*_+$.

[3] See the paragraph after [11, Definition], where 'saturated' is defined only in the finite-dimensional case.

[4] The space of all linear functionals $\mathscr{V} \to \mathbb{R}$.

We now introduce a more primitive notion of models in terms of convex sets and effect modules.

**Definition 7.2.49.** A **state-effect model** is a triple $(K, E, \vDash)$ where $K$ is a convex set, $E$ is an effect module, and $\vDash\colon K \times E \to [0,1]$ is a 'validity' map satisfying:

(SE1) For each $x \in K$, $x \vDash (-)\colon E \to [0,1]$ is a unital module map.

(SE2) For each $a \in E$, $(-) \vDash a\colon K \to [0,1]$ is an affine map.

(SE3) $x \vDash a \le x \vDash b$ for all $x \in K$ implies $a \le b$.

(SE4) $x \vDash a = y \vDash a$ for all $a \in E$ implies $x = y$.

A state-effect model $(K, E, \vDash)$ is seen as an axiomatic model of a (physical) system where $K$ is the set of *states* and $E$ is the set of *effects* (predicates, or yes-no measurements), and for each pair $x \in K$ and $a \in E$, the validity $x \vDash a$ gives the probability that the effect $a$ occurs (is observed) when the system is in the state $x$.

Ludwig studied a similar dual pair $K \times E \to [0,1]$ as an axiomatic model of quantum system, see e.g. [195–198]. Under several axioms on the dual pair he proved the 'embedding' theorem which asserts that the set $K$ of states (which he call 'ensembles') is embedded in a base-norm space, and the set $E$ of effects is embedded in the dual of the base-norm space, hence an order-unit space, see [198, Chapter IV] (see also [197, Th. 3.1] and the re-elaboration by Lami [186, § 1.7]).

Below we will prove that each state-effect model can be embedded into a COM, and moreover that the categories of state-effect models and COMs are equivalent. This may be seen as a variant of Ludwig's embedding theorem. There is however some important difference between our result and Ludwig's original result, which we briefly discuss below. The point is that Ludwig's axioms on the dual pair $K \times E \to [0,1]$ are weaker than ours, i.e. the conditions in Definition 7.2.49. As mentioned in the first paragraph of [198, Chapter IV], the assumptions he used are 'III T 5.1.4' (found in [198, p. 78]), 'APK', and 'ARK' (both found in [198, p. 84]). In our terminology, Ludwig's axioms amount[5] to the following: one has a set $K$ (of states/ensembles), a set $E$ (of effects), and a function $\vDash\colon K \times E \to [0,1]$ such that

(L1) $K$ and $E$ separates each other: $x \vDash a = x \vDash b$ for all $x \in K$ implies $a = b$; and $x \vDash a = y \vDash a$ for all $a \in E$ implies $x = y$.

(L2) There exist $0, 1 \in E$ such that $x \vDash 0 = 0$ and $x \vDash 1 = 1$ for all $x \in X$.

(L3) For each $a \in E$, there exists $a^\perp \in E$ such that $x \vDash a + x \vDash a^\perp = 1$ for all $x \in X$.

(L4) Both $K$ and $E$ are convex sets and the map $\vDash\colon K \times E \to [0,1]$ is affine in each argument separately.

See also [186, § 1.7]. It is easy to see that every state-effect model satisfies these axioms. Although we do not have a concrete counterexample, it is not likely that the converse holds, i.e. that Ludwig's axioms (L1)–(L4) makes $E$ into an effect algebra. In fact, one can show that $E$ is an effect algebra and moreover an effect module, if we add (or replace (L3) with) the following axiom:

(G) If $x \vDash a \le x \vDash b$ for all $x \in K$, then there exists $c \in E$ such that $x \vDash a + x \vDash c = x \vDash b$ for all $x \in K$.

[5] We ignore some minor points such as the use of rationals instead of reals.

This axiom is (essentially) taken from Gudder's *effect-state space* [101, 102]. It allows us to define the difference $b \ominus a = c$, from which we can define the sum as $a \ovee b = (b^\perp \ominus a)^\perp$, see [101] for details. The module structure can be defined from convex structure as $r \cdot a = [\![ r|a\rangle + (1-r)|0\rangle ]\!]$. Moreover (G) implies that states $K$ order-separate effects $E$, making $(K, E, \vDash)$ into a state-effect model. Conversely, any state-effect model satisfies (G), as one can take $c = b \ominus a$ using order-separation. To summarize, $(K, E, \vDash)$ satisfies axioms (L1)–(L4) and (G) if and only if it forms a state-effect model. However, the Ludwig's original axioms (L1)–(L4) are (probably strictly) weaker than state-effect models.

We now prove a categorical equivalence of COMs and state-effect models. We first need definitions of morphisms.

**Definition 7.2.50** (cf. [11, § 3.1])**.** Let $(\mathscr{V}, \mathscr{A}, \langle\,,\rangle)$ and $(\mathscr{W}, \mathscr{B}, \langle\,,\rangle)$ be COMs. A **total morphism** of COMs from $(\mathscr{V}, \mathscr{A}, \langle\,,\rangle)$ to $(\mathscr{W}, \mathscr{B}, \langle\,,\rangle)$ is a pair of a morphism $f_* \colon \mathscr{V} \to \mathscr{W}$ in **pBNS** and $f^* \colon \mathscr{B} \to \mathscr{A}$ in **OUS** that are 'in adjunction', namely $\langle f_*(x), a\rangle = \langle x, f^*(a)\rangle$ for all $x \in \mathscr{V}$ and $a \in \mathscr{B}$. Similarly, a **partial morphism** of COMs is a pair of $f_* \colon \mathscr{V} \to \mathscr{W}$ in $\mathbf{pBNS}_\leq$ and $f^* \colon \mathscr{B} \to \mathscr{A}$ in $\mathbf{OUS}_\leq$ such that $\langle f_*(x), a\rangle = \langle x, f^*(a)\rangle$. We write **COM** for the category of COMs and total morphisms and $\mathbf{COM}_\leq$ for the category COMs and partial morphisms.

**Lemma 7.2.51.** *Let $(\mathscr{V}, \mathscr{A}, \langle\,,\rangle), (\mathscr{W}, \mathscr{B}, \langle\,,\rangle)$ be convex operational models. Let $f_* \colon \mathscr{V} \to \mathscr{W}$ and $f^* \colon \mathscr{B} \to \mathscr{A}$ be positive linear maps in adjunction, i.e. $\langle f_*(x), a\rangle = \langle x, f^*(a)\rangle$ for all $x \in \mathscr{V}$ and $a \in \mathscr{B}$. Then $f_*$ is trace-decreasing if and only if $f^*$ is subunital. Moreover, $f_*$ is trace-preserving if and only if $f^*$ is unital.*

*Proof.* If $f_*$ is trace-decreasing, then for any $x \in \mathscr{V}_+$,

$$\langle x, f^*(u)\rangle = \langle f_*(x), u\rangle = \tau(f_*(x)) \leq \tau(x) = \langle x, u\rangle\,.$$

By order-separation, $f^*(u) \leq u$, i.e. $f^*$ is subunital. Conversely if $f^*$ is subunital, then

$$\tau(f_*(x)) = \langle f_*(x), u\rangle = \langle x, f^*(u)\rangle \leq \langle x, u\rangle = \tau(x)$$

for any $x \in \mathscr{V}_+$, i.e. $f_*$ is trace-decreasing. The trace-preserving/unital case is shown similarly. ∎

**Definition 7.2.52.** Let $(K, E, \vDash)$ and $(L, D, \vDash)$ be state-effect models. A morphism from $(K, E, \vDash)$ to $(L, D, \vDash)$ is a pair of an affine map $f_* \colon K \to L$ and a unital module map $f^* \colon D \to E$ such that $(f_*(x) \vDash a) = (x \vDash f^*(a))$ for all $x \in K$ and $a \in D$. We write **SEM** for the category of state-effect models and morphisms between them.

**Theorem 7.2.53.** *One has an equivalence* $\mathbf{COM} \simeq \mathbf{SEM}$ *of the categories of convex operational models and state-effect models.*

*Proof.* Let $(K, E, \vDash)$ is a state-effect model. By definition, $E$ is order-separated by unital module maps $x \vDash (-) \colon E \to [0, 1]$ for $x \in K$. By Proposition 7.2.43, $E$ is Archimedean. Similarly by Proposition 7.2.44, $K$ is a metric convex set. Therefore we have a functor $\mathbf{SEM} \to \mathbf{MConv} \times \mathbf{AEMod}$ that forgets validity maps $\vDash$. By definition, there is a forgetful functor $\mathbf{COM} \to \mathbf{pBNS} \times \mathbf{OUS}$ too. By Corollary 7.2.38 and Theorem 7.2.20, there are equivalences $\mathbf{MConv} \simeq \mathbf{pBNS}$ and $\mathbf{AEMod} \simeq \mathbf{OUS}$. Therefore there is an equivalence $\mathbf{MConv} \times \mathbf{AEMod} \simeq \mathbf{pBNS} \times \mathbf{OUS}$. Below we will obtain an equivalence $\mathbf{SEM} \simeq \mathbf{COM}$ by lifting the equivalence $\mathbf{MConv} \times \mathbf{AEMod} \simeq \mathbf{pBNS} \times \mathbf{OUS}$ along the forgetful functor. Thus, the core of the proof is to describe how parings $\langle\,,\rangle\colon \mathscr{V} \times \mathscr{A} \to \mathbb{R}$ in $\mathbf{COM}$ induce validities $\vDash\colon K \times E \to [0,1]$ in $\mathbf{SEM}$, and vice versa.

The part of equivalence $\mathbf{pBNS} \times \mathbf{OUS} \overset{\simeq}{\to} \mathbf{MConv} \times \mathbf{AEMod}$ sends pairs $(\mathscr{V}, \mathscr{A})$ to $(\mathrm{B}(\mathscr{V}), [0,1]_{\mathscr{A}})$. Let $(\mathscr{V}, \mathscr{A}, \langle\,,\rangle) \in \mathbf{COM}$ be a COM. Define the validity map by $x \vDash a = \langle x, a\rangle$. It is straightforward to see that $\vDash\colon \mathrm{B}(\mathscr{V}) \times [0,1]_{\mathscr{A}} \to [0,1]$ is a suitable bihomomorphism. The validity $\vDash$ satisfies the appropriate separation property since so does $\langle\,,\rangle$. Therefore $(\mathrm{B}(\mathscr{V}), [0,1]_{\mathscr{A}}, \vDash) \in \mathbf{COM}$. It is also easy to see that morphisms in $\mathbf{COM}$ yield those in $\mathbf{SEM}$, so that one obtains a functor $\mathbf{COM} \to \mathbf{SEM}$.

Let $F\colon \mathbf{MConv} \overset{\simeq}{\to} \mathbf{pBNS}$ and $G\colon \mathbf{AEMod} \overset{\simeq}{\to} \mathbf{OUS}$ be the parts of equivalences (concretely, $F = \mathcal{K} \circ \mathcal{T} \circ \mathcal{L}$ and $G = \mathcal{K} \circ \mathcal{T}$, but the arguments below do not depend on the concrete definitions). Then $F \times G\colon \mathbf{MConv} \times \mathbf{AEMod} \overset{\simeq}{\to} \mathbf{pBNS} \times \mathbf{OUS}$ is a part of equivalence. We will show that $F \times G$ lifts to a functor $\mathbf{SEM} \to \mathbf{COM}$. Let $(K, E, \vDash) \in \mathbf{SEM}$ be a state-effect model. Then $FK$ and $GE$ are a pre-base-norm space and an order-unit space. In order to extend $\vDash\colon K \times E \to [0,1]$ to $\langle\,,\rangle\colon FK \times GE \to \mathbb{R}$, let $v\colon K \to \mathbf{EMod}(E, [0,1])$ be the 'curried' validity map given by $v(x)(a) = (x \vDash a)$. Note that

$$\mathbf{EMod}(E, [0,1]) = \mathbf{AEMod}(E, [0,1]) \cong \mathbf{OUS}(GE, \mathbb{R}) = \mathrm{B}((GE)^*)$$

and $v$ is a map in $\mathbf{MConv}$. Thu by applying $F$ to $v$, we get

$$FK \xrightarrow{Fv} F(\mathbf{EMod}(E, [0,1])) \cong F(\mathrm{B}((GE)^*)) \cong (GE)^*$$

in $\mathbf{pBNS}$, i.e. a trace-preserving positive linear map. By 'uncurrying', we get the paring map $\langle\,,\rangle\colon FK \times GE \to \mathbb{R}$. By the fact that the curried map $FK \to (GE)^*$ is a trace-preserving positive linear map, it follows that $\langle\,,\rangle\colon FK \times GE \to \mathbb{R}$ is bilinear such that $\langle x, a\rangle \geq 0$ for all $x \in (FK)_+$ and $a \in (BE)_+$, and $\langle -, u\rangle = \tau$ for the trace $\tau$ on $FK$ and the order-unit $u$ of $GE$. To see that the separation conditions (CO3) and (CO4) holds, note that the paring $\langle\,,\rangle$ restricted on $\mathrm{B}(FK) \cong K$ and $[0,1]_{GE} \cong E$ satisfies the separation conditions (SE3) and (SE4) of state-effect models, since $\langle\,,\rangle$ extends $\vDash$. To prove (CO3), suppose that $a \in GE$ satisfies $\langle x, a\rangle \geq 0$ for all $x \in (FK)_+$. Let $a = a_1 - a_2$ for $a_1, a_2 \in (GE)_+$. For large enough $n \in \mathbb{N}$ we have $a_1/n, a_2/n \in [0,u]_{GE}$. For all $x \in \mathrm{B}(FK)$, we have $\langle x, a_2/n\rangle \leq \langle x, a_1/n\rangle$. Then $a_2/n \leq a_1/n$ by (SE3), and hence $a = a_1 - a_2 \geq 0$. Next to prove (CO4), suppose that $x \in FK$ satisfies $\langle x, a\rangle = 0$ for all $a \in GE$. There are $r, s \in \mathbb{R}_+$ and $y, z \in \mathrm{B}(FK)$ such that $x = ry - sz$. Then

$$r = \tau(ry) = \langle ry, u\rangle = \langle sz, u\rangle = \tau(sz) = s\,,$$

so we have $x = ry - rz$. If $r = 0$ we are done, and if $r > 0$, then it follows that $\langle x, a\rangle = \langle y, a\rangle$ for all $a \in [0,u]_{GE}$. Thus $y = z$ by (SE4), and hence $x = ry - rz = 0$. We conclude that $(FK, GE, \langle\,,\rangle)$ is a convex operational model. Let $(f_*, f^*)\colon (K, E, \vDash) \to (L, D, \vDash)$ be a morphism of state-effect models. Applying functor $F$ and $G$, we have $Ff_*\colon FK \to FL$ in **pBNS** and $Gf^*\colon GD \to GE$ in **OUS**. By the construction of validities on $(FK, GE)$ and $(FL, GD)$, we have $\langle Ff_*(x), a\rangle = \langle x, Gf^*(a)\rangle$ for all $x \in \mathrm{B}(FK)$ and $a \in [0,u]_{GD}$. Since $FK$ and $GD$ are respectively generated by $\mathrm{B}(FK)$ and $[0,u]_{GD}$, it follows that $\langle Ff_*(x), a\rangle = \langle x, Gf^*(a)\rangle$ holds for all $x \in FK$ and $a \in GD$. Therefore $(Ff^*, Gf_*)$ is a morphism $(FK, GE, \langle\,,\rangle) \to (FL, GD, \langle\,,\rangle)$ in **COM**, and we obtain a functor **SEM** $\to$ **COM**.

We have shown that the functors constituting the equivalences **MConv** $\simeq$ **pBNS** and **AEMod** $\simeq$ **OUS** can lift to the functors **SEM** $\rightleftarrows$ **COM**. It easy to see that the natural isomorphisms constituting the equivalences **MConv** $\simeq$ **pBNS** and **AEMod** $\simeq$ **OUS** can also lift, and thus we obtain an equivalence **SEM** $\simeq$ **COM**. ■

**Example 7.2.54.** We give some examples of convex operational models and the corresponding state-effect models.

(i) Let $\mathscr{V}$ be a pre-base-norm space. Then the dual $\mathscr{V}^*$ is an order-unit space, and the triple $(\mathscr{V}, \mathscr{V}^*, \langle\,,\rangle)$ where $\langle x, \varphi\rangle = \varphi(x)$ is a convex operational model. The corresponding state-effect model consists of the convex set $\mathrm{B}(\mathscr{V})$ and the effect module $[0,u]_{\mathscr{V}^*} = \mathbf{OUS}_{\le}(\mathscr{V}, \mathbb{R})$ of unital positive functionals on $\mathscr{V}$.

(ii) Let $\mathscr{A}$ be an order-unit space. Then the dual $\mathscr{A}^*$ is a base-norm space, and $(\mathscr{A}^*, \mathscr{A}, \langle\,,\rangle)$ is a convex operational model. The corresponding state-effect model consists of the convex set $\mathrm{B}(\mathscr{A}^*) = \mathbf{pBNS}(\mathscr{A}, \mathbb{R})$ of trace-preserving positive functionals on $\mathscr{A}$ and the effect module $[0,u]_{\mathscr{A}}$.

(iii) Let $\mathscr{A}$ be a $W^*$-algebra. Recall from Examples 7.2.21 and 7.2.39 that $\mathscr{A}_{\mathrm{sa}}$ is a order-unit space and $(\mathscr{A}_*)_{\mathrm{sa}}$ is a base-norm space. Thus $((\mathscr{A}_*)_{\mathrm{sa}}, \mathscr{A}_{\mathrm{sa}}, \langle\,,\rangle)$, with $\langle\,,\rangle$ defined in the obvious way, is a convex operational model. The corresponding state-effect model consists of the convex set $\mathrm{B}((\mathscr{A}_*)_{\mathrm{sa}}) = \mathrm{St}(\mathscr{A})$ of normal states on $\mathscr{A}$ and the effect module $[0,1]_{\mathscr{A}}$ of effects in $\mathscr{A}$.

(iv) Let $\mathscr{A}$ be a $C^*$-algebra. Similarly to the previous example, $((\mathscr{A}^*)_{\mathrm{sa}}, \mathscr{A}_{\mathrm{sa}}, \langle\,,\rangle)$ is a convex operational model. The corresponding state-effect model consists of the convex set $\mathrm{B}((\mathscr{A}^*)_{\mathrm{sa}}) = \mathrm{St}(\mathscr{A})$ of (arbitrary) states on $\mathscr{A}$ and the effect module $[0,1]_{\mathscr{A}}$ of effects in $\mathscr{A}$.

Many other examples can be obtained as special cases of the above ones, see Examples 7.2.21 and 7.2.39. We mention an important example, a special case of (iii), which corresponds to the standard Hilbert space model of quantum theory,

(v) Let $\mathscr{H}$ be a Hilbert space. Then $(\mathcal{TC}(\mathscr{H})_{\mathrm{sa}}, \mathcal{B}(\mathscr{H})_{\mathrm{sa}}, \langle\,,\rangle)$, where $\langle T, A\rangle = \mathrm{tr}(TA)$, is a convex operational model. The corresponding state-effect model consists of the convex set $\mathcal{D}en(\mathscr{H})$ of density operators on $\mathscr{H}$ and the effect module $\mathcal{E}f(\mathscr{H})$ of effects on $\mathscr{H}$.

### 7.2.5 Effectuses and convex operational models

Finally we will make a precise explicit connection between effectuses and the convex operational framework. The connection is in some sense bidirectional. We will first prove that the category $\mathbf{pBNS}_{\le}$ of pre-base-norm spaces, the category $\mathbf{OUS}_{\le}^{\mathrm{op}}$ of order-unit spaces, and the category $\mathbf{COM}_{\le}$ of convex operational models are all effectuses, which establishes one way of the connection. There is the other way, starting from effectuses: we will show that for each real effectus $\mathbf{C}$ satisfying a certain separation condition, there is a faithful functor $F\colon \mathbf{C} \to \mathbf{COM}_{\leq}$ into the category $\mathbf{COM}$ of convex operational models (with 'subunital' morphisms). Moreover the functor $F\colon \mathbf{C} \to \mathbf{COM}_{\leq}$ is shown to be a morphism of effectuses. Thus, every suitable real effectus can be embedded into the category of convex operational models, in a way it preserves the effectus structure.

**Proposition 7.2.55.**

(i) *The category* $\mathbf{pBNS}_{\leq}$ *of pre-base-norm spaces is an effectus. The total maps are trace-preserving maps, i.e.* $\mathrm{Tot}(\mathbf{pBNS}_{\leq}) = \mathbf{pBNS}$.

(ii) *The category* $\mathbf{OUS}^{\mathrm{op}}_{\leq}$ *of order-unit spaces is an effectus. The total maps are unital maps, i.e.* $\mathrm{Tot}(\mathbf{OUS}^{\mathrm{op}}_{\leq}) = \mathbf{OUS}^{\mathrm{op}}$.

*Proof.* Note first that $\mathbf{sBNS}_{\leq} \simeq \mathbf{CWMod}_{\leq}$ is an effectus. This follows by Corollary 3.8.8, because $\mathbf{CWMod}_{\leq}$ is a full subcategory of the effectus $\mathbf{WMod}_{\leq}$ (Proposition 3.5.9) that contains the unit $[0,1]$ is closed under finite coproducts. The effectus $\mathbf{sBNS}_{\leq}$ has $\mathbb{R}$ as unit and truth maps $\mathbb{1}_{\mathscr{V}}\colon \mathscr{V} \to \mathbb{R}$ given by traces. Since $\mathbf{pBNS}_{\leq}$ is a full subcategory of $\mathbf{sBNS}_{\leq}$, we again apply Corollary 3.8.8 to prove that $\mathbf{pBNS}_{\leq}$ is an effectus. Clearly $\mathbf{pBNS}_{\leq}$ contains the unit $\mathbb{R}$ and the initial object $0 = \{0\}$. It remains to show that binary coproducts in $\mathbf{sBNS}_{\leq}$ of pre-base-norm spaces are pre-base-norm spaces. Let $(\mathscr{V}, \tau_{\mathscr{V}})$ and $(\mathscr{W}, \tau_{\mathscr{W}})$ be pre-base-norm spaces. Then the coproduct is given by the direct sum $\mathscr{V} \oplus \mathscr{W}$ with trace $\tau\colon \mathscr{V} \oplus \mathscr{W} \to \mathbb{R}$ given by $\tau(x, y) = \tau(x) + \tau(y)$. It is not hard to see that the base seminorm on $\mathscr{V} \oplus \mathscr{W}$ satisfies $\|(x, y)\| = \|x\|_{\mathscr{V}} + \|y\|_{\mathscr{W}}$ where $\|-\|_{\mathscr{V}}$ and $\|-\|_{\mathscr{W}}$ are the base norms of $\mathscr{V}$ and $\mathscr{W}$. Therefore the base seminorm is a norm and hence $\mathscr{V} \oplus \mathscr{W}$ is a pre-base-norm space. Thus $\mathbf{pBNS}$ is an effectus. It is clear that $\mathrm{Tot}(\mathbf{pBNS}_{\leq}) = \mathbf{pBNS}$.

The proof that $\mathbf{OUS}^{\mathrm{op}}_{\leq}$ is an effectus is similar, via the fact that $\mathbf{OUS}^{\mathrm{op}}_{\leq}$ is a full subcategory of the effectus $\mathbf{sOUS}^{\mathrm{op}} \simeq \mathbf{EMod}^{\mathrm{op}}$. ∎

**Proposition 7.2.56.** *The category* $\mathbf{COM}_{\leq}$ *is an effectus. The total maps are total morphisms of convex operational models:* $\mathrm{Tot}(\mathbf{COM}_{\leq}) = \mathbf{COM}$.

*Proof.* We will prove this by lifting the effectus structure along the 'forgetful' functor $\mathbf{COM}_{\leq} \to \mathbf{pBNS}_{\leq} \times \mathbf{OUS}^{\mathrm{op}}_{\leq}$ (the codomain is an effectus, see Proposition 3.8.9). In other words, the effectus structure of $\mathbf{COM}_{\leq}$ can be given 'pointwise' via the structures of $\mathbf{pBNS}_{\leq}$ and $\mathbf{OUS}^{\mathrm{op}}_{\leq}$. Note that the unit of $\mathbf{pBNS}_{\leq}$ is $\mathbb{R}$ and truth maps are traces $\mathbb{1} = \tau\colon \mathscr{V} \to \mathbb{R}$. The unit of $\mathbf{OUS}^{\mathrm{op}}_{\leq}$ is also $\mathbb{R}$ and truth maps are order-units $\mathbb{1} = u\colon \mathbb{R} \to \mathscr{A}$ (via $\mathbb{1}(1) = u$). Then the unit of $\mathbf{COM}_{\leq}$ is a triple $(\mathbb{R}, \mathbb{R}, \langle\,,\rangle)$ with the obvious pairing $\langle r, s\rangle = rs$, and truth maps $\mathbb{1} = (\mathbb{1}_*, \mathbb{1}^*)$ are given by traces and order-units: $\mathbb{1}_* = \tau$ and $\mathbb{1}^* = u$. Similarly, since finite coproducts in $\mathbf{pBNS}_{\leq}$ and $\mathbf{OUS}^{\mathrm{op}}_{\leq}$ are direct sums, it is straightforward to see that the initial object in $\mathbf{COM}_{\leq}$ is $(\{0\}, \{0\}, \langle\,,\rangle)$, and a coproduct of $(\mathscr{V}, \mathscr{A}, \langle\,,\rangle)$ and $(\mathscr{W}, \mathscr{B}, \langle\,,\rangle)$ is $(\mathscr{V} \oplus \mathscr{W}, \mathscr{A} \oplus \mathscr{B}, \langle\,,\rangle)$ with pairing defined by $\langle (x, y), (a, b)\rangle = \langle x, a\rangle + \langle y, b\rangle$. (Put differently, finite coproducts are created by $\mathbf{COM}_{\leq} \to \mathbf{pBNS}_{\leq} \times \mathbf{OUS}^{\mathrm{op}}_{\leq}$.) We can also lift the PCM-structure along $\mathbf{COM}_{\leq} \to \mathbf{pBNS}_{\leq} \times \mathbf{OUS}^{\mathrm{op}}_{\leq}$ — concretely, we can define a sum $\ovee$ of morphisms $\mathbf{COM}_{\le}$ pointwise:

$$(f_*, f^*) \ovee (g_*, g^*) = (f_* + g_*, f^* + g^*)$$

if $f_* + g_*$ is trace-decreasing and $f^* + g^*$ is subunital (by Lemma 7.2.51, if either holds, so do both). By construction, it is easy to see that $\mathbf{COM}_{\le}$ is a finPAC and satisfies (E2) and (E3) from Definition 3.2.1. It remains to verify (E1), i.e. that predicates form effect algebras. Let $(p_*, p^*) \colon (\mathscr{V}, \mathscr{A}, \langle\,,\rangle) \to (\mathbb{R}, \mathbb{R}, \langle\,,\rangle)$ be a predicate in $\mathbf{COM}_{\le}$. Since $p_*$ and $p^*$ are respectively predicates in $\mathbf{pBNS}_{\le}$ and $\mathbf{OUS}_{\le}^{\mathrm{op}}$, there are orthosupplements such that $p_* \ovee (p_*)^{\perp} = \mathbb{1}_* \;(= \tau)$ and $p^* \ovee (p^*)^{\perp} = \mathbb{1}^* \;(= u)$. It is easy to see that $(p_*)^{\perp}$ and $(p^*)^{\perp}$ are in adjunction. Therefore $((p_*)^{\perp}, (p^*)^{\perp})$ is a unique map in $\mathbf{COM}_{\le}$ such that $(p_*, p^*) \ovee ((p_*)^{\perp}, (p^*)^{\perp}) = (\mathbb{1}^*, \mathbb{1}_*)$. Clearly $(p_*, p^*) \perp (\mathbb{1}^*, \mathbb{1}_*)$ implies $(p_*, p^*) = 0$, showing that predicates in $\mathbf{COM}_{\le}$ form effect algebras, and hence $\mathbf{COM}_{\le}$ is an effectus. Clearly $\mathbf{COM}$ is the total part of $\mathbf{COM}_{\le}$. ■

Since $\mathbf{COM} \simeq \mathbf{SEM}$, we obtain:

**Corollary 7.2.57.** $\mathbf{SEM}$ *is an effectus in total form.* ■

Let us describe states and predicates in the effectus $\mathbf{COM}_{\le}$ of convex operational models. A state in $\mathbf{COM}_{\le}$ is a total morphism $\omega \colon (\mathbb{R}, \mathbb{R}, \langle\,,\rangle) \to (\mathscr{V}, \mathscr{A}, \langle\,,\rangle)$, that is, a pair of $\omega_* \colon \mathbb{R} \to \mathscr{V}$ in $\mathbf{pBNS}$ and $\omega^* \colon \mathscr{A} \to \mathbb{R}$ in $\mathbf{OUS}$ such that $\langle \omega_*(r), a\rangle = \langle r, \omega_*(a)\rangle \equiv r \cdot \omega_*(a)$. Since $\omega_*(r) = r \cdot \omega_*(1)$ and $\omega^*(a) = \langle \omega_*(1), a\rangle$, such a pair $(\omega_*, \omega^*)$ is determined by $\omega_*(1)$, which is an element of the base $\mathrm{B}(\mathscr{V})$. Conversely, any element $x \in \mathrm{B}(\mathscr{V})$ induces a state $\omega \colon (\mathbb{R}, \mathbb{R}, \langle\,,\rangle) \to (\mathscr{V}, \mathscr{A}, \langle\,,\rangle)$ by $\omega_*(r) = r \cdot x$ and $\omega^*(a) = \langle x, a\rangle$. This establishes the bijection

$$\mathrm{St}(\mathscr{V}, \mathscr{A}, \langle\,,\rangle) \equiv \mathbf{COM}((\mathbb{R}, \mathbb{R}, \langle\,,\rangle), (\mathscr{V}, \mathscr{A}, \langle\,,\rangle)) \cong \mathrm{B}(\mathscr{V})\,.$$

Reasoning similarly about predicates, we obtain:

$$\mathrm{Pred}(\mathscr{V}, \mathscr{A}, \langle\,,\rangle) \equiv \mathbf{COM}_{\le}((\mathscr{V}, \mathscr{A}, \langle\,,\rangle), (\mathbb{R}, \mathbb{R}, \langle\,,\rangle)) \cong [0,1]_{\mathscr{A}}\,.$$

Thus effectus-theoretic notions of states and predicates capture the view that $\mathscr{V}$ is the 'state space' and $\mathscr{A}$ is the 'effect space' of an convex operational model $(\mathscr{V}, \mathscr{A}, \langle\,,\rangle)$.

Similarly, in the effectus in total form $\mathbf{SEM}$ of state-effect models, one has

$$\mathrm{St}(K, E, \vDash) \cong K \qquad\qquad \mathrm{Pred}(K, E, \vDash) \cong E\,,$$

so that the state functor $\mathrm{St} \colon \mathbf{SEM} \to \mathbf{Conv}$ and predicate functor $\mathrm{Pred} \colon \mathbf{SEM} \to \mathbf{EMod}$ are mere projections.

Next we will show how effectuses may produce convex operational models. Recall that for each object $A$ in an effectus $\mathbf{C}$, there are states $\omega \colon I \to A$, forming a convex set $\mathrm{St}(A)$, and predicates $p \colon A \to I$, forming an effect module $\mathrm{Pred}(A)$. Here the scalars are $\mathcal{S}$ given by morphisms $I \to I$. The validity $\omega \vDash p = p \circ \omega$ defines a dual pair $\mathrm{St}(A) \times \mathrm{Pred}(A) \to \mathcal{S}$. Thus one has a 'generalized state-effect model' $(\mathrm{St}(A), \mathrm{Pred}(A), \vDash)$, where the scalars $\mathcal{S}$ need not be real numbers and the separation conditions need not be satisfied. If $(\mathrm{St}(A), \mathrm{Pred}(A), \vDash)$ is a state-effect model in the sense of Definition 7.2.49, it yields a convex operational model via **COM** $\simeq$ **SEM**. Thus below, we focus on *real* effectuses (Definition 3.4.7), that is, effectuses with $\mathcal{S} = \mathbf{C}(I, I) \cong [0, 1]$. Note that the separation property (Definition 6.2.8) is not enough, since state-effect models involve the 'order-separation' condition. We formulate the order-separation property for effectuses in two ways, as a weak and a strong one.

**Definition 7.2.58.** We say that an effectus $\mathbf{C}$ satisfies the **weak order-separation property** if both

(i) $\mathbf{C}$ satisfies the separation property (see Definition 6.2.8);

(ii) predicates are order-separated by states: for each $p, q \in \mathrm{Pred}(A)$, if $p \circ \omega \leq q \circ \omega$ for all $\omega \in \mathrm{St}(A)$, then $p \leq q$.

An effectus $\mathbf{C}$ satisfies the **strong order-separation property** if morphisms are order-separated by predicates and states: for each $f, g \colon A \to B$, if $p \circ f \circ \omega \leq p \circ g \circ \omega$ for all $\omega \in \mathrm{St}(A)$ and $p \in \mathrm{Pred}(A)$, then $f \leq g$.

**Lemma 7.2.59.** *If* $\mathbf{C}$ *satisfies the strong order-separation property, then* $\mathbf{C}$ *satisfies weak order-separation property.*

*Proof.* Suppose that $f, g \colon A \to B$ satisfy $p \circ f \circ \omega = p \circ g \circ \omega$ for all $\omega \in \mathrm{St}(A)$ and $p \in \mathrm{Pred}(B)$. Then both $p \circ f \circ \omega \leq p \circ g \circ \omega$ and $p \circ f \circ \omega \geq p \circ g \circ \omega$ hold, and hence by order-separation, $f \leq g$ and $f \geq g$. Since $\leq$ is a partial order (see Proposition 3.2.7), we obtain $f = g$. It is obvious that predicates are order-separated by states. ■

We now show that a real effectus with the weak order-separation property indeed yields state-effect models, not only in a functorial way, but also faithfully and in the way it preserves the structure of effectuses, i.e. forms a morphism of effectuses (see Definition 4.2.1).

**Proposition 7.2.60.** *Let* $(\mathbf{C}, I)$ *be a real effectus with the weak order-separation property. Then for each* $A \in \mathbf{C}$*, the triple* $(\mathrm{St}(A), \mathrm{Pred}(A), \vDash)$*, where* $\omega \vDash p = p \circ \omega$*, is a state-effect model, yielding a functor* $F \colon \mathrm{Tot}(\mathbf{C}) \to \mathbf{SEM}$ *by* $FA = (\mathrm{St}(A), \mathrm{Pred}(A), \vDash)$ *and* $Ff = (f_*, f^*)$*. Moreover,* $F$ *is faithful and a morphism of effectuses in total form.*

*Proof.* We know that $\mathrm{St}(A)$ is a convex set, $\mathrm{Pred}(A)$ is an effect module, and $\vDash \colon \mathrm{St}(A) \times \mathrm{Pred}(A) \to [0, 1]$ is a suitable bihomomorphism. By the weak order-separation property, it is easy to see that $\vDash$ satisfies the separation conditions (SE3) and (SE4). Therefore $(\mathrm{St}(A), \mathrm{Pred}(A), \vDash)$ is a state-effect model. For each total map $f \colon A \to B$, the state transformer $f_* \colon \mathrm{St}(A) \to \mathrm{St}(B)$ is an affine map, and the predicate transformer $f^* \colon \mathrm{Pred}(A) \to \mathrm{Pred}(B)$ are a unital module map, satisfying

$$f_*(\omega) \vDash p \quad = \quad p \circ f \circ \omega \quad = \quad \omega \vDash f^*(p)\,.$$

Therefore the functor $F \colon \mathrm{Tot}(\mathbf{C}) \to \mathbf{SEM}$ is well-defined.

To see the faithfulness of $F$, let $f, g \colon A \to B$ be total morphisms in $\mathbf{C}$ such that $(f_*, f^*) = Ff = Fg = (g_*, g^*)$. Then for any $\omega \in \mathrm{St}(A)$ and $p \in \mathrm{Pred}(B)$ we have

$$\omega \circ f \circ p = f_*(\omega) \circ p = g_*(\omega) \circ p = \omega \circ g \circ p\,.$$

By separation $f = g$, showing that $F$ is faithful. Finally we show that $F$ is a morphism of effectuses in total form, i.e. $F$ preserves finite coproducts and the final object. This follows mostly directly from the fact that $\mathrm{St}\colon \mathrm{Tot}(\mathbf{C}) \to \mathbf{Conv}$ and $\mathrm{Pred}\colon \mathrm{Tot}(\mathbf{C}) \to \mathbf{EMod}^{\mathrm{op}}$ preserve finite coproducts and the final object, see § 4.2.1. Below we elaborate the case of binary coproducts only. Let $A + B$ be a coproduct in $\mathbf{C}$. Then

$$\mathrm{St}(A) \xrightarrow{(\kappa_1)_*} \mathrm{St}(A+B) \xleftarrow{(\kappa_2)_*} \mathrm{St}(B)$$
$$\mathrm{Pred}(A) \xrightarrow{(\kappa_1)^*} \mathrm{Pred}(A+B) \xrightarrow{(\kappa_2)^*} \mathrm{Pred}(B)$$

are respectively a coproduct in $\mathbf{Conv}$ and a product in $\mathbf{EMod}$. These coprojections and projections form the coprojections from $(\mathrm{St}(A), \mathrm{Pred}(A), \vDash)$ and $(\mathrm{St}(B), \mathrm{Pred}(B), \vDash)$ into the state-effect model $(\mathrm{St}(A+B), \mathrm{Pred}(A+B), \vDash)$. Now let $(K, E, \vDash)$ be a state-effect models, and $f\colon (\mathrm{St}(A), \mathrm{Pred}(A), \vDash) \to (K, E, \vDash)$ and $g\colon (\mathrm{St}(B), \mathrm{Pred}(B), \vDash) \to (K, E, \vDash)$ be morphisms in $\mathbf{SEM}$. By the universality of $\mathrm{St}(A+B)$ and $\mathrm{Pred}(A+B)$, there are unique mediating morphisms $[f_*, g_*]\colon \mathrm{St}(A+B) \to K$ in $\mathbf{Conv}$ and $\langle f^*, g^*\rangle\colon K \to \mathrm{Pred}(A+B)$ in $\mathbf{EMod}$ (such that $(\kappa_1)^* \circ [f_*, g_*] = f_*$, etc.). We need to show that the pair of $[f_*, g_*]$ and $\langle f^*, g^*\rangle$ form a morphism in $\mathbf{SEM}$. Using normalization, one can write each $\omega \in \mathrm{St}(A+B)$ as:

$$\omega = \langle\!\langle r_1 \cdot \omega_1, r_2 \cdot \omega_2 \rangle\!\rangle = r_1 \cdot (\kappa_1)_*(\omega_1) \ovee r_2 \cdot (\kappa_2)_*(\omega_2)$$

for some $\omega_1 \in \mathrm{St}(A)$ and $\omega_2 \in \mathrm{St}(B)$ and scalars $r_1, r_2 \in \mathcal{S} \cong [0,1]$. By affineness, $[f_*, g_*](\omega) = [\![ r_1 | f_*(x_1)\rangle + r_2 | g_*(x_2)\rangle ]\!]$. It is not hard to see that $\langle f^*, g^*\rangle(a) = [f^*(a), g^*(a)] \in \mathrm{Pred}(A+B)$ for each $a \in E$. Thus:

$$\begin{aligned}
[f_*, g_*](\omega) \vDash a &= [\![ r_1 | f_*(\omega_1)\rangle + r_2 | f_*(\omega_1)\rangle ]\!] \vDash a \\
&= r_1 (f_*(\omega_1) \vDash a) + r_2 (g_*(\omega_2) \vDash a) \\
&= r_1 (\omega_1 \vDash f^*(a)) + r_2 (\omega_2 \vDash g^*(a)) \\
&= f^*(a) \circ (r_1 \cdot \omega_1) \ovee g^*(a) \circ (r_2 \cdot \omega_2) \\
&= [f^*(a), g^*(a)] \circ \langle\!\langle r_1 \cdot \omega_1, r_2 \cdot \omega_2 \rangle\!\rangle \\
&= \omega \vDash \langle f^*, g^*\rangle(a)\,.
\end{aligned}$$

Thus the pair of $[f_*, g_*]$ and $\langle f^*, g^*\rangle$ form a morphism in $\mathbf{SEM}$, which is clearly a unique mediating map for $(f, g)$. ■

Finally we obtain the embedding of an effectus into the category of convex operational models.

**Theorem 7.2.61.** *Let $(\mathbf{C}, I)$ be a real effectus with the weak order-separation property. Then one has a morphism of effectuses $R\colon \mathbf{C} \to \mathbf{COM}_{\leq}$, which is also a faithful functor. Moreover, writing $(\mathscr{V}_A, \mathscr{A}_A, \langle\,,\rangle) = RA$ for the value of the functor at each object $A \in \mathscr{A}$, one has $\mathrm{B}(\mathscr{V}_A) \cong \mathrm{St}(A)$, $\mathrm{B}_{\leq}(\mathscr{V}_A) \cong \mathrm{St}_{\leq}(A)$, and $[0,u]_{\mathscr{A}_A} \cong \mathrm{Pred}(A)$.*

*Proof.* By Theorem 7.2.53 and Proposition 7.2.60, we have a morphism of effectuses in total form:

$$\mathrm{Tot}(\mathbf{C}) \xrightarrow{F} \mathbf{SEM} \simeq \mathbf{COM} = \mathrm{Tot}(\mathbf{COM}_{\leq})$$

By the equivalence between effectuses in partial and total form (Theorem 4.2.10), this extend to a morphism of effectuses $R\colon \mathbf{C} \to \mathbf{COM}_{\leq}$. To see that $R$ is faithful, let $f, g\colon A \to B$ be morphisms in $\mathbf{C}$ such that $Rf = Rg$. Using the fact that $R$ is a morphism of effectuses, we can show $R\langle\!\langle f, (\mathbb{1}f)^{\perp}\rangle\!\rangle = R\langle\!\langle g, (\mathbb{1}g)^{\perp}\rangle\!\rangle$. Since $\langle\!\langle f, (\mathbb{1}f)^{\perp}\rangle\!\rangle$ and $\langle\!\langle g, (\mathbb{1}g)^{\perp}\rangle\!\rangle$ are total morphisms, $\langle\!\langle f, (\mathbb{1}f)^{\perp}\rangle\!\rangle = \langle\!\langle g, (\mathbb{1}g)^{\perp}\rangle\!\rangle$ by faithfulness of $\mathrm{Tot}(R)\colon \mathrm{Tot}(\mathbf{C}) \to \mathrm{Tot}(\mathbf{COM}_{\leq})$. Therefore $f = g$.

It is clear by the construction of $\mathbf{SEM} \simeq \mathbf{COM}$ (see the proof of Theorem 7.2.53) that $\mathrm{B}(\mathscr{V}_A) \cong \mathrm{St}(A)$, $\mathrm{B}_{\leq}(\mathscr{V}_A) \cong \mathrm{St}_{\leq}(A)$, and $[0, u]_{\mathscr{A}_A} \cong \mathrm{Pred}(A)$. ∎

If an effectus satisfies the strongly order-separation, then we can improve the result slightly.

**Corollary 7.2.62.** *Let* $(\mathbf{C}, I)$ *be a real effectus with the strong order-separation property. One can apply Theorem* 7.2.61 *and has a functor* $R\colon \mathbf{C} \to \mathbf{COM}_{\leq}$*. Then for each object* $A \in \mathbf{C}$*, writing* $(\mathscr{V}_A, \mathscr{A}_A, \langle\,,\rangle) = RA$ *for the induced convex operational model, we have that* $\mathscr{V}_A$ *is a base-norm space.*

*Proof.* As shown in Theorem 7.2.61, one has $\mathrm{B}_{\leq}(\mathscr{V}_A) \cong \mathrm{St}_{\leq}(A)$. Note that by the strong order-separation property, substates are order-separated by predicates, that is: for each $\omega_1, \omega_2 \in \mathrm{St}_{\leq}(A)$, if $p \circ \omega_1 \leq p \circ \omega_2$ for all $p \in \mathrm{Pred}(A)$ implies $\omega_1 \leq \omega_2$. By Proposition 7.2.45, this implies that $\mathscr{V}_A$ is a base-norm space. ∎

We give examples of the effectuses that satisfies the assumptions of the embedding theorem, and describe induced convex operational models.

**Example 7.2.63.** The effectus $\mathcal{K}\ell(\mathcal{D}_{\leq})$ is real and strongly order-separated. To see the order-separation, let $f, g\colon X \to \mathcal{D}_{\leq}(Y)$ such that $p \odot f \odot \omega \leq p \odot f \odot \omega$ for any $\omega \in \mathrm{St}(X) \cong \mathcal{D}(X)$ and $p\colon \mathrm{Pred}(Y) \cong [0,1]^Y$. To see this, for each $x \in X$ and $y \in Y$, take the Dirac distribution $\omega = \eta_X(x)$ and the indicator function $p = \mathbb{1}_{\{y\}}$. Then

$$f(x)(y) = \mathbb{1}_{\{y\}} \odot f \odot \eta_X(x) \leq \mathbb{1}_{\{y\}} \odot g \odot \eta_X(x) = g(x)(y)\,.$$

Therefore $f \leq g$, and hence $\mathcal{K}\ell(\mathcal{D}_{\leq})$ is order-separated. Thus Corollary 7.2.62 applies to $\mathcal{K}\ell(\mathcal{D}_{\leq})$. For each $X \in \mathcal{K}\ell(\mathcal{D}_{\leq})$, the base-norm space induced by $\mathrm{St}(X) \cong \mathcal{D}(X)$ is the space $\ell^1_{c,\mathbb{R}}(X)$ of real-valued functions with finite support, see Example 7.2.39(iii). The order-unit space induced by $\mathrm{Pred}(X) \cong [0,1]^X$ is the real $\ell^\infty$-space $\ell^\infty_{\mathbb{R}}(X)$, see Example 7.2.21(ii). Thus the induced convex operational model is $(\ell^1_{c,\mathbb{R}}(X), \ell^\infty_{\mathbb{R}}(X), \langle\,,\rangle)$ with $\langle \varphi, \psi\rangle = \sum_{x \in X} \varphi(x)\psi(x)$.

It works in almost the same way for the Kleisli category $\mathcal{K}\ell(\mathcal{D}^\infty_{\leq})$ of the infinite subdistribution monad. The only difference is that the states are infinite distributions $\mathrm{St}(X) \cong \mathcal{D}^\infty(X)$, and hence one obtains the base-norm space $\ell^1_{\mathbb{R}}(X)$, see Example 7.2.39(ii). The predicates and the yielded order-unit space is the same, $\ell^\infty_{\mathbb{R}}(X)$. Thus one obtains a convex operational model $(\ell^1_{\mathbb{R}}(X), \ell^\infty_{\mathbb{R}}(X), \langle\,,\rangle)$.

**Example 7.2.64.** The effectus $\mathcal{K}\ell(\mathcal{G}_{\leq})$ is also real and strongly order-separated. The argument to show the order-separation is basically the same as above. Suppose that $f, g\colon X \to \mathcal{G}_{\leq}(Y)$ satisfy $p \odot f \odot \omega \leq p \odot f \odot \omega$ for any $\omega \in \mathrm{St}(X) \cong \mathcal{G}(X)$ and $p\colon \mathrm{Pred}(Y) \cong \mathbf{Meas}(Y,[0,1])$. Then for each $x \in X$ and $U \in \Sigma_Y$, taking the Dirac measure $\omega = \eta_X(x)$ and the indicator function $p = \mathbb{1}_U$, one has

$$f(x)(U) = \mathbb{1}_U \circledcirc f \circledcirc \eta_X(x) \leq \mathbb{1}_U \circledcirc g \circledcirc \eta_X(x) = g(x)(U)\,.$$

Thus $f \leq g$. Let $X$ be a measurable space. Then the base-norm space induced by $\mathrm{St}(X) \cong \mathcal{G}(X)$ is the space $ca(X)$ of finite signed measures, see Example 7.2.39(iv). The order-unit space induced by $\mathrm{Pred}(X) \cong \mathbf{Meas}(X,[0,1])$ is the space $\mathcal{L}^\infty_\mathbb{R}(X)$ of bounded real-valued measurable functions, see Example 7.2.21(v). Thus the induced convex operational model is $(ca(X), \mathcal{L}^\infty_\mathbb{R}(X), \langle\,,\rangle)$, with pairing $\langle \mu, f\rangle = \int f\, \mathrm{d}\mu$ done by integration.

**Example 7.2.65.** The effectus $\mathbf{Wstar}^{\mathrm{op}}_{\leq}$ of $W^*$-algebras is real and weakly order-separated. We verify that predicates are order-separated by states. Let $p, q \in \mathrm{Pred}(\mathscr{A}) \cong [0,1]_\mathscr{A}$ be predicates/effects such that $p \circ \omega \leq q \circ \omega$ for all $\omega \in \mathrm{St}(\mathscr{A})$. Then we have $\omega(p) \leq \omega(q)$, i.e. $\omega(q-p) \geq 0$, for all normal states on $\mathscr{A}$. By [232, Lemma 1.7.2], we obtain $q - p \geq 0$ and hence $p \leq q$. Let $\mathscr{A}$ be a $W^*$-algebra. Then the pre-base-norm space induced by $\mathrm{St}(\mathscr{A}) = \mathbf{Wstar}(\mathscr{A}, \mathbb{C})$, is the self-adjoint part of the predual $(\mathscr{A}_*)_{\mathrm{sa}}$, see Example 7.2.39(viii). (Thus it is a base-norm space.) The order-unit space induced by $\mathrm{Pred}(\mathscr{A}) \cong [0,1]_\mathscr{A}$ is the space $\mathscr{A}_{\mathrm{sa}}$ of self-adjoint elements, see Example 7.2.21(vii). Therefore the induced convex operational model is $((\mathscr{A}_*)_{\mathrm{sa}}, \mathscr{A}_{\mathrm{sa}}, \langle\,,\rangle)$, the example we saw in Example 7.2.54(iii).

**Example 7.2.66.** Similarly, the effectus $\mathbf{Cstar}^{\mathrm{op}}_{\leq}$ of $C^*$-algebras is real and weakly order-separated. The order-separation of predicates by states holds by Lemma 7.2.41(ii), since states on $\mathscr{A}$ can be identified with unital positive functionals on the order-unit space $\mathscr{A}_{\mathrm{sa}}$, but see also [167, Theorem 4.3.4]. The pre-base-norm space induced by $\mathrm{St}(\mathscr{A}) = \mathbf{Cstar}(\mathscr{A}, \mathbb{C})$, is the self-adjoint part of the dual space $(\mathscr{A}^*)_{\mathrm{sa}}$, see Example 7.2.39(vii). The order-unit space induced by $\mathrm{Pred}(\mathscr{A}) \cong [0,1]_\mathscr{A}$ is again $\mathscr{A}_{\mathrm{sa}}$. Therefore the induced convex operational model is $((\mathscr{A}^*)_{\mathrm{sa}}, \mathscr{A}_{\mathrm{sa}}, \langle\,,\rangle)$.

## 7.3 Partially $\sigma$-additive structure in effectus theory

Recall that effectuses are finPACs that have extra structures and properties (Definition 3.2.1). Here finPACs (finitely partially additive categories) are the finite variant of Arbib and Manes' PACs (partially additive categories) [7], which are equipped with *countably* additive structure. To clearly distinguish the finite and countable notions of PACs, we will call Arbib and Manes' original version a *partially $\sigma$-additive category* (*$\sigma$-PAC*, for short). Arbib and Manes introduced $\sigma$-PACs to provide a semantics of flow diagrams and programming languages [7, 8, 202]. The reason why countable sums are needed is that programs may contain infinite loop, typically coming from 'while' statement or recursion. More recently, $\sigma$-PACs (and its generalization called *unique decomposition categories*) were also used by Haghverdi and others to provide semantics of linear logic and Geometry of Interaction [3, 113–116].

It is natural to consider an extension of an effectus whose underlying structure is a $\sigma$-PAC, instead of a finPAC. In this section we will study such an extension of effectus, called *$\sigma$-effectus*. As a first step of the study of $\sigma$-effectuses, here we focus on the structure of (sub)states and predicates in a $\sigma$-effectus. Therefore in § 7.3.1 we study the countable extension of effect modules, weight modules, and convex sets. Continuing the work of Section 7.2 in the $\sigma$-additive setting, in § 7.3.2 we study the $\sigma$-additive structure in the context of base-norm and order-unit spaces. In § 7.3.3 we give the definition of $\sigma$-effectus, with some basic results, and examples of $\sigma$-effectuses. Finally in § 7.3.4, we present results about the structure of (sub)states and predicates in a $\sigma$-effectus. Specifically we present state-and-effect triangles for $\sigma$-effectuses (Corollaries 7.3.42 and 7.3.44), and a $\sigma$-effectus version of the result about embedding into convex operational models (Corollary 7.3.46).

Note that in this section we restrict ourselves to the 'real' setting: that is, we consider only effect/weight modules and convex sets over the real unit interval $[0,1] \subseteq \mathbb{R}$. The restriction is mainly for simplicity, and we expect that most results can be extended to general scalars given by a '$\sigma$-effect monoid' (with division, if needed). We leave such generalization for another occasion.

### 7.3.1 Partially $\sigma$-additive structure

We start with partially $\sigma$-additive monoids ($\sigma$-PAMs), which are an extension of partial commutative monoids with countable sums. We then study countable versions of effect algebras/modules, convex sets, and weight modules. We show that both $\sigma$-effect algebras [102] and superconvex sets [176, 177], which are existing notions, can be characterized in terms of the $\sigma$-PAM structure (Proposition 7.3.6 and Theorem 7.3.13). These results are new, as far as the author is aware, and used later in § 7.3.4 to identify the structure of states and predicates in a $\sigma$-effectus.

**Definition 7.3.1.** A **partially $\sigma$-additive monoid** (**$\sigma$-PAM**, for short) is a nonempty set $M$ equipped with a partial operation $\bigovee$ that sends a countable family $(x_j)_{j\in J}$ of elements in $M$ to an element in $M$, satisfying the three axioms below. We write $\bigovee_{j\in J} x_j = \bigovee(x_j)_{j\in J}$ and say that $(x_j)_{j\in J}$ is **summable** if $\bigovee_{j\in J} x_j$ is defined.

**(Partition-associativity axiom)** For each countable family $(x_j)_{j\in J}$ and each countable partition $J = \biguplus_{k\in K} J_k$, the family $(x_j)_{j\in J}$ is summable if and only if $(x_j)_{j\in J_k}$ is summable for each $k \in K$ and $(\bigovee_{j\in J_k} x_j)_{k\in K}$ is summable. In that case, one has:

$$\bigovee_{j\in J} x_j = \bigovee_{k\in K} \bigovee_{j\in J_k} x_j \,.$$

**(Unary sum axiom)** Each family $(x_j)_{j\in J}$ indexed by a singleton, say $J = \{j_1\}$, is summable and satisfies $\bigovee_{j\in J} x_j = x_{j_1}$.

**(Limit axiom)** A countable family $(x_j)_{j\in J}$ is summable whenever for any finite subset $F \subseteq J$, the subfamily $(x_j)_{j\in F}$ is summable.

A function $f\colon M \to N$ between $\sigma$-PAMs is said to be **$\sigma$-additive** if for any summable family $(x_j)_{j\in J}$ in $M$, the family $(f(x_j))_{j\in J}$ is summable in $N$ and moreover $f(\bigovee_{j\in J}(x_j)) = \bigovee_{j\in J} f(x_j)$. A function $f\colon M \times N \to L$ is **$\sigma$-biadditive** if it is $\sigma$-additive in each argument separately. We write **$\boldsymbol{\sigma}$PAM** for the category of partially $\sigma$-additive monoids and $\sigma$-additive maps.

Given a partially $\sigma$-additive monoid $(M, \bigovee)$, we will write $x \ovee y$ for the sum of the family consisting of two elements $x, y \in M$, and 0 for the sum of the empty family. Then $(M, \ovee, 0)$ forms a PCM. Thus there is a functor $\boldsymbol{\sigma}\mathbf{PAM} \to \mathbf{PCM}$ forgetting infinite addition. Here, finite sums defined by the $\sigma$-PAM structure and by the PCM structure coincide:

$$\bigovee_{j\in[n]} x_j = x_1 \ovee \cdots\cdots \ovee x_n \,.$$

The following lemma is useful when one wants to lift a PCM into a $\sigma$-PAM.

**Lemma 7.3.2.** *Let $M$ be a PCM. Let $\bigovee$ be a partial function that sends countable families in $M$ to elements in $M$, in a compatible way with the PCM structure — that is: for each finite family $(x_j)_{j\in F}$, $\bigovee(x_j)_{j\in F}$ is defined if and only if $(x_j)_{j\in F}$ is summable in the PCM $M$, and in that case, $\bigovee(x_j)_{j\in F}$ equals the sum $\bigovee_{j\in F} x_j$ in the PCM $M$. Then $(M, \bigovee)$ is a partially $\sigma$-additive monoid if (and only if) it satisfies the limit axiom and the following 'weak partition-associativity axiom'.*

- *For each countable family $(x_j)_{j\in J}$ and each countable partition $J = \biguplus_{k\in K} J_k$, if $(x_j)_{j\in J}$ is summable, then the family $(x_j)_{j\in J_k}$ is summable for each $k \in K$; $(\bigovee_{j\in J_k} x_j)_{k\in K}$ is summable; and*

$$\bigovee_{j\in J} x_j = \bigovee_{k\in K} \bigovee_{j\in J_k} x_j \,.$$

*Proof.* The unary sum axiom holds by the assumption that $\bigovee$ is compatible with the PCM structure. We only need to show that for each countable family $(x_j)_{j\in J}$ and each countable partition $J = \biguplus_{k\in K} J_k$, if the family $(x_j)_{j\in J_k}$ is summable for each $k \in K$ and $(\bigovee_{j\in J_k} x_j)_{k\in K}$ is summable, then $(x_j)_{j\in J}$ is summable. By the limit axiom, it suffices to show that $(x_j)_{j\in F}$ is summable for each finite subset $F \subseteq J$. By the weak partition-associativity, the sums $y_k := \bigovee_{j\in J_k\cap F} x_j$ and $z_k := \bigovee_{j\in J_k\setminus F} x_j$ are defined, and $y_k \ovee z_k = \bigovee_{j\in J_k} x_j$. Let $G = \{k \in K \mid J_k \cap F \neq \varnothing\}$. Then also

$$\bigovee_{k\in G} \bigovee_{j\in J_k} x_j = \bigovee_{k\in G} y_k \ovee z_k$$

is defined. Since $G$ is finite, the associativity of the PCM implies that the sums

$$\bigovee_{k\in G} y_k = \bigovee_{k\in G} \bigovee_{j\in J_k\cap F} x_j = \bigovee_{j\in F} x_j$$

are all defined, and hence $(x_j)_{j\in F}$ is summable. ∎

Next we show that certain 'complete' PCMs have the structure of $\sigma$-PAMs.

**Lemma 7.3.3.** *Let $M$ be a cancellative positive PCM. The following are equivalent.*

(i) *For any countable family $(x_j)_{j\in J}$, if*

$$\text{the family } (x_j)_{j\in F} \text{ is summable for every finite subset } F \subseteq J\,, \tag{7.5}$$

*then the following supremum exists:*

$$\bigvee\Big\{\bigovee_{j\in F} x_j \;\Big|\; F \subseteq J, F \text{ is finite}\Big\}\,. \tag{7.6}$$

(ii) *Let* $(x_n)_{n\in\mathbb{N}}$ *be a sequence in* $M$ *such that* $(x_n)_{n=0}^N$ *is summable for each* $N \in \mathbb{N}$. *Then the supremum* $\bigvee_{N\in\mathbb{N}} \bigovee_{n=0}^N x_n$ *exists.*

(iii) *The partial order of* $M$ *is* $\omega$*-complete: that is, every ascending sequence* $y_0 \leq y_1 \leq \cdots$ *has a supremum* $\bigvee_{n\in\mathbb{N}} y_n$ *in* $M$.

*Proof.* (i) $\Longrightarrow$ (ii) is obvious.

(ii) $\Longrightarrow$ (iii): Let $y_0 \leq y_1 \leq \cdots$ be an ascending sequence. Define $x_0 = y_0$ and $x_n = y_n \ominus y_{n-1}$ for each $n \in \mathbb{N}_{>0}$. Then $\bigovee_{n=0}^N x_N = y_N$, and hence the supremum $\bigvee_{N\in\mathbb{N}} \bigovee_{n=0}^N x_N = \bigvee_{N\in\mathbb{N}} y_N$ exists.

(iii) $\Longrightarrow$ (i): It holds in general that any $\omega$-complete poset has all suprema of directed countable subsets. To see this, let $U \subseteq M$ be a directed subset that is countable, say $U = \{x_0, x_1, \ldots\}$. By directedness we can find an ascending sequence $x'_0 \leq x'_1 \leq \cdots$ in $U$ such that $x_n \leq x'_n$ for all $n \in \mathbb{N}$. If $M$ is $\omega$-complete, $\bigvee_{n\in\mathbb{N}} x'_n$ exists and is a supremum of $U$. From this fact it is easy to see that (iii) implies (i), since $\{\bigovee_{j\in F} x_j \mid F \subseteq J, F \text{ is finite}\}$ is directed and countable. ■

**Definition 7.3.4.** A cancellative positive PCM is **$\sigma$-orthocomplete** if it satisfies any (hence all) of the equivalent conditions in Lemma 7.3.3. A $\sigma$-orthocomplete effect algebra is called a **$\sigma$-effect algebra**.

Let $M$ be a $\sigma$-orthocomplete cancellative positive PCM. We say that a countable family $(x_j)_{j\in J}$ in $M$ is **summable** if the condition (7.5) holds. In that case, we define the sum of the family by

$$\bigovee_{j\in J} x_j := \bigvee\Big\{\bigovee_{j\in F} x_j \;\Big|\; F \subseteq J, F \text{ is finite}\Big\}$$

**Lemma 7.3.5.** *Every* $\sigma$*-orthocomplete cancellative positive PCM* $M$ *form a* $\sigma$*-PAM via the partial operation* $\bigovee$ *defined above.*

*Proof.* Clearly, the operation $\bigovee$ is compatible with the PCM structure of $M$ in the sense of Lemma 7.3.2. By definition, $\bigovee$ satisfies the limit axiom. Thus it suffices to check the 'weak partition-associativity axiom' of Lemma 7.3.2. Let $(x_j)_{j\in J}$ be a summable family, and $\biguplus_{k\in K} J_k = J$ a countable partition of $J$. Clearly $\bigovee_{j\in J_k} x_j$ is defined for each $k \in K$. We write $y_k := \bigovee_{j\in J_k} x_j$. Let $G \subseteq K$ be a finite subset. By [203, Theorem 2.8] (see also [207, Theorem 4.5 and Corollary 4.6]), the (finite) sum $\bigovee_{k\in G} y_k$ exists and

$$\bigovee_{k\in G} y_k = \bigovee_{j\in\biguplus_{k\in G} J_k} x_j \,.$$

Hence, by definition, $\bigovee_{k\in K} y_k$ exists. Moreover,

$$\begin{aligned}
\bigovee_{k\in K} y_k &= \bigvee\Big\{ \bigovee_{k\in G} y_k \;\Big|\; G\subseteq K, G \text{ is finite}\Big\} \\
&= \bigvee\Big\{ \bigovee_{j\in\biguplus_{k\in G} J_k} x_j \;\Big|\; G\subseteq K, G \text{ is finite}\Big\} \\
&= \bigvee\Big\{ \bigovee_{j\in F} x_j \;\Big|\; G\subseteq K, G \text{ is finite}, F\subseteq \biguplus_{k\in G} J_k, F \text{ is finite}\Big\} \\
&= \bigvee\Big\{ \bigovee_{j\in F} x_j \;\Big|\; F\subseteq J, F \text{ is finite}\Big\} \\
&= \bigovee_{j\in J} x_j\,. \qquad \blacksquare
\end{aligned}$$

**Proposition 7.3.6.** *Let $E$ be an effect algebra. The following are equivalent.*

(i) *$E$ is $\sigma$-orthocomplete, that is, a $\sigma$-effect algebra.*

(ii) *There exists a $\sigma$-PAM structure on $E$ that extends the PCM structure of $E$.*

*Moreover, the $\sigma$-PAM structure in* (ii) *is unique if it exists, and is equal to the one given by Lemma* 7.3.5.

*Proof.* (i) $\Longrightarrow$ (ii) is done by Lemma 7.3.5. We prove the converse, and thus assume that $E$ has a $\sigma$-PAM structure that extends the PCM structure. To prove that $E$ is $\sigma$-orthocomplete, let $(x_j)_{j\in J}$ be a countable family satisfying (7.5). Then the sum $\bigovee_{j\in J} x_j$ exists by the Limit Axiom. We claim that $\bigovee_{j\in J} x_j$ is the supremum (7.6). (Then it also follows that the $\sigma$-PAM structure is unique.) For each finite subset $F\subseteq J$, one has $\bigovee_{j\in J} x_j = (\bigovee_{j\in F} x_j) \ovee (\bigovee_{j\in J\setminus F} x_j)$ and thus $\bigovee_{j\in J} x_j$ is an upper bound of $\bigovee_{j\in F} x_j$ for finite $F\subseteq J$. Let $y$ be another such upper bound. Then for each finite $F\subseteq J$, $\bigovee_{j\in F} x_j \le y$ and hence the sum $(\bigovee_{j\in F} x_j)\ovee y^\perp$ exists. By the Limit Axiom, it follows that the sum $(\bigovee_{j\in J} x_j)\ovee y^\perp$ exists. This means that $\bigovee_{j\in J} x_j \le y$. ■

We define morphisms and categories of $\sigma$-effect algebras.

**Definition 7.3.7.** A unital (resp. subunital) morphism of $\sigma$-effect algebras is a unital (resp. subunital) morphism of effect algebras that is also $\sigma$-additive. These morphisms, with $\sigma$-effect algebras as objects, form (non-full) subcategories $\boldsymbol{\sigma}\mathbf{EA}\hookrightarrow\mathbf{EA}$ and $\boldsymbol{\sigma}\mathbf{EA}_{\le}\hookrightarrow\mathbf{EA}_{\le}$.

It is straightforward to prove the following proposition using the arguments in the proof of Lemma 7.3.3.

**Proposition 7.3.8.** *Let $E, D$ be $\sigma$-effect algebras. A morphism $f\colon E\to D$ in $\mathbf{EA}_{\le}$ is a $\sigma$-additive (hence a morphism in $\boldsymbol{\sigma}\mathbf{EA}_{\le}$) if and only if it is $\omega$-continuous, i.e. preserves suprema of ascending sequences.* ■

We define an extension of $\sigma$-effect algebras with scalar multiplication, i.e. *$\sigma$-effect modules*. Though we can formulate it more generally like effect modules in Definition 3.4.3, for simplicity we will restrict ourselves to $\sigma$-effect modules over $[0,1]$. Note that $[0,1]$ has the structure of a '$\sigma$-effect monoid', i.e. a $\sigma$-effect algebra that is also equipped with multiplication. Countable sums $\sum_j r_j$ in $[0,1]$ are defined iff $\sum_j r_j \le 1$, where we interpret the left hand-side as the usual sum. Moreover multiplication commutes with countable sums: $s \cdot (\sum_j r_j) = \sum_j s r_j$ and $(\sum_j r_j) \cdot t = \sum_j r_j t$.

**Definition 7.3.9.** A **$\sigma$-effect module** is an effect module $E$ (over $[0,1]$) such that

(i) $E$ is $\sigma$-orthocomplete; and

(ii) the $[0,1]$-action $\cdot\colon [0,1] \times E \to E$ is $\sigma$-biadditive.

A unital (resp. subunital) morphism of $\sigma$-effect modules is a unital (resp. subunital) morphism of effect modules that is also $\sigma$-additive, or equivalently, $\omega$-continuous. The definitions yield a category $\boldsymbol{\sigma}\mathbf{EMod}$ (resp. $\boldsymbol{\sigma}\mathbf{EMod}_{\le}$).

Put categorically, a $\sigma$-effect module is a module over $[0,1]$ internal to the monoidal category $\boldsymbol{\sigma}\mathbf{EA}$ with the tensor product [102] of $\sigma$-effect algebras. In fact, we later show that the condition (ii) is redundant — every $\sigma$-orthocomplete effect module over $[0,1]$ (i.e. *$\omega$-effect module* in [138, 142, 149, 151]) is a $\sigma$-effect module; see Proposition 7.3.17. Note however that the proof relies on the properties of the scalars $[0,1]$.

Recall that a convex set (over $[0,1]$) is by definition an algebra for the (finite) distribution monad $\mathcal{D}$. A countable extension of convex sets is thus obtained by replacing $\mathcal{D}$ by the infinite distribution monad $\mathcal{D}^\infty$. Note that $\mathcal{D}^\infty(X)$ consists only of the distributions with countable support (Lemma 2.5.3). Such an extension is known as *superconvex sets* [176, 177].

**Definition 7.3.10.** A **superconvex set** (over $[0,1]$) is an (Eilenberg-Moore) algebra of the monad $\mathcal{D}^\infty$. A superaffine map is a morphism of Eilenberg-Moore $\mathcal{D}^\infty$-algebras. They form a category $\mathbf{SConv} = \mathcal{EM}(\mathcal{D}^\infty)$.

The term '$\sigma$-convex' is reserved for a notion in a topological vector space, which gives prototypical examples of superconvex sets.

**Example 7.3.11.** Let $\mathscr{V}$ be a Hausdorff topological real vector space. A subset $K \subseteq \mathscr{V}$ is said to be **$\sigma$-convex** if for each sequence $(x_n)_n$ in $K$ and $(r_n)_n$ in $[0,1]$ such that $\sum_n r_n = 1$, the series $\sum_{n\in\mathbb{N}} r_n x_n$ converges to some element in $K$. Every $\sigma$-convex subset of $\mathscr{V}$ forms a superconvex set; see [176, Example 1.6] or [177, Proposition 1.2].

Recall from Corollary 4.4.9 and Proposition 4.4.10 that convex sets over $[0,1]$ are categorically equivalent to weight modules over $[0,1]$. It will turn out that such equivalence also exists between superconvex sets and *$\sigma$-weight modules*, an extension of weight modules with countable sums. This can be seen as a characterization of superconvex sets in terms of the $\sigma$-PAM structure.

**Definition 7.3.12.** A **$\sigma$-weight module** is a weight module $X$ (over $[0,1]$) equipped with a $\sigma$-PAM structure $\bigovee$ that extends the PCM structure of $X$ such that

(i) the $[0,1]$-action $\cdot\colon [0,1] \times X \to X$ is $\sigma$-biadditive;

(ii) the weight map $|-|\colon X \to [0,1]$ is $\sigma$-additive;

(iii) a countable family $(x_j)_{j\in J}$ in $X$ is summable whenever $\sum_{j\in J} |x_j| \le 1$ (i.e. $(\,|x_j|\,)_{j\in J}$ is summable in $[0,1]$).

A weight-preserving (resp. weight-decreasing) morphism of $\sigma$-weight modules is a weight-preserving (resp. weight-decreasing) morphism of weight modules that is also $\sigma$-additive. The definitions yield a category $\boldsymbol{\sigma}\mathbf{WMod}$ (resp. $\boldsymbol{\sigma}\mathbf{WMod}_{\leq}$).

**Theorem 7.3.13.** *The equivalence* $\mathbf{WMod} \simeq \mathbf{Conv}$ *from Corollary* 4.4.9 *(for* $M = [0,1]$*) lifts to the categories of* $\sigma$*-weight modules and superconvex sets:*

$$\boldsymbol{\sigma}\mathbf{WMod} \underset{\mathcal{L}}{\overset{\mathrm{B}}{\rightleftarrows}} \mathbf{SConv} \quad (\simeq)$$

*Proof.* Let $X$ be a $\sigma$-weight module. We define a $\mathcal{D}^\infty$-algebra structure

$$[\![-]\!] \colon \mathcal{D}^\infty(\mathrm{B}(X)) \to \mathrm{B}(X) \qquad \text{by} \qquad [\![\textstyle\sum_j r_j|x_j\rangle]\!] = \bigovee_j r_j \cdot x_j\,.$$

The sum in the right-hand side is defined since $\sum_j |r_j \cdot x_j| = \sum_j r_j = 1$, and it is in $\mathrm{B}(X)$ since $|\bigovee_j r_j \cdot x_j| = \sum_j |r_j \cdot x_j| = 1$. In a similar manner to the proof of Proposition 3.6.2, it is easy to verify that $(\mathrm{B}(X), [\![-]\!])$ satisfies the axioms of $\mathcal{D}^\infty$-algebras. It is easy to see that the functor B sends morphisms in $\boldsymbol{\sigma}\mathbf{WMod}$ to those in $\mathbf{SConv}$, so that the functor restricts to $\mathrm{B}\colon \boldsymbol{\sigma}\mathbf{WMod} \to \mathbf{SConv}$.

Let $K$ be a superconvex set. Then we define a $\sigma$-PAM structure on $\mathcal{L}(K)$ as follows: a countable family $((r_j, x_j))_j$ is summable iff $\sum_j r_j \leq 1$, and in that case, writing $t = \sum_j r_j$,

$$\bigovee_j (r_j, x_j) = \begin{cases} (t, [\![\sum_j t\backslash r_j|x_j\rangle]\!]) & \text{if } t \neq 0 \\ (0, *) & \text{if } t = 0\,. \end{cases}$$

Note that by Lemma 4.4.3, the above definition of sums $\bigovee$ agrees with the PCM structure of $\mathcal{L}(X)$ if the family $((r_j, x_j))_j$ is finite. It is clear that $\bigovee$ satisfies the limit axiom. By Lemma 7.3.2, to prove that $\mathcal{L}(X)$ is a $\sigma$-PAM, it suffices to show that the weak partition-associativity holds. Let $((r_j, x_j))_{j\in J}$ be a summable family, that is, $\sum_j r_j \leq 1$. Let $\biguplus_{k\in K} J_k = J$ be a countable partition. Clearly each family $((r_j, x_j))_{j\in J_k}$ is summable, with $|\bigovee_{j\in J_k}(r_j, x_j)| = \sum_{j\in J_k} r_j$. Since $\sum_k |\bigovee_{j\in J_k}(r_j, x_j)| = \sum_k \sum_{j\in J_k} r_j = \sum_j r_j \leq 1$, the family $(\bigovee_{j\in J_k}(r_j, x_j))_{k\in K}$ is summable. Now, writing $t_k = \sum_{j\in J_k} r_j$ and $t = \sum_{k\in K} t_k$, we have

$$\begin{aligned}
\bigovee_{k\in K} \bigovee_{j\in J_k}(r_j, x_j) &= \bigovee_{k\in K}(t_k, [\![\textstyle\sum_{j\in J_k} t_k\backslash r_j|x_j\rangle]\!]) \\
&= \Big(t, \Big[\!\!\Big[\textstyle\sum_{k\in K} t\backslash t_k\big|[\![\sum_{j\in J_k} t_k\backslash r_j|x_j\rangle]\!]\big\rangle\Big]\!\!\Big]\Big) \\
&= \big(t, [\![\textstyle\sum_{k\in K}\sum_{j\in J_k}(t\backslash t_k)\cdot(t_k\backslash r_j)|x_j\rangle]\!]\big) \\
&= (t, [\![\textstyle\sum_{j\in J} t\backslash r_j|x_j\rangle]\!]) \\
&= \bigovee_{j\in J}(r_j, x_j)\,,
\end{aligned}$$

Thus $\mathcal{L}(K)$ has a $\sigma$-PAM structure that extends the PCM structure. Conditions (ii) and (iii) in Definition 7.3.12 hold immediately from the definition. The proof of (i), the $\sigma$-biadditivity of the $[0,1]$-action, is essentially the same as the proof of Lemma 4.4.2, hence omitted. Therefore $\mathcal{L}(K)$ is a $\sigma$-weight module. It is easy to see that for each $f\colon K \to L$ in **SConv**, the map $\mathcal{L}(f)\colon \mathcal{L}(K) \to \mathcal{L}(L)$ preserves countable sums $\bigovee$. Therefore we have a functor $\mathcal{L}\colon \mathbf{SConv} \to \boldsymbol{\sigma}\mathbf{WMod}$.

We have the unit and counit maps $\eta_K\colon K \overset{\cong}{\to} \mathrm{B}(\mathcal{L}(K))$ in **Conv** and $\varepsilon_X\colon \mathcal{L}(\mathrm{B}(X)) \overset{\cong}{\to} X$ in **WMod**, see Section 4.4. It is straightforward to check that these maps respectively preserve superconvex sums and countable partial sums. The map $\varepsilon_X$ moreover reflects summability. It follows that $\eta_K$ is an isomorphism in **SConv** and $\varepsilon_X$ is an isomorphism in $\boldsymbol{\sigma}$**WMod**. Therefore we obtain $\mathbf{SConv} \simeq \boldsymbol{\sigma}\mathbf{WMod}$. ∎

## 7.3.2 Partially $\sigma$-additive structure and convex operational structure

In § 7.2.1 and § 7.2.2 we saw the equivalences between (certain classes of) effect modules and order-unit spaces, and between convex sets, weight modules, and base-norm spaces. In the previous subsection we defined countable extensions of effect modules, convex sets, and weight modules. Then the obvious question is: what are the order-unit and base-norm spaces corresponding to such countable extensions? This will be answered here. It will turn out that such countable structure makes both order-unit and (pre-) base-norm space complete, and hence makes them into Banach spaces. We note that key technical results (e.g. Lemmas 7.3.15 and 7.3.22) are known ones. The contribution here is to collect these known results and put them in the context of the equivalences to effect modules or to convex sets and weight modules.

An ordered vector space $\mathscr{V}$ is said to be **monotone $\sigma$-complete** if it is bounded $\omega$-complete as poset, that is, if every ascending sequence $x_0 \le x_1 \le \cdots$ in $\mathscr{V}$ that is bounded above has a supremum $\bigvee_{n=0}^{\infty} x_n$. Note that the monotone $\sigma$-completeness implies that every bounded-below descending sequence has an infimum, too, since $\mathscr{V}$ is a self-dual poset via negation.

**Lemma 7.3.14.** *A semi-order-unit space $(\mathscr{A}, u)$ is monotone $\sigma$-complete if and only if the unit interval $[0, u]_{\mathscr{A}}$ is $\omega$-complete and hence a $\sigma$-effect algebra (see Lemma* 7.3.3*).*

*Proof.* It is clear that the monotone $\sigma$-completeness of $A$ implies $\omega$-completeness of $[0, u]_{\mathscr{A}}$. Conversely, assume that $[0, u]_{\mathscr{A}}$ is $\omega$-complete. Any ascending sequence in $\mathscr{A}$ bounded above can be shifted and scaled into an ascending sequence in $[0, u]_{\mathscr{A}}$, which has a supremum in $[0, u]_{\mathscr{A}}$. By shifting and scaling the supremum we obtain a supremum of the original sequence, showing that $\mathscr{A}$ is monotone $\sigma$-complete. ∎

**Lemma 7.3.15.** *Let $(\mathscr{A}, u)$ be a monotone $\sigma$-complete semi-order-unit space.*

(i) *The order unit $u$ is Archimedean. Hence $(\mathscr{A}, u)$ is an order-unit space.*

(ii) *Moreover, $(\mathscr{A}, u)$ is a Banach order-unit space (i.e. complete with respect to the order-unit norm).*

*Proof.* (i) (From [260, Lemma 1.1]) The sequence $(u/n)_{n=1}^{\infty}$ is descending and bounded from below, so there is an infimum $\bigwedge_{n=1}^{\infty} u/n \in \mathscr{A}$. Then $\bigwedge_{n=1}^{\infty} u/n = \bigwedge_{n=1}^{\infty} 2u/n = 2 \cdot \bigwedge_{n=1}^{\infty} u/n$ and hence $\bigwedge_{n=1}^{\infty} u/n = 0$. Therefore if $na \le u$ for all $n \in \mathbb{N}$, then $a \le u/n$ for all $n \ge 1$, so that $x \le \bigwedge_{n=1}^{\infty} u/n = 0$.

For a proof of (ii), we refer to [260, Lemma 1.2]. It is also an immediate consequence of [220, Proposition 1.6, Chapter 3], which characterizes Banach order-unit spaces. ∎

**Corollary 7.3.16.** *Let $(\mathscr{A}, u)$ be an semi-order-unit space. If the interval $[0, u]$ is a $\sigma$-effect algebra, then $\mathscr{A}$ is a Banach order-unit space.* ∎

We note that not every Banach order-unit space is monotone $\sigma$-complete. For a counterexample, consider the order-unit space $\mathrm{C}_{\mathbb{R}}([0, 1])$ of continuous functions $f\colon [0, 1] \to \mathbb{R}$, with the constant function $1(x) = 1$ as unit. The order-unit norm of $\mathrm{C}_{\mathbb{R}}([0, 1])$ coincides with the sup norm, and hence $\mathrm{C}_{\mathbb{R}}([0, 1])$ is a Banach space. However $\mathrm{C}_{\mathbb{R}}([0, 1])$ is not monotone $\sigma$-complete.

The following proposition shows that $\sigma$-orthocomplete ($= \omega$-complete) effect modules have quite rich structures. In particular, it is shown that condition (ii) of Definition 7.3.9 is redundant for $\sigma$-effect modules over $[0, 1]$.

**Proposition 7.3.17.** *Every $\sigma$-orthocomplete effect module $E$ is Archimedean, and also a $\sigma$-effect module (Definition 7.3.9). Moreover, $E$ is complete in the canonical 'order-unit' metric (thus, $E$ is a 'Banach effect module' in the sense of [90, 147, 148]).*

*Proof.* By Theorem 7.2.6 we may assume that $E$ is the interval $[0, u]_{\mathscr{A}}$ of some semi-order-unit space $(\mathscr{A}, u)$. Since $E = [0, u]_{\mathscr{A}}$ is $\sigma$-orthocomplete, $\mathscr{A}$ is monotone $\sigma$-complete by Lemma 7.3.14. Then $\mathscr{A}$ is a Banach order-unit space by Lemma 7.3.15. By Theorem 7.2.20, $[0, u]_{\mathscr{A}}$ is Archimedean. Since $[0, u]_{\mathscr{A}}$ is closed in $\mathscr{A}$, it is complete.

It remains to prove that $E = [0, u]_{\mathscr{A}}$ is a $\sigma$-effect module. We need to prove that the scalar multiplication $\cdot\colon [0, 1] \times [0, u]_{\mathscr{A}} \to [0, u]_{\mathscr{A}}$ is $\sigma$-biadditive. By Proposition 7.3.8, it suffices to prove the $\omega$-continuity in each argument.

*$\omega$-continuity in the first argument:* Fix $a \in [0, u]_{\mathscr{A}}$ and we prove that $(-)\cdot a\colon [0, 1] \to [0, u]_{\mathscr{A}}$ is $\omega$-continuous. Let $(r_n)_{n\in\mathbb{N}}$ be an ascending sequence in $[0, 1]$. Clearly $(\bigvee_n r_n) \cdot a$ is a upper bound of $r_n \cdot a$. Let $b \in [0, u]_{\mathscr{A}}$ satisfy $r_n \cdot a \le b$ for all $n \in \mathbb{N}$. Let $N \in \mathbb{N}_{>0}$ be arbitrary. Then there is some $m \in \mathbb{N}$ such that $\bigvee_n r_n < r_m + \frac{1}{N}$, so that

$$\Big(\bigvee\nolimits_n r_n\Big) \cdot a \le (r_m + \frac{1}{N}) \cdot a = r_m \cdot a + \frac{a}{N} \le b + \frac{u}{N} \,.$$

Thus $N \cdot (((\bigvee_n r_n) \cdot a) - b) \le u$. Since $N \in \mathbb{N}_{>0}$ is arbitrary, and $\mathscr{A}$ is Archimedean by Lemma 7.3.15, we obtain $((\bigvee_n r_n) \cdot a) - b \le 0$, that is, $(\bigvee_n r_n) \cdot a \le b$. Therefore $(\bigvee_n r_n) \cdot a = \bigvee_n (r_n \cdot a)$.

*$\omega$-continuity in the second argument:* If $r = 0$, then $0 \cdot (-)\colon [0, u]_{\mathscr{A}} \to [0, u]_{\mathscr{A}}$ is trivially $\omega$-continuous. Fix $r \in (0, 1]$. Then $r \cdot (-)\colon \mathscr{A} \to \mathscr{A}$ is an order isomorphism, with the monotone inverse $r^{-1} \cdot (-)\colon \mathscr{A} \to \mathscr{A}$. Thus $r \cdot (-)\colon \mathscr{A} \to \mathscr{A}$ preserves all suprema, and the restriction $r \cdot (-)\colon [0, u]_{\mathscr{A}} \to [0, u]_{\mathscr{A}}$ is $\omega$-continuous. ∎

We summarize the results above as a categorical equivalence. A positive linear map between monotone $\sigma$-complete ordered vector spaces is said to be **$\sigma$-normal** if it preserves suprema of bounded-above ascending sequences.

**Lemma 7.3.18.** *Let $f\colon \mathscr{A} \to \mathscr{B}$ be a subunital positive linear map between monotone $\sigma$-complete (Banach) order-unit spaces. Then $f$ is $\sigma$-normal if and only if the restriction $f\colon [0, u]_{\mathscr{A}} \to [0, u]_{\mathscr{B}}$ is $\omega$-continuous.*

*Proof.* By a similar reasoning to Lemma 7.3.14. ∎

**Proposition 7.3.19.** *The category* $\boldsymbol{\sigma}$**EMod** *(resp.* $\boldsymbol{\sigma}\mathbf{EMod}_{\leq}$*) is equivalent to the category of monotone σ-complete Banach order-unit spaces and σ-normal unital (resp. subunital) positive linear maps.*

*Proof.* By Proposition 7.3.17, $\boldsymbol{\sigma}$**EMod** is the same as the category of σ-orthocomplete effect modules and ω-continuous unital morphisms of effect modules. The desired equivalence is obtained by restricting the equivalence $\mathbf{EMod} \simeq \mathbf{sOUS}$ from Theorem 7.2.6, using Lemmas 7.3.14, 7.3.15 and 7.3.18. ■

We turn to (pre-)base-norm spaces. We start with proving that for σ-weight modules (and hence for superconvex sets), being cancellative is equivalent to being metric.

**Proposition 7.3.20.** *A σ-weight module is cancellative if and only if it is metric.*

*Proof.* By Lemma 7.2.31, any metric weight module is cancellative. To prove the 'only if', let $X$ be a cancellative σ-weight module. Assume that $d(x, y) = 0$ in the base pseudometric of $X$. By $d(x, y) = 0$, for each $n \in \mathbb{N}$ there are $z_n, w_n \in X$ such that $\frac{1}{2}x \ovee \frac{1}{2}z_n = \frac{1}{2}y \ovee \frac{1}{2}w_n$ and $|z_n| + |w_n| \leq 1/2^{n+1}$. Fix such sequences $(z_n)_{n\in\mathbb{N}}$ and $(w_n)_{n\in\mathbb{N}}$. By $\frac{1}{2}x \ovee \frac{1}{2}z_n = \frac{1}{2}y \ovee \frac{1}{2}w_n$ and $\frac{1}{2}x \ovee \frac{1}{2}z_{n+1} = \frac{1}{2}y \ovee \frac{1}{2}w_{n+1}$, we have

$$\tfrac{1}{4}x \ovee \tfrac{1}{4}z_n \ovee \tfrac{1}{4}y \ovee \tfrac{1}{4}w_{n+1} = \tfrac{1}{4}x \ovee \tfrac{1}{4}z_{n+1} \ovee \tfrac{1}{4}y \ovee \tfrac{1}{4}w_n \,.$$

By cancellation and Lemma 4.4.11, we obtain $z_n \ovee w_{n+1} = z_{n+1} \ovee w_n$. Since $\sum_{n\in\mathbb{N}}|z_n| + \sum_{n\in\mathbb{N}}|w_n| \leq 1$, we have the countable sums:

$$\begin{aligned} z_0 \ovee \Big(\bigovee_{n=1}^{\infty} z_n\Big) \ovee \Big(\bigovee_{n=1}^{\infty} w_n\Big) &= \bigovee_{n=0}^{\infty}(z_n \ovee w_{n+1}) \\ &= \bigovee_{n=0}^{\infty}(z_{n+1} \ovee w_n) = w_0 \ovee \Big(\bigovee_{n=1}^{\infty} z_n\Big) \ovee \Big(\bigovee_{n=1}^{\infty} w_n\Big) \,. \end{aligned}$$

Here the partition-associativity axiom is used. Therefore by cancellation, $z_0 = w_0$. From $\frac{1}{2}x \ovee \frac{1}{2}z_0 = \frac{1}{2}y \ovee \frac{1}{2}w_0$ we obtain $\frac{1}{2}x = \frac{1}{2}y$ and hence $x = y$ by Lemma 4.4.11. ■

We thus obtain an alternative proof to the following result, which was shown in [21, Corollary 2] and [176, 3.5].

**Corollary 7.3.21.** *A superconvex set is cancellative if and only if it is metric.*

*Proof.* By Theorem 7.3.13, Lemma 7.2.11, and Propositions 7.2.36 and 7.3.20. ■

**Lemma 7.3.22.** *Let* $(\mathscr{V}, \tau)$ *is a semi-base-norm space. Assume that the base* $\mathrm{B}(\mathscr{V})$ *is a superconvex set, whose underlying convex structure agrees with the canonical one. Then*

(i) $\mathscr{V}$ *is a Banach pre-base-norm space.*

(ii) $\mathrm{B}(\mathscr{V})$ *is σ-convex (see Example* 7.3.11*) with respect to the base norm, and the superconvex structure in the assumption coincides with the superconvex structure given by σ-convexity.*

*Proof.* (i) By Proposition 7.2.36 and Corollary 7.3.21, $\mathscr{V}$ is a pre-base-norm space. For a proof that $\mathscr{V}$ is complete, see [223, Lemma 3.2] or [90, Proposition 2.4.11 and Lemma 2.4.12(i)].

(ii) follows by [90, Lemma 2.4.12(i)] and by the fact that any $\sigma$-convex set is a superconvex set; see Example 7.3.11. ∎

We write **CSConv** $\hookrightarrow$ **SConv** $= \mathcal{EM}(\mathcal{D}^\infty)$ for the full subcategory of *cancellative* superconvex sets.

**Proposition 7.3.23.** *The forgetful functor* **CSConv** $\to$ **Conv** *is full, faithful, and injective on objects.*

The fullness means that affine maps between cancellative superconvex set are automatically superaffine. In particular, the category **CSConv** is equal to the category of cancellative superconvex sets and affine maps. The injectivity on objects is called the unique extension theorem in [176] and proved also there.

*Proof.* Suppose that $K$ is equipped with two superconvex structures $\alpha, \alpha' \colon \mathcal{D}^\infty(K) \to K$ that are cancellative, such that $\alpha$ and $\alpha'$ agree on finite convex sums, i.e. are equal on $\mathcal{D}(K) \subseteq \mathcal{D}^\infty(K)$. The restriction $\mathcal{D}(K) \to K$ of $\alpha$ or $\alpha'$ makes $K$ into a cancellative convex set. By Corollary 7.2.13, there exists a semi-base-norm space $(\mathscr{V}, \tau)$ with an isomorphism $\varphi\colon K \xrightarrow{\cong} \mathrm{B}(\mathscr{V})$ in **Conv**. Then both composition $\varphi \circ \alpha \circ \mathcal{D}^\infty(\varphi^{-1})$ and $\varphi \circ \alpha' \circ \mathcal{D}^\infty(\varphi^{-1})$ defines a superconvex structure $\mathcal{D}^\infty(\mathrm{B}(\mathscr{V})) \to \mathrm{B}(\mathscr{V})$ that extends the canonical convex structure of $\mathrm{B}(\mathscr{V})$. By Lemma 7.3.22(ii) it follows that $\varphi \circ \alpha \circ \mathcal{D}^\infty(\varphi^{-1}) = \varphi \circ \alpha' \circ \mathcal{D}^\infty(\varphi^{-1})$, so that $\alpha = \alpha'$. This proves that the forgetful functor **CSConv** $\to$ **Conv** is injective on objects. The faithfulness of **CSConv** $\to$ **Conv** is trivial. The fullness is proved by [90, Lemma 2.4.10]. ∎

**Theorem 7.3.24.** *The following categories are equivalent:*

(i) *The category of Banach pre-base-norm spaces whose bases are $\sigma$-convex, and trace-preserving positive linear maps (i.e.* **pBNS***-morphisms).*

(ii) **CSConv***, the category of cancellative superconvex sets and superaffine maps.*

(iii) **C$\boldsymbol{\sigma}$WMod***, the category of cancellative $\sigma$-weight modules and* **$\boldsymbol{\sigma}$WMod***-morphisms.*

*Proof.* Let $(\mathscr{V}, u)$ be a semi-base-norm space. By Lemma 7.3.22, there exists a superconvex structure that extends the canonical convex structure of $\mathrm{B}(\mathscr{V})$ if and only if $\mathscr{V}$ is a Banach pre-base-norm spaces whose base is $\sigma$-convex. By Proposition 7.3.23 **CSConv** can be identified with the full subcategory of **CConv** consisting of convex sets that can extend to superconvex sets. Therefore one obtains an equivalence of the category in (i) and **CSConv** by restricting the equivalence **sBNS** $\simeq$ **CConv** (Corollary 7.2.13). The equivalence **CSConv** $\simeq$ **C$\boldsymbol{\sigma}$WMod** can be obtained by restricting **SConv** $\simeq$ **$\boldsymbol{\sigma}$WMod** (Theorem 7.3.13) using Lemma 7.2.11. ∎

It turns out that imposing *completeness* yields neat equivalences between convex sets, weight modules, and base-norm spaces.

**Lemma 7.3.25.** *Let $(\mathscr{V}, \tau)$ be a pre-base-norm space. The following are equivalent.*

(i) *$\mathscr{V}$ is a Banach base-norm space.*

(ii) $\mathrm{B}(\mathscr{V})$ *is complete with respect to the base norm.*

(iii) $\mathrm{B}_{\leq}(\mathscr{V})$ *is complete with respect to the base norm.*

*Moreover, if any of the conditions holds,* $\mathrm{B}(\mathscr{V})$ *is $\sigma$-convex.*

*Proof.* (ii) $\Longrightarrow$ (i) and '$\mathrm{B}(\mathscr{V})$ is $\sigma$-convex': Since $\mathrm{B}(\mathscr{V})$ is complete, it is closed. This implies that $\mathscr{V}_+$ is closed by [90, Lemma 2.2.14] (the trace $\tau\colon \mathscr{V} \to \mathbb{R}$ is continuous by Lemma 7.2.42(i)). Therefore $\mathscr{V}$ is a base-norm space. Since $\mathrm{B}(\mathscr{V})$ is convex, bounded, and (sequentially) complete, $\mathrm{B}(\mathscr{V})$ is $\sigma$-convex [176, Example 1.6.ii]. Then $\mathrm{B}(\mathscr{V})$ is superconvex and hence by Lemma 7.3.22(i), $\mathscr{V}$ is complete in the base norm.

(i) $\Longrightarrow$ (iii) $\Longrightarrow$ (ii): By the continuity of the trace $\tau\colon \mathscr{V} \to \mathbb{R}$, it is easy to see that if $\mathscr{V}_+$ is closed, so is $\mathrm{B}_{\leq}(\mathscr{V})$, and if $\mathrm{B}_{\leq}(\mathscr{V})$ is closed, so is $\mathrm{B}(\mathscr{V})$. Then we are done, since each complete subset of a metric space is closed, and each closed subset of a complete metric space is complete. ■

We say that a metric (super)convex set or a metric ($\sigma$-)weight module is **complete** if it is complete with respect to the canonical base metric. Then we obtain the following equivalences of categories. This extends the result of Pumplün [223] which showed the equivalences **BBNS** $\simeq$ **CMConv** $\simeq$ **CMSConv**.

**Theorem 7.3.26.** *The following categories are all equivalent.*

(i) **BBNS**, *the category of Banach base-norm spaces and trace-preserving positive linear maps.*

(ii) **CMConv**, *the category of complete metric convex sets and affine maps.*

(iii) **CMSConv**, *the category of complete metric superconvex sets and superaffine maps.*

(iv) **CMWMod**, *the category of complete metric weight modules and* **WMod**-*morphisms.*

(v) **CM$\boldsymbol{\sigma}$WMod**, *the category of complete metric $\sigma$-weight modules and* **$\boldsymbol{\sigma}$WMod**-*morphisms.*

*Proof.* Recall from Corollary 7.2.38 that one has the equivalences **pBNS** $\simeq$ **MConv** $\simeq$ **MWMod**. Let $(\mathscr{V}, \tau)$ be a pre-base-norm space. Then the base metric on $\mathrm{B}(\mathscr{V})$ and $\mathrm{B}_{\leq}(\mathscr{V})$ respectively as a metric convex set and a metric weight module coincide with the metric given by the base norm of $\mathscr{V}$ — metric convex sets and weight modules are defined so that it holds — see Propositions 7.2.28 and 7.2.33. From this and Lemma 7.3.25, it follows that one can restrict the equivalences to **BBNS** $\simeq$ **CMConv** $\simeq$ **CMWMod**. Moreover, again by Lemma 7.3.25, restricting the equivalences of Theorem 7.3.24, one obtains **BBNS** $\simeq$ **CMSConv** $\simeq$ **CM$\boldsymbol{\sigma}$WMod**. ■

### 7.3.3 Definition and examples of $\sigma$-effectuses

We now introduce *$\sigma$-effectuses*, an extension of effectuses with countable sums. It is defined based on Arbib and Manes' *partially $\sigma$-additive categories* (*$\sigma$-PACs*) [7, 8, 202], which are usually simply called *partially additive categories* (*PACs*). We add '$\sigma$' to avoid confusion with finPACs.

We say that a category $\mathbf{C}$ is **enriched over $\sigma$-PAMs** if each homset $\mathbf{C}(A, B)$ is a $\sigma$-PAM and each composition map $\circ\colon \mathbf{C}(B, C) \times \mathbf{C}(A, B) \to \mathbf{C}(A, C)$ is $\sigma$-biadditive.

**Definition 7.3.27.** A **partially $\sigma$-additive category** (**$\sigma$-PAC**, for short) is a category with countable coproducts that is enriched over $\sigma$-PAMs satisfying the following two axioms.

**(Compatible sum axiom)** If a countable family $(f_j\colon A \to B)_{j\in J}$ is compatible, then it is summable.

**(Untying axiom)** If $f, g\colon A \to B$ are summable, then $\kappa_1 \circ f, \kappa_2 \circ g\colon A \to B + B$ are summable too.

Some examples of $\sigma$-PACs are given below in Examples 7.3.33–7.3.36 as $\sigma$-effectuses. Other examples can be found in e.g. [7, 113, 202].

We recall basic results on $\sigma$-PAC.

**Proposition 7.3.28.** *Let $\mathbf{C}$ be a $\sigma$-PAC.*

(i) *The partial projections $\rhd_j\colon \coprod_j A_j \to A_j$ from a countable coproduct are jointly monic.*

(ii) *A countable family $(f_j\colon A \to B)_{j\in J}$ of morphisms is summable if and only if it is compatible, i.e. there exists a morphism $f\colon A \to J \cdot A$ such that $f_j = \rhd_j \circ f$ for all $j \in J$.*

(iii) *If a family $(f_j\colon A \to B)_{j\in J}$ is summable and hence compatible, say via $f\colon A \to J \cdot A$, then $\bigovee_{j\in J} f_j = \nabla \circ f$, where $\nabla\colon J \cdot A \to A$ is the codiagonal.* ■

*Proof.* For the proof we refer to [7, § 3] or [202, § 3.2], though the arguments are basically the same as the case of finPACs in Section 3.1. ■

**Definition 7.3.29.** A $\sigma$-effectus is an effectus $(\mathbf{C}, I)$ that is at the same time a $\sigma$-PAC.

We do not explicitly require that the effectus structure and the $\sigma$-PAC structure are compatible. The compatibility, however, follows automatically, since by Proposition 7.3.28 the $\sigma$-PAM structure of a $\sigma$-PAC is determined by its categorical structure. Explicitly, for each $\sigma$-effectus $\mathbf{C}$, the underlying PCM structure of the $\sigma$-PAC $\mathbf{C}$ agrees with the PCM structure of the effectus $\mathbf{C}$. We note some immediate consequences.

**Lemma 7.3.30.** *Let $\mathbf{C}$ be a $\sigma$-effectus. Let $(f\colon A \to B)_{j\in J}$ be a countable family of morphisms. Then $(f_j)_{j\in J}$ is summable in $\mathbf{C}(A, B)$ if and only if $(\mathbb{1} f_j)_{j\in J}$ is summable in* $\mathrm{Pred}(A) = \mathbf{C}(A, I)$.

*Proof.* By the following equivalences.

| | |
|---|---|
| $(f_j)_{j\in J}$ is summable | |
| $\iff$ $(f_j)_{j\in F}$ is summable for all finite $F \subseteq J$ | (by axioms of $\sigma$-PAM) |
| $\iff$ $(\mathbb{1} f_j)_{j\in F}$ is summable for all finite $F \subseteq J$ | (by Lemma 3.2.4(ii)) |
| $\iff$ $(\mathbb{1} f_j)_{j\in J}$ is summable | (by axioms of $\sigma$-PAM) ■ |

**Proposition 7.3.31.** *Let* $\mathbf{C}$ *be a* $\sigma$*-effectus. Then for each* $A \in \mathbf{C}$*, the set of predicates* $\mathrm{Pred}(A) = \mathbf{C}(A, I)$ *is a* $\sigma$*-effect algebra. Moreover* $A \mapsto \mathrm{Pred}(A)$ *extends to a functor* $\mathrm{Pred}\colon \mathbf{C}^{\mathrm{op}} \to \boldsymbol{\sigma}\mathbf{EA}_{\leq}$*.*

*Proof.* By the definition of an effectus, predicates $\mathrm{Pred}(A) = \mathbf{C}(A, I)$ form an effect algebra. Since $\mathbf{C}$ is a $\sigma$-PAC, there is a $\sigma$-PAM structure on $\mathbf{C}(A, I)$ that extends the PCM structure. Therefore $\mathrm{Pred}(A) = \mathbf{C}(A, I)$ is a $\sigma$-effect algebra by Proposition 7.3.6. The predicate functor $\mathrm{Pred}\colon \mathbf{C}^{\mathrm{op}} \to \mathbf{EA}_{\leq}$ restricts to $\mathrm{Pred}\colon \mathbf{C}^{\mathrm{op}} \to \boldsymbol{\sigma}\mathbf{EA}_{\leq}$ since each predicate transformer $f^*\colon \mathrm{Pred}(B) \to \mathrm{Pred}(A)$ preserves countable sums by the $\sigma$-PAM enrichment. ■

An important consequence from the proposition is that the set of predicates $\mathrm{Pred}(A) = \mathbf{C}(A, I)$ in a $\sigma$-effectus forms a $\omega$-complete poset, see Lemma 7.3.3. Conversely, recall from Lemma 7.3.5 that a suitable PCM with $\omega$-completeness yields a $\sigma$-PAM. We can extend the result and obtain a convenient sufficient condition for an effectus to be a $\sigma$-effectus.

**Lemma 7.3.32.** *Let* $\mathbf{C}$ *be an effectus such that*

(i) $\mathbf{C}$ *has countable coproducts;*

(ii) *every hom-PCM* $\mathbf{C}(A, B)$ *is cancellative;*

(iii) $\mathbf{C}$ *is enriched over* $\omega$*-cpos with respect to the algebraic order — in other words: the algebraic order on each homset* $\mathbf{C}(A, B)$ *is* $\omega$*-complete, and the composition* $\circ$ *of morphisms is* $\omega$*-continuous in each argument separately.*

*Then* $\mathbf{C}$ *is a* $\sigma$*-PAC, and hence a* $\sigma$*-effectus.*

*Proof.* By Lemma 3.2.6 and assumption (ii), each homset $\mathbf{C}(A, B)$ is a cancellative positive PCM. Therefore by Lemmas 7.3.3 and 7.3.5, each homset $\mathbf{C}(A, B)$ forms a $\sigma$-PAM, with countable sums defined as suprema of finite sums. Moreover by Proposition 7.3.8, the composition $\circ$ is $\sigma$-biadditive, so that $\mathbf{C}$ is enriched over $\sigma$-PAMs. It only remains to prove that the compatible sum axiom holds for countable families. Let $(f_j\colon A \to B)_{j \in J}$ be a compatible family, say via $f\colon A \to J \cdot B$. For each finite subset $F \subseteq J$, the family $(f_j\colon A \to B)_{j \in F}$ is compatible via

$$A \xrightarrow{f} J \cdot B \xrightarrow{\rhd_F} F \cdot B$$

where $\rhd_F$ is defined by

$$\rhd_F \circ \kappa_j = \begin{cases} \mathrm{id} & \text{if } j \in F \\ 0 & \text{if } j \notin F\,. \end{cases}$$

By Lemma 3.1.5, $(f_j\colon A \to B)_{j \in F}$ is summable. Therefore the family $(f_j\colon A \to B)_{j \in J}$ is summable by the limit axiom in $\mathbf{C}(A, B)$. ■

We give examples of $\sigma$-effectuses.

**Example 7.3.33.** The category **Pfn** of sets and partial functions is a prototypical example of a $\sigma$-PAC, see [7, 202]. Hence **Pfn** is a $\sigma$-effectus. A countable family $(f_j\colon X \rightharpoonup Y)_j$ of partial functions is summable iff the domains of definition $\mathbb{1} f_j \subseteq X$ are pairwise disjoint. The sum $\bigovee_j f_j\colon X \rightharpoonup Y$ is then defined in a similar manner to the finitary case, see § 3.3.1.

**Example 7.3.34.** It was proved by Panangaden [214] and Haghverdi [113] that the Kleisli category $\mathcal{K}\ell(\mathcal{G}_{\leq})$ of the subprobability Giry monad — called the category **SRel** of stochastic relations in [113, 214] — is a $\sigma$-PAC. For the proof we refer to [214], [215, Chapter 5], or [113, § 3.2]. Therefore $\mathcal{K}\ell(\mathcal{G}_{\leq})$ is a $\sigma$-effectus. A countable family $(f_j \colon X \to \mathcal{G}_{\leq}(Y))_j$ is summable iff $\sum_j f_j(x)(Y) \leq 1$ for all $x \in X$. In that case one defines the sum $\bigovee_j f_j$ by $(\bigovee_j f_j)(x)(U) = \sum_j f_j(x)(U)$, for $U \in \Sigma_Y$.

**Example 7.3.35.** As a discrete version of the previous example, the Kleisli category $\mathcal{K}\ell(\mathcal{D}^{\infty}_{\leq})$ of the infinite subdistribution monad is also a $\sigma$-PAC. This was shown in [135] as an example of a general result about 'partially additive' monads [135, Proposition 4.8]. Since some details are omitted in [135], for the sake of completeness we give another argument to prove that $\mathcal{K}\ell(\mathcal{D}^{\infty}_{\leq})$ is a $\sigma$-PAC. It is well known (see e.g. [63, 119, 135]), and also straightforward to show, that the category $\mathcal{K}\ell(\mathcal{D}^{\infty}_{\leq})$ is enriched over $\omega$-cpos, via the usual pointwise ordering. The category $\mathcal{K}\ell(\mathcal{D}^{\infty}_{\leq})$ is an effectus, and moreover the following hold:

- It has countable coproducts, since any Kleisli category inherits all coproducts that exists in the base category (**Set**, in this case).
- Clearly, each hom-PCM is cancellative.
- It is enriched over $\omega$-cpos with respect to the algebraic order, since the algebraic order coincides with the pointwise order.

Therefore by Lemma 7.3.32, we conclude that $\mathcal{K}\ell(\mathcal{D}^{\infty}_{\leq})$ is a $\sigma$-PAC, and hence a $\sigma$-effectus. A countable family $(f_j \colon X \to \mathcal{D}^{\infty}_{\leq}(Y))_j$ is summable iff $\sum_j \sum_y f_j(x)(y) \leq 1$ for all $x \in X$. In that case, the sum $\bigovee_j f_j$ is defined pointwise: $(\bigovee_j f_j)(x)(y) = \sum_j f_j(x)(y)$.

We note that the Kleisli category $\mathcal{K}\ell(\mathcal{D}_{\leq})$ of the finite subdistribution monad is *not* a $\sigma$-PAC. To see this, recall from Example 7.2.39(iii) that $\mathcal{D}(X)$ is the base of the base-norm space $\ell^1_{c,\mathbb{R}}(X)$. If $\mathcal{K}\ell(\mathcal{D}_{\leq})$ is a $\sigma$-PAC, then it follows that $\mathrm{St}(X) \cong \mathcal{D}(X)$ is a superconvex set (see Theorem 7.3.37(ii) below), and by Lemma 7.3.22 that $\ell^1_{c,\mathbb{R}}(X)$ is complete with respect to the base norm ($= \ell^1$-norm). This is not the case, e.g. for $X = \mathbb{N}$.

**Example 7.3.36.** The opposite $\mathbf{Wstar}^{\mathrm{op}}_{\leq}$ of the category of $W^*$-algebras and subunital normal CP maps is a $\sigma$-PAC. To the best of the author's knowledge this fact was first proved by the author and remarked in his article [37, Remark 4.3]. Since the proof was omitted there, we will give a proof here. In fact, it is a direct consequence of Lemma 7.3.32 combined with the results shown in [37]:

- $\mathbf{Wstar}_{\leq}$ has (arbitrary) products [37, Proposition 2.10]. Hence $\mathbf{Wstar}^{\mathrm{op}}_{\leq}$ has countable coproducts.
- $\mathbf{Wstar}_{\leq}$ is enriched over dcpos with respect to the order $f \leq g \iff g - f$ is CP [37, Theorem 4.3]. Thus so is the opposite $\mathbf{Wstar}^{\mathrm{op}}_{\leq}$, and in particular it is enriched over over $\omega$-cpos. Clearly the order here coincides with the algebraic order (see Remark 3.3.1).

It is clear that each hom-PCM $\mathbf{Wstar}^{\mathrm{op}}_{\leq}(\mathscr{A}, \mathscr{B})$ is cancellative. By Lemma 7.3.32, $\mathbf{Wstar}^{\mathrm{op}}_{\leq}$ is a $\sigma$-PAC. Therefore $\mathbf{Wstar}^{\mathrm{op}}_{\leq}$ is a $\sigma$-effectus.

We note that $\mathbf{Cstar}_{\leq}^{\mathrm{op}}$ is *not* a $\sigma$-PAC. To see this, by Proposition 7.3.31, it is enough to show that predicates in $\mathbf{Cstar}_{\leq}^{\mathrm{op}}$, i.e. effects $[0,1]_{\mathscr{A}}$ in a $C^*$-algebra, do not always form a $\omega$-complete effect algebra. Concretely, take the $C^*$-algebra $\mathrm{C}([0,1])$ of continuous functions $\varphi\colon [0,1] \to \mathbb{C}$. Then $[0,1]_{\mathrm{C}([0,1])}$ is the effect algebra of continuous functions $\varphi\colon [0,1] \to [0,1]$, where the order is the usual one. Clearly, it is not $\omega$-complete.

### 7.3.4 (Sub)states and predicates in a real $\sigma$-effectus

We will investigate the structure of (sub)states and predicates in a $\sigma$-effectus. For simplicity, we will restrict ourselves to *real* $\sigma$-effectuses, i.e. $\sigma$-effectuses $\mathbf{C}$ with $\mathbf{C}(I,I) \cong [0,1]$, see Definition 3.4.7. It still covers our most important examples, namely effectuses for probability and quantum theory.

**Theorem 7.3.37.** *Let* $\mathbf{C}$ *be a real* $\sigma$*-effectus.*

(i) *For each* $A \in \mathbf{C}$*, predicates* $\mathrm{Pred}(A) = \mathbf{C}(A,I)$ *form a* $\sigma$*-effect module, yielding a functor* $\mathrm{Pred}\colon \mathbf{C} \to \boldsymbol{\sigma}\mathbf{EMod}_{\leq}^{\mathrm{op}}$*. By restricting it to total maps, one also has* $\mathrm{Pred}\colon \mathrm{Tot}(\mathbf{C}) \to \boldsymbol{\sigma}\mathbf{EMod}^{\mathrm{op}}$*.*

(ii) *For each* $A \in \mathbf{C}$*, substates* $\mathrm{St}_{\leq}(A) = \mathbf{C}(I,A)$ *form a* $\sigma$*-weight module, yielding a functor* $\mathrm{St}_{\leq}\colon \mathbf{C} \to \boldsymbol{\sigma}\mathbf{WMod}_{\leq}$*. By restricting it to total maps, one also has* $\mathrm{St}_{\leq}\colon \mathrm{Tot}(\mathbf{C}) \to \boldsymbol{\sigma}\mathbf{WMod}$*.*

(iii) *For each* $A \in \mathbf{C}$*, states* $\mathrm{St}(A) = \mathrm{Tot}(\mathbf{C})(I,A)$ *form a superconvex set, yielding a functor* $\mathrm{St}\colon \mathrm{Tot}(\mathbf{C}) \to \mathbf{SConv}$*.*

*Proof.* (i) Recall from Proposition 3.4.5 that one has a functor $\mathrm{Pred}\colon \mathbf{C} \to \mathbf{EMod}_{\leq}^{\mathrm{op}}$. As we saw in Proposition 7.3.31, each $\mathrm{Pred}(A) = \mathbf{C}(A,I)$ is a $\sigma$-effect algebra. Since the action $[0,1] \times \mathrm{Pred}(A) \to \mathrm{Pred}(A)$ is defined via the composition in the $\sigma$-PAM-enriched category $\mathbf{C}$, it is $\sigma$-biadditive. Therefore $\mathrm{Pred}(A)$ is a $\sigma$-effect module. Predicate transformers $f^*\colon \mathrm{Pred}(B) \to \mathrm{Pred}(A)$ are defined via composition, hence they preserves countable sums and are morphisms in $\boldsymbol{\sigma}\mathbf{EMod}_{\leq}$. Therefore one has a functor $\mathrm{Pred}\colon \mathbf{C} \to \mathbf{EMod}_{\leq}^{\mathrm{op}}$. Clearly it can be restricted to $\mathrm{Pred}\colon \mathrm{Tot}(\mathbf{C}) \to \mathbf{EMod}^{\mathrm{op}}$.

(ii) The proof is similar to the previous one, based on Proposition 3.5.4. Note that Lemma 7.3.30 implies that the weight map $|{-}|\colon \mathrm{St}_{\leq}(A) \to [0,1]$ reflects countable summability.

(iii) The set of states is equal to the base of the weight module of substates, i.e. $\mathrm{St}(A) = \mathrm{Tot}(\mathbf{C})(I,A) = \mathrm{B}(\mathrm{St}_{\leq}(A))$. By the previous point, $\mathrm{St}_{\leq}(A)$ is a $\sigma$-effect module, and hence by Theorem 7.3.13, the states $\mathrm{St}(A)$ form a superconvex set. To see the functoriality, note that the state functor is precisely the following composite:

$$\mathrm{St} = \Big(\mathrm{Tot}(\mathbf{C}) \xrightarrow{\mathrm{St}_{\leq}} \boldsymbol{\sigma}\mathbf{WMod} \underset{\simeq}{\xrightarrow{\mathrm{B}}} \mathbf{SConv}\Big)\,. \qquad \blacksquare$$

Our next aim is to establish state-and-effect triangles for $\sigma$-effectuses. To this end, we will first prove that both $\boldsymbol{\sigma}\mathbf{EMod}_{\leq}^{\mathrm{op}}$ and $\boldsymbol{\sigma}\mathbf{WMod}_{\leq}$ are $\sigma$-effectuses, and then prove that they form a dual adjunctions via the substate and predicate functors.

**Lemma 7.3.38.** *An effectus* $\mathbf{C}$ *is a partially* $\sigma$*-additive category, hence a* $\sigma$*-effectus, if (and only if) the following conditions hold.*

(i) $\mathbf{C}$ *has countable coproducts.*

(ii) *For each object* $A$ *and each countable set* $J$*, the partial projections* $\rhd_j\colon J\cdot A\to A$ *from the copower of* $A$ *by* $J$ *(i.e. the* $J$*-fold coproduct) are jointly monic.*

(iii) *Let* $(f_j\colon A\to B)_{j\in J}$ *be a countable family of parallel morphisms. If the family* $(f_j\colon A\to B)_{j\in F}$ *is compatible for each finite subset* $F\subseteq J$*, then* $(f_j\colon A\to B)_{j\in J}$ *is compatible.*

*Proof.* If an effectus $\mathbf{C}$ satisfies (i) and (ii), then we can define a partial sum operation on countable families $(f_j\colon A\to B)_{j\in J}$ by: sum $\bigovee_{j\in J} f_j$ is defined when $(f_j)_j$ is compatible; and in that case $\bigovee_{j\in J} f_j=\nabla\circ f$, where $f\colon A\to J\cdot B$ is a unique morphism that witnesses compatibility of $(f_j)_j$. Note that composition, in each argument, preserves the partial sums defined in this way. Thus to show that $\mathbf{C}$ is a $\sigma$-PAC, we only need to check that each homset $\mathbf{C}(A,B)$ is a $\sigma$-PAM. Note that (iii) says that the limit axiom holds. Since the countable partial sum $\bigovee$ extends the PCM structure of $\mathbf{C}$, it suffices to show that it the 'weak partition-associativity axiom' in Lemma 7.3.2 holds. Let $(f_j\colon A\to B)_{j\in J}$ be a summable family, compatible via $f\colon A\to J\cdot B$. Let $\bigoplus_{k\in K} J_k=J$ be a partition. Then for each $k\in K$, $(f_j\colon A\to B)_{j\in J_k}$ is compatible via

$$A\xrightarrow{f} J\cdot B\xrightarrow{\rhd_{J_k}} J_k\cdot B$$

where $\rhd_{J_k}$ is the obvious projection map. Moreover, the family $(\bigovee_{j\in J_k} f_j\colon A\to B)_{k\in K}$ is compatible via

$$A\xrightarrow{f} J\cdot B\cong\coprod_{k\in K} J_k\cdot B\xrightarrow{\coprod_k\nabla}\coprod_{k\in K} B=K\cdot B\,.$$

It is then straightforward to verify $\bigovee_{j\in J} f_j=\bigovee_{k\in K}\bigovee_{j\in J_k} f_j$. ■

**Proposition 7.3.39.** *The opposite* $\boldsymbol{\sigma}\mathbf{EMod}_{\le}^{\mathrm{op}}$ *of the category of* $\sigma$*-effect modules and subunital maps is a* $\sigma$*-effectus.*

*Proof.* Note that $\boldsymbol{\sigma}\mathbf{EMod}_{\le}^{\mathrm{op}}$ is a subcategory of the effectus $\mathbf{EMod}_{\le}^{\mathrm{op}}$. Thus by Proposition 3.8.7, one proves that $\boldsymbol{\sigma}\mathbf{EMod}_{\le}^{\mathrm{op}}$ (with $[0,1]$ as unit) is an effectus by checking that $\boldsymbol{\sigma}\mathbf{EMod}_{\le}^{\mathrm{op}}$ is an 'sub-effectus' of $\mathbf{EMod}_{\le}^{\mathrm{op}}$. Next we invoke (iii) to show that $\boldsymbol{\sigma}\mathbf{EMod}_{\le}^{\mathrm{op}}$ is a $\sigma$-PAC. The category has countable coproducts since $\boldsymbol{\sigma}\mathbf{EMod}_{\le}$ has products, given by cartesian products $\prod_j E_j$ with the operations defined pointwise. Let $J$ be a countable set. The partial projections $\rhd_j\colon J\cdot E\to E$ in $\boldsymbol{\sigma}\mathbf{EMod}_{\le}^{\mathrm{op}}$ are morphisms $\rhd_j\colon E\to E^J$ in $\boldsymbol{\sigma}\mathbf{EMod}_{\le}$ that send $x\in E$ to the $J$-tuple that has 0 at every coordinate except $x$ at the $j$th coordinate. If $f,g\colon E^J\to J$ in $\boldsymbol{\sigma}\mathbf{EMod}_{\le}$ satisfy $f\circ\rhd_j=g\circ\rhd_j$ for all $j\in J$, then

$$f((x_j)_j)=f(\textstyle\bigovee_j\rhd_j(x_j))=\bigovee_j f(\rhd_j(x_j))=\bigovee_j g(\rhd_j(x_j))=\cdots=g((x_j)_j)\,.$$

Therefore the maps $\rhd_j$ are jointly epic in $\boldsymbol{\sigma}\mathbf{EMod}_{\le}$ and hence jointly monic in the opposite. We have shown that $\boldsymbol{\sigma}\mathbf{EMod}_{\le}^{\mathrm{op}}$ satisfies (i) and (ii) of Lemma 7.3.38. Finally we prove that a countable family $(f_j\colon E \to D)_{j\in J}$ in $\boldsymbol{\sigma}\mathbf{EMod}_{\leq}^{\mathrm{op}}$ is compatible if and only if $(f_j(1))_{j\in J}$ is summable in $E$. It implies, by the limit axiom in $E$, that (iii) of Lemma 7.3.38 holds. Let $(f_j)_{j\in J}$ be a compatible family. Then in $\boldsymbol{\sigma}\mathbf{EMod}_{\leq}$, there exists a map $f\colon D^J \to E$ such that $f \circ \rhd_j = f_j$. Since $(1)_{j\in J} \in D^J$ can be written as $\bigovee_{j\in J} \rhd_j(1)$, it follows that the sum $\bigovee_{j\in J} f_j(1) = f(\bigovee_{j\in J} \rhd_j(1))$ is defined. Conversely, if $(f_j(1))_{j\in J}$ is summable, define a map $\langle\!\langle f_j\rangle\!\rangle_j\colon D^J \to E$ by $\langle\!\langle f_j\rangle\!\rangle_j((a_j)_j) = \bigovee_j f_j(a_j)$. Then $\langle\!\langle f_j\rangle\!\rangle_j\colon D^J \to E$ is a morphism in $\boldsymbol{\sigma}\mathbf{EMod}_{\leq}$ (cf. the proof of Proposition 3.4.10 in § 3.8.3). For example, it preserves the scalar multiplication:

$$\begin{aligned}\langle\!\langle f_j\rangle\!\rangle_j(r\cdot(a_j)_j) &= \langle\!\langle f_j\rangle\!\rangle_j((r\cdot a_j)_j)\\ &= \textstyle\bigovee_j f_j(r\cdot a_j)\\ &= \textstyle\bigovee_j r\cdot f_j(a_j)\\ &= r\cdot \textstyle\bigovee_j f_j(a_j)\\ &= r\cdot \langle\!\langle f_j\rangle\!\rangle_j((a_j)_j)\,.\end{aligned}$$

Then $(f_j)_{j\in J}$ is compatible via $\langle\!\langle f_j\rangle\!\rangle_j$. ■

**Proposition 7.3.40.** *The category $\boldsymbol{\sigma}\mathbf{WMod}_{\leq}$ of σ-weight modules and weight-decreasing maps is a σ-effectus.*

*Proof.* First we show that $\boldsymbol{\sigma}\mathbf{WMod}_{\leq}$ has countable coproducts. For a countable family $(X_\lambda)_{\lambda\in\Lambda}$ of objects, we define the underlying set by

$$\coprod_{\lambda\in\Lambda} X_\lambda = \Big\{(x_\lambda)_\lambda \in \prod_{\lambda\in\Lambda} X_\lambda \;\Big|\; \sum_{\lambda\in\Lambda}|x_\lambda| \leq 1\Big\}$$

and the weight of $(x_\lambda)_\lambda \in \coprod_{\lambda\in\Lambda} X_\lambda$ by $|(x_\lambda)_\lambda| = \sum_{\lambda\in\Lambda}|x_\lambda|$. The weight determines summability in $\coprod_{\lambda\in\Lambda} X_\lambda$. We then define partial σ-addition and $[0,1]$-action pointwise. In much the same way as Lemma 3.5.7, one can verify that $\coprod_{\lambda\in\Lambda} X_\lambda$ is a σ-weight module, and that it is a coproduct with coprojections $\kappa_\lambda\colon X_\lambda \to \coprod_{\lambda\in\Lambda} X_\lambda$ that sends each element $x \in X_\lambda$ to the Λ-tuple with 0 everywhere except $x$ at the λth coordinate.

There are obvious zero maps $0\colon X \to Y$ that sends everything to zero. Hence there are partial projections $\rhd_\lambda\colon \coprod_\lambda X_\lambda \to X_\lambda$, which are given by $\rhd_\lambda((x_\lambda)_\lambda) = x_\lambda$. It is clear that the partial projections $\rhd_\lambda\colon \coprod_\lambda X_\lambda \to X_\lambda$ are jointly monic for each countable $J$.

Let $(f_j\colon X \to Y)_{j\in J}$ be a countable family of morphisms in $\boldsymbol{\sigma}\mathbf{WMod}_{\leq}$. We claim that $(f_j)_j$ is compatible if and only if $\sum_j |f_j(x)| \leq |x|$ for all $x \in X$, which implies that condition (iii) in Lemma 7.3.38 holds. Suppose that $(f_j)_j$ is compatible via $f\colon X \to J\cdot Y$. Then for each $x \in X$, one has $\rhd_j(f(x)) = f_j(x)$, and thus by definition of $\rhd_j$, we have $f(x) = (f_j(x))_j$. As $f$ is weight-decreasing,

$$|x| \geq |f(x)| = |(f_j(x))_j| = \sum\nolimits_j |f_x(x)|\,.$$

Conversely, if $\sum_j |f_j(x)| \leq |x|$ for all $x \in X$, then we can show that the map $f\colon X \to J\cdot Y$ given by $f(x) = (f_j(x))_j$ is a well-defined morphism in $\boldsymbol{\sigma}\mathbf{WMod}_{\leq}$ and that $(f_j)_{j\in J}$ is compatible via $f$.

We have shown that conditions (i)–(iii) of Lemma 7.3.38 hold. Therefore it only remains to show that $\boldsymbol{\sigma}\mathbf{WMod}_{\leq}$ is an effectus. The unit object is $[0,1]$ and the truth maps $\mathbb{1}_X \colon X \to [0,1]$ are defined as weight: $\mathbb{1}_X(x) = |x|$. Then it is straightforward to check that conditions (E′1)–(E′4) in Proposition 3.8.6 hold (cf. the proof of Proposition 3.5.9 in § 3.8.3). ∎

**Theorem 7.3.41.** *There is the following dual adjunction, given by 'homming into* $[0,1]$*':*

$$\boldsymbol{\sigma}\mathbf{EMod}_{\leq}^{\mathrm{op}} \underset{\mathrm{Hom}(-,[0,1])}{\overset{\mathrm{Hom}(-,[0,1])}{\rightleftarrows}} \boldsymbol{\sigma}\mathbf{WMod}_{\leq} \quad (\top)$$

*Proof.* The two functors are well-defined since they are the substate and predicate functor for $\sigma$-effectuses:

$$\begin{aligned} \boldsymbol{\sigma}\mathbf{EMod}_{\leq}(-,[0,1]) &= \boldsymbol{\sigma}\mathbf{EMod}_{\leq}^{\mathrm{op}}([0,1],-) = \mathrm{St}_{\leq} \colon \boldsymbol{\sigma}\mathbf{EMod}_{\leq}^{\mathrm{op}} \to \boldsymbol{\sigma}\mathbf{WMod}_{\leq} \\ \boldsymbol{\sigma}\mathbf{WMod}_{\leq}(-,[0,1]) &= \mathrm{Pred} \colon \boldsymbol{\sigma}\mathbf{WMod}_{\leq} \to \boldsymbol{\sigma}\mathbf{EMod}_{\leq}^{\mathrm{op}} \end{aligned}$$

The adjunction amounts to the following bijective correspondence

$$\frac{X \xrightarrow{f} \boldsymbol{\sigma}\mathbf{EMod}_{\leq}(E,[0,1]) \quad \text{in } \boldsymbol{\sigma}\mathbf{WMod}_{\leq}}{E \xrightarrow{g} \boldsymbol{\sigma}\mathbf{WMod}_{\leq}(X,[0,1]) \text{ in } \boldsymbol{\sigma}\mathbf{EMod}_{\leq}}$$

given by 'swapping arguments': $f(x)(a) = g(a)(x)$. The verification of the correspondence is straightforward, in a similar manner to Proposition 3.7.1. ∎

Combining Proposition 7.3.31 and Theorem 7.3.41, we obtain state-and-effect triangles for $\sigma$-effectuses.

**Corollary 7.3.42.** *For each $\sigma$-effectus* $\mathbf{C}$ *with* $\mathbf{C}(I,I) \cong [0,1]$*, one has the following state-and-effect triangle:*

$$\begin{array}{ccc} \boldsymbol{\sigma}\mathbf{EMod}_{\leq}^{\mathrm{op}} & \underset{\mathrm{Hom}(-,[0,1])}{\overset{\mathrm{Hom}(-,[0,1])}{\rightleftarrows}} \ (\top) & \boldsymbol{\sigma}\mathbf{WMod}_{\leq} \\ & {\scriptstyle \mathbf{C}(-,I) = \mathrm{Pred}} \nwarrow \quad \nearrow {\scriptstyle \mathrm{St}_{\leq} = \mathbf{C}(I,-)} & \\ & \mathbf{C} & \end{array}$$

∎

We will also give a version of state-and-effect triangles using the state functor and the category of superconvex sets.

**Proposition 7.3.43.** *There is the following dual adjunction, given by 'homming into* $[0,1]$*':*

$$\boldsymbol{\sigma}\mathbf{EMod}^{\mathrm{op}} \underset{\mathrm{Hom}(-,[0,1])}{\overset{\mathrm{Hom}(-,[0,1])}{\rightleftarrows}} \mathbf{SConv} \quad (\top)$$

*Proof.* The adjunction can be obtained as the following composition:

$$\boldsymbol{\sigma}\mathbf{EMod}^{\mathrm{op}} \underset{\boldsymbol{\sigma}\mathbf{WMod}_{\leq}(-,[0,1])}{\overset{\boldsymbol{\sigma}\mathbf{EMod}_{\leq}(-,[0,1])}{\rightleftarrows}}_{\top} \boldsymbol{\sigma}\mathbf{WMod} \underset{\mathcal{L}}{\overset{\mathrm{B}}{\rightleftarrows}}_{\simeq} \mathbf{SConv}$$

Here the adjunction on the left is the one from Theorem 7.3.41 restricted to the subcategory of unital/weight-preserving maps, and the adjoint equivalence on the right is from Theorem 7.3.13. We check that the composition gives the desired hom functors. For one composite, we have equality:

$$\boldsymbol{\sigma}\mathbf{EMod}(E,[0,1]) = \mathrm{B}(\boldsymbol{\sigma}\mathbf{EMod}_{\leq}(E,[0,1]))\,.$$

For the other composite, it is not hard to see that each $\mathbf{SConv}$-morphism $f\colon K \to [0,1]$ extends uniquely to a $\boldsymbol{\sigma}\mathbf{WMod}_{\leq}$-morphism $\overline{f}\colon \mathcal{L}(K) \to [0,1]$, yielding a natural isomorphism:

$$\mathbf{SConv}(K,[0,1]) \cong \boldsymbol{\sigma}\mathbf{WMod}_{\leq}(\mathcal{L}(K),[0,1])\,. \qquad \blacksquare$$

**Corollary 7.3.44.** *For each $\sigma$-effectus $\mathbf{C}$ with $\mathbf{C}(I,I) \cong [0,1]$, one has the following state-and-effect triangle:*

$$\begin{array}{ccc}
\boldsymbol{\sigma}\mathbf{EMod}^{\mathrm{op}} & \underset{\mathrm{Hom}(-,[0,1])}{\overset{\mathrm{Hom}(-,[0,1])}{\rightleftarrows}}_{\top} & \mathbf{SConv} \\
{\scriptstyle \mathbf{C}(-,I)\,=\,\mathrm{Pred}} \nwarrow & & \nearrow {\scriptstyle \mathrm{St}\,=\,\mathrm{Tot}(\mathbf{C})(I,-)} \\
& \mathrm{Tot}(\mathbf{C}) &
\end{array}$$

$\blacksquare$

**Remark 7.3.45.** In [142] Jacobs showed that the adjunctions

$$\mathbf{DcEMod}^{\mathrm{op}} \underset{\mathrm{Hom}(-,[0,1])}{\overset{\mathrm{Hom}(-,[0,1])}{\rightleftarrows}}_{\top} \mathbf{Set} \qquad \boldsymbol{\sigma}\mathbf{EMod}^{\mathrm{op}} \underset{\mathrm{Hom}(-,[0,1])}{\overset{\mathrm{Hom}(-,[0,1])}{\rightleftarrows}}_{\top} \mathbf{Meas}$$

yield the monads $\mathcal{D}^{\infty}$ on $\mathbf{Set}$ and $\mathcal{G}$ on $\mathbf{Meas}$, respectively. He then obtained the following adjunctions as a consequence of a general result on monads [142, Theorem 1].

$$\mathbf{DcEMod}^{\mathrm{op}} \underset{\mathrm{Hom}(-,[0,1])}{\overset{\mathrm{Hom}(-,[0,1])}{\rightleftarrows}}_{\top} \mathcal{EM}(\mathcal{D}^{\infty}) \equiv \mathbf{SConv} \qquad \boldsymbol{\sigma}\mathbf{EMod}^{\mathrm{op}} \underset{\mathrm{Hom}(-,[0,1])}{\overset{\mathrm{Hom}(-,[0,1])}{\rightleftarrows}}_{\top} \mathcal{EM}(\mathcal{G})$$

Then one might ask: can the adjunction $\boldsymbol{\sigma}\mathbf{EMod}^{\mathrm{op}} \rightleftarrows \mathbf{SConv}$ in Proposition 7.3.43 be obtained by Jacobs' construction from the following adjunction?

$$\boldsymbol{\sigma}\mathbf{EMod}^{\mathrm{op}} \underset{\mathrm{Hom}(-,[0,1])}{\overset{\mathrm{Hom}(-,[0,1])}{\rightleftarrows}}_{\top} \mathbf{Set}$$

Somewhat surprisingly, this does not work. Note that the adjunction above induces a monad given by:

$$\begin{aligned}
\boldsymbol{\sigma}\mathbf{EMod}(\mathbf{Set}(X,[0,1]),[0,1]) &= \boldsymbol{\sigma}\mathbf{EMod}(\mathbf{Meas}((X,\mathcal{P}(X)),[0,1]),[0,1]) \\
&\cong \mathcal{G}(X,\mathcal{P}(X))\,.
\end{aligned}$$

Here we used the fact that the adjunction $\boldsymbol{\sigma}\mathbf{EMod}^{\mathrm{op}} \rightleftarrows \mathbf{Meas}$ induces the monad $\mathcal{G}$, identifying a set $X$ with a measurable space $(X, \mathcal{P}(X))$ having the full powerset $\sigma$-algebra. To obtain $\boldsymbol{\sigma}\mathbf{EMod}^{\mathrm{op}} \rightleftarrows \mathbf{SConv}$ via Jacobs' construction, we need to show $\mathcal{G}(X, \mathcal{P}(X)) \cong \mathcal{D}^{\infty}(X)$. However, the problem whether probability measures on $(X, \mathcal{P}(X))$ can be identified with discrete probability distributions on $X$ is known to be related with existence of certain large cardinals — real-valued measurable cardinals, see [88, 160] for details. This virtually means that we cannot prove (or disprove) $\mathcal{G}(X, \mathcal{P}(X)) \cong \mathcal{D}^{\infty}(X)$.

If we apply to a $\sigma$-effectus the embedding theorem of an effectus into COMs, shown in § 7.2.5, then the yielded operational convex models have richer structures. This is an immediate consequence from the results we obtained in § 7.3.2.

**Corollary 7.3.46.** *Let $(\mathbf{C}, I)$ be a real $\sigma$-effectus with the weak order-separation property. Let $R\colon \mathbf{C} \to \mathbf{COM}_{\leq}$ be the functor obtained by Theorem 7.2.61. For each object $A \in \mathbf{C}$, write $(\mathscr{V}_A, \mathscr{A}_A, \langle\,,\rangle) = RA$ for the yielded convex operational model. Then*

(i) *$\mathscr{V}_A$ is a Banach pre-base-norm space whose base is $\sigma$-convex; and*

(ii) *$\mathscr{A}_A$ is a monotone $\sigma$-complete Banach order-unit space.*

*Moreover, if $\mathbf{C}$ has the strong order-separation property, then:*

(iii) *$\mathscr{V}_A$ is a Banach base-norm space.*

*Proof.* This follows from Theorem 7.2.61 (or Corollary 7.2.62) combined with Theorem 7.3.37, Proposition 7.3.19, and Theorem 7.3.24. ■

**Example 7.3.47.** We describe the results of this subsection for the $\sigma$-effectuses $\mathcal{K}\ell(\mathcal{D}^{\infty}_{\leq})$, $\mathcal{K}\ell(\mathcal{G}_{\leq})$, and $\mathbf{Wstar}^{\mathrm{op}}_{\leq}$, continuing Examples 7.3.34–7.3.36. This also continues Examples 7.2.63–7.2.66, convex operational models obtained form effectuses.

(i) In the $\sigma$-effectus $\mathcal{K}\ell(\mathcal{D}^{\infty}_{\leq})$, for each set $X$ we have the $\sigma$-effect module $[0,1]^X$ of predicates, the $\sigma$-weight module $\mathcal{D}^{\infty}_{\leq}(X)$ of substates, and the superconvex set $\mathcal{D}^{\infty}(X)$ of states. Since $\mathcal{K}\ell(\mathcal{D}^{\infty}_{\leq})$ is strongly order-separated, we obtain a operational convex model $(\ell^1_{\mathbb{R}}(X), \ell^{\infty}_{\mathbb{R}}(X), \langle\,,\rangle)$, where $\ell^1_{\mathbb{R}}(X)$ is a Banach base-norm space and $\ell^{\infty}_{\mathbb{R}}(X)$ is a monotone $\sigma$-complete Banach order-unit space.

(ii) In the $\sigma$-effectus $\mathcal{K}\ell(\mathcal{G}_{\leq})$, for each measurable space $X$ one has the $\sigma$-effect module $\mathbf{Meas}(X, [0,1])$ of predicates, the $\sigma$-weight module $\mathcal{G}_{\leq}(X)$ of substates, and the superconvex set $\mathcal{G}(X)$ of states. As $\mathcal{K}\ell(\mathcal{G}_{\leq})$ is strongly order-separated, they yields a operational convex model $(ca(X), \mathcal{L}^{\infty}_{\mathbb{R}}(X), \langle\,,\rangle)$, where $ca(X)$ is a Banach base-norm space and $\mathcal{L}^{\infty}_{\mathbb{R}}(X)$ is a monotone $\sigma$-complete Banach order-unit space.

(iii) In the $\sigma$-effectus $\mathbf{Wstar}^{\mathrm{op}}_{\leq}$, for each $W^*$-algebra $\mathscr{A}$ we have the $\sigma$-effect module $[0,1]_{\mathscr{A}}$ of predicates, the $\sigma$-weight module $\mathrm{St}_{\leq}(\mathscr{A}) = \mathbf{Wstar}_{\leq}(\mathscr{A}, \mathbb{C})$ of substates, and the superconvex set $\mathrm{St}(\mathscr{A}) = \mathbf{Wstar}(\mathscr{A}, \mathbb{C})$ of states. Since $\mathbf{Wstar}^{\mathrm{op}}_{\leq}$ is weakly order-separated, we obtain a operational convex model $((\mathscr{A}_*)_{\mathrm{sa}}, \mathscr{A}_{\mathrm{sa}}, \langle\,,\rangle)$, where $(\mathscr{A}_*)_{\mathrm{sa}}$ is a Banach pre-base-norm space and $\mathscr{A}_{\mathrm{sa}}$ is a monotone $\sigma$-complete Banach order-unit space. We note that more is true: $(\mathscr{A}_*)_{\mathrm{sa}}$ is a Banach base-norm space and $\mathscr{A}_{\mathrm{sa}}$ is a monotone (directed) complete Banach order-unit space.

### 7.3.5 Future directions

We conclude the section with two future directions concerning $\sigma$-effectuses.

#### Iteration

We did not discuss one important aspect of partially $\sigma$-additive structure: *iteration.* In any $\sigma$-PAC, given a morphism $f = \langle\!\langle f_1, f_2\rangle\!\rangle\colon A \to A+B$ we can define the *iteration* of $f$ by:

$$\mathrm{Iter}(f) = \bigovee_{n=0}^{\infty} f_2 \circ (f_1)^n \quad : A \longrightarrow B$$

where $(f_1)^n$ denotes the $n$-fold composition of $f_1$. The sum in the right-hand side is always defined [202, Theorem 3.2.24]. The iteration operator Iter satisfies suitable categorical properties such as naturality [113, Proposition 3.1.3], and it makes a $\sigma$-PAC into a *traced monoidal category* [165] with $(+, 0)$ as the monoidal structure [113, Proposition 3.1.4].

As the name explains, the morphism $\mathrm{Iter}(f)\colon A \to B$ represents the process obtained by iterating $f\colon A \to A + B$ until we get an output in $B$. Such iteration plays an important role in program semantics, used for interpreting 'while' loop. Iteration will also play a role in effectus theory. Here is an example.

**Example 7.3.48** (Normalization via iteration)**.** Let $(\mathbf{C}, I)$ be a $\sigma$-effectus and let $\omega\colon I \to A$ be a substate. Writing $s = (\mathbb{1}\omega)^{\perp}$, we have a total map $\langle\!\langle s, \omega\rangle\!\rangle\colon I \to I + A$. The map can be interpreted as a preparation test which may fail (if it goes left, in $I$) or succeed (if it goes right, in $A$). Thus we can interpret the iteration

$$\mathrm{Iter}(\langle\!\langle s, \omega\rangle\!\rangle) = \bigovee_{n=0}^{\infty} \omega \circ s^n \quad : I \longrightarrow A$$

as a process of iterating the test $\langle\!\langle s, \omega\rangle\!\rangle$ until it succeeds. Now we assume that the scalars of $\mathbf{C}$ are commutative and admit division. Then the following holds: *whenever* $\omega$ *is nonzero,* $\mathrm{Iter}(\langle\!\langle s, \omega\rangle\!\rangle)$ *is the normalization of* $\omega$*.* Hence $\mathbf{C}$ has the normalization property.

First we show that $\mathrm{Iter}(\langle\!\langle s, \omega\rangle\!\rangle)$ is a state, i.e. a total map. Let $t := \mathbb{1} \circ \mathrm{Iter}(\langle\!\langle s, \omega\rangle\!\rangle) = \bigovee_{n=0}^{\infty} s^{\perp} \cdot s^n$. Then

$$t = \bigovee_{n=0}^{\infty} s^{\perp} \cdot s^n = s^{\perp} \ovee \Big(\bigovee_{n=0}^{\infty} s^{\perp} \cdot s^n\Big) \cdot s = s^{\perp} \ovee t \cdot s\,.$$

Since $t = t \cdot (s \ovee s^{\perp}) = t \cdot s \ovee t \cdot s^{\perp}$, we obtain $t \cdot s^{\perp} = s^{\perp}$ by cancellation. Because $s^{\perp} = \mathbb{1}\omega$ is nonzero, $t = s^{\perp}/s^{\perp} = 1$. Next we prove:

$$\mathrm{Iter}(\langle\!\langle s, \omega\rangle\!\rangle) \cdot \mathbb{1}\omega = \bigovee_{n=0}^{\infty} \omega \cdot s^n \cdot s^{\perp} = \omega \cdot \bigovee_{n=0}^{\infty} s^{\perp} \cdot s^n = \omega \cdot 1 = \omega\,.$$

We used the commutativity of scalars and $\bigovee_{n=0}^{\infty} s^{\perp} \cdot s^n = 1$ shown above. Finally, to see the uniqueness, let $\rho$ be a state with $\omega = \rho \cdot \mathbb{1}\omega$. Then

$$\rho = \rho \cdot 1 = \rho \cdot \Big(\bigovee_{n=0}^{\infty} s^{\perp} \cdot s^n\Big) = \bigovee_{n=0}^{\infty} \rho \cdot s^{\perp} \cdot s^n = \bigovee_{n=0}^{\infty} \omega \cdot s^n = \mathrm{Iter}(\langle\!\langle s, \omega\rangle\!\rangle)\ .$$

Therefore $\mathrm{Iter}(\langle\!\langle s, \omega \rangle\!\rangle)$ defines the normalization of $\omega$.

We leave further investigation of iteration in effectus theory for future work.

### Relation to synthetic measure theory

Measure and probability theory is based on 'countable structures', such as $\sigma$-algebras and (countably additive) measures. Indeed, a $\sigma$-algebra $\Sigma_X$ on a set $X$ is precisely a $\sigma$-effect subalgebra of $\mathcal{P}(X)$. Moreover probability measures $\mu\colon \Sigma_X \to [0,1]$ are precisely unital morphisms of $\sigma$-effect algebras. Jacobs and Abraham Westerbaan have studied integration in terms of $\sigma$-effect algebras/modules [151]. We thus expect that there is a further connection between $\sigma$-effectuses and measure/probability theory.

As a first step in this direction, we will relate $\sigma$-effectuses to *synthetic measure theory* developed by Kock [175] and further by Ścibior et al. [233]. Specifically, we show that *measure categories* [233] induce $\sigma$-effectuses under suitable additional conditions. We then prove that the measure category of *quasi-Borel spaces* [124] indeed yields a $\sigma$-effectus.

The following definition provides a basic setting for synthetic measure theory.

**Definition 7.3.49** ([233, Definition 4.2]). A **measure category** is a pair $(\mathbf{C}, M)$ consisting of a cartesian closed category $\mathbf{C}$ with countable products and coproducts and a commutative monad $M\colon \mathbf{C} \to \mathbf{C}$ such that

(a) the canonical map $M0 \to 1$ is an isomorphism; and

(b) for each countable family $(A_j)_j$ of objects in $\mathbf{C}$, the canonical map $M \coprod_j A_j \to \prod_j MA_j$ is an isomorphism.

The intuition is that an object $A \in \mathbf{C}$ is an abstract measurable space, and $MA$ consists of measures/distributions on $A \in \mathbf{C}$. A concrete example will be given below by quasi-Borel spaces.

The conditions (a) and (b) says that $M$ turns coproducts into products. It is well known (e.g. [59, 135]) that such a condition on a monad makes the Kleisli category into a category with biproducts. Moreover, the final object 1 in $\mathbf{C}$ yields a suitable 'ground' structure.

**Lemma 7.3.50.** *Let $(\mathbf{C}, M)$ a measure category. Then the Kleisli category $\mathcal{K}\ell(M)$ has countable biproducts, which are given by coproducts in $\mathbf{C}$. Moreover, there is a family of 'ground' maps $\bar{\mp}_A\colon A \to\!\!\!\!\!\to 1$ in $\mathcal{K}\ell(M)$ given by $A \xrightarrow{!} 1 \xrightarrow{\eta} M1$ in $\mathbf{C}$, and the family satisfies* (G1) *and* (G2) *of Definition* 7.1.21.

*Proof.* Straightforward. ∎

In general $\mathcal{K}\ell(M)$ is not a grounded biproduct category, since the conditions (G3) and (G4) of Definition 7.1.21 may fail.

Each category with countable biproducts is enriched over *$\sigma$-additive monoids* (i.e. $\sigma$-PAMs with total addition). To obtain a $\sigma$-effectus from a grounded biproduct category with countable coproducts, we need an additional condition.

**Lemma 7.3.51.** *Let* $(\mathbf{E}, I)$ *be a grounded biproduct category such that* $\mathbf{E}$ *has countable biproducts. Then* $\mathrm{Caus}_{\leq}(\mathbf{E})$ *is a* $\sigma$*-effectus if (and only if)* $\mathbf{E}$ *satisfies the following condition:*

(∗) *For each* $A \in \mathbf{E}$ *and for each sequence* $(p_n \colon A \to I)_{n\in\mathbb{N}}$*, if* $\sum_{n=0}^{N} p_n \leq \bar{\top}_A$ *for all* $N \in \mathbb{N}$*, then* $\sum_{n\in\mathbb{N}} p_n \leq \bar{\top}_A$*.*

*Proof.* We already know that $\mathrm{Caus}_{\leq}(\mathbf{E})$ is an effectus. To prove that it is a $\sigma$-effectus, it suffices to show that it satisfies the conditions (i)–(iii) of Lemma 7.3.38.

Let $\bigoplus_j A_j$ be a countable biproduct in $\mathbf{E}$. By Definition 7.1.21(G2), every coprojection $\kappa_j \colon A_j \to \bigoplus_j A_j \cong A_j \oplus \bigoplus_{k\neq j} A_k$ is causal, which means that $[\bar{\top}]_j = \bar{\top} \colon \bigoplus_j A_j \to I$. If $(f_j \colon A_j \to B)_j$ is a family of subcausal maps, then the cotuple $[f_j]_j \colon \bigoplus_j A_j \to B$ is subcausal because

$$\bar{\top} \circ [f_j]_j = [\bar{\top} \circ f_j]_j \leq [\bar{\top}]_j = \bar{\top}\,.$$

Here the inequality (w.r.t. the algebraic ordering) holds since cotupling

$$[-]_j \colon \prod_j \mathbf{E}(A_j, B) \longrightarrow \mathbf{E}\Big(\bigoplus_j A_j, B\Big)$$

is a monoid (iso)morphism. Therefore $\bigoplus_j A_j$ forms a coproduct in $\mathrm{Caus}_{\leq}(\mathbf{E})$, showing that $\mathrm{Caus}_{\leq}(\mathbf{E})$ satisfies (i).

Since partial projections $\rhd_j$ in $\mathrm{Caus}_{\leq}(\mathbf{E})$ are ordinary projections $\pi_j$ in $\mathbf{E}$, they are jointly monic, i.e. (ii) holds.

Finally we prove (iii) of Lemma 7.3.38. Let $(f_j \colon A \to B)_{j\in J}$ be a countable family of subcausal morphisms. Note that $(f_j)_{j\in J}$ is compatible in $\mathrm{Caus}_{\leq}(\mathbf{E})$ if and only if the sum $\sum_j f_j$ (in $\mathbf{E}(A, B)$) is subcausal. By definition, $\sum_j f_j$ is subcausal if and only if $\bar{\top} \circ \sum_j f_j \leq \bar{\top}$, i.e. $\sum_j \bar{\top} \circ f_j \leq \bar{\top}$. Now assume that $(f_j)_{j\in F}$ is compatible for each finite $F \subseteq J$. Then $\sum_{j\in F} \bar{\top} \circ f_j \leq \bar{\top}$ for each finite $F \subseteq J$. By the assumption (∗) it follows that $\sum_{j\in J} \bar{\top} \circ f_j \leq \bar{\top}$. Therefore $(f_j)_{j\in J}$ is compatible, concluding the proof. ■

By Lemmas 7.3.50 and 7.3.51, we obtain:

**Proposition 7.3.52.** *Let* $(\mathbf{C}, M)$ *be a measure category such that the Kleisli category* $\mathcal{K}\ell(M)$ *satisfies* (G3) *and* (G4) *of Definition* 7.1.21 *and* (∗) *of Lemma* 7.3.51*. Then* $\mathcal{K}\ell(M)$ *is a grounded biproduct category and* $\mathrm{Caus}_{\leq}(\mathcal{K}\ell(\mathcal{Q}))$ *is a* $\sigma$*-effectus.* ■

Now we recall quasi-Borel spaces. They form a measure category, providing a concrete model of synthetic measure theory. Using the lemmas above, we will show that quasi-Borel spaces yield a $\sigma$-effectus.

**Definition 7.3.53** ([124, 233])**.** A **quasi-Borel space** is a set $X$ with a subset $M_X \subseteq X^{\mathbb{R}}$ of functions $\alpha \colon \mathbb{R} \to X$ such that

(a) For each $\alpha \in M_X$ and each (Borel) measurable function $f \colon \mathbb{R} \to \mathbb{R}$, one has $\alpha \circ f \in M_X$.

(b) $M_X$ contains all constant functions $\alpha \colon \mathbb{R} \to X$.

(c) For each family $(\alpha_n)_{n\in\mathbb{N}}$ of members in $M_X$ and each measurable partition $\mathbb{R} = \biguplus_{n\in\mathbb{N}} U_n$, the function $\beta\colon \mathbb{R} \to X$ defined by $\beta(x) = \alpha_n(x)$ for $x \in U_n$ belongs to $M_X$.

A morphism of quasi-Borel spaces $(X, M_X) \to (Y, M_Y)$ is a function $f\colon X \to Y$ such that $\alpha \in M_X$ implies $f \circ \alpha \in M_Y$.

If $(X, \Sigma_X)$ is a standard Borel space, then $X$ can be seen canonically as a quasi-Borel space with $M_X = \mathbf{Meas}(\mathbb{R}, X)$. This defines a full and faithful functor from the category of standard Borel spaces (and measurable functions) to **QBS**.

**Definition 7.3.54** ([233, Definition 4.7])**.** A **measure** on a quasi-Borel space $(X, M_X)$ is a triple $(\Omega, \alpha, \mu)$ where $\Omega$ is a standard Borel space, $\alpha \in \mathbf{QBS}(\Omega, X)$, and $\mu$ is a $\sigma$-finite measure on $\Omega$.

Below we denote by $\overline{\mathbb{R}}_+ = [0, \infty]$ the set of non-negative extended real numbers. Since $\overline{\mathbb{R}}_+$ is a standard Borel space, it can be seen as a quasi-Borel space.

For each measure $(\Omega, \alpha, \mu)$ on $(X, M_X)$ and a morphism $\varphi\colon X \to \overline{\mathbb{R}}_+$ in **QBS**, we can define the *integral*

$$\int \varphi \,\mathrm{d}(\Omega, \alpha, \mu) \coloneqq \int_\Omega (\varphi \circ \alpha) \,\mathrm{d}\mu \quad \in \overline{\mathbb{R}}_+$$

where the right-hand side is the usual (Lebesgue) integral. For two measures $(\Omega, \alpha, \mu)$ and $(\Omega', \alpha', \mu')$ we write $(\Omega, \alpha, \mu) \approx (\Omega', \alpha', \mu')$ and say that they are equivalent if $\int \varphi \,\mathrm{d}(\Omega, \alpha, \mu) = \int \varphi \,\mathrm{d}(\Omega', \alpha', \mu')$ for all $\varphi \in \mathbf{QBS}(X, \overline{\mathbb{R}}_+)$. For $X \in \mathbf{QBS}$ we write

$$\mathcal{Q}(X) = \{\text{measures } (\Omega, \alpha, \mu) \text{ on } (X, M_X)\}/{\approx}$$

for the equivalences classes of measures. Then one can equip $\mathcal{Q}(X)$ with a quasi-Borel structure and moreover the structure of monad. The following theorem summarizes the results on quasi-Borel spaces in [124, 233]. For details we refer to the cited articles.

**Theorem 7.3.55** ([124, 233])**.** *The monad $\mathcal{Q}$ is well-defined. Moreover,* $(\mathbf{QBS}, \mathcal{Q})$ *forms a measure category.* ∎

For the proof of the theorem below, we note that for $f\colon X \to Y$ in **QBS**, the map $\mathcal{Q}(f)\colon \mathcal{Q}(X) \to \mathcal{Q}(Y)$ is defined by $\mathcal{Q}(f)(\Omega, \alpha, \mu) = (\Omega, f \circ \alpha, \mu)$.

Finally we show that the measure category $(\mathbf{QBS}, \mathcal{Q})$ satisfies suitable additional conditions and hence yields a $\sigma$-effectus.

**Theorem 7.3.56.** *The Kleisli category $\mathcal{K}\ell(\mathcal{Q})$ of the monad $\mathcal{Q}$:* $\mathbf{QBS} \to \mathbf{QBS}$ *is a grounded biproduct category, and moreover* $\mathrm{Caus}_\leq(\mathcal{K}\ell(\mathcal{Q}))$ *is a $\sigma$-effectus.*

*Proof.* By Proposition 7.3.52 it suffices to show that $\mathcal{K}\ell(\mathcal{Q})$ satisfies (G3) and (G4) of Definition 7.1.21 and $(*)$ of Lemma 7.3.51.

As noted in [233, § 4.3], one has an isomorphism $\mathcal{Q}(1) \cong \overline{\mathbb{R}}_+$ in **QBS**, which identifies a measure $(\Omega, \alpha, \mu) \in \mathcal{Q}(1)$ with its total mass $\mu(\Omega) \in \overline{\mathbb{R}}_+$. Under this identification, the ground maps $\top_X\colon X \to \mathcal{Q}(1) \cong \overline{\mathbb{R}}_+$ are the constant functions with value 1. Moreover, one can verify that the additive structure on $\mathcal{K}\ell(\mathcal{Q})(X, 1) =$ $\mathbf{QBS}(X, \mathcal{Q}(1)) \cong \mathbf{QBS}(X, \overline{\mathbb{R}}_+)$ coincides with the usual pointwise addition of functions $X \to \overline{\mathbb{R}}_+$. From this fact, (G4) and $(*)$ follow easily.

Finally let us verify (G3). Let $f\colon X \to \mathcal{Q}(Y)$ be a morphism in **QBS** such that $\bar{\top} \odot f = 0$. Here $0\colon X \to \mathcal{Q}(Y)$ is a constant function whose value is (the equivalence class of) the zero measure. Fix $x \in X$ and suppose that $f(x) = (\Omega, \alpha, \mu)$. Since $\bar{\top} = \eta \circ {!}$, we have $\mathcal{Q}(!) \circ f = \bar{\top} \odot f = 0$ and hence

$$0 = \mathcal{Q}(!)(f(x)) = (\Omega, {!} \circ \alpha, \mu) = (\Omega, {!}, \mu) \in \mathcal{Q}(1)\,.$$

Via the identification $\mathcal{Q}(1) \cong \overline{\mathbb{R}}_+$, it follows that $\mu(\Omega) = 0$, i.e. $\mu$ is the zero measure. Therefore $f(x) = (\Omega, \alpha, \mu) \approx 0$ for each $x \in X$, so that $f = 0$. ∎

# Index of Categories and Monads

## Categories

## Monads

# Index of Notation

# Index of Subjects

# Summary

In this thesis we develop a categorical axiomatic approach to quantum theory. In general, we model a theory of physics by a category, where the objects represent types of systems, and the morphisms represent processes between systems. The category is assumed to have certain structures and properties to axiomatize quantum systems and processes.

The approach of this thesis is based on *effectuses*, a class of categories introduced by Jacobs from a categorical logic perspective. The predicates in an effectus form *effect algebras*, which are posets with top ('truth'), bottom ('falsity'), orthosupplement $p^{\perp}$ ('negation of $p$'), and orthosum $p \ovee q$ ('$p$ or $q$'). Effect algebras generalize Boolean algebras and model unsharp quantum logic, axiomatizing *quantum effects* (unsharp measurements).

We present a basic theory of effectuses, together with several leading examples of effectuses. The archetypal example that models quantum theory is the category of $W^*$-algebras with suitable morphisms. Effectuses for classical theories include the category of sets and the Kleisli category of the distribution monad. We study the structure of predicates, states, and substates in an effectus, and we present the duality between predicates and (sub)states as 'state-and-effect' triangles of functors.

There are two formulations of effectus: *partial form* and *total form*. They respectively axiomatize partial and total processes. Effectuses in partial form are defined via partially additive structure. Effectuses in total form involve certain pullback diagrams and can be seen as a generalization of extensive categories. We prove that the two definitions are equivalent in a 2-categorical sense.

We study effectuses further from a logical point of view, systematically using the language of (Grothendieck) fibrations. The fibrational perspective motivates several universal constructions in an effectus, such as *image*, *comprehension*, and *quotient*. Via images and comprehension, we define *sharp* predicates, which corresponds to projection operators in the orthodox Hilbert space formalism of quantum theory. We prove under a mild assumption that sharp predicates form orthomodular lattices.

We then discuss measurements in an effectus. We use the framework of *operational probabilistic theories* which has been developed by Chiribella, D'Ariano and Perinotti, and related to effectus theory by Tull. We study properties of measurements such as repeatability and side-effect-freeness. Repeatable measurements are related to sharp predicates. If measurements are both repeatable and side-effect-free, they are called *Boolean*. Studying these measurements, we obtain a characterization of extensive categories (with final objects) as *Boolean effectuses*, where Boolean measurements are possible for all observables.

We end the thesis with miscellaneous results about effectuses. We establish the relationship between effectuses and biproduct categories, using the technique of *totalization*. Moreover, we show that an effectus satisfying certain separation conditions can be embedded into the category of *convex operational models*, which are axiomatic models of quantum systems based on ordered vector spaces. Finally, we study a natural extension of effectus with countably partially additive structure, following the work of Arbib and Manes.

# Samenvatting (Dutch Summary)

Dit proefschrift zet een categorische axiomatische aanpak uiteen voor de kwantumtheorie. In het algemeen modelleren we een natuurkundige theorie met een categorie, waar de objecten de rol hebben van typen van systemen en de morfismen de rol van processen tussen systemen. De categorie wordt verondersteld bepaalde structuur en eigenschappen te hebben om kwantumsystemen en -processen te axiomatiseren.

De aanpak van dit proefschrift is gebaseerd op *effectussen*, een klasse van categorieën ingevoerd door Jacobs vanuit de zienswijze van de categorische logica. De predicaten in een effectus vormen *effectalgebra's*; dat zijn geordende verzamelingen met een maximum ('waarheid'), minimum ('onwaarheid'), orthosupplement $p^{\perp}$ ('ontkenning van $p$') en orthosom $p \ovee q$ ('$p$ of $q$'). Effectalgebra's generaliseren booleaanse algebra's en modelleren onscherpe kwantumlogica, door het axiomatiseren van *kwantumeffecten* (onscherpe metingen).

Het proefschrift begint met de basis van de effectusleer. We geven verscheidene voorbeelden van effectussen. Het archetypische voorbeeld voor de kwantumtheorie is de categorie van $W^*$-algebra's met bijpassende morfismen. Tussen de effectussen voor klassieke theorieën vindt men onder anderen de categorie van verzamelingen en de Kleislicategorie van de distributiemonad. We bestuderen de structuur van de predicaten, toestanden en (deel)toestanden in een effectus en we zetten een dualiteit uiteen tussen predicaten en (deel)toestanden in de vorm van 'toestand-en-effectdriehoeken' van functoren.

Een effectus heeft twee verschijningvormen: de *partiële* en de *totale vorm*. Ze axiomatiseren, respectievelijk, de partiële en totale processen. Effectussen in partiële vorm worden gedefinieerd aan de hand van een partieel-additieve structuur. Effectussen in totale vorm gaan gepaard met zekere vezelproducten en kunnen gezien worden als generalisatie van extensieve categorieën. We laten zien dat de twee definities overeenkomen in 2-categorische zin.

We bestuderen effectussen vervolgens vanuit een logisch gezichtspunt, waarbij we systematisch de taal van (Grothendieck)vezelingen gebruiken. Dit geeft aanleiding tot enkele universele constructies binnen een effectus, zoals *beeld*, *afscheiding* en *quotiënt*. Middels beeld en afscheiding definiëren we *scherpe* predicaten, die overeenkomen met projectieoperatoren in de het orthodoxe Hilbertruimteformalisme van de kwantumtheorie. We bewijzen dat onder milde omstandigheden scherpe predicaten een orthomodulaire tralies vormen.

Daarna gaan we in op metingen binnen een effectus. We gebruiken het raamwerk van de *operationele probabilistische theorieën*, dat door Chiribella, D'Ariano en Perinotti ontwikkeld is en door Tull verbonden aan de effectusleer. We onderzoeken eigenschappen van metingen zoals herhaalbaarheid en het bijwerkingsloos-zijn. Herhaalbare metingen houden verband met scherpe predicaten. Als metingen zowel herhaalbaar als bijwerkingsloos zijn worden ze *booleaans* genoemd. De studie van deze metingen levert de typering van extensieve categorieën (met finaal object) als *booleaanse effectussen*, waarin alleen booleaanse metingen mogelijk zijn.

We eindigen het proefschrift met allerhande resultaten over effectussen. We leggen een verband tussen effectussen en biproductcategorieën met de techniek van *totalisatie*. Verder laten we zien dat een effectus met zekere scheidingseigenschappen ingebed kan worden in een categorie van *convexe operationele modellen*, wat axiomatische modellen van quantumsystemen zijn gebaseerd op geordende vectorruimten. Tot slot bestuderen we een natuurlijke uitbreiding van de effectus met een partiële aftelbaar-additieve structuur, naar het voorbeeld van het werk van Arbib en Manes.

# Curriculum Vitae

Kenta Cho enrolled in the University of Tokyo, Japan in 2008 and received a bachelor's degree in Information Science in 2012. Continuing to study at the same university, in 2014, he received a master's degree in Computer Science with a thesis on semantics of a quantum programming language supervised by Ichiro Hasuo. Soon after that, he became a PhD student at Radboud University, Nijmegen, the Netherlands under the supervision of Bart Jacobs. There he was engaged in the ERC project *Quantum Computation, Logic, and Security* of Jacobs. In 2018, he started working at National Institute of Informatics, Japan on the ERATO MMSD project led by Ichiro Hasuo.